%% file: Main.tex
\documentclass[a4paper,10pt,twoside]{report}

\input{Inputs/PackagesnStyles.tex}

\graphicspath{{Figures/}}

\input{Inputs/Definitions.tex}

\allowdisplaybreaks

\begin{document}

\newcounter{tblEqCounter}     

\pagestyle{plain}
\hypersetup{pageanchor=false}

\input{Inputs/Cover.tex}

\hypersetup{pageanchor=true}
\pagenumbering{roman}
\setcounter{page}{1}

\input{Inputs/Citation.tex}
\input{Inputs/Acknowledgements.tex}


\input{Inputs/Abstract.tex}
\input{Inputs/Resumo.tex}
\input{Inputs/Preface.tex}

\input{Inputs/TOCnLists.tex}

\pagenumbering{arabic}
\setcounter{page}{1}


\input{Chapters/Intro}

\input{Chapters/Chapter_FM}

\input{Chapters/Chapter_Selection}

\input{Chapters/Chapter_Real}

\input{Chapters/Chapter_Maggie}

\input{Chapters/Chapter_Reno}

\input{Chapters/Chapter_Lavoura}

\input{Chapters/Chapter_Valle}
\input{Chapters/Conclusion}

\appendix 
\input{Appendices/FM-Manual}
\input{Appendices/Pilaftsis}

\input{Appendices/OSS}

\input{Appendices/LSZ}

\input{Appendices/Fermions}

\input{Appendices/WI}

\input{Appendices/Sym}
\input{Appendices/FM-C2HDM}

\input{Appendices/Theorem}

%

\bibliographystyle{h-physrev4}
\bibliography{Inputs/MyReferences}

\end{document}

%% file: Inputs/PackagesnStyles.tex
\usepackage{geometry}	
\usepackage{setspace}

\everydisplay{}

\usepackage{tocloft}
\usepackage{indentfirst}

\makeatletter
\def\hlinewd#1{%
	\noalign{\ifnum0=`}\fi\hrule \@height #1 %
	\futurelet\reserved@a\@xhline}
\makeatother

\usepackage{verbatim}
 
\usepackage{hyperref}
\hypersetup{colorlinks=false}
\usepackage{array}

\usepackage[latin9,utf8]{inputenc} 
\usepackage[T1]{fontenc}
\usepackage{ae}

\usepackage{cancel}
\usepackage{float,hypcap}
\usepackage{changepage,titlesec}
\usepackage{sectsty}
\usepackage{soul}
\usepackage{bbold}
\usepackage{slashed}
\usepackage{tabularx}
\usepackage{amsmath,amsfonts,amssymb}
\usepackage{amsmath,bbm,latexsym,amssymb}
\usepackage{braket}
\usepackage{blkarray}
\usepackage{enumerate}
\usepackage{booktabs,hyperref}
\usepackage{bigints}
\usepackage{cite}
\usepackage{mflogo}
\usepackage{scrextend}
\usepackage{longtable}
\usepackage{enumitem}
\usepackage{pifont}
\usepackage{cases}
\usepackage{colortbl}

\usepackage{multirow}

\usepackage[flushleft]{threeparttable}

\usepackage{verbatim}

\usepackage{fancyvrb}
\usepackage{xcolor}

\usepackage{graphicx}
\usepackage{caption}
\usepackage{subfig}
\usepackage{feynmp-auto}

\usepackage{color}
\definecolor{orange}{rgb}{1,0.5,0}
\definecolor{uglyblue}{RGB}{95,158,160}
\definecolor{newblue}{RGB}{128,0,0}
\definecolor{mygray}{RGB}{129,129,129}
\definecolor{mywhite}{RGB}{255,250,240}
\usepackage[most]{tcolorbox}

\newcommand*{\mywbox}{\tcboxmath[colback=mywhite]}
\newcommand{\Lim}[1]{\raisebox{0.5ex}{\scalebox{0.8}{$\displaystyle \lim_{#1}\;$}}}


%% file: Inputs/Definitions.tex
\def\be{\begin{equation}}
\def\ee{\end{equation}}
\def\ba{\begin{alignedat}}
\def\ea{\end{alignedat}}
\def\bea{\begin{eqnarray}}
\def\eea{\end{eqnarray}}
\newcommand{\bs}{\begin{subequations}}
\newcommand{\es}{\end{subequations}}
\def\ra{\rightarrow}
\def\ppo{\langle}
\def\ppc{\rangle}
\def\vs{\vspace}
\def\hs{\hspace}

\def\n{\noindent}
\def\no{\nonumber\\}
\def\fn{\footnote}

\def\t{\texttt}
\def\ts{\textsc}
\def\vb#1{\vbox to #1 pt{}}
\def\FM{\textsc{FeynMaster}}
\def\FMS{\textsc{FeynMaster} }
\def\FMT{\textsc{FeynMaster} 2}
\def\FMTS{\textsc{FeynMaster} 2 }
\def\FR{\textsc{FeynRules}}
\def\FRS{\textsc{FeynRules} }
\def\FC{\textsc{FeynCalc}}
\def\FCS{\textsc{FeynCalc} }
\def\QG{\textsc{Qgraf}}
\def\QGS{\textsc{Qgraf} }

\newcommand{\newc}{\newcommand}
\newc{\ol}{\overline}
\newc{\wt}{\widetilde}

\newc{\m}{\mathcal}

\newcommand{\beq}{\begin{eqnarray}}
\newcommand{\eeq}{\end{eqnarray}}
\newcommand{\bpmatrix}{\begin{pmatrix}}
\newcommand{\epmatrix}{\end{pmatrix}}

\newcommand{\tb}{\tan\beta}

\renewcommand{\ol}{\text{1l}}

\renewcommand{\eqref}[1]{eq.~(\ref{#1})}

\newcommand{\bc}{\begin{center}}
\newcommand{\ec}{\end{center}}

\def\ra{\rightarrow}
\def\vb#1{\vbox to #1 pt{}}

\def\m#1{m_{#1}}

\def\ltap{\;\centeron{\raise.35ex\hbox{$<$}}{\lower.65ex\hbox{$\sim$}}\;}
\def\gtap{\;\centeron{\raise.35ex\hbox{$>$}}{\lower.65ex\hbox{$\sim$}}\;}
\def\vb#1{\vbox to #1 pt{}}

\usepackage{stackengine}
\stackMath

\newcommand\OSeq{\stackrel{\mathrm{OS}}{=}}
\newcommand\OSSeq{\stackrel{\mathrm{OSS}}{=}}
\newcommand\FJeq{\stackrel{\mathrm{FJ}}{=}}
\newcommand\PReq{\stackrel{\mathrm{PR}}{=}}
\newcommand\CPeq{\stackrel{\mathrm{CP}}{=}}

\newcommand\xxrightarrow[2][]{\mathrel{%
\setbox2=\hbox{\stackon{\scriptstyle#1}{\scriptstyle#2}}%
\stackunder[1pt]{%
\xrightarrow{\makebox[\dimexpr\wd2\relax]{$\scriptstyle#2$}}%
}{%
\scriptstyle#1\,%
}%
}}
\newcommand\xxleftarrow[2][]{\mathrel{%
\setbox2=\hbox{\stackon{\scriptstyle#1}{\scriptstyle#2}}%
\stackunder[1pt]{%
\xleftarrow{\makebox[\dimexpr\wd2\relax]{$\scriptstyle#2$}}%
}{%
\scriptstyle#1\,%
}%
}}

%% file: Inputs/Cover.tex
\thispagestyle {empty}

\def\FontLL{
  \usefont{T1}{cmr}{m}{n}\fontsize{18pt}{18pt}\selectfont}
\def\FontL{
  \usefont{T1}{cmr}{m}{n}\fontsize{17.28pt}{16pt}\selectfont}
\def\FontM{
  \usefont{T1}{cmr}{m}{n}\fontsize{14pt}{14pt}\selectfont}
\def\FontS{
  \usefont{T1}{cmr}{m}{n}\fontsize{12pt}{12pt}\selectfont}
\def\FontT{
  \fontsize{10pt}{10pt}\selectfont}

\noindent \includegraphics[width=5cm]{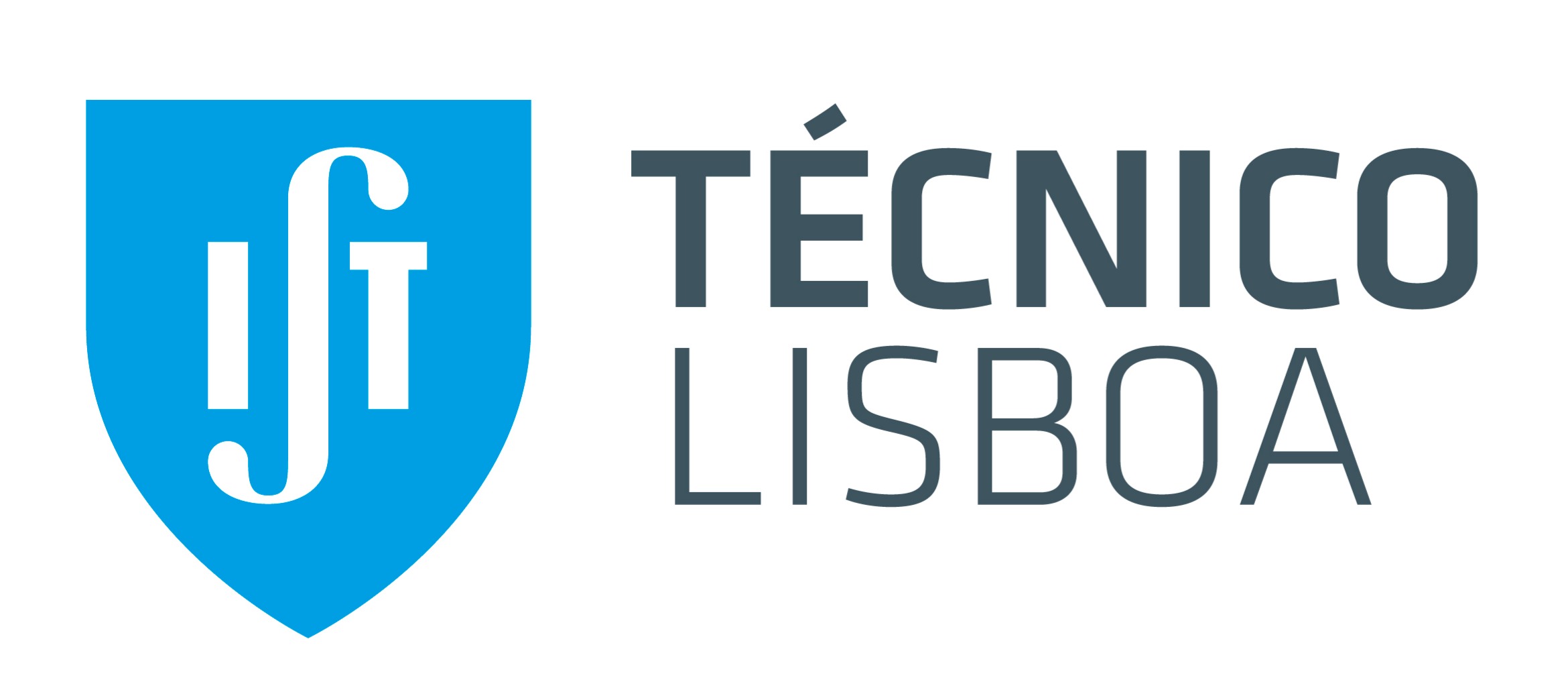}

\begin{center}


{\FontL \textbf{UNIVERSIDADE DE LISBOA}} \\
{\FontL \textbf{INSTITUTO SUPERIOR TÉCNICO}} \\

\vspace{4cm}

{\FontLL \textbf{Multi-Higgs Models:}} \\
{\FontM \textbf{model building, phenomenology and renormalization}} \\

\vspace{1cm}

{\FontM \textbf{Duarte Sarmento de Sousa Machado Fontes}} \\
\end{center}

\vspace{8mm}

{\FontS \textbf{Supervisor}: Doctor Jo\~{a}o Paulo Ferreira da Silva} \\
{\FontS \textbf{Co-Supervisor}: Doctor Jorge Manuel Rodrigues Crispim Rom\~{a}o}

\begin{center}

\vspace{8mm}

{\FontS \textbf{Thesis approved in public session to obtain the PhD Degree in}} \\[2mm]
{\FontL \textbf{Physics}}

\vspace{7mm}

{\FontS \textbf{Jury final classification:}}
{\Large \textbf{Pass with Distinction and Honour}}

\vfill

{\FontM \textbf{2021}}

\end{center}

\cleardoublepage


\thispagestyle {empty}

\noindent \includegraphics[width=5cm]{ist_logo.jpg}

\begin{center}


{\FontL \textbf{UNIVERSIDADE DE LISBOA}} \\
{\FontL \textbf{INSTITUTO SUPERIOR TÉCNICO}} \\

\vspace{1cm}

{\FontLL \textbf{Multi-Higgs Models:}} \\
{\FontM \textbf{model building, phenomenology and renormalization}} \\

\vspace{1cm}

{\FontM \textbf{Duarte Sarmento de Sousa Machado Fontes}} \\
\end{center}

\vspace{5mm}

{\FontS \textbf{Supervisor}: Doctor Jo\~{a}o Paulo Ferreira da Silva} \\
{\FontS \textbf{Co-Supervisor}: Doctor Jorge Manuel Rodrigues Crispim Rom\~{a}o}

\begin{center}

\vspace{4mm}

{\FontS \textbf{Thesis approved in public session to obtain the PhD Degree in}} \\[1mm]
{\FontL \textbf{Physics}}

\vspace{4mm}

{\FontS \textbf{Jury final classification:}}
{\Large \textbf{Pass with Distinction and Honour}}

\vspace{7mm}

{\FontS \textbf{Jury}}
\end{center}

\vspace{-2mm}

{\FontS \textbf{Chairperson}: \\
\hphantom{\hspace{6.5mm}} Doctor M\'{a}rio Jo\~{a}o Martins Pimenta, Instituto Superior Técnico, Universidade de \\[-1mm]
\hphantom{\hspace{20.5mm}} Lisboa} 

\vspace{2mm}

{\FontS \textbf{Members of the Commitee}:

\begin{addmargin}[8mm]{0mm}
Doctor Ansgar Denner, Fakultät für Physik und Astronomie, 
Universität Würzburg, \\[-1mm]
\hphantom{\hspace{12.8mm}} Germany \\
Doctor Jorge Manuel Rodrigues Crispim Rom\~{a}o, Instituto Superior Técnico, Univer- \\[-1mm]
\hphantom{\hspace{12.8mm}} sidade de Lisboa  \\
Doctor José Guilherme Teixeira de Almeida Milhano, Instituto Superior Técnico, \\[-1mm]
\hphantom{\hspace{12.8mm}} Universidade de Lisboa  \\
Doctor Rui Alberto Serra Ribeiro dos Santos, Instituto Superior de Engenharia \\[-1mm]
\hphantom{\hspace{12.8mm}} de Lisboa, Instituto Politécnico de Lisboa
\end{addmargin}
}

\vspace{5mm}

\begin{center}
{\FontS \textbf{Funding Institution:}
Fundação para a Ciência e Tecnologia (FCT)
}

\vfill

{\FontM \textbf{2021}}

\end{center}

\pagebreak

\thispagestyle {empty}
\cleardoublepage

%% file: Inputs/Citation.tex
{\FontT
\null\vskip 3cm

\begin{flushright}



\begin{minipage}{0.55\textwidth}
\begin{flushright}
Il est surprenant que ce qui semble aujourd'hui \\
si facile à discerner, restât aussi embrouillé et \\
aussi voilé aux yeux les plus clairvoyants
\end{flushright}
\end{minipage}\\[3mm]

\vs{-2mm}
\textcolor{mygray}{\rule{7cm}{0.1mm}}
\vs{-1mm}

\begin{footnotesize}
\textit{L'ancient régime et la révolution}, \\[-1mm]
A. de Tocqueville, Flammarion, Paris (1988), p. 116
\end{footnotesize}
\\[20mm]

\begin{minipage}{0.55\textwidth}
\begin{flushright}
Niemand findet leicht als erster etwas Auffälliges: \\
denn es ist den Menschen im allgemeinen nicht \\
gegeben, zu sehen, was ist
\end{flushright}
\end{minipage}\\[3mm]

\vs{-2mm}
\textcolor{mygray}{\rule{7cm}{0.1mm}}
\vs{-1mm}

\begin{footnotesize}
\textit{Erzählungen}, H. von Hofmannsthal, Philipp Reclam \\[-1mm]
jun. GmbH \& Co., Stuttgart (2000), p. 218
\end{footnotesize}
\\[20mm]

%

\begin{minipage}{0.60\textwidth}
\begin{flushright}
Put not yourself into amazement how these things should be: all difficulties are but easy when they are known
\end{flushright}
\end{minipage}\\[3mm]

\vs{-2mm}
\textcolor{mygray}{\rule{7cm}{0.1mm}}
\vs{-1mm}

\begin{footnotesize}
\textit{Measure for Measure}, Shakespeare, 4.2.197-199
\end{footnotesize}
\\[20mm]

\begin{minipage}{0.55\textwidth}
\begin{flushright}
``Speak, then, let us hear the wisdom of Hercule Poirot.''
``Did I not tell you that I was, like you, a very puzzled man? But at least we can face our problem. We can arrange such facts as we have with order and method.''
\end{flushright}
\end{minipage}\\[3mm]

\vs{-2mm}
\textcolor{mygray}{\rule{7cm}{0.1mm}}
\vs{-1mm}

\begin{footnotesize}
\textit{Murder on the Orient Express}, Agatha Christie, \\[-1mm]
HarperPaperbacks, New York (1991), p. 86
\end{footnotesize}
\\[20mm]

\begin{minipage}{0.55\textwidth}
\begin{flushright}
perseverantia opus est
\end{flushright}
\end{minipage}\\[-1mm]

\vs{-2mm}
\textcolor{mygray}{\rule{7cm}{0.1mm}}
\vs{-1mm}

\begin{footnotesize}
\textit{De Sermone Domini in Monte}, Augustinus, \\[-1mm]
Liber Secundus, XXI, 73
\end{footnotesize}
\\[20mm]

%
%
%

\end{flushright}
}

\cleardoublepage

%% file: Inputs/Acknowledgements.tex
\clearpage

\chapter*{Acknowledgements}
\addcontentsline{toc}{chapter}{Acknowledgements}

During these four years of PhD, I had two official supervisors. They were responsible for guiding and supporting me in my investigations. They surely did; but they were not alone. Many others have contributed to the path I ended up taking. I am very grateful to them all.

To my supervisors, Jorge Romão and João P. Silva, who were tireless in their patience, support and motivation, I am most grateful. 
I am also very much indebted to Ansgar Denner and Augusto Barroso, who accepted my endless queries and took plenty of their time to answer them.
A big word of thanks also to Rui Santos, who has always shown very supportive.
I also thank those with whom I directly collaborated in these four years: besides those already mentioned, also Adam Falkowski, Hermès Bélusca-Maito, 
Margarete M\"{u}hlleitner, Jonas Wittbrodt,
Luís Lavoura,
José Valle
and Maximilian Löschner.

I appreciate the support I received both from FCT, as well as from CFTP, especially from Sandra Oliveira and Prof. M. N. Rebelo. I also thank Prof. José Sande Lemos, for always welcoming me in his office and in CENTRA. I am also indebted to those who allowed me to participate in the workshop KIT-NEP at KIT. I also mention all those that allowed my talk about particle physics in the High School D. Dinis, in Lisbon, and all those that were present; I am especially grateful to profs. José António de Oliveira, Maria da Glória Pereira and João Brandão.

I also recall those with whom I had occasional discussions; I recall Matthew D. Schwartz, Luís Lavoura,
Stefan Dittmaier,
Rafael Ribeiro, Diogo Poças,
Vladyslav Shtabovenko, Claude Duhr,
Pietro Grassi, Paolo Gambino, Marcel Krause,
Juan Sebastian Cruz, Nikolai Husung,
Howard Haber, Renato Fonseca, Paulo Nogueira,
Quico Botella,
Pedro Girão and Gustavo Granja.

I am also very grateful to all those who, in one way or another, contributed to FeynMaster. Besides some already mentioned, I recall António P. Lacerda,
Miguel P. Bento, Patrick Blackstone,
Darius Jur\v{c}iukonis,
my sister Isabel
and Sofia Gomes. 

I would also like to thank all those who, albeit in a less direct fashion, contributed to my dedication to this PhD;
I recall Sally Dawson and the High Energy Theory group at Brookhaven,
Jennifer Rittenhouse West,
the Brüder und Schwester von Hl. Benedikt Labre and the Hledik Kommune,
my siblings, brothers- and sister-in-law, and my kind nieces,
my grandparents Isabel and António,
my friends of the classes of philosophy,
and profs. Henrique Leitão, Nuno Ferro and Mário Jorge de Carvalho.

And a very special thanks to my mother, to whom, in a very unsatisfactory attempt to show my gratitude for all she gave me and gives me, I dedicate this thesis.

%% file: Inputs/Abstract.tex
\cleardoublepage

\chapter*{Abstract}
\addcontentsline{toc}{chapter}{Abstract}

The discovery of the Higgs boson in 2012 initiated the exploration of the scalar sector of particle physics. In recent years, several properties of newfound particle have been experimentally explored. However, there are still many open questions; in particular, is the discovered boson the only particle in the scalar sector, as the Standard Model predicts? Or is it just one among others? The possibility of the existence of more than one scalar particle is highly relevant, as it would explain current problems, such as dark matter or matter-antimatter asymmetry.

In this thesis, we investigate several aspects of models with a multiplicity of Higgs bosons. We start by focusing on models with two Higgs doublets.
We claim that one of the most used models in the literature---the so-called real 2-Higgs-Doublet Model (2HDM)---is inconsistent.
We give special emphasis to the model that corrects it in the most simple way, the complex 2HDM; we study its phenomenology in the lowest order of perturbation theory and describe its renormalization in the following order.
Motivated by this renormalization, we analyze in detail tadpole schemes that allow the selection of the true vacuum expectation value in models with spontaneous symmetry breaking, and we propose a theorem regarding the finitude of Green's functions with corrections to the external legs.

Given the complexity of the calculations involved, it is crucial to dispose of an appropriate software. We therefore present \textsc{FeynMaster}, which calculates, among other things, Feynman rules, radiative corrections and counterterms in a simultaneously automatic and flexible way.
We use the potential of \textsc{FeynMaster} to investigate two new projects related to multiple scalar bosons:
first, we consider the class of models with a generic number of scalar doublets and singlets, and investigate radiative corrections to the $Z \to b \bar{b}$ process in this context; then, we explore the phenomenology of a model based on the linear seesaw hypothesis, which, in addition to two scalar doublets, includes a complex singlet.

\vfill
{\Large {\bf Keywords:}} {\large Higgs boson, 2-Higgs-Doublet Model, Multi-Higgs Models, phenomeno- logy, renormalization} 

%% file: Inputs/Resumo.tex
\cleardoublepage

\chapter*{Resumo}
\addcontentsline{toc}{chapter}{Resumo}

O descobrimento do bosão de Higgs em 2012 inaugurou a exploração do sector escalar da física de partículas. Nos últimos anos, várias propriedades da partícula recém-descoberta têm vindo a ser exploradas experimentalmente. No entanto, muitas são ainda as perguntas em aberto; em particular, será essa a única partícula do sector escalar, tal como prevê o Modelo Padrão? Ou será que se trata apenas de uma entre outras? A possibilidade da existência de mais do que uma partícula escalar é altamente relevante, na medida em que permitiria dar resposta a problemas actuais, tais como a matéria escura ou a assimetria matéria-antimatéria.

Nesta tese, investigamos vários aspectos de modelos com uma multiplicidade de bosões de Higgs. Come- çamos por focar modelos com dois dubletos de Higgs.
Reivindicamos que um dos modelos mais usados na literatura -- o chamado \textit{real 2-Higgs-Doublet Model} (2HDM) -- é inconsistente.
Damos especial realce ao modelo que o corrige de forma mais simples, o \textit{complex} 2HDM; estudamos a sua fenomenologia em ordem mais baixa de teoria de perturbações e descrevemos a sua renormalização em ordem seguinte.
Motivados por esta renormalização, analisamos em pormenor esquemas de \textit{tadpole} que permitem a selecção do verdadeiro valor de expectação do vácuo em modelos com quebra espontânea de simetria, e propomos um teorema referente à finitude de funções de Green com correcções às pernas exteriores.

Dada a complexidade dos cálculos envolvidos, é decisivo estar na posse de um programa computacional adequado. Apresentamos, pois, o programa \textsc{FeynMaster}, que calcula, entre muitas outras coisas, regras de Feynman, correcções radiativas e contratermos de forma simultaneamente automática e flexível.
Usamos as potencialidades do programa para investigar dois novos projectos relacionados com múltiplos bosões de Higgs.
Num primeiro, consideramos a classe de modelos com um número genérico de dubletos e singletos escalares, e investigamos correcções radiativas ao processo $Z \to b \bar{b}$ nesse contexto. Num segundo, exploramos a fenomenologia de um modelo baseado na hipótese do \textit{seesaw} linear que, para além de dois dubletos escalares, inclui um singleto carregado.

\vfill
{\Large {\bf Palavras-chave:}} {\large bosão de Higgs, 2-Higgs-Doublet Model, modelos Multi-Higgs, feno- menologia, renormalização}

%% file: Inputs/Preface.tex
\cleardoublepage

\chapter*{Preface}
\addcontentsline{toc}{chapter}{Preface}
\label{chap:Preface}

\vs{-5mm}

Research presented in this thesis has been carried out at Centro de Física Teórica de Partículas (CFTP) in the Physics Department of Instituto Superior Técnico, University of Lisbon, and was supported by Fundação para a Ciência e Tecnologia (FCT) through grants UID/FIS/00777/2013 and SFRH/BD/135698 /2018. 

\vs{2mm}

I declare that this thesis is not substantially the same as any that I have submitted for a degree, diploma or other qualification at any other university, and that no part of it has already been or is concurrently submitted for any such degree, diploma or other qualification.

\vs{2mm}

To the present date, I have contributed to the following works:
\begin{enumerate}

\item Duarte Fontes, Jorge. C. Romão and João. P. Silva, ``A reappraisal of the wrong-sign $hb\overline{b}$ coupling and the study of $h \rightarrow Z \gamma$'', Phys. Rev. D \textbf{90}, 015021 (2014), arXiv: 1406.6080, ref. \cite{Fontes:2014tga},

\item Duarte Fontes, Jorge C. Romão and João P. Silva, ``$h \rightarrow Z \gamma$ in the complex two Higgs doublet model'', JHEP \textbf{12}, 043 (2014), arXiv: 1408.2534, ref. \cite{Fontes:2014xva},

\item Duarte Fontes, Jorge C. Romão, Rui Santos and João P. Silva, ``Large pseudoscalar Yukawa couplings in the complex 2HDM'', JHEP \textbf{06}, 060 (2015), arXiv: 1502.01720, ref. \cite{Fontes:2015mea},

\item Duarte Fontes, Jorge C. Romão, Rui Santos and João P. Silva, ``Undoubtable signs of CP-violation in Higgs boson decays at the LHC run 2'', Phys.Rev. D \textbf{92} no.5, 055014 (2015), arXiv: 1506.06755, ref. \cite{Fontes:2015xva},

\item H. Bélusca-Maïto, A. Falkowski, Duarte Fontes, Jorge C. Romão, João P. Silva, ``Higgs EFT for 2HDM and beyond'', Eur.Phys.J. C \textbf{77} no. 3, 176 (2017), arXiv: 1611.01112, ref. \cite{Belusca-Maito:2016dqe},

\item H. Bélusca-Maïto, A. Falkowski, Duarte Fontes, Jorge C. Romão, João P. Silva, ``CP violation in 2HDM and EFT: the $ZZZ$ vertex'', JHEP \textbf{04}, 002 (2018), arXiv: 1710.05563, ref. \cite{Belusca-Maito:2017iob},

\item Duarte Fontes, M. Mühlleitner, Jorge C. Romão, Rui Santos, João P. Silva, J. Wittbrodt, ``The C2HDM revisited'', JHEP \textbf{02}, 073 (2018), arXiv: 1711.09419, ref. \cite{Fontes:2017zfn},

\item Duarte Fontes, Jorge C. Romão, J. W. F. Valle, ``Electroweak Breaking and Higgs Boson Profile in the Simplest Linear Seesaw Model'', JHEP \textbf{10}, 245 (2019), arXiv: 1908.09587, ref. \cite{Fontes:2019uld},

\item Duarte Fontes, Jorge C. Romão, ``FeynMaster: a plethora of Feynman tools'', Comput. Phys. Commun. \textbf{256}, 107311 (2020), arXiv: 1909.05876, ref. \cite{Fontes:2019wqh},

\item Duarte Fontes, L. Lavoura, Jorge C. Romão, João P. Silva, ``One-loop corrections to the $Zb\bar{b}$ vertex in models with scalar doublets and singlets'', Nucl. Phys. B \textbf{958}, 115131 (2020), arXiv: 1910.11886, ref. \cite{Fontes:2019fbz},

\item Duarte Fontes, Maximilian Löschner, Jorge C. Romão, João P. Silva, ``Leaks of CP violation in the real two-Higgs doublet model'', Eur. Phys. J. C \textbf{81} no. 6, 541 (2021)  arXiv: 2103:05002, ref. \cite{Fontes:2021znm},

\item Duarte Fontes, Jorge C. Romão, ``Renormalization of the C2HDM with FeynMaster 2'', JHEP \textbf{06}, 016 (2021), arXiv: 2103.06281, ref. \cite{Fontes:2021iue}.

\end{enumerate}
%
%
%

All these works have been published in refereed journals.
%
%
Refs. \cite{Fontes:2014tga,Fontes:2014xva,Fontes:2015mea,Fontes:2015xva,Belusca-Maito:2016dqe,Belusca-Maito:2017iob} are prior to this thesis and have not been included in it.
Significant content of refs. \cite{Fontes:2017zfn,Fontes:2019uld,Fontes:2019wqh,Fontes:2019fbz,Fontes:2021znm,Fontes:2021iue} has been included in this thesis, according to:
\begin{itemize}

\item chapter \ref{Chap-FM} (and appendix \ref{App-FM-Manual}) contains a significant part of ref. \cite{Fontes:2019wqh}, done in collaboration with Jorge C. Romão,

\item chapter \ref{Chap-Real} contains most part of ref. \cite{Fontes:2021znm}, done in collaboration with Maximilian Löschner, Jorge C. Romão and João P. Silva,

\item chapter \ref{Chap-Maggie} contains a significant part of ref. \cite{Fontes:2017zfn}, done in collaboration with  M. Mühlleitner, Jorge C. Romão, Rui Santos, João P. Silva and J. Wittbrodt,

\item chapter \ref{Chap-Reno} (and appendices
	\ref{App-OSS},
	\ref{App-Fermions},
	\ref{App-Sym} and
	\ref{App-FM-C2HDM}) contains most part of ref. \cite{Fontes:2021iue}, done in collaboration with Jorge C. Romão,

\item chapter \ref{Chap-Lavou} contains a significant part of ref. \cite{Fontes:2019fbz}, done in collaboration with L. Lavoura, Jorge C. Romão and João P. Silva,

\item chapter \ref{Chap-Valle} contains a significant part of ref. \cite{Fontes:2019uld}, done in collaboration with Jorge C. Romão and J. W. F. Valle.

\end{itemize}

%% file: Inputs/TOCnLists.tex
\cleardoublepage

\tableofcontents

\clearpage

\listoffigures
\addcontentsline{toc}{chapter}{List of Figures}

\clearpage

\listoftables
\addcontentsline{toc}{chapter}{List of Tables}

\null\newpage

\chapter*{\huge List of Abbreviations}
\addcontentsline{toc}{chapter}{List of Abbreviations}

\vs{-5mm}

\begin{normalsize}
\begin{center}
\begin{longtable}{@{\hspace{0mm}} >{\raggedright\arraybackslash\bf}p{1.6cm} >{\raggedright\arraybackslash}p{8cm} }	
1PI & one-particle irreducible\\[2mm]
2HDM & 2-Higgs-Doublet Model\\[2mm]
BFB & bounded from below\\[2mm]
BR & Branching Ratio\\[2mm]
BSM & Beyond the Standard Model\\[2mm]
C2HDM & Complex 2-Higgs-Doublet Model\\[2mm]
cMSSM & complex Minimal Supersymmetric Standard Model\\[2mm]
CKM & Cabibbo–Kobayashi–Maskawa\\[2mm]
CP & Charge Parity\\[2mm]
{\large $\CPeq$} & valid assuming CP conservation\\[2mm]
dof & degrees of freedom\\[2mm]
EDM & Electric Dipole Moment\\[2mm]
eq. & equation\\[2mm]
EW & electroweak\\[2mm]
fb & fentobarn\\[2mm]
FCNC & Flavour Changing Neutral Currents\\[2mm]
FJTS & Fleischer-Legerlehner tadpole scheme\\[2mm]
{\large $\FJeq$} & valid in the FJTS\\[2mm]
fig. & figure\\[2mm]
GeV & Giga electron Volt\\[2mm]
GF & Green's function\\[2mm]
h.c. & hermitian conjugate\\[2mm]
IR & infrared\\[2mm]				
LHC & Large Hadron Collider\\[2mm]
l.h.s. & left-hand side\\[2mm]
LO & Leading Order\\[2mm]
LS & Lepton-Specific\\[2mm]
$\overline{\text{MS}}$ & modified minimal subtraction\\[2mm]
$\stackrel{\overline{\mathrm{MS}}}{=}$ & valid in $\overline{\text{MS}}$\\[2mm]
MHM & Multi-Higgs Models\\[2mm]
NDR & naive dimensional regularization\\[2mm]
NLO & Next-to-Leading Order\\[2mm]
NP & New Physics\\[2mm]
OS & on-shell\\[2mm]
OSS & on-shell subtraction\\[2mm]
{\large $\OSSeq$} & valid in OSS\\[2mm]
pb & picobarn\\[2mm]
PDF & Portable Document Format\\[2mm]
PRTS & parameter renormalized tadpole scheme\\[2mm]
{\large $\PReq$} & valid in the PRTS\\[2mm]
QCD & Quantum Chromodynamics\\[2mm]
QED & Quantum Electrodynamics\\[2mm]
ref. & reference\\[2mm]
r.h.s. & right-hand side\\[2mm]
SM & Standard Model\\[2mm]
SQED & Scalar Quantum Electrodynamics\\[2mm]
SSB & Spontaneous Symmetry Breaking\\[2mm]
TeV & Tera electron Volt\\[2mm]
UV & ultraviolet\\[2mm]
vev & vacuum expectation value\\[2mm]
WI & Ward identity\\[-20mm]
\end{longtable}
\end{center}
\end{normalsize}

\cleardoublepage

%% file: Chapters/Intro.tex
\chapter*{Introduction}
\addcontentsline{toc}{chapter}{Introduction}
\label{chap:Intro}

\vs{-5mm}

\textit{1) The Higgs boson: the last element of the Standard Model to be observed}
\vs{1.5mm}

Nine years ago, the Large Hadron Collider (LHC) at CERN announced what was immediatly recognized as a spectacular milestone in particle physics: the discovery of a scalar boson.
After decades of exploring other sectors \cite{Altarelli:1989hv,Bardin:1997xq,Bardin:1999gt,Schael:2013ita}, never had the fundamental scalar sector been directly probed---until 2012.
Observed by the LHC collaborations ATLAS \cite{Aad:2012tfa} and CMS \cite{Chatrchyan:2012ufa} with a mass close to 125 GeV, the newfound particle was compatible with the Higgs boson, predicted in the '60s by several authors \cite{Higgs:1964pj,Englert:1964et,Guralnik:1964eu,Kibble:1967sv} and constituting the last piece of the Standard Model (SM) of particle physics \cite{Weinberg:1967tq,Salam:1968rm,Glashow:1970gm}.
Besides being a fundamental element to understand spontaneous symmetry breaking (SSB)
and mass generation, the Higgs boson is also indispensable to ensure the unitarity of high-energy scatterings between longitudinal $W$ and $Z$ gauge bosons \cite{LlewellynSmith:1973yud,Bell:1973ex,Cornwall:1973tb,Romao:2016ien}.

\begin{figure}[h!]
\centering
\includegraphics[width=0.45\linewidth]{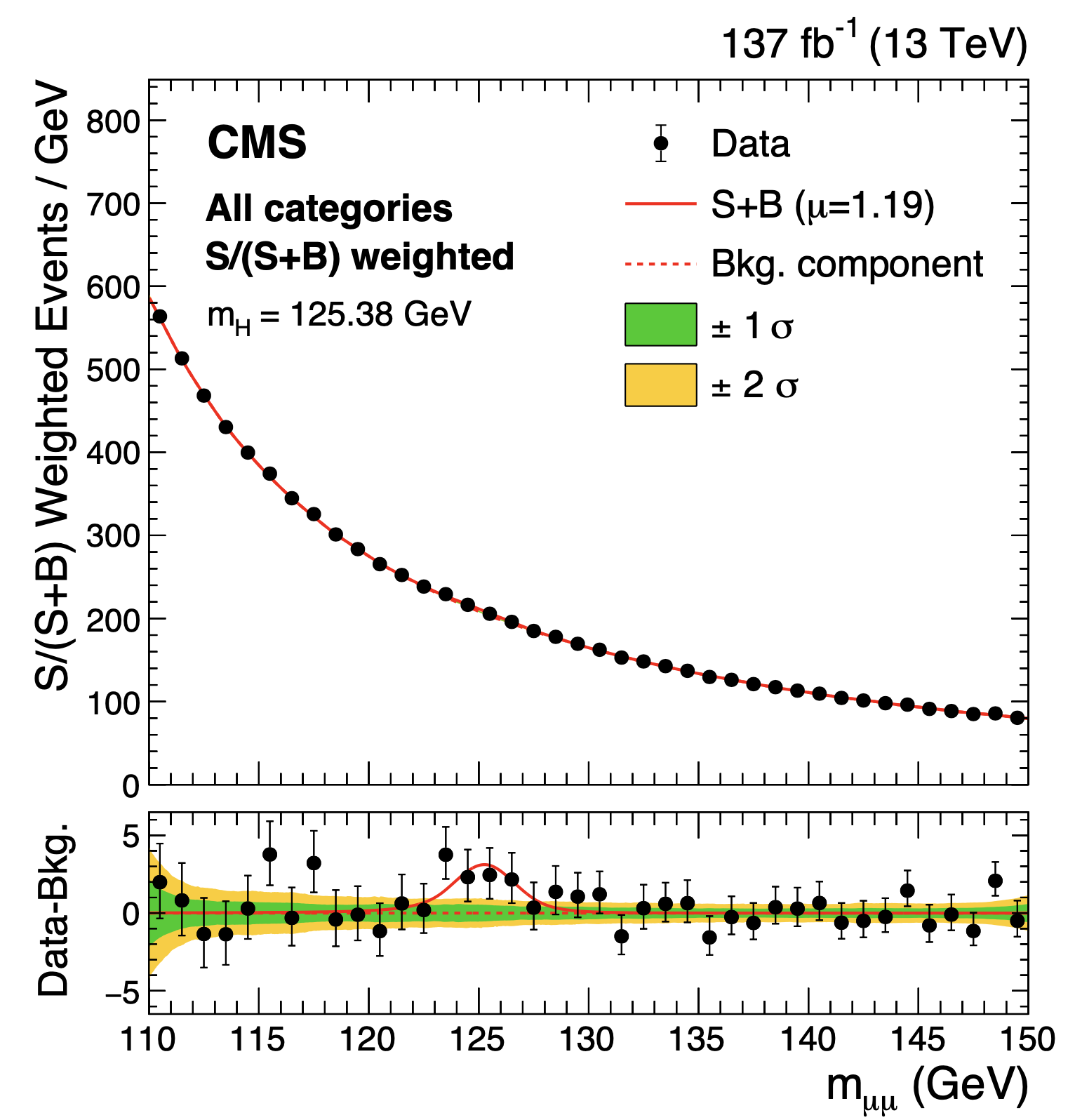}
\includegraphics[width=0.45\linewidth]{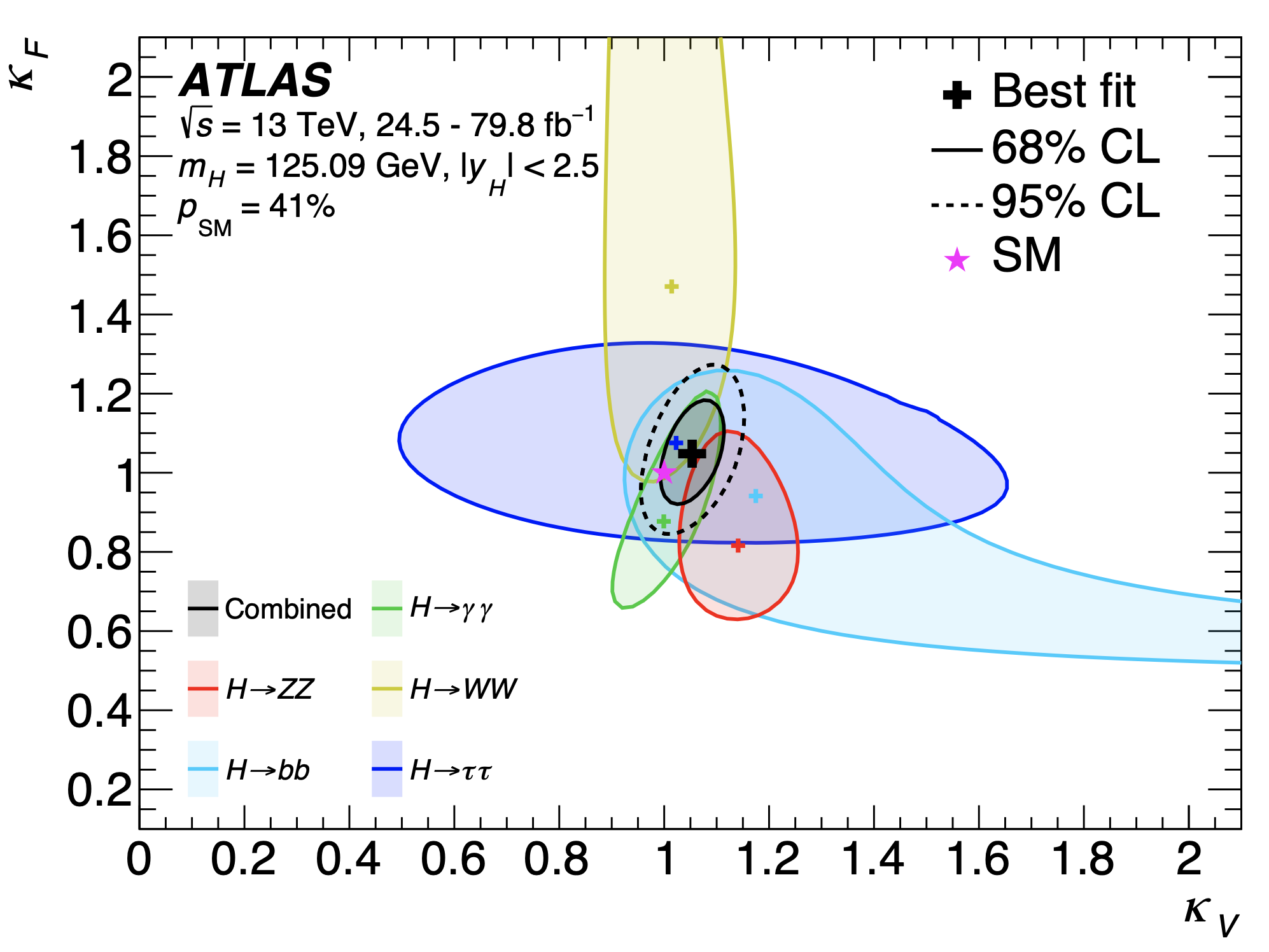}
\caption[
Left: the distribution of the $m_{\mu\mu}$ invariant mass;
the lower panel displays what is left after subtraction of the background, with the red line representing the best fit of the $h \to \mu^- \mu^+$ signal contribution for $m_h =125.38$ GeV.
Right: negative log-likelihood contours at 68\% and 95\% confidence level in the plane formed by the coupling modifiers concerning the Higgs-vector-boson ($\kappa_V$) and Higgs-fermion ($\kappa_F$) couplings, for the individual decay modes; also shown (in black) is their combination. The cross represents the best-fit value for each measurement, whereas the star indicates the SM hypothesis.
]
{Left: the distribution of the $m_{\mu\mu}$ invariant mass;
the lower panel displays what is left after subtraction of the background, with the red line representing the best fit of the $h \to \mu^- \mu^+$ signal contribution for $m_h =125.38$ GeV.
Right: negative log-likelihood contours at 68\% and 95\% confidence level in the plane formed by the coupling modifiers relative to the Higgs-vector-boson ($\kappa_V$) and Higgs-fermion ($\kappa_F$) couplings, for the individual decay modes; also shown (in black) is their combination. The cross represents the best-fit value for each measurement, whereas the star indicates the SM hypothesis.
Figures taken from refs. \cite{Sirunyan:2020two} and \cite{Aad:2019mbh}.
}
\label{Chap-Intro:fig:XarExp2}
\end{figure}

The new particle was discovered at the LHC through decays to gauge bosons, from data with an integrated luminosity of about $10 \, \mathrm{fb}^{-1}$ obtained between 2010 and 2012, corresponding to a center-of-mass energy of 7 and 8 TeV (constituting the LHC Run-1) \cite{Agashe:2014kda}.
%
During the months following the discovery, an extensive set of measurements of both the Higgs boson mass \cite{ATLAS:2012klq,Chatrchyan:2012jja}
and its couplings to $W$ and $Z$ bosons \cite{ATLAS:2013nma,Chatrchyan:2013mxa} was carried out.
\begin{figure}[h!]
\centering
\includegraphics[width=0.45\linewidth]{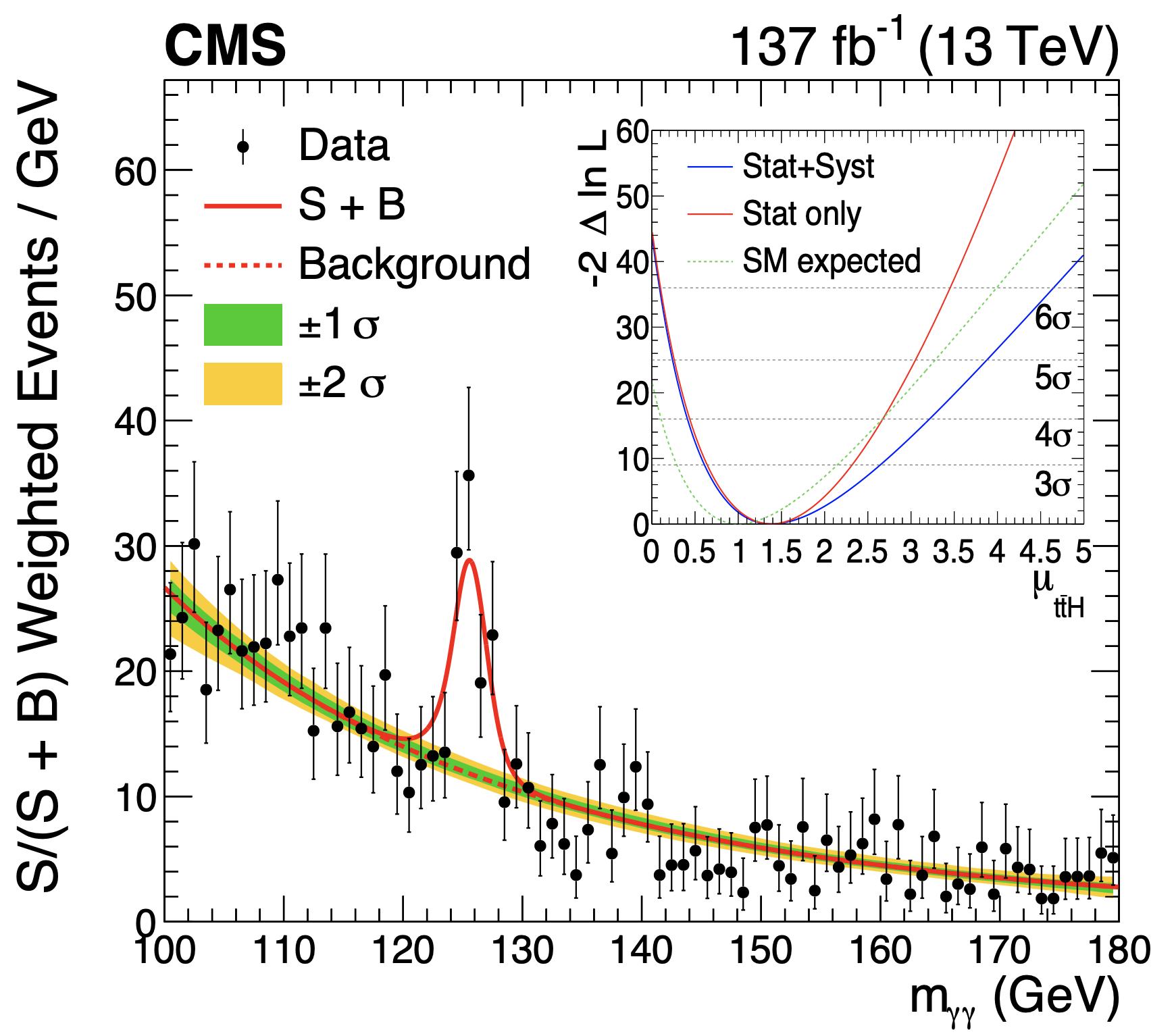}
\includegraphics[width=0.45\linewidth]{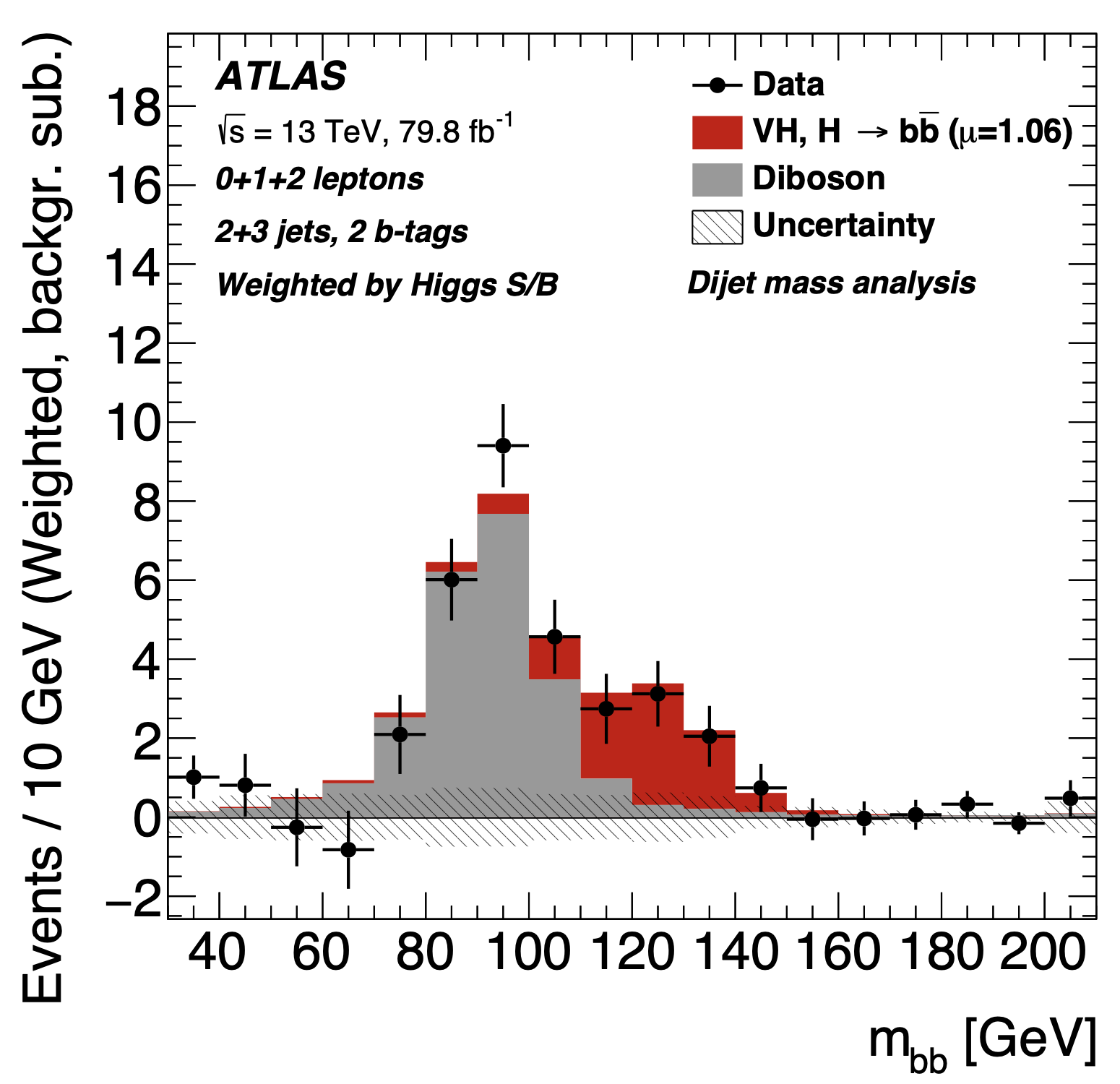}
\caption[
Left: distribution of the invariant mass of the diphoton system for selected events (black points) in $t\bar{t}h$ specific topologies. The curves in solid red and dashed red represent the fitted signal + background and the background only, respectively. The green and yellow bands cover the $\pm 1\sigma$ and $\pm 2\sigma$ uncertainties in the fitted background, respectively. The inner panel shows the measurement’s likelihood as a function of the strength of signal.
Right: the distribution of the $m_{bb}$ invariant mass summed over all channels and regions, showing the $Z \to bb$ (gray) and $h \to bb$ signal (red)
]
{Left: distribution of the invariant mass of the diphoton system for selected events (black points) in $t\bar{t}h$ specific topologies. The curves in solid red and dashed red represent the fitted signal + background and the background only, respectively. The green and yellow bands cover the $\pm 1\sigma$ and $\pm 2\sigma$ uncertainties in the fitted background, respectively. The inner panel shows the measurement’s likelihood as a function of the strength of signal.
Right: the distribution of the $m_{bb}$ invariant mass summed over all channels and regions, showing the $Z \to b\bar{b}$ (gray) and $h \to b\bar{b}$ signal (red).
Figures taken from refs. \cite{Sirunyan:2020sum} and \cite{Aaboud:2018zhk}.
}
\label{Chap-Intro:fig:XarExp1}
\end{figure}
%
%
%
%
%
%
Its CP parity was also probed, with the result that the possibility of a pure CP-odd state was definitely excluded \cite{Chatrchyan:2012jja,Khachatryan:2014kca,Aad:2013xqa}.
%
A direct observation of Yukawa couplings was essential to understand the nature of the new scalar \cite{Bass:2021acr}; this was allowed by data collected between 2015 and 2018 with $139 \mathrm{fb}^{-1}$ at a center-of-mass energy of 13 TeV (Run-2)\cite{Sirunyan:2018koj}.
The first direct observation of the Higgs-top Yukawa coupling was obtained by ATLAS and CMS collaborations with Run-2 data in 2018 \cite{Aaboud:2018urx,Sirunyan:2018hoz}; a distribution of the invariant mass of the diphoton system for events corresponding to $t\bar{t}h$ topologies can be seen on the left panel of fig. \ref{Chap-Intro:fig:XarExp1}.
The observation of the decay channel $h \to b\bar{b}$ also in 2018 \cite{Aaboud:2018zhk,Sirunyan:2018kst} constituted another experimental breakthrough, for although it corresponds to the largest branching ratio of the Higgs boson, it involves a huge background \cite{Heinemeyer:2013tqa}; on the right panel of fig. \ref{Chap-Intro:fig:XarExp1}, a distribution of the $m_{bb}$ invariant mass is shown.
Both ATLAS and CMS have also employed the Run-2 data to search for the $h \to \mu^- \mu^+$ channel, and thus inaugurate the exploration of the couplings to the second fermion generation;
%
%
the first results were divulged in summer 2020 \cite{Aad:2020xfq,Sirunyan:2020two}; the distribution of the $m_{\mu\mu}$ invariant mass can be seen on the left panel of fig. \ref{Chap-Intro:fig:XarExp2}.

Several analyses are still being finalized, but many Run-2 results are complete \cite{Bass:2021acr}. All in all, the experimental results concerning the Higgs boson obtained so far comply with the SM expectation \cite{Aad:2019mbh,Sirunyan:2018koj}; this behaviour is illustrated on the right panel of fig. \ref{Chap-Intro:fig:XarExp2}, which shows the so-called coupling modifiers $\kappa$ (which quantify deviations to SM predictions)
for the couplings of the Higgs boson to fermions and to vector bosons.
In fact, as the experimental precision grows, the results get closer and closer to those determined by the SM \cite{Zyla:2020zbs}

\vs{1.5mm}
\textit{2) Beyond the Standard Model: there must be something else}
\vs{1.5mm}

\begin{figure}[h!]
\centering
\includegraphics[width=0.55\linewidth]{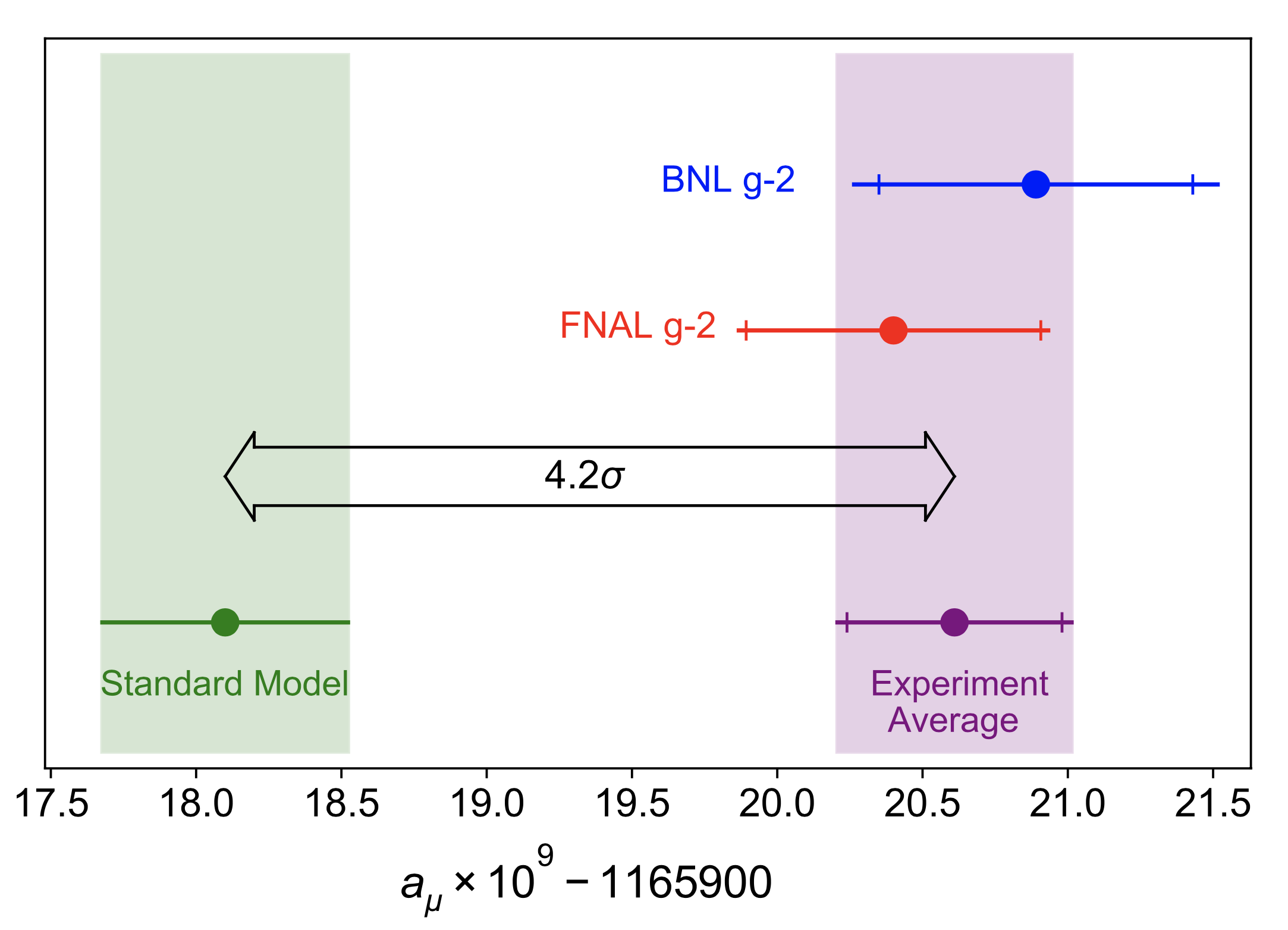}
\caption[
From top to bottom: experimental values of $a_{\mu}$ from experiment E821 at Brookhaven National Laboratory, the new FNAL measurement, and the combined average.
]
{
From top to bottom: experimental values of $a_{\mu}$ from experiment E821 at Brookhaven National Laboratory, the new FNAL measurement, and the combined average. The value recommended by ref. \cite{Aoyama:2020ynm} for the SM is also shown. Figure taken from ref. \cite{Abi:2021gix}.
}
\label{Chap-Intro:fig:XarExp3}
\end{figure}

And yet, the SM cannot be the ultimate theory of particle physics.
Despite the tremendous success of the SM in explaining the observed phenomena (in particular regarding the Higgs boson), it is by now clear that there must be physics beyond the SM (BSM).
The most striking indications come from inability of the SM to properly explain the baryon asymmetry of the Universe \cite{Trodden:1998qg}, as well as to accommodate both dark matter
and neutrino masses \cite{Zyla:2020zbs}.
In addition, a new measurement of the muon anomalous magnetic moment
$a_{\mu} \equiv (g_{\mu}-2)/2$ has been recently announced by the FNAL Muon g-2 collaboration \cite{Abi:2021gix}, supporting the claim for BSM physics.
This measurement is compatible with a previous one, obtained by the E821 experiment at Brookhaven National Laboratory \cite{Bennett:2006fi}. Fig. \ref{Chap-Intro:fig:XarExp3} shows that the combination of results deviates by $4.2 \sigma$ from the SM prediction, described in the white paper of the Muon g-2 Theory Initiative \cite{Aoyama:2020ynm}.
%
%
%
%
%

The quest for New Physics is thus highly relevant.
Among the different directions that can be followed, one of them addresses an essential question: how many elementary scalar particles are there? After all, there is no fundamental reason why the scalar sector should be restricted to a single Higgs doublet, as predicted by the SM. Moreover, although the neutral scalar observed at the LHC is compatible with the Higgs boson of the SM, it can also correspond to just one particle of a larger set of scalar bosons \cite{Zyla:2020zbs}.
%
%
So, what if there is an extended scalar sector?

\vs{1.5mm}
\textit{3) Multiple Higgs bosons: an exciting possibility}
\vs{1.5mm}

The possibility of Multi-Higgs bosons is extremely exciting; it constitutes the backbone of this thesis, as a follow-up to work developed earlier  \cite{Fontes:2014tga,Fontes:2014xva,Fontes:2015mea,Fontes:2015xva,Belusca-Maito:2016dqe,Belusca-Maito:2017iob}. 
From the experimental side, searches for additional Higgs bosons have been performed by both ATLAS \cite{Morvaj:2019ldy} and CMS \cite{Tao:2019hpy} collaborations (a recent report on the searches for extra scalar particles can be found in ref. \cite{Zyla:2020zbs}).
From the theoretical side, there have been countless studies on models with an extended scalar sector, also known as Multi-Higgs Models (for reviews, see refs. \cite{Gunion:1989we,Ivanov:2017dad} and references therein).
Multi-Higgs Models allow for extremely rich phenomenologies, including for example the possibility of charged scalar bosons.

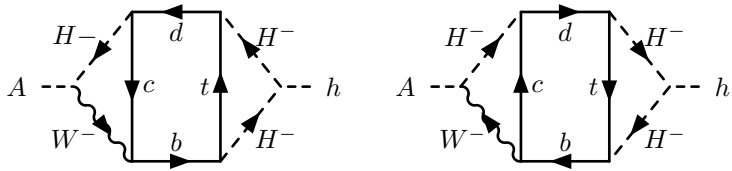
\begin{figure}[!htb] 
\vspace{-8mm}
\hspace{6mm}
\centering
\subfloat{ 
\begin{fmffile}{Intro1} 
\begin{fmfgraph*}(100,100) 
\fmfset{arrow_len}{3mm} 
\fmfset{arrow_ang}{20} 
\fmfleft{nJ1} 
\fmflabel{$A$}{nJ1} 
\fmfright{nJ2} 
\fmflabel{$h$}{nJ2} 
\fmf{dashes,tension=25}{nJ1,nJ1J4J1} 
\fmf{dashes,tension=25}{nJ2,nJ2J7J6}
\fmftop{yt1}
\fmftop{yt2}
\fmfbottom{yb1}
\fmfbottom{yb2}
\fmf{phantom,tension=8}{yt1,J12J9J2}
\fmf{phantom,tension=8}{yt2,J16J11J5}
\fmf{phantom,tension=8}{yb1,J10J13J3}
\fmf{phantom,tension=8}{yb2,J14J15J8}
\fmf{scalar,label=$H{-}$,label.side=right,tension=6,label.dist=1thick}{J12J9J2,nJ1J4J1} 
\fmf{photon,label=$W^{-}$,label.side=right,tension=6,label.dist=1thick}{nJ1J4J1,J10J13J3}
\fmf{phantom_arrow,tension=0}{nJ1J4J1,J10J13J3}
\fmf{scalar,label=$H^{-}$,label.side=right,tension=6,label.dist=1thick}{nJ2J7J6,J16J11J5} 
\fmf{scalar,label=$H^{-}$,label.side=right,tension=6,label.dist=1thick}{J14J15J8,nJ2J7J6} 
\fmf{fermion,label=$c$,label.side=left,tension=0.1,label.dist=2thick}{J12J9J2,J10J13J3} 
\fmf{fermion,label=$d$,label.side=left,tension=0.1,label.dist=2thick}{J16J11J5,J12J9J2} 
\fmf{fermion,label=$b$,label.side=left,tension=0.1,label.dist=2thick}{J10J13J3,J14J15J8} 
\fmf{fermion,label=$t$,label.side=left,tension=0.1,label.dist=2thick}{J14J15J8,J16J11J5} 
\end{fmfgraph*} 
\end{fmffile} 
} 
\hspace{9mm} 
\subfloat{ 
\begin{fmffile}{Intro2} 
\begin{fmfgraph*}(100,100) 
\fmfset{arrow_len}{3mm} 
\fmfset{arrow_ang}{20} 
\fmfleft{nJ1} 
\fmflabel{$A$}{nJ1} 
\fmfright{nJ2} 
\fmflabel{$h$}{nJ2} 
\fmf{dashes,tension=25}{nJ1,nJ1J3J2} 
\fmf{dashes,tension=25}{nJ2,nJ2J5J8}
\fmftop{yt1}
\fmftop{yt2}
\fmfbottom{yb1}
\fmfbottom{yb2}
\fmf{phantom,tension=8}{yt1,J10J11J1}
\fmf{phantom,tension=8}{yt2,J12J15J6}
\fmf{phantom,tension=8}{yb1,J14J9J4}
\fmf{phantom,tension=8}{yb2,J16J13J7}
\fmf{scalar,label=$H^{-}$,label.side=left,tension=6,label.dist=1thick}{nJ1J3J2,J10J11J1} 
\fmf{photon,label=$W^{-}$,label.side=right,tension=6,label.dist=1thick}{nJ1J3J2,J14J9J4}
\fmf{phantom_arrow,tension=0}{J14J9J4,nJ1J3J2}
\fmf{scalar,label=$H^{-}$,label.side=left,tension=6,label.dist=1thick}{J12J15J6,nJ2J5J8} 
\fmf{scalar,label=$H^{-}$,label.side=left,tension=6,label.dist=1thick}{nJ2J5J8,J16J13J7} 
\fmf{fermion,label=$c$,label.side=right,tension=0.1,label.dist=2thick}{J14J9J4,J10J11J1} 
\fmf{fermion,label=$d$,label.side=right,tension=0.1,label.dist=2thick}{J10J11J1,J12J15J6} 
\fmf{fermion,label=$b$,label.side=right,tension=0.1,label.dist=2thick}{J16J13J7,J14J9J4} 
\fmf{fermion,label=$t$,label.side=right,tension=0.1,label.dist=2thick}{J12J15J6,J16J13J7} 
\end{fmfgraph*} 
\end{fmffile} 
}
\vspace{-5mm}
\caption{Feynman diagrams contributing to CP violation in the scalar sector of the real 2HDM.}
\label{Intro:fig:jarl-pair}
\end{figure}

Particularly appealing inside Multi-Higgs scenarios is the simplest one that can provide a new source of CP violation---as required by the three Sakharov criteria for baryogenesis~\cite{Sakharov:1967dj}---, the so-called 2-Higgs-Doublet Model (2HDM)~\cite{Lee:1973iz} (for a review, see ref.~\cite{Branco:2011iw}).
Several variants of 2HDM can be considered;
the most common one prescribes a softly-broken $\mathbb{Z}_2$ symmetry and enforces CP conservation in the scalar sector.
This model is sometimes referred to as the ``real 2HDM'', because the
parameter in the potential that softly-brakes $\mathbb{Z}_2$ is taken to be real (since CP conservation is imposed in the potential).
In this thesis, we argue that this variant of 2HDM is theoretically unsound. The reason is that, in order to comply with experimentally observed CP violation \cite{Zyla:2020zbs}, the real 2HDM allows for CP violation in quark interactions, present as a complex phase in the Cabibbo-Kobayashi-Maskawa (CKM) matrix \cite{Cabibbo:1963yz,Kobayashi:1973fv}. At higher orders in perturbation theory, ultraviolet (UV) divergent CP-violation effects coming from that phase end up leaking into the scalar sector (see fig. \ref{Intro:fig:jarl-pair}), which however lacks CP-violating counterterms to absorb such divergences. The real 2HDM thus reveals itself as non-renormalizable, which means that it is not a (consistent) model.

The inconsistency of the \textit{real} 2HDM can be solved by allowing the scalar sector to be in general CP-violating. In that case, one obtains the \textit{complex} 2HDM (C2HDM). This model was originally discussed in ref.~\cite{Ginzburg:2002wt}, having been later developed and used by many authors~\cite{Khater:2003wq, ElKaffas:2006gdt, WahabElKaffas:2007xd, ElKaffas:2007rq, Osland:2008aw, Grzadkowski:2009iz, Arhrib:2010ju, Barroso:2012wz, Inoue:2014nva, Cheung:2014oaa, Fontes:2014xva, Fontes:2015mea, Chen:2015gaa, Fontes:2015xva, Belusca-Maito:2017iob, Fontes:2017zfn, Basler:2017uxn, Aoki:2018zgq}.
Special attention is devoted to the C2HDM in this thesis, for several reasons already suggested, viz.: it explores the exciting scenario of Multi-Higgs bosons in a simple fashion, it provides an extra source of CP-violation required for baryogenesis---thus introducing the interesting hypothesis according to which the neutral scalars have mixed CP parities---and it corrects in a simple and alegant way the flaws of the real 2HDM. Furthermore, a systematic investigation of next-to-leading-order corrections of the model is still barely explored, which motivates an undertaking in this direction.

Although less popular than the real 2HDM, the C2HDM has aroused quite some interest in the last few months. For example, a 2-loop renormalization group evolution of the model was discussed in ref.~\cite{Oredsson:2019mni};
ref.~\cite{Wang:2019pet} used the model to study phase transition dynamics and gravitational wave signals;
ref.~\cite{Boto:2020wyf} investigated a basis-independent treatment of the model; studies on electric dipole moments (EDM) of the C2HDM have been put forward in refs.~\cite{Cheung:2020ugr, Altmannshofer:2020shb};
ref.~\cite{Azevedo:2020vfw} discussed the impact of the discovery of a new scalar particle on the parameter space of the model;
ref.~\cite{Huang:2020zde} considered CP-violating gauge-scalar interactions;
ref.~\cite{Low:2020iua} investigated Higgs alignment and signatures of CP violation in the context of the model;
and ref. \cite{Azevedo:2020fdl} derived constraints on the model from the phenomenology of light scalars.
As we mentioned in passing, the three neutral scalar bosons predicted by the C2HDM contain in general a mixture of a CP-even and a CP-odd component. Notably, the state corresponding to the scalar observed at the LHC could have a CP-odd component. Although the possibility of a pure CP-odd state is excluded (as discussed above), the scenario where the Higgs boson has a partial CP-odd factor is still allowed \cite{Zyla:2020zbs}.
This possibility has been experimentally investigated in couplings of the Higgs boson: both in couplings to vector bosons \cite{Aad:2016nal,Sirunyan:2019twz,Aaboud:2017vzb,Sirunyan:2019nbs}, as well as in couplings to fermions \cite{Berge:2011ij,Sirunyan:2019wxt,Aad:2020ivc}. Here, we shall investigate the phenomenology of the C2HDM, focusing on scenarios where the neutral scalar can display a fascinating CP-character. In particular, we will show that the neutral scalar particles can couple in a pure CP-even fashion to some fermions and in a pure CP-odd fashion to other fermions.


\vs{1.5mm}
\textit{4) Going up to one-loop level: a necessary condition for precision physics}
\vs{1.5mm}

From the initial discussion, it became clear that the LHC has not yet detected any significant deviations between the observed properties of the Higgs boson and the SM predictions. Hence, if discrepancies are to be detected (and we presented a compelling case in favor of this hypothesis), they shall be very subtle. It is therefore of paramount importance that \textit{precise predictions} from the theory side are put forward, so as to properly interpret small experimental signals of BSM physics. 
In other words, it is compelling to go beyond the leading-order (LO) predictions of the model and include the next-to-leading-order (NLO) in perturbation theory.
NLO corrections are indispensable also for a correct determination of the parameter space of models and an accurate distinction between different BSM models.
They require the inclusion of one-loop electroweak corrections, which in turn require the one-loop electroweak renormalization of the model at stake.%
\fn{Hereafter, and unless stated otherwise, all references to renormalization mean the one-loop electroweak renormalization.}

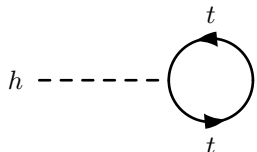
\begin{figure}[htp] 
\vspace{-3mm}
\centering 
\begin{fmffile}{Intro3} 
\begin{fmfgraph*}(80,70) 
\fmfset{arrow_len}{3mm} 
\fmfset{arrow_ang}{20} 
\fmfleft{nJ1} 
\fmflabel{$h$}{nJ1} 
\fmfright{o1} 
\fmf{dashes,tension=1.3}{nJ1,J2J1nJ1} 
\fmf{fermion,label=$t$,right=1}{J2J1nJ1,o1,J2J1nJ1} 
\end{fmfgraph*} 
\end{fmffile}
\vspace{-5mm}
\caption{Feynman diagram modifying the tree-level minimum of the potential of the SM.}
\label{Intro:fig:my-tad}
\end{figure}

The tandem composed of these two topics---one-loop corrections and renormalization---constitutes another major pillar of this thesis.
We shall devote special attention to the selection of the true vacuum expectation value (vev), a central aspect in such tandem. In theories with SSB, a non-null vev must be considered in order for fields to correspond to small excitations around the minimum of the potential. When higher orders in perturbation theory are included, the minimum of the potential is in general modified (see fig. \ref{Intro:fig:my-tad}), so that the true vev no longer corresponds to the tree-level one.
It turns out that there is more than one consistent method to select the true vev, in such a way that the different alternatives have significant consequences for the renormalization of the theory. This discussion becomes especially interesting whenever the vev can be complex, which occurs in models with a CP-violating potential, such as the C2HDM.

Despite the aforementioned recent works on this model, its renormalization is still missing in the literature.%
\fn{Except for ref. \cite{Fontes:2021iue}, included in this thesis.}
We fill this gap in the present thesis, thus paving the way for an investigation of precise predictions in the C2HDM. As shall be seen, the existence of CP violation in the scalar sector of the model leads to a quite unique process of renormalization, since it requires the introduction of several non-physical parameters. This will have as an interesting consequence that there will be more independent counterterms than independent renormalized parameters, which means that there will be several possible combinations of the former for the same set of the latter.

Still in the context of one-loop corrections and renormalization, we discuss several other topics, such as the impact of CP-violation in fermionic counterterms (and the possibility of complex mass counterterms), 
the renormalization of fields whose mass is a dependent parameter and the role of loop corrections to external propagators. Concerning the latter, we put forward a theorem, which states that, for a certain process, the inclusion of Feynman diagrams containing non-renormalized loop corrections to external propagators will render the process finite if and only if a) the process does not exist at lower orders and b) the corrections to external propagators are set on-shell. The theorem is completely general, leading to an interesting corollary for 3-particle processes.

\vs{1.5mm}
\textit{5) Flexibility: the key for a mastery of Feynman tools}
\vs{1.5mm}

Given the complexity of NLO and renormalization calculations, the use of computational tools is today virtually indispensable. In the specific context of model building and NLO calculations, there is currently a myriad of software addressing one or several of the following tasks (e.g. refs. \cite{Belyaev:2012qa,Belanger:2003sd,Denner:2016kdg,Cullen:2011ac,Cullen:2014yla,Lorca:2004fg,Degrande:2014vpa,Mertig:1990an,Shtabovenko:2016sxi,Shtabovenko:2020gxv,Christensen:2008py,Semenov:2014rea,Alloul:2013bka,Hahn:1998yk,Hahn:2000kx,Kublbeck:1990xc,Pukhov:1999gg,Nogueira:1991ex,Semenov:1996es,Semenov:1998eb,Tentyukov:1999is,Wang:2004du,Alwall:2014hca,Feng:2021kha}):
\vs{1.0mm}
\begin{center}
	\quad a) generation of Feynman rules; \qquad b) generation and drawing of Feynman diagrams;\\
	c) generation of amplitudes; \hs{3mm} d) loop calculations; \hs{3mm} e) algebraic calculations; \hs{3mm} f) renormalization.
\end{center}
\vs{1.0mm}

However, despite the undisputed quality of some of the referred software---which perform almost all {tasks} of the above list---, they usually do not combine an automatic character with the possibility of manipulating the final analytical expressions in a practical way. And although interfaces between different software exist, they tend not to be free of constraints, since the notation changes between software and a conversion is not totally automatic. It would thus be desirable to have a single software that could perform all the above listed tasks, and at the same time allowing the user to handle the final results.

In this thesis, a new such software is introduced. \ts{FeynMaster} \cite{Fontes:2019wqh} is a single program, written in both \t{\ts{Python}} and \t{\ts{Mathematica}}, that combines \ts{FeynRules}~\cite{Christensen:2008py,Alloul:2013bka}, \QG~\cite{Nogueira:1991ex} and \ts{FeynCalc}~\cite{Mertig:1990an,Shtabovenko:2016sxi,Shtabovenko:2020gxv}
to perform \textit{all} the referred tasks in a flexible and consistent way.
It has a hybrid character concerning automatization: not only does it automatically generate the results, but it also allows the user to act upon them. This feature is extremely useful, since very often in research one is not interested in obtaining a rigid list of final expressions, but in handling them at will. This is made possible in \ts{FeynMaster} due to the creation of notebooks for both \ts{FeynRules} and \ts{FeynCalc}, in which a multiplicity of different tools enables the user to manipulate the results.

\vs{1.5mm}
\textit{6) Structure of the thesis}
\vs{1.5mm}

The thesis is organized as follows.
We start with the presentation of \FM, which lies at the root of virtually all subsequent chapters. This we do in chapter \ref{Chap-FM}.
Then, chapter \ref{Chap-Selec} is devoted to the topic of the selection of the true vev, briefly described above. Here, we also introduce the technique of renormalization, explaining in detail several aspects that may be less clear in the literature.%
\fn{This chapter shows for the first time a very personal view on
tadpoles and renormalization, and it defends positions which may
not be shared by all my collaborators.}
The chapter is thus highly relevant for understanding our subsequent criticism of the real 2HDM, as well as the renormalization of the C2HDM.
It is to the former that we turn in chapter \ref{Chap-Real}, where we argue that there are irremovable infinities in the model, as a consequence of assuming CP conservation in the scalar sector while allowing CP violation through quark interactions.

As a simple way to heal the pathologies of the real 2HDM, we investigate in detail the C2HDM, which constitutes the main subject of the following two  chapters.
More specifically, in chapter \ref{Chap-Maggie} we study the C2HDM at tree-level, describing the theory in detail and exploring interesting cases of its phenomenology.
In chapter \ref{Chap-Reno}, we investigate the model for up-to-one-loop level purposes. We show that this requires a different treatment of the model immediatly at tree-level; we perform the electroweak one-loop renormalization, and compare the behaviour of selected combinations of independent counterterms in specific NLO processes, which are ensured to be gauge independent via a simple prescription.

The final chapters of the thesis are devoted to more general Multi-Higgs models. Chapter \ref{Chap-Lavou} is devoted to one-loop corrections to the $Z\bar{b}b$ vertex in extensions of the SM with an arbitrary numbers of scalar particles. We shall ascertain the reasonableness of certain approximations usually performed in the literature, like the neglect of diagrams with neutral scalars. Finally, in chapter \ref{Chap-Valle} we examine a simple realization of the linear seesaw mechanism---which configures two Higgs doublets and a complex singlet---, exploring the parameter space of the model.

After an overall Conclusion, we present several appendices.
Appendix \ref{App-FM-Manual} contains the updated version of the manual of \FM, just as it can be found online.
Appendix \ref{App-Pilaftsis} clarifies how a model by Pilaftsis \cite{Pilaftsis:1998pe} is theoretically sound, thus not suffering from the inconsitencies of the real 2HDM.
Appendix \ref{App-OSS} provides a simple description of the on-shell subtraction (OSS) scheme and investigates the scenarios where the mass of a particle is a dependent parameter.
These scenarios are also the main object of appendix \ref{App-LSZ}, but now in the context of the LSZ formula.
Appendix \ref{App-Fermions} is devoted to CP violation in fermionic 2-point functions and its influence on counterterms, which is particulatly relevant for the renormalization of the C2HDM.
Appendix \ref{App-WI} derives the Ward identity that can be used to simplify the renormalization of the electric charge, and we prove explicitly how that simplification is realized. 
Then, in appendix \ref{App-Sym}, we show how counterterms can be fixed through symmetry relations (again very useful for the renormalization of the C2HDM).
Appendix \ref{App-FM-C2HDM} presents \FMS as a rather convenient tool to renormalize the C2HDM.
Finally, appendix \ref{App-Theorem} is devoted to the theorem which we alluded to.

%% file: Chapters/Chapter_FM.tex
\chapter{FeynMaster}
\label{Chap-FM}

\vs{-5mm}

%
In this chapter, we start by presenting an overview of \FM, after which we
compare the software with similar ones.
The latest version (\FMT) can be downloaded online at:
\begin{center}
\url{https://porthos.tecnico.ulisboa.pt/FeynMaster/}.
\end{center}
The current version of the manual is included in this thesis in appendix \ref{App-FM-Manual}, and an application of the program to the renormalization of the C2HDM can be found in appendix \ref{App-FM-C2HDM}.

\section{Overview}
\label{Chap-FM:overview}

\begin{figure}[h!]
\centering
\includegraphics[width=0.8\textwidth]{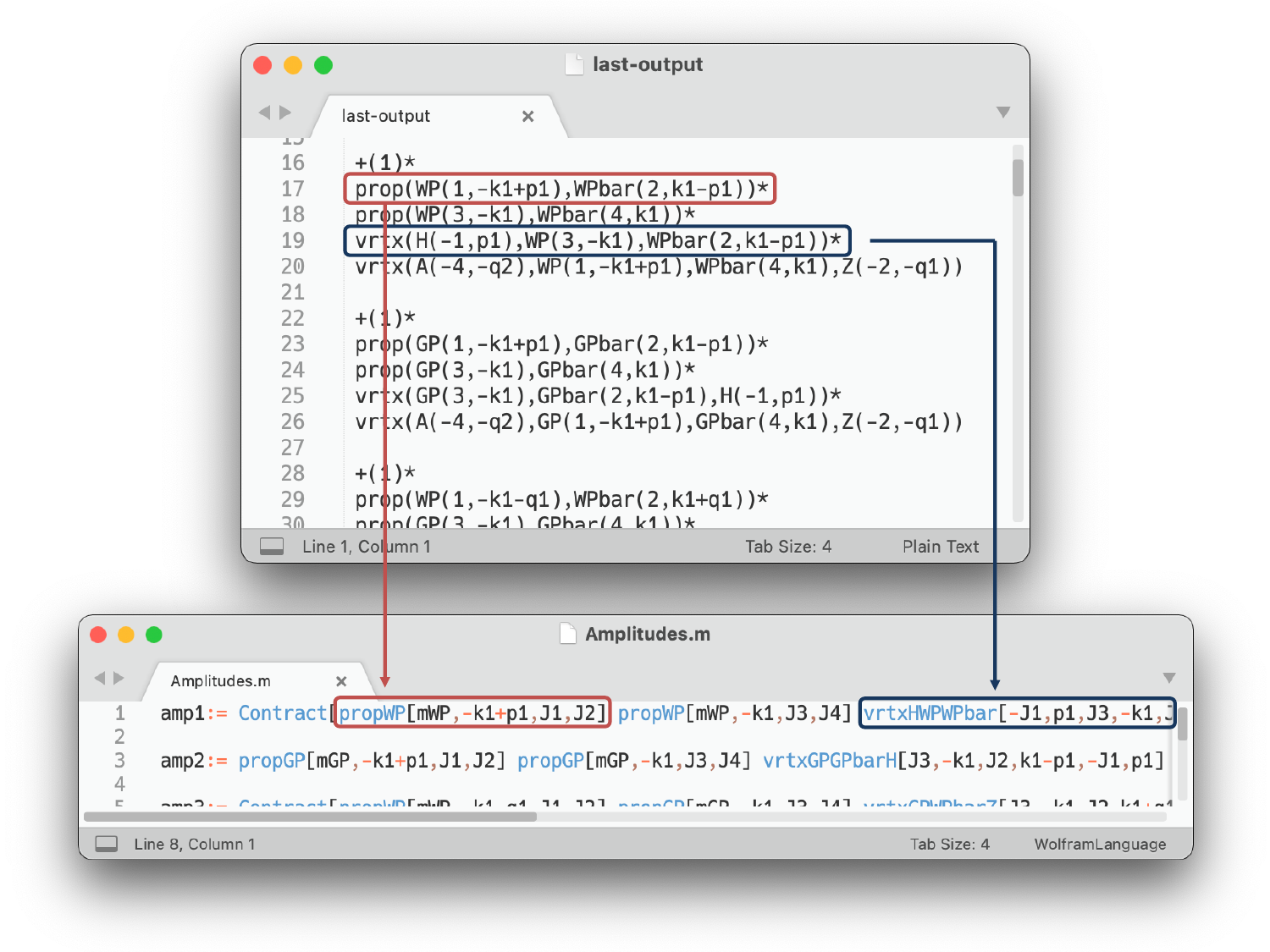}
\vspace{-7mm}
\caption{Conversion of the \QGS output (top) to a file with the amplitudes for \textsc{FeynCalc} (bottom).}
\label{Chap-FM:fig:Lacerda-basic}
\end{figure}

The original motivation for \FMS was an automatic generation of one-loop amplitudes that could be managed with \ts{FeynCalc}~\cite{Mertig:1990an,Shtabovenko:2016sxi,Shtabovenko:2020gxv}.
The latter is an excellent tool to perform calculations and manipulate the results, due to its flexibility: for one thing, it is contained inside the general framework of \t{\textsc{Mathematica}}, which allows all kinds of manipulations; for another, \textsc{FeynCalc} itself contains a multiplicity of very useful functions to deal with one-loop amplitudes.
It turns out that the output of \QG~\cite{Nogueira:1991ex} contains all the information required to write one-loop amplitudes. One just needs to reorganize that information; this task is performed by \FM, as illustrated in fig. \ref{Chap-FM:fig:Lacerda-basic}.

Yet, an automatic conversion from \QGS to \FCS is only really useful if the Feynman rules contributing to the amplitudes (e.g. \t{propWP} and \t{vrtxHWPWPbar} in the definition of \t{amp1} in fig. \ref{Chap-FM:fig:Lacerda-basic}) are also automatically generated. \FMS thus resorts to \ts{FeynRules}~\cite{Christensen:2008py,Alloul:2013bka}, which yields the complete set of Feynman rules for a given Lagrangian. \FMS then writes those rules in a way that \FCS can read them, and uses the information of the \FRS output to automatically generate the model file for \QG. In this way, by defining a Lagrangian, one has automatic access to both Feynman rules and amplitudes, all of them defined in the user-friendly setup of \FC.

Another crucial ingredient of \FMS is the automatic generation of counterterms. Given that one-loop amplitudes are in general divergent, their study is severely deficient unless one can render them finite. This requires the calculation of the total counterterm for the process under investigation, which in turn requires the determination of the complete set of individual counterterms contributing to it.
\FMS not only automatically determines the complete set of counterterms for all the possible processes, but also automatically calculates the total counterterm for the process at stake in a particular subtraction scheme (modified minimal subtraction, $\overline{\text{MS}}$). As a matter of fact, \FMS can automatically calculate the total counterterm not only for one process, but also for a sequence of processes. This happens in such a way that, in each process of the sequence, a different individual counterterm of the theory is determined. Thus, by choosing an appropriate sequence, a model can be fully and automatically renormalized with a single \FMS run.

\begin{figure}[t]
\centering
\includegraphics[width=0.99\textwidth]{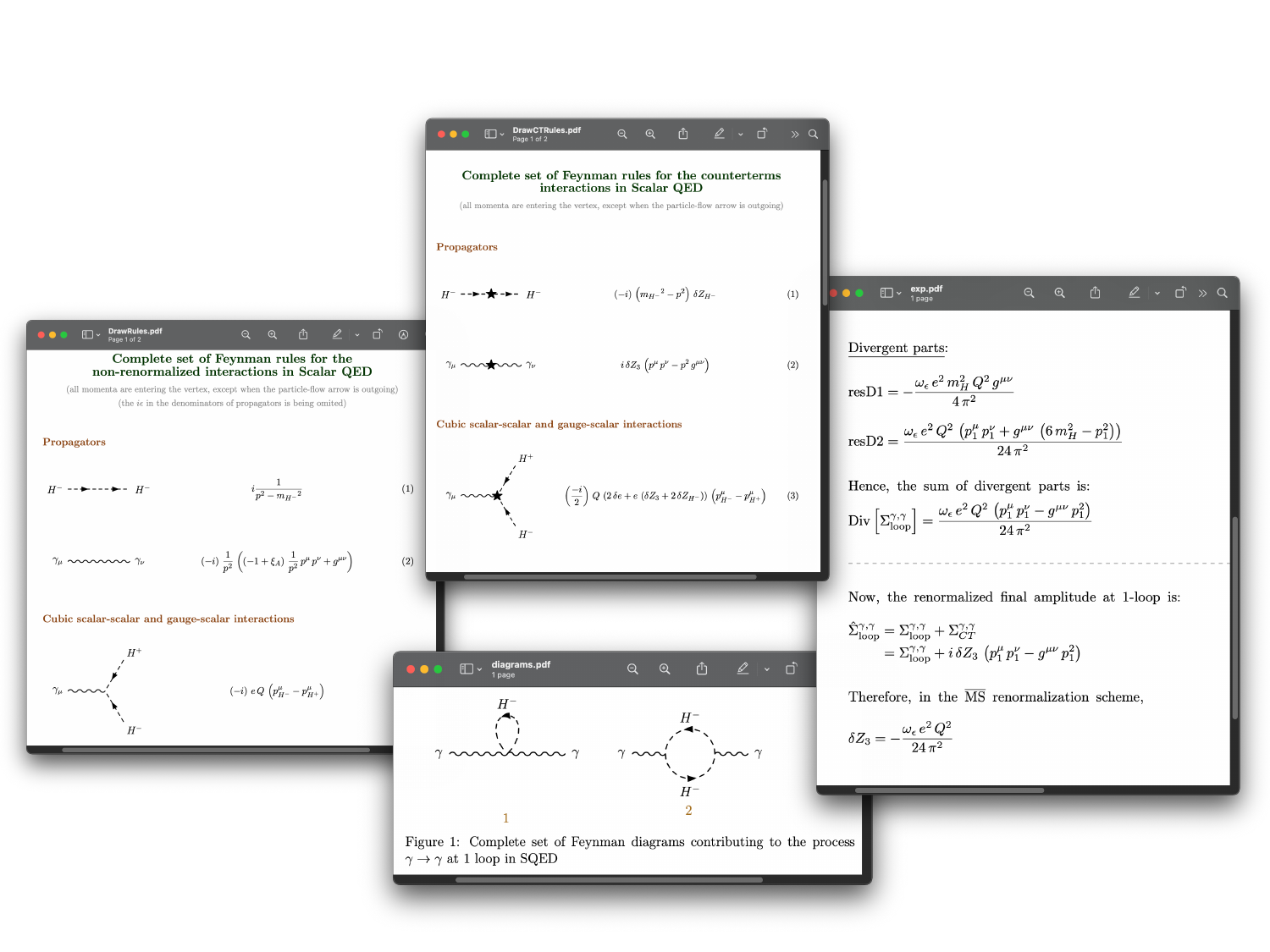}
\vspace{-3mm}
\caption{Set of PDF outputs automatically generated by \FMS in scalar QED: the Feynman rules for the tree-level interactions (left), the Feynman rules for the counterterms (top), the expressions and the counterterms in $\overline{\text{MS}}$ (right), the Feynman diagrams (bottom).}
\label{Chap-FM:fig:SQED-bouquet}
\end{figure}

All of these steps are rendered more intuitive thanks to the visual interface of \FM. Instead of simply generating the necessary elements for \FCS to perform the different calculations, \FMS also automatically prints three different types of PDF files. First, it prints the Feynman rules, both for the tree-level interactions and the counterterms; for each rule, the Feynman diagram (drawn with \t{feynmf} \cite{Ohl:1995kr}) and the Feynman rule are presented side-by-side.
Second, \FMS draws all the Feynman diagrams contributing to the process at stake. Finally, it prints the results of the \FCS calculations, which includes not only the amplitudes, but also counterterms. An illutration of these PDF outputs can be seen in fig. \ref{Chap-FM:fig:SQED-bouquet}, for scalar QED.
The files are all written in \LaTeX, which allows an immediate inclusion of their content in a paper.

\begin{figure}[h!]
\centering
\includegraphics[width=0.95\linewidth]{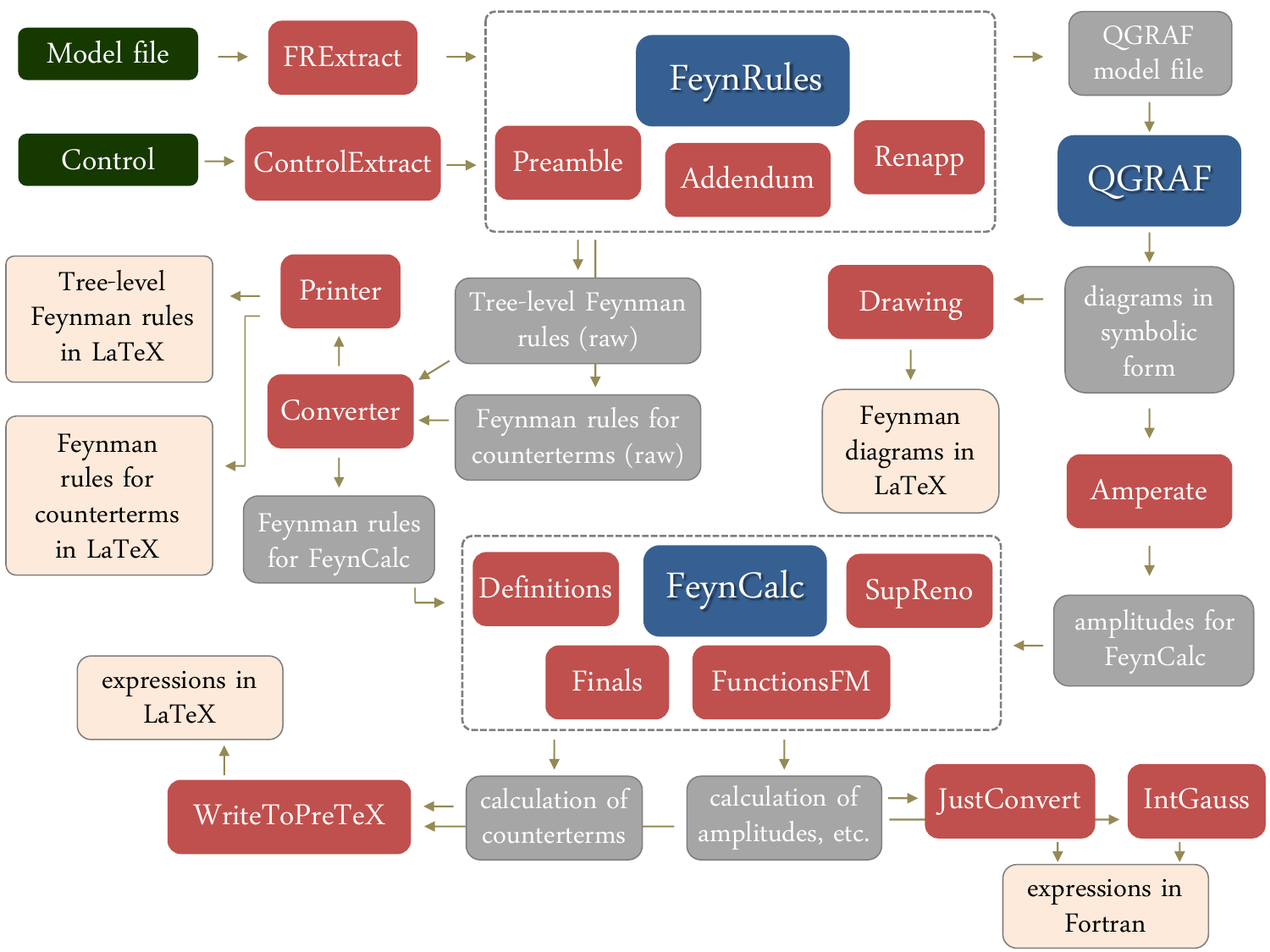}
\caption{Articulation of the different components of  \textsc{FeynMaster}. All of them are governed by the \t{\textsc{Python}} subroutine \t{General}, which is not shown. See text for details.}
\label{Chap-FM:fig:FM-scheme}
\end{figure}

We show in fig. \ref{Chap-FM:fig:FM-scheme} the articulation of the different components that constitute \FMS as a whole.
The different colors represent different types of components: in green, the two files that trigger the total run; in blue, the three software that \textsc{FeynMaster} is based on; in gray, intermediate files that chain the whole process; in beige, the main outputs; finally, in red, the subroutines composing \textsc{FeynMaster}. These subroutines are characterized in table \ref{Chap-FM:tab:subrou}.
\begin{table}[!h]%
\begin{normalsize}
\normalsize
\begin{center}
\begin{tabular}
{@{\hspace{3mm}}
>{\raggedright\arraybackslash}p{2.7cm}
>{\raggedright\arraybackslash}p{2.5cm}
>{\raggedright\arraybackslash}p{8.8cm}
@{\hspace{3mm}}}
\hlinewd{1.1pt}
Subroutine & Language & Purpose \\
\hline
\\[-3.5mm]
\t{Addendum} & \t{\textsc{Mathematica}} &  generate the \QGS model and rules for \textsc{FeynCalc} \\[3mm]
\t{Amperate} & \t{\textsc{Python}} &  convert the \QGS output into amplitudes \\[3mm]
\t{ControlExtract} & \t{\textsc{Python}} & extract information from the \t{Control} file \\[3mm]
\t{Converter} & \t{\textsc{Python}} & convert notations \\[3mm]
\t{Definitions} & \t{\textsc{Mathematica}} & make definitions and replacement rules for \textsc{FeynCalc}  \\[3mm]
\t{Drawing} & \t{\textsc{Python}} & draw Feynman diagrams \\[3mm]
\t{Finals} & \t{\textsc{Mathematica}} &  handle the final expressions of \textsc{FeynCalc} \\[3mm]
\t{FRExtract} & \t{\textsc{Python}} & extract information from the model file \\[3mm]
\t{FunctionsFM} & \t{\textsc{Mathematica}} & provide extra functions for  \textsc{FeynRules} and  \textsc{FeynCalc}\\[3mm]
\t{General} & \t{\textsc{Python}} & coordinate all the remaining subroutines\\[3mm]
\t{IntGauss} & \t{\textsc{Fortran}} & perform gaussian integrations \\[3mm]
\t{JustConvert} & \t{\textsc{Python}} & support \t{Converter}   \\[3mm]
\t{Preamble} & \t{\textsc{Mathematica}} & make definitions and lists for \textsc{FeynRules} \\[3mm]
\t{Printer} & \t{\textsc{Python}} & draw and print the Feynman rules \\[3mm]
\t{Renapp} & \t{\textsc{Mathematica}} & generate the counterterms of the model \\[3mm]
\t{SupReno} & \t{\textsc{Mathematica}} &  perform $\overline{\text{MS}}$ renormalization \\[3mm]
\t{WriteToPreTeX} & \t{\textsc{Mathematica}} & generate and print the final expressions \\
\hlinewd{1.1pt}
\end{tabular}
\end{center}
\vspace{-4mm}
\end{normalsize}
\vs{-1mm}
\caption{Subroutines composing \textsc{FeynMaster 2}.}
\label{Chap-FM:tab:subrou}
\vspace{-2mm}
\end{table}

Finally, as already suggested in the Introduction, one of the most striking advantages of \FMS is its flexibility. More than just producing the outputs we just mentioned, \FMS allows the user to manipulate intermediate and final results, due to the user-friendly features of \textsc{FeynCalc} and \t{\textsc{Mathematica}}. This flexible character will be described in more detail in appendices \ref{App-FM-Manual} and \ref{App-FM-C2HDM}.

\section{Comparison with other software}
\label{Chap-FM:sec:Compa}

We now provide a brief explanation of how \textsc{FeynMaster} relates to other frameworks which also study particle physics processes at the one-loop level. Among those more closely related to \textsc{FeynMaster}, we select four: \textsc{GRACE}~\cite{Belanger:2003sd}, \textsc{FormCalc}~\cite{Hahn:1998yk}, \textsc{GoSam}~\cite{Cullen:2011ac,Cullen:2014yla} 
and \textsc{NLOCT}~\cite{Degrande:2014vpa} (for reviews comparing different software, see refs. \cite{Harlander:1998dq,Denner:2019vbn}).

Before entering the comparison, let us consider a preliminary aspect. As we just noted, one of the major advantages of \FMS is its flexibility. However, software based on other frameworks (such as \t{\textsc{FORM}}~\cite{Kuipers:2012rf}) can be much faster than \textsc{FeynMaster}. Actually, if one is interested in obtaining the numerical result for a one-loop process with hundreds of diagrams in a well known model, other software can be preferable. If, however, one wants to use a slightly different model, or ascertain some properties of the analytical results, or make sure to have no problems with interfaces between different software, or simply obtain a printed list of the Feynman rules, then \textsc{FeynMaster} is an interesting alternative. In this sense, the purpose of \textsc{FeynMaster} is not compete with the five aforementioned programs, but rather to enrich the flexibility of \textsc{FeynCalc} by creating a unified platform around it that performs a vast list of tasks.


Let us now consider the referred programs. \textsc{GRACE} is a software that aims at numerically computing cross sections. It generates Feynman rules from a model, generates and draws Feynman diagrams, and writes the differential cross section in a \t{\textsc{Fortran}} file. Unlike \FM, it is not restricted to 2 particles in the final state; it employs other software for the numerical integration and event generation; it also addresses renormalization, and calculates counterterms in the on-shell renormalization scheme. However, it cannot calculate one-loop processes that do not exist at tree-level, such as $h\to\gamma\gamma$. And although it performs algebraic calculations, it does not seem to allow a large margin for the user to manipulate analytical expressions.

\textsc{FormCalc} is a program that calculates tree-level and one-loop Feynman diagrams based on both \t{\textsc{FORM}} and \t{\textsc{Mathematica}}. It combines the speed of the former and the vast amount of instructions of the latter. It gets the amplitudes from \textsc{FeynArts}~\cite{Hahn:2000kx} and generates an output to numerically evaluate the one-loop integrals with \textsc{LoopTools}. Although \textsc{FormCalc} is excellent for fast results, it lacks the flexibility to perform simple manipulations involving Dirac algebra or Lorentz contractions.

\textsc{GoSam} is designed to an automated calculation of one-loop amplitudes for multi-particle processes. Like \textsc{FeynMaster}, it uses \textsc{QGRAF} to generate the Feynman diagrams in symbolic form, and also uses both \t{\textsc{Python}} and \t{feynmf} to draw the diagrams. Like \textsc{GRACE}, \textsc{GoSam} produces a \t{\textsc{Fortran}} file required to perform the evaluation of one-loop matrix elements. But unlike \textsc{FeynMaster}, it uses helicity amplitudes formalism, and makes renormalization for QCD corrections only. And although it generates one-loop amplitudes, it seems to be mostly focused on numerical evaluation, and not on allowing a manipulation of intermediate or final expressions.


\textsc{NLOCT} determines the UV counterterms for any Lagrangean in an automated way. It requires an interaction with both \textsc{FeynRules} and \textsc{FeynArts}, and works only in the Feynman gauge. It not only obtains the Feynman rules for the counterterms, but also computes the analytical expressions for these counterterms in the on-shell renormalization scheme.

In conclusion, although some of software considered here are more powerful than \textsc{FeynMaster} in some specific tasks (e.g. \textsc{FormCalc} for fast results), \textsc{FeynMaster} exploits the strength of \textsc{FeynCalc} to allow a practical manipulation of the analytical expressions. In addition, \textsc{FeynMaster} condenses in itself a multiplicity of tasks, so that there is no need to convert (and hence there are no obstacles in converting) between different software. All of this renders \textsc{FeynMaster} a truly flexible framework.

%% file: Chapters/Chapter_Selection.tex
\chapter{Selection of the true vev}
\label{Chap-Selec}

\vs{-5mm}

In this chapter, we present an introduction to the selection of the true vev, which is closely related to a discussion on tadpole schemes.
Tadpole schemes have been discussed in recent years in the context of the renormalization of models with extended scalar sectors~\cite{Krause:2016oke,Krause:2016xku,Kanemura:2017wtm,Krause:2017mal,Altenkamp:2017ldc,Krause:2018wmo,Dudenas:2018wlr,Denner:2018opp,Krause:2019oar,Dao:2019nxi,Altenkamp:2019wht,Denner:2016etu,Denner:2019xti,Denner:2017vms,Krause:2019qwe} (a recent work~\cite{Dudenas:2020ggt} is devoted to this discussion, focusing especially on gauge dependence).
And indeed, they constitute an important aspect in the treatment of any gauge theory with SSB. However, we believe that a clear description of this topic is still missing in the literature. Actually, from our point of view, the language normally used to address tadpole schemes tends to be misleading. We thus found it relevant to clarify this discussion.

We do so in such a way that we do not restrict the discussion to a particular model. The goal, in fact, is to introduce the topic of tadpole schemes in a general gauge theory with SSB. Yet, we shall give special emphasis to the SM, given its universally known character and its preponderance in modern particle physics. In addition, the SM provides an ideal framework for an introduction to the selection of the true vev, since it is simple enough to allow a clear description of the problem, but simultaneously complex enough to reveal several important details.

%
We extend the discussion to models with two Higgs doublets, as an example of an extended scalar sector.
As shall be seen, the introduction of a second scalar doublet slightly  increases the complexity of the problem, since there will be more than one vev. As a consequence, some features not present in the case of the SM will show up, and thus must be conveniently clarified. The discussion can then be easily generalized to a model with a more elaborated scalar sector, since the inclusion of more vevs does not constitute a significative addition of complexity.



The chapter is organized as follows: in section \ref{Chap-Selec:sec:SM-tree}, we perform a tree-level description of the SM, showing how the true vev is selected in this simple context; then, in section \ref{Chap-Selec:sec:SMutol}, we consider the SM up to one-loop level, and clarify several aspects related to the selection of the true vev; finally, in section \ref{Chap-Selec:sec:2HDM}, we discuss the case of the 2HDM.

\section{Standard Model: tree-level}
\label{Chap-Selec:sec:SM-tree}

We essentially focus one sector of the SM (the potential, $V$); we also resort to other sectors, which we cannot introduce in detail.
Reviews of the SM can be found in refs. \cite{Denner:1991kt,Romao:2012pq,Denner:2019vbn}. We closely follow the notation of the latter; this reference (\!\!\cite{Romao:2012pq}) introduced several parameters---all dubbed $\eta$, with different indices---to allow for different sign conventions used in the literature. Throughout this thesis, and unless stated otherwise, we use the sign convention defined by taking all $\eta$'s positive.%
\fn{Refs. \cite{Denner:1991kt,Denner:2019vbn} choose $\eta = \eta_{\theta} = -1$, and the remaining $\eta$'s positive. In general, we follow the conventions of ref. \cite{Romao:2012pq} (although we use $g_1$ and $g_2$ instead of $g'$ and $g$, respectively).
Note that the $\lambda$ paremeter of ref. \cite{Romao:2012pq} is four times smaller than that of refs. \cite{Denner:1991kt,Denner:2019vbn}, and the unitary transformations that diagonalize the fermions in ref. \cite{Romao:2012pq} are the dagger of those in refs. \cite{Denner:1991kt,Denner:2019vbn}.}

Let us then focus on potential of the SM. We write it as:
\be
V = - \mu^2 \Phi^{\dagger} \Phi + \lambda \left(\Phi^{\dagger} \Phi\right)^2,
\label{Chap-Selec:eq:SMpot}
\ee
where $\mu$ and $\lambda$ real positive constants, and $\Phi$ is the Higgs doublet, which can be parameterized as:%
\fn{The parameterization $\bar{v} + h$ is based on section 3.1.6 of ref. \cite{Denner:2019vbn}. We took $\bar{v}$ to be real, positive and in the lower component of the doublet without loss of generality.}
\be
\Phi(x)
=
\begin{pmatrix} G^+(x) \\ \dfrac{1}{\sqrt{2}}[\bar{v} + h(x) + i G_0(x)] \end{pmatrix}.
\label{Chap-Selec:eq:doublet}
\ee
Here, $h$ is the Higgs field, while $G_0$ and $G^{+}$ are the neutral and charged would-be Goldstone fields, respectively. Besides, $\bar{v}$ is a real and positive parameter, which turns out to be non-arbitrary. In fact, perturbation theory
\textit{fixes} its value.
%
The reason is that fields used in perturbation theory must represent small excitations around the state of minimum energy of the theory.%
%
\fn{For details, cf. e.g. section 28.1 of ref. \cite{Schwartz:2013pla} or section 4.1.1 of ref. \cite{Bohm:2001yx}.}
In other words, the state around which the small excitations of fields take place---i.e. the state where they vanish, let us identify it with $|\Omega_m\rangle$---must correspond to the state of minimum energy. As a consequence, $\bar{v}$ must be precisely given by the value that ensures that, at the state $|\Omega_m\rangle$ (i.e. for vanishing fields), the theory has its minimum energy (in short, its minimum).

The state $|\Omega_m\rangle$ is called the \textit{vacuum}, and the Green's functions (GFs) used in perturbation theory correspond to the \textit{vacuum expectation value} (vev) of a time-ordered products of fields. Generically,
\be
G^{\phi_1...\phi_n}(x_1, ..., x_n)
=
\langle \Omega_m | T\{ \phi_1(x_1) \, . . . \, \phi_n(x_n) \}| \Omega_m \rangle,
\label{Chap-Selec:eq:GF-vev}
\ee
where $T$ represents the time-ordering operator; in what follows, we use the short notation $\langle \phi_1(x_1) \, . . .$  $\phi_n(x_n) \rangle$ instead of $\langle \Omega_m | T\{ \phi_1(x_1) ... \phi_n(x_n)\}| \Omega_m \rangle$, for simplicity.
In this way, all GFs correspond to vevs. However, we usually associate the name `vev' to non-null 1-point functions (i.e. non-null GFs with just one field or, more generally, with just one multiplet). As we now show, the Higgs doublet $\Phi$ configures one such vev; and since it is the only multiplet of the theory with this property, we associate the term `vev' to the vacuum expectation value of $\Phi$, $\langle \Phi(x) \rangle$.

Going back to $\bar{v}$: as we saw, to determine its value we must find the minimum of the theory, which in turn is given by the minimum of the potential. Since we are considering a tree-level description, the minimum of the potential can be determined classically, by taking derivatives of the potential in order to the fields. Or, what is even simpler, we can take derivatives in order to the doublet $\Phi$. Then, if we recall that $\mu$ and $\lambda$ are both positive, we realize that the minimum of the potential is verified if and only if $\bar{v}$ obeys:
\be
\bar{v} = v,
\label{Chap-Selec:eq:pert}
\ee
where $v$ is defined as:
\be
v := \sqrt{\dfrac{\mu^2}{\lambda}}.
\label{Chap-Selec:eq:THE-TRUE}
\ee
The quantity $v$ is \textit{the true vev}: it is the value of $\bar{v}$ that corresponds to the minimum of the potential.%
\fn{The notion `true vev', as well as equivalent notions such as `proper vev', `correct value', `true minimum', are common in the literature \cite{Fleischer:1980ub,Denner:2019vbn,Denner:1991kt,Krause:2016oke,Peskin:1995ev,Krause:2017mal,Barroso:2012mj,Schwartz:2013pla,Cheng:1985bj}.}
What we just found is that perturbation theory requires $\bar{v}$ (i.e. the quantity $\bar{v}$ that we introduced in eq. \ref{Chap-Selec:eq:doublet}) to be equal to the true vev $v$, defined in eq. \ref{Chap-Selec:eq:THE-TRUE}.
To better illustrate this point, suppose that we did not choose eq. \ref{Chap-Selec:eq:pert}, but rather $\bar{v} = v' \neq v$. 
In that case, the Lagrangian $\mathcal{L}$ would have a linear term in $h$,
\be
\mathcal{L} \, \ni \, -V \, \ni \, t \, h,
\label{Chap-Selec:eq:my-min}
\ee
with $t$ defined by
\be
t := \bar{v} \left(\mu^2 - \lambda \bar{v}^2\right),
\label{Chap-Selec:eq:tad-tree-Lag}
\ee 
as can be verified by inserting eq. \ref{Chap-Selec:eq:doublet} in eq. \ref{Chap-Selec:eq:SMpot} and expanding the terms. Since we would be choosing  $\bar{v} \neq v$, then $t$ would not vanish, which would imply that $h$ would have a non-null vev, $\langle h(x) \rangle = t \neq 0$.
But that would be inconsistent: the fields used in perturbation theory (like $h$) must vanish when evaluated at $|\Omega_m\rangle$, which in turn must describe the state of minimum energy. If $t\neq0$, nothing of this holds.
Note that eq. \ref{Chap-Selec:eq:my-min} corresponds to the Feynman diagram and rule%
\fn{In this diagram, the cross means that there is no momentum corresponding to that point.}
\be
\begin{minipage}[h]{.40\textwidth}
\vspace{5mm}
\begin{picture}(0,42)
\begin{fmffile}{Chap-Selec-1616} 
\begin{fmfgraph*}(50,70) 
\fmfset{arrow_len}{3mm} 
\fmfset{arrow_ang}{20} 
\fmfleft{nJ1} 
\fmfright{nJ2}
\fmflabel{$h$}{nJ1}
\fmflabel{$\vspace{-0.01mm}\hspace{-4.5mm}\times$}{nJ2}
\fmf{dashes,tension=1}{nJ1,nJ2} 
\end{fmfgraph*} 
\end{fmffile} 
\end{picture}
\vspace{-5mm}
\end{minipage}
\hspace{-35mm}
i \, t.
\label{Chap-Selec:eq:tree-tad}
\ee
The presence of such structure always indicates that we are selecting the false vev, i.e. choosing a value for $\bar{v}$ 
which does not lead to the state of minimum energy---in which case perturbation theory is not consistent.
We dub both the term on the l.h.s. of eq. \ref{Chap-Selec:eq:tad-tree-Lag} and the Feynman diagram in eq. \ref{Chap-Selec:eq:tree-tad} a \textit{tree-level tadpole}. The tree-level tadpole always corresponds to the linear term in the scalar field.
And the conclusion is this: if the theory generates a tree-level tadpole, we are not selecting the true vev, i.e. we are not choosing $\bar{v}$ to be equal to $v$. Conversely, if we do make this choice (eq. \ref{Chap-Selec:eq:pert}), the tree-level tadpole vanishes:
\be
\bar{v} = v 
\ \
\Longrightarrow
\ \ 
t = 0.
\ee
As we want to have a consistent perturbation theory, we select the true vev;
this implies, as predicted, that the vacuum expectation value of $\Phi$ is non-vanishing:
\be
\langle \Phi(x) \rangle
=
\begin{pmatrix} 0 \\ \frac{v}{\sqrt{2}} \end{pmatrix}.
\label{Chap-Selec:eq:doublet-vev}
\ee

The selection of the true vev is straightforward at tree-level, as we just saw. But this will not be the case at higher orders.
To properly introduce that scenario, we consider some aspects that shall prove useful.
We start by noting that, with eq. \ref{Chap-Selec:eq:pert}, eq. \ref{Chap-Selec:eq:doublet} becomes:
\be
\Phi(x)
=
\begin{pmatrix} G^+(x) \\ \dfrac{1}{\sqrt{2}}[v + h(x) + i G_0(x)] \end{pmatrix}.
\label{Chap-Selec:eq:doublet-2}
\ee
Then, the masses of the particles become proportional to $v$ \cite{Denner:2019vbn}:%
\begin{gather}
m_{\mathrm{W}}=\dfrac{1}{2} g_2 v,
\quad
m_{\mathrm{Z}} = \dfrac{1}{2} \sqrt{g_1^2 + g_2^2} \, v,
\quad
m_h^2 = 2 \mu^2 = 2 \lambda v^2,
\nonumber \\
m_{f,t,i} = \dfrac{v}{\sqrt{2}} \sum_{j,k}
\Big(U_{t_{\mathrm{L}}}^{\dagger}\Big)_{ij} \big(Y_t\big)_{jk} \Big(U_{t_{\mathrm{R}}}\Big)_{ki}.
\label{Chap-Selec:eq:15}
\end{gather}
Here, $m_{\mathrm{W}}$, $m_{\mathrm{Z}}$ and $m_h$ are the masses of the $W$, $Z$ and Higgs bosons, respectively; $g_1$ and $g_2$  are the $\mathrm{U(1)_Y}$ and $\mathrm{SU(2)_L}$ couplings, respectively;
$m_{f,t,i}$ is the mass of the fermion $i$ of type $t$ (where $t=\{d,u,l\}$, for the down-type quarks, up-type quarks and leptons, respectively), $Y_t$ represents the Yukawa matrix for fermions of type $t$ and $U_{t_{\mathrm{L}}}^{\dagger}$ and $U_{t_{\mathrm{R}}}$ are the unitary matrices that lead the fermions of type $t$ to the mass basis (for details, cf. ref. \cite{Denner:2019vbn}).

We also introduce the weak mixing angle $\theta_{\text{w}}$, such that
\be
c_{\text{w}} = \dfrac{m_{\mathrm{W}}}{m_{\mathrm{Z}}},
\label{Chap-Selec:eq:cw}
\ee
and
\be
s_{\text{w}} = \dfrac{e}{g_2},
\label{Chap-Selec:eq:sw}
\ee
where we used the short notation $c_{\text{w}} = \cos(\theta_{\text{w}})$, $s_{\text{w}} = \sin(\theta_{\text{w}})$, and where $e$ is the electric charge.
Besides, we can use the relations in eq. \ref{Chap-Selec:eq:15}, as well as
\be
e = \dfrac{g_1 g_2}{\sqrt{g_1^2+g_2^2}}, \qquad V = U_{u_{\mathrm{L}}}^{\dagger} U_{d_{\mathrm{L}}},
\label{Chap-Selec:eq:extra}
\ee
where $V$ is the CKM matrix, to replace the original set of independent parameters of the SM,
\be
g_1, g_2, \lambda, \mu^2, Y_d, Y_u, Y_l,
\label{Chap-Selec:eq:setoriginal}
\ee
by the (also independent among themselves, but related to the former) parameters \cite{Denner:1991kt}:
\be
e, m_{\mathrm{W}}, m_{\mathrm{Z}}, m_h, m_{f}, V.
\label{Chap-Selec:eq:setsecond}
\ee
%
%
Whichever the set chosen, the independent parameters must be experimentally determined. The set of eq. \ref{Chap-Selec:eq:setsecond} is thus more convenient, since all the parameters contained in it have an intuitive physical meaning (i.e. they can be easily related to experimental quantities) \cite{Denner:1991kt,Bohm:2001yx}.
Finally, using eqs. \ref{Chap-Selec:eq:THE-TRUE}, \ref{Chap-Selec:eq:15} and \ref{Chap-Selec:eq:sw}, we can write $\mu^2$ and $\lambda$ as:
\be
\mu^2 = \dfrac{m_h^2}{2},
\quad
\lambda = \dfrac{m_h^2 \, e^2}{8 \, s_{\text{w}}^2 \, m_{\mathrm{W}}^2}.
\label{Chap-Selec:eq:lambda1}
\ee
If, however, we decided to ignore the fact that $t=0$, we could use eq. \ref{Chap-Selec:eq:tad-tree-Lag} (instead of \ref{Chap-Selec:eq:pert}) to obtain:
\be
\mu^2 = \dfrac{m_h^2}{2} + \dfrac{3 \, t \, e}{4 \, s_{\text{w}} \, m_{\mathrm{W}}},
\qquad
\lambda = \dfrac{m_h^2 \, e^2}{8 \, s_{\text{w}}^2 \, m_{\mathrm{W}}^2} + \dfrac{t \, e^3}{16 \, s_{\text{w}}^3 \, m_{\mathrm{W}}^3},
\label{Chap-Selec:eq:lambda2}
\ee
which, at this point, is just a more complicated (and rather useless) way of writing eq. \ref{Chap-Selec:eq:lambda1}, since $t=0$. Later on, though, it shall prove useful.

\section{Standard Model: up to one-loop level}
\label{Chap-Selec:sec:SMutol}

%

%


\subsection{Finite $S$-matrix elements}

When we go up to one-loop level, the parameters of the theory no longer have an intuitive physical meaning.
As a matter of fact, they have no meaning at all: they are UV divergent.
In what follows, these parameters are dubbed \textit{bare} parameters and are identified with the subscript ``$(0)$''.
In non-abelian gauge theories with SSB like the SM, the divergences end up cancelling in physical predictions of $S$-matrix elements 
\cite{tHooft:1971akt,tHooft:1971qjg,tHooft:1972tcz,Lee:1972fj,Lee:1974zg,Lee:1973fn,Lee:1973rb}, so that the latter become meaningful.
Such cancellation can be carried out through different approaches (cf. e.g. ref. \cite{Denner:1991kt} and references therein); in this thesis, we follow the \textit{counterterm approach}.
Here, in order to obtain finite $S$-matrix elements, we \textit{renormalize} the bare parameters.%
\fn{An excellent introduction to renormalization can be found in ref. \cite{Schwartz:2013pla}.}
This involves equating them with the sum of \textit{renormalized} (i.e. finite) quantities and (infinite) \textit{counterterms}; taking the Higgs boson mass as an example, we say
\be
m_{h(0)}^2 = m_h^2 + \delta m_h^2,
\label{Chap-Selec:eq:mHexample}
\ee
where the first term on the r.h.s represents the renormalized parameter, and the second represents the counterterm. Then, the renormalization is completed by fixing the counterterms.%
\fn{\label{Chap-Selec:note:LSZ}In this chapter, we do not aim at fixing the counterterms, as we do not intend to renormalize the theory. Several examples of counterterm-fixing can be found in chapter \ref{Chap-Reno}, where we renormalize the C2HDM. 
Finally, note that the attainment of finite $S$-matrix elements also requires the renormalization of LSZ factors, which we discuss in appendix \ref{App-LSZ}.}

\subsection{Independent parameters}
\label{Chap-Selec:sec:indep}

The Higgs boson mass used in eq. \ref{Chap-Selec:eq:mHexample} belongs to the set of parameters in eq. \ref{Chap-Selec:eq:setsecond}. To ensure that all $S$-matrix elements are finite, only one set of bare independent parameters needs to be renormalized;
we can choose either the bare version of eq. \ref{Chap-Selec:eq:setoriginal},
\be
g_{1(0)}, \, g_{2(0)}, \, \lambda_{(0)}, \, \mu_{(0)}^2, \, Y_{d(0)}, \, Y_{u(0)}, \, Y_{l(0)},
\label{Chap-Selec:eq:setoriginal0}
\ee
or the bare version of eq. \ref{Chap-Selec:eq:setsecond},
\be
e_{(0)}, \, m_{\mathrm{W}(0)}, \, m_{\mathrm{Z}(0)}, \, m_{h(0)}, \, m_{f(0)}, \, V_{(0)}.
\label{Chap-Selec:eq:setsecond0}
\ee
Given their physical relevance, we choose the latter.\fn{Note, however, that the renormalized parameters like $m_h$ will only be physical (i.e. they will only correspond to observables) if on-shell subtraction conditions are used; more details can be found in section \ref{Chap-Selec:sec:compar} below.} This means that the parameters therein contained must be renormalized (in particular, they must be split into renormalized parameter and counterterm, just like in eq. \ref{Chap-Selec:eq:mHexample}), while those in eq. \ref{Chap-Selec:eq:setoriginal0} need not be subject to that process.
Indeed, if we choose the set of parameters in eq. \ref{Chap-Selec:eq:setsecond0} as independent, then the parameters in eq. \ref{Chap-Selec:eq:setoriginal0} are \textit{de}pendent---that is, they can always be written in terms of the independent parameters of eq. \ref{Chap-Selec:eq:setsecond0}. And since the latter will be renormalized, the ones in eq. \ref{Chap-Selec:eq:setoriginal0} do not have to be submitted to this process. Yet, although one does not have to renormalize dependent parameters, it may be convenient to do so, as we now clarify.

\subsection{Renormalizing for convenience}
\label{Chap-Selec:sec:conv}

Let us consider the cosine of the weak angle. Given eq. \ref{Chap-Selec:eq:cw}, and since we are taking eq. \ref{Chap-Selec:eq:setsecond} as the set of independent parameters, it follows that $c_{\text{w}}$ is a dependent parameter.
In the context of the up-to-one-loop theory, eq. \ref{Chap-Selec:eq:cw} is written in terms of bare quantities,
\be
c_{{\text{w}}(0)} = \dfrac{m_{\mathrm{W}(0)}}{m_{\mathrm{Z}(0)}}.
\label{Chap-Selec:eq:cw0}
\ee
Now, as a dependent parameter, $c_{{\text{w}}(0)}$ does not need to be renormalized. To say it clearly, $S$-matrix elements will not be divergent if we do not renormalize $c_{{\text{w}}(0)}$. 
In that case, though, we need to replace it everywhere by $m_{\mathrm{W}(0)}/m_{\mathrm{Z}(0)}$, and then renormalize $m_{\mathrm{W}(0)}$ and $m_{\mathrm{Z}(0)}$ (if we just kept on using $c_{{\text{w}}(0)}$, we would run into problems, since $c_{{\text{w}}(0)}$ is a divergent quantity).
In this method, $c_{{\text{w}}(0)}$ vanishes from the theory, since it was replaced everywhere.
Consequently, there is no such thing as a renormalized version of eq. \ref{Chap-Selec:eq:cw0}.

But this is one reason why, although it is not necessary, it is \textit{convenient} to renormalize $c_{{\text{w}}(0)}$: one may find it convenient to have eq. \ref{Chap-Selec:eq:cw0} valid for the renormalized parameters as well (and not only for the bare parameters). 
Then, we introduce a new parameter, $c_{\text{w}}$, such that
\be
c_{\text{w}} = \dfrac{m_{\mathrm{W}}}{m_{\mathrm{Z}}},
\label{Chap-Selec:eq:cwnovo}
\ee
where $m_{\mathrm{W}}$ and $m_{\mathrm{Z}}$ are the renormalized masses of the $W$ and the $Z$ bosons, respectively.
To see how $c_{{\text{w}}(0)}$ and $c_{\text{w}}$ are related, we just expand the r.h.s. of eq. \ref{Chap-Selec:eq:cw0}:
\be
\begin{split}
c_{{\text{w}}(0)} &= \dfrac{m_{\mathrm{W}(0)}}{m_{\mathrm{Z}(0)}} = \dfrac{\sqrt{m_{\mathrm{W}}^2 + \delta m_{\mathrm{W}}^2}}{\sqrt{m_{\mathrm{Z}}^2 + \delta m_{\mathrm{Z}}^2}} = \left( m_{\mathrm{W}} + \dfrac{\delta m_{\mathrm{W}}^2}{2 m_{\mathrm{W}}}\right) \left( \dfrac{1}{m_{\mathrm{Z}}} - \dfrac{\delta m_{\mathrm{Z}}^2}{2 m_{\mathrm{Z}}^3}\right) + \mathcal{O}(\delta^2) \\
&=  \dfrac{m_{\mathrm{W}}}{m_{\mathrm{Z}}} + \dfrac{\delta m_{\mathrm{W}}^2}{2 m_{\mathrm{W}} m_{\mathrm{Z}}} - \dfrac{m_{\mathrm{W}} \, \delta m_{\mathrm{Z}}^2}{2 m_{\mathrm{Z}}^3} + \mathcal{O}(\delta^2).
\label{Chap-Selec:eq:cw0yet}
\end{split}
\ee
So, using eq. \ref{Chap-Selec:eq:cwnovo}, and neglecting second order terms, we conclude that $c_{{\text{w}}(0)}$ and $c_{\text{w}}$ are related by:
\be
c_{{\text{w}}(0)} = c_{\text{w}} + \dfrac{\delta m_{\mathrm{W}}^2}{2 m_{\mathrm{W}} m_{\mathrm{Z}}} - \dfrac{m_{\mathrm{W}} \, \delta m_{\mathrm{Z}}^2}{2 m_{\mathrm{Z}}^3}.
\label{Chap-Selec:eq:cwexp}
\ee
This allows us to define a counterterm $\delta c_{\text{w}}$, such that
\be
c_{{\text{w}}(0)} = c_{\text{w}} + \delta c_{\text{w}},
\label{Chap-Selec:eq:cwren}
\ee
with
\be
\delta c_{\text{w}} = \dfrac{\delta m_{\mathrm{W}}^2}{2 m_{\mathrm{W}} m_{\mathrm{Z}}} - \dfrac{m_{\mathrm{W}} \, \delta m_{\mathrm{Z}}^2}{2 m_{\mathrm{Z}}^3} .
\label{Chap-Selec:eq:cwCT}
\ee
These equations illustrate another motivation for the renormalization of $c_{{\text{w}}(0)}$: if we replace it using eq. \ref{Chap-Selec:eq:cw0} and renormalize $m_{\mathrm{W}(0)}$ and $m_{\mathrm{Z}(0)}$, we end up having more counterterms than if we renormalize $c_{{\text{w}}(0)}$ (compare eqs. \ref{Chap-Selec:eq:cwexp} and \ref{Chap-Selec:eq:cwren}).
Of course, $\delta c_{\text{w}}$ \textit{is} given by eq. \ref{Chap-Selec:eq:cwCT}, so that we always end up having the same number of counterterms. Nonetheless, the manipulation of expressions can become quite easier if we use the l.h.s. of \ref{Chap-Selec:eq:cwCT} instead of the r.h.s.%
\fn{In this simple case, the difference in number of counterterms in not significative, so that this discussion does not seem very relevant. Yet, the difference can be very significative in more complicated cases.}

But this reveals another important consequence of the renormalization of $c_{\text{w}}$, namely: the counterterm $\delta c_{\text{w}}$ is fixed.%
\fn{The renormalization condition \ref{Chap-Selec:eq:cwCT} follows from eqs. \ref{Chap-Selec:eq:cwren} and \ref{Chap-Selec:eq:cwnovo}. Had we chosen $c_{{\text{w}}(0)} = c_{\text{w}}^{\prime} + \delta c_{\text{w}}^{\prime}$, with $c_{\text{w}}^{\prime} \neq c_{\text{w}}$, then we would have freedom to fix the finite parts of $\delta c_{\text{w}}^{\prime}$ (the divergent parts of $\delta c_{\text{w}}^{\prime}$ and $\delta c_{\text{w}}$ must necessarily be equal, as both $c_{\text{w}}^{\prime}$ and $c_{\text{w}}$ are assumed to be finite). In that case, though, by comparing with eq. \ref{Chap-Selec:eq:cw0}, we would necessarily have $c_{\text{w}}^{\prime} =  c_{\text{w}} + \delta c_{\text{w}} - \delta c_{\text{w}}^{\prime}$, with $c_{\text{w}}$ and $\delta c_{\text{w}}$ given by eqs. \ref{Chap-Selec:eq:cwnovo} and \ref{Chap-Selec:eq:cwCT}, respectively. Hence, the freedom in defining the finite parts of $\delta c_{\text{w}}^{\prime}$ would result in the introduction of yet another variable---the alternative renormalized parameter $c_{\text{w}}^{\prime}$---, corresponding to a fixed (and in general complicated) expression. The choice adopted in eqs. \ref{Chap-Selec:eq:cwren}, \ref{Chap-Selec:eq:cwnovo} and \ref{Chap-Selec:eq:cwCT} is not only simpler, but has the additional advantage that the finite parts of the renormalized parameters are independent of the finite parts of counterterms.}
In fact, if we define eqs. \ref{Chap-Selec:eq:cwren} and \ref{Chap-Selec:eq:cwnovo}, and given eq. \ref{Chap-Selec:eq:cw0}, we have no freedom to choose $\delta c_{\text{w}}$: it is (and must be) given by eq. \ref{Chap-Selec:eq:cwCT}. This is a simple consequence of the fact that $c_{{\text{w}}(0)}$ is a dependent variable; and, as a dependent variable, it is fixed---that is, fixed by eq. \ref{Chap-Selec:eq:cw0}. So, the same way $c_{\text{w}}$ is a dependent, fixed variable, so $\delta c_{\text{w}}$ is a dependent, fixed counterterm.

What we just did with the cosine of the weak angle applies to all dependent parameters. That is, a bare dependent parameter does not need to be renormalized, although it can be renormalized for convenience. If one decides not to renormalize it, one must replace it by the corresponding independent parameters, and then renormalize the latter. Alternatively, one can renormalize the bare dependent parameter itself---introducing a renormalized version for it, as well as a counterterm. Yet, just as the bare parameter is dependent and fixed, so will the renormalized parameter and the counterterm be dependent and fixed.

%
%

\subsection{Finite Green's functions}

What we have been describing so far---renormalization of parameters of the theory---is motivated by the need to generate finite $S$-matrix elements, as we noted. If, in addition, we intend to obtain finite GFs (and not simply finite $S$-matrix elements), the logic we apply to the parameters in eq. \ref{Chap-Selec:eq:setoriginal0} should be extended to fields as well. In that case, fields are also identified as bare fields and are split into renormalized ones through a counterterm; for example,
\be
h_{(0)} (x) = h(x) + \dfrac{1}{2} \delta Z_h \, h(x),
\label{Chap-Selec:eq:mHexample2}
\ee
where $h$ is the renormalized Higgs field and $\delta Z_h$ the counterterm. Therefore, assuming that we want to get finite GFs, we must renormalize both the parameters and the GFs.
%

\subsection{The true vev}

What about the vev?
Just like at tree-level, we are interested in selecting the true vev;
and just like at tree-level, the true vev is the one for which the linear terms in the Higgs field vanish. However, since we are considering the theory up to one-loop level, the true vev is no longer the same as it was at tree-level. 
That is to say, the true vev is no longer the same as it was in the original, bare theory. In fact, the bare vev---i.e. the vev that is the true vev in the bare theory (the vev which we have been identifying with $v$ in section \ref{Chap-Selec:sec:SM-tree}, but now identify with $v_{(0)}$ in the context of the up-to-one-loop theory)---is characterized by the fact that it obeys the bare minimum relation,
\be
v_{(0)} = \sqrt{\dfrac{\mu_{(0)}^2}{\lambda_{(0)}}}.
\label{Chap-Selec:eq:fixed0}
\ee
But since we are now considering the theory up to one-loop order, we can consider a conceptually different vev: the true vev of the up-to-one-loop theory, which we identify in the following with $v$.
The new quantity $v$ corresponds to the true minimum of the \textit{effective} potential. The effective potential, $V_{\mathrm{eff}}$, includes not only the bare terms (which, as we saw, can be split into renormalized terms and counterterms), but also the one-loop terms.
This suggests that, in order to select the true vev at up-to-one-loop order,
we must ensure the vanishing of linear terms in the Higgs field in the effective potential. It is to this vanishing that we now turn.

The first thing to notice is that we want such vanishing to happen \textit{before} we renormalize the theory. In other words, we want to select the true vev before we initiate the process which a) expands bare parameters and fields into renormalized ones and counterterms, and b) fixes the different counterterms. Although the two processes---selection of the true vev and renormalization---are close to each other (as shall be discussed in detail), the former is performed before the latter.
We thus consider the bare doublet, $\Phi_{(0)}$.
We follow the same stategy as in tree-level, by parameterizing the doublet with a generic quantity $\bar{v}$:
\be
\Phi_{(0)}(x) =\begin{pmatrix} G_{(0)}^+(x) \\
\dfrac{1}{\sqrt{2}}[\bar{v} + h_{(0)}(x) + i G_{(0)}(x)] \end{pmatrix}.
\label{Chap-Selec:eq:doublet-base-new}
\ee
From what we have been discussing, the selection of the true vev of the up-to-one-loop theory will happen if and only if $\bar{v} = v$. However, as we are about to see, there is more than one consistent way to obtain this equality. These different ways are called \textit{tadpole schemes}, and we are about to study two of them in great detail. To antecipate a little, one of the tadpole schemes will split $\bar{v}$ in two terms, while the other scheme will equate it immediatly with $v$ in a single term. In both schemes, we will have $\bar{v} = v$, i.e. the true up-to-one-loop vev will be selected. Yet, to account for the different approaches, we keep the generic parameter $\bar{v}$ for now.

Whichever the method used to select the true vev, the linear terms in the Higgs field in the effective potential must vanish.
If we insert $\Phi_{(0)}$ into the bare version of the potential,
we obtain only part of the effective potential: we still lack the one-loop quantities. This can be easily solved by defining coefficients for the one-loop terms. Since we are interested in the linear terms in the Higgs field only, we just have to be concerned with the coefficient of that term, which we identify with $T_h$.
Then, we find that the totality of linear terms in the Higgs field in the effective potential, $V_{\mathrm{eff}}^{h_{(0)}^1}$, are such that:
\be
\mathcal{L}_{\text{eff}} \ni -V_{\text{eff}}^{h_{(0)}^1} = \bar{v} \left(\mu_{(0)}^2 - \lambda_{(0)} \bar{v}^2 \right) h_{(0)} + T_h \, h_{(0)}.
\label{Chap-Selec:eq:VrutolHnew}
\ee
Hence, the linear terms in the Higgs field vanish if and only if we have:
\be
\bar{v}  \left(\mu_{(0)}^2 - \lambda_{(0)} \bar{v} ^2 \right) = - T_h.
\label{Chap-Selec:eq:condition}
\ee
This is the key condition for the selection of the up-to-one-loop true vev:
if the quantity $\bar{v}$ introduced in eq. \ref{Chap-Selec:eq:doublet-base-new} obeys eq. \ref{Chap-Selec:eq:condition}, then it corresponds to the true up-to-one-loop true vev $v$.
As we suggested, eq. \ref{Chap-Selec:eq:condition} can be verified by manipulating $\bar{v}$ in different ways, the so-called tadpole schemes.
We now proceed to characterize separately the two tadpole schemes we alluded to; we will explain in detail how the selection of the true vev is interconnected with the renormalization of the theory.
Before that, however, three final notes.

First, since the r.h.s. of eq. \ref{Chap-Selec:eq:condition} is of one-loop order, so will the l.h.s. also be. This means that, when we renormalize the theory---and thus split the bare field $h_{(0)}$ according to eq. \ref{Chap-Selec:eq:mHexample2}---, the contributions of the counterterm $\delta Z_h$ to eq. \ref{Chap-Selec:eq:VrutolHnew} can be neglected; antecipating that, we can replace the bare field $h_{(0)}$ in eq. \ref{Chap-Selec:eq:VrutolHnew} by the renormalized one, $h$.
In that case, $T_h$ is the one-loop linear term in $h$. That is, $T_h$ is the \textit{one-loop tadpole},%
\fn{Throughout this thesis, the hatched circle represents by default a one-loop non-renormalized generic GF.}
\be
\begin{minipage}[h]{.40\textwidth}
\begin{picture}(0,42)
\begin{fmffile}{Chap-Selec-6} 
\begin{fmfgraph*}(50,70) 
\fmfset{arrow_len}{3mm} 
\fmfset{arrow_ang}{20} 
\fmfleft{nJ1} 
\fmfright{nJ2}
\fmflabel{$h$}{nJ1}
\fmf{dashes,tension=1}{nJ1,nJ2} 
\fmfv{decor.shape=circle,decor.filled=hatched,decor.size=9thick}{nJ2}
\end{fmfgraph*} 
\end{fmffile} 
\end{picture}
\vspace{-10mm}
\end{minipage}
\hs{-35mm}
i \, T_h.
\label{Chap-Selec:eq:loop-tad}
\vs{1mm}
\ee

Second, as we have been suggesting, the selection of the true vev is a procedure which is prior to the procedure of renormalization.  
In particular, counterterms (which result from the identification of bare parameters or fields with renormalized ones and counterterms) only show up after the true vev has been selected. Nonetheless, and as we shall see, it can be useful to think of the l.h.s. of eq. \ref{Chap-Selec:eq:condition} \textit{as a counterterm}, the so-called tadpole counterterm $\delta t$,
\be
\delta t : = \bar{v}  \left(\mu_{(0)}^2 - \lambda_{(0)} \bar{v} ^2 \right).
\label{Chap-Selec:eq:consistent}
\ee
Then, the first term on the r.h.s. of eq. \ref{Chap-Selec:eq:VrutolHnew} is simply $\delta t \, h$, and is represented as:
\be
\begin{minipage}[h]{.40\textwidth}
\begin{picture}(0,42)
\begin{fmffile}{Chap-Selec-9} 
\begin{fmfgraph*}(50,70) 
\fmfset{arrow_len}{3mm} 
\fmfset{arrow_ang}{20} 
\fmfleft{nJ1} 
\fmfright{nJ2}
\fmflabel{$h$}{nJ1}
\fmf{dashes,tension=1}{nJ1,nJ2} 
\fmfv{decor.shape=pentagram,decor.filled=full,decor.size=6thick}{nJ2}
\end{fmfgraph*} 
\end{fmffile} 
\end{picture}
\vs{-10mm}
\end{minipage}
\hspace{-35mm}
i \, \delta t,
\label{Chap-Selec:eq:tad-counter}
\ee
in which case eq. \ref{Chap-Selec:eq:condition} can be rewritten as
\be
\delta t = - T_h,
\label{Chap-Selec:eq:allagree}
\ee
which in turn can be diagrammatically represented as:
\be
\begin{minipage}[h]{.40\textwidth}
\begin{picture}(0,42)
\begin{fmffile}{Chap-Selec-9b} 
\begin{fmfgraph*}(50,70) 
\fmfset{arrow_len}{3mm} 
\fmfset{arrow_ang}{20} 
\fmfleft{nJ1} 
\fmfright{nJ2}
\fmflabel{$h$}{nJ1}
\fmf{dashes,tension=1}{nJ1,nJ2} 
\fmfv{decor.shape=pentagram,decor.filled=full,decor.size=6thick}{nJ2}
\end{fmfgraph*} 
\end{fmffile} 
\end{picture}
\vspace{-10mm}
\end{minipage}
\hspace{-38mm}
= \ \ -
\hs{5mm}
\begin{minipage}[h]{.40\textwidth}
\begin{picture}(0,42)
\begin{fmffile}{Chap-Selec-6b} 
\begin{fmfgraph*}(50,70) 
\fmfset{arrow_len}{3mm} 
\fmfset{arrow_ang}{20} 
\fmfleft{nJ1} 
\fmfright{nJ2}
\fmflabel{$h$}{nJ1}
\fmf{dashes,tension=1}{nJ1,nJ2} 
\fmfv{decor.shape=circle,decor.filled=hatched,decor.size=9thick}{nJ2}
\end{fmfgraph*} 
\end{fmffile} 
\end{picture}
\vspace{-10mm}
\end{minipage}
\hspace{-38mm} .
\label{Chap-Selec:eq:bonecos}
\ee
On the other hand, as we will discuss in detail in section \ref{Chap-Selec:sec:reno}, the name counterterm for $\delta t$ can be misleading, precisely because $\delta t$ is not a regular counterterm (we insist, the selection of the true vev is prior to renormalization).
However, since $\delta t$ will contribute to the counterterms of other GFs, we keep the name. 

Finally, both tadpole schemes will obey eq. \ref{Chap-Selec:eq:condition}, which means that, in of both of them, eq. \ref{Chap-Selec:eq:VrutolHnew} is zero. And this in turn implies that, in both of them, there are no linear terms in $h$.
But there is an important detail here. What we have been saying is that linear terms in $h$ will always cancel out; and by `linear terms' we mean terms which are linear \textit{only} in $h$ (i.e. 1-point functions in $h$, just as $T_h h$).
But this does not mean that a term like $T_h \, h G^+ G^-$ cannot exist. In fact, all we ensure with eq. \ref{Chap-Selec:eq:condition} is that 1-point functions in $h$ vanish; but since $T_h \, h G^+ G^-$ is not a 1-point function in $h$, there is nothing preventing it from appearing---even if it is proportional to $T_h$.

In other words, our goal never was that $T_h$ by itself is zero, or that it vanishes from the theory, but rather that it vanishes from the Higgs 1-point function (because we wanted the Higgs 1-point function to vanish).
We realized that this required eq. \ref{Chap-Selec:eq:condition} to be true, which ensures that the only two quantities that contribute to the Higgs 1-point function $\big(\bar{v} [\mu_{(0)}^2 - \lambda_{(0)} \bar{v}^2]$ and $T_h\big)$ precisely cancel.
But suppose that, in another GF, $\bar{v} [\mu_{(0)}^2 - \lambda_{(0)} \bar{v}^2]$ contributes, but in such a way that there is no $T_h$ added to it.
That is to say, in that GF we do not have something like:
\be
\bar{v} [\mu_{(0)}^2 - \lambda_{(0)} \bar{v}^2] + T_h,
\ee
but rather like:
\be
\bar{v} [\mu_{(0)}^2 - \lambda_{(0)} \bar{v}^2].
\label{Chap-Selec:eq:twist}
\ee
In that GF, then, $\bar{v} [\mu_{(0)}^2 - \lambda_{(0)} \bar{v}^2]$ will not cancel: it will contribute to the GF. On the other hand, eq. \ref{Chap-Selec:eq:condition} is always valid, as it is the assurance that we have selected the true vev. Hence, eq. \ref{Chap-Selec:eq:twist} becomes:
\be
-T_h.
\ee
This means that, in that GF, there \textit{will be} a one-loop tadpole. Only, there is \textit{nothing wrong with this}, since this does not spoil \ref{Chap-Selec:eq:condition}, i.e. it does not spoil the fact that there are no 1-point functions in $h$. Actually, both tadpole schemes will see one-loop tadpoles appearing in GFs (other than 1-point functions).

We should not let terminology confuse us: if by tadpole we mean a term linear only in $h$ (a 1-point function in $h$), then both schemes will have no tadpoles; if by tadpole we mean a one-loop tadpole contributing to GFs other than 1-point functions, then tadpoles will emerge in both schemes. In order to avoid misunderstandings, we shall name \textit{proper tadpoles} to the former (i.e. to 1-point functions in $h$) and \textit{broad tadpoles} to the latter (i.e. to one-loop tadpoles contributing to GFs other than 1-point functions).

%

We now introduce the two schemes, following ref.~\cite{Denner:2019vbn} closely.

\subsection{Fleischer-Jegerlehner tadpole scheme}
\label{Chap-Selec:sec:FJTS}

It is in this scheme, introduced by Fleischer and Jegerlehner in 1981~\cite{Fleischer:1980ub}, that $\bar{v}$ is split into two terms,
\be
\bar{v} \FJeq v_{(0)} + \Delta v,
\label{Chap-Selec:eq:FJbasic}
\ee
where $\FJeq$ means that the equality is valid in the Fleischer-Jegerlehner tadpole scheme (FJTS). Recall that $v_{(0)}$ is the bare vev, so that it obeys eq. \ref{Chap-Selec:eq:fixed0}---an equation which we will use repeatedly, and which can in particular be used to derive the relations
\be 
\mu_{(0)}^2 = \dfrac{m_{h(0)}^2}{2},
\qquad
\lambda_{(0)} \FJeq \dfrac{m_{h(0)}^2 \, e_{(0)}^2}{8 \, s_{{\text{w}}(0)}^2 \, m_{\mathrm{W}(0)}^2},
\label{Chap-Selec:eq:lambdaFJ}
\ee
which is just the bare version of eq. \ref{Chap-Selec:eq:lambda1}. Moreover, in this scheme, the bare masses are defined in terms of the bare vev:
\begin{gather}
m_{\mathrm{W}(0)} \FJeq \dfrac{1}{2} \dfrac{e_{(0)}}{s_{\mathrm{w}(0)}} v_{(0)},
\qquad
m_{\mathrm{Z}(0)} \FJeq \dfrac{1}{2} \sqrt{g_{1(0)}^2 + g_{2(0)}^2} v_{(0)},
\qquad
m_{h(0)}^2 \FJeq 2 \mu_{(0)}^2 \FJeq 2 \lambda_{(0)} v^2_{(0)}, \nonumber\\[3mm]
m_{f,t,i(0)} \FJeq  \dfrac{v_{(0)}}{\sqrt{2}}
\sum_{j,k} \Big(U_{t_{\mathrm{L}}(0)}^{\dagger}\Big)_{ij} \big(Y_{t(0)}\big)_{jk} \Big(U_{t_{\mathrm{R}}(0)}\Big)_{ki}. 
\label{Chap-Selec:eq:FJmasses}
\end{gather}
In its turn, $\Delta v$ in eq. \ref{Chap-Selec:eq:FJbasic} will be responsible for cancelling the one-loop proper tadpole, $T_h \, h$. To see how this comes up, we can insert the bare Higgs doublet (eq. \ref{Chap-Selec:eq:doublet-base-new}) in the bare potential using eq. \ref{Chap-Selec:eq:FJbasic}, and consider only the terms which are linear in $\Delta v$. We find:
\be
\begin{split}
- V_{(0)}^{\Delta v^1} 
= - \Delta v \lambda_{(0)} \bigg[
2 h_{(0)} v_{(0)}^{2} + &v_{(0)}
\left(2 \, G_{(0)}^{-} G_{(0)}^{+} + 3 h_{(0)}^{2} + G_{0(0)}^{2}\right)
\\[-2mm]
& \hs{20mm}
+ h_{(0)} \left(2 \, G_{(0)}^{-} G_{(0)}^{+} + h_{(0)}^{2} + G_{0(0)}^{2}\right)
\bigg]. 
\label{Chap-Selec:eq:linearDv}
\end{split}
\ee
The fact that $\Delta v$ will cancel the one-loop tadpole implies that it is formally of one-loop order, so that we can neglect $\mathcal{O}(\Delta v^2)$. Moreover, since $\Delta v$ should cancel the one-loop term linear only in $h$, the term responsible for such cancellation should also be linear only in $h$. That is, using the definition \ref{Chap-Selec:eq:allagree}, $\delta t$ should be multiplying the linear term in $h_{(0)}$ in eq. \ref{Chap-Selec:eq:linearDv}, which leads to the identification:
\be
\delta t \FJeq -2 \, \Delta v \, \lambda_{(0)} v_{(0)}^2.
\label{Chap-Selec:eq:solve}
\ee
This is consistent with eq. \ref{Chap-Selec:eq:consistent}; to see it explicitly,
\be
\delta t = \bar{v} \left(\mu_{(0)}^2 - \lambda_{(0)} \bar{v}^2 \right) \FJeq \left(v_{(0)}+\Delta v\right) \left(\mu_{(0)}^{2} - \lambda_{(0)} \left(v_{(0)}+\Delta v\right)^{2}\right) = -2 \, \Delta v \, \lambda_{(0)} v_{(0)}^{2},
\ee
%

where we neglected $\mathcal{O}(\Delta v^2)$. Using now eq. \ref{Chap-Selec:eq:FJmasses} to solve for $\Delta_v$ in eq. \ref{Chap-Selec:eq:solve} in the one-loop approximation fixes it to:
\be
\Delta v=-\dfrac{\delta t}{m_h^{2}}.
\label{Chap-Selec:eq:Dv}
\ee
Now, to begin to ascertain the impact of these equations on the renormalization of the theory, note that, because $\Delta v$ is of one-loop order, we can replace all the bare parameters and fields in eq. \ref{Chap-Selec:eq:linearDv} for their renormalized equivalents. 
As for $v_{(0)}$, we use eq. \ref{Chap-Selec:eq:FJmasses} to replace it by $\frac{2 \, s_{\mathrm{w}(0)} \, m_{\mathrm{W}(0)}}{e_{(0)}}$ and then we replace these parameters for their renormalized versions.
Therefore, using eqs. \ref{Chap-Selec:eq:FJmasses} and \ref{Chap-Selec:eq:Dv}, we can rewrite eq. \ref{Chap-Selec:eq:linearDv} as:
\be
-V^{\delta t}_{\text{FJTS}}=\delta t \, h + \frac{\delta t \, e}{4 \, m_{\mathrm{W}} s_{\text{w}}}\left[2 \, G^{-} G^{+} + 3 h^{2} + G_0^{2}\right] + \frac{\delta t \, e^{2}}{8 \, m_{\mathrm{W}}^{2} s_{\text{w}}^{2}} h \left[2 \, G^{-} G^{+} + h^{2} + G_0^{2}\right].
\label{Chap-Selec:eq:VFJ}
\ee
This equation represents the totality of terms in the effective potential that, in the FJTS, are proportional to $\delta t$.
%
%
%
%
The first term on the r.h.s. will precisely cancel the one-loop proper tadpole (i.e. the one-loop 1-point function) of $h$, just as eq. \ref{Chap-Selec:eq:allagree} predicts. So, in the end of the day, there will be no proper tadpoles in the theory, as intended.

However, we see in eq. \ref{Chap-Selec:eq:VFJ} that broad tadpoles will be around. For example, the term $\frac{3}{4} \frac{\delta t \, e}{m_{\mathrm{W}} s_{\text{w}}} h^2$ will contribute to the Higgs 2-point function; and the $\delta t$ (which is a one-loop tadpole, recall eq. \ref{Chap-Selec:eq:allagree}) contained in it is an example of what we called a broad tapole. Indeed, it is a one-loop tadpole contributing to a GF other than a 1-point function (in this case, the 2-point function of the Higgs boson).
The reason why it may be useful to think of $\delta t$ as a counterterm is that we can include the new term $\frac{3}{4} \frac{\delta t \, e}{m_{\mathrm{W}} s_{\text{w}}} h^2$ in the set of counterterms of the Higgs 2-point function. Such set then becomes:%
\fn{The factor $1/2$ in the last term (instead of $1/4$) is due to the $h^2$.}
\begin{equation}
\noindent
\begin{minipage}[h]{.40\textwidth}
\vspace{-10mm}
\begin{picture}(0,42)
\hspace{10mm}
\begin{fmffile}{Chap-Selec-10b} 
\begin{fmfgraph*}(70,70) 
\fmfset{arrow_len}{3mm}
\fmfset{arrow_ang}{20}
\fmfleft{nJ1}
\fmfright{nJ2}
\fmflabel{$h$}{nJ1}
\fmflabel{$h$}{nJ2}
\fmf{dashes,tension=3}{nJ1,nJ1nJ2}
\fmf{dashes,tension=3}{nJ1nJ2,nJ2}
\fmfv{decor.shape=pentagram,decor.filled=full,decor.size=6thick}{nJ1nJ2}
\end{fmfgraph*} 
\end{fmffile} 
\end{picture}
\vspace{-20mm}
\end{minipage}
\hspace{-14mm}
  i   \,  \bigg[ \left(p^2 - m_h^2\right) \delta Z_h - \delta m_h^2 + \dfrac{3}{2} \dfrac{\delta t \, e}{m_{\mathrm{W}} s_{\text{w}}} \bigg].
\label{Chap-Selec:eq:Hprop-counter}
\end{equation}
It turns out that eq. \ref{Chap-Selec:eq:Dv} allows an interesting reorganization of the counterterms.
Let us start by noting that the cubic Higgs coupling is given by
\be
\vspace{5mm}
\begin{minipage}[h]{.40\textwidth}
\begin{picture}(0,60)
\begin{fmffile}{Chap-Selec-2} 
\begin{fmfgraph*}(60,60)
\fmfset{arrow_len}{3mm}
\fmfset{arrow_ang}{20}
\fmfleft{nJ1}
\fmfright{nJ2,nJ4}
\fmflabel{$h$}{nJ1}
\fmflabel{$h$}{nJ2}
\fmflabel{$h$}{nJ4}
\fmf{dashes,tension=3}{nJ1,nJ2nJ4nJ1}
\fmf{dashes,tension=3}{nJ2,nJ2nJ4nJ1}
\fmf{dashes,tension=3}{nJ4,nJ2nJ4nJ1}
\end{fmfgraph*}
\end{fmffile} 
\end{picture}
\end{minipage}
\hspace{-30mm}
- \frac{3}{2} i \dfrac{m_h^2 \, e}{m_{\mathrm{W}} s_{\text{w}}},
\label{Chap-Selec:eq:cubic}
\ee
and the zero-momentum Higgs propagator by
\be
\begin{minipage}[h]{.40\textwidth}
\begin{picture}(0,42)
\begin{fmffile}{Chap-Selec-3}
\begin{fmfgraph*}(70,70)
\fmfset{arrow_len}{3mm}
\fmfset{arrow_ang}{20}
\fmfleft{nJ1}
\fmfright{nJ2}
\fmflabel{$h$}{nJ1}
\fmflabel{$h$}{nJ2}
\fmf{dashes,label=\small $\hspace{12mm} p=0$,label.side=left,tension=3,label.dist=2thick}{nJ1,nJ1nJ2}
\fmf{dashes,tension=3}{nJ1nJ2,nJ2}
\end{fmfgraph*}
\end{fmffile}
\end{picture}
\vs{-10mm}
\end{minipage}
\hspace{-25mm} 
i \frac{1}{- m_h^2} .
\label{Chap-Selec:eq:0mom}
\ee
Then, combining these equations with eq. \ref{Chap-Selec:eq:tad-counter}, we find%
\fn{We draw small black circles on the dashed lines to highlight the presence of a tree-level propagator.}
\be
\begin{minipage}[h]{.40\textwidth}
\begin{picture}(0,120)
\begin{fmffile}{Chap-Selec-11}
\begin{fmfgraph*}(70,115)
\fmfset{arrow_len}{3mm}
\fmfset{arrow_ang}{20}
\fmfleft{nJ1}
\fmfright{nJ2}
\fmftop{nJ3}
\fmflabel{$h$}{nJ1}
\fmflabel{$h$}{nJ2}
\fmf{dashes,tension=3}{nJ1,nJ1nJ2}
\fmf{dashes,tension=0.1}{nJ1nJ2,x}
\fmf{dashes,tension=0.1}{x,y}
\fmf{dashes,tension=0.1}{y,nJ3}
\fmf{dashes,tension=3}{nJ1nJ2,nJ2}
\fmfv{decor.shape=circle,decor.filled=70,decor.size=1thick}{x}
\fmfv{decor.shape=circle,decor.filled=70,decor.size=1thick}{y}
\fmfv{decor.shape=pentagram,decor.filled=full,decor.size=6thick}{nJ3}
\end{fmfgraph*}
\end{fmffile}
\end{picture}
\end{minipage}
\hspace{-25mm} 
- \frac{3}{2} i \dfrac{\delta t \, e}{m_{\mathrm{W}} s_{\text{w}}},
\label{Chap-Selec:eq:redtadcount}
\vs{-15mm}
\ee
and, using eq. \ref{Chap-Selec:eq:bonecos},
\be
\begin{minipage}[h]{.40\textwidth}
\begin{picture}(0,120)
\begin{fmffile}{Chap-Selec-12}
\begin{fmfgraph*}(70,115)
\fmfset{arrow_len}{3mm}
\fmfset{arrow_ang}{20}
\fmfleft{nJ1}
\fmfright{nJ2}
\fmftop{nJ3}
\fmflabel{$h$}{nJ1}
\fmflabel{$h$}{nJ2}
\fmf{dashes,tension=3}{nJ1,nJ1nJ2}
\fmf{dashes,tension=0.1}{nJ1nJ2,x}
\fmf{dashes,tension=0.1}{x,y}
\fmf{dashes,tension=0.1}{y,nJ3}
\fmf{dashes,tension=3}{nJ1nJ2,nJ2}
\fmfv{decor.shape=circle,decor.filled=70,decor.size=1thick}{x}
\fmfv{decor.shape=circle,decor.filled=70,decor.size=1thick}{y}
\fmfv{decor.shape=circle,decor.filled=hatched,decor.size=9thick}{nJ3}
\end{fmfgraph*}
\end{fmffile}
\end{picture}
\end{minipage}
\hspace{-25mm} 
\frac{3}{2} i \dfrac{\delta t \, e}{m_{\mathrm{W}} s_{\text{w}}}.
\label{Chap-Selec:eq:redtadloop}
\vs{-15mm}
\ee
In order to realize the importance of these results, note that the most general structure of a renormalized (connected) 2-point GF at one-loop level is given by:
\vspace{-4mm}
\begin{equation}
\begin{minipage}[h]{.40\textwidth}
\vspace{10mm}
\begin{picture}(0,42)
\begin{fmffile}{Chap-Selec-13} 
\begin{fmfgraph*}(40,70) 
\fmfset{arrow_len}{3mm}
\fmfset{arrow_ang}{20}
\fmfleft{nJ1}
\fmfright{nJ2}
\fmf{plain,tension=3}{nJ1,nJ1nJ2}
\fmf{plain,tension=3}{nJ1nJ2,nJ2}
\fmfv{decor.shape=circle,decor.filled=full,decor.size=6thick}{nJ1nJ2}
\end{fmfgraph*} 
\end{fmffile} 
\end{picture}
\end{minipage}
\hspace{-43mm}
=
\hspace{1mm}
\begin{minipage}[h]{.40\textwidth}
\vspace{10mm}
\begin{picture}(0,42)
\begin{fmffile}{Chap-Selec-14} 
\begin{fmfgraph*}(40,70) 
\fmfset{arrow_len}{3mm}
\fmfset{arrow_ang}{20}
\fmfleft{nJ1}
\fmfright{nJ2}
\fmf{plain,tension=3}{nJ1,nJ1nJ2}
\fmf{plain,tension=3}{nJ1nJ2,nJ2}
\fmfv{decor.shape=circle,decor.filled=hatched,decor.size=7thick}{nJ1nJ2}
\end{fmfgraph*} 
\end{fmffile} 
\end{picture}
\end{minipage}
\hspace{-43mm}
+
\hspace{2mm}
\begin{minipage}[h]{.40\textwidth}
\vspace{10mm}
\begin{picture}(0,42)
\begin{fmffile}{Chap-Selec-15} 
\begin{fmfgraph*}(40,70) 
\fmfset{arrow_len}{3mm}
\fmfset{arrow_ang}{20}
\fmfleft{nJ1}
\fmfright{nJ2}
\fmf{plain,tension=3}{nJ1,nJ1nJ2}
\fmf{plain,tension=3}{nJ1nJ2,nJ2}
\fmfv{decor.shape=pentagram,decor.filled=full,decor.size=5thick}{nJ1nJ2}
\end{fmfgraph*} 
\end{fmffile} 
\end{picture}
\end{minipage}
\hspace{-43mm}
+
\hspace{2mm}
\begin{minipage}[h]{.40\textwidth}
\vspace{-10.5mm}
\begin{picture}(0,100)
\begin{fmffile}{Chap-Selec-16}
\begin{fmfgraph*}(40,70)
\fmfset{arrow_len}{3mm}
\fmfset{arrow_ang}{20}
\fmfleft{nJ1}
\fmfright{nJ2}
\fmftop{nJ3}
\fmf{plain,tension=3}{nJ1,nJ1nJ2}
\fmf{dashes,tension=0.1}{nJ1nJ2,x}
\fmf{dashes,tension=0.1}{x,y}
\fmf{dashes,tension=0.1}{y,nJ3}
\fmf{plain,tension=3}{nJ1nJ2,nJ2}
\fmfv{decor.shape=circle,decor.filled=70,decor.size=0.8thick}{x}
\fmfv{decor.shape=circle,decor.filled=70,decor.size=0.8thick}{y}
\fmfv{decor.shape=circle,decor.filled=hatched,decor.size=7thick}{nJ3}
\end{fmfgraph*}
\end{fmffile}
\end{picture}
\end{minipage}
\hspace{-43mm}
+
\hspace{2mm}
\begin{minipage}[h]{.40\textwidth}
\vspace{-10.5mm}
\begin{picture}(0,100)
\begin{fmffile}{Chap-Selec-17}
\begin{fmfgraph*}(40,70)
\fmfset{arrow_len}{3mm}
\fmfset{arrow_ang}{20}
\fmfleft{nJ1}
\fmfright{nJ2}
\fmftop{nJ3}
\fmf{plain,tension=3}{nJ1,nJ1nJ2}
\fmf{dashes,tension=0.1}{nJ1nJ2,x}
\fmf{dashes,tension=0.1}{x,y}
\fmf{dashes,tension=0.1}{y,nJ3}
\fmf{plain,tension=3}{nJ1nJ2,nJ2}
\fmfv{decor.shape=circle,decor.filled=70,decor.size=0.8thick}{x}
\fmfv{decor.shape=circle,decor.filled=70,decor.size=0.8thick}{y}
\fmfv{decor.shape=pentagram,decor.filled=full,decor.size=6thick}{nJ3}
\end{fmfgraph*}
\end{fmffile}
\end{picture}
\end{minipage}
\hspace{-45mm}.
\label{Chap-Selec:eq:novosbonecos}
\end{equation}

\vspace{-8mm}

This is a universal relation (valid for every tadpole scheme), where the straight lines represents any type of particle.
The terms on the r.h.s represent respectively the (non-renormalized)  one-particle irreducible (1PI) one-loop GF, the corresponding counterterm, the reducible diagram with a one-loop tadpole, and the corresponding counterterm.
Now, due to eq. \ref{Chap-Selec:eq:allagree} (diagrammatically represented in eq. \ref{Chap-Selec:eq:bonecos}), the last two terms of the r.h.s. of eq. \ref{Chap-Selec:eq:novosbonecos} necessarily cancel. We just found this explicitly for the Higgs 2-point function in eqs. \ref{Chap-Selec:eq:redtadcount} and \ref{Chap-Selec:eq:redtadloop}. As a consequence, the complete 2-point function at one-loop includes only the first two terms of the r.h.s. of eq. \ref{Chap-Selec:eq:novosbonecos}, that is, the 1PI terms: the non-renormalized one-loop function and its counterterm, respectively.
Moreover, the counterterm (i.e. the second term on the r.h.s. of eq. \ref{Chap-Selec:eq:novosbonecos}) in general includes a broad tadpole, as we saw in eq. \ref{Chap-Selec:eq:Hprop-counter} for the Higgs 2-point function. Hence, we can split it explicitly in two parts:
\vspace{-7mm}
\begin{equation}
\begin{minipage}[h]{.40\textwidth}
\vspace{10mm}
\begin{picture}(0,42)
\begin{fmffile}{Chap-Selec-213b} 
\begin{fmfgraph*}(40,70) 
\fmfset{arrow_len}{3mm}
\fmfset{arrow_ang}{20}
\fmfleft{nJ1}
\fmfright{nJ2}
\fmf{plain,tension=3}{nJ1,nJ1nJ2}
\fmf{plain,tension=3}{nJ1nJ2,nJ2}
\fmfv{decor.shape=pentagram,decor.filled=full,decor.size=5thick}{nJ1nJ2}
\end{fmfgraph*} 
\end{fmffile} 
\end{picture}
\end{minipage}
\hspace{-40mm}
=
\hspace{2mm}
\begin{minipage}[h]{.40\textwidth}
\vspace{10mm}
\begin{picture}(0,42)
\begin{fmffile}{Chap-Selec-214b} 
\begin{fmfgraph*}(40,70) 
\fmfset{arrow_len}{3mm}
\fmfset{arrow_ang}{20}
\fmfleft{nJ1}
\fmfright{nJ2}
\fmf{plain,tension=3,label={\footnotesize \hspace{7mm} $\delta t$-part},label.distance=-60thick,label.side=left}{nJ1,nJ1nJ2}
\fmf{plain,tension=3}{nJ1nJ2,nJ2}
\fmfv{decor.shape=pentagram,decor.filled=full,decor.size=5thick}{nJ1nJ2}
\end{fmfgraph*} 
\end{fmffile} 
\end{picture}
\end{minipage}
\hspace{-40mm}
+
\hspace{2mm}
\begin{minipage}[h]{.40\textwidth}
\vspace{10mm}
\begin{picture}(0,42)
\begin{fmffile}{Chap-Selec-215b} 
\begin{fmfgraph*}(40,70) 
\fmfset{arrow_len}{3mm}
\fmfset{arrow_ang}{20}
\fmfleft{nJ1}
\fmfright{nJ2}
\fmf{plain,tension=3,label={\footnotesize \hspace{7mm} without $\delta t$},label.distance=-60thick,label.side=left}{nJ1,nJ1nJ2}
\fmf{plain,tension=3}{nJ1nJ2,nJ2}
\fmfv{decor.shape=pentagram,decor.filled=full,decor.size=5thick}{nJ1nJ2}
\end{fmfgraph*} 
\end{fmffile} 
\end{picture}
\end{minipage}
\hspace{-45mm}
\vspace{-5mm}.
\label{Chap-Selec:eq:my-kind-exp}
\end{equation}
\vspace{-7mm}

In the Higgs 2-point function, however, we found a curious result:
the $\delta t$-part (given by the last term in eq. \ref{Chap-Selec:eq:Hprop-counter}) precisely cancels eq. \ref{Chap-Selec:eq:redtadcount}.
%
%
%
This means that, instead of letting the last two terms on the r.h.s. of eq. \ref{Chap-Selec:eq:novosbonecos} cancel (so that the complete 2-point function at one-loop includes only the first two terms of that side), we can let the broad tadpole contained in the second term cancel against the fourth term, in which case the complete 2-point function at one-loop is given by the first term, the third term and the second without the broad tadpole. That is, in the FJTS, instead of 
\vspace{-7mm}
\begin{equation}
\begin{minipage}[h]{.40\textwidth}
\vspace{10mm}
\begin{picture}(0,42)
\begin{fmffile}{Chap-Selec-13b} 
\begin{fmfgraph*}(40,70) 
\fmfset{arrow_len}{3mm}
\fmfset{arrow_ang}{20}
\fmfleft{nJ1}
\fmfright{nJ2}
\fmf{plain,tension=3}{nJ1,nJ1nJ2}
\fmf{plain,tension=3}{nJ1nJ2,nJ2}
\fmfv{decor.shape=circle,decor.filled=full,decor.size=6thick}{nJ1nJ2}
\end{fmfgraph*} 
\end{fmffile} 
\end{picture}
\end{minipage}
\hspace{-42mm}
\FJeq
\hspace{3mm}
\begin{minipage}[h]{.40\textwidth}
\vspace{10mm}
\begin{picture}(0,42)
\begin{fmffile}{Chap-Selec-14b} 
\begin{fmfgraph*}(40,70) 
\fmfset{arrow_len}{3mm}
\fmfset{arrow_ang}{20}
\fmfleft{nJ1}
\fmfright{nJ2}
\fmf{plain,tension=3}{nJ1,nJ1nJ2}
\fmf{plain,tension=3}{nJ1nJ2,nJ2}
\fmfv{decor.shape=circle,decor.filled=hatched,decor.size=7thick}{nJ1nJ2}
\end{fmfgraph*} 
\end{fmffile} 
\end{picture}
\end{minipage}
\hspace{-42mm}
+
\hspace{3mm}
\begin{minipage}[h]{.40\textwidth}
\vspace{10mm}
\begin{picture}(0,42)
\begin{fmffile}{Chap-Selec-15b} 
\begin{fmfgraph*}(40,70) 
\fmfset{arrow_len}{3mm}
\fmfset{arrow_ang}{20}
\fmfleft{nJ1}
\fmfright{nJ2}
\fmf{plain,tension=3,label={\footnotesize \hspace{7mm} $\delta t$-part},label.distance=-60thick,label.side=left}{nJ1,nJ1nJ2}
\fmf{plain,tension=3}{nJ1nJ2,nJ2}
\fmfv{decor.shape=pentagram,decor.filled=full,decor.size=5thick}{nJ1nJ2}
\end{fmfgraph*} 
\end{fmffile} 
\end{picture}
\end{minipage}
\hspace{-42mm}
+
\hspace{3mm}
\begin{minipage}[h]{.40\textwidth}
\vspace{10mm}
\begin{picture}(0,42)
\begin{fmffile}{Chap-Selec-715b} 
\begin{fmfgraph*}(40,70) 
\fmfset{arrow_len}{3mm}
\fmfset{arrow_ang}{20}
\fmfleft{nJ1}
\fmfright{nJ2}
\fmf{plain,tension=3,label={\footnotesize \hspace{7mm} without $\delta t$},label.distance=-60thick,label.side=left}{nJ1,nJ1nJ2}
\fmf{plain,tension=3}{nJ1nJ2,nJ2}
\fmfv{decor.shape=pentagram,decor.filled=full,decor.size=5thick}{nJ1nJ2}
\end{fmfgraph*} 
\end{fmffile} 
\end{picture}
\end{minipage}
\hspace{-42mm},
\label{Chap-Selec:eq:myoldeq}
\vspace{-8mm}
\end{equation}
we can write
\vspace{-5mm}
\begin{equation}
\begin{minipage}[h]{.40\textwidth}
\vspace{10mm}
\begin{picture}(0,42)
\begin{fmffile}{Chap-Selec-13c} 
\begin{fmfgraph*}(40,70) 
\fmfset{arrow_len}{3mm}
\fmfset{arrow_ang}{20}
\fmfleft{nJ1}
\fmfright{nJ2}
\fmf{plain,tension=3}{nJ1,nJ1nJ2}
\fmf{plain,tension=3}{nJ1nJ2,nJ2}
\fmfv{decor.shape=circle,decor.filled=full,decor.size=6thick}{nJ1nJ2}
\end{fmfgraph*} 
\end{fmffile} 
\end{picture}
\end{minipage}
\hspace{-42mm}
\FJeq
\hspace{3mm}
\begin{minipage}[h]{.40\textwidth}
\vspace{10mm}
\begin{picture}(0,42)
\begin{fmffile}{Chap-Selec-14c} 
\begin{fmfgraph*}(40,70) 
\fmfset{arrow_len}{3mm}
\fmfset{arrow_ang}{20}
\fmfleft{nJ1}
\fmfright{nJ2}
\fmf{plain,tension=3}{nJ1,nJ1nJ2}
\fmf{plain,tension=3}{nJ1nJ2,nJ2}
\fmfv{decor.shape=circle,decor.filled=hatched,decor.size=7thick}{nJ1nJ2}
\end{fmfgraph*} 
\end{fmffile} 
\end{picture}
\end{minipage}
\hspace{-42mm}
+
\hspace{3mm}
\begin{minipage}[h]{.40\textwidth}
\vspace{10mm}
\begin{picture}(0,42)
\begin{fmffile}{Chap-Selec-15c} 
\begin{fmfgraph*}(40,70) 
\fmfset{arrow_len}{3mm}
\fmfset{arrow_ang}{20}
\fmfleft{nJ1}
\fmfright{nJ2}
\fmf{plain,tension=3,label={\footnotesize \hspace{7mm} without $\delta t$},label.distance=-60thick,label.side=left}{nJ1,nJ1nJ2}
\fmf{plain,tension=3}{nJ1nJ2,nJ2}
\fmfv{decor.shape=pentagram,decor.filled=full,decor.size=5thick}{nJ1nJ2}
\end{fmfgraph*} 
\end{fmffile} 
\end{picture}
\end{minipage}
\hspace{-42mm}
+
\hspace{3mm}
\begin{minipage}[h]{.40\textwidth}
\vspace{5mm}
\begin{picture}(0,57)
\begin{fmffile}{Chap-Selec-16c}
\begin{fmfgraph*}(40,70)
\fmfset{arrow_len}{3mm}
\fmfset{arrow_ang}{20}
\fmfleft{nJ1}
\fmfright{nJ2}
\fmftop{nJ3}
\fmf{plain,tension=3}{nJ1,nJ1nJ2}
\fmf{dashes,tension=0.1}{nJ1nJ2,x}
\fmf{dashes,tension=0.1}{x,y}
\fmf{dashes,tension=0.1}{y,nJ3}
\fmf{plain,tension=3}{nJ1nJ2,nJ2}
\fmfv{decor.shape=circle,decor.filled=70,decor.size=0.8thick}{x}
\fmfv{decor.shape=circle,decor.filled=70,decor.size=0.8thick}{y}
\fmfv{decor.shape=circle,decor.filled=hatched,decor.size=7thick}{nJ3}
\end{fmfgraph*}
\end{fmffile}
\end{picture}
\end{minipage}
\hspace{-42mm} .
\label{Chap-Selec:eq:myneweq}
\vspace{-8mm}
\end{equation}
The two equations are equivalent. In particular, both include a broad tadpole: in eq. \ref{Chap-Selec:eq:myoldeq}, it corresponds to the second term on the r.h.s., while in \ref{Chap-Selec:eq:myneweq} it corresponds to the third term on the r.h.s.

But why would we want to use eq. \ref{Chap-Selec:eq:myneweq} instead of \ref{Chap-Selec:eq:myoldeq}? It turns out that, with software such as \FM, it is much easier to calculate a reducible Feynman diagram with a one-loop tadpole than to derive a term like the last one in eq. \ref{Chap-Selec:eq:Hprop-counter}.
In fact, by using eq. \ref{Chap-Selec:eq:myneweq}, we do not have to go through all the process involved in obtaining the last term in eq. \ref{Chap-Selec:eq:Hprop-counter}---which involves separating $\bar{v}$ into $v_{(0)}$ and $\Delta v$ as in eq. \ref{Chap-Selec:eq:FJbasic}, identifying $\Delta v$ with the quotient in eq. \ref{Chap-Selec:eq:Dv} and looking in eq. \ref{Chap-Selec:eq:VFJ} for terms contributing to the GF we are after. Instead, we can just use Feynman rules to calculate the reducible diagram with a one-loop tadpole.%

Actually, this applies to all the other GFs as well: not only to 2-point functions for particles other than the Higgs, but also to GFs with more than 2 particles. That this is true can be easily proven if we realize that, inserting eqs. \ref{Chap-Selec:eq:mHexample2}, \ref{Chap-Selec:eq:FJbasic} and \ref{Chap-Selec:eq:Dv} inside eq. \ref{Chap-Selec:eq:doublet-base-new}, 
%
%
%
then $- \frac{\delta t}{m_h^2}$ and $h$ always appear together. Therefore, for every term in the Lagrangian with $- \frac{\delta t}{m_h^2}$, there is a term with $h$ instead. So, for every $n$-point function with a $\delta t$ contribution, there is a $(n+1)$-point function with a Higgs field $h$ instead of the $\delta t$; and with that Higgs field, we can always form a reducible diagram (with a one-loop tadpole) which, using eqs. \ref{Chap-Selec:eq:loop-tad}, \ref{Chap-Selec:eq:allagree} and \ref{Chap-Selec:eq:0mom}, precisely equals the $n$-point function with a broad tadpole. For example, for the $W$ 2-point function (in the FJTS), we have:
%
%
%
\begin{equation}
\noindent
\begin{minipage}[h]{.40\textwidth}
\vspace{-10mm}
\begin{picture}(0,42)
\hspace{10mm}
\begin{fmffile}{Chap-Selec-18} 
\begin{fmfgraph*}(70,70) 
\fmfset{arrow_len}{3mm}
\fmfset{arrow_ang}{20}
\fmfleft{nJ1}
\fmfright{nJ2}
\fmflabel{$W_{\mu}^+$}{nJ1}
\fmflabel{$W_{\nu}^+$}{nJ2}
\fmf{photon,tension=3}{nJ1,nJ1nJ2}
\fmf{phantom_arrow,tension=0}{nJ1,nJ1nJ2}
\fmf{photon,tension=3}{nJ1nJ2,nJ2}
\fmf{phantom_arrow,tension=0}{nJ1nJ2,nJ2}
\fmfv{decor.shape=pentagram,decor.filled=full,decor.size=6thick}{nJ1nJ2}
\end{fmfgraph*} 
\end{fmffile} 
\end{picture}
\vspace{-20mm}
\end{minipage}
\hspace{-12mm}
i \,  \bigg[ \delta Z_{W} \, p^{ \mu} \, p^{ \nu} +  \left( \delta m_{\mathrm{W}}^2 -\left[p^2 -m_{\mathrm{W}}^2 \right] \delta Z_{W} - \dfrac{\delta t \, e \, m_{\mathrm{W}}}{s_{\text{w}} \, m_h^2}\right) \, g^{\mu \nu} \bigg] 
\label{Chap-Selec:eq:Wprop-counter}
\vs{-3mm}
\end{equation}
and
%
%
%
\begin{equation}
\begin{minipage}[h]{.40\textwidth}
\vspace{15mm}
\begin{picture}(0,80)
\begin{fmffile}{Chap-Selec-19}
\begin{fmfgraph*}(70,110)
\fmfset{arrow_len}{3mm}
\fmfset{arrow_ang}{20}
\fmfleft{nJ1}
\fmfright{nJ2}
\fmftop{nJ3}
\fmflabel{$W_{\mu}^+$}{nJ1}
\fmflabel{$W_{\nu}^+$}{nJ2}
\fmf{photon,tension=3}{nJ1,nJ1nJ2}
\fmf{phantom_arrow,tension=0}{nJ1,nJ1nJ2}
\fmf{dashes,tension=0.1}{nJ1nJ2,x}
\fmf{dashes,tension=0.1}{x,y}
\fmf{dashes,tension=0.1}{y,nJ3}
\fmf{photon,tension=3}{nJ1nJ2,nJ2}
\fmf{phantom_arrow,tension=0}{nJ1nJ2,nJ2}
\fmfv{decor.shape=circle,decor.filled=70,decor.size=1thick}{x}
\fmfv{decor.shape=circle,decor.filled=70,decor.size=1thick}{y}
\fmfv{decor.shape=circle,decor.filled=hatched,decor.size=9thick}{nJ3}
\end{fmfgraph*}
\end{fmffile}
\end{picture}
\end{minipage}
\hspace{-25mm} 
- i \dfrac{\delta t \, e \, m_{\mathrm{W}}}{s_{\text{w}} \, m_h^2} \, g^{\mu \nu}.
\vspace{-15mm}
\end{equation}
In the same way, for the Higgs 3-point function (in the FJTS), we have:
\vspace{5mm}
\begin{equation}
\begin{minipage}[h]{.40\textwidth}
\vspace{-5mm}
\begin{picture}(0,80)
\begin{fmffile}{Chap-Selec-269}
\begin{fmfgraph*}(70,70)
\fmfset{arrow_len}{3mm}
\fmfset{arrow_ang}{20}
\fmfleft{nJ1}
\fmfright{nJ2,nJ4}
\fmflabel{$h$}{nJ1}
\fmflabel{$h$}{nJ2}
\fmflabel{$h$}{nJ4}
\fmf{dashes,tension=3}{nJ1,nJ2nJ4nJ1}
\fmf{dashes,tension=3}{nJ2,nJ2nJ4nJ1}
\fmf{dashes,tension=3}{nJ4,nJ2nJ4nJ1}
\fmfv{decor.shape=pentagram,decor.filled=full,decor.size=6thick}{nJ2nJ4nJ1}
\end{fmfgraph*}
\end{fmffile}
\end{picture}
\end{minipage}
\hspace{-30mm} 
i \dfrac{3 \, e^2 \delta t}{4\, m_{\mathrm{W}}^2 \, s_{\text{w}}^2 }  + ...
\label{Chap-Selec:eq:h3ct}
\vspace{5mm}
\end{equation}
where we showed only the term with the broad tadpole, and
\vspace{5mm}
\begin{equation}
\begin{minipage}[h]{.40\textwidth}
\vspace{-5mm}
\begin{picture}(0,80)
\begin{fmffile}{Chap-Selec-21}
\begin{fmfgraph*}(70,70)
\fmfset{arrow_len}{3mm}
\fmfset{arrow_ang}{20}
\fmfleft{nJ1} 
\fmflabel{$h$}{nJ1} 
\fmfright{nJ2,nJ4} 
\fmflabel{$h$}{nJ2} 
\fmflabel{$h$}{nJ4}
\fmftop{nJ3}
\fmf{dashes,tension=25}{nJ1,nJ1nJ2nJ4J4} 
\fmf{dashes,tension=25}{nJ2,nJ1nJ2nJ4J4} 
\fmf{dashes,tension=25}{nJ4,nJ1nJ2nJ4J4}
\fmf{dashes,tension=0.1}{nJ1nJ2nJ4J4,x}
\fmf{dashes,tension=0.1}{x,y}
\fmf{dashes,tension=0.08}{y,nJ3}
\fmfv{decor.shape=circle,decor.filled=70,decor.size=1thick}{x}
\fmfv{decor.shape=circle,decor.filled=70,decor.size=1thick}{y}
\fmfv{decor.shape=circle,decor.filled=hatched,decor.size=9thick}{nJ3}
\end{fmfgraph*}
\end{fmffile}
\end{picture}
\end{minipage}
\hspace{-30mm} 
i \dfrac{3 \, e^2 \delta t}{4\, m_{\mathrm{W}}^2 \, s_{\text{w}}^2 } .
\end{equation}

\vspace{5mm}

In sum: in the FJTS, we can account for all the broad tadpoles---the new terms generated through $\Delta v$ contributing to GFs other than the 1-point function---by including all possible reducible diagrams with one-loop tadpoles in all possible GFs. Note that it is easy to predict which GF will contain broad tadpoles: considering eqs. \ref{Chap-Selec:eq:FJbasic} and \ref{Chap-Selec:eq:Dv} again, we see that there will be a term with $- \frac{\delta t}{m_h^2}$ for every term with $v_{(0)}$. It follows that broad tadpoles show up for every GF whose tree-level contribution contains a $v_{(0)}$, which includes all the 2-point functions and some 3-point functions.


\subsection{Parameter-renormalized tadpole scheme}
\label{Chap-Selec:sec:SM-PRTS}

We now consider the other scheme, introduced in ref. \cite{Denner:1991kt} and dubbed parameter-renormalized tadpole scheme (PRTS) in ref. \cite{Denner:2018opp}.
Here, $\bar{v}$ is not separated in two, but is immediately taken as the true vev $v$:
\be
\bar{v} \PReq v.
\label{Chap-Selec:PR-base}
\ee
Besides, there is no mention of $v_{(0)}$ in the PRTS. In particular, the bare masses are defined in terms of $v$:
\begin{gather}
m_{\mathrm{W}(0)} \PReq \dfrac{1}{2} \dfrac{e_{(0)}}{s_{\mathrm{w}(0)}} v,
\quad
m_{\mathrm{Z}(0)} \PReq \dfrac{1}{2} \sqrt{g_{1(0)}^2 + g_{2(0)}^2} v,
\nonumber\\[1mm]
m_{f,t,i(0)} \PReq  \dfrac{v}{\sqrt{2}}
\sum_{j,k}\Big(U_{t_{\mathrm{L}}(0)}^{\dagger}\Big)_{ij} \big(Y_{t(0)}\big)_{jk} \Big(U_{t_{\mathrm{R}}(0)}\Big)_{ki}. 
\label{Chap-Selec:eq:PRmasses}
\end{gather}
To obtain the bare Higgs mass, we isolate the quadratic terms in $h_{(0)}$ in the bare potential; we find:
\be
m_{h(0)}^2 \PReq - \mu_{(0)}^2 + 3 \lambda_{(0)} v^2.
\label{Chap-Selec:eq:MHPR}
\ee 
Finally, we can combine eqs. \ref{Chap-Selec:eq:consistent} and \ref{Chap-Selec:PR-base} to obtain:
%
%
\be
\delta t \PReq \, v \left( \mu_{(0)}^2 - \lambda_{(0)} v^2 \right).
\label{Chap-Selec:eq:deltatPR}
\ee
%
%
%
Like before, proper tadpoles vanish: eq. \ref{Chap-Selec:eq:bonecos} is valid. Yet, just like before, broad tadpoles (again, one-loop tadpoles like $\delta t$ contributing to GFs other than 1-point functions) will be around after renormalization. To see it explicitly, note that we can use eqs. \ref{Chap-Selec:eq:MHPR} and \ref{Chap-Selec:eq:deltatPR}, as well as the first relation of eq. \ref{Chap-Selec:eq:PRmasses}, to conclude that $\delta t$ shows up inside $\mu^2_{(0)}$ and $\lambda_{(0)}$ only:
\be
\mu^2_{(0)} \PReq
\dfrac{m_{h(0)}^2}{2}
+
\dfrac{3 \, \delta t \, e_{(0)}}{4 \, s_{{\text{w}}(0)} m_{\mathrm{W}(0)}},
\qquad
\lambda_{(0)} \PReq
\dfrac{m_{h(0)}^2 \, e_{(0)}^2}{8 \, s_{{\text{w}}(0)}^2 m_{\mathrm{W}(0)}^2}
+
\dfrac{\delta t \, e_{(0)}^3}{16 \, s_{{\text{w}}(0)}^3 m_{\mathrm{W}(0)}^3}.
\label{Chap-Selec:eq:lambdaPR}
\ee
Then, it is easy to show that the totality of terms proportional to $\delta t$ in the effective Lagrangian is:
\be
\begin{split}
\mathcal{L}^{\delta t}_{\mathrm{PRTS}} = \delta t \, h
+ \frac{1}{4} \frac{\delta t \, e}{m_{\mathrm{W}} s_{\text{w}}}
&
\left[2 \, G^{-} G^{+} + G_0^{2}\right] - \frac{1}{8} \frac{\delta t \, e^{2}}{m_{\mathrm{W}}^{2} s_{\text{w}}^{2}} h \left[2 \, G^{-} G^{+}
+ h^{2} + G_0^{2}\right]
\\
& -\frac{1}{64} \frac{\delta t \, e^{3}}{m_{\mathrm{W}}^{3} s_{\text{w}}^{3}}\left[4 \left(G^{-} G^{+}\right)^{2} + 4 \, G^{-} G^{+}\left(h^{2}+G_0^{2}\right)+\left(h^{2}+G_0^{2}\right)^{2}\right].
\label{Chap-Selec:eq:myfinal}
\end{split}
\ee
Hence, while the first term on the r.h.s. cancels the one-loop 1-point function of the Higgs boson, all the other terms represent broad tadpoles: the second and third terms contribute to the 2-point function of the charged and neutral would-be Goldstone bosons, respectively, the fourth term to the 3-point function containing one Higgs and two charged would-be Goldstones, and so on.
Some final notes should be considered.

First, in the PRTS, eq. \ref{Chap-Selec:eq:deltatPR} can be seen as the relation that defines $v$. Or, as we could also state it, the true vev is defined as the value $v$  which obeys eq. \ref{Chap-Selec:eq:deltatPR}, since it is this relation that ensures that the theory is free from proper tadpoles.


Second, equation \ref{Chap-Selec:eq:myfinal} implies that, in the PRTS, $\delta t$ appears only inside terms of the potential; what is more, it does not contribute to the 2-point function of any physical particle.%
\fn{\label{Chap-Selec:note:explain}It is easy to see that both these statements hold in the PRTS for a general scalar sector. Concerning the circumstance that 2-point functions of physical particles never contain broad tadpoles, this is a consequence of the definition of the bare mass of the physical particles. The bare mass, indeed, is defined as the collection of terms that contribute to the bilinear term of the field at stake; so, even if broad tadpoles are included in those terms (and in general they are), they are subsumed under the definition of the bare mass, and thus do not show up contributing to the 2-point function.
That broad tadpoles show up exclusively in the potential (i.e. contribute exclusively to GFs coming from the potential) will be shown below.}
In other words, although broad tadpoles are in general present in the PRTS, they are only present in a very specific set, namely: the GFs that come from the potential and that do not correspond to a 2-point function of a physical particle.%
\fn{And, of course, that do not correspond to 1-point functions, by definition of broad tadpole.}
So, for example, there are broad tadpoles in the Higgs 3-point function, but none in the Higgs 2-point function; and there are broad tadpoles in the 2-point function of would-be Goldstone bosons, but none in the 2-point functions of gauge bosons or fermions. All in all, then, in the PRTS there are no broad tadpoles contributing to any 2-point functions of physical particles (which will be relevant when we discuss renormalization in the on-shell subtraction scheme).

Third, one cannot in general calculate broad tadpoles in the PRTS through reducible diagrams with one-loop tadpoles, like in the FJTS, since there is no equivalent to eq. \ref{Chap-Selec:eq:Dv} in the PRTS.%
\fn{This does not mean that broad tadpoles in the PRTS can never be calculated through reducible diagrams with one-loop tadpoles. In some particular cases, that may be possible. In the SM, such scenario happens in the 2-point functions of non-physical scalars. A simple explanation is that, in these cases, the broad tadpole contribution in the PRTS is precisely equal to that in the FJTS: compare eqs. \ref{Chap-Selec:eq:VFJ} and \ref{Chap-Selec:eq:myfinal}. In any event, eq. \ref{Chap-Selec:eq:novosbonecos} continues to be valid in the PRTS, and the last two terms of its r.h.s. still cancel.} As a consequence, it is mandatory to write the terms in the potential in terms of $\delta t$ and track down all the terms in which it shows up.
This task can be simplified by realizing that, at tree-level, eq. \ref{Chap-Selec:eq:tad-tree-Lag} allows to antecipate eq. \ref{Chap-Selec:eq:deltatPR}: one just needs to promote the tree-level true vev and the tree-level tadpole to the up-to-one-loop vev and $\delta t$, respectively.
In fact, the terms which will have $\delta t$ in the up-to-one-loop theory are precisely those with a tree-level tadpole in the tree-level theory, and only those.%
\fn{This allows to explain why, in the PRTS, broad tadpoles show up exclusively in the potential (recall note \ref{Chap-Selec:note:explain}).
Indeed, the equations for the tree-level tadpoles (the minimization equations) exclusively involve parameters of the potential.}
And this is the advantage of writing eq. \ref{Chap-Selec:eq:lambda2} instead of \ref{Chap-Selec:eq:lambda1} at tree-level: it allows to anticipate eq. \ref{Chap-Selec:eq:lambdaPR} in the PRTS.

Finally, note that eq. \ref{Chap-Selec:eq:lambdaPR} implies eq. \ref{Chap-Selec:eq:deltatPR}. This means that the relations in eq. \ref{Chap-Selec:eq:lambdaPR} are sufficient to ensure that no proper tadpoles will be around. Therefore, the strategy outlined in the previous paragraph---rewriting the tree-level relations in terms of the tree-level tadpole and then, when considering the up-to-one-loop theory, promoting the $t$ to $\delta t$---is sufficient to ensure that the true vev is chosen.

\subsection{Comparison between the two tadpole schemes}
\label{Chap-Selec:sec:compar}

\subsubsection{Fundamental differences}
\label{Chap-Selec:sec:diffs}

There are two fundamental differences between the two tadpoles schemes we presented. The first one has to do with \textit{the way} $\bar{v}=v$ \textit{is verified}---and not with the value of $\bar{v}$. In reality, as we saw, both schemes end up having $\bar{v}=v$; this implies that there are proper tadpoles in neither of them, although this comes at the price of having broad tadpoles in both. But although both obey $\bar{v}=v$, they do it in different ways: the PRTS takes that relation from scratch; the FJTS, by contrast, picks $\bar{v}=v_{(0)} + \Delta v$. Naturally, this difference has consequences, which we now describe in detail for the two schemes.

In the PRTS, the true vev $v$ is used. This $v$ is no longer fixed by the tree-level relation \ref{Chap-Selec:eq:THE-TRUE}, but rather by eq. \ref{Chap-Selec:eq:deltatPR}.
As a consequence, not every tree-level relation can be immediately taken as a bare relation. In other words, when passing from the tree-level to the up-to-one-loop level, it is not always true that, for a certain tree-level relation, one can simply convert all the parameters involved to their bare versions, assuming that the relation written in terms of bare parameters will also hold. This is because tree-level relations which assumed the vanishing of the tree-level tadpole must be rewritten without this assumption (as we did in eq. \ref{Chap-Selec:eq:lambda2}). Then, the relation will be valid for bare parameters if and only if the tree-level tadpole is identified with the tadpole counterterm $\delta t$ (see eq. \ref{Chap-Selec:eq:lambdaPR}).
%
It turns out that broad tadpoles show up in GFs of the potential only, and never in 2-point functions of physical particles. Yet, they must be calculated by explicitly tracing down $\delta t$. 

The FJTS is clearly different. By splitting $\bar{v}$ into $\bar{v}=v_{(0)} + \Delta v$, 
it introduces broad tadpoles in every Feynman rule which has a $v_{(0)}$ in the bare theory---in particular, in 2-point functions of physical particles. On the other hand, broad tadpoles can be accounted for by considering one-loop tadpole insertions in all possible GFs. Furthermore, since the FJTS uses $v_{(0)}$ (the bare vev, obeying eq. \ref{Chap-Selec:eq:fixed0}), the tree-level relations can be immediately taken as bare relations, and there is no need to be careful with tree-level tadpoles; we saw an explicit example of this in equations \ref{Chap-Selec:eq:lambda1} and \ref{Chap-Selec:eq:lambdaFJ}.

On top of all this, there is a second fundamental difference between the two schemes, which is \textit{the way bare parameters} are defined. In the PRTS, they are defined in terms of $v$ through eqs. \ref{Chap-Selec:eq:PRmasses} and \ref{Chap-Selec:eq:MHPR}, whereas in the FJTS they are defined in terms of $v_{(0)}$ through eq. \ref{Chap-Selec:eq:FJmasses}. This difference leads to an interesting discussion concerning gauge dependence,
which we now introduce.

\subsubsection{Gauge dependence}
\label{Chap-Selec:sec:gauge-dep-app}

Let us consider a generic parameter $x$, renormalized according to:
\be
x_{(0)} = x + \delta x,
\label{Chap-Selec:eq:x}
\ee
and let us assume that $x_{(0)}$ depends on the vev. We are interested in ascertaining the gauge dependence of the variables involved in this equation: not only in the context of both tadpole schemes (PRTS and FJTS), but also in the context of two common subtraction schemes: the \textit{on-shell subtraction} (OSS) scheme and the \textit{modified minimal subtraction} ($\overline{\text{MS}}$) scheme.%
\fn{Subtraction schemes are different prescriptions to define the finite part of a certain counterterm. For more details, see section \ref{Chap-Reno:sec:calculation-CTs}.}

%

In the PRTS, the vev which $x_{(0)}$ depends on is $v$. But if we consider eq. \ref{Chap-Selec:eq:deltatPR}, we see that $v$ is related not only to the parameters $\mu_{(0)}^2$ and $\lambda_{(0)}$, but also to $\delta t$; and while the former are gauge independent by construction, the latter is defined by a one-loop process through eq. \ref{Chap-Selec:eq:allagree}, which turns out to be gauge dependent~\cite{Degrassi:1992ff}. It follows that $x_{(0)}$, by depending on $v$, will also be gauge dependent. Now, in OSS, the parameters are renormalized in such a way that they (i.e. the renormalized parameters) correspond to observables---which implies that they are gauge independent.%
\fn{\label{Chap-Selec:note:indirect}Care should be taken with cases of `indirect' OSS renormalization. For example, the renormalization of mixing angles can be determined through symmetry relations (cf. appendix \ref{App-Sym}). In this case, the parameter counterterms are fixed by field counterterms, which in turn may happen to be fixed through OSS. In this `indirect' (since parameter counterterms are not directly fixed via OSS, but rather through OSS-fixed field counterterms) OSS renormalization, the renormalized parameters will in general not correspond to observables, and hence will in general be gauge dependent.}
In that case, $x$ is gauge independent, and $\delta x$ compensates the gauge dependence of $x_{(0)}$\cite{Denner:2016etu}.
In fact, if $\xi$ represents a generic gauge parameter, it follows from eq. \ref{Chap-Selec:eq:x} that:
\be
\dfrac{\partial x_{(0)}}{\partial \xi}  = \dfrac{\partial x}{\partial \xi} + \dfrac{\partial \, \delta x}{\partial \xi},
\label{Chap-Selec:eq:xxi}
\ee
so that, in OSS,
\bs
\bea
\dfrac{\partial x}{\partial \xi} &\OSSeq& 0, \label{Chap-Selec:eq:joker}\\
\quad \Longrightarrow \quad
 \dfrac{\partial x_{(0)}}{\partial \xi} &\OSSeq& - \dfrac{\partial \, \delta x}{\partial \xi}, \\
\quad \text{where}  \quad \dfrac{\partial x_{(0)}}{\partial \xi} &\stackrel{\textrm{PR}}{\neq}& 0.
\eea
\es
In $\overline{\text{MS}}$, however (and still in the context of the PRTS), there is nothing forcing the renormalized parameter $x$ to be gauge independent. Therefore, both $x$ and $\delta x$ will in general be gauge dependent.

The situation changes if we consider the FJTS. Here, the vev which $x_{(0)}$ depends on is $v_{(0)}$. This vev, in contrast to $v$, is a bare quantity and, as such, gauge independent. Consequently, $x_{(0)}$ is also gauge independent. We now prove that this implies that the three quantities present in eq. \ref{Chap-Selec:eq:x} are gauge independent in both OSS and $\overline{\text{MS}}$.

That this is true in OSS follows immediately from eqs. \ref{Chap-Selec:eq:xxi}, \ref{Chap-Selec:eq:joker} and 
\be
\dfrac{\partial x_{(0)}}{\partial \xi} \FJeq 0.
\label{Chap-Selec:eq:FJx0}
\ee
Note, by the way, that for everything to be consistent, the GFs in the FJTS must include the broad tadpoles. This applies, in particular, to 2-point functions (recall eq.
\ref{Chap-Selec:eq:myneweq})---which is crucial to ensure the gauge independence of parameter counterterms in OSS. Indeed, parameter counterterms (like mass counterterms) in OSS are defined in terms of 2-point functions evaluated in the on-shell (OS) limit; but these are in general gauge independent if and only if the reducible diagrams with one-loop tadpole are included in them~\cite{Degrassi:1992ff}; that is, 
\vspace{-9mm}
\be
\dfrac{\partial}{\partial \xi}
\left(\vphantom{\begin{array}{c} \\ \\ \\ \end{array}}\right.
\hspace{-1mm}
\begin{minipage}[h]{.40\textwidth}
\vspace{10mm}
\begin{picture}(0,42)
\begin{fmffile}{Chap-Selec-14d} 
\begin{fmfgraph*}(40,70) 
\fmfset{arrow_len}{3mm}
\fmfset{arrow_ang}{20}
\fmfleft{nJ1}
\fmfright{nJ2}
\fmf{plain,tension=3}{nJ1,nJ1nJ2}
\fmf{plain,tension=3}{nJ1nJ2,nJ2}
\fmfv{decor.shape=circle,decor.filled=hatched,decor.size=7thick}{nJ1nJ2}
\end{fmfgraph*} 
\end{fmffile} 
\end{picture}
\end{minipage}
\hspace{-43mm}
+
\hspace{1mm}
\begin{minipage}[h]{.40\textwidth}
\vspace{2mm}
\begin{picture}(0,100)
\begin{fmffile}{Chap-Selec-16d}
\begin{fmfgraph*}(40,70)
\fmfset{arrow_len}{3mm}
\fmfset{arrow_ang}{20}
\fmfleft{nJ1}
\fmfright{nJ2}
\fmftop{nJ3}
\fmf{plain,tension=3}{nJ1,nJ1nJ2}
\fmf{dashes,tension=0.1}{nJ1nJ2,x}
\fmf{dashes,tension=0.1}{x,y}
\fmf{dashes,tension=0.1}{y,nJ3}
\fmf{plain,tension=3}{nJ1nJ2,nJ2}
\fmfv{decor.shape=circle,decor.filled=70,decor.size=0.8thick}{x}
\fmfv{decor.shape=circle,decor.filled=70,decor.size=0.8thick}{y}
\fmfv{decor.shape=circle,decor.filled=hatched,decor.size=7thick}{nJ3}
\end{fmfgraph*}
\end{fmffile}
\end{picture}
\end{minipage}
\hspace{-45mm} 
\left.\vphantom{\begin{array}{c} \\ \\ \\ \end{array}}\right)
\OSeq 0,
\ee
\vspace{-12mm}

(where $\OSeq$ means that the OS limit is considered), in such a way that

\vspace{-12mm}
\be
\dfrac{\partial}{\partial \xi}
\,
\Bigg(
\begin{minipage}[h]{.40\textwidth}
\vspace{10mm}
\begin{picture}(0,42)
\begin{fmffile}{Chap-Selec-14e} 
\begin{fmfgraph*}(40,70) 
\fmfset{arrow_len}{3mm}
\fmfset{arrow_ang}{20}
\fmfleft{nJ1}
\fmfright{nJ2}
\fmf{plain,tension=3}{nJ1,nJ1nJ2}
\fmf{plain,tension=3}{nJ1nJ2,nJ2}
\fmfv{decor.shape=circle,decor.filled=hatched,decor.size=7thick}{nJ1nJ2}
\end{fmfgraph*} 
\end{fmffile} 
\end{picture}
\end{minipage}
\hspace{-44mm}
\Bigg)
\OSeq
\hspace{1mm}
- \dfrac{\partial}{\partial \xi}
\,
\left(\vphantom{\begin{array}{c} \\ \\ \\ \end{array}}\right.
\begin{minipage}[h]{.40\textwidth}
\vspace{2mm}
\begin{picture}(0,100)
\begin{fmffile}{Chap-Selec-16e}
\begin{fmfgraph*}(40,70)
\fmfset{arrow_len}{3mm}
\fmfset{arrow_ang}{20}
\fmfleft{nJ1}
\fmfright{nJ2}
\fmftop{nJ3}
\fmf{plain,tension=3}{nJ1,nJ1nJ2}
\fmf{dashes,tension=0.1}{nJ1nJ2,x}
\fmf{dashes,tension=0.1}{x,y}
\fmf{dashes,tension=0.1}{y,nJ3}
\fmf{plain,tension=3}{nJ1nJ2,nJ2}
\fmfv{decor.shape=circle,decor.filled=70,decor.size=0.8thick}{x}
\fmfv{decor.shape=circle,decor.filled=70,decor.size=0.8thick}{y}
\fmfv{decor.shape=circle,decor.filled=hatched,decor.size=7thick}{nJ3}
\end{fmfgraph*}
\end{fmffile}
\end{picture}
\end{minipage}
\hspace{-45mm}
\left.\vphantom{\begin{array}{c} \\ \\ \\ \end{array}}\right)
\neq
0.
\label{Chap-Selec:eq:yetanother}
\vspace{-7mm}
\ee
%
%
This means that, if we did not include the reducible diagrams with the one-loop tadpole, we would end up with a gauge dependent 2-point function, and hence a gauge dependent parameter counterterm---which is precisely what happens in the PRTS (recall that, in the PRTS, there are no broad tadpoles contributing to the 2-point functions of physical particles).

Going back to the FJTS, we now turn to $\overline{\text{MS}}$. Here, the counterterm $\delta x$ is defined as
proportional to $\Delta_{\varepsilon}$, which contains a divergence (see section \ref{Chap-Reno:sec:calculation-CTs} for details). 
On the other hand, $x$ is finite (by definition of renormalized parameter), so that it is independent of $\Delta_{\varepsilon}$. Hence,%
\fn{We have been assuming that $x_{(0)}$ depends on the vev; as we saw, this implies that, in the PRTS, $x_{(0)}$ is gauge dependent. If, however, $x_{(0)}$ does not depend on the vev, eq. \ref{Chap-Selec:eq:FJMS} also holds in the PRTS.}
\be
\dfrac{\partial x_{(0)}}{\partial \xi} \FJeq 0
\quad \Longrightarrow \quad 
\dfrac{\partial x}{\partial \xi} \FJeq - \dfrac{\partial \, \delta x}{\partial \xi}
\quad \Longrightarrow \quad 
\begin{cases}
\dfrac{\partial x}{\partial \xi} \stackrel{\mathrm{FJ},\overline{\mathrm{MS}}}{=} 0 \\[4mm]
 \dfrac{\partial \, \delta x}{\partial \xi} \stackrel{\mathrm{FJ},\overline{\mathrm{MS}}}{=} 0
\end{cases}.
\label{Chap-Selec:eq:FJMS}
\ee
%

In conclusion, renormalized parameters and their counterterms are gauge independent in the FJTS: both in OSS and $\overline{\text{MS}}$.
The PRTS, by contrast, is plagued by gauge dependences: only the renormalized parameter $x$ is gauge independent, and only if it is renormalized through OSS.%
\fn{Recall that, as mentioned in note \ref{Chap-Selec:note:indirect}, counterterms fixed in the FJTS via an `indirect' OSS renormalization will in general be gauge dependent.
As it is clear, the parameter counterterms will in general contribute to $S$-matrix elements, so that one should in principle ensure that both are gauge independent, in order to obtain gauge independent $S$-matrix elements (nonetheless, see discussion in section \ref{Chap-Reno:sec:gauge-dep}). In the PRTS, we just saw that, using OSS, the parameter counterterms are in general gauge dependent. 
However, if all parameters are renormalized using a momentum-subtraction scheme (like OSS), the gauge dependence ends up cancelling in renormalized functions (and hence in physical predictions). In fact, a momentum-dependent subtraction scheme is such that the counterterms are defined as a function of a one-loop integral evaluated at a specific momentum (we ignore terms of derivatives, which are irrelevant for the argument). Accordingly, a renormalized GF corresponds to the non-renormalized function evaluated at a general momentum, minus the same non-renormalized function evaluated at a specific momentum. Now, the gauge dependence of the non-renormalized function is related to that of the reducible diagrams with the one-loop tadpole (recall eq. \ref{Chap-Selec:eq:yetanother}). But since those diagrams do not depend on the momentum, we conclude that the gauge dependence cancels in the renormalized GF \cite{Denner:2016etu}. I thank Ansgar Denner for clarifications on this aspect.}

\subsubsection{Observable differences}
\label{Chap-Selec:sec:obs}

Are there physical (i.e. observable) differences between tadpole schemes?
In a first approach, one might think that, in this respect, tadpole schemes are like subtraction schemes.
As is well known, although observables in a renormalized theory must be independent of subtraction schemes, such independence must only be verified if one had an exact (i.e. including all-orders in perturbation theory) calculation. More precisely, if one would perform an exact calculation in two different schemes, the relations between observables in the two cases would be equivalent.
Yet, calculations performed at a certain finite order in perturbation theory (i.e. including only a certain number of loops) will in general depend on the subtraction scheme.%
\fn{The differences between calculation performed in two different schemes, though, are of higher order\cite{Bohm:2001yx}.}

The question now is whether the same happens in calculations with different tadpole schemes. The answer is no: differences in calculation at a certain finite order between consistent tadpole schemes cancel exactly. That is, one does not need to get the exact result to obtain equivalent descriptions between two tadpole schemes: finite order calculations are also equivalent. As matter of fact, we are about to see that the relevant connected Green functions are the same.%
\fn{It follows that, in the FJTS, when one considers processes that do not come from terms in the potential, the broad tadpoles showing up in the renormalized connected one-loop amplitude must cancel between themselves. The reason is that, on the one hand, and as we just saw, there cannot be differences at finite orders in connected GFs calculated in different tadpole schemes; on the other hand, there are no broad tadpoles in the PRTS outside of the potential.}

\subsection{The need to select the true vev}
\label{Chap-Selec:sec:need}

In the previous sections, we have been concerned with selecting the true vev of the up-to-one-loop theory. We introduced two different ways to achieve such selection, and we compared them in detail.

What we should now realize is that we really \textit{did not need} to perform such selection. To say it clearly, we did not need to choose the true vev of the up-to-one-loop theory \cite{Collins:1984xc,Appelquist:1973ms,Bohm:2001yx,Denner:2016etu}.%
\fn{I am particularly indebted to Ansgar Denner for several clarifications on this topic.}
This is in contrast to what happened at tree-level. As we saw in section \ref{Chap-Selec:sec:SM-tree}, indeed, the selection of the true vev at tree-level was a necessary condition for perturbation theory to be consistent. To recap, fields used in perturbation theory should correspond to small excitations around the (true) minimum of theory. So, if we were to choose a vev which would not ensure this basic condition, perturbation theory would not make sense; the poles of the propagators would in general not have corresponded to the physical masses, so that $S$-matrix elements could not be calculated. 
Conversely, if the correct (or true) vev is selected, in such a way that the minimum of the potential is chosen, then perturbation theory is sound, i.e. well-defined.

The point we should now stress is this: as long as perturbation theory is well-defined at tree-level (i.e. at lowest order), it will stay well-defined at higher orders.
The reason is simply that higher orders correspond to small corrections (i.e. small perturbations) relative to the lowest order; consequently, if 
the fields were correctly defined at the lowest order, then perturbation theory is not spoiled at higher orders. This holds, in particular, for the vev. In fact, the true vev of the up-to-one-loop theory differs from the true vev of the tree-level theory by higher orders, as it is obvious. So, the consistency of perturbation theory is not spoiled if we choose a vev that differs from the true vev by higher orders. In other words, precisely because higher order are small, perturbation theory is consistent both when we select the true tree-level level, and when we choose vevs that differ from it by higher order effects.

But this seems to lead to a problem. Indeed, suppose that, when we consider the theory up-to-one-loop level, we use the bare vev. In that case, the one-loop proper tadpole (i.e. one-loop 1-point function of the Higgs boson) will not vanish. It turns out that the one-loop proper tadpole is a divergent GF. Hence, our theory will have a divergence, which is not removed.

Yet, that is really not a problem. Divergences in the theory constitute a problem when they lead to divergent predictions. This is because the theory should describe the physical (i.e. observable) reality, which is obviously free from divergences. So, if the theory makes predictions which are divergent, then it is deeply sick. But the point is that 1-point functions do not correspond to any prediction; they are not like 2-point functions (which are intrinsically related to a physical---i.e. observable---mass), nor like 3-point functions (which are intrinsically related to a physical---i.e. observable---coupling). In 1-point functions, there is really nothing physical, i.e. observable. Therefore, the fact that they are divergent does not imply that the theory will end up making divergent observable predictions.

As we have been suggesting, physical meaning is obtained from GFs with more than one external leg. However, connected GFs with more than one external leg \textit{do not} depend on the selection of the vev. This was explicitly proven in ref. \cite{Denner:2016etu}; in what follows, we reproduce their proof.
We consider two different generating functionals for the SM: $Z[J]$, defined in terms of the field $h$, and $Z^{\prime}[J]$, defined in terms of the field $h^{\prime}$, in such a way that the two fields are related by:
\be
h^{\prime} = h - \omega
\quad
\Leftrightarrow
\quad
h = h^{\prime} + \omega,
\label{Chap-Selec:eq:trick-change}
\ee
with $\omega$ an arbitrary constant. Then, and ignoring the remaining fields of the theory,%
\fn{We omit the dependence of both the current ($J$) and the fields ($h$ and $h^{\prime}$) on the space-time coordinate ($x$).}
\bs
\bea
Z[J] &=& \int \mathcal{D} h \, \mathrm{exp}
\Big\{
i \int d^4 x 
\big[
\mathcal{L}(h) + J \, h
\big]
\Big\},
\label{Chap-Selec:eq:Z-i}
\\
Z^{\prime}[J] &=& \int \mathcal{D} h^{\prime} \, \mathrm{exp}
\Big\{
i \int d^4 x 
\big[
\mathcal{L}(h^{\prime} + \omega) + J \, h^{\prime}
\big]
\Big\}.
\label{Chap-Selec:eq:Z-ii}
\eea
\es
Note that we wrote $\mathcal{L}(h^{\prime} + \omega)$ in $Z^{\prime}[J]$, because the theory is the same as the one described by $Z[J]$, and $\mathcal{L}(h^{\prime} + \omega) = \mathcal{L}(h)$ due to eq. \ref{Chap-Selec:eq:trick-change}. Now, since the path integral measure is invariant under a constant shift, we can use eq. \ref{Chap-Selec:eq:trick-change} to rewrite eq. \ref{Chap-Selec:eq:Z-ii} as:
\be
Z^{\prime}[J] = \int \mathcal{D} h \, \mathrm{exp}
\Big\{
i \int d^4 x 
\big[
\mathcal{L}(h) + J \times \left(h - \omega\right)
\big]
\Big\}.
\label{Chap-Selec:eq:Z-iii}
\ee
Therefore, the generating functionals of the connected GFs $W[J] = \log Z[J]$ and $W^{\prime}[J] = \log Z^{\prime}[J]$ are related via:
\be
W^{\prime}[J] = W[J] - i \, \omega \int d^4x \, J(x).
\ee

As a consequence, the connected GFs of the two descriptions are related by:
%
%
\bs
\begin{numcases}{}
\dfrac{\delta^n W^{\prime}}{i \delta J(x_1)} \bigg|_{J=0}
= 
\dfrac{\delta^n W}{i \delta J(x_1)} \bigg|_{J=0} - \omega,
& for n=1, \\[2mm]
\dfrac{\delta^n W^{\prime}}{i \delta J(x_1) ... i \delta J(x_n)} \bigg|_{J=0}
= 
\dfrac{\delta^n W}{i \delta J(x_1) ... i \delta J(x_n)} \bigg|_{J=0},
& for $n > 1$.
\label{Chap-Selec:eq:connec:two}
\end{numcases}
\es
In the end, then, the connected GFs with more than one external leg ($n > 1$) do not depend on the arbitrary constant $\omega$. 
For this reason, even if two descriptions of the same theory decide to expand their field around different vevs (so that the fields differ by a constant), the connected GF with more than one external leg will be the same.%
\fn{It should be clear that this also holds at tree-level; the only requirement is that the propagator is defined with the correct mass.
To understand this aspect, suppose the $\varphi^4$ theory, $\mathcal{L} = \frac{1}{2} \partial_{\mu} \varphi \partial^{\mu} \varphi - \frac{1}{2} m^2 \varphi^2 - \frac{\lambda}{4} \varphi^4$, for $\varphi$ a real scalar field, and for $m$ and $\lambda$ real positive parameters. With this choice of parameters, there is no SSB, so that the field configuration that leads to the minimum happens for the constant solution $\varphi(x)=0$.
Now, in the most obvious description, perturbation theory is performed with the field $\varphi$ itself, which describes small excitations around the vacuum $\varphi(x)=0$. The mass of the propagator is $m$. However, one can also define $\varphi^{\prime} = \varphi - \omega$ (for an arbitrary constant $\omega$) and describe the perturbation theory using $\varphi^{\prime}$. In this case, all the terms containing $\omega$ should be considered as vertices (even the 1-point functions and the 2-point functions), which means that the mass of the propagator of $\varphi^{\prime}$ is still $m$. Then, if the symmetry factors are taken care of, and if different powers of $\omega$ are treated as formally independent (so that a proper expansion in orders of $\omega$ is performed), one can verify that the dependences on $\omega$ vanish in all connected GF with more than one external leg. Note that, in both descriptions (the one using $\varphi$ and the one using $\varphi^{\prime}$), the propagator has the same mass $m$, which must always be determined from the true minimum of the theory; that is to say, the mass of a propagator must always correspond to the coefficient of the bilinear terms of the field that describes small excitations around the true minimum (in this sense, it is mandatory to select the true vev at tree-level). As long as this is the case, perturbation theory can be consistently defined by fields that differ from each other by a constant, as we just saw. I thank Ansgar Denner for clarifications on this point.}
In sum, when condering the theory up to one-loop level, we really do not need to select the true vev to that order. This means that eq. \ref{Chap-Selec:eq:allagree}---which is a necessary condition to ensure that the true vev of the up-to-one-loop theory is selected---is really not necessary to have a consistent description of the theory.%
\fn{Contrary to what is sometimes mentioned in the literature \cite{Gambino:1999ai,Dudenas:2020ggt}, eq. \ref{Chap-Selec:eq:allagree} is not needed for a correct definition of the Legendre transformation that yields the generating functional of the 1PI GFs.}

It is relevant to ascertain what happens if eq. \ref{Chap-Selec:eq:allagree} does not hold (i.e. if the true vev of the up-to-one-loop theory is not chosen). Suppose that, in the theory considered up to one-loop level, we select the bare vev. In that case, there will be no tadpole counterterm whatsoever; in fact, if we choose $\bar{v} = v_0$ in eq. \ref{Chap-Selec:eq:consistent}, then $\delta t = 0$, due to eq. \ref{Chap-Selec:eq:fixed0}. Accordingly, we see from eq. \ref{Chap-Selec:eq:novosbonecos} that the complete 2-point function becomes:
\vs{-1mm}
\begin{equation}
\begin{minipage}[h]{.40\textwidth}
\vspace{10mm}
\begin{picture}(0,42)
\begin{fmffile}{Chap-Selec-113c} 
\begin{fmfgraph*}(40,70) 
\fmfset{arrow_len}{3mm}
\fmfset{arrow_ang}{20}
\fmfleft{nJ1}
\fmfright{nJ2}
\fmf{plain,tension=3}{nJ1,nJ1nJ2}
\fmf{plain,tension=3}{nJ1nJ2,nJ2}
\fmfv{decor.shape=circle,decor.filled=full,decor.size=6thick}{nJ1nJ2}
\end{fmfgraph*} 
\end{fmffile} 
\end{picture}
\end{minipage}
\hspace{-42mm}
\stackrel{\bar{v} = v_0}{=}
\hspace{3mm}
\begin{minipage}[h]{.40\textwidth}
\vspace{10mm}
\begin{picture}(0,42)
\begin{fmffile}{Chap-Selec-114c} 
\begin{fmfgraph*}(40,70) 
\fmfset{arrow_len}{3mm}
\fmfset{arrow_ang}{20}
\fmfleft{nJ1}
\fmfright{nJ2}
\fmf{plain,tension=3}{nJ1,nJ1nJ2}
\fmf{plain,tension=3}{nJ1nJ2,nJ2}
\fmfv{decor.shape=circle,decor.filled=hatched,decor.size=7thick}{nJ1nJ2}
\end{fmfgraph*} 
\end{fmffile} 
\end{picture}
\end{minipage}
\hspace{-42mm}
+
\hspace{3mm}
\begin{minipage}[h]{.40\textwidth}
\vspace{10mm}
\begin{picture}(0,42)
\begin{fmffile}{Chap-Selec-115c} 
\begin{fmfgraph*}(40,70) 
\fmfset{arrow_len}{3mm}
\fmfset{arrow_ang}{20}
\fmfleft{nJ1}
\fmfright{nJ2}
\fmf{plain,tension=3,label={\footnotesize \hspace{7mm} without $\delta t$},label.distance=-60thick,label.side=left}{nJ1,nJ1nJ2}
\fmf{plain,tension=3}{nJ1nJ2,nJ2}
\fmfv{decor.shape=pentagram,decor.filled=full,decor.size=5thick}{nJ1nJ2}
\end{fmfgraph*} 
\end{fmffile} 
\end{picture}
\end{minipage}
\hspace{-42mm}
+
\hspace{3mm}
\begin{minipage}[h]{.40\textwidth}
\vspace{5mm}
\begin{picture}(0,57)
\begin{fmffile}{Chap-Selec-116c}
\begin{fmfgraph*}(40,70)
\fmfset{arrow_len}{3mm}
\fmfset{arrow_ang}{20}
\fmfleft{nJ1}
\fmfright{nJ2}
\fmftop{nJ3}
\fmf{plain,tension=3}{nJ1,nJ1nJ2}
\fmf{dashes,tension=0.1}{nJ1nJ2,x}
\fmf{dashes,tension=0.1}{x,y}
\fmf{dashes,tension=0.1}{y,nJ3}
\fmf{plain,tension=3}{nJ1nJ2,nJ2}
\fmfv{decor.shape=circle,decor.filled=70,decor.size=0.8thick}{x}
\fmfv{decor.shape=circle,decor.filled=70,decor.size=0.8thick}{y}
\fmfv{decor.shape=circle,decor.filled=hatched,decor.size=7thick}{nJ3}
\end{fmfgraph*}
\end{fmffile}
\end{picture}
\end{minipage}
\hspace{-45mm} .
\label{Chap-Selec:eq:myneweq-2}
\vs{-7mm}
\end{equation}
But this is precisely what we found for the FJTS, eq. \ref{Chap-Selec:eq:myneweq}. Actually, this is hardly a surprise. To see it in detail: in the FJTS, the field we used was defined by reference to the up-to-one-loop true vev, which we wrote as $\bar{v} = v_{(0)} + \Delta v$; in eq. \ref{Chap-Selec:eq:myneweq-2}, by contrast, the field was defined by reference to the tree-level true vev $v_{(0)}$. So, the fields in the two description differ by a constant $\Delta v$. 
But as we just saw in eq. \ref{Chap-Selec:eq:connec:two}, if we define two fields that differ by a constant, the complete connected 2-point function does not depend on that constant. Hence, and since the constant $\Delta v$ is directly proportional to the tadpole counterterm (eq. \ref{Chap-Selec:eq:Dv}), we conclude that, in the FJTS, the complete connected 2-point function does not depend on the tadpole counterterm. Or, which is the same, the terms in the complete 2-point function that depend on $\delta t$ must cancel:
%
%
%
\begin{equation}
\begin{minipage}[h]{.40\textwidth}
\vspace{10mm}
\begin{picture}(0,42)
\begin{fmffile}{Chap-Selec-614b} 
\begin{fmfgraph*}(40,70) 
\fmfset{arrow_len}{3mm}
\fmfset{arrow_ang}{20}
\fmfleft{nJ1}
\fmfright{nJ2}
\fmf{plain,tension=3,label={\footnotesize \hspace{7mm} $\delta t$-part},label.distance=-60thick,label.side=left}{nJ1,nJ1nJ2}
\fmf{plain,tension=3}{nJ1nJ2,nJ2}
\fmfv{decor.shape=pentagram,decor.filled=full,decor.size=5thick}{nJ1nJ2}
\end{fmfgraph*} 
\end{fmffile} 
\end{picture}
\end{minipage}
\hspace{-44mm}
\FJeq
\hspace{1mm}
-
\hspace{1mm}
\begin{minipage}[h]{.40\textwidth}
\vspace{-10.5mm}
\begin{picture}(0,100)
\begin{fmffile}{Chap-Selec-617}
\begin{fmfgraph*}(40,70)
\fmfset{arrow_len}{3mm}
\fmfset{arrow_ang}{20}
\fmfleft{nJ1}
\fmfright{nJ2}
\fmftop{nJ3}
\fmf{plain,tension=3}{nJ1,nJ1nJ2}
\fmf{dashes,tension=0.1}{nJ1nJ2,x}
\fmf{dashes,tension=0.1}{x,y}
\fmf{dashes,tension=0.1}{y,nJ3}
\fmf{plain,tension=3}{nJ1nJ2,nJ2}
\fmfv{decor.shape=circle,decor.filled=70,decor.size=0.8thick}{x}
\fmfv{decor.shape=circle,decor.filled=70,decor.size=0.8thick}{y}
\fmfv{decor.shape=pentagram,decor.filled=full,decor.size=6thick}{nJ3}
\end{fmfgraph*}
\end{fmffile}
\end{picture}
\end{minipage}
\hspace{-44mm}.
\label{Chap-Selec:eq:novosbonecos-2}
\vs{-7mm}
\end{equation}
A similar relation holds also for 3-point functions. In conclusion, the FJTS (where the true vev at up-to-one-loop level was selected due to the introduction of the quantity $\Delta v$) is equivalent to a scheme where the bare vev is selected. Note that, in this case (i.e. if the bare vev is chosen), one still finds the advantages of the FJTS which result from the use of that vev in the bare relations---namely, the fact that tree-level relations can be immediatly converted into bare relations, as well as the advantages concerning gauge independence.

Despite of this, one can argue that, when considering the theory up-to-one-loop level, it is convenient to select the true up-to-one-loop vev. The advantage would be that, with that choice, eq. \ref{Chap-Selec:eq:bonecos} holds, which implies that the last two terms of the r.h.s. of eq. \ref{Chap-Selec:eq:novosbonecos} cancel, which in turn implies that one does not have to consider reducible diagrams with one-loop tadpole insertions (third term of the r.h.s. of eq. \ref{Chap-Selec:eq:novosbonecos}).
However, as we discussed at the end of section \ref{Chap-Selec:sec:FJTS}, we believe that there is no inconvenience in such reducible diagrams. As a matter of fact, we find it highly preferable to let the terms with tadpole counterterms cancel (as in eqs. \ref{Chap-Selec:eq:myneweq} and \ref{Chap-Selec:eq:myneweq-2}).

\subsection{Renormalization}
\label{Chap-Selec:sec:reno}

\subsubsection{Renormalized tadpoles?}

Let us go back to the case where we select the true vev, as we have been discussing in the previous sections. This is by far the most common approach in the literature, and usually goes by the name of
%
%
\textit{tadpole renormalization}%
~\cite{
Krause:2016oke,Krause:2017mal,Altenkamp:2017ldc,Denner:2017vms,Krause:2018wmo,Dudenas:2018wlr,Denner:2018opp,Krause:2019oar,Dao:2019nxi,Altenkamp:2019wht,Denner:2016etu,Denner:2019xti,Denner:2019vbn,Dudenas:2020ggt}.
Some authors~\cite{Altenkamp:2019wht} even define $t_{(0)}$ and $t$ (corresponding to the bare and the renormalized tadpole, respectively), such that:
\be
t_{(0)} = t + \delta t.
\label{Chap-Selec:eq:tadeq}
\ee
Then, in the FJTS, $t_{(0)}$ is taken to be:
\be
t_{(0)} \FJeq v_{(0)} \left(\mu_{(0)}^2 - \lambda_{(0)} v_{(0)}^2 \right),
\label{Chap-Selec:eq:t0FJ}
\ee
which is zero due to eq. \ref{Chap-Selec:eq:fixed0}. Then, using eqs. \ref{Chap-Selec:eq:allagree} and \ref{Chap-Selec:eq:tadeq},
\be
t \FJeq T_h.
\label{Chap-Selec:eq:tFJ}
\ee
By contrast, in the PRTS, $t_{(0)}$ is taken to be:
\be
t_{(0)} \PReq v \left(\mu_{(0)}^2 -  \lambda_{(0)} v^2 \right),
\label{Chap-Selec:eq:t0PR}
\ee
which is not zero, but rather equal to $\delta t$, through eq. \ref{Chap-Selec:eq:deltatPR}. Using eq. \ref{Chap-Selec:eq:tadeq}, it then follows that:
\be
t \PReq 0.
\ee
%
%
\subsubsection{A misleading name}
\label{Chap-Selec:sec:misleading}

Although this interpretation---according to which tadpoles were renormalized---is sometimes presented, we find it highly misleading, for two main reasons.

First, renormalization is a process through which unphysical divergences are removed from the theory. In the previous sections, however, our focus was never divergences, but only tadpoles. Our goal was never to remove the divergences of tadpoles, but rather to remove tadpoles altogether. Even in section \ref{Chap-Selec:sec:need}, where we realized that it is not imperative to select the true up-to-one-loop vev (and hence to remove proper tapoles), the discussion was centered on the selection of the vev; indeed, the main question was `which vev to select?', and not `how to remove divergences?'.


Second, a very simple argument against the thesis according to which we renormalized something is the fact that the quantities that we are left with after the `renormalization' are not finite at all. Indeed, in an attempt to be consistent with the idea of tadpole renormalization, eq. \ref{Chap-Selec:eq:tadeq} is introduced. This equation just looks like a regular renormalization equation, similar to eqs. \ref{Chap-Selec:eq:mHexample} and \ref{Chap-Selec:eq:cwren}, with a non-renormalized term ($t_{(0)}$), a renormalized one ($t$) and a counterterm ($\delta t$). But this equation is misleading in both tadpole schemes: in the FJTS, and as we just saw in eq. \ref{Chap-Selec:eq:tFJ}, the `renormalized' parameter $t$ is equal to $T_h$, which is the one-loop proper tadpole and therefore a divergent quantity. In the PRTS, the parameter we are left with is not even the renormalized one, but rather the quantity given in eq. \ref{Chap-Selec:eq:t0PR}, which is also divergent.
The same applies, by the way, to the vevs: as we are about to discuss in detail, both $v_{(0)}$ and $v$ are divergent.%
\fn{Moreover, the parameter $\Delta v$ introduced in the FJTS in eq. \ref{Chap-Selec:eq:FJbasic} is no counterterm; rather, it is a parameter that eliminates one-loop proper tadpoles. Contrary to counterterms, indeed, it allows to renormalize nothing: neither the tadpole (recall eq. \ref{Chap-Selec:eq:tFJ}), nor the vev. As a matter of fact, the quantities $v_{(0)}$, $\Delta v$ and $v = v_{(0)} + \Delta v$ are all divergent.}
Therefore, by talking about renormalization, one can be mislead and conclude that the remaining parameters really are renormalized, i.e. finite.

One can argue that eq. \ref{Chap-Selec:eq:condition} (or, equivalently, eq. \ref{Chap-Selec:eq:allagree}) is just a renormalization condition for the tadpole, and that other renormalization conditions would be possible, according to what we discussed in section \ref{Chap-Selec:sec:need}. But this section (\ref{Chap-Selec:sec:need}), by clearly demonstrating that one does not have to get rid of divergences of 1-point functions, shows that the discussion of which vev to select
is \textit{not a question of renormalization}. That is to say, eq. \ref{Chap-Selec:eq:condition} is not a renormalization condition among others, but a selection of vev among others.

All this should convince us that we should not use the term `renormalization' for the subject underlying the different tadpole schemes. In our point of view, the correct description is \textit{selection of the true vev}. Or, if the true vev is not selected (as suggested in section \ref{Chap-Selec:sec:need}), the description would be simply \textit{selection of the vev}.
%
%
Unfortunately for the sake of clarity, we were led to identify the l.h.s. of eq. \ref{Chap-Selec:eq:condition} with a counterterm, since terms with broad tadpoles can be included in the counterterms of GFs, as in eqs. \ref{Chap-Selec:eq:Wprop-counter} and \ref{Chap-Selec:eq:h3ct}. In any event, we find counterproductive the introduction of eq. \ref{Chap-Selec:eq:tadeq}, and we believe that the notion `tadpole counterterm' should be used with caution.

Finally, this also means that, unless one actually uses the relations of the theory to choose the vev as an independent parameter, it makes no sense to include the vev---or the tadpole, for that matter---in the list of independent parameters of the theory, as if it needed to be renormalized. As we saw in section \ref{Chap-Selec:sec:indep}, only independent parameters need to be renormalized; and the vev is \textit{not} chosen as independent parameter in the usual approaches to the SM (in particular, it is not independent neither in eq. \ref{Chap-Selec:eq:setoriginal0} nor in eq. \ref{Chap-Selec:eq:setsecond0}).


\subsubsection{Actual renormalization}
\label{Chap-Selec:sec:actual}

The present discussion becomes more interesting when we realize that, \textit{after} the selection of the true vev, one \textit{does} speak of renormalization. Indeed, as we just saw, after we eliminated proper tadpoles through the choice of the true vev, the parameters we are left with are divergent. Actually, not only the broad tadpoles are divergent (they correspond to one-loop GFs), but the vev used to define the bare masses as well. This is not surprising, since these are precisely \textit{bare} masses, so that there is no reason why the vevs involved in their definition---$v_{(0)}$ in the FJTS and $v$ in the PRTS---should be finite.%
\fn{The fact that $v$ is the true vev of the theory might lead us to think that it has some physical meaning, in which case the vev used to define the masses in the PRTS would need to be finite. But this is incorrect: $v$ has no physical meaning, i.e. there is no observable associated to $v$.}
These vevs have no physical meaning whatsoever: not only can they be gauge dependent (as $v$ in the PRTS is), but they can also be divergent---and both $v_{(0)}$ in the FJTS and $v$ in the PRTS are, in fact, divergent. In the following, we call the vev used to define the masses the bare-mass-defining-vev, or bmd-vev for short. In the FJTS, then, the bmd-vev is $v_{(0)}$, while in the PRTS it is $v$.

Now, as we suggested, unless one uses the relations of the theory to replace one of the independent parameters in eq. \ref{Chap-Selec:eq:setsecond0} by the bmd-vev, the latter will be a dependent variable. In that case, then, the bmd-vev is just like the cosine of the weak angle discussed in section \ref{Chap-Selec:sec:conv}: on the one hand, it is fixed (in this case, by the mass relations); on the other hand, it does not need to be renormalized.
If one decides not to renormalize it, one must replace it everywhere by the parameters it depends on, and the latter must be renormalized, just as we saw for $c_{\text{w}}$. However, and as we also saw in that case, it may be convenient to renormalize the bmd-vev.%
\fn{\label{Chap-Selec:note:relevant}In the literature, what usually happens is that the bmd-vev is not renormalized, whereas the sine and cosine of the weak angle are renormalized for convenience~\cite{Denner:1991kt,Denner:2019vbn}. What this means is that, while the bmd-vev is replaced by the quantities it depends on and the latter are renormalized (so that the bmd-vev does not show up anymore in the theory), the bare parameters $s_{{\text{w}}(0)}$ and $c_{{\text{w}}(0)}$ are replaced by renormalized parameters and counterterms, and these renormalized parameters and counterterms in general contribute to the Feynman rules.}
%
%
%
%
To see this in detail, we consider again separately the different tadpole schemes.

We start with the FJTS. Based on eq. \ref{Chap-Selec:eq:FJmasses}, we can write:
\be
v_{(0)} \FJeq \dfrac{2 \, m_{\mathrm{W}(0)} \, s_{{\text{w}}(0)}}{e_{(0)}}.
\label{Chap-Selec:eq:vFJbare}
\ee
Note that $m_{\mathrm{W}(0)}$ and $e_{(0)}$ belong to eq. \ref{Chap-Selec:eq:setsecond0}, so that they must be renormalized, while $s_{{\text{w}}(0)}$ must not (although we renormalize it for convenience, as we did with $c_{{\text{w}}(0)}$ in section \ref{Chap-Selec:sec:conv}). To proceed to the renormalization of these parameters, we define the counterterms $\delta m_{\mathrm{W}}^{2 \, \mathrm{FJ}}$, $\delta Z_e^{\mathrm{FJ}}$ and $\delta s_{\text{w}}^{\mathrm{FJ}}$ such that:
\be
m_{\mathrm{W}(0)}^{2} \FJeq m_{\mathrm{W}}^2 + \delta m_{\mathrm{W}}^{2 \, \mathrm{FJ}},
\qquad 
e_{(0)} \FJeq e \left(1 + \delta Z_e^{\mathrm{FJ}}\right),
\qquad
s_{{\text{w}}(0)} \FJeq s_{\text{w}} + \delta s_{\text{w}}^{\mathrm{FJ}}.
\label{Chap-Selec:eq:somedefs}
\ee
At this point, if we want to have an equivalent to eq. \ref{Chap-Selec:eq:vFJbare} for renormalized quantities, we must introduce the parameter $v_{\mathrm{R}}$:
\be
v_{\mathrm{R}} = \dfrac{2 \, m_{\mathrm{W}} \, s_{\text{w}}}{e}.
\label{Chap-Selec:eq:vFJreno}
\ee
Just like in the case of the cosine of the weak mixing angle, we can relate $v_{(0)}$ and $v_{\mathrm{R}}$ by expanding the r.h.s. of eq. \ref{Chap-Selec:eq:vFJbare}. Then, if we define a counterterm $\delta Z_v^{\mathrm{FJ}}$ obeying the relation
\be
v_{(0)} \FJeq v_{\mathrm{R}} \left(1 + \delta Z_v^{\mathrm{FJ}}\right),
\label{Chap-Selec:eq:renov}
\ee
we conclude that:
\be
\delta Z_v^{\mathrm{FJ}} = \dfrac{\delta m_{\mathrm{W}}^{2 \, \mathrm{FJ}}}{2 m_{\mathrm{W}}^2} + \dfrac{\delta s_{\text{w}}^{\mathrm{FJ}}}{s_{\mathrm{w}}} - \delta Z_e^{\mathrm{FJ}}.
\label{Chap-Selec:eq:deltaZvFJ}
\ee
Some aspects should not be neglected here. First, contrary to what usually happens (e.g. in eq. \ref{Chap-Selec:eq:somedefs}), we cannot define the renormalized quantity in eq. \ref{Chap-Selec:eq:renov} without index---since the parameter $v$ is already attributed to the true vev; we thus defined $v_{\mathrm{R}}$. Second, this parameter $v_{\mathrm{R}}$ is indeed renormalized and thus finite. Third, the counterterm $\delta Z_v^{\mathrm{FJ}}$ is fixed and exactly equal to the totality of counterterms that showed up at the expansion of eq. \ref{Chap-Selec:eq:vFJbare}. Finally, it is only when we renormalize the bmd-vev that we can employ the relation eq. \ref{Chap-Selec:eq:vFJreno} (it is a matter of course that one can always assume that eq. \ref{Chap-Selec:eq:vFJreno} is valid and not mention explicitly any counterterm $\delta Z_v^{\mathrm{FJ}}$; in doing so, however, eqs. \ref{Chap-Selec:eq:renov} and \ref{Chap-Selec:eq:deltaZvFJ} are implicitly used).

In the PRTS, the situation is different. Here, the bmd-vev is $v$, not $v_{(0)}$; we can use eq. \ref{Chap-Selec:eq:PRmasses} to write
\be
v \PReq  \dfrac{2 \, \, m_{\mathrm{W}(0)} \, \, s_{{\text{w}}(0)}}{e_{(0)}},
\label{Chap-Selec:eq:vPRbare}
\ee
Then, proceeding in a similar way as before, we now introduce the relation
\be
v \PReq v_{\mathrm{R}} \left(1 + \delta Z_v^{\mathrm{PR}}\right),
\label{Chap-Selec:eq:vev-ren-PR}
\ee
which ends up leading to:
\be
\delta Z_v^{\mathrm{PR}} = \dfrac{\delta m_{\mathrm{W}}^{2 \, \mathrm{PR}}}{2 m_{\mathrm{W}}^2} + \dfrac{\delta s_{\text{w}}^{\mathrm{PR}}}{s_{\mathrm{w}}} - \delta Z_e^{\mathrm{PR}}.
\label{Chap-Selec:eq:deltaZvPR}
\ee
By comparing eqs. \ref{Chap-Selec:eq:deltaZvFJ} and \ref{Chap-Selec:eq:deltaZvPR}, we realize the importance of using the superscripts FJ or PR: the two equations are different (since the counterterms involved are calculated in different ways in the different tadpole schemes), although they would look the same without the superscripts.

\subsubsection{An alternative approach}
\label{Chap-Selec:sec:alternative}

We have been describing a general approach to the treatment of the theory when considered up to one-loop level, according to which one is faced with two different tasks (in the following order): first, one chooses which vev to select (and, possibly, through which way or scheme to perform that selection); then, one takes care of renormalizing the theory.

An alternative approach is also possible. It aims \textit{ab initio} at a single goal: removing all divergences of the 1PI GFs. In that case, all one is really concerned about are divergent parts (finite parts can be addressed in a later moment).
Just as in our approach, an essential question is: how should one treat the vev?
In the alternative approach we are now discussing, one usually considers a counterterm for the vev, as if the vev was a normal parameter. Then, three questions become relevant:
\begin{enumerate}
\item Is such counterterm really necessary (in order for all divergences of the 1PI GFs to be removed)?
\item If so, what value should it have?
\item How is such counterterm related to other counterterms?
\end{enumerate}
In simple models, such as the linear sigma model, one finds that a counterterm for the vev is necessary, although it is the same as the counterterm for the scalar field \cite{Peskin:1995ev,Bohm:2001yx}. Consequently, the counterterms available in the theory before SSB are enough to remove all divergences (cf. also ref. \cite{Srednicki:2007qs}).
In more complex theories, such as the SM, the situation is more subtle, due to gauge-fixing terms. A vev counterterm is also necessary, but it will only be equal to the counterterm for the scalar field in some specific gauges \cite{Chankowski:1991md,Sperling:2013eva,Grimus:2018rte,Dudenas:2020ggt}; in general, one needs a vev counterterm that is independent of the counterterm for the scalar field.

Although such an alternative approach is legitimate, we find it less transparent, and we will not consider it in the following.

\section{2HDM}
\label{Chap-Selec:sec:2HDM}

So far, we have been exploring properties of tadpole schemes in the SM, where there is just a single Higgs doublet. New features appear when one considers additional doublets. We discuss the simplest case, that of two Higgs doublets. We start with the case where CP is assumed to be conserved in the potential, and then address the case with general CP violation.
Although the former configures a description which, as we shall argue in chapter \ref{Chap-Real}, is not theoretically sound, the discussion of vev selection (and tadpole schemes) in that case is nonetheless relevant, for two reasons: first, it has been repeatedly considered in recent literature \cite{Krause:2016oke,Krause:2016xku,Kanemura:2017wtm,Krause:2017mal,Altenkamp:2017ldc,Krause:2018wmo,Dudenas:2018wlr,Denner:2018opp,Krause:2019oar,Dao:2019nxi,Altenkamp:2019wht,Denner:2016etu,Denner:2019xti,Denner:2017vms,Krause:2019qwe}; second, a model with two Higgs doublets and CP conservation, by itself, is not unsound;%
\fn{The problem with the real 2HDM, as shall be seen, is that it enforces CP conservation in a part of the theory, while allowing CP violation in another.}
hence, the discussion of the selection of the vev in that case is not only legitimate, but also instructive.
We do not provide an exhaustive treatment neither in the CP-conserving case nor in the CP-violating; rather, we illustrate only the relevant new features concerning tadpole schemes.

\subsection{CP conservation}

%
%

We begin with the tree-level description.
Supposing two Higgs doublets $\Phi_1$ and $\Phi_2$, we have:
%
\be
\langle\Phi_1\rangle
=
\begin{pmatrix} 0 \\ \dfrac{v_1}{\sqrt{2}} \end{pmatrix},
\qquad
\langle\Phi_2\rangle
=
\begin{pmatrix} 0 \\ \dfrac{v_2}{\sqrt{2}} \end{pmatrix},
\label{Chap-Selec:eq:doublet2H}
\ee
with $v_1$ and $v_2$ real vevs.
It is usual to define the variable $\beta$ such that:
\be
\tan \beta = \dfrac{v_2}{v_1},
\label{Chap-Selec:eq:tanbtree}
\ee
as well as the total vev $v$, such that:
\be
v^2 = v_1^2 + v_2^2,
\label{Chap-Selec:eq:totalvevtree}
\ee
with:
\be
v = \dfrac{2 \, m_{\mathrm{W}} \, s_{\text{w}}}{e}.
\ee

Let us now consider what happens when we consider the theory up to one-loop level. The case with two scalar doublets is a simple generalization of that with just one scalar doublet. In the SM, when considering the theory up-to-one-loop level, we discussed how to select the true vev, so that all proper tadpoles would vanish from the theory. With two scalar doublets, there are two vevs, but the principle is the same. Specifically, one may select the true vevs---identified in the following (in the context of the theory considered up to one-loop level) with $v_1$ and $v_2$---, which again guarantees that no proper tadpoles show up. 


\subsubsection{FJTS}

The FJTS for a general Higgs sector was studied in detail in ref.~\cite{Denner:2016etu}. Here, we present but a brief sketch of its application to a theory with two scalar doublets. As a trivial extension of what we found with one doublet, the bare doublets $\Phi_{1(0)}$ and $\Phi_{2(0)}$ are written in terms of $\bar{v}_1$ and $\bar{v}_2$, respectively. Each one is split into two terms:
\be
\bar{v}_1 = v_{1(0)} + \Delta v_1, 
\qquad
\bar{v}_2 = v_{2(0)} + \Delta v_2,
\label{Chap-Selec:eq:mysplit}
\ee
where $v_{1(0)}$ and $v_{2(0)}$ correspond to the bare vevs (the vevs that were the true vevs in the bare theory) and $\Delta v_1$ and $\Delta v_2$ to the quantities that will cancel the one-loop tadpole terms. Similarly to what we had in eq.~\ref{Chap-Selec:eq:FJmasses}, the bare masses of the particles of the theory are defined in terms of the bare total vev,
\be
v_{(0)} = \sqrt{v_{1(0)}^2 + v_{2(0)}^2} \, ,
\ee
an expression that directly follows from eq. \ref{Chap-Selec:eq:totalvevtree}.
Concerning $\Delta v_1$ and $\Delta v_2$, one needs to rotate them into the mass basis (i.e. the basis where the scalar fields are well-defined) in order for them to cancel the one-loop tadpole terms of the physical scalar fields.
By defining the CP-even physical scalar fields with the usual notation $h$ and $H$ (with mass $m_h$ and $m_H$, respectively),
one can accordingly define the rotated versions of $\Delta v_1$ and $\Delta v_2$
as $\Delta v_h$ and $\Delta v_H$. Then, one can prove the relations:%
\fn{For details, cf. section \ref{Chap-Reno:sec:FJTS-C2HDM}.}
\be
\Delta v_h = \dfrac{T_h}{m_h^2},
\qquad 
\Delta v_H = \dfrac{T_H}{m_H^2},
\ee
where $T_h$ and $T_H$ are the one-loop tadpoles of $h$ and $H$, respectively.
Following the same line of thought as in the SM, one concludes that all the broad tadpoles in the FJTS can be consistently accounted for by including all the possible reducible diagrams with one-loop tadpoles in all possible GFs.%
\fn{In theories with two scalar doublets, the variable $\beta$ usually represents not only a relation between the two vevs, but also a mixing angle that establishes the connection between two different bases of fields (cf. eq. \ref{Chap-Real:eq:HiggsBasis} below). Now, it is sometimes read in the literature \cite{Krause:2017mal} that, when using the FJTS, one has to be careful to distinguish between the two meanings of $\beta$. The point is that, if a certain term involves $\beta$ in the sense of a relation between the two vevs, then such term follows from the expansion of the scalar doublets; in that case, the splitting in eq. \ref{Chap-Selec:eq:mysplit} will generate---besides the term at stake---also terms with $\Delta v_1$ or $\Delta v_2$. The same will not happen for terms involving $\beta$ in the sense of a mixing angle. As a consequence---so the argument goes---, one must be careful to distinguish the two scenarios. However, as already noted, all the terms with $\Delta v_1$ or $\Delta v_2$---that is, all the terms with broad tadpoles---can be accounted for by including all the possible reducible diagrams with one-loop tadpoles in all possible GFs. In this case, there is no need to distinguish the two different meanings of $\beta$.}
The generalization of the FJTS from one scalar doublet to two is thus quite simple.

The use of $\beta$ in the context of the FJTS is also straightforward. As we saw, in the FJTS the bare relations can be immediately derived from the tree-level ones, since the bare vevs are used in the definitions. Then, the bare version of eq. \ref{Chap-Selec:eq:tanbtree} in the FJTS is:%
\fn{A couple of points should be stressed here. First, note that the definition of $\beta_{(0)}$ is \textit{not} modified by $\Delta v_1$ and $\Delta v_2$. Recall that these variables were only introduced in the scalar doublets in order to cancel the one-loop tadpoles, as we saw in detail in the case of the SM. In that case, it would make no sense to change the definition of $\beta_{(0)}$ to include such quantities. Actually, if we did so, we would be absorbing some of the broad tadpoles in the definition of $\beta_{(0)}$---which would render a consistent use of the FJTS very complicated. Indeed, one would not be able to include in all possible GFs all the reducible diagrams with one-loop tadpoles, as some of them would be absorbed inside $\beta_{(0)}$. It follows that, in the FJTS, even if $\bar{v}_1$ and $\bar{v}_2$ are each split into two terms, the definition of $\beta_{(0)}$ is still given by eq. \ref{Chap-Selec:eq:tanbbareFJTS}.
Second, it should be clear that using $\beta_{(0)}$ and $v_{(0)}$ is simply an alternative to using $v_{1(0)}$ and $v_{2(0)}$. And just as in the SM, in the FJTS, the renormalization of $v_{(0)}$ (discussed in section \ref{Chap-Selec:sec:actual}) has nothing to do with $\Delta v$ and the elimination of proper tadpoles (discussed in section \ref{Chap-Selec:sec:FJTS}), the same happens with two scalar doublets in the FJTS: the renormalization of $v_{1(0)}$ and $v_{2(0)}$ is independent of $\Delta v_i$ and the elimination of proper tadpoles. Therefore, since $\beta_{(0)}$ and $v_{(0)}$ are just an alternative to $v_{1(0)}$ and $v_{2(0)}$, the same conclusion follows for them, namely: in the FJTS, the renormalization of $\beta_{(0)}$ and $v_{(0)}$ is independent of $\Delta v_i$ and the elimination of proper tadpoles.}
\be
\tan \beta_{(0)} \FJeq \dfrac{v_{2(0)}}{v_{1(0)}}.
\label{Chap-Selec:eq:tanbbareFJTS}
\ee

When renormalizing theories with two scalar doublets, the variable $\beta$ is usually taken as an independent parameter, whereas $v_1$, $v_2$ and $v$ are not. As such, $\beta$ must be renormalized, according to:
\be
\beta_{(0)} = \beta + \delta \beta,
\ee
whereas $v_1$, $v_2$ and $v$ must not. And while one could always renormalize these variables for convenience (as we did with $c_{\text{w}}$ above), there is no real advantage in doing so. Thus, immediately before renormalization, one should use the dependence relations of the theory to replace those variables for independent parameters. Specifically, one should use eq. \ref{Chap-Selec:eq:vFJbare}, which also allows one to write
\be
v_{1(0)} \FJeq \dfrac{2 \, m_{\mathrm{W}(0)} \, s_{{\text{w}}(0)}}{e_{(0)}} \cos \beta_{(0)},
\qquad
v_{2(0)} \FJeq \dfrac{2 \, m_{\mathrm{W}(0)} \, s_{{\text{w}}(0)}}{e_{(0)}} \sin \beta_{(0)},
\ee
and then renormalize $e$, $m_{\mathrm{W}}$, $s_{\text{w}}$ and $\beta$ as usual.%
\fn{
\label{Chap-Selec:note:alternative}
Alternatively, should one had decided to renormalize $v_1$ and $v_2$ for convenience, one could define $v_{1\mathrm{R}}$ and $v_{2\mathrm{R}}$, such that
\be
v_{1(0)} \FJeq v_{1\mathrm{R}} \left( 1 + \delta Z_{v_1}^{\mathrm{FJ}} \right),
\qquad 
v_{2(0)} \FJeq v_{2\mathrm{R}} \left( 1 + \delta Z_{v_2}^{\mathrm{FJ}} \right), 
\ee
so that the renormalized $\beta$ would obey the relation $\tan \beta = v_{2\mathrm{R}} / v_{1\mathrm{R}}$. Then, together with the definitions:
\be
v_{\mathrm{R}}^2 = v_{1\mathrm{R}}^2 + v_{2\mathrm{R}}^2,
\qquad
v_{(0)} \FJeq v_{\mathrm{R}} \left( 1 + \delta Z_v^{\mathrm{FJ}}\right),
\ee
this would lead to the relations:
\be
\delta \beta = \dfrac{v_{1\mathrm{R}} \, v_{2\mathrm{R}}}{v_{\mathrm{R}}^2} \left(\delta Z_{v_2}^{\mathrm{FJ}} - Z_{v_1}^{\mathrm{FJ}}\right), \\
\qquad 
\delta Z_v^{\mathrm{FJ}} = \dfrac{v_{1\mathrm{R}}^2 \delta Z_{v_1}^{\mathrm{FJ}} + v_{2\mathrm{R}}^2 \delta Z_{v_2}^{\mathrm{FJ}}}{v_{\mathrm{R}}^2}.
\ee
}

\subsubsection{PRTS}
\label{Chap-Selec:sec:2HDM-PRTS}

In the PRTS with two scalar doublets, the procedure is similar to that of the SM: for one thing, the quantities $\bar{v}_1$ and $\bar{v}_2$ are immediately taken as the true vevs; besides, the bare relations that depend on the vevs are written in terms of the true vevs. This means that, just as the SM tree-level relation for the mass of the $W$-boson was $m_{\mathrm{W}} = \frac{1}{2} g_{2} v$, but in such a way that the PRTS bare version of this relation was $m_{\mathrm{W}(0)} \PReq \frac{1}{2} g_{2(0)} v$ (with the true vev $v$, not with the bare vev $v_{(0)}$), in the same way the tree-level $\beta$ obeys eq. \ref{Chap-Selec:eq:tanbtree}, but the PRTS bare version of this relation is:
\be
\tan \beta_{(0)} \PReq \dfrac{v_2}{v_1}.
\ee
Assuming---as in the FJTS---that $\beta$ is chosen as an independent parameter, and $v_1$, $v_2$ and $v$ as dependent ones, one can use eq. \ref{Chap-Selec:eq:vPRbare} to write:
\be
v_{1} \PReq \dfrac{2 \, \, m_{\mathrm{W}(0)} \, \, s_{{\text{w}}(0)}}{e_{(0)}} \cos \beta_{(0)},
\qquad 
v_{2} \PReq \dfrac{2 \, \, m_{\mathrm{W}(0)} \, \, s_{{\text{w}}(0)}}{e_{(0)}} \sin \beta_{(0)},
\ee
and then renormalize $e$, $m_{\mathrm{W}}$, $s_{\text{w}}$ and $\beta$ as usual.%
\fn{
We follow here an equivalent derivation to the one in note \ref{Chap-Selec:note:alternative}, but now for the PRTS.
Should one had decided to renormalize $v_1$ and $v_2$ for convenience, one could define $v_{1\mathrm{R}}$ and $v_{2\mathrm{R}}$, such that
\be
\label{Chap-Selec:eq:PRTS-vev-reno}
v_{1} \PReq v_{1\mathrm{R}} \left( 1 + \delta Z_{v_1}^{\mathrm{PR}} \right),
\qquad
v_{2} \PReq v_{2\mathrm{R}} \left( 1 + \delta Z_{v_2}^{\mathrm{PR}} \right), 
\ee
so that the renormalized $\beta$ would obey the relation $\tan \beta = v_{2\mathrm{R}} / v_{1\mathrm{R}}$. Then, together with the definitions:
\be
v_{\mathrm{R}}^2 = v_{1\mathrm{R}}^2 + v_{2\mathrm{R}}^2,
\qquad 
v \PReq v_{\mathrm{R}} \left( 1 + \delta Z_v^{\mathrm{PR}}\right),
\ee
this would lead to the relations:
\be
\delta \beta = \dfrac{v_{1\mathrm{R}} \, v_{2\mathrm{R}}}{v_{\mathrm{R}}^2} \left(\delta Z_{v_2}^{\mathrm{PR}} - Z_{v_1}^{\mathrm{PR}}\right),
\qquad 
\delta Z_v^{\mathrm{PR}} = \dfrac{v_{1\mathrm{R}}^2 \delta Z_{v_1}^{\mathrm{PR}} + v_{2\mathrm{R}}^2 \delta Z_{v_2}^{\mathrm{PR}}}{v_{\mathrm{R}}^2}.
\ee
}

Finally, recall that, in the PRTS, one cannot in general account for all the broad tadpoles (which, in the PRTS, only show up for terms in the potential) by considering reducible diagrams with one-loop tadpoles. As described at the end of section \ref{Chap-Selec:sec:SM-PRTS}, one can get a consistent theory by rewriting the tree-level relations assuming that tadpoles do not vanish, and then, when considering the theory up to one-loop level, identify the bare tadpoles with their corresponding tadpole counterterms. Only, while in the SM only two tree-level relations needed to be modified in order to account for all the broad tadpoles (recall eqs. \ref{Chap-Selec:eq:lambda2} and \ref{Chap-Selec:eq:lambdaPR}), in a model with two Higgs doublets there will be more (cf. e.g. eqs. 2.19 of ref. \cite{Altenkamp:2017ldc}).

\subsection{CP violation}
\label{Chap-Selec:sec:CPV}

The existence of CP violation leads to generally complex vevs (or, which is equivalent, phases besides the real vevs), which constitute the main novelty when compared to the scenarios studied above. We discuss in detail the FJTS in section \ref{Chap-Reno:sec:FJTS-C2HDM}.

Here, we restrict ourselves to a very brief comment on the role of phases of the vevs in the PRTS.
Similarly to what we did before, when we consider the theory up to one-loop level, we write the expectation values of the two bare doublets as:
\be
\langle\Phi_{1(0)}\rangle= 
\begin{pmatrix}
0 \\
e^{i \bar{\zeta}_1} \frac{\bar{v}_1}{\sqrt{2}}
\end{pmatrix},
\qquad
\langle\Phi_{2(0)}\rangle= 
\begin{pmatrix}
0 \\
e^{i \bar{\zeta}_2} \frac{\bar{v}_2}{\sqrt{2}}
\end{pmatrix},
\label{Chap-Selec:eq:ExpecPR}
\ee
where the quantities $\bar{\zeta}_1$, $\bar{\zeta}_2$, $\bar{v}_1$ and $\bar{v}_2$ are real. As usual in the PRTS, these quantities are identified with the true quantities (without bars):
\be
\bar{\zeta}_1 = \zeta_1,
\quad
\bar{\zeta}_2 = \zeta_2,
\quad
\bar{v}_1 = v_1,
\quad
\bar{v}_2 = v_2.
\ee
Hence, just as in the CP-conserving case one had true vevs ($v_1$ and $v_2$), in the CP-violating one also has true phases ($\zeta_1$ and $\zeta_2$). And, just as in the previous models, the values of these true quantities (which are in general divergent) are fixed such that there are no proper tadpoles. This can be ensured in the usual way: rewriting the tree-level relations in terms of bare tadpoles and, when considering the theory in the context of one-loop level, replacing the bare tadpoles by the corresponding counterterms (just as was done above in the SM in eqs. \ref{Chap-Selec:eq:lambda2} and \ref{Chap-Selec:eq:lambdaPR}).
The values of $v_1$, $v_2$, $\zeta_1$ and $\zeta_2$ in those relations thus correspond to the true values.%
\fn{The true phases $\zeta_1$ and $\zeta_2$ will in general be non-null. However, one can always use the rephasing freedom of the theory to absorb the (true) values of $\zeta_1$ and $\zeta_2$ in the remaining parameters of the theory.}

\section{Summary}
\label{Chap-Selec:sec:summary}

We discussed the selection of the true vev in models with SSB, focusing especially on the SM.
Whereas this selection is trivial when the theory is taken at tree-level, it becomes more involved when one goes up to one-loop level.
On the other hand, although such selection is necessary in the first case,
it is not so in the second. In fact, as we showed, one can find a consistent description of the theory using connected GFs, even if divergent one-loop 1-point functions are not removed.
Nevertheless, we focused on the selection of the true vev of the up-to-one-loop theory.

We described in detail two different ways to implement such selection: the tadpole schemes FJTS and PRTS. We have seen that, by assuring that the theory is free from one-loop 1-point functions (identified as proper tadpoles), both schemes end up introducing one-loop tadpoles contributing to other GFs (identified as broad tadpoles).
After comparing the two schemes, it becomes patent that the FJTS has significant advantages over the PRTS, notably: the triviality to convert tree-level relations to bare relations, the easiness to account for all the broad tadpoles of the theory, and the simplicity to obtain gauge independent parameters and counterterms.
We found that the FJTS is actually equivalent to an approach where the bare vev is used. Thus, and given the comodity with which, using tools such as \FM, reducible diagrams with one-loop tadpoles are calculated, one can simply ignore the selection of the true vev, and use the bare vev throughout.%
\fn{Recall that we are treating theories only up to one-loop level. The elimination of proper tadpoles at higher orders may be preferable.}

We argued that, whichever the vev chosen in the context of the up-to-one-loop theory, such selection is not a question of renormalization, but rather an undertaking that precedes renormalization. As we noted, despite the potential convenience in identifying a certain quantity with a tadpole counterterm, 
the identification of the selection of the vev with `tadpole renormalization' is misleading, among other reasons because neither vevs nor tadpoles become finite (i.e. renormalized) due to the selection of the true vev. We explained that, when one performs the renormalization of the theory (after the true vev has been selected), the vevs can be treated as regular parameters. As such, they must be renormalized should they be taken as independent, and they can be renormalized for convenience otherwise.

Finally, we discussed the case of models with two Higgs doublets. We have seen that the selection of the true vev in the SM can be easily extended to that case. We analyzed how the parameter $\beta$ (related to the two vevs) should be considered in both the FJTS and the PRTS. We also commented on the case with CP violation, where one also has true phases besides true vevs.

%% file: Chapters/Chapter_Real.tex
\chapter{Is there a real 2HDM?}
\label{Chap-Real}

\vs{-5mm}

In this chapter, we address the claim raised in the Introduction according to which the so-called real 2HDM is a theoretically unsound model. In section~\ref{Chap-Real:sec:short}, we start by discussing the inconsistency of requiring CP conservation in the potential of the real 2HDM, while allowing for CP violation elsewhere. We argue that, at sufficient high order in perturbation theory, there will be divergences in CP-violating 1-point and 2-point functions that cannot be removed by the counterterms provided by the theory.
We show in section~\ref{Chap-Real:sec:toymodel} that this is precisely what happens in a toy model suffering from the same problem as the real 2HDM. Then, in section \ref{Chap-Real:sec:probing}, we investigate the consistency of the real 2HDM itself. There, we present the result for the leading pole of the three-loop 1-point function of the alleged CP-odd physical field of the real 2HDM, together with the details of the calculation.

\section{Shortcomings of the real 2HDM}
\label{Chap-Real:sec:short}

Let us consider a $\mathrm{SU(2)_L} \otimes \mathrm{U(1)_Y}$ gauge theory with
two Higgs-doublets $\Phi_1$ and $\Phi_2$,
with the same hypercharge $1/2$,
and with real vevs $v_1$ and $v_2$, such that:
\begin{equation}
\langle \Phi_1 \rangle
=
\left(
\begin{array}{c}
0\\
\dfrac{v_1}{\sqrt{2}}
\end{array}
\right),
\qquad
\langle \Phi_2 \rangle
=
\left(
\begin{array}{c}
0\\
\dfrac{v_2}{\sqrt{2}}
\end{array}
\right).
\label{Chap-Real:vev}
\end{equation}
The most general 2HDM scalar potential may be written as
\begin{flalign}
& \hs{3mm} V
= 
m_{11}^2 \Phi_1^\dagger \Phi_1 + m_{22}^2 \Phi_2^\dagger \Phi_2
- \left[ m_{12}^2 \Phi_1^\dagger \Phi_2 + \mathrm{h.c.} \right]
+ \dfrac{\lambda_1}{2} (\Phi_1^\dagger\Phi_1)^2
+ \dfrac{\lambda_2}{2} (\Phi_2^\dagger\Phi_2)^2
+ \lambda_3 (\Phi_1^\dagger\Phi_1) (\Phi_2^\dagger\Phi_2)&
\nonumber\\[2pt]
& \ \ + \lambda_4 (\Phi_1^\dagger\Phi_2) (\Phi_2^\dagger\Phi_1)
+ \left[
\dfrac{\lambda_5}{2} (\Phi_1^\dagger\Phi_2)^2
+ \mathrm{h.c.}
\right] + \left[
\lambda_6 (\Phi_1^\dagger\Phi_1) (\Phi_1^\dagger\Phi_2)
+ \lambda_7 (\Phi_2^\dagger\Phi_2) (\Phi_1^\dagger\Phi_2)
+ \mathrm{h.c.}
\right].&
\label{Chap-Real:VH1}
\end{flalign}
%
The coefficients
$m_{11}^2$, $m_{22}^2$, and $\lambda_1,\cdots,\lambda_4$
are real, while
$m_{12}^2$, $\lambda_5$, $\lambda_6$ and $\lambda_7$
may be complex.

When extending this model to the fermion sector, one finds
flavour changing neutral scalar interactions, which are very
strongly constrained by experiments on neutral meson systems.
This problem can be solved by imposing a $\mathbb{Z}_2$ symmetry:
$\Phi_1 \rightarrow \Phi_1$; $\Phi_2 \rightarrow - \Phi_2$
\cite{Glashow:1976nt,Paschos:1976ay}.
If the symmetry is exact, both the quadratic term $m_{12}^2$ 
and the quartic terms $\lambda_6$ and $\lambda_7$ vanish.
This has the consequence that the model has no decoupling limit; 
that is, one cannot smoothly approach the SM limit by rendering the masses of the new
particles (i.e. the particles resultant from the presence of the second scalar doublet) arbitrarily large.
Such a decoupling is a desirable feature,
especially since the couplings probed by current
LHC data are consistent with the SM predictions,
within errors of order 20\% \cite{Khachatryan:2016vau}.
Decoupling is recovered by reintroducing
$m_{12}^2 \neq 0$ \cite{Gunion:2002zf},
which,
because it breaks softly the $\mathbb{Z}_2$ symmetry,
does not affect the renormalizability of the theory.%
\fn{Cf. e.g. ref. \cite{Cheng:1985bj} page 187 ff. and references therein.}

Most articles addressing this model then decide
that both the vevs $v_1$ and $v_2$, as well as the parameters $m_{12}^2$ and $\lambda_5$, are \textit{all real}, arguing that CP conservation in the scalar sector has been imposed (choice 1).
One would then proceed to discuss the various implementations of the
$\mathbb{Z}_2$ symmetry in the fermion sector, and perform a variety of
fits to experiment.
Among these, one must fit the well measured CP violation with origin
in the Cabibbo-Kobayashi-Maskawa (CKM) matrix
\cite{Cabibbo:1963yz,Kobayashi:1973fv},
accommodated by the complex Yukawa couplings (choice 2).

Choice 1 and choice 2 are both assumed by the so-called real 2HDM.
Only, they are incompatible.

In fact, either the CP symmetry is applied to the whole Lagrangian,
in which case the Yukawa couplings are real (which means that one cannot account for the observed CP violation); or, else, the CP symmetry is not applied anywhere, in which case not only the Yukawa couplings are in general complex, but $m_{12}^2$ is also in general complex.%
\fn{
We mention only $m_{12}^2$ (and not $\lambda_5$) simply because, in the basis of $\Phi_1$ and $\Phi_2$ where $v_1$ and $v_2$ are real, the minimization equations lead to a relation between the phase of $m_{12}^2$ and the phase of $\lambda_5$, so that the latter can be seen as dependent from the former. For details, see section \ref{Chap-Maggie:section:pot} below.}
A compromising scenario---allowing complex Yukawa matrices but excluding $\textrm{Im}\left( m_{12}^2\right)$---leads to a \textit{non-renormalizable theory}.
At sufficiently high loop level, indeed, the CP violation
in the quark sector will leak into the scalar sector,
through a divergent contribution that cannot be absorbed by
a $\textrm{Im}\left( m_{12}^2\right)$ counterterm---which was excluded from the theory from the start.

So why do all articles fitting the real 2HDM ``model'' to experiment
ignore this problem?
Because the divergent contribution can only be shown to happen 
in at least
three loops.
However,
precisely because they are divergent, the problem cannot be ignored
if one wishes to use a theoretically sound model.

Given the fact that
the problem seems to occur due to (the lack of)
$\textrm{Im}\left(m_{12}^2\right)$,
one could be tempted to assume that such a dimension two operator
could not affect renormalizability.
And indeed, it cannot affect renormalizability due to the circumstance that it softly-breaks the $\mathbb{Z}_2$ symmetry.
But the problem with CP symmetry being
invoked is \textit{not}
that it is broken by $m_{12}^2$ (real or complex); rather,
it is (hardly) broken
by the dimension four Yukawa couplings.

It is true that one can look at the real 2HDM as a limiting case of
the $\mathbb{Z}_2$ 2HDM, softly broken by a complex $m_{12}^2$.
This model is known as the complex 2HDM (C2HDM), and will be considered in detail in the following chapters.
In that case, one can choose any tree-level values for the parameters,
and, in particular, set $\textrm{Im}\left( m_{12}^2 \right) = 0$
at tree-level.
In that context, 
setting $\textrm{Im}\left( m_{12}^2 \right) = 0$ at tree-level
does not constitute a problem,
since the theory does have its counterterm and is renormalizable (cf. chapter \ref{Chap-Reno}).
Is this the same as the real 2HDM? \textit{No, it is not}:
setting $\textrm{Im}\left( m_{12}^2 \right) = 0$ in the C2HDM means that
we are studying a specific corner of tree-level parameter space
of a more general model. The real 2HDM, where there is no
$\textrm{Im}\left( m_{12}^2 \right)$ nor its counterterm, is not a consistent model.

There is a more physical way to state the non-renormalization problem.
In any 2HDM, there are three physical neutral scalars.
%
In the real 2HDM, the (proclaimed) lack of CP violation in the scalar sector leads to the separation of the three neutral scalars into
one single CP-odd state ($A$) and two CP-even states (usually denoted by $h$ for the lightest and $H$ for the heaviest).
If the CKM CP violation seeps into the scalar sector,
there should be divergent contributions to the $h$-$A$ and $H$-$A$
2-point functions, as well as to the $A$ 1-point function.
Since such terms are absent from the real scalar sector at tree-level,
there are no counterterms to absorb those infinities,
and the theory is formally inconsistent.

At this point, a hasty reader could argue that Pilaftsis put forward a model \cite{Pilaftsis:1998pe} where he (allegedly) shows to no such problem exists. In Pilaftsis's model, CP-conservation at tree-level in the scalar potential leads him to define the CP-odd state $A$ and the CP-even states $h$ and $H$; yet, CP violation coming from another sector of the theory seeps into the scalar sector at one-loop, thus rendering the $h$-$A$ and $H$-$A$ 2-point functions and the $A$ 1-point function divergent; however, the theory ``fortunately'' happens to generate counterterms for those GFs \cite{Pilaftsis:1998pe}, so that the theory is consistent.

The problem with this argument is that the model by Pilaftsis has a small, but crucial difference regarding the real 2HDM, namely: in Pilaftsis's model, 
the fact that all parameters of the potential are real
\textit{is a consequence of a basis choice}.
In the real 2HDM, by contrast, there is no basis where the generally complex quantities $v_1$, $v_2$, $m_{12}^2$ and $\lambda_5$ can all be rendered real.
%
We discuss Pilaftsis's model in detail in appendix \ref{App-Pilaftsis}, where we show that it is a sound model.
In the following sections, we argue that the real 2HDM is theoretically unsound.


\section{A theoretically unsound toy model}
\label{Chap-Real:sec:toymodel}

We start by considering a toy model, which suffers from the same inconsistency as the real 2HDM. In both models, CP conservation is enforced in a particular sector of an otherwise CP-violating theory.
As a consequence, CP-violating radiative effects end up contaminating
the alleged CP-conserving sector, thus leading to divergences
that cannot be absorbed by the counterterms.
The major feature of our toy model is that such divergences show up
immediately at one-loop order.
Therefore, it constitutes a simple materialization of the
same theoretical pathology that we claim to be present
at three-loop or above in the real 2HDM.

The present model is inspired in the model by Pilaftsis which we alluded to, and which we discuss in detail in appendix \ref{App-Pilaftsis}.%
\fn{Only (we insist), whereas the model discussed in appendix \ref{App-Pilaftsis} is sound and consistent, the model we discuss in this section is as unsound and inconsistent as the real 2HDM.}
Consider two Abelian gauge symmetries $\mathrm{U(1)_Q}$ and $\mathrm{U(1)_B}$,
with gauge bosons $A_{\mu}$ and $B_{\mu}$,
respectively.
Suppose also four complex scalars, $\Phi_1$,
$\Phi_2$, $\chi_L$ and $\chi_R$, with charges 
\begin{gather}
Q(\Phi_1) = 0, \ \
Q(\Phi_2) = 0, \ \
Q(\chi_L) = 1, \ \
Q(\chi_R) = 1,
\\
B(\Phi_1) = 1, \ \
B(\Phi_2) = 1, \ \
B(\chi_L) = -\dfrac{1}{5}, \ \
B(\chi_R) = \dfrac{4}{5},
\end{gather}
where $Q$ and $B$ represent the conserved charges of $\mathrm{U(1)_Q}$ and $\mathrm{U(1)_B}$,
respectively.
A discrete symmetry $D$ is imposed on the fields, under which:
\be
\Phi_1 \stackrel{D}{\to} - \Phi_1, \ \ 
\Phi_2 \stackrel{D}{\to} \Phi_2, \ \
\chi_L \stackrel{D}{\to} - \chi_L, \ \
\chi_R \stackrel{D}{\to} \chi_R.
\ee
However, $D$ is allowed to be softly broken.
The complete renormalizable Lagrangian can then be written in four terms,
\be
\mathcal{L} = \mathcal{L}_{kin} + \mathcal{L}_{\Phi} + \mathcal{L}_{\chi} + \mathcal{L}_{\Phi\chi},
\label{Chap-Real:eq:main}
\ee
where $\mathcal{L}_{kin} $ represents the kinetic
terms\footnote{We assume no $A_\mu$-$B_\mu$ kinetic mixing.} and
\bs
\bea
- \mathcal{L}_{\Phi} &=& \mu_1^2 \Phi_1^* \Phi_1 +
\mu_2^2 \Phi_2^* \Phi_2 + \mu^2 \Phi_1^* \Phi_2 +
(\mu^2)^* \, \Phi_2^* \Phi_1
+ \lambda_1 {\left ( \Phi_1^{*} \Phi_1 \right )}^2
+ \lambda_2 {\left ( \Phi_2^{*} \Phi_2 \right )}^2
+ \lambda_{34} \, \Phi_1^{*} \Phi_1 \Phi_2^{*} \Phi_2
\nonumber\\
&& + \lambda_5 { \left (  \Phi_1^{*} \Phi_2 \right )}^2
+ \lambda_5^* {\left ( \Phi_2^{*} \Phi_1 \right )}^2,
\label{Chap-Real:eq:pot} \\[2mm]
- \mathcal{L}_{\chi} &=&  m_L^2 \, \chi_L \chi_L^*
+ m_R^2 \, \chi_R \chi_R^* + \rho_1 (\chi_L^* \chi_L)^2 + \rho_2 (\chi_R^* \chi_R)^2
+  \rho_{34} \,  \chi_L^* \chi_L  \chi_R^* \chi_R,\label{Chap-Real:eq:7} \\[2mm]
- \mathcal{L}_{\Phi\chi} &=&  f_1 \, \Phi_1  \chi_L \chi_R^*
+ f_1^* \, \Phi_1^*  \chi_L^* \chi_R
+ f_2 \, \Phi_2  \chi_L \chi_R^*
+ f_2^* \, \Phi_2^*  \chi_L^* \chi_R
+ g_1 \, \Phi_1^{*} \Phi_1 \chi_L^* \chi_L
+ g_2 \, \Phi_2^{*} \Phi_2\chi_L^* \chi_L
\nonumber \\
&&  + \, g_3 \, \Phi_1^{*} \Phi_1 \chi_R^* \chi_R
+ \, g_4 \, \Phi_2^{*} \Phi_2 \chi_R^* \chi_R.
\label{Chap-Real:eq:Yuk}
\eea
\es
The parameters $\mu^2$, $\lambda_5$, $f_1$ and $f_2$ are
in general complex, while the remaining ones are real by
construction. The terms involving $\mu^2$ and $f_2$
break the symmetry $D$ softly.
It is easy to show that the conditions for CP conservation are:
\bs
\label{Chap-Real:eq:conds}
\bea
\text{Im} \left[ \mu^2 \, f_1 \, f_2^* \right] &=& 0, \label{Chap-Real:eq:9a}\\
\text{Im}  \Big[ \lambda_5 \, f_1^2  \,(f_2^*)^2 \Big] &=& 0, \label{Chap-Real:eq:9b}\\
\text{Im}  \Big[ \lambda_5^* \, (\mu^2)^2 \Big] &=& 0.
\label{Chap-Real:eq:9c}
\eea
\es
Mimicking the usual real 2HDM treatment,
we take $\langle \Phi_1 \rangle$ and $\langle \Phi_2 \rangle$ real
and parameterize
\be
\label{Chap-Real:eq:6}
\Phi_1 = 
\tfrac{1}{\sqrt{2}} \left( v_1 + H_1 + i A_1 \right),
\qquad
\Phi_2 = 
\tfrac{1}{\sqrt{2}} \left( v_2 + H_2 + i A_2 \right),
\ee
where 
$v_1$, $v_2$ are real and non-negative,
and
$H_1$, $H_2$, $A_1$, and $A_2$ are real fields.
The vevs $v_1$ and $v_2$ break spontaneously the gauge symmetry $\mathrm{U(1)_B}$.
Recall that $\mu^2$, $\lambda_5$, $f_1$, $f_2$ are in general complex.
But suppose we force CP to be conserved in $\mathcal{L}_{\Phi}$
by the ad-hoc imposition that
$\mu^2$ and $\lambda_5$ are real.
This we do in the following. As we will show, though, it will lead to irremovable divergences at one-loop.

We start by determining the minimization (or tadpole) equations.
These are:\footnote{The tadpole equations for $A_1$ and $A_2$ are trivially zero.}
\bs
\label{Chap-Real:eq:tree-tads}
\bea
0 &=& t_{H_1} := \dfrac{\partial \mathcal{L}_{\Phi}}{\partial
H_1}\bigg|_{< >=0}
=
-v_1 \left( \mu_1^2 + \dfrac{v_2}{v_1} \mu ^2
+  v_1^2 \lambda_1 + \dfrac{1}{2} v_2^2 \lambda _{34}
+  v_2^2 \, \lambda _5\right)\hs {-1mm}, \\
0 &=& t_{H_2}  := \dfrac{\partial \mathcal{L}_{\Phi}}{\partial
H_2}\bigg|_{< >=0}
=
-v_2 \left( \mu_2^2 + \dfrac{v_1}{v_2} \mu ^2
+  v_2^2 \lambda_2 + \dfrac{1}{2} v_1^2 \lambda _{34}
+  \, v_1^2 \, \lambda _5\right) \hs {-1mm},
\eea
\es
where  $t_{H_1}, t_{H_2}$ represent the tree-level
tadpoles for $H_1$, $H_2$, respectively, and
$< >=0$ means that the expectation
values of all fields on the r.h.s. of
eqs.~(\ref{Chap-Real:eq:6}) are set to zero.
Recall that we are taking $\mu^2$ and $\lambda_5$ as
real parameters. For that reason, the mass matrices for $H_1$ and
$H_2$, on the one hand, and for $A_1$ and $A_2$, on the other,
can be separately diagonalized. We thus define the
angles $\theta$ and $\beta$ such that:
\bs
\bea
\left(\begin{array}{c}
H_{1} \\
H_{2}
\end{array}\right)
&=&
\left(\begin{array}{cc}
c_{\theta} & -s_{\theta} \\
s_{\theta} & c_{\theta}
\end{array}\right)\left(\begin{array}{l}
h \\
H
\end{array}\right),
\\
\left(\begin{array}{c}
A_{1} \\
A_{2}
\end{array}\right)
&=&
\left(\begin{array}{cc}
c_{\beta} & -s_{\beta} \\
s_{\beta} & c_{\beta}
\end{array}\right)\left(\begin{array}{c}
G_0 \\
A
\end{array}\right),
\eea
\es
where $h$ and $H$ are the CP-even states, and
$A$ and $G_0$ are the CP-odd states, where
$G_0$ is the massless would-be Goldstone boson.\footnote{$G_0$
is eaten by the longitudinal component of $B_{\mu}$ in the
unitary gauge through the Higgs mechanism.
The fact that $G_0$ is massless forces $\beta$ to
obey the relation $\tan \beta = \frac{v_2}{v_1}$.}
As for the $\mathcal{L}_{\chi}$ sector, the mass matrix is given by
\be
-\mathcal{L}^{\chi}_{\text{mass}}=
\left(\begin{array}{cc}
\chi_L^* & \chi_R^*
\end{array}\right)
\left(\begin{array}{cc}
a & b^* \\
b & c
\end{array}\right)
\left(\begin{array}{c}
\chi_L \\
\chi_R
\end{array}\right),
\label{Chap-Real:eq:mymass}
\ee
with
\be
a = \frac{1}{2} g_1 v_1^2+\frac{1}{2} g_2 v_2^2+m_L^2,
\qquad
b = \dfrac{f_1 v_1+ f_2 v_2}{\sqrt{2}},
\qquad
c = \frac{1}{2} g_3 v_1^2+\frac{1}{2} g_4 v_2^2+m_R^2\, ,
\ee
where $a$ and $c$ are real, while $b$ is in principle complex.
However, we can rephase $\chi_R$ through
$\chi_R \to e^{i \, \text{arg}(b)} \chi_R$
so that it absorbs the phase of $b$.
In the new basis, then, $b$ is real, which implies that 
the mass matrix in eq.~(\ref{Chap-Real:eq:mymass}) is
symmetric.\footnote{If it were hermitian, one would
need a unitary matrix to diagonalize it, instead of
an orthogonal one. Moreover, note that 
$b$ real forces the relation
$f_1^* = f_1 + \left( f_2 - f_2^*\right)  \tan{\beta}$.}
We thus need an orthogonal matrix with a new angle
$\phi$ to diagonalize the states:
\be
\left(\begin{array}{c}
\chi_L \\
\chi_R
\end{array}\right)
=
\left(\begin{array}{cc}
c_{\phi} & -s_{\phi} \\
s_{\phi} & c_{\phi}
\end{array}\right)\left(\begin{array}{c}
\chi_1 \\
\chi_2
\end{array}\right),
\ee
where $\chi_1$ and $\chi_2$ are the (complex)
diagonalized states with (real) masses $M_1$ and $M_2$.
The model was easily implemented on \textsc{FeynMaster}, which allowed to determine the Feynman rules, counterterms and one-loop results.

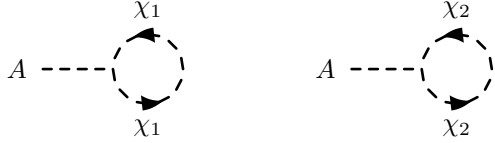
\begin{figure}[h!] 
\hspace{6mm}
\centering 
\subfloat{ 
\begin{fmffile}{Chap-Real-200} 
\begin{fmfgraph*}(70,40) 
\fmfset{arrow_len}{3mm} 
\fmfset{arrow_ang}{20} 
\fmfleft{nJ1} 
\fmflabel{$A$}{nJ1} 
\fmfright{nJ2} 
\fmf{dashes,tension=2}{nJ1,nJ1J1J4} 
\fmf{phantom,tension=3}{nJ2,nJ2J3J2} 
\fmf{scalar,label=$\chi_1$,right=1}{nJ2J3J2,nJ1J1J4} 
\fmf{scalar,label=$\chi_1$,right=1}{nJ1J1J4,nJ2J3J2} 
\end{fmfgraph*} 
\end{fmffile} 
} 
\hspace{9mm} 
\subfloat{ 
\begin{fmffile}{Chap-Real-201} 
\begin{fmfgraph*}(70,40) 
\fmfset{arrow_len}{3mm} 
\fmfset{arrow_ang}{20} 
\fmfleft{nJ1} 
\fmflabel{$A$}{nJ1} 
\fmfright{nJ2} 
\fmf{dashes,tension=2}{nJ1,nJ1J1J4} 
\fmf{phantom,tension=3}{nJ2,nJ2J2J3} 
\fmf{scalar,label=$\chi_2$,right=1}{nJ2J2J3,nJ1J1J4} 
\fmf{scalar,label=$\chi_2$,right=1}{nJ1J1J4,nJ2J2J3} 
\end{fmfgraph*} 
\end{fmffile} 
}
\caption{Feynman diagrams contributing to
the one-loop tadpole of the CP-odd state.}
\label{Chap-Real:fig:mytads}
\end{figure}

When considering the theory up to one-loop level,
the quantities of the original theory are identified as bare quantities. We decide to select the true vev of the up-to-one-loop theory through the PRTS. Recalling the discussion of sections \ref{Chap-Selec:sec:SM-PRTS} and \ref{Chap-Selec:sec:2HDM-PRTS}, it follows from the set of eqs. \ref{Chap-Real:eq:tree-tads} that:
\bs
\label{Chap-Real:eq:loop-tads}
\bea
\delta t_{H_1} &:=& -v_1 \bigg( \mu_{1(0)}^2 + \dfrac{v_2}{v_1} \mu_{(0)} ^2 +  v_1^2 \lambda_{1(0)} + \dfrac{1}{2} v_2^2 \lambda _{34(0)} + v_2^2 \, \lambda_{5(0)}\bigg),  \\
\delta t_{H_2} &:=& -v_2 \bigg( \mu_{2(0)}^2 + \dfrac{v_1}{v_2} \mu_{(0)} ^2 +  v_2^2 \lambda_{2(0)} + \dfrac{1}{2} v_1^2 \lambda_{34(0)} + v_1^2 \, \lambda_{5(0)}\bigg),
\eea
\es
where $v_1$ and $v_2$ are the true vevs, and where the tadpole counterterms $\delta t_{H_1}$ and $\delta t_{H_2}$ are such that
$\delta t_{H_1}  = -T_{H_1}$ and
$\delta t_{H_2}  = -T_{H_2}$, with $T_{H_1}$ and  $T_{H_2}$
being the one-loop tadpole for $H_1$ and $H_2$, respectively.
%
%
Note that, since we imposed CP conservation in
$\mathcal{L}_{\Phi}$, there are no tadpole counterterms
for the CP-odd fields.
Specifically, in the mass basis,
\be
\delta t_{A} = 0.
\label{Chap-Real:eq:tads-mass-odd}
\ee
But it is easy to see that this is inconsistent in the PRTS.
Indeed, \textit{there is} a one-loop tadpole for $A$,
whose diagrams are represented in fig.~\ref{Chap-Real:fig:mytads}.
The sum of diagrams is divergent.
In fact,
\be
\left(T_{A}\right)\Big|_{\varepsilon} =
- \dfrac{1}{\varepsilon}
\dfrac{{c_{\phi }}  \, {s_{\phi }} \,
\left( M_1^2 - M_2^2 \right)  \,
\text{Im}[f_2]}{4 \, \sqrt{2} \, \pi^2 \, c_{\beta}},
\label{Chap-Real:eq:taddivs}
\ee
in $d=4-\varepsilon$ dimensions,
where $T_{A}$ represents the one-loop tadpole for
$A$ and $\big|_{\varepsilon}$ means that we consider
only the terms proportional to inverse powers of $\varepsilon$.%
\fn{\label{Chap-Real:note:uu}Note that, in the paper \cite{Fontes:2021znm}, the alternative convention $d=4-2\varepsilon$ is used. We changed the results to the convention $d=4-\varepsilon$, because this convention is the one that is used throughout the thesis.}

The origin of the problem lies in the fact that we
imposed $\mu^2$ and $\lambda_5$ to be real.
To clarify this point, let us provisionally
take these parameters to be complex, as they originally were.
By rewriting eq.~(\ref{Chap-Real:eq:pot}) in terms of bare quantities,
and separating the real and imaginary parts of
$\mu_{(0)}^2$ and $\lambda_{5(0)}$,
the terms proportional to these parameters are:
\be
\begin{split}
- \mathcal{L}_{\Phi_{(0)}}
&\ni
\mu_{(0)}^2 \, \Phi_{1(0)}^* \Phi_{2(0)}
+ (\mu_{(0)}^2)^* \, \Phi_{2(0)}^* \Phi_{1(0)} +
\lambda_{5(0)} \, \left(\Phi_{1(0)}^* \Phi_{2(0)}\right)^2
+ \lambda_{5(0)}^* \, \left(\Phi_{2(0)}^* \Phi_{1(0)}\right)^2
\\
&= \text{Re} [\mu_{(0)}^2]
(\Phi_{1(0)}^* \Phi_{2(0)} + \Phi_{2(0)}^* \Phi_{1(0)}) +
\text{Re} [\lambda_{5(0)}] \Big\{\left(\Phi_{1(0)}^* \Phi_{2(0)}\right)^2
+ \left(\Phi_{2(0)}^* \Phi_{1(0)}\right)^2\Big\} \\
& \hspace{1mm} +
i \, \text{Im} [\mu_{(0)}^2] \left(\Phi_{1(0)}^* \Phi_{2(0)}
- \Phi_{2(0)}^* \Phi_{1(0)}\right) + i \,
\text{Im} [\lambda_{5(0)}] \Big\{\left(\Phi_{1(0)}^*
\Phi_{2(0)}\right)^2 - \left(\Phi_{2(0)}^* \Phi_{1(0)}\right)^2\Big\}.
\end{split}
\label{Chap-Real:eq:decide}
\ee
As a consequence, when we set
$\text{Im} [\mu_{(0)}^2] = \text{Im} [\lambda_{5(0)}] = 0$,
we are not including in the model the terms of the
last line of eq.~(\ref{Chap-Real:eq:decide}).
Naturally, since such terms are not in the model,
there is no counterterm for the parameters involved therein.
That is, there is neither $\text{Im} [\delta \mu^2]$
nor $\text{Im} [\delta \lambda_5]$.\footnote{The situation
would not be different if we decided
to exclude any other term from the theory. For example,
had we decided not to include the term
proportional to $\lambda_1$ in the model,
there would be no counterterm $\delta \lambda_1$.}
Now, it is a matter of course that this would not be a problem
if the fact that we did not include the terms in the last
line of eq.~(\ref{Chap-Real:eq:decide}) would follow from a symmetry that forbade them.
In other words, should there be a symmetry in the theory that
proscribed those terms, they could logically not be included;
and since the symmetry would prevent any GFs
generated by such terms from showing up, there would never be
divergences involved therein, so that the absence of
counterterms for them would never be a problem.
So, for example, if CP was a symmetry of the theory,
it would preclude those terms, in which case the absence
of $\text{Im} [\delta \mu^2]$ and $\text{Im} [\delta \lambda_5]$
would not be inconsistent.

However, CP is \textit{not} a symmetry of theory:
even if we try to impose it in the $\mathcal{L}_{\Phi}$ sector,
it still is violated in the $\mathcal{L}_{\Phi \chi}$
sector through the phases of $f_1$ and $f_2$,
as eqs. \ref{Chap-Real:eq:9a} and \ref{Chap-Real:eq:9b} show.
So, there is no CP symmetry forbidding the terms in
the last line of eq. \ref{Chap-Real:eq:decide}.
As a consequence, even if we exclude them, CP-violating radiative effects
can nonetheless contribute to the GFs involved therein.
Such GFs will in general be divergent;
but since the last line of eq. \ref{Chap-Real:eq:decide}
were not included in the theory, there will in general
not be enough counterterms to absorb them.

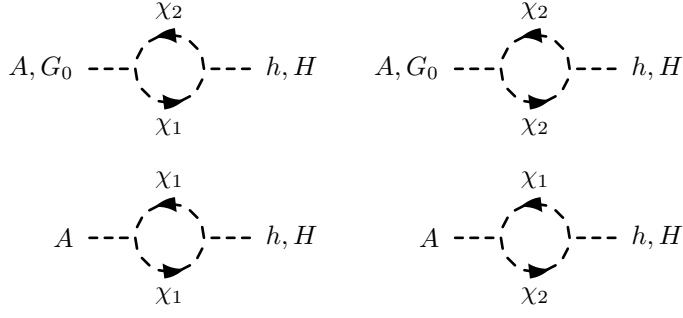
\begin{figure}[htp] 
\centering 
\subfloat{ 
\begin{fmffile}{Chap-Real-3} 
\begin{fmfgraph*}(60,30) 
\fmfset{arrow_len}{3mm} 
\fmfset{arrow_ang}{20} 
\fmfleft{nJ1} 
\fmflabel{$A,G_0$}{nJ1} 
\fmfright{nJ2} 
\fmflabel{$h,H$}{nJ2} 
\fmf{dashes,tension=3}{nJ1,nJ1J2J3} 
\fmf{dashes,tension=3}{nJ2,nJ2J1J4} 
\fmf{scalar,label=$\chi_1$,right=1}{nJ1J2J3,nJ2J1J4} 
\fmf{scalar,label=$\chi_2$,right=1}{nJ2J1J4,nJ1J2J3} 
\end{fmfgraph*} 
\end{fmffile} 
} 
\hspace{20mm} 
\subfloat{ 
\begin{fmffile}{Chap-Real-4} 
\begin{fmfgraph*}(60,30) 
\fmfset{arrow_len}{3mm} 
\fmfset{arrow_ang}{20} 
\fmfleft{nJ1} 
\fmflabel{$A,G_0$}{nJ1} 
\fmfright{nJ2} 
\fmflabel{$h,H$}{nJ2} 
\fmf{dashes,tension=3}{nJ1,nJ1J1J4} 
\fmf{dashes,tension=3}{nJ2,nJ2J3J2} 
\fmf{scalar,label=$\chi_2$,right=1}{nJ2J3J2,nJ1J1J4} 
\fmf{scalar,label=$\chi_2$,right=1}{nJ1J1J4,nJ2J3J2} 
\end{fmfgraph*} 
\end{fmffile} 
}
\\[8mm]
%
\subfloat{ 
\begin{fmffile}{Chap-Real-1} 
\begin{fmfgraph*}(60,30) 
\fmfset{arrow_len}{3mm} 
\fmfset{arrow_ang}{20} 
\fmfleft{nJ1} 
\fmflabel{$A$}{nJ1} 
\fmfright{nJ2} 
\fmflabel{$h,H$}{nJ2} 
\fmf{dashes,tension=3}{nJ1,nJ1J1J4} 
\fmf{dashes,tension=3}{nJ2,nJ2J3J2} 
\fmf{scalar,label=$\chi_1$,right=1}{nJ2J3J2,nJ1J1J4} 
\fmf{scalar,label=$\chi_1$,right=1}{nJ1J1J4,nJ2J3J2} 
\end{fmfgraph*} 
\end{fmffile} 
} 
\hspace{20mm} 
\subfloat{ 
\begin{fmffile}{Chap-Real-2} 
\begin{fmfgraph*}(60,30) 
\fmfset{arrow_len}{3mm} 
\fmfset{arrow_ang}{20} 
\fmfleft{nJ1} 
\fmflabel{$A$}{nJ1} 
\fmfright{nJ2} 
\fmflabel{$h,H$}{nJ2} 
\fmf{dashes,tension=3}{nJ1,nJ1J1J4} 
\fmf{dashes,tension=3}{nJ2,nJ2J2J3} 
\fmf{scalar,label=$\chi_1$,right=1}{nJ2J2J3,nJ1J1J4} 
\fmf{scalar,label=$\chi_2$,right=1}{nJ1J1J4,nJ2J2J3} 
\end{fmfgraph*} 
\end{fmffile} 
}
\vs{5mm}
\caption{Feynman diagrams contributing to the
one-loop CP-violating 2-point functions.}
\label{Chap-Real:fig:myf}
\end{figure}

We have already seen one example of GF
whose divergences cannot be removed: the one-loop 1-point
function $T_{A}$. Other examples are the one-loop
CP-violating 2-point functions $\Sigma^{G_0 h}$,
$\Sigma^{G_0 H}$, $\Sigma^{A h}$ and $\Sigma^{A H}$
for the scalar-pseudoscalar mixing of $G_0 \, h$,
$G_0 \, H$, $A\, h$ and $A\, H$, respectively.
Their Feynman diagrams are represented in fig.~\ref{Chap-Real:fig:myf}. 
%
%
%
There simply is no
counterterm for these functions, which nonetheless
are divergent. Their divergent parts are:
\bs
\label{Chap-Real:eq:divs}
\bea
&& \Sigma^{G_0 h}(k^2)\Big|_{\varepsilon}
= \Sigma^{AH}(k^2)\Big|_{\varepsilon}
= - \dfrac{1}{\varepsilon}
\, \sin(\beta  - \theta) \,
\frac{\text{Im}\left[f_2\right]
\left( f_1 + f_2 \tan \beta \right)}{8 \pi ^2},
\\[3mm]
&& \Sigma^{Ah}(k^2)\Big|_{\varepsilon}
= - \Sigma^{G_0 H}(k^2)\Big|_{\varepsilon}
=
- \dfrac{1}{\varepsilon} \, \cos(\beta  - \theta)
\, \frac{\text{Im}\left[f_2\right]
\left( f_1 + f_2 \tan \beta \right)}{8 \pi ^2}.
\eea
\es
In conclusion, the fact that we imposed
$\mu^2$ and $\lambda_5$ 
to be real leads to several
divergences that cannot be removed from the theory.

There are two ways to heal this model: either CP is
imposed as a whole, or it is not imposed at all.
In the first case, all the three relations in
eqs. \ref{Chap-Real:eq:conds} should be verified, which implies
that there is a basis where $\mu^2$, $\lambda_5$,
$f_1$ and $f_2$ are all real. In this scenario, therefore,
CP-violating GFs are precluded, which implies,
in particular, that no divergent CP-violating GFs will appear at any order.
This is consistent with what we obtained in
eqs. \ref{Chap-Real:eq:taddivs} and \ref{Chap-Real:eq:divs},
which vanish in the limit of real $f_1$ and $f_2$.
In the second case, $\mu^2$, $\lambda_5$, $f_1$ and $f_2$
are in general complex parameters, which implies
that their counterterms are also in general complex.
Since CP is violated, there are no scalar states with
well-defined CP, and GFs will in general
be CP-violating.
The model is renormalizable as long as all the terms
compatible with the symmetries are included.
Finally, note that, in such a CP-violating scenario,
there may be regions of the parameter space in which 
$\lambda_5$ and $\mu^2$ are real, and $f_1$ and $f_2$ complex.
But this is a completely different situation from that where one
builds a theory taking \textit{ab initio} $\lambda_5$ and
$\mu^2$ real, while $f_1$ and $f_2$ in general complex.
In fact, while the former situation corresponds to a
particular solution of a consistent, renormalizable theory,
the latter suffers from the inconsistencies we have shown.

\section{Probing the real 2HDM}
\label{Chap-Real:sec:probing}

The previous section ought to convince us that the real 2HDM, by imposing CP conservation in the potential sector but allowing it elsewhere, will suffer from the same inconsistencies as the toy model we just presented.
Our goal now is to investigate this inconsistency in the real 2HDM itself. We start by presenting the potential sector of the model, followed by a short analysis of the CP violation originating from the CKM matrix. Then, in section \ref{Chap-Real:sec:3looptad}, we perform a three-loop calculation, whose details are given in section \ref{Chap-Real:sec:details}.

\subsection{The potential sector}

As suggested in section \ref{Chap-Real:sec:short}, the scalar potential of the real 2HDM reads:
\begin{gather}
V_r
=\, 
m_{11}^2 \Phi_1^\dagger \Phi_1 + m_{22}^2 \Phi_2^\dagger \Phi_2
- m_{12}^2 \left[\Phi_1^\dagger \Phi_2 + \Phi_2^\dagger \Phi_1 \right]
+ \dfrac{\lambda_1}{2} (\Phi_1^\dagger\Phi_1)^2
+ \dfrac{\lambda_2}{2} (\Phi_2^\dagger\Phi_2)^2
+ \lambda_3 (\Phi_1^\dagger\Phi_1) (\Phi_2^\dagger\Phi_2)
\no
+\, \lambda_4 (\Phi_1^\dagger\Phi_2) (\Phi_2^\dagger\Phi_1)
+ 
\dfrac{\lambda_5}{2}
\left[
(\Phi_1^\dagger\Phi_2)^2 + (\Phi_2^\dagger\Phi_1)^2
\right],
\label{Chap-Real:Vreal}
\end{gather}
with all parameters real. The doublets can be parameterized as
\begin{align}
\Phi_1 = 
\begin{pmatrix}
\phi_1^+ \\
\frac{1}{\sqrt{2}}(v_1 + \rho_1 + i \eta_1)
\end{pmatrix},
\hspace{3mm} \Phi_2 = 
\begin{pmatrix}
\phi_2^+ \\
\frac{1}{\sqrt{2}}(v_2 + \rho_2 + i \eta_2)
\end{pmatrix},
\label{Chap-Real:parametrizacao-PHIs}
\end{align}
with $v_1 , v_2$ real constants, $\rho_1, \rho_2, \eta_1, \eta_2$ real fields, and $\phi_1^+$ and $\phi_2^+$ complex fields. The vevs $v_1$ and $v_2$ obey eq. \ref{Chap-Selec:eq:tanbtree}, i.e. $\tan \beta = v_2/v_1$.
In that case, if one inserts eq. \ref{Chap-Real:parametrizacao-PHIs} inside eq. \ref{Chap-Real:Vreal} and expands the terms, one finds that, since $m_{12}^2$ and $\lambda_5$ are real, there are no linear terms for the $\eta$ fields, just as there are no bilinear terms mixing $\eta$ fields with $\rho$ fields. 
Accordingly, the $\eta$ fields are identified as CP-odd, and the $\rho$ fields as CP-even.
The requirement that the linear terms in the $\rho$ fields vanish leads to two minimization (or tadpole) conditions; since there were no linear terms in the $\eta$ fields to start with, there are no such conditions for these fields. In what follows, we use the short notation $s_x \equiv \sin(x)$,
$c_x \equiv \cos(x)$ for a generic angle $x$.

The Higgs basis \cite{Lavoura:1994fv,Botella:1994cs}
$\{ {\cal H}_1, {\cal H}_2 \}$ can be defined by introducing the orthogonal matrix $R_H$, such that:
\beq
\left( \begin{array}{c} {\cal H}_1 \\ {\cal H}_2 \end{array} \right) =
R^{\mathrm{T}}_H \left( \begin{array}{c} \Phi_1 \\
    \Phi_2 \end{array} \right) \equiv
\left( \begin{array}{cc} c_\beta & s_\beta \\ - s_\beta &
    c_\beta \end{array} \right) \left( \begin{array}{c} \Phi_1 \\
    \Phi_2 \end{array} \right) .
\label{Chap-Real:eq:HiggsBasis}
\eeq
The doublets in the Higgs basis are written as
\begin{align}
{\cal H}_1 = 
\begin{pmatrix}
G^+ \\
\frac{1}{\sqrt{2}}(v + h_1^{\mathrm{H}} + i G_0)
\end{pmatrix},
\hspace{3mm}
{\cal H}_2 = 
\begin{pmatrix}
H^+ \\
\frac{1}{\sqrt{2}}(h_2^{\mathrm{H}} + i A)
\end{pmatrix},
\label{Chap-Real:eq:parametrizacaoHs}
\end{align}
with $G^+, H^+$ complex fields, and $h_1^{\mathrm{H}}, h_2^{\mathrm{H}}, A, G_0$ real fields, and $v$ is a real parameter defined as in eq. \ref{Chap-Selec:eq:totalvevtree}, i.e. $v = \sqrt{v_1^2 + v_2^2}$.
Expanding the terms of the potential, one finds that $G^+, G_0, H^+, A$ are already mass states (i.e. they are already diagonalized): $G^+$ and $G_0$ correspond to the charged and neutral would-be Goldstone bosons, respectively, $H^+$ corresponds to a charged scalar boson and $A$ to the CP-odd physical state. The states $h_1^{\mathrm{H}}$ and $h_2^{\mathrm{H}}$ can be diagonalized introducing the angle $\alpha$, such that:
\be
\begin{pmatrix}
H \\
h
\end{pmatrix}
= 
\begin{pmatrix}
c_{\alpha} & s_{\alpha}\\
-s_{\alpha} & c_{\alpha}
\end{pmatrix}
\begin{pmatrix}
\rho_1 \\
\rho_2
\end{pmatrix}
= 
\begin{pmatrix}
c_{\alpha-\beta} & s_{\alpha-\beta} \vspace{0.7mm}\\
-s_{\alpha-\beta} & c_{\alpha-\beta}
\end{pmatrix}
\begin{pmatrix}
h_1^{\mathrm{H}} \vspace{0.7mm}\\
h_2^{\mathrm{H}}
\end{pmatrix},
\label{Chap-Real:my-alpha}
\ee
where $H$ and $h$ are the CP-even mass states.

\subsection{CP violation from the CKM matrix}

In the real 2HDM, just as in the SM, CP violation arises from the complex
Yukawa couplings.
When the quark fields are rotated into their mass basis,
all CP violation phases are contained in the 
CKM matrix \cite{Cabibbo:1963yz,Kobayashi:1973fv}.
However, each quark field can still be rephased at will,
thus moving the CP-violating phase around the various entries
of the CKM matrix.
The only rephasing invariant quantity is
\cite{Jarlskog:1985ht,Jarlskog:1985cw,Dunietz:1985uy}
\be
I^{\alpha i}_{\beta  j}
= \textrm{Im}\left(
V_{\alpha i} V_{\beta j} V_{\alpha j}^\ast V_{\beta i}^\ast
\right)\, ,
\label{Chap-Real:eq:Jarlskog_1}
\ee
where $\alpha \neq \beta$ and $i \neq j$ (we use the notation of ref. \cite{Branco:1999fs}, where Greek letters $\alpha, \beta, \gamma, \dots$ refer to up-type quarks $u_\alpha = u, c, t$,
while Roman letters $i, j, k, \dots$
refer to down-type quarks $d_i = d, s, b$).
There are nine distinct four quark combinations with different flavours:
three combinations for the down-type quarks---$(ds)$, $(db), (sb)$---times three for the up-type quarks---$(uc)$, $(ut)$, and $(ct)$.
Using the unitarity of the CKM matrix,
the following symmetries hold
\be
I^{\alpha i}_{\beta  j} = I^{\beta  j}_{\alpha i}
=
- I^{\alpha j}_{\beta  i} = - I^{\beta i}_{\alpha  j},
\ee
showing that indeed there is only one independent CP-violating quantity.
The antisymmetry with respect to interchange of same quark-type indices
is easiest to see in the form 
\be
I^{\alpha i}_{\beta  j}
= J\, \sum_{\gamma, k} \epsilon_{\alpha \beta \gamma}\, \epsilon_{i j k}\, ,
\label{Chap-Real:eq:Jarlskog_2}
\ee
where
$J$ is the Jarlskog invariant
\cite{Jarlskog:1985ht,Jarlskog:1985cw,Dunietz:1985uy},
defined for example as $J=I^{ud}_{cs}$.%
%

As emphasized by Khriplovich and Pospelov \cite{Pospelov:1991zt}
and by Booth \cite{Booth:1993af}
in the context of the electric dipole moments (EDM)
of the $W$ and the electron, the antisymmetry
of $I^{\alpha i}_{\beta j}$ is very powerful.
In fact, any CP-violating amplitude from a fermion box diagram will appear as the product
$I^{\alpha i}_{\beta  j}$ with some amplitude
\be
\mathcal{A}(m_{u_\alpha},m_{u_\beta},m_{d_i},m_{d_j})\, .
\label{Chap-Real:eq:A_ms}
\ee
It turns out that, when all contributions are summed over ($\alpha$, $\beta$, $i$, and $j$),
all terms in $\mathcal{A}(m_{u_\alpha},m_{u_\beta},m_{d_i}$, $m_{d_j})$
symmetric under $\alpha \leftrightarrow \beta$
or $i \leftrightarrow j$ will not contribute.
A much more involved analysis along these lines
was used in refs. \cite{Pospelov:1991zt,Booth:1993af}
to show that the SM electroweak contributions to the
EDMs
of the $W$  and the electron vanish in the two-loop
and three-loop approximation, respectively.

\subsection{Three-loop tadpole for $A$ in the real 2HDM}
\label{Chap-Real:sec:3looptad}

\begin{figure}[htp] 
\vspace{-5mm}
\hspace{6mm}
\centering
\subfloat{
\begin{fmffile}{Chap-Real-three-loop-tadpole-J1}
\fmfset{arrow_len}{3mm} 
\fmfset{arrow_ang}{20} 
\begin{fmfgraph*}(120,85) 
\fmfleft{l} 
\fmflabel{$A$}{l} 
\fmfright{r}
\fmftop{t}
\fmfbottom{b}
\fmf{phantom}{vrb,r,vrt}
\fmf{phantom}{vlb,l,vlt}
\fmf{phantom,tension=1.5}{vlt,t,vrt}
\fmf{phantom,tension=1.5}{vlb,b,vrb}
\fmf{phantom}{r,vr,vl,l}
\fmf{dashes,tension=2}{l,vl}
\fmffreeze
\fmf{fermion,label.side=left,label.dist=3,label=$u_\alpha$}{vlb,vl,vlt}
\fmf{fermion,label.side=left,label.dist=3,label=$d_j$}{vlt,vrt}
\fmf{fermion,left=0.5,label.side=left,label.dist=4,label=$u_\beta$}{vrt,vrb}
\fmf{fermion,label.side=left,label.dist=4,label=$d_i$}{vrb,vlb}
\fmf{dashes_arrow,right=0.5,label.side=left,label.dist=2,label=$H^-$}{vlb,vlt}
\fmf{photon,right=0.5,label.side=left,label.dist=3,label=$W^-$}{vrt,vrb}
\fmf{phantom_arrow,right=0.5,tension=1}{vrt,vrb}
\end{fmfgraph*} 
\end{fmffile}
}
\hspace{9mm} 
\subfloat{ 
\begin{fmffile}{Chap-Real-three-loop-tadpole-J2}
\fmfset{arrow_len}{3mm} 
\fmfset{arrow_ang}{20} 
\begin{fmfgraph*}(120,85) 
\fmfleft{l} 
\fmflabel{$A$}{l} 
\fmfright{r}
\fmftop{t}
\fmfbottom{b}
\fmf{phantom}{vrt,r,vrb}
\fmf{phantom}{vlt,l,vlb}
\fmf{phantom,tension=1.5}{vlb,t,vrb}
\fmf{phantom,tension=1.5}{vlt,b,vrt}
\fmf{phantom}{r,vr,vl,l}
\fmf{dashes,tension=2}{l,vl}
\fmffreeze
\fmf{fermion,label.side=right,label.dist=3,label=$u_\alpha$}{vlb,vl,vlt}
\fmf{fermion,label.side=right,label.dist=3,label=$d_i$}{vlt,vrt}
\fmf{fermion,right=0.5,label.side=right,label.dist=4,label=$u_\beta$}{vrt,vrb}
\fmf{fermion,label.side=right,label.dist=4,label=$d_j$}{vrb,vlb}
\fmf{dashes_arrow,left=0.5,label.side=right,label.dist=2,label=$H^-$}{vlb,vlt}
\fmf{photon,left=0.5,label.side=right,label.dist=3,label=$W^-$}{vrt,vrb}
\fmf{phantom_arrow,left=0.5,tension=1}{vrt,vrb}
\end{fmfgraph*} 
\end{fmffile} 
}
\vspace{-2mm}
\caption{Example of pair of Feynman diagrams where $J$ factorizes (for fixed $\alpha,\beta,i,j$). They differ only by the direction of fermion flow.
}
\label{Chap-Real:fig:jarl-pair}
\end{figure}
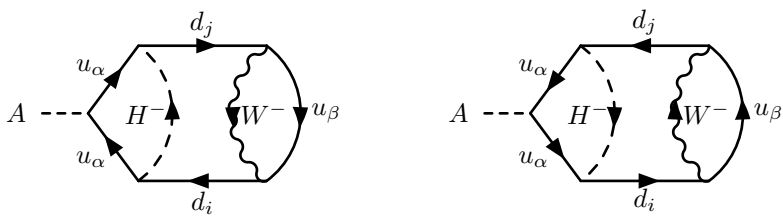

We now ascertain whether quark-induced CP-violating effects contaminate the
otherwise CP-conserving scalar sector of the real 2HDM via radiative corrections. Since the quantity that represents quark-induced CP violation in a convention-independent way is the Jarlskog invariant $J$, we are looking for radiative corrections to the 2HDM which contain this quantity.
As the simplest check, we have looked for diagrams proportional to $J$ contributing to the $A$ tadpole.%
\fn{Strictly speaking, and as we saw in section \ref{Chap-Selec:sec:need}, even if a 1-point function ends up being divergent, that does not necessarily imply that the theory is inconsistent. On the other hand, just as in the toy model discussed in section \ref{Chap-Real:sec:toymodel}, the inconsistency of the real 2HDM should manifest itself in divergences contained both in 1-point and 2-point CP-violating GFs. Hence, if we find divergent CP-violating effects in 1-point functions, that is a clear indication that the 2-point functions will be sick, with disastrous consequences for the real 2HDM.}
This can only happen in amplitudes with a) at least four vertices containing each a factor of $V_{u_\alpha d_j}$ and b) a vertex to couple to $A$.
Consequently, the first possible appearance of $J$ in
$A$-tadpoles is at three loops.
An example of a pair of diagrams yielding the
Jarlskog invariant is shown in fig.~\ref{Chap-Real:fig:jarl-pair}.
%
%
%

It goes without saying that the complexity of a calculation at three-loop level---even the most simple one, like the tadpole of $A$---cannot be compared to the complexity of the one-loop calculations of section \ref{Chap-Real:sec:toymodel}.
For a start, the complete calculation of three-loop tadpole of $A$ would require the renormalization of this process up to that order; unfortunately, such enterprise is not available in the literature and is beyond the scope of this thesis.
It is clear that, in the context of the PRTS, and just as in the toy model of section \ref{Chap-Real:sec:toymodel}, there is no tadpole counterterm $\delta t_{A}$ at the one-loop level, since there is no tadpole equation for $A$ at tree-level. However, this does not imply that there is no $\delta t_{A}$ at three-loops; in fact, one can form a multiplicity of diagrams that can be paired to generate $J$, and that in principle contribute to the total three-loop tadpole counterterm of $A$. Therefore, even if we are able to detect divergences in the non-renormalized diagrams, that would not immediatly demonstrate the unsoundness of the theory.

Nonetheless, we propose to tackle this problem by addressing a significant part of the calculation. Specifically, we look for the leading-pole contributions (i.e. terms proportional to $1/\varepsilon^3$) of the non-renormalized three-loop tadpole of A.
We generate all three-loop amplitudes that contain a fixed set of quarks $\{u_\alpha, u_\beta, d_i, d_j\}$;%
\fn{Other contributions are CP-conserving operators and, thus, irrelevant to our discussion.}
the total set of those amplitudes is dubbed $\left(T_A\right)^{\alpha i}_{\beta j} \big|_{\textrm{3L}}$ in the following.
Our calculation was carried out in three independent ways (two numeric; one analytical); the details are shown in section \ref{Chap-Real:sec:details} below. The result is%
\footnote{Notice that the angle $\beta$
in $s_\beta c_\beta$ is the angle in eq. \ref{Chap-Selec:eq:tanbtree},
while in all other instances of eq. \ref{Chap-Real:eq:tadpole},
$\beta$ refers to the up-type quark being considered.
Here and henceforth, which $\beta$ is meant should be
clear from the context. Finally, note that the results differ from those of ref. \cite{Fontes:2021znm} due to the definition of $\varepsilon$ (recall note \ref{Chap-Real:note:uu}).
}
\begin{fmffile}{Chap-Real-tadpole-A0}
\vspace{-5mm}
\be
\left(T_A\right)^{\alpha i}_{\beta j} \big|_{\textrm{3L},\varepsilon}
=
-i \,\Big( \;
\begin{gathered}
\vspace{-4pt}
\begin{fmfgraph*}(50,70)
\fmfright{t}
\fmfleft{b}
\fmf{dashes,label=$A$,label.dist=3}{b,t}
\fmfblob{20}{t}
\end{fmfgraph*}
\end{gathered}
\hs{5mm}
\Big)^{u_\alpha d_i}_{u_\beta d_j} \bigg|_{\textrm{3L},\varepsilon}
=
\frac{e^5}{\varepsilon^3 \, s_{\mathrm{w}}^5 \, m_{\mathrm{W}}^3 \, s_{\beta} \, c_{\beta}}
M^{\alpha i}_{\beta j}\, I^{\alpha i}_{\beta j}
+\mathcal{O}(\varepsilon^{-2}),
\label{Chap-Real:eq:tadpole}
\vspace{-5mm}
\ee
\end{fmffile}%
where 3L indicates three-loops, where there is no sum over repeated indices, and $M^{\alpha i}_{\beta j}$ is given by:
\be
M^{\alpha i}_{\beta j}
=
(m_{u_\alpha}^2 - m_{u_\beta}^2) (m_{d_i}^2 - m_{d_j}^2) (m_{u_\alpha}^2 - m_{d_i}^2 + m_{u_\beta}^2 - m_{d_j}^2)\, .
\label{Chap-Real:eq:M}	
\ee
%

Remarkably,
when summing over all different sets
of up- and down-type quark contributions,
the leading pole vanishes exactly.
Note that
both $M^{\alpha i}_{\beta j}$ and $I^{\alpha i}_{\beta j}$
are antisymmetric under $\alpha \leftrightarrow \beta$
(or $i \leftrightarrow j$).
As a consequence, the vanishing of 
eq. \ref{Chap-Real:eq:tadpole} is not due to the simple symmetry reasons
mentioned in connection with eq. \ref{Chap-Real:eq:A_ms}; rather, it is the specific form of the mass term
$M^{\alpha i}_{\beta j}$ in eq. \ref{Chap-Real:eq:M} which makes this possible.

We cannot see how one would have guessed from the start
such peculiar cancellation. We resonate with Khriplovich and Pospelov's remark in the context of EDM that:
``We cannot get rid of the feeling that this simple
result (...) should have a simple transparent explanation.
Unfortunately, we have not been able to find it.''
In our case, it remains uncertain whether the cancellation we just found has a physical origin, or if should be interpreted as accidental.
It is possible that the next orders in 
$1/\varepsilon$ are non-vanishing, even after the sums in $\left(T_A\right)^{\alpha i}_{\beta j} \big|_{\textrm{3L},\varepsilon}$ are performed. Otherwise, $\delta t_{A}$ would only become relevant at the four-loop level.

\subsection{Details of the calculation}
\label{Chap-Real:sec:details}

In this final section, we discuss our derivation of eq.  \ref{Chap-Real:eq:tadpole}.
One caveat in our calculation is the treatment of amplitudes
with an uneven number of $\gamma$-matrices together
with $\gamma_5$.
We chose to work in naive dimensional regularisation (NDR),
with the expectation that the leading $\varepsilon$-poles do
not depend on the choice of a $\gamma_5$-scheme.

As mentioned before, at least three generations of quarks are necessary to generate a CP-violating tadpole.
We focused on a particular set of diagrams, $S^{\{dcbt\}}$, corresponding to the set of all the three-loop tadpole diagrams for $A$ containing the quarks $d$, $c$, $b$, $t$.
We started by generating the amplitudes for $S^{\{dcbt\}}$ in an $R_\xi$-gauge.
We did this through two independent software: \FMS
and \textsc{FeynArts} \cite{Hahn:2000kx}.%
\footnote{Although \FMS is not prepared to draw the diagrams, it can correctly write the amplitudes. \FMS was used by myself, whereas \textsc{FeynArts} was used by Maximilian Löschner. It is worth emphasizing that the three-loop tadpole amplitudes generated with \FMS and \textsc{FeynArts} coincide.}
At three loops, there are 360 amplitudes containing the quarks $d,c,b$ and $t$. However, 120 among them involve two closed loops of fermions, which means that they can never factorize the Jarlskog invariant $J$; therefore, since the tadpole for $A$ violates CP, and since all the CP violation in the real 2HDM must be proportional to $J$, those diagrams must sum up to zero. We checked this explicitly using \FCS \cite{Mertig:1990an,Shtabovenko:2016sxi,Shtabovenko:2020gxv}.

We then focused on the remaining 240 diagrams.
After simplifying the Lorentz and Dirac algebra of the 240 diagrams with \FC, another set of 32 diagrams (such as the ones with two internal $W$-boson loops) immediately vanishes in NDR due to the chirality of the interactions involved. This eventually left us with 208 diagrams, which can be categorized as follows:
\begin{enumerate}
\item For the first group, diagrams can be generated by connecting $A$ to any of the fermion lines in any of the diagrams in fig.~\ref{Chap-Real:fig:category-1}; this gives 8 possibilities. If we add the corresponding ones with reversed fermion flow, we obtain 16 diagrams.
If we now consider all possible vector boson and scalar insertions, namely $\{HW,WH,HG,GH,HH,GW,WG,GG\}$, we obtain 128 diagrams. Yet, diagrams where $A$ is connected to a line with an attached $W$-loop vanish. Hence, the first group effectively contains 112 diagrams.
\begin{figure}[htp] 
\vspace{-5mm}
\hspace{6mm}
\centering 
\subfloat{ 
\begin{fmffile}{Chap-Real-three-loop-tadpole-C1a}
\fmfset{arrow_len}{3mm} 
\fmfset{arrow_ang}{20} 
\fmfframe(0,10)(0,10){
\begin{fmfgraph*}(90,70) 
\fmfleft{lb,lt} 
\fmfright{rb,rt}
\fmf{fermion,label.side=left,label=$c$}{lb,lt}
\fmf{fermion,label.side=left,label=$d$}{lt,rt}
\fmf{fermion,label.side=left,label=$t$}{rt,rb}
\fmf{fermion,label.side=left,label=$b$}{rb,lb}
\fmf{dashes_arrow,right=0.8,label.side=left,label.dist=3,label=$H^-$}{lb,lt}
\fmf{photon,right=0.7,label.side=left,label.dist=3,label=$W^-$}{rt,rb}
\fmf{phantom_arrow,right=0.7,tension=1}{rt,rb}
\end{fmfgraph*}
}
\end{fmffile} 
} 
\hspace{9mm} 
\subfloat{ 
\begin{fmffile}{Chap-Real-three-loop-tadpole-C1b}
\fmfset{arrow_len}{3mm} 
\fmfset{arrow_ang}{20} 
\fmfframe(0,10)(0,10){ 
\begin{fmfgraph*}(90,70)
\fmfleft{lb,lt} 
\fmfright{rb,rt}
\fmf{fermion,label.side=left,label=$d$}{lb,lt}
\fmf{fermion,label.side=left,label=$t$}{lt,rt}
\fmf{fermion,label.side=left,label=$b$}{rt,rb}
\fmf{fermion,label.side=left,label=$c$}{rb,lb}
\fmf{dashes_arrow,left=0.8,label.side=right,label.dist=2,label=$H^-$}{lt,lb}
\fmf{photon,right=0.7,label.side=left,label.dist=3,label=$W^-$}{rt,rb}
\fmf{phantom_arrow,left=0.7,tension=1}{rb,rt}
\end{fmfgraph*} 
}
\end{fmffile}
}
\caption{Attaching $A$ to the fermion lines of these diagrams generates part of diagrams of the first category.}
\label{Chap-Real:fig:category-1}
\end{figure}
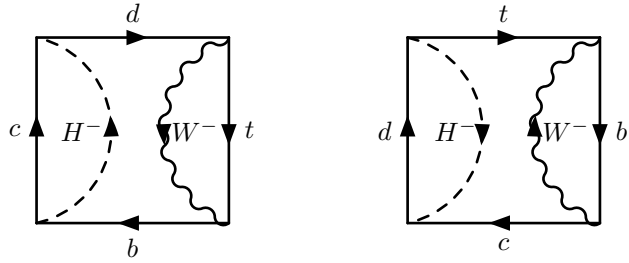

\item An example of the second group is shown in fig.~\ref{Chap-Real:fig:category-2}. From this diagram, and from the one with reversed fermion flow, we get 8 diagrams by cyclic permutations of the fermions.
The $A$ boson can be connected to either $\{WH,HW,GH,HG\}$, and we can have either a $W$-, $H$-, or $G$-boson in the r.h.s. of the diagram. This gives $3 \times 4 \times 8 = 96$ diagrams.
\begin{figure}[htp] 
\hspace{6mm}
\centering 
\begin{fmffile}{Chap-Real-three-loop-tadpole-C2a}
\fmfset{arrow_len}{3mm} 
\fmfset{arrow_ang}{20} 
\fmfframe(0,10)(0,10){
\begin{fmfgraph*}(115,83) 
\fmfleft{l} 
\fmflabel{$A$}{l} 
\fmfright{r}
\fmftop{t}
\fmfbottom{b}
\fmf{phantom}{vrb,r,vrt}
\fmf{phantom}{vlb,l,vlt}
\fmf{phantom,tension=1.5}{vlt,t,vrt}
\fmf{phantom,tension=1.5}{vlb,b,vrb}
\fmf{phantom}{r,vr,vl,l}
\fmf{dashes,tension=2}{l,vl}
\fmffreeze
\fmf{photon,label.side=left,label=$W^-$}{vl,vlt}
\fmf{phantom_arrow}{vl,vlt}
\fmf{dashes_arrow,label.side=left,label=$H^-$}{vlb,vl}
\fmf{fermion,label.side=left,label=$d$}{vlt,vrt}
\fmf{fermion,label.side=right,label=$t$}{vrt,vrb}
\fmf{fermion,label.side=left,label=$b$}{vrb,vlb}
\fmf{fermion,label.side=right,label=$c$}{vlb,vlt}
\fmf{photon,left=0.5,label.side=left,label=$W^-$}{vrt,vrb}
\fmf{phantom_arrow,left=0.5}{vrt,vrb}
\end{fmfgraph*} 
}
\end{fmffile} 
\vspace{-3mm}
\caption{The second set of relevant diagrams is generated from permutations of the fermions in this diagram and by replacing the $W$-insertions with a would-be Goldstone boson.}
\label{Chap-Real:fig:category-2}
\end{figure}
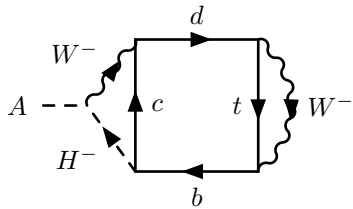
\end{enumerate}
Our goal, then, was to use \textsc{Fiesta} \cite{Smirnov:2015mct} to numerically evaluate the most divergent part of the 208 diagrams (in the Feynman gauge). In order to obtain an accurate result with \textsc{Fiesta}, we needed to decompose our integrals. Here, the \FCS function \texttt{ApartFF} played an essential role.
All integrals could be easily decomposed, except for one, namely: an integral with a scalar product in the numerator and five different propagator factors.
This type of integral yielded large error estimates in \textsc{Fiesta}, so that the results could no be trusted.
We thus resorted to \textsc{Fire}~\cite{Smirnov:2019qkx}, whose  integration-by-parts identities have proven very useful.
The intermediate steps that were needed to link \FC, \textsc{Fire} and \textsc{Fiesta} were performed by two independent sets of private codes.

For the numerical input values of the scalar sector, one should choose a point in parameter space which does not violate any theoretical or experimental constraints.
The theoretical bounds include boundedness from below, perturbative
unitarity~\cite{Kanemura:1993hm,Akeroyd:2000wc,Ginzburg:2005dt} as well
as electroweak precision measurements using the oblique parameters S,
T and U~\cite{Branco:2011iw}.  The experimental constraints include the
exclusion bounds from Higgs searches at LHC that were verified using
\textsc{HiggsBounds5}~\cite{Bechtle:2013wla,Bechtle:2020pkv} and the signal
strengths for the SM-like Higgs boson were forced to be within
2$\sigma$ of the fits given in~\cite{Khachatryan:2016vau,Aad:2019mbh}.
Among the points that pass all constraints, we pick the following one:
\begin{gather}
\alpha = -0.83797, \ \
\beta = 0.73908, \ \
m_{H^{\pm}} = 581.18 \, \, \text{GeV}, \ \
m_H = 592.81 \, \, \text{GeV}, \ \ 
m_{A} = 597.44 \, \, \text{GeV}, \no
m_{12}^2 = 19.458 \, \, \text{TeV}, \ \
m_{\mathrm{W}}  = 80.358 \, \, \text{GeV}, \ \
m_u  = 2.2 \times 10^{-3} \, \, \text{GeV}, \ \
m_d  = 4.8 \times 10^{-3} \, \, \text{GeV}, \no
m_c  = 1.4464 \, \, \text{GeV}, \ \
m_s  = 0.093 \, \, \text{GeV}, \ \
m_t  = 172.5 \, \, \text{GeV}, \ \
m_b  = 4.8564 \, \, \text{GeV}, \no
e  = 0.30812, \ \
\sin \theta_{\mathrm{w}} = 0.47206.
\label{Chap-Real:eq:mypoint}
\end{gather}
The result for the most divergent part of $S^{\{dcbt\}}$ is:%
\be
\left(i T_{A}\right)^{cd}_{tb} \big|_{\textrm{3L},\varepsilon}
= 19140.8 \, \, (\text{GeV})^3
 \times \dfrac{1}{\varepsilon^3}  J + \mathcal{O}(\varepsilon^{-2}).
\label{Chap-Real:eq:TAdiv1}
\ee
\n We ascertained whether sets of diagrams with different combinations of quarks could possibly cancel with each other.
For example, we considered the set of diagrams $S^{\{dcst\}}$, defined as identical to $S^{\{dcbt\}}$, except that the $b$ quark is replaced by an $s$ quark. Using the same point in parameter space (eq.~\ref{Chap-Real:eq:mypoint}), the result for the most divergent part of $S^{\{dcst\}}$ is
\be
\left(i T_{A}\right)^{cd}_{ts} \big|_{\textrm{3L},\varepsilon}
 = - 7.30728 \, \, (\text{GeV})^3
\times \dfrac{1}{\varepsilon^3} \, J + \mathcal{O}(\varepsilon^{-2}).
\label{Chap-Real:eq:TAdiv2}
\ee
Clearly, the numbers differ, which might lead one to believe that the sum over all quarks combinations would likely yield a non-zero result.

In the meantime, however, our attention was drawn to \textsc{TVID}~\cite{Bauberger:2017nct,Bauberger:2019heh}, a software that allows to determine \textit{analytical} expressions for the most divergent part of three-loop vacuum integrals. The authors discuss a set of three master integrals into which any three-loop tadpole diagram can be decomposed.
Using the same decomposition as in the numerical evaluations, we could find such master-integrals, which eventually led us to the analytical result shown in eq. \ref{Chap-Real:eq:tadpole}. We checked that the analytic results coincided with our previous numerical findings.%
\fn{This calculation also showed that the evaluation of scalar three-loop integrals with up to five propagators and different mass scales via \textsc{Fiesta} yields accurate results for the leading poles. This might be the first such stress test on this package.}
Since we now disposed of an analytical expression, we were able to sum over all possible quark combinations, which lead us to the surprising result described at the end of section \ref{Chap-Real:sec:3looptad}.

\section{\label{Chap-Real:sec:conclusions}Summary}

We argued that the real 2HDM is a theoretically unsound model, as the simultaneous enforcement of CP conservation in the potential and allowance of CP violation in another sector leads to divergences that cannot be removed by the counterterms.
Because such divergences cannot show up at two-loop level and below, the unsoundness of the model has been by and large ignored in the literature.
But the problem cannot be dismissed.
In order to highlight its inescapable nature, we introduced a simple toy model, characterized by the same inconsistency as the real 2HDM.
There, and as we showed, the irremovable divergences (that are expected at least at three-loops in the real 2HDM) show up immediately at one-loop level.
This simple example ought to make the point: the real 2HDM will suffer from the same kind of pathology.
We addressed this claim by calculating the leading pole of the three-loop 1-point function of the $A$ field in the real 2HDM.
We showed that, surprisingly, the pole vanishes exactly after summing all contributions.
This does not mean that the model is sound after all, but only that its unsoundness is to be found either at lower order in $1/\varepsilon$, or upon two-loop renormalization, or possibly at four-loop order.

%% file: Chapters/Chapter_Maggie.tex
\chapter{C2HDM: phenomenology at LO}
\label{Chap-Maggie}

\vs{-5mm}

%

In section \ref{Chap-Real:sec:short}, we noted that the real 2HDM can be given sense if it is understood not as model by itself, but rather as a particular case of a sound model, namely: the C2HDM. It is to this model that we now turn our attention. In this chapter, we investigate its phenomenology at LO, and in the next one we discuss its renormalization at NLO.

Here, we are particularly interested in ascertaining the CP parity of both the discovered Higgs boson and further scalars yet undiscovered. Due to its simplicity, the C2HDM is the ideal benchmark model to test the CP quantum numbers of the scalars at the LHC.
In principle, these quantum numbers could be ascertained not only through the Yukawa couplings (i.e. couplings between scalars and fermions), but also through the couplings between scalars and gauge bosons.
Concerning the latter, both ATLAS \cite{Aad:2015mxa} and CMS \cite{Khachatryan:2016tnr} collaborations have probed the CP-nature of the Higgs boson couplings to gauge bosons~\cite{Choi:2002jk, Buszello:2002uu, Godbole:2007cn}, by using correlations in the momentum distributions of leptons that were produced in the decays of the Higgs boson to gauge bosons;
the most general CP-violating $hVV$ coupling was used, and limits were set on the anomalous couplings. Yet, such anomalous couplings can only appear at loop level in the C2HDM (the tree-level couplings are similar to those of the SM), which means they are rather small.%
\fn{Recently (and after this chapter was concluded), Huang, Morais and Santos \cite{Huang:2020zde} investigated the one-loop CP-violating gauge-scalar interactions in the C2HDM, and concluded that they may be within reach of machines planned for the near-future.}
Therefore, in this model, only the Yukawa couplings can reasonably lead to direct observations of CP violation. Yukawa couplings have been experimentally ascertained in some particular cases, e.g. using asymmetries to study the top Yukawa coupling \cite{Gunion:1996xu, Boudjema:2015nda, AmorDosSantos:2017ayi}, or using the decays of tau leptons to investigate the tau Yukawa coupling~\cite{Berge:2008wi, Berge:2008dr, Berge:2011ij, Berge:2014sra, Berge:2015nua}.


However, there are other ways to probe CP violation besides analyzing the couplings of the scalar particles. As proposed in refs. \cite{Mendez:1991gp, Branco:1999fs, Fontes:2015xva}, several combinations of three simultaneously observed Higgs decay modes can constitute an undoubtable sign of CP violation. The argument can be summarized as follows: if CP were conserved, a decay of the type $h_i \rightarrow h_j Z$ would imply opposite CP parities for $h_i$ and $h_j$; on the other hand, a Higgs boson decaying into a pair of gauge bosons has to be CP-even;%
\fn{Strictly speaking, a CP-odd scalar can decay to a pair of gauge bosons at loop level. However, and in the context of a CP-conserving limit of the C2HDM, it was shown that the width of $h \to ZZ$ is several orders of magnitude smaller than the corresponding tree-level one \cite{Arhrib:2006rx, Bernreuther:2010uw}.}
hence, the combination of the decays $h_i \rightarrow Z Z$, $h_j \rightarrow Z Z$ and $h_j \rightarrow h_i Z$ is a clear sign of CP violation.
In ref. \cite{Fontes:2015xva}, seven classes of decays were proposed, some of which indicate CP violation for any extension of the SM.
%
%
We focus on those that include the already observed decay of the SM-like Higgs boson (denoted in the following by $h_{125}$) to gauge bosons, $h_{125} \to ZZ$. Furthermore, decays of the type $h_j \to ZZ$ and $h_j \to h_i Z$ were the subject of searches during LHC Run-1, and will proceed during Run-2.%
\fn{Note that this chapter was finished in 2017. Throughout the chapter, we keep to the original phrasing, and assume that the Run-2 lies in the future.}
Therefore, if a new Higgs boson $h_j$ is observed in both the final states $ZZ$ and $h_{125} Z$, the scalar sector is immediately established to be CP-violating.

In this chapter, we start by describing the C2HDM. Contrary to what will happen in chapter \ref{Chap-Reno}, here we are only interested in a tree-level description, which we present in detail in section \ref{Chap-Maggie:sec:model}. Then, in section \ref{Chap-Maggie:sec:c2hdmspace}, we investigate the parameter space of the model, which is restricted by several theoretical and experimental constraints; we will focus on the CP-character of $h_{125}$, and we will propose benchmark points with remarkable properties. In section~\ref{Chap-Maggie:sec:measure}, we try to find a relation between variables that signal CP violation, on the one hand, and production rates of the CP-violating classes described above, on the other.
Finally, motivated by those classes, we discuss Higgs-to-Higgs decays in the C2HDM in section \ref{Chap-Maggie:sec:higgstohiggs}.

\section{The C2HDM (for a tree-level description)}
\label{Chap-Maggie:sec:model}

\n The complete Lagrangian of the C2HDM can be written as a sum of the partial Lagrangians for the different sectors of the theory:
\be
\mathcal{L}_{\text{C2HDM}}
=
\mathcal{L}_{\text{Gauge}}
+
\mathcal{L}_{\text{Fermion}}
+
\mathcal{L}_{\text{Higgs}}
+
\mathcal{L}_{\text{Yukawa}}
+
\mathcal{L}_{\text{GF}}
+
\mathcal{L}_{\text{Ghost}}.
\label{Chap-Maggie:eq:myLag}
\ee
The terms $\mathcal{L}_{\text{Gauge}}$ and $ \mathcal{L}_{\text{Fermion}}$ are just those of the SM, while $\mathcal{L}_{\text{GF}}$ and $\mathcal{L}_{\text{Ghost}}$ can be easily derived from the SM ones (cf. e.g. ref.~\cite{Romao:2012pq}).
Here, we address the Higgs and the Yukawa partial Lagrangians. The former can be split into kinetic terms and potential,
\be
\mathcal{L}_{\text{Higgs}}
=
\mathcal{L}_{\text{Higgs}}^{\text{kin}}
- V.
\label{Chap-Maggie:eq:LHiggs}
\ee
We start by studying in detail the potential $V$, in section \ref{Chap-Maggie:section:pot}. Then, in sections \ref{Chap-Maggie:section:kin} and \ref{Chap-Maggie:section:Yukawa}, we address the scalar kinetic terms and the Yukawa Lagrangian, respectively.

\subsection{The potential}
\label{Chap-Maggie:section:pot}

The C2HDM has an explicitly
CP-violating scalar potential, with a  softly broken $\mathbb{Z}_2$ symmetry $\Phi_1 \ra \Phi_1, \Phi_2 \ra -\Phi_2$
written as
\beq
V &=& m_{11}^2 |\Phi_1|^2 + m_{22}^2 |\Phi_2|^2
- \left(m_{12}^2 \, \Phi_1^\dagger \Phi_2 + \text{h.c.}\right)
+ \frac{\lambda_1}{2} (\Phi_1^\dagger \Phi_1)^2 +
\frac{\lambda_2}{2} (\Phi_2^\dagger \Phi_2)^2 \nonumber \\
&& + \lambda_3
(\Phi_1^\dagger \Phi_1) (\Phi_2^\dagger \Phi_2) + \lambda_4
(\Phi_1^\dagger \Phi_2) (\Phi_2^\dagger \Phi_1) +
\left[\frac{\lambda_5}{2} (\Phi_1^\dagger \Phi_2)^2 + \text{h.c.}\right] .
\label{Chap-Maggie:eq:pot}
\eeq
The hermiticity of the Lagrangian obliges all parameters to be real except $m_{12}^2$ and $\lambda_5$, which are in general complex.
In what follows, we define:
\be
m_{12 \mathrm{R}}^2 \equiv \mathrm{Re}\left(m_{12}^2\right),
\qquad
m_{12 \text{I}}^2 \equiv \mathrm{Im}\left(m_{12}^2\right),
\qquad
\lambda_{5 \mathrm{R}} \equiv \mathrm{Re}\left(\lambda_{5}\right),
\qquad
\lambda_{5 \text{I}} \equiv \mathrm{Im}\left(\lambda_{5}\right).
\label{Chap-Maggie:eq:my-re-and-im}
\ee

We write each of the doublets $\Phi_1$ and $\Phi_2$ as in eq. \ref{Chap-Real:parametrizacao-PHIs}, namely: as an expansion around
the real vevs $v_{1}$ and $v_{2}$,
in terms of the charged complex fields ($\phi_i^+$)
and the real neutral fields ($\rho_i$ and $\eta_i$).%
\fn{\label{Chap-Maggie:note:vevs}The vevs $v_1$ and $v_2$ are in general complex. However, when considering the theory solely at tree-level (i.e. not aiming at the one-loop renormalization), one can always go to the basis of $\Phi_1$ and $\Phi_2$ where the vevs are real. This is what we do in this chapter. Finally, note that, since $m_{12}^2$ and $\lambda_5$ are in general complex, there are bilinear terms mixing the $\rho$ fields and the $\eta$ fields.}
Just like in eq. \ref{Chap-Real:parametrizacao-PHIs}, then, the doublets read
\beq
\Phi_1 = \left(
\begin{array}{c}
\phi_1^+ \\
\tfrac{1}{\sqrt{2}}(v_1 + \rho_1 + i \eta_1)
\end{array}
\right),
\qquad 
\Phi_2 = \left(
\begin{array}{c}
\phi_2^+ \\
\tfrac{1}{\sqrt{2}} (v_2 + \rho_2 + i \eta_2)
\end{array}
\right).
\label{Chap-Maggie:eq:2hdmdoubletexpansion}
\eeq
We also define the real parameters $v$ and $\beta$ as in eqs. \ref{Chap-Selec:eq:tanbtree} and \ref{Chap-Selec:eq:totalvevtree}, which we rewrite here as:
\be
v^2 \equiv v_1^2 + v_2^2,
\qquad
\tan \beta \equiv \dfrac{v_2}{v_1},
\qquad
v_1 = v \, c_{\beta},
\qquad
v_2 = v \, s_{\beta}.
\label{Chap-Maggie:eq:preliminar}
\ee
The minimum conditions for the potential read:
\bs
\label{Chap-Maggie:eq:min-eqs}
\beq
m_{11}^2 v_1 + \frac{\lambda_1}{2} v_1^3 + \frac{\lambda_{345}}{2} v_1
v_2^2 &=& m_{12 \mathrm{R}}^2 v_2 , \label{Chap-Maggie:eq:mincond1} \\
m_{22}^2 v_2 + \frac{\lambda_2}{2} v_2^3 + \frac{\lambda_{345}}{2} v_1^2
v_2 &=& m_{12 \mathrm{R}}^2 v_1 , \label{Chap-Maggie:eq:mincond2} \\
2\, m_{12 \text{I}}^2 &=& v_1 v_2 \lambda_{5 \text{I}}
, 
\label{Chap-Maggie:eq:mincond3}
\eeq
\es
where we used the short notation $\lambda_{345} \equiv \lambda_3 + \lambda_4 + \lambda_{5 \mathrm{R}}$. If we define $\phi (m_{12}^2)$ and $\phi (\lambda_5)$ as
\begin{equation}
m_{12}^2 = |m_{12}^2|\, e^{i\, \phi (m_{12}^2)},
\qquad \qquad \lambda_5 = |\lambda_5|\, e^{i\, \phi (\lambda_5)},
\label{Chap-Maggie:eq-phases}
\end{equation}
we can use eq. \ref{Chap-Maggie:eq:mincond3} to conclude that these two
phases are related. Indeed, we can re-write eq. \ref{Chap-Maggie:eq:mincond3} as
\begin{equation}
2 \, m_{12 \mathrm{R}}^2 \, \tan \phi (m_{12}^2) = v_1 v_2 \, \lambda_{5 \mathrm{R}} \, \tan \phi (\lambda_5) \, .
\end{equation}
The condition $\phi (\lambda_5) \neq 2\, \phi (m_{12}^2)$ (together with $v_1$ and $v_2$ real) ensures that the two phases cannot be removed simultaneously \cite{Gunion:2002zf,Ginzburg:2002wt}; otherwise, we are in the CP-conserving limit of the model.%
\fn{Recall that we already used the rephasing freedom of $\Phi_1$ and $\Phi_2$ to choose a basis where $v_1$ and $v_2$ are real (cf. note \ref{Chap-Maggie:note:vevs}), which means that there is no longer freedom to rephase $\phi (\lambda_5)$ or $\phi (m_{12}^2)$ away.}

The Higgs basis is still defined by eq. \ref{Chap-Real:eq:HiggsBasis}. However, the doublets in that basis are now written as:
\beq
{\cal H}_1 = \left( \begin{array}{c} G^+ \\ \frac{1}{\sqrt{2}} (v + H^0
    + i G^0) \end{array} \right) \quad \mbox{and} \qquad
{\cal H}_2 = \left( \begin{array}{c} H^+ \\ \frac{1}{\sqrt{2}} (R_2
    + i I_2) \end{array} \right),
\label{Chap-Maggie:eq:Doublets}
\eeq
with $H^0$, $R_2$ and $I_2$ real fields.
If we identify $\rho_3 \equiv I_2$, the neutral Higgs mass eigenstates $h_1$, $h_2$ and $h_3$ can be obtained by
\beq
\left( \begin{array}{c} h_1 \\ h_2 \\ h_3 \end{array} \right)
=
R
\left( \begin{array}{c} \rho_1 \\ \rho_2 \\ \rho_3 \end{array} \right)
=
T^{\mathrm{T}}
\left( \begin{array}{c} H^0 \\ R_2 \\ I_2 \end{array} \right)
,
\label{Chap-Maggie:eq:c2hdmrot}
\eeq
where $R$ is an orthogonal matrix that we write as \cite{ElKaffas:2007rq,Arhrib:2010ju}:
\be
\begin{split}
R = R_3 \, R_2 \, R_1 &=
\left(
\begin{array}{ccc}
1 & 0 & 0 \\
0 & c_3 & s_3 \\
0 & -s_3 & c_3 
\end{array}
\right)
\left(
\begin{array}{ccc}
c_2 & 0 & s_2 \\
0 & 1 & 0 \\
-s_2 & 0 & c_2 
\end{array}
\right)
\left(
\begin{array}{ccc}
c_1 & s_1 & 0 \\
-s_1 & c_1 & 0 \\
0 & 0 & 1
\end{array}
\right)
\\
&=
\left(
\begin{array}{ccc}
c_1 c_2 & s_1 c_2 & s_2\\
-(c_1 s_2 s_3 + s_1 c_3) & c_1 c_3 - s_1 s_2 s_3  & c_2 s_3\\
- c_1 s_2 c_3 + s_1 s_3 & -(c_1 s_3 + s_1 s_2 c_3) & c_2 c_3
\end{array}
\right),
\label{Chap-Maggie:matrixR}
\end{split}
\ee
with $s_i = \sin{\alpha_i}$, $c_i = \cos{\alpha_i}$ ($i = 1, 2, 3$), and
where the matrix $T$, used in ref.~\cite{Branco:2011iw} for the expression of the oblique radiative corrections, is defined in such a way that:
\beq
T^{\mathrm{T}} = R \, \left( \begin{array}{cc} R_H & 0 \\ 0 & 1 \end{array}
\right)
=
R
\left( \begin{array}{ccc}
c_\beta & - s_\beta & 0\\
s_\beta & c_\beta & 0\\
0 & 0 & 1
\end{array}
\right),
\eeq
where we used eq. \ref{Chap-Real:eq:HiggsBasis}.
%
The mass matrix of the neutral scalar states,
\beq
({\cal M}_n^2)_{ij} = \left\langle \frac{\partial^2 V}{\partial \rho_i
  \partial \rho_j} \right\rangle ,
\label{Chap-Maggie:eq:c2hdmmassmat}
\eeq
is diagonalized by construction through the matrix $R$, i.e.
\beq
R {\cal M}_n^2 R^{\mathrm{T}} = \mbox{diag} (m_{1}^2, m_{2}^2, m_{3}^2) ,
\eeq
which in turn implies
\beq
{\cal M}_n^2= R^{\mathrm{T}} \, \mbox{diag} (m_{1}^2, m_{2}^2, m_{3}^2) \, R ,
\label{Chap-Maggie:eq:target}
\eeq
where $m_i$ represents the mass of $h_i$, such that $m_1 < m_2 < m_3$.
We choose as independent parameters of the potential sector of the C2HDM, $\{p^{\mathrm{V}}\}$, the following list:
\be
\{p^{\mathrm{V}}\}
=
\{ \alpha_1, \, \alpha_2, \, \alpha_3, \, \beta, \, m_{1}, \, m_{2}, \,  m_{\mathrm{H}^{+}}, \, \mu^2 \},
\label{Chap-Maggie:eq:indeps}
\ee
where $m_{\mathrm{H}^{+}}$ is the mass of the physical charged scalar $H^+$, and where we define:
\be
\mu^2 \equiv
\dfrac{m_{12 \mathrm{R}}^2}{c_{\beta} \, s_{\beta}}.
\label{Chap-Maggie:eq:mu}
\ee
With this choice, the mass of the heaviest neutral scalar is a dependent parameter. In fact, the matrix ${\cal M}_n^2$ is such that 
\be
\dfrac{\left(\mathcal{M}_n^{2}\right)_{13}}{\left(\mathcal{M}_n^{2}\right)_{23}} = \tan \beta ;
\label{Chap-Maggie:eq:rel-Mn}
\ee
since the r.h.s. of eq. \ref{Chap-Maggie:eq:target} must also obey this relation, we obtain:%
\fn{The parameter space points will have to comply with $m_{3} > m_{2}$.}
\be
m_{3}^2 = \frac{m_{1}^2\, R_{13} (R_{12} \tan{\beta} - R_{11})
+ m_{2}^2\ R_{23} (R_{22} \tan{\beta} - R_{21})}{R_{33} (R_{31} - R_{32} \tan{\beta})}.
\label{Chap-Maggie:m3_derived}
\ee
Finally, the $\lambda$'s in eq. \ref{Chap-Maggie:eq:pot} can be rewritten as:
\bs
\label{Chap-Maggie:eq:lambdas}
\bea
v^2\, \lambda_1 &=&
- \frac{1}{c_{\beta}^2}
\left[- m_1^2\, c_1^2 c_2^2 - m_2^2 (c_3 s_1 + c_1 s_2 s_3)^2
- m_3^2\, (c_1 c_3 s_2 - s_1 s_3)^2 + \mu^2\, s_{\beta}^2
\right],
\\
v^2\, \lambda_2 &=&
- \frac{1}{s_{\beta}^2}
 \left[
- m_1^2\, s_1^2 c_2^2 - m_2^2\, (c_1 c_3 - s_1 s_2 s_3)^2
- m_3^2\, (c_3 s_1 s_2 + c_1 s_3)^2  + \mu^2\, c_{\beta}^2
\right],
\\
v^2\, \lambda_3 &=&
\frac{1}{s_{\beta} c_{\beta}}
\left[
\left(
m_1^2\, c_2^2
+ m_2^2\, (s_2^2 s_3^2 - c_3^2)
+ m_3^2\, (s_2^2 c_3^2 - s_3^2)
\right) c_1 s_1
\right.
\nonumber \\
& &
\hspace{15ex}
\left.
+\,
(m_3^2 - m_2^2) (c_1^2 - s_1^2) s_2 c_3 s_3
\right]
- \mu^2 + 2 m_{\mathrm{H}^{+}}^2,
\\
v^2\, \lambda_4 &=&
m_1^2\, s_2^2 + ( m_2^2\,  s_3^2 + m_3^2\, c_3^2) c_2^2
+ \mu^2 - 2 m_{\mathrm{H}^{+}}^2,
\\
v^2\, \lambda_{5 \mathrm{R}}
&=&
- m_1^2\, s_2^2 - (m_2^2\, s_3^2 + m_3^2\, c_3^2) c_2^2 + \mu^2,
\\
v^2\, \lambda_{5 \text{I}}
&=&
\frac{2}{s_{\beta}}
c_2
\left[
(- m_1^2 + m_2^2\, s_3^2 + m_3^2\, c_3^2) c_1 s_2
+ (m_2^2 - m_3^2) s_1 s_3 c_3
\right].
\eea
\es

\subsection{Scalar kinetic sector}
\label{Chap-Maggie:section:kin}

Concerning the term $\mathcal{L}_{\text{Higgs}}^{\text{kin}}$ in eq. \ref{Chap-Maggie:eq:LHiggs}, and for the purposes of this chapter, it is sufficient to parameterize the couplings between the physical neutral scalars $h_i$ and the massive gauge bosons $V=W,Z$.%
\fn{More details can be found in section \ref{Chap-Reno:section:kin}. The full set of couplings is contained in the web page~\cite{C2HDM_FR}, and can also be obtained with \FMS (which was not yet available by the time this chapter was completed).}
Those couplings are given by
\beq
i \, g_{\mu\nu} \, c(h_i VV) \, g_{h_{\mathrm{SM}} VV} , \label{Chap-Maggie:eq:gaugecoupdef}
\eeq
where $h_{\mathrm{SM}}$ denotes the SM Higgs boson, so that the couplings $g_{h_{\mathrm{SM}} VV}$ are
\beq
g_{h_{\mathrm{SM}} VV} = \Bigg\{
\begin{array}{ll}
e \, m_{\mathrm{W}}/s_{\mathrm{w}}, & \textrm{for} \ \ V=W, \\[1.6mm]
e \, m_{\mathrm{Z}} /(s_{\mathrm{w}} c_{\mathrm{w}}), & \textrm{for} \ \ V=Z,
\end{array}
\eeq
and the coefficients $c(h_i VV)$ can be written as
\beq
c(h_i VV) = T_{1i} = c_\beta R_{i1} + s_\beta R_{i2}, \label{Chap-Maggie:eq:c2dhmgaugecoup}
\eeq
where the elements of matrix $R$ are given in eq. \ref{Chap-Maggie:matrixR}.

\subsection{Yukawa sector}
\label{Chap-Maggie:section:Yukawa}

As mentioned in section \ref{Chap-Real:sec:short}, the existence of two Higgs doublets is such that the Yukawa Lagrangian in general leads to flavour-changing neutral currents at tree-level. A simple way to avoid this is to ensure that each right-handed fermionic singlet couples to only one Higgs doublet. This in turn can be accomplished if the $\mathbb{Z}_2$ symmetry is extended to the fermion fields \cite{Glashow:1976nt,Paschos:1976ay}, such that:
\be
\bar{q}_{\mathrm{L}} \to (-1)^a \, \bar{q}_{\mathrm{L}},
\hs{4.5mm}
\bar{n}_{\mathrm{R}} \to (-1)^b \, \bar{n}_{\mathrm{R}},
\hs{4.5mm}
p_{\mathrm{R}} \to (-1)^c \, p_{\mathrm{R}},
\hs{4.5mm}
\bar{L}_{\mathrm{L}} \to (-1)^d \, \bar{L}_{\mathrm{L}},
\hs{4.5mm}
l_{\mathrm{R}} \to (-1)^e \, l_{\mathrm{R}}.
\label{Chap-Reno:eq:coeffs-a-to-e}
\ee
Here, $a, b, c, d, e$ are general powers, $q_{\mathrm{L}} = (p_{\mathrm{L}} \, n_{\mathrm{L}})^{\mathrm{T}}$  and $L_{\mathrm{L}} = (\nu_{\mathrm{L}} \, l_{\mathrm{L}})^{\mathrm{T}}$ are the quark and lepton left-handed $\mathrm{SU(2)_L}$ doublets, respectively, and $p_{\mathrm{R}}$, $n_{\mathrm{R}}$ and $l_{\mathrm{R}}$ are the up-type quark, down-type quark and lepton right-handed singlets, respectively. There are four different combinations of the powers $a$ to $e$, and therefore four different types of C2HDM: Type I, Type II, Lepton-Specific (LS) and Flipped.
For each type of C2HDM, we show in table \ref{Chap-Maggie:tab:coeffs_models}
\begin{table}[!h]
\begin{normalsize}
\normalsize
\begin{center}
\begin{tabular}
{@{\hspace{3mm}} >{\raggedright\arraybackslash}p{2.9cm} >{\raggedleft\arraybackslash}p{0.8cm}
>{\raggedleft\arraybackslash}p{0.8cm}
>{\raggedleft\arraybackslash}p{0.8cm}
>{\raggedleft\arraybackslash}p{0.8cm}
>{\raggedleft\arraybackslash}p{0.8cm}
>{\raggedleft\arraybackslash}p{0.2cm}
>{\raggedleft\arraybackslash}p{0.8cm}
>{\raggedleft\arraybackslash}p{0.8cm}
>{\raggedleft\arraybackslash}p{0.8cm}
@{\hspace{3mm}}}
\hlinewd{1.1pt}
Type of C2HDM & $a$ & $b$ & $c$ & $d$ & $e$ & & $p_{\mathrm{R}}$ & $n_{\mathrm{R}}$& $l_{\mathrm{R}}$ \\
\hline
Type I & $0$ & $1$ & $1$ & $0$ & $1$ & & $\Phi_2$ & $\Phi_2$ & $\Phi_2$ \\
Type II & $0$ & $0$ & $1$ & $0$ & $0$ & & $\Phi_2$ & $\Phi_1$ & $\Phi_1$ \\
Lepton-Specific & $0$ & $1$ & $1$ & $0$ & $0$ & & $\Phi_2$ & $\Phi_2$ & $\Phi_1$ \\
Flipped & $0$ & $0$ & $1$ & $0$ & $1$ & & $\Phi_2$ & $\Phi_1$ & $\Phi_2$ \\
\hlinewd{1.1pt}
\end{tabular}
\end{center}
\vspace{-5mm}
\end{normalsize}
\caption{For each type of C2HDM, the powers $a, b, c, d, e$ of eq. \ref{Chap-Reno:eq:coeffs-a-to-e} (columns 2-6) and the Higgs doublet which the right-handed fermionic singlets $p_{\mathrm{R}}$, $n_{\mathrm{R}}$ and $l_{\mathrm{R}}$ couple to (columns 7-9).}
\label{Chap-Maggie:tab:coeffs_models}
\end{table}
\normalsize
the values of the different powers $a$ to $e$, as well as the Higgs doublet which the different fermionic singlets couple to.

After SSB and diagonalization of the fields, the part of the Yukawa Lagrangean corresponding to the interactions between the physical neutral scalars and the fermions can be written as:
\beq
{\cal L}_{\text{Yukawa}}^{h} = - \sum_{i=1}^3 \frac{m_f}{v} \bar{\psi}_f \left[ c^e(h_i
ff) + i c^o(h_i ff) \gamma_5 \right] \psi_f h_i, \label{Chap-Maggie:eq:yuklag}
\eeq
where $\psi_f$ denote the fermion fields with mass $m_f$, and where $c^e(h_i ff)$ and $c^o (h_i ff)$ represent the coefficients of the CP-even and of the CP-odd part of the Yukawa coupling, respectively. These coefficients are presented in table~\ref{Chap-Maggie:tab:yukcoup} in the form $c^e(h_i ff) + i c^o(h_i ff) \gamma_5$ for the different types of C2HDM and for the different types of fermions (up-type quarks, down-type quarks and leptons).
\begin{table}[h!]
\begin{center}
\begin{tabular}{rccc} \toprule
& up-type quarks & down-type quarks & leptons \\ \midrule
Type I & $\dfrac{R_{i2}}{s_\beta} - i \dfrac{R_{i3}}{t_\beta} \gamma_5$
& $\dfrac{R_{i2}}{s_\beta} + i \dfrac{R_{i3}}{t_\beta} \gamma_5$ &
$\dfrac{R_{i2}}{s_\beta} + i \dfrac{R_{i3}}{t_\beta} \gamma_5$ \\[4mm]
Type II & $\dfrac{R_{i2}}{s_\beta} - i \dfrac{R_{i3}}{t_\beta} \gamma_5$
& $\dfrac{R_{i1}}{c_\beta} - i t_\beta R_{i3} \gamma_5$ &
$\dfrac{R_{i1}}{c_\beta} - i t_\beta R_{i3} \gamma_5$ \\[4mm]
LS & $\dfrac{R_{i2}}{s_\beta} - i \dfrac{R_{i3}}{t_\beta} \gamma_5$
& $\dfrac{R_{i2}}{s_\beta} + i \dfrac{R_{i3}}{t_\beta} \gamma_5$ &
$\dfrac{R_{i1}}{c_\beta} - i t_\beta R_{i3} \gamma_5$ \\[4mm]
Flipped & $\dfrac{R_{i2}}{s_\beta} - i \dfrac{R_{i3}}{t_\beta} \gamma_5$
& $\dfrac{R_{i1}}{c_\beta} - i t_\beta R_{i3} \gamma_5$ &
$\dfrac{R_{i2}}{s_\beta} + i \dfrac{R_{i3}}{t_\beta} \gamma_5$ \\ \bottomrule
\end{tabular}
\caption{Yukawa couplings of the Higgs
bosons $h_i$ in the C2HDM, divided by the corresponding SM Higgs couplings. The expressions correspond to
$c^e(h_i ff) +i c^o (h_i ff) \gamma_5$ from
eq.~\ref{Chap-Maggie:eq:yuklag}, and the elements of the matrix $R$ can be found in eq. \ref{Chap-Maggie:matrixR}.}
\label{Chap-Maggie:tab:yukcoup}
\end{center}
\vs{-7mm}
\end{table}
%
%
%

\section{Parameter space \label{Chap-Maggie:sec:c2hdmspace}}

\subsection{Experimental and Theoretical Restrictions}
\label{Chap-Maggie:sec:rest}

The C2HDM was implemented as a model class in
\textsc{ScannerS}~\cite{Coimbra:2013qq,ScannerS}.
The relevant theoretical and experimental bounds are either built
in the code or acessible via interfaces with other codes. We have
imposed all available constraints on the model and performed a parameter scan. The resulting viable points are the basis for our phenomenological analyses.

The theoretical bounds included in \textsc{ScannerS} are boundness from below and
perturbative unitarity~\cite{Kanemura:1993hm, Akeroyd:2000wc,Ginzburg:2003fe}.
Contrary to the SM, in the 2HDM coexisting minima can occur at tree-level; therefore,  we also force the minimum to be global~\cite{Ivanov:2015nea}, precluding the possibility of vacuum decay. The points generated comply with electroweak precision measurements, making
use of the oblique parameters S, T and U~\cite{Branco:2011iw}; we ask for a
$2\sigma$ compatibility of S, T and U with the SM fit presented in~\cite{Baak:2014ora} (the full correlation among these parameters is taken into account).

The charged sector of the C2HDM has exactly the same couplings as the charged sector of the so-called real 2HDM, so that the exclusion bounds on the $m_{\mathrm{H}^{+}}-t_\beta$ plane in the C2HDM can be imported from that scenario.
The most constraining bounds on this plane come from the measurements of $B \to X_s \gamma$ \cite{Deschamps:2009rh,Mahmoudi:2009zx,Hermann:2012fc,Misiak:2015xwa,Misiak:2017bgg}.
$2\sigma$ bounds on this plane were obtained in~\cite{Misiak:2017bgg} and force the charged Higgs mass to be
$m_{\mathrm{H}^{+}} > 580 \mbox{ GeV}$ for models Type II and Flipped,
almost independently of $\tan \beta$.
Due to the structure of the charged Higgs couplings to fermions, in models Type I and LS the bound has a strong dependence on $\tan \beta$. In fact, for $\tan \beta \approx 1$ the bound is about $400  \mbox{ GeV}$, while the LEP bound derived from $e^+ e^- \to H^+ H^-$~\cite{Abbiendi:2013hk}
(approximately $100  \mbox{ GeV}$) is recovered for $\tan \beta \approx 1.8$.
We further apply the flavour constraints from $R_b$~\cite{Haber:1999zh,Deschamps:2009rh} (which are revisited in chapter \ref{Chap-Lavou}).
All the constraints are checked as $2\sigma$ exclusion bounds on the $m_{\mathrm{H}^{+}}-t_\beta$ plane.

The SM-like Higgs boson (which, recall, we denote by $h_{125}$) has a mass
$m_{h_{125}} = 125.09  \mbox{GeV}$ \cite{Aad:2015zhl}. We exclude points of the parameter space with
the discovered Higgs signal built by two nearly degenerate Higgs boson states, by forcing the non-SM scalar masses to be outside the mass window
$m_{h_{125}} \pm 5$~GeV. Compatibility with the exclusion bounds from Higgs searches is checked with the
\textsc{HiggsBounds4} code \cite{Bechtle:2008jh,Bechtle:2011sb,Bechtle:2013wla},
while the individual signal strengths for the SM-like
Higgs boson are forced to be within $2 \sigma$ of the fits presented in ref. \cite{Khachatryan:2016vau}.
Branching ratios and decay widths of all Higgs bosons are calculated with the
\textsc{C2hdm\_Hdecay} code \cite{Fontes:2017zfn}, which is an implementation of the C2HDM model into \textsc{Hdecay v6.51} \cite{Djouadi:1997yw,Butterworth:2010ym}.
The code has state-of-the-art QCD corrections and
off-shell decays (off-shell decays of one scalar into two are not included). The Higgs boson production cross sections
 via gluon fusion and $b$-quark
fusion are calculated with \textsc{SusHiv1.6.0} \cite{Harlander:2012pb,Harlander:2016hcx} (which is interfaced with \textsc{ScannerS}) at next-to-next-to-leading
order in QCD. As the neutral scalars have no definite CP,
we needed to combine the CP-odd and the CP-even contributions by
summing them incoherently; for details, cf. ref. \cite{Fontes:2017zfn}.

%

The C2HDM is a model with explicit CP violation in the scalar sector.
There are several experiments that constrain
the amount of CP violation in the model; for details, cf. ref. \cite{Fontes:2017zfn} and references therein.
The most restrictive bounds on the CP-phase~\cite{Inoue:2014nva}
(see also refs. \cite{Buras:2010zm, Cline:2011mm, Jung:2013hka, Shu:2013uua,
Brod:2013cka,Basler:2017uxn}) originate from the ACME~\cite{Baron:2013eja} results on the ThO molecule electric dipole moment (EDM), which allows to determine the electron EDM, $d_e$. By the time this chapter was concluded, the ACME bound on the electron EDM was $|d_e|< 9.3 \times 10^{-29} \, e \, \textrm{cm}$ (90\% confidence) \cite{Baron:2013eja}.%
\fn{In the meantime, a more stringent constraint was determined by ACME, $|d_e|< 1.1 \times 10^{-29} \, e \, \textrm{cm}$ \cite{Andreev:2018ayy}. We briefly analyzed its impact on the results of this chapter; although it obviously restricts even more the parameter space, it does not affect the main conclusions derived here.}
We require the points in parameter space to conform to this bound.

In our scan, one of the Higgs bosons $h_i$ is identified with $h_{125}$.
One of the other neutral Higgs bosons is varied between $30\mathrm{GeV}\leq m_{i}<1\mathrm{ TeV}$, while the third neutral Higgs boson is not an independent parameter (recall eq. \ref{Chap-Maggie:m3_derived}) and is calculated by \textsc{ScannerS}, but its mass is forced to be in the same interval. In Type II and Flipped, the charged Higgs boson mass is forced to be in the range $580 \mbox{ GeV } \le m_{\mathrm{H}^{+}} < 1 \mbox{ TeV }$, 
while in Type I and LS we choose $ 80 \mbox{ GeV } \le m_{\mathrm{H}^{+}} < 1 \mbox{ TeV }$.
Taking into account all the constraints, in order to optimize
the scan, we have chosen the following regions for the remaining input parameters:
$0.8 \le \tan \beta  \le 35$, $- \frac{\pi}{2} \le \alpha_{1,2,3} < \frac{\pi}{2}$ and
$ 0 \mbox{ GeV}^2 \le m_{12 \mathrm{R}}^2 < 500 000 \mbox{ GeV}^2$.%
\fn{We have generated a sample with
just the theoretical constraints and the number of points with a negative $m_{12 \mathrm{R}}^2$
is of the order of 1 in 10 million. When we further impose the experimental constraints, all such points vanish.}


\subsection{\label{Chap-Maggie:subsec:constraints}Constraints on the parameter space: $c^e$ versus $c^o$}

\begin{figure}[h]
\centering
\includegraphics[width=0.57\linewidth]{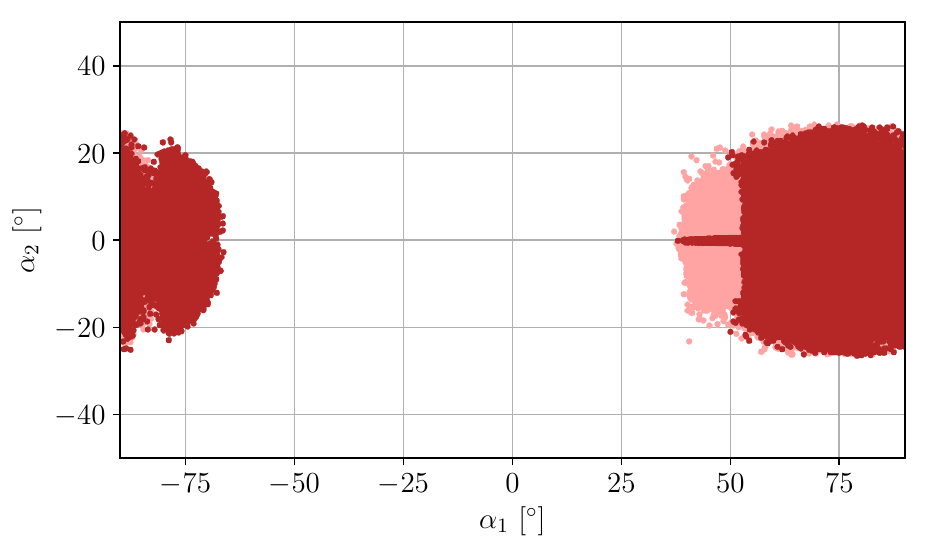}
\includegraphics[width=0.37\linewidth]{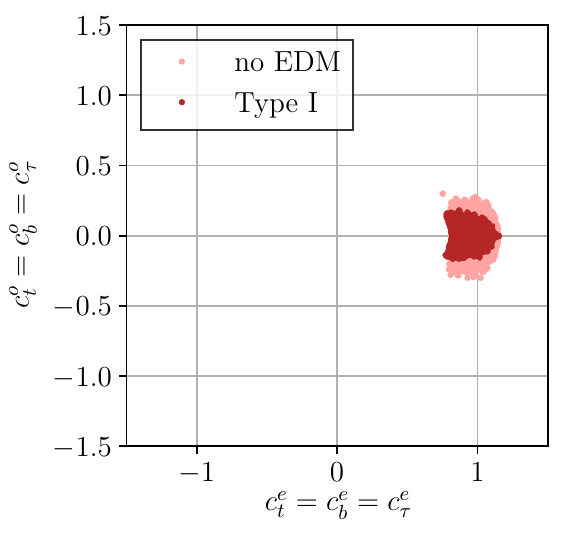}
\caption{Type I, $h_1=h_{125}$, for sample 1 (dark) and sample 2 (light).
Left panel: mixing angles $\alpha_1$ and $\alpha_2$. Right panel: Yukawa couplings.} \label{Chap-Maggie:fig:t1c2hdmyuks}
\end{figure}

In this subsection, we confront the C2HDM parameter space with all the restrictions presented above.
Our aim is a double one: on the one hand, we wish to ascertain what regions of the remaining parameter are still allowed; on the other, we want to study the CP-nature of the 125 GeV scalar, encoded in the coefficients
\be
c_f^e \equiv c^e(h_{125} f f),
\qquad
c_f^o \equiv c^o(h_{125}ff)
\ee
of eq.~\ref{Chap-Maggie:eq:yuklag}.
These coefficients allow to test the CP content of $h_{125}$.
We know that $h_{125}$
must have some CP-even content, because it couples at tree-level to $ZZ$.
However, in a theory with CP violation in the scalar sector (such as the C2HDM), $h_{125}$ could have a mixed CP nature.
This possibility can be probed in the couplings to fermions in a variety of ways.
The simplest case occurs if $c^e_f c^o_f \neq 0$, meaning that, in the coupling to some fermion, there are both CP-even and CP-odd components  (thus establishing CP violation).
A more peculiar and rather interesting case occurs if $h_{125}$ (or some other mass eigenstate) has a pure scalar component to a given type of fermion ($f$) and a pure pseudoscalar component to a different type of fermion ($g$). This would render $c^e_f c^o_f = 0 = c^e_g c^o_g$, whereas $c^e_f c^o_g \neq 0$. In the C2HDM, and as we will now show, this peculiar scenario is no longer available for Type I, but it is possible in all other three types.
%
%
For what follows, we consider two different samples of points:
\begin{equation*}
\textrm{sample 1: applying all constraints},
\qquad
\textrm{sample 2: applying all constraints except EDM}.
\end{equation*}
We start by investigating the Type I in fig. \ref{Chap-Maggie:fig:t1c2hdmyuks}, assuming $h_1=h_{125}$.
The left panel displays the mixing angles $\alpha_2$
and $\alpha_1$,
while the right one displays the CP-odd ($c^o_f$) and the CP-even ($c^e_f$) components of the Yukawa couplings of the Higgs boson $h_{125}=h_1$.
Both plots clearly show that the EDM constraints have little effect on $|\alpha_2|$, which can go up to 25$^\circ$ when all constraints
are taken into account.
%
The right plot also demonstrates that the so-called wrong-sign
regime, $c_b^e < 0$, is in conflict with the Type I constraints.
%
%
This is consistent with what was shown previously in refs. \cite{Ferreira:2014naa, Ferreira:2014dya}.

\begin{figure}[h]
\centering
\includegraphics[width=0.55\linewidth]{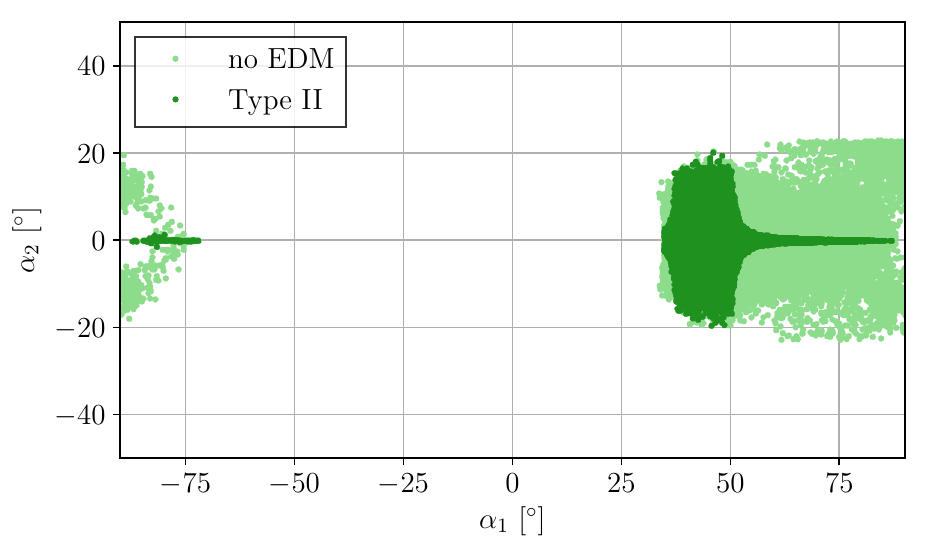}
\vs{-3mm}
\caption{Type II, $h_1=h_{125}$: mixing angles $\alpha_1$ and $\alpha_2$ for sample 1 (dark) and sample 2 (light).}
\label{Chap-Maggie:fig:mixangledist}
\end{figure}

\begin{figure}[h]
\centering
\includegraphics[width=0.9\linewidth]{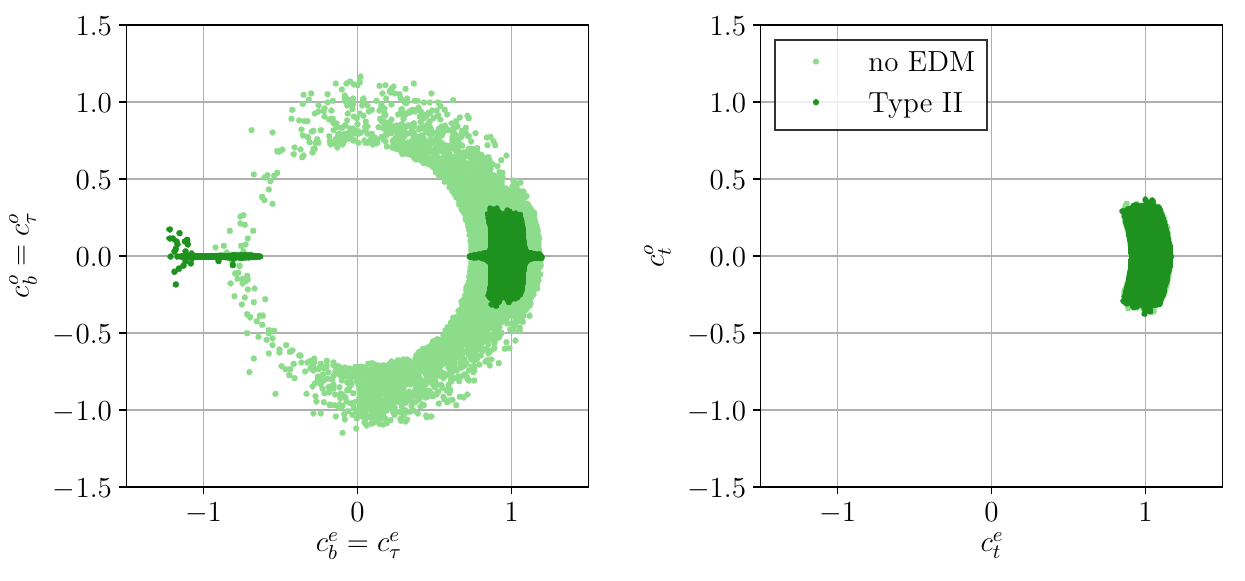}
\vspace{-2mm}
\caption{Type II, $h_{125}=h_1$: Yukawa couplings to bottom quarks and tau leptons (left) and
top quarks (right) for sample 1 (dark) and sample 2
(light). \label{Chap-Maggie:fig:t2c2hdmyuks_H1}}
\end{figure}

In fig.~\ref{Chap-Maggie:fig:mixangledist}, we present again $\alpha_2$ versus $\alpha_1$, but now for Type II (still assuming $h_1=h_{125}$).
It is clear that the EDM constraints (applied in sample 1) strongly reduce $|\alpha_2|$ to small values.
%
%
%
%
%
The phenomenological implications of such reduction are demonstrated in
fig. \ref{Chap-Maggie:fig:t2c2hdmyuks_H1},
which shows $c^o_f$ versus
$c^e_f$, not only for bottom quarks and tau leptons (left), but also for top quarks (right).
As can be inferred from the left panel,
the Higgs data alone still allow for vanishing scalar
couplings to down-type quarks ($c^e_b=0$), as discussed
in ref. \cite{Fontes:2015mea}.
The inclusion of the EDM constraints, however, clearly
rules out this possibility.
Nevertheless, the wrong-sign regime for down-type Yukawa couplings, $c_b^e < 0$, is still possible. The electron EDM has no discernable effect on the allowed coupling to up-type quarks, as can be read off from the right panel. Recall that these results were obtained for $h_1=h_{125}$.

\begin{figure}[h]
\centering
\includegraphics[width=0.9\linewidth]{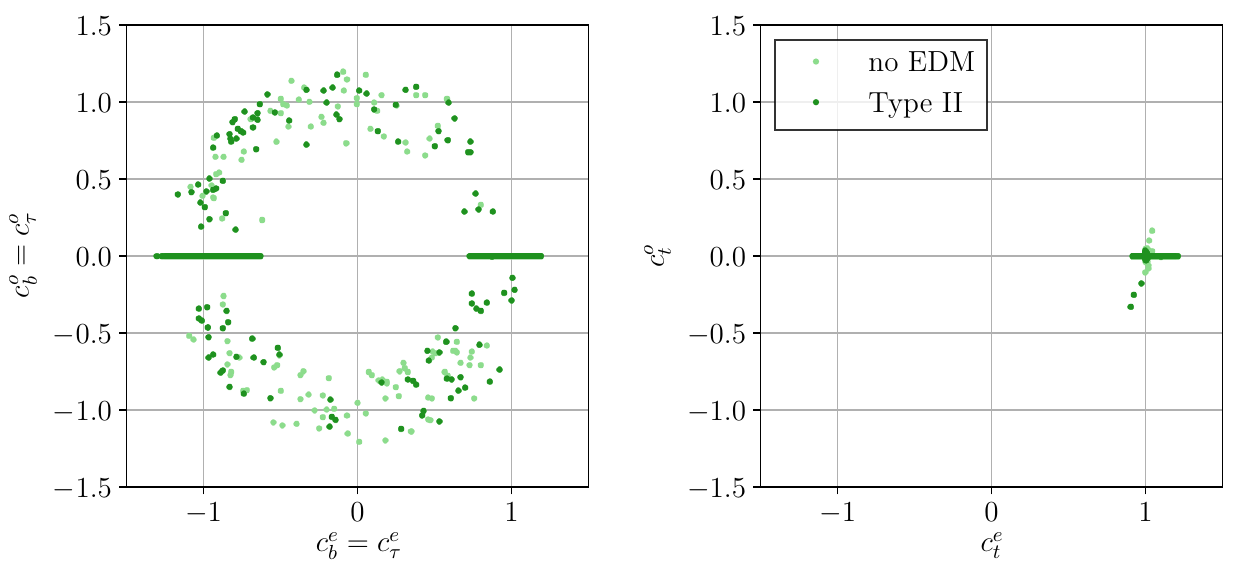}
\vs{-2mm}
\caption{Type II, $h_{125}=h_2$: Yukawa couplings to bottom quarks (left) and
top quarks (right) for sample 1 (dark) and sample 2
(light).
\label{Chap-Maggie:fig:t2c2hdmyuks_H2}}
\vs{-1mm}
\end{figure}

The situation changes (still in the context of Type II) when we consider $h_2 = h_{125}$, as shown in fig.~\ref{Chap-Maggie:fig:t2c2hdmyuks_H2}.
Here, one can find realizations of the peculiar possibility we alluded to above, where the top coupling is mostly CP-even ($c_t^e \simeq 1$),
while the bottom coupling is mostly CP-odd ($c_b^o \simeq 1$).%
\fn{It is noteworthy that EDM precludes all such points when $h_{125}=h_1$ (as fig. \ref{Chap-Maggie:fig:t2c2hdmyuks_H1} shows), but allows them when $h_{125}=h_2$.}
One can also find scenarios where $c^e_b c^o_b \neq 0$.
To more properly investigate both cases, we propose three benchmark points, described in table~\ref{Chap-Maggie:tab:benchII}.
\begin{table}[h!]
\centering
\begin{tabular}{lccc}
\toprule
Type II & BP2m & BP2c & BP2w\\
\midrule
$m_{1}$ & 94.187 & 83.37 & 84.883 \\
$m_{2}$ & 125.09 & 125.09 & 125.09 \\
$m_{\mathrm{H}^{+}}$ & 586.27 & 591.56 & 612.87 \\
$m_{12 \mathrm{R}}^2$ & 24017 & 7658 & 46784 \\
$\alpha_1$ & -0.1468 & -0.14658 & -0.089676 \\
$\alpha_2$ & -0.75242 & -0.35712 & -1.0694 \\
$\alpha_3$ & -0.2022 & -0.10965 & -0.21042 \\
$\tan\beta$ & 7.1503 & 6.5517 & 6.88 \\
\midrule
$m_{3}$ & 592.81 & 604.05 & 649.7 \\
$c^e_b=c^e_\tau$ & 0.0543 & 0.7113 & -0.6594 \\
$c^o_b=c^o_\tau$ & 1.0483 & 0.6717 & 0.6907 \\
\bottomrule
\end{tabular}
\caption{Benchmark points with large pseudoscalar Yukawa couplings
in Type II, $h_{125}=h_2$. Lines 1-8 contain the input parameters, whereas lines 9-11 display the derived third Higgs boson mass and the relevant Yukawa couplings (multiplied by $\mathrm{sgn}(c(h_{125}VV))$).
}
\label{Chap-Maggie:tab:benchII}
\vs{-3mm}
\end{table}
The first one, BP2m, has maximal $c^o_b$ with nearly vanishing
$c^e_b$; since $c^e_t$ is always $\simeq1$, $c^e_t c^o_b$
is maximal here. The other two points, BP2c and BP2w, both have
maximal $c^e_b c^o_b$; the former describes the correct sign regime ($c_b^e > 0$), whereas the latter describes the wrong-sign one ($c_b^e < 0$).

%

\begin{figure}[h!]
\centering
\includegraphics[width=0.86\linewidth]{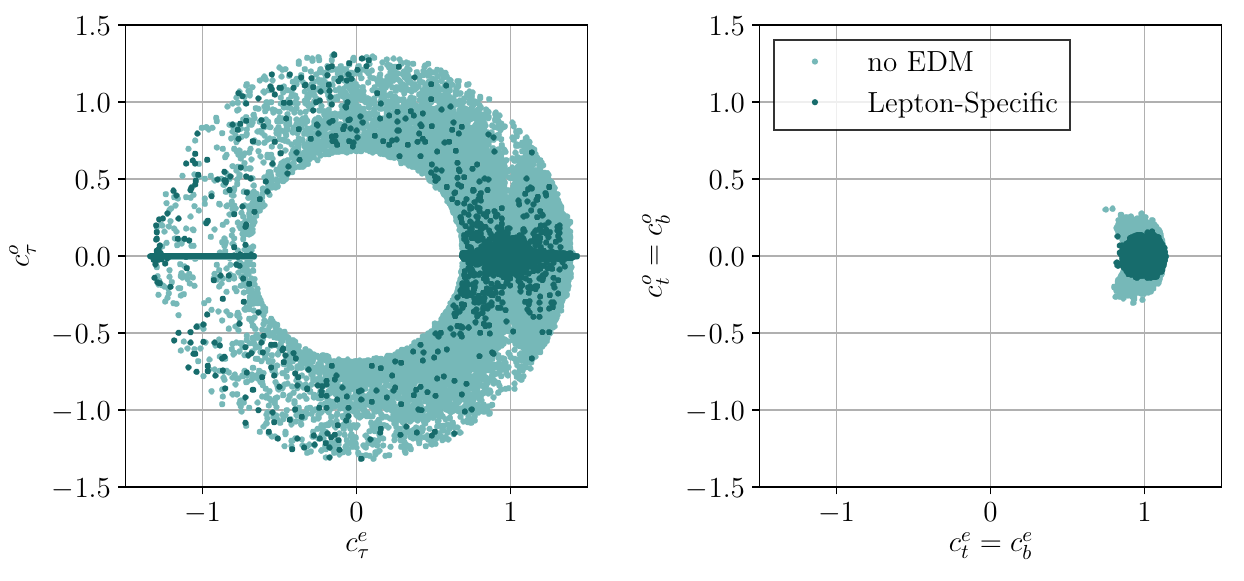}
\caption{LS: Yukawa couplings to charged-leptons (left) and
bottom and top quarks (right) for sample 1 (dark) and sample 2
(light). \label{Chap-Maggie:fig:lsc2hdmyuks}}
\end{figure}

\begin{figure}[h]
\centering
\includegraphics[width=0.86\linewidth]{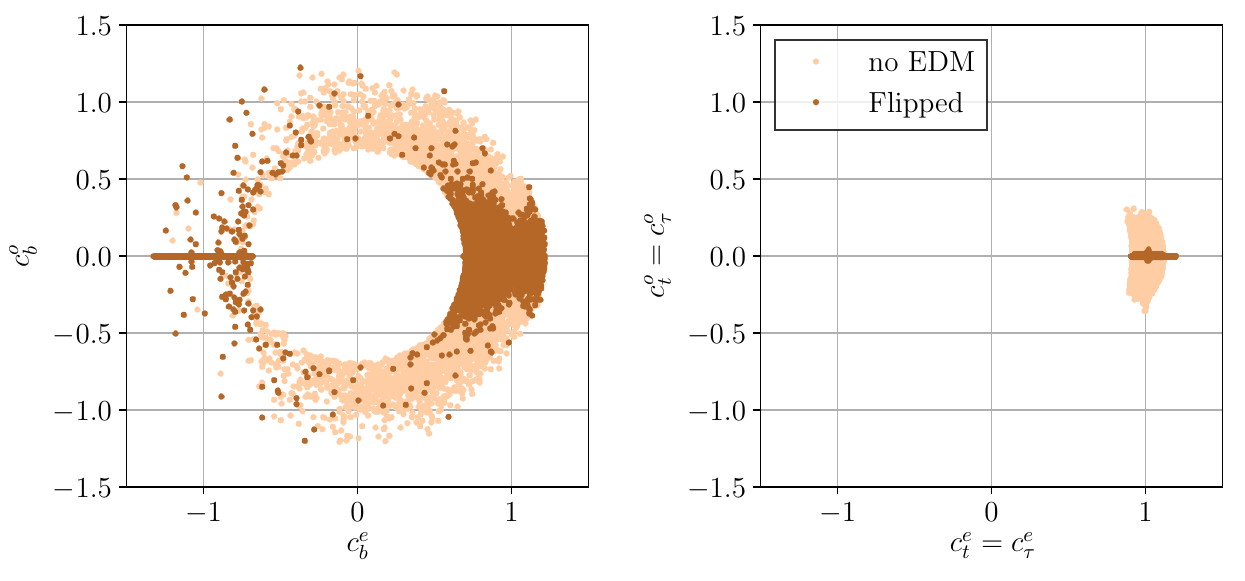}
\caption{Flipped: Yukawa couplings to bottom quarks (left) and
charged-leptons and top quarks (right) for sample 1 (dark) and sample 2
(light). \label{Chap-Maggie:fig:fc2hdmyuks}}
\end{figure}

If the situation in Type II is already interesting, it is even more so in the other two Yukawa types.
In the LS, the down-type quark couplings are tied to the up-type couplings (recall table \ref{Chap-Maggie:tab:coeffs_models}); the right panel of fig. \ref{Chap-Maggie:fig:lsc2hdmyuks} shows that they are heavily constrained to lie close to the SM (fully CP-even) solution.
However, the left panel shows that the charged-lepton couplings can still be fully CP-odd, despite EDM constraints.
Hence, and remarkably, $h_{125}$ can couple in a fully CP-even way to quarks, and in a fully CP-odd way to charged leptons.
Fig. \ref{Chap-Maggie:fig:fc2hdmyuks} shows that the Flipped model has an equivalent scenario, where the role of the CP-odd coupling is now played by the down-type quarks. 
%
%
%
As will be discussed below, such large CP-odd components are still viable
in both LS and Flipped models due to cancellations
between the various diagrams entering the EDM calculation (this is
also true for Type II when $h_2 = h_{125}$).
%
%
%
We present benchmark points for the LS and Flipped types in table~\ref{Chap-Maggie:tab:benchLSF}, where we follow a similar strategy to that in table \ref{Chap-Maggie:tab:benchII}.%
\fn{In contrast to the Type II, all benchmark points (except BPLSw) have $h_1=h_{125}$. We chose BPLSw with $h_2=h_{125}$, since this mass ordering is a consequence of maximizing $c^e_\tau c^o_\tau$ close to the wrong-sign limit of the LS model.}
\begin{table}[h!]
  \centering
\begin{tabular}{lcccclccc}
    \cmidrule[\heavyrulewidth]{1-4}\cmidrule[\heavyrulewidth]{6-9}
    LS & BPLSm & BPLSc & BPLSw & \hspace{2cm}& Flipped & BPFm & BPFc & BPFw\\
    \cmidrule{1-4}\cmidrule{6-9}
    $m_{1}$ & 125.09 & 125.09 & 91.619 &&$m_{1}$ & 125.09 & 125.09 & 125.09 \\
    $m_{2}$ & 138.72 & 162.89 & 125.09 &&$m_{2}$ & 154.36 & 236.35 & 148.75 \\
    $m_{\mathrm{H}^{+}}$ & 180.37 & 163.40 & 199.29 &&$m_{\mathrm{H}^{+}}$ & 602.76 & 589.29 & 585.35 \\
    $m_{12 \mathrm{R}}^2$ & 2638 & 2311 & 1651 &&$m_{12 \mathrm{R}}^2$ & 10277 & 8153 & 42083 \\
    $\alpha_1$ & -1.5665 & 1.5352 & 0.0110 && $\alpha_1$ & -1.5708 & 1.5277 & -1.4772 \\
    $\alpha_2$ & 0.0652 & -0.0380 & 0.7467 && $\alpha_2$ & -0.0495 & -0.0498 & 0.0842 \\
    $\alpha_3$ & -1.3476 & 1.2597 & 0.0893 && $\alpha_3$ & 0.7753 & 0.4790 & -1.3981 \\
    $\tan\beta$ & 15.275 & 17.836 & 9.870 && $\tan\beta$ & 18.935 & 14.535 & 8.475 \\
    \cmidrule{1-4}\cmidrule{6-9}
    $m_{3}$ & 206.49 & 210.64 & 177.52 && $m_{3}$ & 611.27 & 595.89 & 609.82 \\
    $c^e_\tau$ & -0.0661 & 0.6346 & -0.7093 && $c^e_b$ & -0.0003 & 0.6269 & -0.7946 \\
    $c^o_\tau$ & 0.9946 & 0.6780 & -0.6460 && $c^o_b$ & -0.9369 & 0.7239 & 0.7130 \\
    \cmidrule[\heavyrulewidth]{1-4}\cmidrule[\heavyrulewidth]{6-9}
  \end{tabular}
  \caption{Benchmark points with large pseudoscalar Yukawa couplings in the LS and Flipped types. Lines 1-8 contain the input parameters, whereas lines 9-11 display the derived third Higgs mass and the relevant Yukawa couplings (multiplied by $\mathrm{sgn}(c(h_{125}VV))$).
}
  \label{Chap-Maggie:tab:benchLSF}
\end{table}
%
%
%
%
%
%
%
%
%
%

\subsection{\label{Chap-Maggie:subsec:H3}The case $h_3=h_{125}$}

\begin{figure}[h]
\centering
\includegraphics[width=0.8\linewidth]{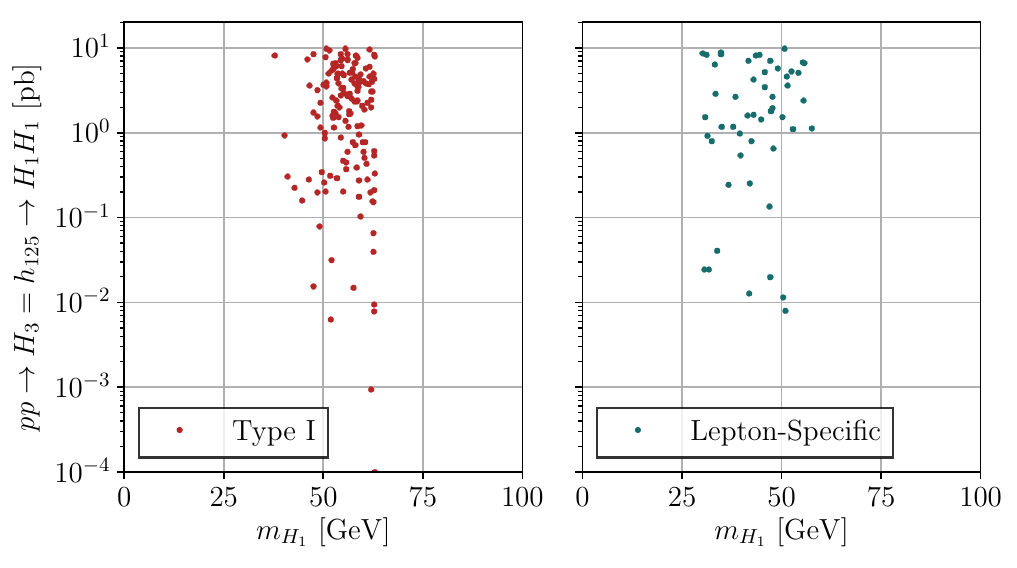}
\vs{-3mm}
\caption{Rate for $pp \to h_3(=h_{125}) \to h_1 h_1$ in Type I (left)
and LS (right).\label{Chap-Maggie:fig:rates_HsmEqH3_HlHl}}
\end{figure}

We now turn to another interesting possibility in the C2HDM, which associates $h_{125}$ to the heaviest Higgs boson, $h_3$.
This possibility is excluded for Type II and Flipped, since
the $B$-physics constraints impose that the charged Higgs boson must be
quite heavy, $m_{H^+}>580$ GeV.%
\fn{This poses a problem with the electroweak precision tests, especially with the T parameter. Indeed, the way to accommodate the experimental bounds on T is to have a spectrum that has some degree of degeneracy. Requiring $m_{H^+}>580$ GeV implies that the other Higgs boson masses cannot all be below 125 GeV. Thus, $m_{H^+}>580$ GeV is not compatible with $m_{2}< 125$ GeV.}
However, $h_3=h_{125}$ is feasible for Type I and LS,
as in these cases  $B$-physics constraints only impose $m_{H^+}>100$ GeV.%
\fn{As explained in section \ref{Chap-Maggie:sec:rest}, this bound could be slightly higher for low $\tan\beta$.}
The situation here is quite fascinating,
because it highlights a complementarity
between LHC and the old LEP results.
To understand this aspect, note that the Feynman rules for the cubic interactions between $Z$ and neutral scalars are:
\be
[h_k Z_\mu Z_\nu]:\ \
i g_{\mu \nu} \frac{e}{s_{\mathrm{w}} c_{\mathrm{w}}} m_{\mathrm{Z}}\, c(h_k Z Z),
\qquad
\quad
[h_i h_j Z_\mu]:\ \
i \frac{e}{2 s_{\mathrm{w}} c_{\mathrm{w}}} (p_i - p_j)_\mu\, c(h_i h_j Z),
\ee
where $p_i$ and $p_j$ are the incoming momenta of particles $h_i$ and $h_j$, respectively, $c(h_k Z Z)$ was defined in eq. \ref{Chap-Maggie:eq:gaugecoupdef} and given in eq.~\ref{Chap-Maggie:eq:c2dhmgaugecoup}, and
\begin{equation}
c(h_i h_j Z) = \epsilon_{ijk}\,  c(h_k Z Z).
\end{equation}
Now, in the SM, $c(h_k Z Z)=1$ and $c(h_i h_j Z) = 0$.
Thus, if we equate $h_3=h_{125}$, the LHC signal $h_{125} \to ZZ$ forces $c(h_3 Z Z) \sim 1$, which in turn implies $c(h_1 h_2 Z) \sim 1$. In that case, if we had $m_{3} > m_{1} + m_{2}$, the decay $Z \to h_1 h_2$ would have been seen at LEP \cite{Schael:2006cr}. Since it was not seen, we must have $m_{3} < m_{1} + m_{2} \leq 2 \, m_{2}$. This ordering, however, is still compatible with $m_{3} > 2 \, m_{1}$, which would thus render the decay $h_3 \to h_1 h_1$ possible. In sum, then, although LEP precludes $h_3 \to h_2 h_2$ and $h_3 \to h_1 h_2$, it allows $h_3 \to h_1 h_1$. It turns out that this decay is still possible in Type I and LS, as shown in fig.~\ref{Chap-Maggie:fig:rates_HsmEqH3_HlHl}.%
\fn{The figures that follow obey an old convention, according to which $H_i \equiv h_i$ and $M_{H_i} \equiv m_i$.}
In fact, the rates can go
up to about 10 pb and therefore have excellent prospects of being probed
at the LHC.
%
%
%
%

\section{Measures of CP violation}
 \label{Chap-Maggie:sec:measure}

We now investigate manifestations of CP violation in the C2HDM.
On the one hand, they can be probed through a number of variables---such as the CP-violation phase $\phi (\lambda_5)$, defined in eq. \ref{Chap-Maggie:eq-phases}.
On the other hand, and as discussed in refs. \cite{Branco:1999fs, Fontes:2015xva}, there are several combinations of Higgs decays that constitute a clear sign of CP violation in any extension of the SM.
%
%
This leads us to the following question: are there variables that probe CP violation from a theoretical point of view, on the one hand,
and that can be related to combination of decays which would establish CP violation experimentally, on the other?

\begin{figure}[h!]
\centering
\includegraphics[width=0.55\linewidth]{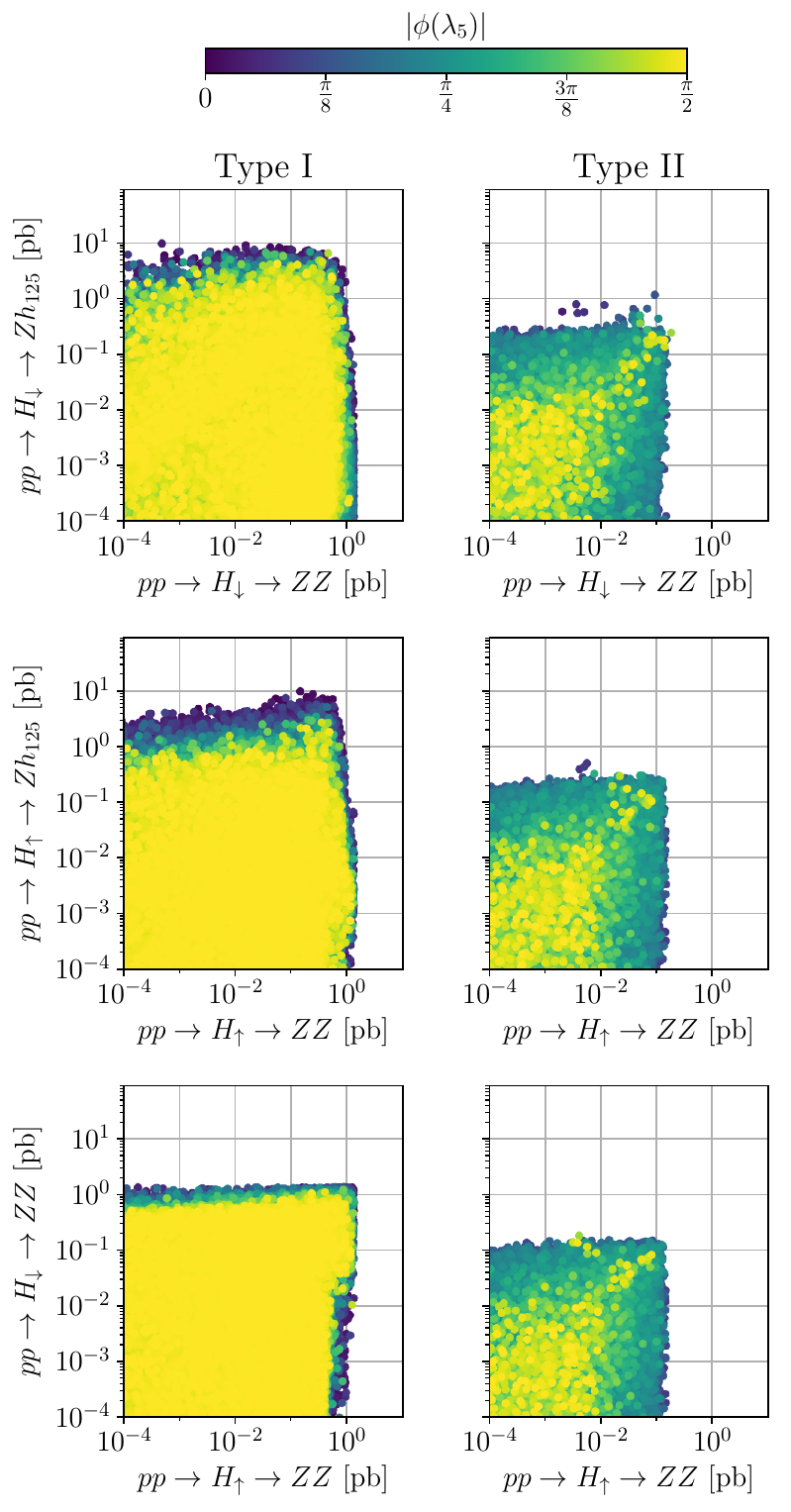}
\caption{Set of CP-violating processes as a function of the CP violation phase $|\phi (\lambda_5)|$
(see colour code) for
Type I (left column) and Type II (right column).
In the first row we show $pp \to h_\downarrow \to Z h_{125}$ against $pp \to h_\downarrow \to ZZ$, in the second row
we have $pp \to h_\uparrow \to Z h_{125}$ against $pp \to h_\uparrow \to ZZ$ and in the third row we plot
$pp \to h_\downarrow \to ZZ$ against $pp \to h_\uparrow \to ZZ$. Note that the yellow points are superimposed
on the darker points - there are dark points underneath the yellow points.} \label{Chap-Maggie:fig:CPvarangle}
\end{figure}

Let us start by considering $\phi (\lambda_5)$ as CP-probing variable. We introduce the following notation to designate the neutral scalars other than $h_{125}$:%
\fn{The figures that follow obey an old convention, according to which $H_\downarrow \equiv h_\downarrow$ and $H_\uparrow \equiv h_\uparrow$.}
\begin{equation*}
h_\downarrow \, \Leftrightarrow \, \text{the lightest scalar},
\qquad
h_\uparrow \, \Leftrightarrow \, \text{the heaviest scalar},
\end{equation*}
and in such a way that their mass can be below or above 125 GeV. With this notation, we present in fig. \ref{Chap-Maggie:fig:CPvarangle} three classes of CP-violating processes, as a function of $|\phi (\lambda_5)|$.
In the first row, we show $pp \to h_\downarrow \to Z h_{125}$  against $pp \to h_\downarrow \to ZZ$,
in the second row
$pp \to h_\uparrow \to Z h_{125}$ against $pp \to h_\uparrow \to ZZ$, and in the third row $pp \to h_\downarrow \to ZZ$ against $pp \to h_\uparrow \to
ZZ$.
%
%
%
%
%
The plots in the left column concern Type I, whereas those on the right one concern Type II.

There are two main features in the plots.
First, there is no correlation between the magnitude
of $\phi (\lambda_5)$ and the rates: the points with larger
$\phi (\lambda_5)$ are almost evenly spread throughout the plot.%
\fn{It is not that we expected that large values of the
CP-violating phase would correspond to large rates; in fact, whereas maximal CP violation is attained for specific sets of values of the angles, the rates are a complicated combination of all parameters of the model. In any event, a correlation could be found, but fig. \ref{Chap-Maggie:fig:CPvarangle} rejects such hypothesis.}
Second, the rates in Type I are almost one order of magnitude above the ones for Type II. This can be explained by the fact that there are some constraints (like $b \to s \gamma$ or the EDM, as we will see below)
acting more strongly on Type II than on Type I.
%

Since $\phi (\lambda_5)$ allows no correlation to the classes of decays presented, we look for other variables that may probe CP violation.
The sets of variables proposed in the literature are basically of two types: the ones where the CP-violating variables appear in a product of squares, and the ones where they appear in a sum of squares \cite{Mendez:1991gp, Khater:2003ym}.%
\fn{Also interesting is the relation between invariants and CP symmetry, explored in refs. \cite{Lavoura:1994fv} and \cite{Botella:1994cs}.}
While the former can be zero even if the model is CP-violating, the latter
is zero only when there is no CP violation in the model (if CP is conserved, both variables are zero).
We introduce $\xi_V$, a multiplicative variable defined through the couplings between gauge bosons and neutral scalars as\cite{Mendez:1991gp}
\begin{equation}
\xi_V=27[g_{h_1VV} \, g_{h_2VV} \, g_{h_3VV}]^2 = 27\prod_{i=1}^3[\cos\beta R_{i1}+\sin\beta R_{i2}]^2.
 \label{Chap-Maggie:Eq:xi-v}
\end{equation}
Variables can also be built with the scalar and pseudoscalar components of the Yukawa couplings. In fact, if a model has CP-violating scalars at tree-level, its Yukawa couplings have the general form $c^e_f + i c^o_f \gamma_5$. Hence, as discussed at the beginning of subsection~\ref{Chap-Maggie:subsec:constraints},
variables of the type $c^e c^o$ clearly signal CP violation in the model.
We thus define the normalized multiplicative variables \cite{Khater:2003ym}
\begin{equation} \label{Chap-Maggie:Eq:gamma_tb}
\gamma_t=1024\prod_{i=1}^3[R_{i2}\,R_{i3}]^2, \quad
\gamma_b=1024\prod_{i=1}^3[R_{i1}\,R_{i3}]^2,
\end{equation}
as measures of CP violation in the up- and down-quark sectors, respectively.
The corresponding normalized sum variables are defined as \cite{Khater:2003ym}
\begin{equation}
\zeta_t=2\sum_{i=1}^3[R_{i2}\,R_{i3}]^2, \quad \zeta_b=2\sum_{i=1}^3[R_{i1}\,R_{i3}]^2 ,
\end{equation}
again for the up- and down-quark sectors, respectively. All the variables that we just introduced take values between 0 and 1 (i.e. $0 \le \xi_V, \gamma_t, \gamma_b, \zeta_t, \zeta_b \le 1$).

In fig.~\ref{Chap-Maggie:fig:CPvar1}, we investigate these variables:
\begin{figure}[h!]
\centering
\includegraphics[width=0.75\linewidth]{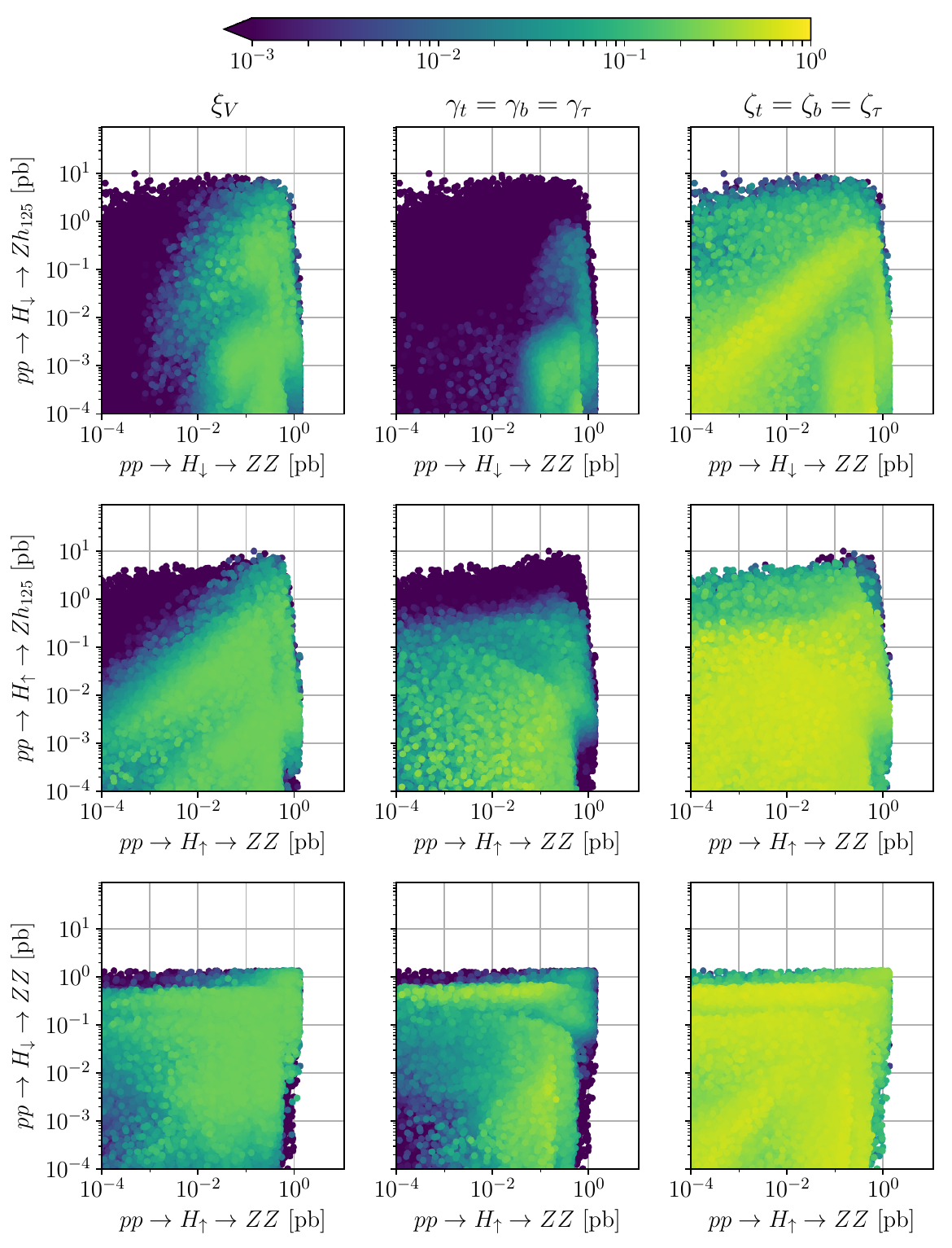}
\caption{Classes of CP-violating processes as a function of the CP-violating variables (see colour code)
for the Type I C2HDM.
In the first row we show $pp \to h_\downarrow \to Z h_{125}$ against $pp \to h_\downarrow \to ZZ$, in the second row
we have $pp \to h_\uparrow \to Z h_{125}$ against $pp \to h_\uparrow \to ZZ$ and in the third row we plot
$pp \to h_\downarrow \to ZZ$ against $pp \to h_\uparrow \to ZZ$.
In each column, we show the variable that is being probed. The darker points are underneath the lighter ones.} \label{Chap-Maggie:fig:CPvar1}
\end{figure}
we present the three classes that signal CP violation as a function of such variables, for Type I.
In the first row, we show $pp \to h_\downarrow \to Z h_{125}$ against $pp \to h_\downarrow \to ZZ$, in the second row $pp \to h_\uparrow \to Z h_{125}$ against $pp \to h_\uparrow \to ZZ$, and in the third row $pp \to h_\downarrow \to ZZ$ against $pp \to h_\uparrow \to ZZ$. In each column, we show the variable that is being probed.

%
The general picture is that there is no striking correlation between the large values for the variables (more yellow points) and the large cross sections for each process. There is a quite even spread of yellow points for the sum variable $\zeta_f$; the reason is that, even if the product of scalar/pseudoscalar components of the SM-like Higgs boson is very constrained, any of the products of the other two Higgs bosons can be very large, in which case one obtains a large value for the sum in $\zeta_f$.
Therefore, variables of this (sum) type can always be large, and we will not show them in the remaining plots.
Regarding the (multiplicative) variables $\gamma_f$, we can see some structure in the plots: there are cases where more yellow points are clustered closer to the maximum values of the rates.

\begin{figure}[h!]
\centering
\includegraphics[width=0.75\linewidth]{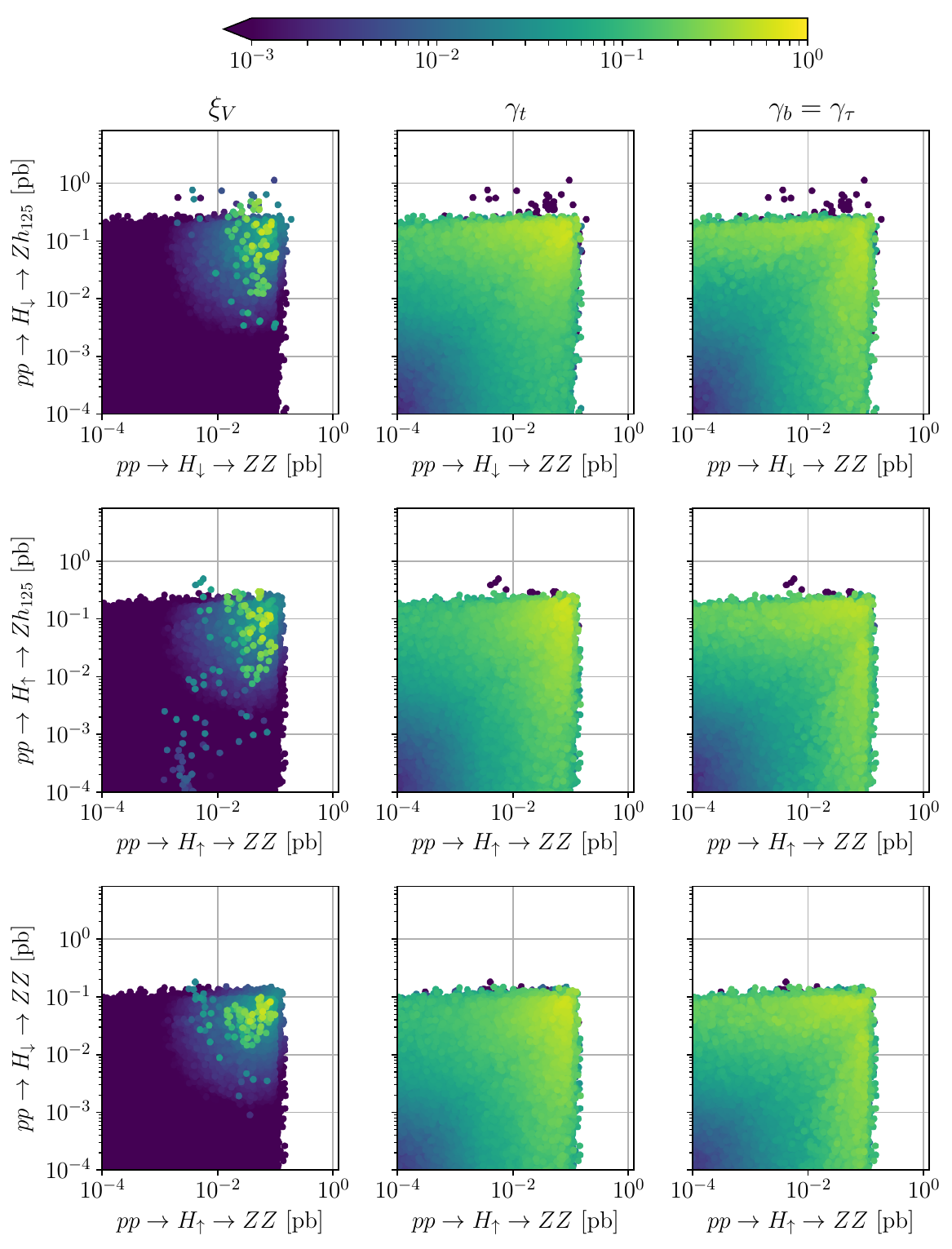}
\caption{Same as fig.~\ref{Chap-Maggie:fig:CPvar1}, but for Type II.} \label{Chap-Maggie:fig:CPvar2}
\end{figure}

In fig.~\ref{Chap-Maggie:fig:CPvar2}, we present an equivalent set of plots, but now for Type II.
In this case, the distribution of the yellow points is more structured and more clustered in specific regions. In fact, the yellow points tend to cluster more in the regions where the rates are larger. This behaviour is sharper for the variable $\xi_V$ (first column), where all yellow points are in the parameter region where the rates are maximal. For the remaining variables, the distribution of yellow points is again not so structured. Still, all variables are larger when the rates are both large, and are much smaller when both rates are smaller. However, in all plots, there are always points with small values of the CP-violating variables and large values of the rates. Hence, although the variables show us a trend, they are not conclusive as a measure of CP violation in the scalar sector.

\begin{figure}[h!]
\centering
\includegraphics[width=0.75\linewidth]{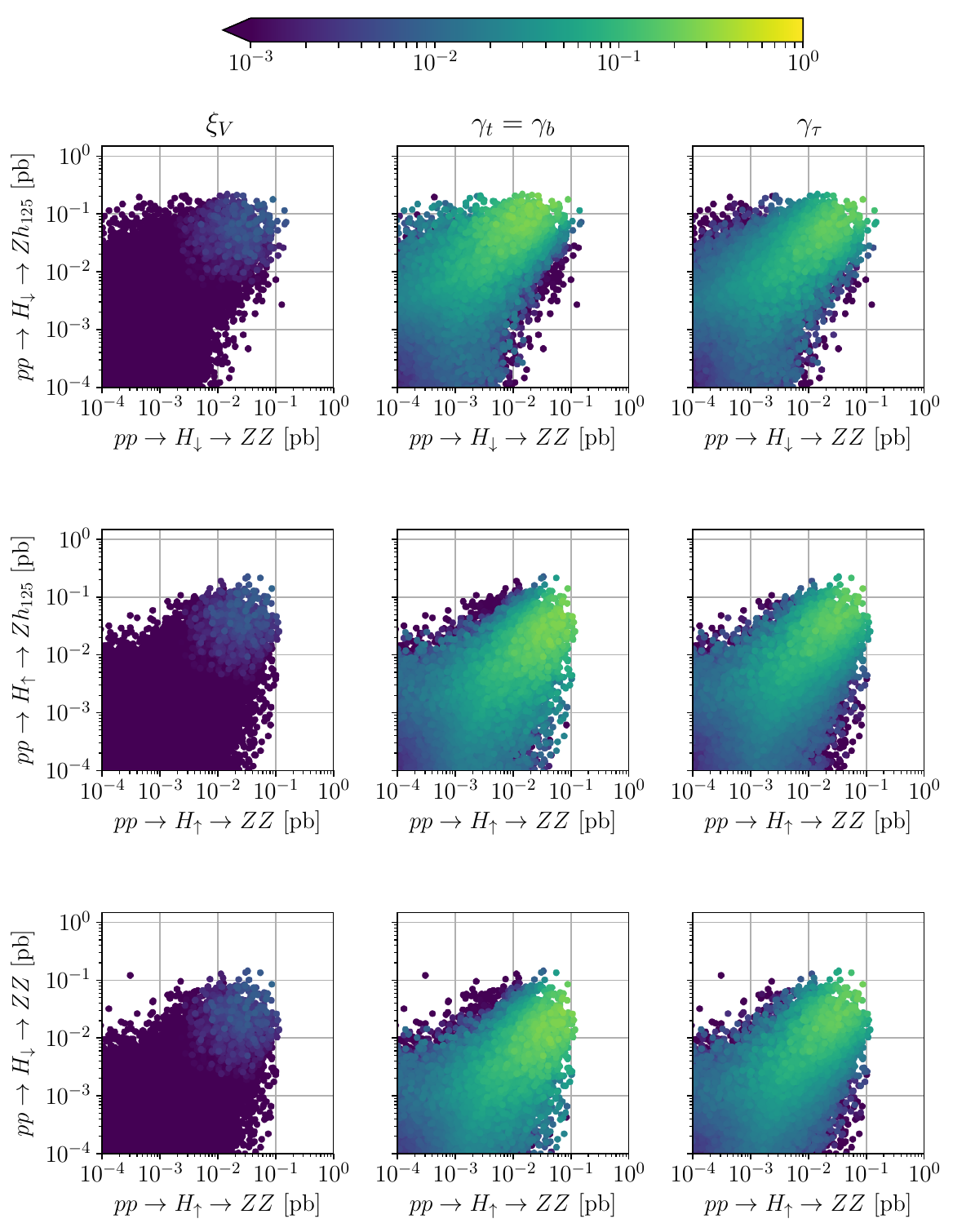}
\caption{Same as fig.~\ref{Chap-Maggie:fig:CPvar2}, but with signal strengths
within 5\% of the SM values, for Type II.} \label{Chap-Maggie:fig:CPvar3}
\end{figure}

In fig.~\ref{Chap-Maggie:fig:CPvar3},
we show the same plots, but with signal strengths
within 5\% of the SM values (instead of just within $2 \sigma$ of the fits of ref. \cite{Khachatryan:2016vau}).
This gives us a hint on what to expect regarding future experimental results. There is a clear effect in reducing the rates, but not in the distribution of yellow points. The main difference is that now no yellow points appear in the first column---which means that the points with very large rates were excluded. The distribution of points in the other columns did not change significantly, but the points with the higher rates were also excluded as for the first row. In conclusion, there seems to be an overall reduction in the parameter space of the model leading to smaller rates.

\begin{figure}[h]
\centering
\includegraphics[width=0.75\linewidth]{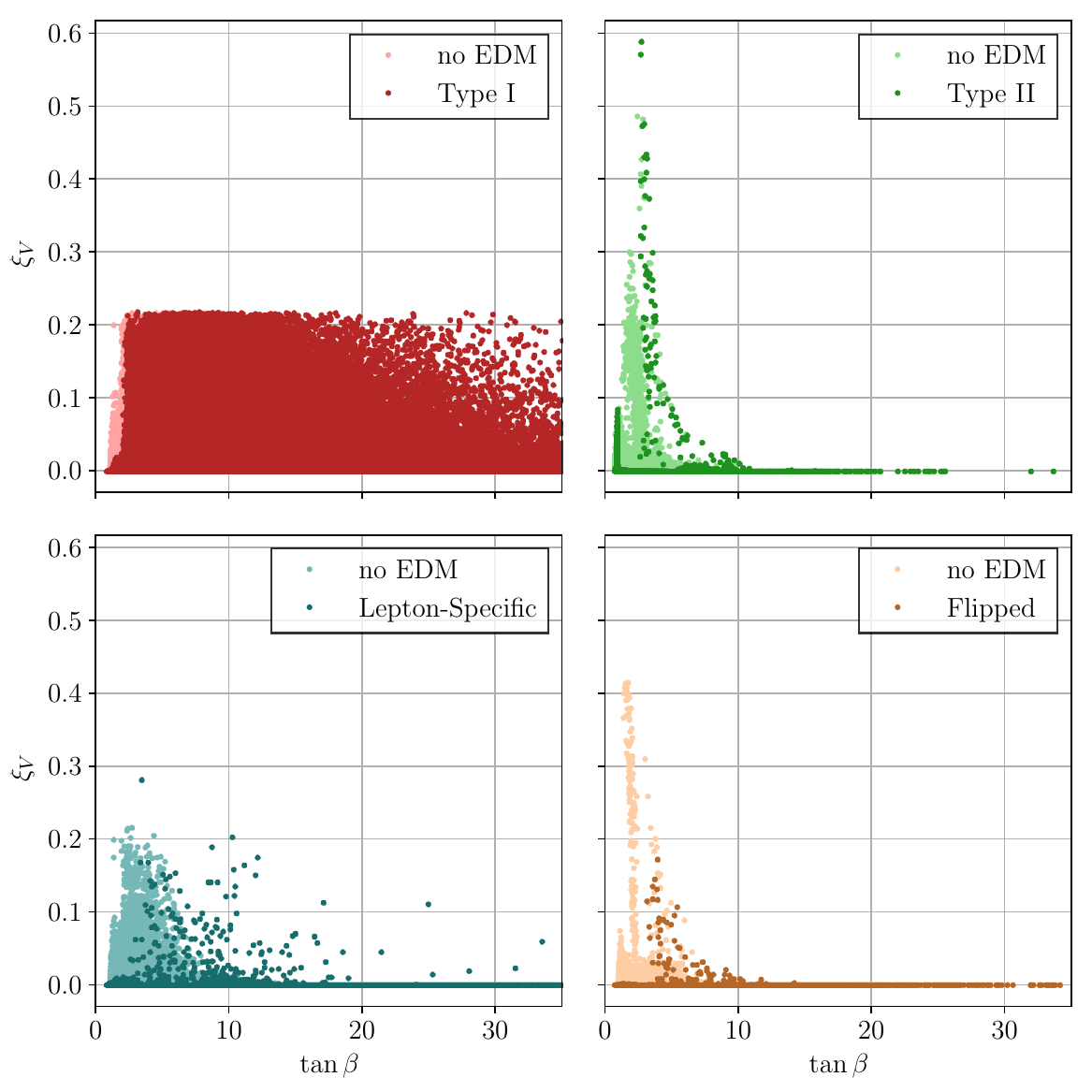}
\caption{The CP-violating parameter $\xi_V$ as a function of $\tan \beta$ for
Type I (top left), Type II (top right), LS (bottom left) and
Flipped (bottom right).
The lighter points have passed all the constraints except for the EDM bounds while the darker
points have passed all constraints.} \label{Chap-Maggie:fig:xitanbeta}
\end{figure}

In fig.~\ref{Chap-Maggie:fig:xitanbeta}, we show the CP-violating parameter $\xi_V$ as a function of $\tan \beta$ for Type I (top left), Type II (top right), LS (bottom left) and Flipped (bottom right).
%
%
%
In Type I, there is not much difference between the two samples of points (with and without EDM constraints), and there are no special regions regarding the allowed values of $\tan \beta$.
Also, the maximum value for $\xi_V$ is around 0.2 almost independently of $\tan \beta$.
As for Type II, the results are much more striking: even after EDM constraints, we end with two almost straight lines---one for $\tan \beta \approx 1$ and another for $\xi_V \approx 0$---, as well as a region around $\tan \beta \approx 3$ permitting values of $\xi_V$ up to 0.6. This means that $\tan \beta$ can only be large when we approach the CP-conserving limit $\xi_V = 0$.%
\fn{Except for a few points, which lie in the wrong-sign regime.}
%
The situation in Flipped is similar to Type II, with a lower maximum value of $\xi_V \sim 0.2$ after imposing the EDM constraints.

\begin{figure}[h]
\centering
\includegraphics[width=0.8\linewidth]{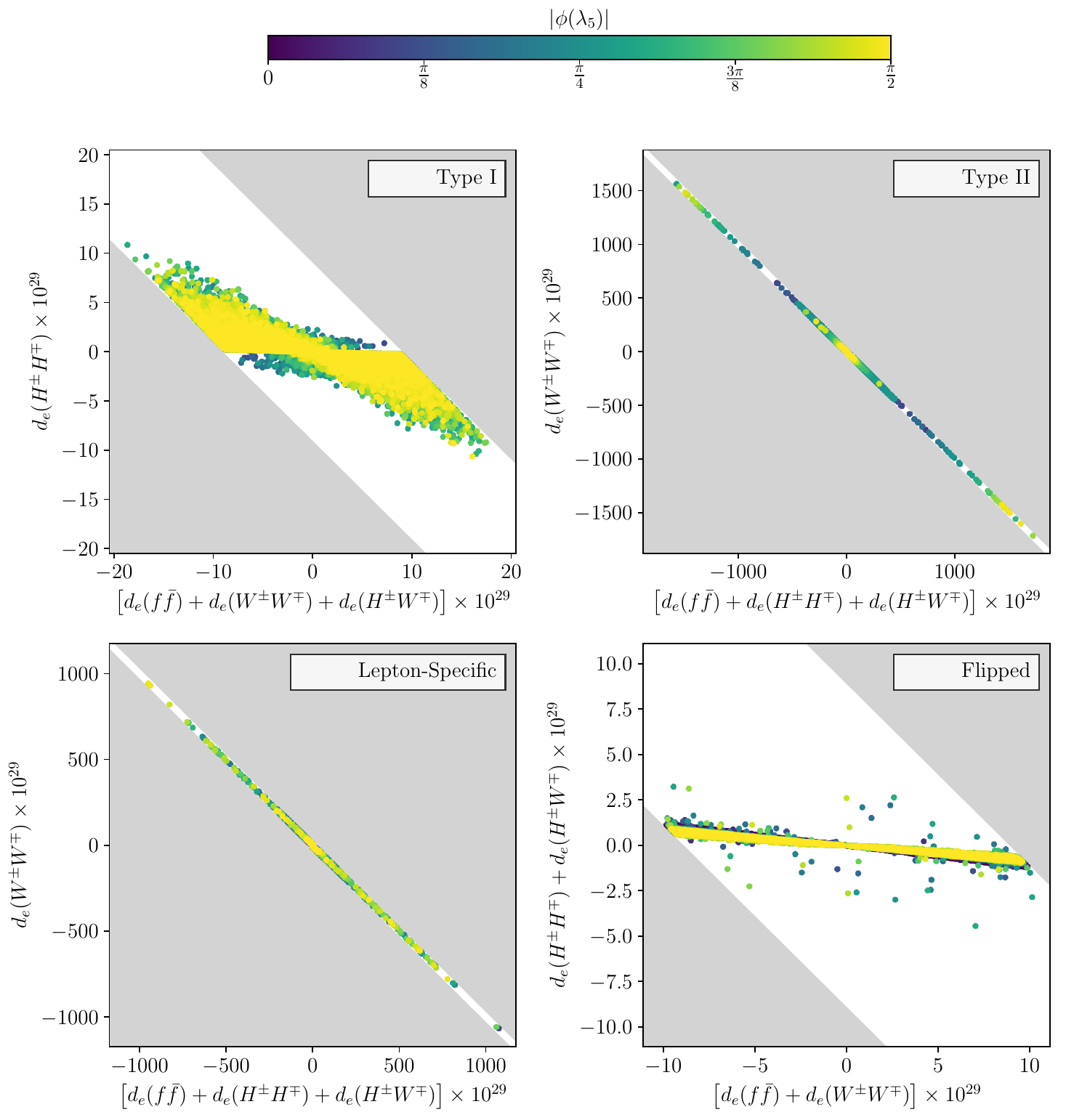}
\caption{Contributions to the EDM according to their relative sign in Type I (top left),
Type II (top right), LS (bottom left) and Flipped (bottom right).
The colour code represents the absolute value of the magnitude of $\phi (\lambda_5)$.
The gray shaded regions represent the parameter space excluded by the EDM constraints only. The coloured
points are the ones that passed all the constraints, that is, EDMs plus theoretical and all other
experimental constraints.} \label{Chap-Maggie:fig:EDMs}
\end{figure}

In fig. \ref{Chap-Maggie:fig:EDMs}, we show the individual contributions to the EDM coming from $W$-loops, fermion-loops, charged Higgs loops and charged Higgs plus $W$-loops.
For each type of C2HDM, we have grouped the contributions to the EDM according to their relative sign; for example, in Type II the contributions of the $W$-loops (y-axis) have a sign which is opposite to that of the sum of the contributions of the fermion loops, charged Higgs loops and charged Higgs plus $W$-loops (x-axis).
The gray shaded region represents the parameter space excluded by the EDM constraints only, and the colour code represents the absolute value of the magnitude of $\phi (\lambda_5)$.

In that figure, \ref{Chap-Maggie:fig:EDMs}, the first important difference between the models is that the maximum value of the individual contributions is around two orders of magnitude smaller in Type I than in Type II.
This implies that Type I is significantly less constrained by the EDM bounds.
Therefore,  the remaining constraints play a much more important role in Type I than in Type II.
This also leads to the distribution of values for the CP-violating phase in the figure: in Type II, large values of $\phi (\lambda_5)$ prefer regions where either the EDM contributions are very small (i.e. there are cancellations between loops in the individual contributions), or where huge cancellations between different contributions occur; this is in contrast to Type I, where large values of the CP-violating phase can be found all over the allowed region.
The Flipped case behaves roughly like Type I, while the
LS case behaves like Type II. This is very different from
the usual behaviour of the Yukawa types. The reason is that observables are usually dominated by quark effects, which yield similar behaviour in
LS /Type I and Flipped/Type II; however, since we are considering the EDM of the electron, the lepton Yukawa couplings are the most important ones, and those are equal in Flipped/Type I and LS/Type II (recall table \ref{Chap-Maggie:tab:coeffs_models}).

\section{Higgs-to-Higgs decays}
\label{Chap-Maggie:sec:higgstohiggs}

\begin{figure}[h!]
\centering
\includegraphics[width=0.75\linewidth]{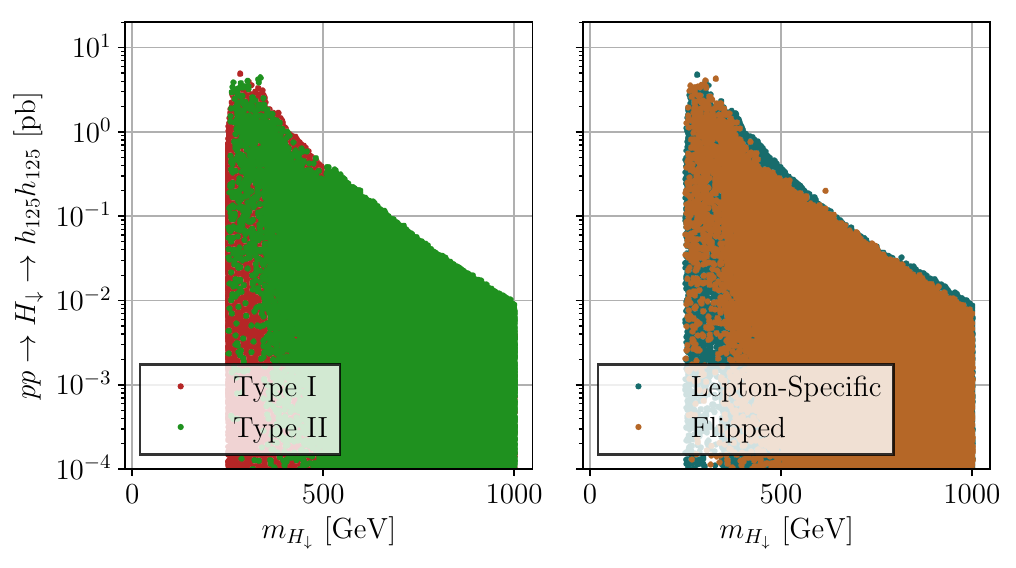}
\includegraphics[width=0.75\linewidth]{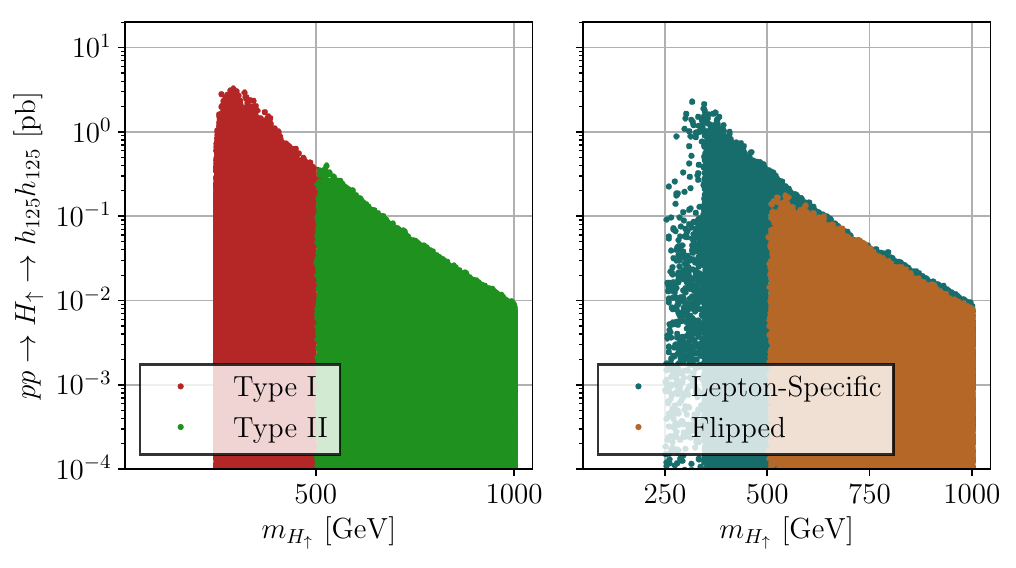}
\caption{Rates for the processes
$pp \to h_\downarrow \to h_{125} h_{125}$ (top row) and
$pp \to h_\uparrow \to h_{125} h_{125}$ (bottom row)
as a function of the respective mass for all C2HDM types.} \label{Chap-Maggie:fig:HHsmHsm}
\end{figure}

In the previous section, we discussed the correlation CP-probing variables and the some of the classes of decays that indicate CP violation. These particular classes were chosen not only because they can yield large rates, but also because there are currently searches being performed for these channels at the LHC that have already started during Run-1.
There are, however, other classes of decays that can probe CP violation; they involve Higgs-to-Higgs decays and led to the proposal of benchmarks for Run-2 \cite{Fontes:2015xva, deFlorian:2016spz}.%
\fn{For example,
the combination
$h_{\downarrow \uparrow} \to h_{125} h_{125}, \quad h_{\downarrow \uparrow}  \to h_{125} Z, \quad h_{125} \to ZZ$, 
as well as
$h_{125} \to h_{\downarrow \uparrow} h_{\downarrow \uparrow}, \quad h_{125}
\to h_{\downarrow \uparrow} Z, \quad h_{\downarrow \uparrow} \to ZZ$,
are a clear sign of CP violation \cite{Fontes:2015xva}.}
%
%
This leads us to investigate Higgs-to-Higgs decays.
In fig.~\ref{Chap-Maggie:fig:HHsmHsm}, we present the rates for the processes $pp \to h_\downarrow \to h_{125} h_{125}$ (top row) and $pp \to h_\uparrow \to h_{125} h_{125}$ (bottom row)
as a function of the respective mass, for all C2HDM types.
In all cases, the rates decrease as one increases the decaying scalar mass. In the four types, the  $pp \to h_\downarrow \to h_{125} h_{125}$
rates can be quite large, reaching about 4 pb in all types. The maximum values are similar in Type I and LS for  $pp \to h_\uparrow \to h_{125} h_{125}$.
By contrast, for Type II and Flipped, the largest rates in $pp \to
h_\uparrow \to h_{125} h_{125}$ decrease by about an order of
magnitude; in these cases, indeed, the heavier neutral scalar cannot be
much lighter than the charged Higgs boson, which is heavy to comply with $B$-physics constraints.

\begin{figure}[h!]
\centering
\includegraphics[width=0.75\linewidth]{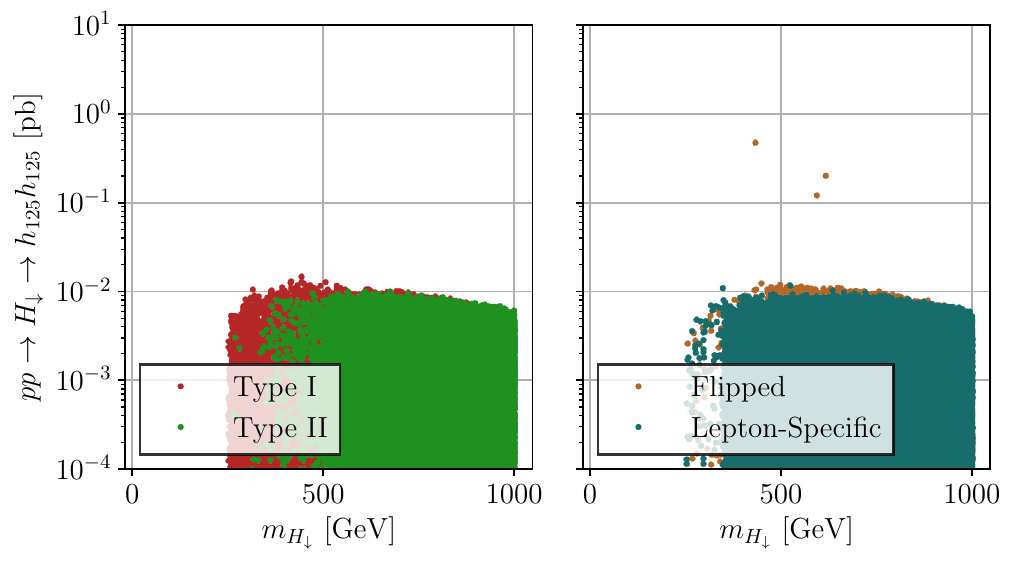}
\includegraphics[width=0.75\linewidth]{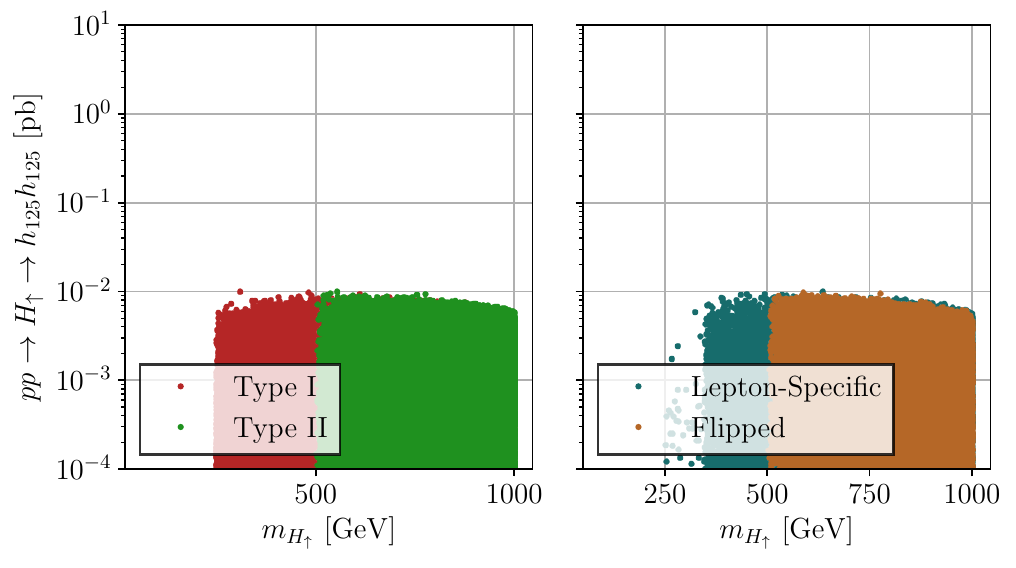}
\caption{Rates for the processes
$pp \to h_\downarrow \to h_{125} h_{125}$ (top row) and
$pp \to h_\uparrow \to h_{125} h_{125}$ (bottom row) as a
function of the respective mass for the four C2HDM types (same as fig.~\ref{Chap-Maggie:fig:HHsmHsm})
but with the extra condition
$\sigma (pp \to h_\downarrow \to ZZ) < 1 \,
\textrm{fb}$ for the top plots and $\sigma (pp \to h_\uparrow \to ZZ) < 1 \, \textrm{fb}$
for the bottom plots.} \label{Chap-Maggie:fig:HHsmHsmlow}
\end{figure}

In order to understand how relevant the searches for the two scalar final states are, we show in fig.~\ref{Chap-Maggie:fig:HHsmHsmlow} the same rates as in the previous fig.~\ref{Chap-Maggie:fig:HHsmHsm}, but with the extra condition   $\sigma (pp \to h_\downarrow \to ZZ) < 1 \, \textrm{fb}$ for the top plots and $ \sigma (pp \to h_\uparrow \to ZZ) < 1 \, \textrm{fb}$ for the lower plots.
It is clear that, with the extra
restriction on the $ZZ$ final state, the cross sections now barely
reach 10 fb for the two decay scenarios and for all types.
Hence, although possible, it will be very hard to detect the new scalars in the $h_{125} h_{125}$ final state if they are not detected in the $ZZ$ final state.

\begin{figure}[h!]
\centering
\includegraphics[width=0.75\linewidth]{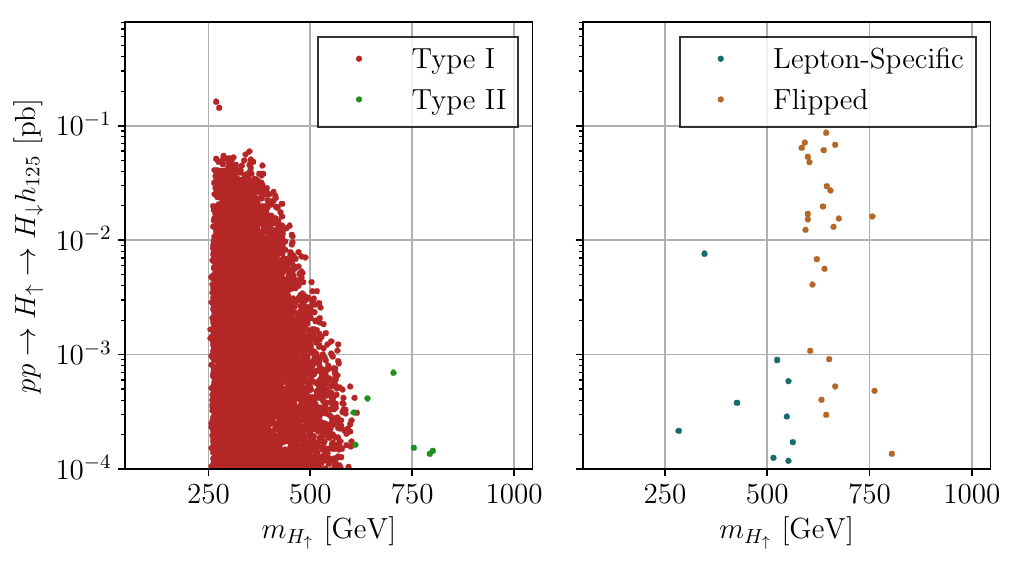}
\vs{-3mm}
\caption{Rates for the process $pp \to h_\uparrow \to h_\downarrow h_{125}$ as a function
of the heavier Higgs boson mass, for all C2HDM types.} \label{Chap-Maggie:fig:HHlHsm}
\end{figure}
\begin{figure}[h!]
\centering
\includegraphics[width=0.75\linewidth]{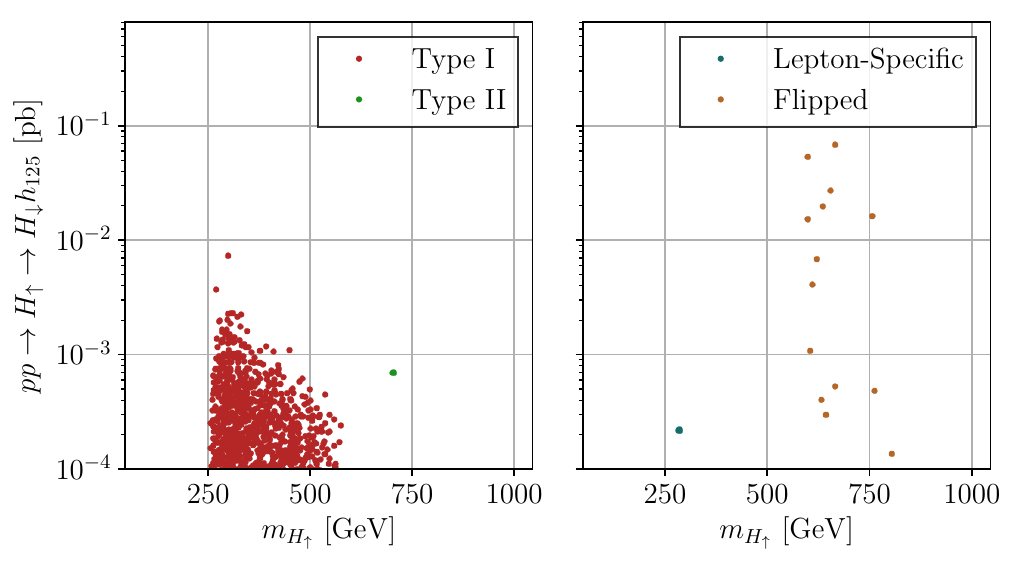}
\vs{-1mm}
\caption{Rates for the process $pp \to h_\uparrow \to h_\downarrow h_{125}$ as a function of
the heavier Higgs boson mass, for all C2HDM types (same as fig.~\ref{Chap-Maggie:fig:HHlHsm}) but with the extra condition
$\sigma (pp \to h_\uparrow \to ZZ) < 1 \, \textrm{fb}$.} \label{Chap-Maggie:fig:HHlHsmlow}
\end{figure}

In fig.~\ref{Chap-Maggie:fig:HHlHsm}, we show the rates for the process $pp \to h_\uparrow \to h_\downarrow h_{125}$ as a function of the heavier Higgs mass, for all C2HDM types.
In this channel, the rates can reach at most about 100 fb,
and only for Type I and Flipped.
The  rates for the $h_\downarrow h_{125}$ final state with the extra condition
$ \sigma (pp \to h_\uparrow \to ZZ) < 1 \, \textrm{fb}$
are shown in fig.~\ref{Chap-Maggie:fig:HHlHsmlow}.
The maximum rates (for low masses) are now reduced by about a factor of 5
for Type I.
However, the rates do not decrease much for the Flipped C2HDM,
and some signal at the LHC
could point to this C2HDM type.
Finally, although $h_\uparrow \to h_\downarrow h_{125}$ appears hard to detect in these models,
it is a clear sign of non-minimal models and should therefore be a priority for the LHC searches.
%
%
%



\vs{-3mm}

\section{Summary}
\label{Chap-Maggie:sec:concl}

We analyzed in detail the LO phenomenology of the C2HDM, focusing especially on signs of CP violation.
We have shown that large CP-odd Yukawa couplings of $h_{125}$ are still possible in all Yukawa types except Type I.
We stressed that two rather peculiar scenarios are allowed in the C2HDM: the one in which $h_{125}$ couples like a scalar to some fermions and like a pseudoscalar to others, and the one in which the scalar and pseudoscalar couplings of $h_{125}$ to some fermion are of similar size.
We proposed benchmark points for both these pictures.

Based on a previous work where we presented classes of three decays that are a clear sign of CP violation, we looked for correlations between the rates of the processes in each class (on the one hand) and CP-probing variables (on the other).
We have shown that the CP-violating phase $\phi (\lambda_5)$ is not a relevant variable. We then tested the correlation with other
variables proposed in the literature. The conclusion is that
some correlation can be seen between the rates and the variables. This is particularly true for the variable $\xi_V$, especially for Type II. However, in most cases, there is almost no correlation between the high rates of the CP-violating processes and the proposed CP-violating variables. The results also tell us that measuring small rates should not be interpreted as a sign of a small CP-violating angle.

Finally, we studied scenarios where the Higgs bosons decay to two other scalars. The search for a scalar decaying into two $h_{125}$ Higgs bosons was already performed during Run-1, whereas the search for $h_i \to h_j h_{125}$ has not started yet. It is clear from the results that it will be much harder to probe CP violation with classes of decays that involve Higgs-to-Higgs decays. However,
the observation of such decays would be a first step to reconstruct the Higgs potential. Hence, these searches should be a priority for LHC searches.

%% file: Chapters/Chapter_Reno.tex
\chapter{C2HDM: renormalization at NLO}
\label{Chap-Reno}

\vs{-5mm}


We now turn to the renormalization of the C2HDM.
As noted in the Introduction, the renormalization is essential to explore precise predictions in the model. Precise predictions, in turn, are the key to probe the SM and BSM models, since they are indispensable for a) a sound interpretation of the observed results, b) a correct determination of the parameter space of models and c) a proper distinction between different BSM models.
The literature devoted to the renormalization of models with an extended Higgs sector is vast (for the SM, reviews can be found in refs. \cite{Aoki:1982ed, Hollik:1988ii,Denner:1991kt,Denner:2019vbn,Freitas:2020kcn}).
For example, the ``real 2HDM'' has been subject to several studies~\cite{Santos:1996vt, Kanemura:2004mg, Kanemura:2015mxa, Krause:2016oke, Altenkamp:2017ldc, Denner:2016etu, Denner:2017vms, Denner:2018opp};
the next-to-minimal 2HDM has been considered in ref.~\cite{Krause:2017mal}, and
the scalar sector of a variant of the Minimal Supersymmetric extension of the SM with complex parameters (cMSSM) was explored e.g. in refs.~\cite{Pilaftsis:1996ac,Pilaftsis:1997dr,Pilaftsis:1998pe,Pilaftsis:1998dd,Pilaftsis:1999qt,Carena:2000ks,Choi:2000wz,Frank:2006yh,Fowler:2009ay,Heinemeyer:2010mm,Fritzsche:2011nr,Fritzsche:2013fta}.

To our knowledge, this is the first time that the renormalization of a model with explicit CP violation in the scalar sector is put forward. We will show that this leads to a unique process of renormalization, as one is forced to introduce several parameters that can be rephased away in the context of the renormalized parameters, but must be considered anyway in order to ensure the generation of all the necessary counterterms. As a consequence, there will be more independent counterterms than independent renormalized parameters. Notably, different combinations of counterterms can be taken as independent for the same set of independent renormalized parameters.

The chapter is organized as follows. 
In section \ref{Chap-Reno:sec:model}, we describe the C2HDM in such a way that we aim at the one-loop renormalization of the model, presenting different combinations of independent parameters.
Section \ref{Chap-Reno:sec:UTOL} introduces the treatment of the theory when considered at up to one-loop level, clarifying how the theory can contain counterterms for quantities that do not show up in the set of renormalized quantities.
Sections \ref{Chap-Reno:sec:FJTS-C2HDM} and \ref{Chap-Reno:sec:calculation-CTs} proceed with the aforementioned treatment: the former describes the selection of the true vev, whereas the latter is devoted to the counterterms of the C2HDM.
%
Finally, we present our results in section \ref{Chap-Reno:sec:num-res-main}, and a summary in section \ref{Chap-Reno:sec:conclu}.

\section{The C2HDM (for one-loop renormalization)}
\label{Chap-Reno:sec:model}

Here, we follow a similar procedure to that of section \ref{Chap-Maggie:sec:model}; in particular, we start by studying the potential $V$ in section \ref{Chap-Reno:section:pot}, 
then the scalar kinetic terms in section \ref{Chap-Reno:section:kin} and finally the Yukawa Lagrangian in section \ref{Chap-Reno:section:Yukawa}.
However, we shall now use a general parameterization, which implies that we will introduce several parameters that are absent in the usual tree-level description of chapter \ref{Chap-Maggie}.%
\fn{By the usual tree-level description, we mean the tree-level description that does not aim at the one-loop renormalization.}
As it will become clear, this is required in order to ensure the one-loop renormalization of the theory.

\subsection{The potential}
\label{Chap-Reno:section:pot}

The potential is still given by eq. \ref{Chap-Maggie:eq:pot}.
However, we parameterize the Higgs doublets in a more general way. Indeed, we admit that their vevs are in general complex, and in general relatively complex when compared to the scalar fields. So, instead of eq. \ref{Chap-Maggie:eq:2hdmdoubletexpansion}, we write:
\be
\Phi_1 =
\left(
\begin{array}{c}
\phi_1^+\\
\tfrac{1}{\sqrt{2}} (v_1 \, e^{i \zeta_1}+ \rho_1 + i \eta_1)
\end{array}
\right),
\hspace{5ex}
\Phi_2 =
\left(
\begin{array}{c}
\phi_2^+\\
\tfrac{1}{\sqrt{2}} (v_2 \, e^{i \zeta_2}+ \rho_2 + i \eta_2)
\end{array}
\right),
\label{Chap-Reno:eq:doublets-first}
\ee
where $\zeta_i$ are (real) phases and, as before, $v_i$ are real parameters, $\phi_i^+$ complex fields and $\rho_i$ and $\eta_i$ real fields ($i=1,2$).
The writing of the vevs with modulus and phase allows us to define the real parameters $v$ and $\beta$ according to eq. \ref{Chap-Maggie:eq:my-re-and-im}.
Finally, note that the phases $\zeta_1$ and $\zeta_2$ constitute the first example of parameters which we introduce in this chapter and which were absent in the previous one.

\subsubsection{Neutral linear terms: minimum equations}

Just as in section \ref{Chap-Selec:sec:SM-tree}, one must start by assuring that the true minimum of the theory at tree-level is selected. This corresponds to the requirement that
no tree-level tadpoles $t_i$ for the neutral scalar fields $\phi_{n,i}$ exist, i.e. that the terms of the potential which are linear in those fields vanish.%
\fn{The tadpoles for the charged fields $\phi_{c,i}$ are trivially zero.}
Hence,
\be
t_i \equiv \left\langle \frac{\partial V}{\partial \phi_{n,i}} \right\rangle
=
0.
\ee
This leads to the minimum equations
\bs
\bea
{{m}_{11}^{2}} &=&  \dfrac{1}{4} \bigg[  \left( -{\lambda_1} + \lambda_{345} \right)  \, {v}^2 \, \cos(2 \, {\beta}) + 4 \, {{m}_{12 \mathrm{R}}^{2}} \, \tan({\beta}) \, \sec({\zeta_1} - {\zeta_2}) \nonumber \\[-4mm]
&& \hs{30mm} - {v}^2 \,  \left( {\lambda_1} + \lambda_{345} + 2 \, {\lambda_{5 \text{I}}} \, \sin({\beta})^2 \, \tan({\zeta_1} - {\zeta_2}) \right)  \bigg], \\[1mm]
{{m}_{22}^{2}} &=&  \dfrac{1}{4} \bigg[  \left( {\lambda_2} -  \lambda_{345} \right)  \, {v}^2 \, \cos(2 \, {\beta}) + 4 \, {{m}_{12 \mathrm{R}}^{2}} \, \cot({\beta}) \, \sec({\zeta_1} - {\zeta_2}) \nonumber \\[-4mm]
&& \hs{30mm} - {v}^2 \,  \left( {\lambda_2} +  \lambda_{345} + 2 \, {\lambda_{5 \text{I}}} \, \cos({\beta})^2 \, \tan({\zeta_1} - {\zeta_2}) \right) \bigg], \\[1mm]
{{m}_{12 \text{I}}^{2}} &=&  {{m}_{12 \mathrm{R}}^{2}} \, \tan({\zeta_1} - {\zeta_2}) + \dfrac{1}{4} \bigg[{v}^2 \, \sec({\zeta_1} - {\zeta_2}) \, \sin(2 \, {\beta}) \,  \Big( {\lambda_{5 \text{I}}} \, \cos(2 \,  \left( {\zeta_1} - {\zeta_2} \right) ) \nonumber \\[-2mm]
&& \hs{64mm} - {\lambda_{5 \mathrm{R}}} \, \sin(2 \,  \left( {\zeta_1} - {\zeta_2} \right) ) \Big) \bigg],
\eea
\label{Chap-Reno:eq:min}
\es
where, as before, we define $\lambda_{345} \equiv \lambda_3 + \lambda_4 + \lambda_{5 \mathrm{R}}$.
Eqs. \ref{Chap-Reno:eq:min} reduce to \ref{Chap-Maggie:eq:min-eqs} for $\zeta_1 = \zeta_2 = 0$.

\subsubsection{Charged bilinear terms}
\label{Chap-Reno:section:charged}

The mass matrix of the charged scalars, defined by
\be
({\cal M}_c^{2})_{ij} = \left\langle \frac{\partial^2 V}{\partial \phi_{c,i}^{*} \,
  \partial \phi_{c,j}} \right\rangle \;,
\label{Chap-Reno:eq:chargedmassmat}
\ee
is hermitian, so that one needs in general a unitary matrix to diagonalize it. We then define the unitary matrix $X$ such that:%
\fn{\label{Chap-Reno:note:overall}The most general parameterization of $X$ includes an overall phase, which we ignore (in doing so, we force $\det X$ to be 1). It can be shown that such phase is not necessary; for details, cf. appendix \ref{App-Sym}.}
\be
S_c
=
X \, \phi_c
\quad
\Leftrightarrow
\quad
\begin{pmatrix}
G^+ \\
H^+
\end{pmatrix}
=
\begin{pmatrix}
e^{i(-\zeta_a - \zeta_b)} \cos \chi &
- e^{i(-\zeta_a + \zeta_b)} \sin \chi \\
e^{i(\zeta_a - \zeta_b)} \sin \chi &
e^{i(\zeta_a + \zeta_b)} \cos \chi
\end{pmatrix}
\begin{pmatrix}
\phi_1^+ \\
\phi_2^+
\end{pmatrix},
\label{Chap-Reno:eq:charged-param-original}
\ee
where $\zeta_a$ and $\zeta_b$ are (real) phases, $\chi$ a (real) angle, and the fields $S_c$ (with $S_c=(G^+ \, \, H^+)^{\mathrm{T}}$) are the charged states in the mass basis, with $G^+$ corresponding to the charged would-be Goldstone boson, as usual.
By definition, $X$ is such that:
\be
X {\cal M}_c^{2} X^{\dagger} = {\cal D}_c^2 \equiv 
\text{diag}(0,m_{\mathrm{H}^{+}}^2),
\label{Chap-Reno:eq:charged-mass-diag}
\ee
(with $m_{\mathrm{H}^{+}}$ representing again the mass of $H^+$), which in turn implies
\be
{\cal M}_c^{2} = X^{\dagger}  {\cal D}_c^2 X.
\label{Chap-Reno:eq:identity-charged}
\ee
Recall that, in chapter \ref{Chap-Maggie}, we found a relation (eq. \ref{Chap-Maggie:eq:rel-Mn}) that fixed a certain parameter (eq. \ref{Chap-Maggie:m3_derived}). Something similar happens here. In fact, there are two non-trivial relations between the elements of $\mathcal{M}_c^{2}$, namely,
\be
\dfrac{\left(\mathcal{M}_c^{2}\right)_{11}}{\left(\mathcal{M}_c^{2}\right)_{22}} = \tan(\beta)^2,
\qquad
\dfrac{\left(\mathcal{M}_c^{2}\right)_{11}}{\left(\mathcal{M}_c^{2}\right)_{12}} = - \tan(\beta) e^{i(\zeta_2 - \zeta_1)},
\label{Chap-Reno:eq:charged-rels}
\ee
which must also be verified for the r.h.s. of eq. \ref{Chap-Reno:eq:identity-charged}, thus leading to two identities:
\be
\chi = - \beta,
\qquad
\zeta_b = \dfrac{\zeta_1 - \zeta_2}{2}.
\label{Chap-Reno:eq:charged-conditions}
\ee

\subsubsection{Neutral bilinear terms}
\label{Chap-Reno:sec:neutral-bi}

The mass matrix of the neutral scalars, defined by
\be
({\cal M}_n^{2})_{ij}
=
\left\langle \frac{\partial^2 V}{\partial \phi_{n,i} \,
  \partial \phi_{n,j}} \right\rangle ,
\label{Chap-Reno:eq:c2hdmmassmat}
\ee
is symmetric, which means that one needs an orthogonal matrix to diagonalize it. Hence, we define the orthogonal matrix $Q$ such that:
\be
S_n = Q \, \phi_n
\quad
\Leftrightarrow
\quad
\begin{pmatrix}
h_1\\
h_2\\
h_3\\
G_0
\end{pmatrix}
= Q
\begin{pmatrix}
\rho_1\\
\rho_2\\
\eta_1\\
\eta_2
\end{pmatrix},
\label{Chap-Reno:eq:newmain}
\ee
where $S_n$ (with $S_n = (h_1 \,\, h_2 \,\, h_3 \,\, G_0)^{\mathrm{T}}$) are the neutral states in the mass basis, with $G_0$ corresponding to the neutral would-be Goldstone boson.%
\fn{A striking difference compared to chapter \ref{Chap-Maggie} is the conjoint diagonalization of the 4 neutral states (compare to eq. \ref{Chap-Maggie:eq:c2hdmrot}).}
%
Concerning $Q$, since it is a $4 \times 4$ orthogonal matrix, one in general needs 6 angles to parameterize it, which we take to be $\alpha_0, \alpha_1, \alpha_2, \alpha_3, \alpha_4, \alpha_5$, such that:
\be
Q
=
Q_5 \, Q_4 \, Q_3 \, Q_2 \, Q_1 \, Q_0,
\label{Chap-Reno:eq:Q1}
\ee
with
\bea
&&
Q_5
=
\begin{pmatrix}
1 & 0 & 0 & 0\\
0 & c_{5} & 0 & s_{5}\\
0 & 0 & 1 & 0\\
0 & -s_{5} & 0 & c_{5}
\end{pmatrix},
\ \
Q_4
=
\begin{pmatrix}
c_{4} & 0 & 0 & s_{4}\\
0 & 1 & 0 & 0\\
0 & 0 & 1 & 0\\
-s_{4} & 0 & 0 & c_{4}
\end{pmatrix}
\ \
Q_3
=
\begin{pmatrix}
1 & 0 & 0 & 0\\
0 & c_{3} & s_{3} & 0\\
0& -s_{3} & c_{3} & 0 \\
0 & 0 & 0 & 1
\end{pmatrix},
\nonumber \\[3mm]
&&
Q_2
=
\begin{pmatrix}
 c_{2} & 0 & s_{2} & 0\\
0 & 1 & 0 & 0\\
-s_{2} & 0 & c_{2} & 0 \\
0 & 0 & 0 & 1
\end{pmatrix},
\ \
Q_1
=
\begin{pmatrix}
c_{1} & s_{1} & 0 & 0\\
-s_{1} & c_{1} & 0 & 0\\
0 & 0 & 1 & 0\\
0 & 0 & 0 & 1
\end{pmatrix}
\ \
Q_0
=
\begin{pmatrix}
1 & 0 & 0 & 0\\
0 & 1 & 0 & 0\\
0 & 0 & -s_{0} & c_{0}\\
0 & 0& c_{0} & s_{0}
\end{pmatrix},
\label{Chap-Reno:eq:Q2}
\eea
with $s_i = \sin \alpha_i$, $c_i = \cos \alpha_i$ ($i = \{0,1,2,3,4,5\}$).
%
By definition, $Q$ is such that:
\be
Q {\cal M}_n^{2} Q^{\mathrm{T}} = {\cal D}_n^2 \equiv
\text{diag}(m_1^2,m_2^2,m_3^2,0),
\label{Chap-Reno:eq:neutral-mass-diag}
\ee
where, just as in chapter \ref{Chap-Maggie}, $m_i$ represents the mass of $h_i$, such that $m_1 < m_2 < m_3$.
In order to find relations in this (neutral) case, it is convenient to rewrite the doublets as:
\be
\Phi_1 = e^{i \zeta_1}
\left(
\begin{array}{c}
\phi_1^{+ \, \prime}\\
\tfrac{1}{\sqrt{2}} (v_1 + \rho_1^{\prime} + i \eta_1^{\prime})
\end{array}
\right),
\hspace{5ex}
\Phi_2 = e^{i \zeta_2}
\left(
\begin{array}{c}
\phi_2^{+ \, \prime}\\
\tfrac{1}{\sqrt{2}} (v_2 + \rho_2^{\prime} + i \eta_2^{\prime})
\end{array}
\right),
\label{Chap-Reno:eq:alternativebasis}
\ee
where the neutral fields with primes are related to the original ones (in eq. \ref{Chap-Reno:eq:doublets-first}) via:
\be
\phi_n = Z \, \phi_n^{\prime}
\quad
\Leftrightarrow
\quad
\begin{pmatrix}
\rho_1\\
\rho_2\\
\eta_1\\
\eta_2
\end{pmatrix}
=
\begin{pmatrix}
c_{\zeta_1} & 0 & -s_{\zeta_1} & 0\\
0 & c_{\zeta_2} & 0 & -s_{\zeta_2}\\
s_{\zeta_1} & 0 & c_{\zeta_1} & 0\\
0 & s_{\zeta_2} & 0 & c_{\zeta_2}
\end{pmatrix}
\begin{pmatrix}
\rho_1^{\prime}\\
\rho_2^{\prime}\\
\eta_1^{\prime}\\
\eta_2^{\prime}
\end{pmatrix}.
\ee
Then,
\begin{gather}
({\cal M}_n^{2 })_{ij}
=
\left\langle \frac{\partial^2 V}{\partial \phi_{n,i} \,
  \partial \phi_{n,j}} \right\rangle
=
({\cal M}_n^{2\prime})_{kl}
\dfrac{\partial \phi_{n,k}'}{\phi_{n,i}}
\dfrac{\partial \phi_{n,l}'}{\phi_{n,j}}
=
({\cal M}_n^{2 \prime})_{kl}
Z^{\mathrm{T}}_{ki} Z^{\mathrm{T}}_{lj}
=
(Z {\cal M}_n^{2 \prime} Z^{\mathrm{T}})_{ij},
\label{Chap-Reno:eq:trick-neutral}
\end{gather}
where we defined:
\be
({\cal M}_n^{2\prime})_{ij}
=
\left\langle \frac{\partial^2 V}{\partial \phi_{n,i}^{\prime} \,
  \partial \phi_{n,j}^{\prime}} \right\rangle.
\label{Chap-Reno:eq:neutral-mass-matrix-primes}
\ee
%
%
There are simple relations between the elements of ${\cal M}_n^{2\prime}$:
\be
\begin{pmatrix}
\dfrac{\left(\mathcal{M}_n^{2\prime}\right)_{13}}{\left(\mathcal{M}_n^{2\prime}\right)_{23}}
&
\dfrac{\left(\mathcal{M}_n^{2\prime}\right)_{14}}{\left(\mathcal{M}_n^{2\prime}\right)_{23}}
&
\dfrac{\left(\mathcal{M}_n^{2\prime}\right)_{23}}{\left(\mathcal{M}_n^{2\prime}\right)_{24}}
\\[5mm]
\dfrac{\left(\mathcal{M}_n^{2\prime}\right)_{13}}{\left(\mathcal{M}_n^{2\prime}\right)_{14}}
&
\dfrac{\left(\mathcal{M}_n^{2\prime}\right)_{13}}{\left(\mathcal{M}_n^{2\prime}\right)_{24}}
&
\dfrac{\left(\mathcal{M}_n^{2\prime}\right)_{14}}{\left(\mathcal{M}_n^{2\prime}\right)_{24}}
\\[5mm]
\dfrac{\left(\mathcal{M}_n^{2\prime}\right)_{33}}{\left(\mathcal{M}_n^{2\prime}\right)_{34}}
&
\dfrac{\left(\mathcal{M}_n^{2\prime}\right)_{33}}{\left(\mathcal{M}_n^{2\prime}\right)_{44}}
&
\dfrac{\left(\mathcal{M}_n^{2\prime}\right)_{34}}{\left(\mathcal{M}_n^{2\prime}\right)_{44}}
\end{pmatrix}
=
\begin{pmatrix}
\tan(\beta)
&
-1
&
-\tan(\beta)
\\
-\tan(\beta)
&
-\tan^2(\beta)
&
\tan(\beta)
\\
-\tan(\beta)
&
\tan^2(\beta)
&
-\tan(\beta)
\label{Chap-Reno:eq:neutral-rels}
\end{pmatrix}.
\ee
On the other hand, eqs. \ref{Chap-Reno:eq:neutral-mass-diag} and \ref{Chap-Reno:eq:trick-neutral} imply:
\be
{\cal M}_n^{2 \prime} = (Q Z)^{\mathrm{T}} {\cal D}_n^2 \, Q Z,
\label{Chap-Reno:eq:identity-neutral}
\ee
which means that the elements of the r.h.s. of this equation must obey the same nine relations of eq. \ref{Chap-Reno:eq:neutral-rels}.
We thus have nine relations, involving in general the twelve parameters:
\be
\alpha_0, \, \alpha_1, \, \alpha_2, \, \alpha_3, \, \alpha_4, \, \alpha_5, \, \zeta_1, \, \zeta_2, \, \beta, \, m_1^2, \, m_2^2, \, m_3^2.
\label{Chap-Reno:eq:params-aux}
\ee
It turns out that only four relations are independent, which means that one can only fix four of the parameters of eq. \ref{Chap-Reno:eq:params-aux}.
Obviously, there are many choices---or combinations---for the possible set of dependent parameters.%
\fn{In section \ref{Chap-Reno:section:charged} above, we could also have considered combinations of independent parameters. However, we did not want to take $\zeta_1$, $\zeta_2$ or $\beta$ as dependent, since they also show up in eq. \ref{Chap-Reno:eq:params-aux}; moreover, the phase $\zeta_a$ turns out not to be present in the relations of section \ref{Chap-Reno:section:charged}. Hence, there were two relations to fix two parameters ($\chi$ and $\zeta_b$), so that there were no combinations.}
In this work, we consider the combinations $C_i$ ($i=1,2,3,4$) described in table \ref{Chap-Reno:tab:combos}.
\begin{table}[h!]
\centering
\begin{tabular}
{
@{\hspace{-0.8mm}}
>{\centering}p{3cm}
>{\centering}p{0.5cm}
>{\centering}p{0.5cm}
>{\centering}p{0.5cm}
>{\centering\arraybackslash}p{0.5cm}@{\hspace{3mm}}
}
\hlinewd{1.1pt}
Combination & \multicolumn{4}{c}{Dependent parameters}\\
\hline\\[-4mm]
$C_1$ & $m_3^2,$ & $\zeta_1,$ & $\alpha_0,$ & $\alpha_4$\\[1mm]
$C_2$ & $m_3^2,$ & $\zeta_1,$ & $\alpha_0,$ & $\alpha_5$\\[1mm]
$C_3$ & $m_3^2,$ & $\zeta_1,$ & $\alpha_4,$ & $\alpha_5$\\[1mm]
$C_4$ & $m_3^2,$ & $\alpha_0,$ & $\alpha_4,$ & $\alpha_5$\\[0.5mm]
\hlinewd{1.1pt}
\end{tabular}
\vspace{-3mm}
\caption{The dependent parameters associated to the different combinations $C_i$.}
\label{Chap-Reno:tab:combos}
\end{table}
\normalsize
Some notes are in order here.

First, $m_3^2$ is chosen as a dependent parameter in all combinations; this is not accidental, and can be justified as follows. In this chapter, and until here, we have been looking at tree-level;%
\fn{The tree-level description we have been doing so far in this chapter---which aims at one-loop renormalization and thus includes several parameters that were absent in chapter \ref{Chap-Maggie}---should not be confused with what we call the usual tree-level description, which does not aim at renormalization and which was discussed in chapter \ref{Chap-Maggie}.}
when we consider the model up to one-loop level, we shall separate the parameters (meanwhile identified with bare parameters) into renormalized ones and counterterms, and we shall find that there can be a clear relation between the renormalized parameters, on the one hand, and the parameters of the usual tree-level description of chapter \ref{Chap-Maggie}, on the other. 
Now, as we saw in section \ref{Chap-Maggie:section:pot}, $m_3^2$ is taken as a dependent parameter in the usual tree-level description; this means that, if we want to exploit the aforementioned clear relation, $m_3^2$ should be a dependent parameter in the renormalized parameters of the model considered up to one-loop level. In order for that to happen, then, it must necessarily be taken as dependent at tree-level.

This also allows to explain the four combinations chosen in table \ref{Chap-Reno:tab:combos}. Indeed, if we wish to obtain the clear relation we have been aluding to, the parameters taken as independent in chapter \ref{Chap-Maggie} (recall eq. \ref{Chap-Maggie:eq:indeps}) must also be taken as independent here; in particular, $\alpha_1, \alpha_2, \alpha_3, \beta, m_1^2, m_2^2$ must always be taken as independent. As a consequence, besides $m_3^2$ (which is always dependent), the only parameters available to be dependent are $\alpha_0$, $\alpha_4$, $\alpha_5$, $\zeta_1$ and $\zeta_2$.
We decide to take $\zeta_2$ as independent, which we will justify below. Therefore, since there are only four dependence relations, there are only four possible combinations of dependent counterterms: precisely those of table \ref{Chap-Reno:tab:combos}.

Finally, the system of equations at stake is non-linear, and a rather complex one. We shall only solve it when we consider the model up to one-loop level.%
\fn{By then, and as suggested, the parameters involved in the system of equations are identified as bare parameters, and are split in renormalized parameters and counterterms (cf. section \ref{Chap-Reno:sec:UTOL}). The equations are then solved separately for the former and the latter.} 
For now, the dependent parameters must be understood simply as functions of the independent parameters.

\subsubsection{Combined sectors}

We have already derived some dependence conditions that resulted from relations we found within the squared mass matrices. Specifically, we found relations between the elements of $\mathcal{M}_c^{2}$, as well as relations between the elements of $\mathcal{M}_n^{2\prime}$. The former lead to the two conditions in eq. \ref{Chap-Reno:eq:charged-conditions}, whereas the latter to the dependence relations implied in table \ref{Chap-Reno:tab:combos}.

But there are still relations we have not yet used, which come from two equalities: that of eq. \ref{Chap-Reno:eq:identity-charged}
and that of eq. \ref{Chap-Reno:eq:identity-neutral}.
%
These lead to six independent relations, that we may use to rewrite the six $\lambda_i$ ($i=1,2,3,4,5\mathrm{R},5\text{I})$ in terms of other parameters.%
\fn{These expressions, though, are too long to be written here.}

\subsection{Scalar kinetic sector}
\label{Chap-Reno:section:kin}

We now consider the first term of the r.h.s. of eq. \ref{Chap-Maggie:eq:LHiggs},
\be
\mathcal{L}_{\text{scalar}}^{\text{kin}}
=
\left( D_{\mu} \Phi_1\right)^{\dagger} \left( D_{\mu} \Phi_1\right)
+
\left( D_{\mu} \Phi_2\right)^{\dagger} \left( D_{\mu} \Phi_2\right),
\label{Chap-Reno:eq:kin}
\ee
where the covariant derivative is defined by
\be
D_{\mu}
=
\partial_{\mu}
+
i g_2 \dfrac{\tau_a}{2} W^a_{\mu}
+
i g_1 Y B_{\mu}.
\ee
Here, $g_1$ and $g_2$ are the gauge couplings of the $\mathrm{U(1)_Y}$ and $\mathrm{SU(2)_L}$ gauge groups, respectively, with $B_{\mu}$ and $W^a_{\mu}$ ($a = 1,2,3$) being the corresponding gauge fields, and $Y$ and $\tau_a$ the corresponding group generators, respectively.%
\fn{As usual, we are following the conventions of ref.~\cite{Romao:2012pq} with all $\eta$'s positive.}
After SSB, the physical gauge fields $W^{\pm}_{\mu}$, $A_{\mu}$ and $Z_{\mu}$ are obtained from the original fields through the relations:
\be
W^{\pm}_{\mu} = \dfrac{W^1_{\mu} \mp i W^2_{\mu}}{\sqrt{2}},
\qquad
\begin{pmatrix}
A_{\mu} \\
Z_{\mu}
\end{pmatrix}
=
\begin{pmatrix}
c_{\text{w}} & s_{\text{w}} \\
-s_{\text{w}} & c_{\text{w}} 
\end{pmatrix}
\begin{pmatrix}
B_{\mu} \\
W^3_{\mu}
\end{pmatrix},
\label{Chap-Reno:eq:gauge-rot-tree}
\ee
with $s_{\text{w}} = \sin\theta_{\text{w}}$, $c_{\text{w}} = \cos\theta_{\text{w}}$, where $\theta_{\text{w}}$ is the weak mixing parameter, which is such that:
\be
c_{\text{w}} = \dfrac{g_2}{\sqrt{g_1^2 + g_2^2}}.
\label{Chap-Reno:eq:mycw}
\ee
Expanding the bilinear terms in the gauge fields in eq. \ref{Chap-Reno:eq:kin}, one easily finds that the photon $A_{\mu}$ is massless, whereas the squared masses of the $W^{+}_{\mu}$ and $Z_{\mu}$ bosons are, respectively,
\be
m_{\mathrm{W}}^2 = \dfrac{1}{4} \, g_2^2 \, v^2,
\qquad
m_{\mathrm{Z}}^2 = \dfrac{1}{4} \left(g_1^2 + g_2^2 \right) v^2.
\label{Chap-Reno:eq:gauge-masses}
\ee
Finally, defining the electric charge as
\be
e = \dfrac{g_1 g_2}{\sqrt{g_1^2 + g_2^2}},
\label{Chap-Reno:eq:charge}
\ee
we can take $v$, $g_1$ and $g_2$ as dependent parameters, which are then written as:%
\fn{The weak mixing parameter will be a dependent parameter itself, through the relation $c_{\text{w}} = m_{\mathrm{W}}/m_{\mathrm{Z}}$ (which follows from eqs. \ref{Chap-Reno:eq:mycw} and \ref{Chap-Reno:eq:gauge-masses}), since $m_{\mathrm{W}}$ and $m_{\mathrm{Z}}$ will be taken as independent.}
%
\be
v = \dfrac{2 \, m_{\mathrm{W}} \, s_{\text{w}}}{e},
\qquad
g_1 = \dfrac{e}{c_{\text{w}}},
\qquad
g_2 = \dfrac{e}{s_{\text{w}}}.
\label{Chap-Reno:eq:dep-params-gauge}
\ee

\subsection{Yukawa sector}
\label{Chap-Reno:section:Yukawa}

The different types of C2HDM were introduced in section \ref{Chap-Maggie:section:Yukawa}. Here, we restrict ourselves to the Type II model. While in section \ref{Chap-Maggie:section:Yukawa} we were interested in a specific part of the physical Yukawa Lagrangian for a generic type of C2HDM (eq. \ref{Chap-Maggie:eq:yuklag}), here we are interested in the complete original Yukawa Lagrangian for a specific type of C2HDM. To write it, it is convenient to parameterize the Higgs doublets according to:
\be
\Phi_i =
\begin{pmatrix}
\phi_i^+ \\
\phi_i^0
\end{pmatrix}
\,
, 
\quad
\Phi_i^* =
\begin{pmatrix}
\phi_i^- \\
\phi_i^{0*}
\end{pmatrix},
\ee
in which case we have, in the Type II C2HDM:
%
\be
- \mathcal{L}_{\text{Yukawa}}
=
\begin{pmatrix}
\bar{p}_{\mathrm{L}} & \bar{n}_{\mathrm{L}}
\end{pmatrix}
Y_d
\begin{pmatrix}
\phi_1^+ \\
\phi_1^0
\end{pmatrix}
n_{\mathrm{R}}
\, + \,
\begin{pmatrix}
\bar{p}_{\mathrm{L}} & \bar{n}_{\mathrm{L}}
\end{pmatrix}
Y_u 
\begin{pmatrix}
\phi_2^{0*} \\
-\phi_2^-
\end{pmatrix}
p_{\mathrm{R}}
\, + \,
\begin{pmatrix}
\bar{\nu}_{\mathrm{L}} & \bar{l}_{\mathrm{L}}
\end{pmatrix}
Y_l 
\begin{pmatrix}
\phi_1^+ \\
\phi_1^0
\end{pmatrix}
l_{\mathrm{R}}
\, + \, 
\text{h.c.},
\label{Chap-Reno:eq:LYukawa-original}
\ee
where $Y_d$, $Y_u$ and $Y_l$ are the Yukawa matrices for the down-type quarks, up-type quarks and leptons, respectively.%
\fn{In this equation, the $\mathrm{SU(2)_L}$ product is shown explicitly, but the sum over fermion generations is left implicit by the matrix notation. Had we written it explicitly, the quantities  $\bar{p}_{\mathrm{L}}, \bar{n}_{\mathrm{L}}, Y_d, n_{\mathrm{R}}$ in the first term on the r.h.s. of eq. \ref{Chap-Reno:eq:LYukawa-original}, for example, would have been $\bar{p}_{L_i}, \bar{n}_{L_i}, Y_d^{ij}, n_{R_j}$, respectively.}
We can, however, rewrite this equation in a more meaningful way.
We start by noting that the quarks are rotated to the mass basis through unitary transformations
\be
\bar{p}_{\mathrm{L}} = \bar{u}_{\mathrm{L}} U_{u_{\mathrm{L}}}^\dagger,
\hspace{8mm}
\bar{n}_{\mathrm{L}} = \bar{d}_{\mathrm{L}} U_{d_{\mathrm{L}}}^\dagger,
\hspace{8mm}
p_{\mathrm{R}} = U_{u_{\mathrm{R}}} u_{\mathrm{R}},
\hspace{8mm}
n_{\mathrm{R}} = U_{d_{\mathrm{R}}} d_{\mathrm{R}} ,
\label{Chap-Reno:eq:quarks-rot}
\ee
in such a way that the interaction with the vevs of the Higgs doublets generates the mass terms, that is,
\be
-\mathcal{L}_{\text{Yukawa}}^{\text{mass}}
=
\bar{d}_{\mathrm{L}} \, U_{d_{\mathrm{L}}}^{\dagger} Y_d U_{d_{\mathrm{R}}} \,  d_{\mathrm{R}} \, \langle \phi_1^0 \rangle
+ \bar{u}_{\mathrm{L}} \, U_{u_{\mathrm{L}}}^{\dagger} Y_u U_{u_{\mathrm{R}}} \, u_{\mathrm{R}} \, \langle \phi_2^{0*} \rangle
+ \bar{l}_{\mathrm{L}} \, Y_l \, l_{\mathrm{R}}  \, \langle \phi_1^0 \rangle
\, + \, 
\text{h.c.} \, .
\ee
On the other hand, the mass terms must also obey:
\be
- \mathcal{L}_{\text{Yukawa}}^{\text{mass}}
=
\bar{d}_{\mathrm{L}} \, M_d \, d_{\mathrm{R}}
+ \bar{u}_{\mathrm{L}} \, M_u \, u_{\mathrm{R}}
+ \bar{l}_{\mathrm{L}} \, M_l \, l_{\mathrm{R}}
\, + \, 
\text{h.c.} \, ,
\ee
with
$M_d = \text{diag}(m_d, m_s, m_b)$,
$M_u = \text{diag}(m_u, m_c, m_t)$,
$M_l = \text{diag}(m_e, m_{\mu}, m_{\tau})$.
Then, noting that
\be
\langle \phi_1^0 \rangle = \dfrac{v}{\sqrt{2}} c_{\beta} \, e^{i \zeta_1},
\qquad
\langle \phi_2^{0*} \rangle = \dfrac{v}{\sqrt{2}} s_{\beta} \, e^{-i \zeta_2},
\ee
we find:%
\fn{We assume there are no right-handed neutrinos, which implies we can take $Y_l$ to be diagonal without loss of generality.}
\be
Y_d = \dfrac{\sqrt{2}}{v \, c_{\beta} \, e^{i \zeta_1}} U_{d_{\mathrm{L}}} M_d U_{d_{\mathrm{R}}}^{\dagger},
\qquad
Y_u = \dfrac{\sqrt{2}}{v \, s_{\beta} \, e^{-i \zeta_2}} U_{u_{\mathrm{L}}} M_u U_{u_{\mathrm{R}}}^{\dagger},
\qquad
Y_l = \dfrac{\sqrt{2}}{v \, c_{\beta} \, e^{i \zeta_1}} M_l,
\ee
which allows us to write eq. \ref{Chap-Reno:eq:LYukawa-original} as
\begin{align}
&
-\mathcal{L}_{\text{Yukawa}}
=
\dfrac{\sqrt{2}}{v}
\Bigg[
\dfrac{1}{c_{\beta} \, e^{i \zeta_1}}
\begin{pmatrix}
\bar{u}_{\mathrm{L}} V & \bar{d}_{\mathrm{L}}
\end{pmatrix}
M_d
\begin{pmatrix}
\phi_1^+ \\
\phi_1^0
\end{pmatrix}
d_{\mathrm{R}}
\, + \,
\dfrac{1}{s_{\beta} \, e^{-i \zeta_2}}
\begin{pmatrix}
\bar{u}_{\mathrm{L}} & \bar{d}_{\mathrm{L}} V^{\dagger}
\end{pmatrix}
M_u
\begin{pmatrix}
\phi_2^{0*} \\
-\phi_2^-
\end{pmatrix}
u_{\mathrm{R}}
\nonumber
\\[1mm]
&
\hspace{40mm}
+
\dfrac{1}{c_{\beta} \, e^{i \zeta_1}}
\begin{pmatrix}
\bar{\nu}_{\mathrm{L}} & \bar{l}_{\mathrm{L}}
\end{pmatrix}
M_l 
\begin{pmatrix}
\phi_1^+ \\
\phi_1^0
\end{pmatrix}
l_{\mathrm{R}}
\Bigg]
\, + \, 
\text{h.c.} \, ,
\label{Chap-Reno:eq:LYukawa-final}
\end{align}
where $V= U_{u_{\mathrm{L}}}^{\dagger} U_{d_{\mathrm{L}}}$ is the CKM matrix, as usual.

Finally, note that the fermion masses $m_f$ are in general complex parameters, since the Yukawa matrices are general complex matrices.%
\fn{Here, $f$ represents the physical down-type quarks, up-type quarks and leptons, whose masses are respectively contained in $M_d$, $M_u$ and $M_l$.}
Usually, one performs a chiral rotation (that is, a rotation of the Weyl spinors) in order to render the masses real.%
\fn{Cf. e.g. section 29.3.2 of ref.~\cite{Schwartz:2013pla}.}
But the circumstance that the masses are in general complex means that, when the theory is considered up to one-loop level, the counterterms for the masses will also be in general complex. And although this is an irrelevant detail in most models, it becomes most relevant whenever there is CP violation in fermionic 2-point functions, as in the present model at one-loop level.
We discuss in detail complex mass counterterms---as well as their relation with CP violation in fermionic 2-point functions---in appendix \ref{App-Fermions}.
For now, we assume $m_f$ to be in general complex.

\subsection{Parameters}

The relations introduced in the previous sections allow us to replace the original sets of parameters in the Yukawa and Higgs sectors, respectively given by
\bs
\label{Chap-Reno:eq:original-params}
\bea
\{p^{\text{Y}}_{\text{ori.}}\}
&=&
\{ 
Y_d, Y_u, Y_l
\},
\\
\{p^{\mathrm{H}}_{\text{ori.}}\}
&=&
\{ 
g_1, \, g_2, \, m_{11}^2, \, m_{22}^2, \, m_{12 \mathrm{R}}^2, \, m_{12 \text{I}}^2, \,
\lambda_1, \, \lambda_2, \, \lambda_3, \, \lambda_4, \, \lambda_{5\mathrm{R}}, \, \lambda_{5\text{I}}
\},
\eea
\es
by new sets of parameters:
\bs
\label{Chap-Reno:eq:new-params}
\bea
\label{Chap-Reno:eq:new-params-Y}
\{p^{\text{Y}}_{\text{new}}\}
&=&
\{ 
m_f, V
\},
\\[1mm]
\label{Chap-Reno:eq:new-params-H}
\begin{blockarray}{c}
\\[-1.4mm]
\{p^{\mathrm{H}}_{C_1}\}\\[1.5mm]
\{p^{\mathrm{H}}_{C_2}\}\\[1.5mm]
\{p^{\mathrm{H}}_{C_3}\}\\[1.5mm]
\{p^{\mathrm{H}}_{C_4}\}
\end{blockarray}
\hspace{-1.5mm}
&=&
\{e, \, m_{\mathrm{W}}, \, m_{\mathrm{Z}},
\, \alpha_1, \, \alpha_2, \, \alpha_3,
\, \beta, \, m_1, \, m_2, \, m_{\mathrm{H}^{+}},
\, \mu^2, \, \zeta_2, \, \zeta_a,
\begin{blockarray}{c}
\\[-1.4mm]
\alpha_5\},\\[1.5mm]
\alpha_4\},\\[1.5mm]
\alpha_0\},\\[1.5mm]
\zeta_1\},
\end{blockarray}
\eea
\es
where $\mu^2$ is defined in eq. \ref{Chap-Maggie:eq:mu}.
We thus see that, in the Higgs sector, there are four possibilities, $\{p^{\mathrm{H}}_{C_i}\}$ ($i=1,2,3,4$), respectively associated to the combinations $C_i$ introduced in section \ref{Chap-Reno:sec:neutral-bi}.
For a given $C_i$, the parameters in the new total set (including parameters from the Yukawa and Higgs sectors) are all independent, and are more convenient than the ones in eqs. \ref{Chap-Reno:eq:original-params}.
One can then express the first four terms of the r.h.s. of eq. \ref{Chap-Maggie:eq:myLag} in terms of the parameters of eqs. \ref{Chap-Reno:eq:new-params}.

\section{Going up to one-loop level}
\label{Chap-Reno:sec:UTOL}

So far, and as we mentioned, we have been discussing the theory at tree-level, i.e. in lowest order (LO) in perturbation theory. When we include the next order---that is, when we consider the theory up to one-loop level, or at the next-to-leading order (NLO)---, we must start by modifying the notation.%
\fn{\label{Chap-Reno:note:utol}The notion `NLO' does not refer to one-loop level only, but to both tree-level and one-loop level. For the sake of clarity, and as we have been doing so far, we often use the notion `up to one-loop' instead. Both notions are equivalent and are used indifferently in this thesis.
Note that what we identify here as up-to-one-loop theory is then part (namely, the part which includes the tree-level and one-loop level) of what is usually identified as the effective theory.}
Just as in section \ref{Chap-Selec:sec:SMutol}, the theory we have been considering is to be identified with a \textit{bare} theory, and the parameters and fields therein contained as \textit{bare} parameters and fields. Moreover, we make sure to select the true vev of the theory up to one-loop level; such selection is discussed in section \ref{Chap-Reno:sec:FJTS-C2HDM}.

To recap what we saw in section \ref{Chap-Selec:sec:SMutol}, the bare parameters and fields are identified with an index ``$(0)$'' and they can renormalized. To do so, one starts by splitting them into renormalized quantities and their corresponding counterterms; for example, for a generic bare parameter $p_{(0)}$ and a generic bare field $\psi_{(0)}$,
\be
p_{(0)} = p + \delta p,
\qquad
\psi_{(0)} = \psi + \dfrac{1}{2} \delta Z_{\psi} \, \psi.
\label{Chap-Reno:eq:generic-expansion}
\ee
Here, $p$ represents the renormalized parameter and $\delta p$ the corresponding counterterm, and $\psi$ the renormalized field and $\delta Z_{\psi}$ the corresponding counterterm.%
\fn{The renormalized quantities in the context of the theory up to one-loop level (which have no index whatsoever) should not be confused with the bare quantities in the context of the theory at tree-level (which, in that context, did not have the index ``$(0)$'').
Note also that $\delta p$ is dubbed counterterm for the parameter (or parameter counterterm) and $\delta Z_{\psi}$ is dubbed counterterm for the field (or field counterterm).
}
The renormalization is completed by the calculation of the counterterms, which is performed in section \ref{Chap-Reno:sec:calculation-CTs}.%
\fn{\label{Chap-Reno:note:LSZ}As already suggested in note \ref{Chap-Selec:note:LSZ}, renormalized LSZ factors (present in the LSZ reduction formula) must also be calculated in order to obtain correct $S$-matrix elements. However, they become trivial when fields are renormalized in the on-shell subtraction scheme, so that they can be ignored. For details, see appendix \ref{App-LSZ}.}

As we saw, in order to ensure that all $S$-matrix elements are UV finite, one \textit{must} renormalize the independent parameters. The first step is choosing an independent set of parameters.%
\fn{As discussed in section \ref{Chap-Selec:sec:conv}, dependent parameters need not be renormalized, although they can be renormalized for convenience.}
%
Besides those of eq. \ref{Chap-Reno:eq:new-params-Y} we will consider the four possibilities of independent sets of eq. \ref{Chap-Reno:eq:new-params-H}. All these parameters are taken as bare parameters in the context of the up-to-one-loop theory, and thus read:
\begingroup\makeatletter\def\f@size{9.5}\check@mathfonts
\def\maketag@@@#1{\hbox{\m@th\normalsize\normalfont#1}}
\makeatother
\bs
\label{Chap-Reno:eq:bare-params}
\begin{flalign}
\label{Chap-Reno:eq:bare-params-Y}
&\{p^{\text{Y}}_{\text{new}(0)}\}
=
\{ 
m_{f(0)}, V_{(0)}
\},
\\[2mm]
&
\label{Chap-Reno:eq:bare-params-H}
\hs{-2mm}
\begin{blockarray}{c}
\\[-1.2mm]
\{p^{\mathrm{H}}_{C_1(0)}\}\\[1.5mm]
\{p^{\mathrm{H}}_{C_2(0)}\}\\[1.5mm]
\{p^{\mathrm{H}}_{C_3(0)}\}\\[1.5mm]
\{p^{\mathrm{H}}_{C_4(0)}\}
\end{blockarray}
\hspace{-3mm}
=
\hspace{-1mm}
\{e_{(0)}, m_{\mathrm{W}(0)},  m_{\mathrm{Z}(0)},
 \alpha_{1(0)},  \alpha_{2(0)},  \alpha_{3(0)},
 \beta_{(0)},  m_{1(0)},  m_{2(0)},  m_{\mathrm{H}^{+}(0)},
 \mu^2_{(0)},  \zeta_{2(0)}, \zeta_{a(0)},
\begin{blockarray}{c}
\\[-1.2mm]
\alpha_{5(0)}\},\\[1.5mm]
\alpha_{4(0)}\},\\[1.5mm]
\alpha_{0(0)}\},\\[1.5mm]
\zeta_{1(0)}\}.
\end{blockarray}
\end{flalign}
\es
\endgroup
Afterwards, and as suggested above, these bare parameters are separated into renormalized parameter and parameter counterterm, and the latter must be calculated.

On the other hand, if one intends to obtain finite Green's functions (GFs) and not only finite $S$-matrix elements, fields must be renormalized as well. This can be done in the symmetric form of the theory \cite{Hollik:1988ii,Bohm:1986rj}---i.e. before the SSB of the symmetry---, in which case it is enough to consider one field counterterm for each multiplet \cite{Denner:2019vbn}. Alternatively, one can renormalize the fields in the mass basis \cite{Aoki:1982ed,Denner:1991kt,Denner:2019vbn}. Here, counterterms for the mixing of fields (mixed fields counterterms) must in general be considered whenever fields mix at loop-level.

In the present section, we start by identifying the bare quantities of the theory with the sum of renormalized ones and counterterms (we do not do this for gauge parameters, nor ghost fields, since it would be irrelevant for the calculation of $S$-matrix elements~\cite{Denner:2019vbn}).%
\fn{Actually, in linear gauges (such as the Feynman gauge), it can be shown that one can simply refrain from applying any renormalization transformation whatsoever to the Gauge Fixing Lagrangian $\mathcal{L}_{\text{GF}}$
\cite{tHooft:1972qbu,Ross:1973fp,Baulieu:1983tg}.
That is, all the parameters and fields therein are assumed to be already renormalized.}
Then, in section \ref{Chap-Reno:sec:simple}, we review the need to consider a general parameterization, and show that the renormalized parameters can be described by a simpler parameterization.

\subsection{Expansion of the bare quantities}
\label{Chap-Reno:sec:expa}

In eq. \ref{Chap-Reno:eq:generic-expansion} above, we considered generic expansions. We now need to apply them to the specific content of the C2HDM. 
Concerning the parameters, we include not only those of eqs. \ref{Chap-Reno:eq:bare-params}, but also some dependent parameters, for convenience (recall section \ref{Chap-Selec:sec:conv}).
Some renormalized parameters will be set equal to non-obvious quantities, which will be justified below.
As for the fields, we choose those in the mass basis.


We start with the gauge boson masses and fields. For the masses, we have:
\be
m_{{\mathrm{W}}(0)}^2 = m_{\mathrm{W}}^2 + \delta m_{\mathrm{W}}^2,
\qquad
m_{\mathrm{Z}(0)}^2 = m_{\mathrm{Z}}^2 + \delta m_{\mathrm{Z}}^2,
\label{Chap-Reno:eq:mass-gauge-expa}
\ee
while for the fields:
\be
W_{\mu(0)}^{+} =
\Big( 1 + \dfrac{1}{2} \delta Z_W \Big) W_{\mu}^{+},
\qquad
\begin{pmatrix}
A_{\mu(0)} \\ Z_{\mu(0)}
\end{pmatrix}
=
\begin{pmatrix}
1 + \dfrac{1}{2} \delta Z_{AA} & \dfrac{1}{2} \delta Z_{AZ} \\
\dfrac{1}{2} \delta Z_{ZA} & 1 + \dfrac{1}{2} \delta Z_{ZZ}
\end{pmatrix}
\begin{pmatrix}
A_{\mu}
\\
Z_{\mu}
\end{pmatrix}.
\label{Chap-Reno:eq:field-gauge-expa}
\ee
%


In the fermions, the masses obey
\be
m_{f,i(0)} = m_{f,i} + \delta m_{f,i},
\label{Chap-Reno:eq:mass-fermion-expa}
\ee
in such a way that $m_{f,i}$ are real, and we define:%
\fn{For details, cf. appendix \ref{App-Fermions}.}
\be
\delta m_{f,i}^{\mathrm{R}} \equiv \delta m_{f,i},
\qquad
\delta m_{f,i}^{\mathrm{L}} \equiv \delta m_{f,i}^*.
\label{Chap-Reno:eq:mass-fermion-expa-2}
\ee
The fields obey:
\be
f_{i(0)}^{\mathrm{L}} =
\sum_j \Big( \delta_{ij} + \dfrac{1}{2} \delta Z_{ij}^{f,\mathrm{L}} \Big) f_j^{\mathrm{L}},
\qquad
f_{i(0)}^{\mathrm{R}} =
\sum_j \Big( \delta_{ij} + \dfrac{1}{2} \delta Z_{ij}^{f,\mathrm{R}} \Big) f_j^{\mathrm{R}}.
\label{Chap-Reno:eq:field-fermion-expa}
\ee


Considering now the scalar sector, we have for the masses:
\be
m_{\mathrm{H}^{+}(0)}^2 = m_{\mathrm{H}^{+}}^2 + \delta m_{\mathrm{H}^{+}}^2,
\qquad
m_{1(0)}^2 = m_1^2 + \delta m_1^2,
\qquad
m_{2(0)}^2 = m_2^2 + \delta m_2^2.
\label{Chap-Reno:eq:mass-scalar-expa}
\ee
We renormalize $m_{3(0)}^2$ for convenience:%
\be
m_{3(0)}^2 = m_{3\mathrm{R}}^2 + \delta m_{3}^2.
\label{Chap-Reno:eq:mass-scalar-expa-2}
\ee
The renormalized squared mass in this case explicitly includes an index $\mathrm{R}$ (from `renormalized').%
\fn{Not to be confused with the index $\mathrm{R}$ expressing `real', as in eq. \ref{Chap-Maggie:eq:my-re-and-im}.}
The reason why this is necessary in this case is that, since $m_{3(0)}$ is a dependent parameter, it is fixed by other parameters. As a consequence, when the bare quantities are identified with a renormalized quantity plus a counterterm, the renormalized squared mass $m_{3\mathrm{R}}^2$ will also be fixed by other parameters, which means that it cannot be set equal to the \textit{pole} squared mass---which we identify in the following as $m_{3\mathrm{P}}^2$. Moreover, $\delta m_{3}^2$ will be a dependent counterterm.
All this implies that $h_3$ plays a special role in theory, which is described in detail in appendix \ref{App-OSS}.

For the scalar fields,
\bs
\label{Chap-Reno:eq:field-scalar-expa}
\bea
\label{Chap-Reno:eq:charged-diag-bare-exp}
\begin{pmatrix}
G_{(0)}^{+} \\ H_{(0)}^{+}
\end{pmatrix}
&=&
\begin{pmatrix}
1 + \dfrac{1}{2} \delta Z_{G^{+}G^{+}} & \dfrac{1}{2} \delta Z_{G^{+}H^{+}} \\
\dfrac{1}{2} \delta Z_{H^{+}G^{+}} & 1 + \dfrac{1}{2} \delta Z_{H^{+}H^{+}}
\end{pmatrix}
\begin{pmatrix}
G^{+} \\ H^{+}
\end{pmatrix},
\\
\begin{pmatrix}
h_{1(0)} \\ h_{2(0)} \\ h_{3(0)} \\ G_{0(0)}
\end{pmatrix}
&=&
\begin{pmatrix}
1 + \frac{1}{2} \delta Z_{h_1h_1} & \frac{1}{2} \delta Z_{h_1h_2} & \frac{1}{2} \delta Z_{h_1h_3} & \frac{1}{2} \delta Z_{h_1G_0}\\
\frac{1}{2} \delta Z_{h_2h_1} & 1 + \frac{1}{2} \delta Z_{h_2h_2} & \frac{1}{2} \delta Z_{h_2h_3} & \frac{1}{2} \delta Z_{h_2G_0}\\
\frac{1}{2} \delta Z_{h_3h_1} & \frac{1}{2} \delta Z_{h_3h_2} & 1 + \frac{1}{2} \delta Z_{h_3h_3}  & \frac{1}{2} \delta Z_{h_3G_0}\\
\frac{1}{2} \delta Z_{G_0h_1} & \frac{1}{2} \delta Z_{G_0h_2} & \frac{1}{2} \delta Z_{G_0h_3} & 1 + \frac{1}{2} \delta Z_{G_0G_0}
\end{pmatrix}
\begin{pmatrix}
h_{1} \\ h_{2} \\ h_{3} \\ G_{0}
\end{pmatrix}.
\eea
\es
The mixing parameters are such that:%
\fn{Recall that $\alpha_{0(0)}$, $\alpha_{4(0)}$ and $\alpha_{5(0)}$ will be independent according to the combination $C_i$ chosen.}
\bs
\label{Chap-Reno:eq:mix-angles-bare-exp}
\begin{gather}
\label{Chap-Reno:eq:my-dza}
\beta_{(0)} = \beta + \delta \beta,
\qquad
\zeta_{a(0)} = \delta \zeta_a,
\\
\alpha_{1(0)} = \alpha_1 + \delta \alpha_{1},
\qquad
\alpha_{2(0)} = \alpha_2 + \delta \alpha_{2},
\qquad
\alpha_{3(0)} = \alpha_3 + \delta \alpha_{3},
\\
\alpha_{0(0)} = \beta + \delta \alpha_{0},
\qquad
\alpha_{4(0)} = \delta \alpha_{4},
\qquad
\alpha_{5(0)} = \delta \alpha_{5}.
\label{Chap-Reno:eq:consoante}
\end{gather}
\es
The electric charge and the CKM matrix obey:
\be
e_{(0)}
= e + \delta Z_e e,
\qquad
V_{ij(0)} = V_{ij} + \delta V_{ij}.
\ee
As for the remaining parameters that are taken as independent ($\zeta_{1(0)}$ will be independent in the combination $C_4$), we have:
\be
\mu_{(0)}^2 = \mu^2 + \delta \mu^2,
\qquad
\zeta_{1(0)} = \delta \zeta_1,
\qquad
\zeta_{2(0)} = \delta \zeta_2.
\label{Chap-Reno:eq:remaining-bare-exp}
\ee

For convenience, we also renormalize several dependent parameters:%
\fn{
\label{Chap-Reno:eq:note-dep-CT}
In the combinations $C_i$ for which the parameters $\alpha_{0(0)}$, $\alpha_{4(0)}$, $\alpha_{5(0)}$ and $\zeta_{1(0)}$ are dependent, we also renormalize them for convenience (according to eqs. \ref{Chap-Reno:eq:consoante} and \ref{Chap-Reno:eq:remaining-bare-exp}).
We do not renormalize $\chi$, $\zeta_b$, $g_1$, $g_2$ or $v$. This means that, before renormalization, they must be replaced by the corresponding expressions (cf. eqs. \ref{Chap-Reno:eq:charged-conditions} and \ref{Chap-Reno:eq:dep-params-gauge}), and the latter must be renormalized. For details, cf. section \ref{Chap-Selec:sec:SMutol}.}
\begin{gather}
s_{{\text{w}}(0)} = s_{\text{w}} + \delta s_{\text{w}}, 
\quad
c_{{\text{w}}(0)} = c_{\text{w}} + \delta c_{\text{w}},
\nonumber
\\
m_{11(0)}^2 = m_{11}^2 + \delta m_{11}^2,
\quad
m_{22(0)}^2 = m_{22}^2 + \delta m_{22}^2,
\qquad
m_{12\text{I}(0)}^2 = m_{12\text{I}}^2 + \delta m_{12\text{I}}^2,
\nonumber
\\
\lambda_{i(0)} = \lambda_i + \delta \lambda_i,
\qquad
X^{\dagger}_{ij(0)} = X^{\dagger}_{ij} + \delta X^{\dagger}_{ij},
\qquad
Q_{ij(0)} = Q_{ij} + \delta Q_{ij}.
\label{Chap-Reno:eq:unnec}
\end{gather}

\subsection{General \textit{vs.} simple parameterization}
\label{Chap-Reno:sec:simple}

When studying the theory at tree-level, we considered the most general parameterization of the Higgs doublets, as well as of the diagonalization of the scalar states---recall section \ref{Chap-Reno:section:pot}.
This general information was required in order to have general counterterms (which are necessary to absorb the UV divergences at one-loop). But it is \textit{not} required to describe the renormalized parameters. Indeed, if the only reason we considered such a general parameterization was to account for the generality of the counterterms, there is no need to apply such general description to the renormalized parameters as well. 

To better understand the claim we are making, consider for example the phase $\zeta_a$. At tree-level, we had the freedom to rephase the fields $G^+$ and $H^+$ to absorb $\zeta_a$. Had we used that freedom, though, we would have run into problems; for when we were to consider the theory up to one-loop level, we would have no counterterm $\delta \zeta_a$ (as there would have been no bare parameter $\zeta_{a(0)}$ to start with). It turns out that such counterterm is necessary for the absorption of the divergences of the theory, so that it cannot be discarded.
%
%
That is to say, we could not use the freedom to rephase $\zeta_a$ away at tree-level.
But once in the up-to-one-loop theory, and given that $\delta \zeta_a$ was already generated, there is nothing stopping us from using the rephasing freedom. We can thus rephase the \textit{renormalized} parameter $\zeta_a$ away in the up-to-one-loop theory; and when we do so, the counterterm $\delta \zeta_a$ (that was meanwhile generated because we did not rephase $\zeta_a$ away at tree-level) will not vanish.

To see this explicitly, we may consider a simplified version of eq. \ref{Chap-Reno:eq:charged-param-original}, where we take $\zeta_b=\chi=0$ (which are irrelevant for the argument):
\be
\begin{pmatrix}
\phi_1^+ \\
\phi_2^+
\end{pmatrix}
=
\begin{pmatrix}
e^{i \zeta_a} & 0\\
0 & e^{-i \zeta_a} 
\end{pmatrix}
\begin{pmatrix}
G^+ \\
H^+
\end{pmatrix}.
\label{Chap-Reno:eq:ex:tree}
\ee
It is obvious that, in this tree-level relation, one is free to rephase $\zeta_a$ away through $G^+ \to e^{-i \zeta_a} G^+$, $H^+ \to e^{i \zeta_a} H^+$. Yet, as we want to generate a counterterm $\delta \zeta_a$, we do not use that rephasing freedom. So, when considering the theory up to one-loop level, we take the bare version of eq. \ref{Chap-Reno:eq:ex:tree};
then, expanding the charged scalar states in the mass basis and $\zeta_{a(0)}$ into renormalized quantities and counterterms, according to eq. \ref{Chap-Reno:eq:charged-diag-bare-exp} and
\be
\zeta_{a(0)} = \zeta_a + \delta \zeta_a,
\label{Chap-Reno:eq:310}
\ee
we obtain, to first order,
\be
\begin{pmatrix}
\phi_{1(0)}^+ \\
\phi_{2(0)}^+
\end{pmatrix}
=
\begin{pmatrix}
e^{i \zeta_a} \left( i \delta \zeta_a + \delta Z_{G^{+}G^{+}} + 1 \right) &
e^{i \zeta_a} \delta Z_{G^{+}H^{+}} \\
e^{-i \zeta_a} \delta Z_{H^{+}G^{+}} &
e^{-i \zeta_a} \left( -i \delta \zeta_a + \delta Z_{H^{+}H^{+}} + 1 \right)
\end{pmatrix}
\begin{pmatrix}
G^{+} \\ H^{+}
\end{pmatrix}.
\label{Chap-Reno:eq:ex:first}
\ee
Since $\delta \zeta_a$ was already generated, the renormalized parameter $\zeta_a$ can be rephased away through
$G^+ \to e^{-i \zeta_a} G^+$, 
$H^+ \to e^{i \zeta_a} H^+$,
$Z_{G^{+}H^{+}} = e^{2 i \zeta_a} Z_{G^{+}H^{+}}^{\prime}$,
$Z_{H^{+}G^{+}} = e^{-2 i \zeta_a} Z_{H^{+}G^{+}}^{\prime}$,
yielding:
\be
\begin{pmatrix}
\phi_{1(0)}^+ \\
\phi_{2(0)}^+
\end{pmatrix}
=
\begin{pmatrix}
i \delta \zeta_a + \delta Z_{G^{+}G^{+}} + 1 &
\delta Z_{G^{+}H^{+}}^{\prime} \\
\delta Z_{H^{+}G^{+}}^{\prime} &
-i \delta \zeta_a + \delta Z_{H^{+}H^{+}} + 1
\end{pmatrix}
\begin{pmatrix}
G^{+} \\ H^{+}
\end{pmatrix}.
\label{Chap-Reno:eq:ex:second}
\ee
This simple example illustrates several relevant points.
First, it shows that one must start with a general description including the tree-level parameter $\zeta_a$, for otherwise the counterterm $\delta \zeta_a$ will not be generated. 
Second, after $\delta \zeta_a$ was generated, the renormalized parameter $\zeta_a$ can be rephased away, and in such a way that  $\delta \zeta_a$ does not vanish.
Third, instead of starting by splitting $\zeta_{a(0)}$ into a non-zero $\zeta_a$ plus a counterterm (as in eq. \ref{Chap-Reno:eq:310}), and afterwards rephasing $\zeta_a$ away, we can assume beforehand that we will work in a basis for the renormalized parameters where $\zeta_a=0$, which lets us equate $\zeta_{a(0)}$ immediately with the counterterm only. This is what we did in eq. \ref{Chap-Reno:eq:my-dza} above.
%
%
%
%
%

Finally, what we obtained for the renormalized parameters looks just the same as that which we would have obtained at tree-level should we not have been aiming at renormalizing the theory.
As a matter of fact, if we were not concerned with the generation of counterterms, but rather in a mere tree-level description (as in chapter \ref{Chap-Maggie}), we would not have needed $\zeta_a$, so that we could have rephased it away in eq. \ref{Chap-Reno:eq:ex:tree}. In that case, we would have obtained the same which we obtained in eq. \ref{Chap-Reno:eq:ex:second} for the renormalized terms (i.e. excluding counterterms).%
\fn{Care should be taken with the notion `the same', because in one case what is at stake are renormalized quantities in the context of the up-to-one-theory, whereas in the other case there are only tree-level quantities in a tree-level theory. Hence, they are formally different. Nonetheless, the physical description in both cases is equivalent.}
This also holds for the other parameters that were not rephased away for the sake of renormalization only. Therefore, the basis for the renormalized parameters (i.e. the basis that we can and shall use for the renormalized parameters) is considerably simpler than the general one used in section \ref{Chap-Reno:section:pot}, and is just the same as the one usually employed when considering the theory solely at tree-level---as in chapter \ref{Chap-Maggie}. Comparing the parameterization used in that chapter with the general parameterization used in section \ref{Chap-Reno:section:pot}, we conclude that, for the renormalized parameters in the up-to-one-loop theory, all the fermion masses are real, and%
\fn{The reason for identifying the renormalized phases $\zeta_{1\mathrm{R}}$ and $\zeta_{2\mathrm{R}}$ with the index R is given in note \ref{Chap-Reno:note:thezetas}.}
\be
\zeta_{1\mathrm{R}} =
\zeta_{2\mathrm{R}} =
\zeta_a =
\alpha_4 =
\alpha_5 =
0,
\qquad
\alpha_0 = \beta.
\label{Chap-Reno:eq:key}
\ee
As a consequence, the relations one obtains for the renormalized parameters in the context of the up-to-one-loop theory are precisely those that were derived in chapter \ref{Chap-Maggie}. We checked this explicitly for different expressions: the minimum equations (eqs. \ref{Chap-Maggie:eq:min-eqs}), the relation for $m_{3\mathrm{R}}^2$ (eq. \ref{Chap-Maggie:m3_derived}), as well as the relations for the $\lambda_i$ (eqs. \ref{Chap-Maggie:eq:lambdas}).
Moreover, given eq. \ref{Chap-Reno:eq:key}, the renormalized matrix $Q$ becomes:
\be
Q
=
Q_{321} \, Q_{\beta},
\label{Chap-Reno:eq:Qsimple1}
\ee
with
\be
Q_{321}
=
Q_3 \, Q_2 \, Q_1
=
\begin{pmatrix}
& & &0\\
& R & &0\\
& & & 0\\
0&0&0& 1
\end{pmatrix},
\qquad
Q_{\beta}
=
\begin{pmatrix}
1&0&0&0\\
0&1&0&0\\
0&0&- s_{\beta} & c_{\beta}\\
0&0&c_{\beta} & s_{\beta}
\end{pmatrix},
\label{Chap-Reno:eq:Qsimple2}
\ee
with $R$ given in eq. \ref{Chap-Maggie:matrixR}.
The original parameterization of $Q$ in eqs. \ref{Chap-Reno:eq:Q1} and \ref{Chap-Reno:eq:Q2} thus allows a simple connection between the matrix $Q$ for the renormalized parameters and the matrix $R$.
The matrix $X$ becomes orthogonal, and is simply:
\be
X
=
\begin{pmatrix}
c_{\beta} & s_{\beta} \\
- s_{\beta} & c_{\beta}
\end{pmatrix},
\label{Chap-Reno:eq:X-reno}
\ee
which is just matrix $R_H^T$ of eq. \ref{Chap-Real:eq:HiggsBasis}.

Finally, the total independent renormalized parameters become:
\bs
\label{Chap-Reno:eq:reno-params}
\bea
\{p^{\text{Y}}_{\text{reno.}}\}
&=&
\{ 
m_{f}, V
\},
\\
\label{Chap-Reno:eq:reno-Higgs}
\{p^{\mathrm{H}}_{\text{reno.}}\}
&=&
\{
e, \, m_{\mathrm{W}}, \, m_{\mathrm{Z}},
\, \alpha_1, \, \alpha_2, \, \alpha_3,
\, \beta, \, m_1, \, m_2, \, m_{\mathrm{H}^{+}}, \, \mu^2
\}.
\eea
\es
These parameters are precisely those which are usually taken as independent in the theory considered solely at tree-level (recall eq. \ref{Chap-Maggie:eq:indeps}).
Moreover, the set of renormalized parameters of the Higgs sector, $\{p^{\mathrm{H}}_{\text{reno.}}\}$, is the same whichever the set of bare parameters $\{p^{\mathrm{H}}_{C_i(0)}\}$ taken as the independent (recall eqs. \ref{Chap-Reno:eq:bare-params-H}).
Actually, not only it is the same, but it contains less parameters than each of the sets $\{p^{\mathrm{H}}_{C_i(0)}\}$, since the renormalized versions of the last three independent bare parameters of each set were rephased away. 
On the other hand, for each bare independent parameter, there is an independent counterterm. 
Thus, whichever the combination of independent bare parameters chosen, \textit{there will be more independent counterterms than independent renormalized parameters}.
In fact, identifying $\{\delta p^{\mathrm{H}}_{C_i}\}$ with the set of independent counterterms of the Higgs sector in the combination $C_i$, we have:
\be
\label{Chap-Reno:eq:CT-params-H}
\hs{-2mm}
\begin{blockarray}{c}
\\[-1.2mm]
\{\delta p^{\mathrm{H}}_{C_1}\}\\[1.5mm]
\{\delta p^{\mathrm{H}}_{C_2}\}\\[1.5mm]
\{\delta p^{\mathrm{H}}_{C_3}\}\\[1.5mm]
\{\delta p^{\mathrm{H}}_{C_4}\}
\end{blockarray}
\hspace{-3mm}
=
\hspace{-1mm}
\{\delta e, \delta m_{\mathrm{W}}, \delta m_{\mathrm{Z}},
 \delta \alpha_{1}, \delta \alpha_{2}, \delta \alpha_{3},
 \delta \beta, \delta m_{1}, \delta m_{2}, \delta m_{\mathrm{H}^{+}}, \delta \mu^2, \delta \zeta_{2}, \delta \zeta_{a},
\begin{blockarray}{c}
\\[-1.2mm]
\delta \alpha_{5}\},\\[1.5mm]
\delta \alpha_{4}\},\\[1.5mm]
\delta \alpha_{0}\},\\[1.5mm]
\delta \zeta_{1}\}.
\end{blockarray}
\ee
Something similar happens in the Yukawa sector, as the phases of the renormalized fermion masses were rephased away. This means that, while the counterterms $\delta m_f$ are complex, the renormalized masses $m_f$ are real.
In summary, the presence of more independent counterterms than independent renormalized parameters is a consequence of CP violation, which forces the introduction of several parameters; these are required for the one-loop renormalization of the model, but their renormalized versions can be rephased away.

\section{Selection of the true vev}
\label{Chap-Reno:sec:FJTS-C2HDM}

We studied in detail the selection of the true vev in chapter \ref{Chap-Selec}. We realized that the most convenient way to perform such selection---the so-called Fleischer-Jegerlehner tadpole scheme, or FJTS---turns out to be equivalent to an approach where the tree-level vev is selected. We verified this explicitly for the SM.
In this section, we perform the selection of the true vev in the C2HDM using the FJTS. The complexity over the SM is increased not only by the existence of more fields, but also due to the presence of phases. Nonetheless, we will reach the same conclusion as before, namely: the FJTS can be mimicked by selecting the bare vev and by resorting to connected GFs.

We start by considering the bare version of eq. \ref{Chap-Reno:eq:doublets-first}, using the notation of bars introduced in chapter \ref{Chap-Selec}:
\be
\Phi_{1(0)} =
\left(
\begin{array}{c}
\phi_{1(0)}^+\\
\tfrac{1}{\sqrt{2}} (\bar{v}_1 \, e^{i \bar{\zeta}_1}+ \rho_{1(0)} + i \eta_{1(0)})
\end{array}
\right),
\quad
\Phi_{2(0)} =
\left(
\begin{array}{c}
\phi_{2(0)}^+\\
\tfrac{1}{\sqrt{2}} (\bar{v}_2 \, e^{i \bar{\zeta}_2}+ \rho_{2(0)} + i \eta_{2(0)})
\end{array}
\right).
\label{Chap-Reno:eq:doublets-bare-first}
\ee
As we are interested in selecting the true vev, the quantities with a bar correspond to \textit{true} quantities, in the sense that they guarantee that no proper tadpoles (i.e. 1-point GFs) show up in the up-to-one-loop theory. In the FJTS, they are split as follows:
\be
\bar{v}_1 = v_{1(0)} + \Delta v_1, 
\qquad
\bar{v}_2 = v_{2(0)} + \Delta v_2,
\qquad
\bar{\zeta}_1 = \zeta_{1(0)} + \Delta \zeta_1,
\qquad
\bar{\zeta}_2 = \zeta_{2(0)} + \Delta \zeta_2.
\label{Chap-Reno:eq:mybars}
\ee
Thus, each quantity with a bar is split into two terms: the bare one and an extra quantity. These extra quantities are identified in what follows as the $\Delta$ quantities and are responsible for the elimination of proper tadpoles (they are introduced with this purpose). To see this, it is convenient to write eq. \ref{Chap-Reno:eq:doublets-bare-first} in an alternative formulation:
\be
\Phi_{1(0)} =
\left(
\begin{array}{c}
\phi_{1(0)}^+\\
\tfrac{1}{\sqrt{2}} (\bar{v}_{\rho_1} + \rho_{1(0)} + i \bar{v}_{\eta_1} + i \eta_{1(0)})
\end{array}
\right),
\quad
\Phi_{2(0)} =
\left(
\begin{array}{c}
\phi_{2(0)}^+\\
\tfrac{1}{\sqrt{2}} (\bar{v}_{\rho_2} + \rho_{2(0)} + i \bar{v}_{\eta_2} + i \eta_{2(0)})
\end{array}
\right),
\label{Chap-Reno:eq:doublets-bare-second}
\ee
with $\bar{v}_{\rho_1}$, $\bar{v}_{\rho_2}$, $\bar{v}_{\eta_1}$, $\bar{v}_{\eta_2}$ real parameters, such that:
\be
\label{Chap-Reno:eq:mybars-new}
\begin{split}
& \bar{v}_{\rho_1} = v_{\rho_1(0)} + \Delta v_{\rho_1}, 
\qquad
\bar{v}_{\eta_1} = v_{\eta_1(0)} + \Delta v_{\eta_1}, 
\\
& \bar{v}_{\rho_2} = v_{\rho_2(0)} + \Delta v_{\rho_2}, 
\qquad
\bar{v}_{\eta_2} = v_{\eta_2(0)} + \Delta v_{\eta_2}.
\end{split}
\ee
Comparing with eqs. \ref{Chap-Reno:eq:doublets-bare-first} and \ref{Chap-Reno:eq:mybars}, we obtain:
\be
\begin{split}
&\Delta v_{\rho_1} = \Delta v_1 \, \cos \zeta_{1(0)} - v_{1(0)} \, \Delta \zeta_1 \sin \zeta_{1(0)},
\hspace{7mm}
\Delta v_{\eta_1} = \Delta v_1 \, \sin \zeta_{1(0)} + v_{1(0)} \, \Delta \zeta_1 \, \cos \zeta_{1(0)},
\\
&\Delta v_{\rho_2} = \Delta v_2 \, \cos \zeta_{2(0)} - v_{2(0)} \, \Delta \zeta_2 \sin \zeta_{2(0)},
\hspace{7mm}
\Delta v_{\eta_2} = \Delta v_2 \, \sin \zeta_{2(0)} + v_{2(0)} \, \Delta \zeta_2 \, \cos \zeta_{2(0)}.
\end{split}
\ee
We neglected terms of order $\Delta^2$, since the $\Delta$ quantities are of one-loop order (as we are about to show). This also means that the bare quantities can be replaced by their renormalized versions.%
\fn{We shall use this argument in the equations that follow.}
Hence, using the fact that we choose a basis for the renormalized parameters such that the renormalized phases $\zeta_{1\mathrm{R}}$ and $\zeta_{2\mathrm{R}}$ vanish (eq. \ref{Chap-Reno:eq:key}), we get:%
\fn{\label{Chap-Reno:note:thezetas}%
The phases $\zeta_{1\mathrm{R}}$ and $\zeta_{2\mathrm{R}}$ are the renormalized phases, such that $\zeta_{1(0)} = \zeta_{1\mathrm{R}} + \delta \zeta_1$
and
$\zeta_{2(0)} = \zeta_{2\mathrm{R}} + \delta \zeta_2$.
We use the index $\mathrm{R}$ to distinguish them from the true phases, 
just as we did with the vev in chapter \ref{Chap-Selec}.
Finally, note that we keep the bare vevs for convenience in eq. \ref{Chap-Reno:eq:auxFJTS}.}
\be
\begin{pmatrix}
\Delta v_{\rho_1} \\
\Delta v_{\rho_2} \\
\Delta v_{\eta_1} \\
\Delta v_{\eta_2}
\end{pmatrix}
=
\begin{pmatrix}
\Delta v_1 \\
\Delta v_2 \\
v_{1(0)} \, \Delta \zeta_1 \\
v_{2(0)} \, \Delta \zeta_2
\end{pmatrix}.
\label{Chap-Reno:eq:auxFJTS}
\ee
%

Now, when the Lagrangian is expanded with eq. \ref{Chap-Reno:eq:doublets-bare-second}, the $\Delta v_{\phi_{n}}$ of eq. \ref{Chap-Reno:eq:mybars-new} must ensure that proper tadpoles vanish. In order for this to happen, the terms with $\Delta v_{\phi_{n,j}}$ that contribute to the 1-point function of a certain neutral scalar field $\phi_{n,i}$ must precisely cancel the one-loop 1-point function in that field. That is, if $T_{\phi_{n,i}}$ represents the one-loop tadpole for $\phi_{n,i}$ (i.e. the one-loop contribution to the 1-point function in $\phi_{n,i}$), we must have:
\be
T_{\phi_{n,i}} = - \sum_j
\dfrac{\partial^2 \mathcal{L}}{\partial \phi_{n,i} \, \partial \Delta v_{\phi_{n,j}}} \bigg|_{\phi=0, \, \Delta v = 0}
\Delta v_{\phi_{n,j}}
.
\label{Chap-Reno:eq:sacred}
\ee
Using the fact that every $\Delta v_{\phi_{n,j}}$ always shows up together with the corresponding field $\phi_{n,j}$ (cf. eqs. \ref{Chap-Reno:eq:doublets-bare-second} and \ref{Chap-Reno:eq:mybars-new}), we can write:
\be
T_{\phi_{n,i}} 
=
\sum_j \dfrac{\partial^2 V}{\partial \phi_{n,i} \, \partial \phi_{n,j}} \bigg|_{\phi=0, \, \Delta v = 0} \Delta v_{\phi_{n,j}} 
=
\sum_j ({\cal M}_n^{2})_{ij} \Delta v_{\phi_{n,j}} ,
\ee
where the mass matrix ${\cal M}_n^{2}$ obeys eq. \ref{Chap-Reno:eq:neutral-mass-diag}.
Then, going to the diagonal basis,
\be
\begin{pmatrix}
T_{h_1}\\
T_{h_2}\\
T_{h_3}\\
T_{G_0}
\end{pmatrix}
=
Q
\begin{pmatrix}
T_{\rho_1}\\
T_{\rho_2}\\
T_{\eta_1}\\
T_{\eta_2}
\end{pmatrix}
=
Q \,
{\cal M}_n^{2} \,
Q^T \,
Q
\begin{pmatrix}
\Delta v_{\rho_1}\\
\Delta v_{\rho_2}\\
\Delta v_{\eta_1}\\
\Delta v_{\eta_2}
\end{pmatrix}
=
{\cal D}_n^2
\begin{pmatrix}
\Delta v_{h_1}\\
\Delta v_{h_2}\\
\Delta v_{h_3}\\
\Delta v_{G_0}
\end{pmatrix},
\ee
which implies:
\be
\Delta v_{h_1} = \dfrac{T_{h_1}}{m_1^2},
\qquad
\Delta v_{h_2} = \dfrac{T_{h_2}}{m_2^2},
\qquad
\Delta v_{h_3} = \dfrac{T_{h_3}}{m_{3\mathrm{R}}^2},
\qquad
\Delta v_{G_0} = 0,
\label{Chap-Reno:eq:DelandTrel}
\ee
where we used the fact that $T_{G_0} = 0$, which is a consequence of the Goldstone theorem. Some comments are in order.

First, just as in the SM, by obeying eq. \ref{Chap-Reno:eq:sacred} the $\Delta$ quantities---in the physical basis, $\Delta v_{h_1}$, $\Delta v_{h_2}$, $\Delta v_{h_3}$, $\Delta v_{G_0}$---ensure that no proper tadpoles (1-point GFs) show up in the theory. In other words, they guarantee that the true vev up to one-loop level is selected. Yet, just as in the SM, since they are contained inside the scalar doublets, they will contribute to other GFs of theory. For example, although no 1-point GF for $h_2$ remains in the theory (that is, no proper tadpole for $h_2$ remains), $\Delta v_{h_2}$ will contribute to some 2-point and 3-point functions; and given eq. \ref{Chap-Reno:eq:DelandTrel}, the one-loop tadpole $T_{h_2}$ will contribute to those GFs; just as in chapter \ref{Chap-Selec}, we dub these one-loop tadpole structures contributing to GFs other than the 1-point GFs \textit{broad tadpoles}.
In this way, $\Delta v_{h_2}$ ensures that no proper tadpole for $h_2$ exists, but introduces broad tadpoles in the theory. Broad tadpoles are thus a consequence of the selection of the true vev and must be considered for consistency.

Second, also as in the description of the FJTS in chapter \ref{Chap-Selec}, all broad tadpoles can be accounted for by considering all possible one-loop tadpole insertions in all possible GFs. Indeed,
as suggested above, each renormalized scalar field will have the corresponding $\Delta$ quantity added to it: $\Delta v_{h_1} + h_1$, $\Delta v_{h_2} + h_2$, and so on. This means that, for every term in the Lagrangian with (say) $\Delta v_{h_2}$, there will be a term with $h_2$ instead. As a consequence, for every $n$-point function with a $\Delta v_{h_2}$ contribution, there is a $(n+1)$-point function with a $h_2$ field contribution instead; and with that field, a reducible diagram with a one-loop tadpole can be formed. But it is easy to see that such reducible diagram precisely equals the $n$-point function with $\Delta v_{h_2}$: one just needs to consider the definition of the one-loop tadpole for $h_2$,
\be
\begin{minipage}[h]{.40\textwidth}
\vspace{5mm}
\begin{picture}(0,42)
\begin{fmffile}{1616} 
\begin{fmfgraph*}(50,70) 
\fmfset{arrow_len}{3mm} 
\fmfset{arrow_ang}{20} 
\fmfleft{nJ1} 
\fmfright{nJ2}
\fmflabel{$h_2$}{nJ1}
\fmf{dashes,tension=1}{nJ1,nJ2} 
\fmfv{decor.shape=circle,decor.filled=hatched,decor.size=9thick}{nJ2}
\end{fmfgraph*} 
\end{fmffile} 
\end{picture}
\vspace{-5mm}
\end{minipage}
\hspace{-30mm}
i \, T_{h_2},
\label{Chap-Reno:eq:loop-tad}
\ee
and the zero-momentum $h_2$ propagator,
\be
\begin{minipage}[h]{.40\textwidth}
\vspace{5mm}
\begin{picture}(0,42)
\begin{fmffile}{1179}
\begin{fmfgraph*}(70,70)
\fmfset{arrow_len}{3mm}
\fmfset{arrow_ang}{20}
\fmfleft{nJ1}
\fmfright{nJ2}
\fmflabel{$h_2$}{nJ1}
\fmflabel{$h_2$}{nJ2}
\fmf{dashes,label=\small $\hspace{12mm} p=0$,label.side=left,tension=3,label.dist=2thick}{nJ1,nJ1nJ2}
\fmf{dashes,tension=3}{nJ1nJ2,nJ2}
\end{fmfgraph*}
\end{fmffile}
\end{picture}
\vspace{-5mm}
\end{minipage}
\hspace{-22mm}
 i \frac{1}{- m_2^2},
\label{Chap-Reno:eq:0mom}
\ee
and compare with eq. \ref{Chap-Reno:eq:DelandTrel}.
%
%
The same argument obviously applies to the remaining broad tadpoles. In sum, although broad tadpoles must be considered for consistency, in the FJTS they can be accounted for by including all possible one-loop tadpole insertions in all possible GFs.

Third, the alternative just proposed (calculating reducible diagrams with one-loop tadpoles instead of tracing down the terms with $\Delta$ showing up in the Lagrangian) will be the one adopted in the following. The reason is simply that, as argued in chapter \ref{Chap-Selec}, the calculation of reducible diagrams with one-loop tadpoles is trivial with software such as \FM;%
\fn{For details, cf. appendix \ref{App-FM-C2HDM}.}
by contrast, the task of expanding the Lagrangian and tracing down all the terms with $\Delta$ is quite cumbersome; and since the two methods are equivalent, we opt for the former.

But it is clear that this option (calculating reducible diagrams with one-loop tadpoles) is precisely the same as the option that a) selects the bare vev instead of the true up-to-one-loop vev and b) considers connected GFs. In fact, by opting for calculating reducible diagrams with one-loop tadpoles in the FJTS, all GFs that \textit{may} receive an insertion of a one-loop tadpole \textit{must} receive such insertion.
%
In general, then, non-renormalized 2-point functions in the FJTS will get contributions from two different types of terms:
\vspace{4mm}
\be
\begin{minipage}[h]{.40\textwidth}
\vspace{13mm}
\begin{picture}(0,42)
\begin{fmffile}{777} 
\begin{fmfgraph*}(60,80) 
\fmfset{arrow_len}{3mm}
\fmfset{arrow_ang}{20}
\fmfleft{nJ1}
\fmfright{nJ2}
\fmf{plain,tension=3}{nJ1,nJ1nJ2}
\fmf{plain,tension=3}{nJ1nJ2,nJ2}
\fmfv{decor.shape=circle,decor.filled=hatched,decor.size=9thick}{nJ1nJ2}
\end{fmfgraph*} 
\end{fmffile} 
\end{picture}
\end{minipage}
\hspace{-31mm}
+
\hspace{4mm}
\begin{minipage}[h]{.40\textwidth}
\vspace{-3mm}
\begin{picture}(0,100)
\begin{fmffile}{888}
\begin{fmfgraph*}(60,90)
\fmfset{arrow_len}{3mm}
\fmfset{arrow_ang}{20}
\fmfleft{nJ1}
\fmfright{nJ2}
\fmftop{nJ3}
\fmf{plain,tension=3}{nJ1,nJ1nJ2}
\fmf{dashes,tension=0.1}{nJ1nJ2,x}
\fmf{dashes,label=$h_i$,label.dist=4,label.side=right,tension=0.1}{x,y}
\fmf{dashes,tension=0.06}{y,nJ3}
\fmf{plain,tension=3}{nJ1nJ2,nJ2}
\fmfv{decor.shape=circle,decor.filled=70,decor.size=1thick}{x}
\fmfv{decor.shape=circle,decor.filled=70,decor.size=1thick}{y}
\fmfv{decor.shape=circle,decor.filled=hatched,decor.size=9thick}{nJ3}
\end{fmfgraph*}
\end{fmffile}
\end{picture}
\end{minipage}
\hspace{-35mm}
,
\vspace{-10mm}
\label{Chap-Reno:eq:FJTS-2p}
\ee
where the full lines represent any type of particle and the dashed lines represent $h_1$, $h_2$, $h_3$. The same will happen for non-renormalized 3-point functions, which then include:
\vspace{3mm}
\be
\begin{minipage}[h]{.40\textwidth}
\vspace{9mm}
\begin{picture}(0,42)
\begin{fmffile}{777b} 
\begin{fmfgraph*}(70,70) 
\fmfset{arrow_len}{3mm}
\fmfset{arrow_ang}{20}
\fmfleft{nJ1}
\fmfright{nJ2,nJ4}
\fmf{plain,tension=3}{nJ1,nJ1nJ2}
\fmf{plain,tension=3}{nJ1nJ2,nJ2}
\fmf{plain,tension=3}{nJ1nJ2,nJ4}
\fmfv{decor.shape=circle,decor.filled=hatched,decor.size=9thick}{nJ1nJ2}
\end{fmfgraph*} 
\end{fmffile} 
\end{picture}
\end{minipage}
\hspace{-30mm}
+
\hspace{5mm}
\begin{minipage}[h]{.40\textwidth}
\vspace{-10mm}
\begin{picture}(0,100)
\begin{fmffile}{889}
\begin{fmfgraph*}(70,70)
\fmfset{arrow_len}{3mm}
\fmfset{arrow_ang}{20}
\fmfleft{nJ1} 
\fmfright{nJ2,nJ4} 
\fmftop{nJ3}
\fmf{plain,tension=25}{nJ1,nJ1nJ2nJ4J4} 
\fmf{plain,tension=25}{nJ2,nJ1nJ2nJ4J4} 
\fmf{plain,tension=25}{nJ4,nJ1nJ2nJ4J4}
\fmf{dashes,tension=0.1}{nJ1nJ2nJ4J4,x}
\fmf{dashes,tension=0.1,label=$h_i$,label.dist=3,label.side=left}{x,y}
\fmf{dashes,tension=0.08}{y,nJ3}
\fmfv{decor.shape=circle,decor.filled=70,decor.size=1thick}{x}
\fmfv{decor.shape=circle,decor.filled=70,decor.size=1thick}{y}
\fmfv{decor.shape=circle,decor.filled=hatched,decor.size=9thick}{nJ3}
\end{fmfgraph*}
\end{fmffile}
\end{picture}
\end{minipage}
\hspace{-35mm}
.
\label{Chap-Reno:eq:FJTS-3p}
\ee
\vspace{-2mm}

These two equations are nothing but the non-renormalized contributions for 2-point and 3-point connected GFs in an approch where the bare vev is selected (i.e. where all the $\Delta$ quantities are simply ignored). 
The conclusion is thus the same as in the SM: although the $\Delta$ quantities are used in the FJTS to select the true vev, they do not contribute to connected GFs (with more than one external leg). Hence, if one decides to calculate the 2-point and 3-point functions as in eqs. \ref{Chap-Reno:eq:FJTS-2p} and \ref{Chap-Reno:eq:FJTS-3p}, all the discussion between eqs. \ref{Chap-Reno:eq:mybars} and \ref{Chap-Reno:eq:DelandTrel} concerning the $\Delta$ quantities is not necessary and can be ignored.
In other words, one does not need to take the trouble of selecting the true vev up to one-loop level: by ignoring $\Delta$ quantities and by including the reducible diagrams with one-loop tadpoles, one obtains the same predictions as in the FJTS. This was already noted with generality in ref. \cite{Denner:2016etu}, but was now explicitly verified in a non-trivial case with CP violation.%
\fn{It is worth mentioning that nowhere in the present section did we need to mention tadpole counterterms. As discussed in section \ref{Chap-Selec:sec:reno}, this notion can lead to misunderstandings, so that it is preferable to avoid it.}

\section{Counterterms}
\label{Chap-Reno:sec:calculation-CTs}

As mentioned in section \ref{Chap-Reno:sec:UTOL}, the renormalization of the theory is completed by fixing the counterterms. 
Independent counterterms are \textit{a priori} not fixed; rather, they must be fixed through independent \textit{renormalization conditions}.
%
%
%
%
Now, counterterms in general contain a divergent part and a finite part; whichever the renormalization condition used to fix a certain counterterm, its divergent part will always be the same. But the same is not true for the finite part, which in general depends on the prescription used. The different prescriptions that can be used to define the finite part of counterterms are called \textit{subtraction schemes}.
In what follows, most counterterms will be fixed either through the modified minimal subtraction ($\overline{\text{MS}}$) scheme or through the on-shell subtraction (OSS) scheme.
%
The former is simpler: one selects a process that depends on the counterterm to be calculated; then, the renormalization condition simply states that the terms of the renormalized one-loop amplitude proportional to $\Delta_{\varepsilon}$ are zero. Here, $\Delta_{\varepsilon}$ is 
\be
\Delta_{\varepsilon} = \dfrac{2}{\varepsilon} - \gamma_{\text{E}} + \ln 4 \pi,
\label{Chap-Reno:Delta-eps}
\ee
where $\varepsilon = 4-d$ ($d$ being the dimension) and $\gamma_{\text{E}}$ is the Euler-Mascheroni constant. The counterterm at stake will then be proportional to $\Delta_{\varepsilon}$ only. When $\overline{\text{MS}}$ is used to calculate a parameter counterterm,
the corresponding renormalized parameter will have no clear physical meaning.
By contrast, OSS is defined precisely so that all renormalized parameters have a direct physical meaning \cite{Denner:2019vbn}; the renormalization conditions in this case are less direct, though, and are discussed in detail in appendix \ref{App-OSS}.
In what follows, we adopt the conventions therein defined.
In section \ref{Chap-Reno:sec:calc}, we fix all of the counterterms of the theory, proceeding by sectors.
%
Finally, in section \ref{Chap-Reno:sec:gauge-dep}, we discuss the gauge dependence of the counterterms.

\subsection{Calculation of counterterms}
\label{Chap-Reno:sec:calc}

\subsubsection{Gauge masses and fields}
\label{Chap-Reno:sec:CT-gauge}

For gauge fields, the  renormalized up-to-one-loop 2-point functions are, in the Feynman gauge,%
\fn{Although 2-point functions configure two momentum arguments, we exploit momentum conservation to write only one argument, always positive.
Note also that our notation for signs is different from that of ref.~\cite{Denner:2019vbn}: where we use a plus sign (in eqs. \ref{Chap-Reno:eq:GammaRenGauge}), they use a minus.
Moreover, ref.~\cite{Denner:2019vbn} also uses the relative sign between the non-renormalized transverse one-loop functions and the corresponding counterterms (in eqs. \ref{Chap-Reno:eq:SigmaRenoDecompGauge} below) opposite to ours. These differences led us to introduce the variable $\eta_g$ in appendix \ref{App-WI} (cf. eqs. \ref{App-WI:eq:etag} and \ref{App-WI:eq:ZACT}), so that ref.~\cite{Denner:2019vbn} uses $\eta_g=-1$, whereas $\eta_g=1$ is used in this thesis. Finally, note that ref. \cite{Fontes:2021iue} implicitly uses the $\eta_g=-1$.}
\bs
\label{Chap-Reno:eq:GammaRenGauge}
\bea
\hat{\Gamma}^{W^{+} W^{-}}_{\mu \nu} (k) &=& -g_{\mu \nu} \left(k^2 - m_{\mathrm{W}}^2\right) + \hat{\Sigma}_{\mu \nu}^{W^{+} W^{-}} (k),
\label{Chap-Reno:eq:GammaRenGaugeCharged}
\\
\hat{\Gamma}^{V V^{\prime}}_{\mu \nu} (k) &=& -g_{\mu \nu} \, \delta_{V V^{\prime}} \left(k^2 - m_V^2\right) + \hat{\Sigma}_{\mu \nu}^{V V^{\prime}} (k),
\label{Chap-Reno:eq:GammaRenGaugeNeutral}
\eea
\es
where $V, V^{\prime}=\{\gamma, Z\}$. They can be decomposed into transversal (T) and longitudinal (L) components:
\bs
\label{Chap-Reno:eq:GammaRenoGauge}
\bea
\hat{\Gamma}_{\mu \nu}^{W^+W^-}(k) &=& \left(g_{\mu \nu}-\frac{k_{\mu} k_{\nu}}{k^{2}}\right) \hat{\Gamma}_{\mathrm{T}}^{W^+W^-}(k^2)+\frac{k_{\mu} k_{\nu}}{k^{2}} \hat{\Gamma}_{\mathrm{L}}^{W^+W^-}(k^2),
\label{Chap-Reno:eq:GammaRenoGaugeCharged}
\\
\hat{\Gamma}_{\mu \nu}^{V V^{\prime}}(k) &=& \left(g_{\mu \nu}-\frac{k_{\mu} k_{\nu}}{k^{2}}\right) \hat{\Gamma}_{\mathrm{T}}^{V V^{\prime}}(k^2)+\frac{k_{\mu} k_{\nu}}{k^{2}} \hat{\Gamma}_{\mathrm{L}}^{V V^{\prime}}(k^2),
\label{Chap-Reno:eq:GammaRenoGaugeNeutral}
\eea
\es
which can also be applied to the particular case of the renormalized one-loop GFs:%
\fn{Eqs. \ref{Chap-Reno:eq:GammaRenoGauge} and \ref{Chap-Reno:eq:SigmaRenoGauge} also hold for non-renormalized GFs.}
\bs
\label{Chap-Reno:eq:SigmaRenoGauge}
\bea
\hat{\Sigma}_{\mu \nu}^{W^+W^-}(k) &=& \left(g_{\mu \nu}-\frac{k_{\mu} k_{\nu}}{k^{2}}\right) \hat{\Sigma}_{\mathrm{T}}^{W^+W^-}(k^2)+\frac{k_{\mu} k_{\nu}}{k^{2}} \hat{\Sigma}_{\mathrm{L}}^{W^+W^-}(k^2),
\label{Chap-Reno:eq:SigmaRenoGaugeCharged}
\\
\hat{\Sigma}_{\mu \nu}^{V V^{\prime}}(k) &=& \left(g_{\mu \nu}-\frac{k_{\mu} k_{\nu}}{k^{2}}\right) \hat{\Sigma}_{\mathrm{T}}^{V V^{\prime}}(k^2) + \frac{k_{\mu} k_{\nu}}{k^{2}} \hat{\Sigma}_{\mathrm{L}}^{V V^{\prime}}(k^2).
\label{Chap-Reno:eq:SigmaRenoGaugeNeutral}
\eea
\es
Now, expanding the Lagrangian using eqs. \ref{Chap-Reno:eq:mass-gauge-expa} and \ref{Chap-Reno:eq:field-gauge-expa}, we find:
\bs
\label{Chap-Reno:eq:SigmaRenoDecompGauge}
\bea
\hs{-10mm}
\hat{\Sigma}_{\mathrm{T}}^{W^+W^-} (k^2) &=& \Sigma_{\mathrm{T}}^{W^+W^-}(k^2) - \left(k^{2}-m_{\mathrm{W}}^{2}\right) \delta Z_{W} + \delta m_{\mathrm{W}}^{2},
\label{Chap-Reno:eq:SigmaRenoDecompGaugeCharged}
\\
\hs{-10mm}
\hat{\Sigma}_{\mathrm{T}}^{V V^{\prime}}(k^2) &=& \Sigma_{\mathrm{T}}^{V V^{\prime}}(k^2) - \frac{1}{2}\left(k^{2}-m_{V}^{2}\right) \delta Z_{V V^{\prime}}   - \frac{1}{2}\left(k^{2}-m_{V^{\prime}}^{2}\right) \delta Z_{V^{\prime} V} + \delta_{V V^{\prime}} \delta m_{V}^{2}.
\label{Chap-Reno:eq:SigmaRenoDecompGaugeNeutral}
\eea
\es
Having seen this, we start by considering the charged fields. The OSS renormalization conditions are:%
\fn{Note that the residue of the propagator of a gauge field (of mass $m$, projected onto the state $\varepsilon^{\sigma}$) is defined with a plus sign:
$\hat{R}^{\mu} =  \Lim{k^2 \to m^2} i \, g^{\mu \nu} (k^2 - m^2) \, \widetilde{\operatorname{Re}} \left[ i \hat{\Gamma}_{\nu \sigma}^{-1}(k) \right] \varepsilon^{\sigma}$ (compare with eq. \ref{App-OSS:eq:resi}).}
\bs
\bea
\left.\widetilde{\operatorname{Re}} \left[\hat{\Gamma}_{\mu \nu}^{W^+W^-}(k)\right] \varepsilon^{\nu}(k)\right|_{k^{2}=m_{\mathrm{W}}^{2}} &=& 0, \\
\lim_{k^{2} \rightarrow m_{\mathrm{W}}^{2}} \frac{(-1)}{k^{2}-m_{\mathrm{W}}^{2}} \widetilde{\operatorname{Re}} \left[\hat{\Gamma}_{\mu \nu}^{W^+W^-}(k)\right] \varepsilon^{\nu}(k) &=& \varepsilon_{\mu}(k).
\eea
\es
Inserting eq. \ref{Chap-Reno:eq:GammaRenoGaugeCharged}, we get respectively:\fn{In this way, the longitudinal components disappear in OSS. This is consistent with the fact that they are non-physical, in the sense that they do not contribute to the 2-point function in the OS limit.}
\bs
\label{Chap-Reno:eq:AuxGaugeCharged}
\bea
\widetilde{\operatorname{Re}} \, \hat{\Gamma}_{\mathrm{T}}^{W^+W^-}\left(m_{\mathrm{W}}^{2}\right) &=& 0,
\label{Chap-Reno:eq:AuxGaugeCharged1}
\\
\left.\quad \widetilde{\operatorname{Re}} \, \frac{\partial \hat{\Gamma}_{\mathrm{T}}^{W^+W^-}\left(k^{2}\right)}{\partial k^{2}}\right|_{k^{2} = m_{\mathrm{W}}^{2}} &=& -1.
\label{Chap-Reno:eq:AuxGaugeCharged2}
\eea
\es
Then, using eqs.
\ref{Chap-Reno:eq:GammaRenGaugeCharged},
\ref{Chap-Reno:eq:GammaRenoGaugeCharged},
\ref{Chap-Reno:eq:SigmaRenoGaugeCharged}, \ref{Chap-Reno:eq:SigmaRenoDecompGaugeCharged} in eqs. \ref{Chap-Reno:eq:AuxGaugeCharged1} and \ref{Chap-Reno:eq:AuxGaugeCharged2}, we get respectively:
\bs
\bea
\delta m_{\mathrm{W}}^{2} &=& - \widetilde{\operatorname{Re}} \, \Sigma_{\mathrm{T}}^{W^+W^-}\left(m_{\mathrm{W}}^{2}\right), \\
\delta Z_{W} &=& \left.\widetilde{\operatorname{Re}} \frac{\partial \Sigma_{\mathrm{T}}^{W^+W^-}\left(k^{2}\right)}{\partial k^{2}}\right|_{k^{2}=m_{\mathrm{W}}^{2}}.
\eea
\es
In a similar way for the neutral fields, the OSS renormalization conditions read:%
\fn{We thus write in a compact form the OSS conditions for the mass counterterms (eq. \ref{Chap-Reno:eq:OSS-gn1} for $V^{\prime} = V$), for the off-diagonal field counterterms (eq. \ref{Chap-Reno:eq:OSS-gn1} for $V^{\prime} \neq V$) and for the diagonal field counterterms (eq. \ref{Chap-Reno:eq:OSS-gn2}).
Moreover, note that $\widetilde{\operatorname{Re}} \, \hat{\Gamma}_{\mu \nu}^{V^{\prime} V}(k)$ $=$ $\widetilde{\operatorname{Re}} \, \hat{\Gamma}_{\mu \nu}^{V V^{\prime}}(k)$, as a consequence of the Lorentz invariance of the renormalized up-to-one-loop action.}
\bs
\label{Chap-Reno:eq:OSS-gauge-neutral}
\bea
\label{Chap-Reno:eq:OSS-gn1}
\left.\widetilde{\operatorname{Re}} \left[\hat{\Gamma}_{\mu \nu}^{V V^{\prime}}(k)\right] \varepsilon^{\nu}(k)\right|_{k^{2}=m_{V}^{2}} &=& 0,
\label{Chap-Reno:eq:my48a}
\\
\label{Chap-Reno:eq:OSS-gn2}
\lim _{k^{2} \rightarrow m_{V}^{2}} \frac{(-1)}{k^{2}-m_{V}^{2}} \widetilde{\operatorname{Re}} \left[\hat{\Gamma}_{\mu \nu}^{V V}(k)\right] \, \varepsilon^{\nu}(k) &=& \varepsilon_{\mu}(k).
\eea
\es
Inserting eq. \ref{Chap-Reno:eq:GammaRenoGaugeNeutral}, we get respectively:%
\bs
\label{Chap-Reno:eq:auxgaugeneutral}
\bea
\widetilde{\operatorname{Re}} \, \hat{\Gamma}_{\mathrm{T}}^{V V^{\prime}}\left(m_{V}^{2}\right) &=& 0,
\label{Chap-Reno:eq:AuxGaugeNeutral1}
\\
\left.\quad \widetilde{\operatorname{Re}} \, \frac{\partial \hat{\Gamma}_{\mathrm{T}}^{V V}\left(k^{2}\right)}{\partial k^{2}}\right|_{k^{2} = m_{V}^{2}} &=& -1.
\label{Chap-Reno:eq:AuxGaugeNeutral2}
\eea
\es
Using eqs.
\ref{Chap-Reno:eq:GammaRenGaugeNeutral},
\ref{Chap-Reno:eq:GammaRenoGaugeNeutral},
\ref{Chap-Reno:eq:SigmaRenoGaugeNeutral}, \ref{Chap-Reno:eq:SigmaRenoDecompGaugeNeutral}, we obtain:
\bs
\bea
\delta m_{\mathrm{Z}}^{2} &=& - \widetilde{\operatorname{Re}} \, \Sigma_{\mathrm{T}}^{Z Z}\left(m_{\mathrm{Z}}^{2}\right),\\
\begin{pmatrix}
\delta Z_{A A} & \delta Z_{A Z}\\
\delta Z_{Z A} & \delta Z_{Z Z}
\end{pmatrix}
&=&
\begin{pmatrix}
\left.\widetilde{\operatorname{Re}} \dfrac{\partial \Sigma_{\mathrm{T}}^{A A}\left(k^{2}\right)}{\partial k^{2}}\right|_{k^{2}=0}
&
2 \widetilde{\operatorname{Re}} \dfrac{\Sigma_{\mathrm{T}}^{A Z}\left(m_{\mathrm{Z}}^{2}\right)}{m_{\mathrm{Z}}^{2}} \\
-2 \dfrac{\Sigma_{\mathrm{T}}^{A Z}(0)}{m_{\mathrm{Z}}^{2}}
&
\left.\widetilde{\operatorname{Re}} \dfrac{\partial \Sigma_{\mathrm{T}}^{Z Z}\left(k^{2}\right)}{\partial k^{2}}\right|_{k^{2}=m_{\mathrm{Z}}^{2}}
\end{pmatrix}.
\label{Chap-Reno:eq:my50b}
\eea
\es

Finally, if one neglects the absorptive parts of loop integrals (through the operator $\widetilde{\operatorname{Re}}$), the one-loop processes introduce no new imaginary parts.%
\fn{For details, cf. appendix \ref{App-OSS}.
It goes without saying that the absorptive parts of loop integrals are neglected in the calculation of counterterms only; the loop integrals of non-renormalized one-loop diagrams contributing to a certain process have in general absorptive parts.}
In that case, the hermiticity of the Lagrangian implies the hermiticity of the up-to-one-loop action. Specifically in the case of up-to-one-loop 2-point functions of the gauge bosons of the C2HDM, this means:
\be
\widetilde{\operatorname{Re}} \, \Gamma_{\mu \nu}^{W^+W^-}(k)
=
\left[
\widetilde{\operatorname{Re}} \, \Gamma_{\mu \nu}^{W^+W^-}(k)
\right]^{*},
\qquad
\widetilde{\operatorname{Re}} \, \Gamma_{\mu \nu}^{V V^{\prime}}(k)
=
\left[\widetilde{\operatorname{Re}} \, \Gamma_{\mu \nu}^{V V^{\prime}}(k)\right]^{*},
\ee
which implies, in particular,%
\fn{As the different orders are formally independent, the hermiticity of the up-to-one-loop 2-point function implies the hermiticity of the one-loop 2-point function, which in turn leads to the hermiticity of its independent components (transversal and longitudinal).}
\be
\widetilde{\operatorname{Re}} \, \Sigma_{\mathrm{T}}^{W^+W^-}(k^2)
=
\left[
\widetilde{\operatorname{Re}} \, \Sigma_{\mathrm{T}}^{W^+W^-}(k^2)
\right]^{*},
\qquad
\widetilde{\operatorname{Re}} \, \Sigma_{\mathrm{T}}^{V V^{\prime}}(k^2)
=
\left[\widetilde{\operatorname{Re}} \, \Sigma_{\mathrm{T}}^{V V^{\prime}}(k^2)\right]^{*}.
\ee
As a consequence, all the counterterms for gauge boson masses and fields are real.

\subsubsection{Fermion masses and fields}
\label{Chap-Reno:sec:fermionsCT}

In the case of fermionic 2-point functions, we have:
\be
\hat{\Gamma}^{\bar{f} f}_{i j} (p) = \delta_{i j} \left(\slashed{p} - m_{f,i}\right) + \hat{\Sigma}^{\bar{f} f}_{i j} (p),
\label{Chap-Reno:eq:GammaRenFermion}
\ee
and we can use the parameterization \cite{Denner:2019vbn}:
\be
\hat{\Gamma}_{i j}^{\bar{f} f}(p) = \slashed{p} \frac{1-\gamma_{5}}{2} \hat{\Gamma}_{i j}^{f, \mathrm{L}}(p^2)+ \slashed{p} \frac{1+\gamma_{5}}{2} \hat{\Gamma}_{i j}^{f, \mathrm{R}}(p^2)+\frac{1-\gamma_{5}}{2} \hat{\Gamma}_{i j}^{f, \mathrm{l}}(p^2)+\frac{1+\gamma_{5}}{2} \hat{\Gamma}_{i j}^{f, \mathrm{r}}(p^2),
\label{Chap-Reno:eq:GammaRenoFermion}
\ee
which, in the particular case of the renormalized one-loop GFs, reads:
\be
\hat{\Sigma}_{i j}^{\bar{f} f}(p) = \slashed{p} \frac{1-\gamma_{5}}{2} \hat{\Sigma}_{i j}^{f, \mathrm{L}}(p^2)+ \slashed{p} \frac{1+\gamma_{5}}{2} \hat{\Sigma}_{i j}^{f, \mathrm{R}}(p^2)+\frac{1-\gamma_{5}}{2} \hat{\Sigma}_{i j}^{f, \mathrm{l}}(p^2)+\frac{1+\gamma_{5}}{2} \hat{\Sigma}_{i j}^{f, \mathrm{r}}(p^2).
\label{Chap-Reno:eq:SigmaRenoFermion}
\ee

Expanding the Lagrangian using eqs. \ref{Chap-Reno:eq:mass-fermion-expa} to \ref{Chap-Reno:eq:field-fermion-expa}, we find%
\bs
\label{Chap-Reno:eq:SigmaRenoDecompFermion}
\bea
\hat{\Sigma}_{i j}^{f, \mathrm{L}}(p^2) &=& \Sigma_{i j}^{f, \mathrm{L}}(p^2)+\frac{1}{2}\left(\delta Z_{i j}^{f, \mathrm{L}}+\delta Z_{i j}^{f, \mathrm{L}^{\dagger}}\right), \\
\hat{\Sigma}_{i j}^{f, \mathrm{R}}(p^2) &=& \Sigma_{i j}^{f, \mathrm{R}}(p^2)+\frac{1}{2}\left(\delta Z_{i j}^{f, \mathrm{R}}+\delta Z_{i j}^{f, \mathrm{R}^{\dagger}}\right), \\
\hat{\Sigma}_{i j}^{f, \mathrm{l}}(p^2) &=& \Sigma_{i j}^{f, \mathrm{l}}(p^2)-\frac{1}{2}\left(m_{f, i} \delta Z_{i j}^{f, \mathrm{L}}+m_{f, j} \delta Z_{i j}^{f, \mathrm{R}^{\dagger}}\right)-\delta_{i j} \, \delta m_{f, i}^{\mathrm{L}}, \\
\hat{\Sigma}_{i j}^{f, \mathrm{r}}(p^2) &=& \Sigma_{i j}^{f, \mathrm{r}}(p^2)-\frac{1}{2}\left(m_{f, i} \delta Z_{i j}^{f, \mathrm{R}}+m_{f, j} \delta Z_{i j}^{f, \mathrm{L} \dagger}\right)-\delta_{i j} \, \delta m_{f, i}^{\mathrm{R}},
\eea
\es
so that, using eqs. \ref{Chap-Reno:eq:GammaRenFermion}, \ref{Chap-Reno:eq:SigmaRenoFermion} and \ref{Chap-Reno:eq:SigmaRenoDecompFermion}, we obtain:
\bs
\label{Chap-Reno:eq:GammaRenoDecompFermion}
\bea
\hat{\Gamma}_{i j}^{f, \mathrm{L}}(p^2) &=& \delta_{ij} + \Sigma_{i j}^{f, \mathrm{L}}(p^2)+\frac{1}{2}\left(\delta Z_{i j}^{f, \mathrm{L}}+\delta Z_{i j}^{f, \mathrm{L}^{\dagger}}\right), \\
\hat{\Gamma}_{i j}^{f, \mathrm{R}}(p^2) &=& \delta_{ij} + \Sigma_{i j}^{f, \mathrm{R}}(p^2)+\frac{1}{2}\left(\delta Z_{i j}^{f, \mathrm{R}}+\delta Z_{i j}^{f, \mathrm{R}^{\dagger}}\right), \\
\hat{\Gamma}_{i j}^{f, \mathrm{l}}(p^2) &=& - \delta_{ij} m_{f,i} + \Sigma_{i j}^{f, \mathrm{l}}(p^2)-\frac{1}{2}\left(m_{f, i} \delta Z_{i j}^{f, \mathrm{L}}+m_{f, j} \delta Z_{i j}^{f, \mathrm{R}^{\dagger}}\right)-\delta_{i j} \, \delta m_{f, i}^{\mathrm{L}}, \\
\hat{\Gamma}_{i j}^{f, \mathrm{r}}(p^2) &=& - \delta_{ij} m_{f,i} + \Sigma_{i j}^{f, \mathrm{r}}(p^2)-\frac{1}{2}\left(m_{f, i} \delta Z_{i j}^{f, \mathrm{R}}+m_{f, j} \delta Z_{i j}^{f, \mathrm{L} \dagger}\right)-\delta_{i j} \, \delta m_{f, i}^{\mathrm{R}}.
\eea
\es

The OSS renormalization conditions are (no summation in the indices is implicit):%
\fn{The conditions for the spinor $\bar{u}(p)$ do not need to be considered, since they end up containing no new information if the hermiticity of the up-to-one-loop action is used.}
\bs
\bea
\left[
\widetilde{\operatorname{Re}} \left\{\hat{\Gamma}_{i j}^{\bar{f} f}(p)\right\} u_{j}(p)
\right]
\Big|_{\slashed{p} \, u_{j}(p) = m_{f, j} u_{j}(p)} &=& 0, \\
\lim_{\slashed{p} \, u_{i}(p) \to m_{f, i} u_{i}(p)}
\,
\frac{\slashed{p}+m_{f,i}}{p^{2}-m_{f, i}^{2}}\widetilde{\operatorname{Re}} \left[\hat{\Gamma}_{i i}^{\bar{f} f}(p)\right] u_{i}(p) &=& u_{i}(p).
\label{Chap-Reno:eq:fermion-res-cond}
\eea
\es

Inserting eq. \ref{Chap-Reno:eq:GammaRenoFermion}, we have respectively:%
\fn{The derivation of eq. \ref{Chap-Reno:eq:AuxFermionsB} is subtle and requires eqs. \ref{Chap-Reno:eq:AuxFermionsA}. We thank Ansgar Denner for clarifications on this aspect.}
\bs
\label{Chap-Reno:eq:AuxFermions}
\begin{flalign}
&
m_{f, j} \, \widetilde{\operatorname{Re}} \, \hat{\Gamma}_{i j}^{f, \mathrm{L}}\left(m_{f, j}^{2}\right)+\widetilde{\operatorname{Re}} \, \hat{\Gamma}_{i j}^{f, \mathrm{r}}\left(m_{f, j}^{2}\right)=0,
\qquad
m_{f, j} \, \widetilde{\operatorname{Re}} \, \hat{\Gamma}_{i j}^{f, \mathrm{R}}\left(m_{f, j}^{2}\right)+\widetilde{\operatorname{Re}} \, \hat{\Gamma}_{i j}^{f, \mathrm{l}}\left(m_{f, j}^{2}\right)=0, 
\label{Chap-Reno:eq:AuxFermionsA}
\\
&
\widetilde{\operatorname{Re}} \, \Bigg\{\hat{\Gamma}_{i i}^{f, \mathrm{R}}\left(m_{f, i}^{2}\right)+\hat{\Gamma}_{i i}^{f, \mathrm{L}}\left(m_{f, i}^{2}\right)+2 \frac{\partial}{\partial p^{2}} \bigg[m_{f, i}^{2}\left(\hat{\Gamma}_{i i}^{f, \mathrm{R}}\left(p^{2}\right)+\hat{\Gamma}_{i i}^{f, \mathrm{L}}\left(p^{2}\right)\right)
\hs{25mm} \nonumber \\[-3mm]
&
\hs{55mm} +m_{f, i}\left(\hat{\Gamma}_{i i}^{f, \mathrm{r}}\left(p^{2}\right)+\hat{\Gamma}_{i i}^{f, \mathrm{l}}\left(p^{2}\right)\right)\bigg]\bigg|_{p^{2}=m_{f, i}^{2}}\Bigg\}=2.
\label{Chap-Reno:eq:AuxFermionsB}
\end{flalign}
\es
In this case, however, the equations leave some freedom. We show in detail in appendix \ref{App-Fermions} how to derive the counterterms. The results are:
\bs
\label{Chap-Reno:eq:reno-ferm}
\begin{flalign}
&\delta m_{f,i}^{\mathrm{L}}
=
\dfrac{1}{2} \widetilde{\operatorname{Re}} 
\left[
m_{f,i} \, \Sigma_{ii}^{f, \mathrm{L}}(m_{f,i}^2)
+ m_{f,i} \, \Sigma_{ii}^{f, \mathrm{R}}(m_{f,i}^2)
+ 2 \, \Sigma_{ii}^{f, \mathrm{l}}(m_{f,i}^2)
\right], \\[3mm]
&\delta m_{f,i}^{\mathrm{R}}
=
\dfrac{1}{2} \widetilde{\operatorname{Re}} 
\left[
m_{f,i} \, \Sigma_{ii}^{f, \mathrm{L}}(m_{f,i}^2)
+ m_{f,i} \, \Sigma_{ii}^{f, \mathrm{R}}(m_{f,i}^2)
+ 2 \, \Sigma_{ii}^{f, \mathrm{r}}(m_{f,i}^2)
\right], \\[5mm]
&\delta Z_{i i}^{f, \mathrm{L}}
=
-\widetilde{\operatorname{Re}} \, \Sigma_{i i}^{f, \mathrm{L}}\left(m_{f, i}^{2}\right) - m_{f, i} \frac{\partial}{\partial p^{2}} \widetilde{\operatorname{Re}}\Big[m_{f, i}\left(\Sigma_{i i}^{f, \mathrm{L}}(p^2)+\Sigma_{i i}^{f, \mathrm{R}}(p^2)\right) \nonumber \\[-2mm]
&\hs{80mm}+ \Sigma_{i i}^{f, \mathrm{l}}(p^2)+\Sigma_{i i}^{f, \mathrm{r}}(p^2)\Big]\Big|_{p^{2}=m_{f, i}^{2}},
\label{Chap-Reno:eq:reno-ferm-diag-L}
\\
&\delta Z_{i i}^{f, \mathrm{R}}
=
-\widetilde{\operatorname{Re}} \, \Sigma_{i i}^{f, \mathrm{R}}\left(m_{f, i}^{2}\right) - m_{f, i} \frac{\partial}{\partial p^{2}} \widetilde{\operatorname{Re}}\Big[m_{f, i}\left(\Sigma_{i i}^{f, \mathrm{L}}(p^2)+\Sigma_{i i}^{f, \mathrm{R}}(p^2)\right) \nonumber \\[-2mm]
&\hs{80mm}+\Sigma_{i i}^{f, \mathrm{l}}(p^2)+\Sigma_{i i}^{f, \mathrm{r}}(p^2)\Big]\Big|_{p^{2}=m_{f, i}^{2}},
\label{Chap-Reno:eq:reno-ferm-diag-R}
\\
&\delta Z_{i j}^{f, \mathrm{L}}
\stackrel{i \neq j}{=}
\frac{2}{m_{f, i}^{2}-m_{f, j}^{2}} \widetilde{\operatorname{Re}}\Big[m_{f, j}^{2} \Sigma_{i j}^{f, \mathrm{L}}\left(m_{f, j}^{2}\right)+m_{f, i} m_{f, j} \Sigma_{i j}^{f, \mathrm{R}}\left(m_{f, j}^{2}\right) \nonumber \\[-2mm]
&\hs{60mm}+m_{f, i} \Sigma_{i j}^{f, \mathrm{l}}\left(m_{f, j}^{2}\right)+m_{f, j} \Sigma_{i j}^{f, \mathrm{r}}\left(m_{f, j}^{2}\right)\Big],
\label{Chap-Reno:eq:reno-fermB-pre}
\\
&\delta Z_{i j}^{f, \mathrm{R}}
\stackrel{i \neq j}{=}
\frac{2}{m_{f, i}^{2}-m_{f, j}^{2}} \widetilde{\operatorname{Re}}\Big[m_{f, j}^{2} \Sigma_{i j}^{f, \mathrm{R}}\left(m_{f, j}^{2}\right)+m_{f, i} m_{f, j} \Sigma_{i j}^{f, \mathrm{L}}\left(m_{f, j}^{2}\right)
\nonumber \\[-2mm]
&\hs{60mm}+m_{f, j} \Sigma_{i j}^{f, \mathrm{l}}\left(m_{f, j}^{2}\right)+m_{f, i} \Sigma_{i j}^{f, \mathrm{r}}\left(m_{f, j}^{2}\right)\Big].
\label{Chap-Reno:eq:reno-fermB}
\end{flalign}
\es

Finally, in the case of fermions, the hermiticity of the up-to-one-loop action implies:
\be
\widetilde{\operatorname{Re}} \,\Gamma_{ij}^{\bar{f} f}(p) = \gamma^0  \left[ \widetilde{\operatorname{Re}} \, \Gamma_{ij}^{\bar{f} f} (p) \right]^{\dagger} \gamma^0, 
\ee
so that, in particular,
\begin{gather}
\widetilde{\operatorname{Re}} \,\Sigma_{ij}^{f,\mathrm{L}}(p^2) =
\left[
\widetilde{\operatorname{Re}} \,\Sigma_{ji}^{f,\mathrm{L}}(p^2)
\right]^*,
\qquad
\widetilde{\operatorname{Re}} \,\Sigma_{ij}^{f,\mathrm{R}}(p^2)
=
\left[
\widetilde{\operatorname{Re}} \,\Sigma_{ji}^{f,\mathrm{R}}(p^2)
\right]^*,
\no
\widetilde{\operatorname{Re}} \,\Sigma_{ij}^{f,\mathrm{l}}(p^2)
=
\left[
\widetilde{\operatorname{Re}} \,\Sigma_{ji}^{f,\mathrm{r}}(p^2)
\right]^*.
\label{Chap-Reno:eq:apD:30c}
\end{gather}
It follows that $\delta m_{f,i}^{\mathrm{L}}$, $\delta m_{f,i}^{\mathrm{R}}$, $\delta Z_{i j}^{f, \mathrm{L}}$, $\delta Z_{i j}^{f, \mathrm{R}}$ (for $i \neq j$) are in general complex, whereas $\delta Z_{i i}^{f, \mathrm{L}}$ and $\delta Z_{i i}^{f, \mathrm{R}}$ are real.

\subsubsection{Scalar masses and fields}
\label{Chap-Reno:sec:CT-scalar}

In this case, we simply have:%
\fn{For the neutral scalar fields, $\hat{\Gamma}^{S_n S_n^{\prime}} (k^2) = \hat{\Gamma}^{S_n^{\prime} S_n} (k^2)$ by construction.}
\bs
\label{Chap-Reno:eq:GammaRenScalar}
\bea
\hat{\Gamma}^{S_c S_c^{\prime}} (k^2) &=& \delta_{S_c S_c^{\prime}} \left(k^2 - m_{S_c}^2\right) + \hat{\Sigma}^{S_c S_c^{\prime}} (k^2),
\label{Chap-Reno:eq:GammaRenScalarCharged}
\\
\hat{\Gamma}^{S_n S_n^{\prime}} (k^2) &=& \delta_{S_n S_n^{\prime}} \left(k^2 - m_{S_n}^2\right) + \hat{\Sigma}^{S_n S_n^{\prime}} (k^2), 
\label{Chap-Reno:eq:GammaRenScalarNeutral}
\eea
\es
where $S_c, S_c^{\prime} = \{H^+, G^+\}$, $S_n, S_n^{\prime} = \{h_1, h_2, h_3, G_0\}$, and $m_{S_n}^2 = m_{3\mathrm{R}}^2$ for $S_n = h_3$.%
\fn{Moreover, we are taking $m_{G_0} = m_{G^{+}} = 0$ (and also $\delta m_{G_0}^{2} = \delta m_{G^{+}}^{2} = 0$ in eq. \ref{Chap-Reno:eq:SigmaRenoDecompScalar}).
The would-be Goldstone boson fields are not physical, so that they cannot be OS; this means that, strictly speaking, one cannot apply the OSS renormalization conditions to GFs with external would-be Goldstone boson fields. Yet, since it will prove convenient, we extend these conditions by supposing that those fields are physical, with masses $m_{G_0} = m_{G^{+}} = 0$.}
Expanding the Lagrangian with eqs. \ref{Chap-Reno:eq:mass-scalar-expa} to \ref{Chap-Reno:eq:field-scalar-expa}, we find:
\bs
\label{Chap-Reno:eq:SigmaRenoDecompScalar}
\bea
\hat{\Sigma}^{S_c S_c^{\prime}}(k^2) &=& \Sigma^{S_c S_c^{\prime}}(k^2)  +\frac{1}{2}\left(k^{2}-m_{S_c^{\prime}}^{2}\right) \delta Z_{S_c^{\prime} S_c} \nonumber \\[-1mm]
&& \hs{35mm} + \frac{1}{2}\left(k^{2}-m_{S_c}^{2}\right) \delta Z_{S_c S_c^{\prime}}^* -\delta_{S_c S_c^{\prime}} \, \delta m_{S_c}^{2},
\label{Chap-Reno:eq:SigmaRenoDecompScalarCharged}\\
\hat{\Sigma}^{S_n S_n^{\prime}}(k^2) &=& \Sigma^{S_n S_n^{\prime}}(k^2)  +\frac{1}{2}\left(k^{2}-m_{S_n^{\prime}}^{2}\right) \delta Z_{S_n^{\prime} S_n} \nonumber \\[-1mm]
&& \hs{35mm} + \frac{1}{2}\left(k^{2}-m_{S_n}^{2}\right) \delta Z_{S_n S_n^{\prime}}
-\delta_{S_n S_n^{\prime}} \, \delta m_{S_n}^{2}.
\label{Chap-Reno:eq:SigmaRenoDecompScalarNeutral}
\eea
\es
We start by considering the charged fields. The OSS renormalization conditions are:
\bs
\bea
\left.\widetilde{\operatorname{Re}} \, \hat{\Gamma}^{S_c^{*} S_c^{\prime}}(k)\right|_{k^{2}=m_{S_c}^{2}} &=& 0,\\
\lim _{k^{2} \to m_{S_c}^{2}} \frac{1}{k^{2}-m_{S_c}^{2}} \widetilde{\operatorname{Re}} \, \hat{\Gamma}^{S_c^* S_c}(k) &=& 1,
\eea
\es
Inserting eq. \ref{Chap-Reno:eq:GammaRenScalarCharged}, we get respectively:
\bs
\label{Chap-Reno:eq:auxscalarcharged}
\bea
\widetilde{\operatorname{Re}} \, \hat{\Sigma}^{S_c^{*} S_c^{\prime}}\left(m_{S_c}^{2}\right) &=& 0,
\label{Chap-Reno:eq:AuxScalarCharged1}
\\
\left.\quad \widetilde{\operatorname{Re}} \, \frac{\partial \hat{\Sigma}^{S_c^* S_c}\left(k^{2}\right)}{\partial k^{2}}\right|_{k^{2} = m_{S_c}^{2}} &=& 0,
\label{Chap-Reno:eq:AuxScalarCharged2}
\eea
\es
and, using eq. \ref{Chap-Reno:eq:SigmaRenoDecompScalarCharged}, we obtain:
\bs
\bea
\hs{-5mm}
\delta m_{\mathrm{H}^{+}}^2 &=& \widetilde{\operatorname{Re}} \, \Sigma^{H^+H^-}(m_{\mathrm{H}^{+}}^2), \\
\hs{-5mm}
\begin{pmatrix}
\delta Z_{G^{+}G^{+}} & \delta Z_{G^{+}H^{+}}\\
\delta Z_{H^{+}G^{+}} & \delta Z_{H^{+}H^{+}}
\end{pmatrix}
&=&
\begin{pmatrix}
-\left.\widetilde{\operatorname{Re}} \, \dfrac{\partial \Sigma^{G^{+}G^{+}}\left(k^{2}\right)}{\partial k^{2}}\right|_{k^{2}=0}
&
- 2 \,\widetilde{\operatorname{Re}}  \dfrac{\Sigma^{H^{+}G^{+}}\left(m_{\mathrm{H}^{+}}^2\right)}{m_{\mathrm{H}^{+}}^2}
\\[4mm]
2 \, \widetilde{\operatorname{Re}}  \dfrac{\Sigma^{G^{+}H^{+}}\left(0\right)}{m_{\mathrm{H}^{+}}^2}
&
-\left.\widetilde{\operatorname{Re}} \, \dfrac{\partial \Sigma^{H^{+}H^{+}}\left(k^{2}\right)}{\partial k^{2}}\right|_{k^{2}=m_{\mathrm{H}^{+}}^2}
\end{pmatrix}.
\eea
\es
The case of the neutral fields is more subtle, and is considered in detail in appendix \ref{App-OSS}. The independent counterterms are:
\bs
\begin{gather}
\delta m_1^2 = \widetilde{\operatorname{Re}} \, \Sigma^{h_1h_1}(m_1^2), \qquad
\delta m_2^2 = \widetilde{\operatorname{Re}} \, \Sigma^{h_2h_2}(m_2^2),
\\[5mm]
\delta Z_{S_n S_n} = -\left.\widetilde{\operatorname{Re}} \, \frac{\partial \Sigma^{S_nS_n}\left(k^{2}\right)}{\partial k^{2}}\right|_{k^{2}=m_{S_n}^{2}},
\qquad
\delta Z_{S_n^{\prime} S_n} \stackrel{S_n^{\prime} \neq S_n}{=}  2 \, \frac{\widetilde{\operatorname{Re}}  \, \Sigma^{S_n S_n^{\prime}}\left(m_{S_n}^{2}\right)}{m_{S_n^{\prime}}^2 - m_{S_n}^2},
\label{Chap-Reno:eq:CT-scalar-neutral}
\end{gather}
\es
and the pole squared mass $m_{3\mathrm{P}}^2$ is given by:
\be
m_{3\mathrm{P}}^2 = m_{\mathrm{3R}}^2 - \widetilde{\operatorname{Re}} \, \Sigma^{h_3h_3}(m_{3\mathrm{R}}^2) + \delta m_3^2.
\ee

The hermiticity of the up-to-one-loop action implies in this case:
\be
\widetilde{\operatorname{Re}} \, \Gamma^{S_c^{\prime} S_c}(k)
=
\left[\widetilde{\operatorname{Re}} \, \Gamma^{S_c S_c^{\prime}}(k)\right]^*,
\quad
\widetilde{\operatorname{Re}} \, \Gamma^{S_n S_n^{\prime}}(k)
=
\left[\widetilde{\operatorname{Re}} \, \Gamma^{S_n S_n^{\prime}}(k)\right]^*.
\ee
so that, in particular,
\be
\widetilde{\operatorname{Re}}  \, \Sigma^{S_c^{\prime} S_c}\left(k^2\right)
=
\left[\widetilde{\operatorname{Re}}  \, \Sigma^{S_c S_c^{\prime}}\left(k^2\right)\right]^*,
\quad
\widetilde{\operatorname{Re}}  \, \Sigma^{S_n S_n^{\prime}}\left(k^2\right)
=
\left[\widetilde{\operatorname{Re}}  \, \Sigma^{S_n S_n^{\prime}}\left(k^2\right)\right]^*.
\label{Chap-Reno:eq:ConsHermSigmaScalar}
\ee
Therefore, the counterterms $\delta Z_{G^{+}H^{+}}$ and $\delta Z_{H^{+}G^{+}}$ are in general complex, while the remaining counterterms of this subsection are real.

\subsubsection{Electric charge}
\label{Chap-Reno:sec:charge}

The counterterm for the electric charge is also fixed through OSS; this involves considering the interaction between electrons and photon, on the one hand, and requiring that, for OS electrons and for the \textit{Thomson limit} (i.e. for vanishing photon momentum), the full (renormalized) interaction is given by the tree-level one, on the other.
Or, which is equivalent, the OSS scheme imposes that higher orders do not contribute to the interaction $A_{\mu}\bar{e}e$ in the particular limit just mentioned.
Besides the vertex (or 1PI) functions, the full interaction in general involves LSZ factors---c.f. appendix \ref{App-LSZ}. However, as discussed in that appendix, the field counterterms in OSS are chosen in such a way that those factors become trivial, and can be ignored.
%
The OSS condition then becomes:
\be
\bar{u}(p) \, \,
i \hat{\Gamma}^{A\bar{e}e}_{\mu}(0,-p,p) \Big|_{\text{1L}}
\, u(p) = 0,
\label{Chap-Reno:eq:OSS-vert-cond}
\ee
where 1L indicates the one-loop contribution.
The reason why this condition allows to fix $\delta Z_e$ is that $\hat{\Gamma}^{A\bar{f}f}_{\mu}(0,-p,p) \Big|_{\text{1L}}$, which is a renormalized GF, depends on
that counterterm.%
\fn{Cf. eqs. \ref{App-WI:eq:vert:boneco} and \ref{App-WI:eq:vert:conta}.}
Hence, $\delta Z_e$  becomes fixed (i.e. it is fixed precisely so that eq. \ref{Chap-Reno:eq:OSS-vert-cond} is verified). However, that results in a quite complicated expression. It turns out that there is a Ward identity that substantially simplifies the result, as it leads to a simple expression that relates $\delta Z_e$ to other counterterms. A detailed derivation of both the Ward identity and the expression for $\delta Z_e$ are provided in appendix \ref{App-WI}. The final result is:
\be
\delta Z_{e} = -\frac{1}{2} \delta Z_{A A} + \frac{s_{\text{w}}}{c_{\text{w}}} \frac{1}{2} \delta Z_{Z A}=\left.\frac{1}{2} \frac{\partial \Sigma_{\mathrm{T}}^{A A}\left(k^{2}\right)}{\partial k^{2}}\right|_{k^{2}=0} + \frac{s_{\text{w}}}{c_{\text{w}}} \frac{\Sigma_{\mathrm{T}}^{A Z}(0)}{m_{\mathrm{Z}}^{2}}.
\label{Chap-Reno:eq:deltaZe}
\ee

\subsubsection{CKM matrix}
\label{Chap-Reno:sec:CKM}

For the CKM matrix, we follow the prescription originally presented in ref.~\cite{Denner:1990yz}. We fix the expression in the Feynman gauge, which will be justified in section \ref{Chap-Reno:sec:gauge-dep} below. We then have:
\be
\delta V_{i j}=\frac{1}{4} \left.\bigg[\left(\delta Z_{i k}^{u, \mathrm{L}}-\delta Z_{i k}^{u, \mathrm{L} \dagger}\right) V_{k j}-V_{i k}\left(\delta Z_{k j}^{d, \mathrm{L}}-\delta Z_{k j}^{d, \mathrm{L} \dagger}\right)\bigg]\right|_{\xi=1},
\ee
where $\xi=1$ represents the Feynman gauge.

\subsubsection{Mixing parameters}
\label{Chap-Reno:sec:RenoMixingParameters}

The renormalization of mixing parameters has been recurrently discussed in recent years \cite{Kanemura:2004mg, Kanemura:2015mxa, Krause:2016oke, Altenkamp:2017ldc, Denner:2016etu, Denner:2017vms, Denner:2018opp, Krause:2017mal}, especially in ref.~\cite{Denner:2018opp}.%
\fn{Mixing parameters usually reduce to mixing \textit{angles}. Yet, in the present model (and due to CP violation), there are not only mixing angles (like $\beta$), but also mixing phases (like $\zeta_a$). Hence, we promote the designation `mixing angles' to `mixing parameters'.}
In the latter, several methods for fixing counterterms for mixing parameters are proposed and analysed.
In the present work, we consider one of those methods, which involves fixing the independent counterterms for mixing parameters using symmetry relations. This method is simple and leads to well behaved results, as shall be discussed below.
A description of the procedure, as well as the derivation of the expressions for the counterterms for mixing parameters of the C2HDM, can be found in appendix \ref{App-Sym}.
In this section, we present the results therein derived. For the charged sector,
\bs
\label{Chap-Reno:eq:rela-charged-real}
\bea
{\delta \beta}
&=& \dfrac{1}{4}
\operatorname{Re}
\left.\Big[\delta Z_{G^+H^+} - \delta Z_{H^+G^+}\Big]
\right|_{\xi=1},
\\
\delta \zeta_a
&=&
-\dfrac{1}{2} \cot(2 \beta) \operatorname{Im} \left[\delta Z_{G^+H^+} \right]\big|_{\xi=1}.
\eea
\es
For the neutral sector, the set of independent counterterms for mixing parameters depends on the combination $C_i$ (recall eq. \ref{Chap-Reno:eq:CT-params-H}). There are three counterterms which are independent in all the combinations:
\bs
\begin{flalign}
&
\delta \alpha_1
=
\dfrac{1}{4} \sec(\alpha_2) \left. \Big[  \cos(\alpha_3) \left( \delta Z_{h_1h_2} - \delta Z_{h_2h_1} \right)  +  \sin(\alpha_3) \left( \delta Z_{h_3h_1} -\delta Z_{h_1h_3} \right)  \Big] \right|_{\xi=1},
\\[3mm]
&
\delta \alpha_2
= 
\left.
\dfrac{1}{4}  \sin(\alpha_3) \Big[ \delta Z_{h_1h_2} - \delta Z_{h_2h_1} + \cot(\alpha_3) \left( \delta Z_{h_1h_3} - \delta Z_{h_3h_1} \right)  \Big] \right|_{\xi=1},
\\[3mm]
&
\delta \alpha_3
=
\dfrac{1}{4} \bigg[\delta Z_{h_2h_3} - \delta Z_{h_3h_2} -   \cos(\alpha_3) \tan(\alpha_2)  \left( \delta Z_{h_1h_2} - \delta Z_{h_2h_1} \right) \nonumber\\[-3mm]
& \hs{50mm} + \sin(\alpha_3) \tan(\alpha_2) \left( \delta Z_{h_1h_3} - \delta Z_{h_3h_1} \right)  \bigg] \bigg|_{\xi=1}.
\end{flalign}
\es
Then, the combinations $C_1$, $C_2$ and $C_3$ have $\delta \alpha_5$, $\delta \alpha_4$ and $\delta \alpha_0$ as independent, respectively, given by:%
\fn{The notation $\stackrel{C_i}{=}$ clarifies which combination $C_i$ the expression is applicable to. For example, eq. \ref{Chap-Reno:eq:indep-dalfa5} is only valid if $C_1$ is chosen; in other combinations, $\delta \alpha_5$ will be a dependent counterterm, so that eq. \ref{Chap-Reno:eq:indep-dalfa5} is not valid.
Finally, note that $\delta \alpha_5$, $\delta \alpha_4$ and $\delta \alpha_0$ are all dependent counterterms in the combination $C_4$.}
\bs
\label{Chap-Reno:eq:indep-dalfas}
\begin{flalign}
\label{Chap-Reno:eq:indep-dalfa5}
&
\delta \alpha_5
\stackrel{C_1}{=}
\dfrac{1}{4} \bigg[\delta Z_{h_2G_0} -\delta Z_{G_0h_2} + \tan(\alpha_3) \left( \delta Z_{G_0h_3} - \delta Z_{h_3G_0} \right)
\bigg]
\bigg|_{\xi=1}
,
\\[3mm]
&
\label{Chap-Reno:eq:indep-dalfa4}
\delta \alpha_4
\stackrel{C_2}{=}
\dfrac{1}{4} \bigg[\delta Z_{h_1G_0} -\delta Z_{G_0h_1} + \sec(\alpha_3) \tan(\alpha_2) \left( \delta Z_{G_0h_3} - \delta Z_{h_3G_0} \right)  \bigg]
\bigg|_{\xi=1},
\\[3mm]
&
\label{Chap-Reno:eq:indep-dalfa0}
\delta \alpha_0
\stackrel{C_3}{=}
\dfrac{1}{4} \sec(\alpha_2) \sec(\alpha_3)  \big( \delta Z_{G_0h_3} - \delta Z_{h_3G_0} \big)
\big|_{\xi=1}.
\end{flalign}
\es

\subsubsection{Remaining parameters}
\label{Chap-Reno:sec:remaining}

By now, all the field counterterms are determined; as for the independent parameter counterterms, considering eq. \ref{Chap-Reno:eq:CT-params-H} we see that we still need to fix $\delta \mu^2$ and $\delta \zeta_2$ (both required for all combinations $C_i$) and, in the specific case of the combination $C_4$, also $\delta \zeta_1$.
%
%
We calculate the latter and $\delta \mu^2$ in the $\overline{\text{MS}}$ scheme.
%
%
As described above, this requires selecting a process where the counterterm to be fixed intervenes.
For $\delta \mu^2$, we choose $h_3 \to H^+H^-$. Here, the total counterterm is such that:
\be
\vs{3mm}
\begin{minipage}[h]{.35\textwidth}
\begin{picture}(0,70)
\begin{fmffile}{h3HPHMCTb} 
\begin{fmfgraph*}(70,70) 
\fmfset{arrow_len}{3mm} 
\fmfset{arrow_ang}{20} 
\fmfleft{nJ1} 
\fmfright{nJ2,nJ4} 
\fmf{dashes,label=$h_3$,label.side=left,tension=3,label.dist=3thick}{nJ1,nJ2nJ4J2}
\fmf{scalar,label=$H^-$,label.side=right,tension=3}{nJ2nJ4J2,nJ2} 
\fmf{scalar,label=$H^+$,label.side=left,tension=3}{nJ2nJ4J2,nJ4} 
\fmfv{decor.shape=pentagram,decor.filled=full,decor.size=6thick}{nJ2nJ4J2}
\end{fmfgraph*} 
\end{fmffile}
\end{picture}
\end{minipage}
\hs{-30mm}
\ni 
\hs{1mm}
i \dfrac{e}{2 m_{\mathrm{W}} s_{\text{w}} c_{\beta} s_{\beta}}
\left(s_{\beta} R_{31} + c_{\beta} R_{32}\right) \delta \mu^2
.
\vspace{-2mm}
\ee
Then, by requiring the renormalized 3-point function to be such that the terms proportional to $\Delta_{\varepsilon}$ are zero,
\vspace{2mm}
\be
i \hat{\Gamma}^{h_3 H^+ H^-}\Big|_{\Delta_{\varepsilon}}
=
\hs{2mm}
\left.
\begin{bmatrix}
\begin{minipage}[h]{.35\textwidth}
\begin{picture}(0,70)
\begin{fmffile}{h3HPHM1L} 
\begin{fmfgraph*}(70,70) 
\fmfset{arrow_len}{3mm} 
\fmfset{arrow_ang}{20} 
\fmfleft{nJ1} 
\fmfright{nJ2,nJ4} 
\fmf{dashes,label=$h_3$,label.side=left,tension=3,label.dist=3thick}{nJ1,nJ2nJ4J2}
\fmf{scalar,label=$H^-$,label.dist=3,tension=3}{nJ2nJ4J2,nJ2} 
\fmf{scalar,label=$H^+$,label.dist=3,label.side=right,tension=3}{nJ2nJ4J2,nJ4} 
\fmfv{decor.shape=circle,decor.filled=hatched,decor.size=11thick}{nJ2nJ4J2}
\end{fmfgraph*} 
\end{fmffile}
\end{picture}
\end{minipage}
\hs{-23mm}
+
\begin{minipage}[h]{.35\textwidth}
\begin{picture}(0,70)
\begin{fmffile}{h3HPHM1LB} 
\begin{fmfgraph*}(70,70)
\fmfset{arrow_len}{3mm}
\fmfset{arrow_ang}{20}
\fmfleft{nJ1} 
\fmfright{nJ2,nJ4} 
\fmftop{nJ3}
\fmf{dashes,label=$h_3$,label.side=right,tension=35}{nJ1,nJ1nJ2nJ4J4} 
\fmf{scalar,label=$H^-$,label.side=left,label.dist=3,tension=25}{nJ1nJ2nJ4J4,nJ2} 
\fmf{scalar,label=$H^+$,label.side=right,label.dist=3,tension=25}{nJ1nJ2nJ4J4,nJ4}
\fmf{dashes,tension=0.1}{nJ1nJ2nJ4J4,x}
\fmf{dashes,label=$h_i$,label.dist=3,label.side=left,tension=0.1}{x,y}
\fmf{dashes,tension=0.08}{y,nJ3}
\fmfv{decor.shape=circle,decor.filled=70,decor.size=1thick}{x}
\fmfv{decor.shape=circle,decor.filled=70,decor.size=1thick}{y}
\fmfv{decor.shape=circle,decor.filled=hatched,decor.size=9thick}{nJ3}
\end{fmfgraph*}
\end{fmffile}
\end{picture}
\end{minipage}
\hs{-23mm}
+
\hs{1mm}
\begin{minipage}[h]{.35\textwidth}
\begin{picture}(0,70)
\begin{fmffile}{h3HPHMCT} 
\begin{fmfgraph*}(70,70) 
\fmfset{arrow_len}{3mm} 
\fmfset{arrow_ang}{20} 
\fmfleft{nJ1} 
\fmfright{nJ2,nJ4} 
\fmf{dashes,label=$h_3$,label.side=left,tension=4,label.dist=3thick}{nJ1,nJ2nJ4J2}
\fmf{scalar,label=$H^-$,label.side=right,label.dist=3,tension=3}{nJ2nJ4J2,nJ2} 
\fmf{scalar,label=$H^+$,label.side=left,label.dist=3,tension=3}{nJ2nJ4J2,nJ4} 
\fmfv{decor.shape=pentagram,decor.filled=full,decor.size=6thick}{nJ2nJ4J2}
\end{fmfgraph*} 
\end{fmffile}
\end{picture}
\end{minipage}
\hs{-28mm}
\hphantom{.}
\end{bmatrix}
\right|_{\Delta_{\varepsilon}}
= 0 ,
\ee
%
%
%
%
we fix $\delta \mu^2$. Concerning $\delta \zeta_1$ in $C_4$, we choose $h_3 \to  \bar{\tau} \tau$. Knowing that:
\be
\vs{3mm}
\begin{minipage}[h]{.35\textwidth}
\begin{picture}(0,70)
\begin{fmffile}{h3tTCTb} 
\begin{fmfgraph*}(70,70) 
\fmfset{arrow_len}{3mm} 
\fmfset{arrow_ang}{20} 
\fmfleft{nJ1} 
\fmfright{nJ2,nJ4} 
\fmf{dashes,label=$h_3$,label.side=left,tension=3,label.dist=3thick}{nJ1,nJ2nJ4J2}
\fmf{fermion,label=$\tau$,label.side=right,tension=3}{nJ2nJ4J2,nJ2} 
\fmf{fermion,label=$\bar{\tau}$,tension=3}{nJ4,nJ2nJ4J2} 
\fmfv{decor.shape=pentagram,decor.filled=full,decor.size=6thick}{nJ2nJ4J2}
\end{fmfgraph*} 
\end{fmffile}
\end{picture}
\end{minipage}
\hs{-30mm}
\ni 
\hs{1mm}
i \, \delta \zeta_1 \left(\dfrac{e \, m_{\tau} \, R_{33} \tan \beta}{2 m_{\mathrm{W}} s_{\text{w}}} + ...\right),
\ee
where the ellipses represent terms proportional to $\gamma_5$, 
we fix $\delta \zeta_1$ in $C_4$ by imposing:
\be
\vs{3mm}
i \hat{\Gamma}^{h_3  \bar{\tau} \tau}\Big|_{\Delta_{\varepsilon}}
=
\hs{2mm}
\left.
\begin{bmatrix}
\begin{minipage}[h]{.35\textwidth}
\begin{picture}(0,70)
\begin{fmffile}{h3tT1L} 
\begin{fmfgraph*}(70,70) 
\fmfset{arrow_len}{3mm} 
\fmfset{arrow_ang}{20} 
\fmfleft{nJ1} 
\fmfright{nJ2,nJ4} 
\fmf{dashes,label=$h_3$,label.side=left,tension=3,label.dist=3thick}{nJ1,nJ2nJ4J2}
\fmf{fermion,label=$\tau$,tension=3}{nJ2nJ4J2,nJ2} 
\fmf{fermion,label=$\bar{\tau}$,tension=3}{nJ4,nJ2nJ4J2} 
\fmfv{decor.shape=circle,decor.filled=hatched,decor.size=11thick}{nJ2nJ4J2}
\end{fmfgraph*} 
\end{fmffile}
\end{picture}
\end{minipage}
\hs{-28mm}
+
\hs{5mm}
\begin{minipage}[h]{.35\textwidth}
\begin{picture}(0,70)
\begin{fmffile}{h3tTCT} 
\begin{fmfgraph*}(70,70) 
\fmfset{arrow_len}{3mm} 
\fmfset{arrow_ang}{20} 
\fmfleft{nJ1} 
\fmfright{nJ2,nJ4} 
\fmf{dashes,label=$h_3$,label.side=left,tension=3,label.dist=3thick}{nJ1,nJ2nJ4J2}
\fmf{fermion,label=$\tau$,label.side=right,tension=3}{nJ2nJ4J2,nJ2} 
\fmf{fermion,label=$\bar{\tau}$,tension=3}{nJ4,nJ2nJ4J2} 
\fmfv{decor.shape=pentagram,decor.filled=full,decor.size=6thick}{nJ2nJ4J2}
\end{fmfgraph*} 
\end{fmffile}
\end{picture}
\end{minipage}
\hs{-28mm}
\hphantom{.}
\end{bmatrix}
\right|_{\Delta_{\varepsilon}}
\stackrel{C_4}{=} 0 .
\label{Chap-Reno:eq:delta-zeta1}
\ee

As for $\delta \zeta_2$, we checked that its divergent part is related to that of $\delta \zeta_1$ via:%
\fn{The divergent part of $\delta \zeta_2$ can be easily calculated by requiring e.g. the process $h_3 \to \bar{u} u$ to be finite.}
\be
\delta \zeta_2 \big|_{\Delta_{\varepsilon}} = -\cot^2(\beta) \,  \delta \zeta_1 \big|_{\Delta_{\varepsilon}}.
\label{Chap-Reno:eq:rel-z1-z2-divs}
\ee
Then, similarly to what can be done with symmetry relations (cf. appendix \ref{App-Sym}), we fix $\delta \zeta_2$ as a whole by requiring that eq. \ref{Chap-Reno:eq:rel-z1-z2-divs} also holds for the finite parts. That is, we fix $\delta \zeta_2$ by imposing:%
\fn{Eq. \ref{Chap-Reno:eq:rel-z1-z2-divs} thus provides a simple prescription to fix $\delta \zeta_2$. This simplicity is the reason why we chose this counterterm as independent in all the four combinations discussed above.}
%
%
\be
\delta \zeta_2 = -\cot^2(\beta) \, \delta \zeta_1 \big|_{\xi=1}.
\label{Chap-Reno:eq:rel-z1-z2-fin}
\ee
Note that this condition applies to all four combinations $C_i$. Accordingly, $\delta \zeta_2$ will have a general finite part in the first three combinations, whereas in $C_4$ it will be proportional to $\Delta_{\varepsilon}$ only.

\subsection{Gauge dependence}
\label{Chap-Reno:sec:gauge-dep}


Physical observables cannot depend on the gauge chosen, which implies that $S$-matrix elements must be gauge independent.
In a renormalized up-to-one-loop process, both the (non-renormalized) one-loop diagrams and the counterterms are in general gauge dependent.
Now, it can be shown that, if parameter counterterms are gauge independent, the gauge dependences of the one-loop diagrams and those of the field counterterms precisely cancel, thus rendering the $S$-matrix elements gauge independent \cite{Bohm:2001yx}.
Conversely, gauge dependent parameter counterterms in general lead to gauge dependent $S$-matrix elements. 
Motivated by this, several authors have insisted in recent years on the importance of finding gauge independent parameter counterterms \cite{Krause:2016oke,Krause:2017mal,Denner:2016etu,Kanemura:2017wtm}.

However, as noted in ref.~\cite{Denner:2018opp}, the circumstance that parameter counterterms are gauge dependent does not by itself ruin the gauge independence of $S$-matrix elements. In fact, all that is required for $S$-matrix elements to be gauge independent is that parameter counterterms are \textit{considered} gauge independent. In order to ensure the gauge independence of $S$-matrix elements, indeed, it does not really matter whether the parameter counterterms are truly gauge independent or not; all what is required is that they are considered gauge independent, i.e. that they do not change when the gauge is changed \cite{Denner:2018opp,Bohm:2001yx}.%
\fn{We thank Ansgar Denner for clarifications on this point.
}
But this can be easily implemented by calculating a gauge dependent counterterm in a particular gauge. When calculating observables in other gauges, one must stick to the value of the parameter counterterm in the gauge it was calculated in. Then, the gauge dependence of field counterterms cancels the gauge dependence of the one-loop diagrams, and $S$-matrix elements will be gauge independent.
In face of this, there does not seem to be a clear advantage of true gauge independence in parameter counterterms over considered gauge independence. In both scenarios, $S$-matrix elements end up being gauge-independent, and both lead to correct relations between physical observables.%
\fn{It goes without saying that, when using considered gauge independent counterterms, one has to make sure that no inconsistencies are introduced (i.e. one must make sure not to change the gauge in the gauge dependent parameter counterterms). In this sense, truly gauge independent counterterms may be preferable.
It has been claimed that true gauge independence is a desirable property, as it allows a ``more physical interpretation'' of non-observable parameters \cite{Freitas:2002um}. However, a non-observable parameter is always non-physical anyway; the fact that it is truly gauge independent does not render it any more physical (in whatever physically meaningful way).}

Now, this procedure is obviously irrelevant when parameter counterterms are truly gauge independent. In the FJTS, this happens for parameter counterterms fixed in either the OSS or the $\overline{\text{MS}}$ schemes (recall section \ref{Chap-Selec:sec:gauge-dep-app}).
However, the method described in the previous paragraph is very convenient for the remaining independent parameter counterterms, which in this model (and according to our choices) consist of those for the CKM matrix elements, those for the independent mixing parameters, and $\delta \zeta_2$ (sections \ref{Chap-Reno:sec:CKM}, \ref{Chap-Reno:sec:RenoMixingParameters} and \ref{Chap-Reno:sec:remaining} above, respectively).
We thus applied the method in those cases, following the suggestion proposed in ref.~\cite{Denner:2018opp} of selecting the Feynman gauge as the gauge in which the counterterms are calculated, which avoids artificially large parameters.


\section{Results}
\label{Chap-Reno:sec:num-res-main}

We now study the behaviour of the different combinations $C_i$ introduced above. This agenda is quite novel in studies of renormalization, due to the mere existence of combinations. In fact, there is usually no such thing as different combinations of independent parameters, as the number of independent renormalized parameters usually equals the number of independent counterterms. The studies of the renormalization of models thus tend to be focused either on the choice of different subtraction schemes for the counterterms, or on the choice of different renormalization scales.
%
%

In our case, the fact that there are more independent counterterms than independent renormalized parameters means that, for the same set of the latter, there can be different combinations of the former. Our goal, then, is to compare those combinations, ascertaining how the choice of one or the other combination (i.e. the choice of one or the other set of independent counterterms) affects the predictions of the theory at NLO.%
\fn{In this paper, we restrict ourselves to the four combinations $C_i$ implied in eq. \ref{Chap-Reno:eq:CT-params-H} and, for a certain combination, to the subtraction schemes proposed in section \ref{Chap-Reno:sec:calculation-CTs}.}
%
To do that, we consider three specific NLO processes: the decays of $h_2$ to $ZZ$, $h_1 Z$ and $h_1 h_1$.
%

In the present section, we start by describing both the computational tools and the simulation procedure used to study these decays. Then, we discuss how the decays are influenced by the different combinations $C_i$. Finally, after presenting the expressions for the one-loop decay widths, we show numerical results that allow to compare the different combinations.

\subsection{Computational tools and simulation procedure}

The generation of Feynman rules for both the tree-level and the counterterm interactions, the drawing of the Feynman rules and diagrams, and the calculation of one-loop amplitudes, counterterms and one-loop decay widths were all performed with \FMT. These different tasks are discussed in detail in appendix \ref{App-FM-C2HDM}.
\FMTS was also used for two additional tasks.
The first was a confirmation that the model is renormalized; by considering processes of the different sectors, we numerically checked that all the counterterms calculated in section \ref{Chap-Reno:sec:calc} lead to finite results.
The second task was a conversion of the results to \t{\ts{Fortran}}, which were then numerically evaluated using \ts{LoopTools}~\cite{Hahn:1998yk}.%
\fn{This conversion is highly non-trivial, involving millions of lines of code; for details, see appendix \ref{App-FM-C2HDM}.}

For the scatter plots of section \ref{Chap-Reno:sec:num-res} below, we 
essentially followed the procedure outlined in section \ref{Chap-Maggie:sec:rest}, with four differences.
First, we assumed that $h_1$ corresponds to the SM-like Higgs boson, such that $m_1 = 125 \, \, \textrm{GeV}$.
Second, for the experiments concerning the different signals for the SM-like Higgs boson, we used the more recent reference \cite{Aad:2019mbh}. 
Third, we used a more recent version of \textsc{HiggsBounds}, namely \textsc{HiggsBounds5} \cite{Bechtle:2020pkv}.
Finally, for the EDM, we considered the most recent result from the ACME collaboration, $|d_e| < 1.1 \times 10^{-29} e \mbox{ cm}$ ($90\%$ confidence level) \cite{Andreev:2018ayy}.

\subsection{The influence of the different combinations}
\label{Chap-Reno:sec:influ}

The renormalized NLO amplitude for the process $j$ (with $j = \{ h_2 \to ZZ, h_2 \to h_1Z, h_2 \to h_1h_1\}$),
which we represent as $\hat{\mathcal{M}}_j$, can generically be written as:
\be
\hat{\mathcal{M}}_j = \mathcal{M}^{\mathrm{tree}}_j + \hat{\mathcal{M}}^{\mathrm{loop}}_j,
\ee
where $\mathcal{M}^{\mathrm{tree}}_j$ and $\hat{\mathcal{M}}^{\mathrm{loop}}_j$ respectively represent the tree-level amplitude and the renormalized one-loop amplitude, the latter of which can in turn be split according to:
\be
\hat{\mathcal{M}}^{\mathrm{loop}}_j =
 \mathcal{M}^{\mathrm{loop}}_j + \mathcal{M}^{\mathrm{CT}}_j,
\ee
where the terms on the right-hand side represent the non-renormalized one-loop amplitude and the counterterms, respectively.
Here, $\mathcal{M}^{\mathrm{CT}}_j$ includes all the counterterms that, after expanding the bare quantities into renormalized plus counterterm, end up contributing to the process $j$. Now, the set of counterterms that constitute $\mathcal{M}^{\mathrm{CT}}_j$ is just a consequence of such an expansion, and is the same for all the combinations $C_i$. The differences between the combinations begin
when the counterterms are to be given an expression,
since the latter in general depend on the combination. For example, suppose that $\delta \zeta_1$ contributes to $\mathcal{M}^{\mathrm{CT}}_j$; recalling eq. \ref{Chap-Reno:eq:CT-params-H}, it is clear that the expression for this counterterm depends on $C_i$. Indeed, in $C_4$, $\delta \zeta_1$ is an independent counterterm, so that it is determined by eq. \ref{Chap-Reno:eq:delta-zeta1}. In the remaining combinations, by contrast, $\delta \zeta_1$ is dependent, which means that it is determined by a function (which in general depends on the particular combination at stake) of independent counterterms. As a consequence, $\mathcal{M}^{\mathrm{CT}}_j$ will in general have different values according to the combination $C_i$ chosen. In this way, the different $C_i$ in general lead to different predictions for the renormalized NLO amplitudes.

At this point, it is worth noting that, in a renormalized function such that all its counterterms are fixed through OSS or other momentum-dependent subtraction schemes, the dependence on the renormalization scale $\mu_{\mathrm{R}}$ drops out.%
\fn{A simple way to see this is to note that, in a one-loop integral calculated through dimensional regularization, the UV divergent term ($\frac{2}{\varepsilon}$) and the renormalization scale dependent term ($\ln\mu_{\mathrm{R}}^2$) always come together ($\frac{2}{\varepsilon} + \ln\mu_{\mathrm{R}}^2$). Now, a momentum-dependent subtraction scheme is such that the counterterms are defined as a function of a one-loop integral evaluated at a specific momentum. Thus, since neither $\frac{2}{\varepsilon}$ nor $\ln\mu_{\mathrm{R}}^2$ depend on the momentum, the counterterms will contain both. And just as they (i.e. the counterterms) ensure that the renormalized function is UV finite (i.e. ensure that the $\frac{2}{\varepsilon}$ of the counterterms cancel against that of the original one-loop integral), they also ensure that it does not depend on the renormalization scale (i.e. the $\ln\mu_{\mathrm{R}}^2$ cancel).
A final note: the renormalization scale $\mu_{\mathrm{R}}$ should not be confused with the parameter $\mu^2$ introduced in eq. \ref{Chap-Maggie:eq:mu}.
}
Conversely, if there is at least a counterterm which is fixed in $\overline{\text{MS}}$, such counterterm will not cancel the dependence of the non-renormalized function on the renormalization scale, so that the renormalized function at stake will depend on $\mu_{\mathrm{R}}$.
In section \ref{Chap-Reno:sec:calc}, all independent counterterms were fixed through a momentum-dependent subtraction scheme, except for $\delta \mu^2$ and (whenever it is independent) $\delta \zeta_1$.%
\fn{Whenever $\delta \zeta_1$ is fixed in $\overline{\text{MS}}$, $\delta \zeta_2$ will also be (cf. eq. \ref{Chap-Reno:eq:rel-z1-z2-fin}).} 
These exceptions will be quite relevant for the impact of the different combinations on the three processes $j$ introduced above.

In order to better illustrate such impact, we show in table \ref{Chap-Reno:tab:CTs-per-process} all the independent counterterms contributing to the different $\mathcal{M}^{\mathrm{CT}}_{j}$, for the four combinations $C_i$.%
\fn{These counterterms are fixed according to the prescriptions presented in section \ref{Chap-Reno:sec:calc}. We should mention that $\delta \zeta_2$ contributes to all processes, in all combinations; and since it is an independent counterterm, it should in principle be included in table \ref{Chap-Reno:tab:CTs-per-process}. However, we fixed it by reference to $\delta \zeta_1$ (recall eq. \ref{Chap-Reno:eq:rel-z1-z2-fin}), which is only independent in $C_4$. In the remaining combinations, the counterterms which $\delta \zeta_1$ depends on are already accounted for in the table. Hence, for the sake of clarity, we do not show $\delta \zeta_2$.}
\begin{table}[!h]%
\begin{normalsize}
\normalsize
\begin{center}
\begin{tabular}
{@{\hspace{0.1mm}}
>{\centering\arraybackslash}m{1.8cm} >{\centering\arraybackslash}m{8.7cm}
>{\centering\arraybackslash}m{0.4cm}
>{\centering\arraybackslash}m{0.4cm}
>{\centering\arraybackslash}m{0.4cm}
>{\centering\arraybackslash}m{0.4cm}
@{\hspace{2mm}}}
\hlinewd{1.1pt}
Process $j$ & Common to all the $C_i$ & $C_1$ & $C_2$ & $C_3$ & $C_4$ \\
\hline\\[-4mm]
{\vspace{3mm}
$h_2 \to ZZ$} &
\multirow{2}{9.3cm}{
\centering
$
{\delta Z_e},  \,  {\delta m_{\mathrm{W}}^2},  \,  {\delta m_{\mathrm{Z}}^2},  \,  {\delta \alpha_1},  \,  {\delta \alpha_2}, \, {\delta \alpha_3}, \, {\delta \beta},
$ \\[0.3mm]
$
{\delta Z_{ZZ}},  \,   {\delta Z_{h_1h_2}},  \,  {\delta Z_{h_2h_2}},  \,  {\delta Z_{h_3h_2}} 
$
}
& $\delta \alpha_5$ & $\delta \alpha_4$ & $\delta \alpha_0$ & $\delta \zeta_1$
\\[7mm]
{\vspace{3mm}
$h_2 \to h_1 Z$} &
\multirow{2}{9.3cm}{
\centering
$
{\delta Z_e},  \,  {\delta m_{\mathrm{W}}^2},  \,  {\delta m_{\mathrm{Z}}^2},  \,  {\delta \alpha_1},  \,  {\delta \alpha_2},  \,  {\delta \alpha_3}, \, {\delta \beta}, \, {\delta Z_{ZZ}},$ \\[0.3mm]
${\delta Z_{h_1h_1}},  \,  {\delta Z_{h_2h_2}},  \,  {\delta Z_{h_3h_1}},  \,  {\delta Z_{h_3h_2}},  \,  {\delta Z_{G_0h_1}},  \,  {\delta Z_{G_0h_2}} 
$
} & $\delta \alpha_5$ & $\delta \alpha_4$ & $\delta \alpha_0$
& $\delta \zeta_1$
\\[7mm]
{\vspace{3mm}
$h_2 \to h_1 h_1$} 
&
\multirow{3}{9.3cm}{
\centering
${\delta Z_e},  \,  {\delta m_{\mathrm{W}}^2},  \,  {\delta m_{\mathrm{Z}}^2},  \,  {\delta \alpha_1},  \,  {\delta \alpha_2},  \,  {\delta \alpha_3}, \, {\delta \beta}, \, {\delta m_1^2}, $\\[0.3mm]
$ {\delta m_2^2}, \,  {\delta \mu^2}, \, {\delta Z_{h_1h_1}},  \,  {\delta Z_{h_1h_2}},  \,  {\delta Z_{h_2h_1}}, \, {\delta Z_{h_2h_2}}, $\\[0.3mm]
${\delta Z_{h_3h_1}},  \,  {\delta Z_{h_3h_2}},  \,  {\delta Z_{G_0h_1}},  \,  {\delta Z_{G_0h_2}}$
}
& $\delta \alpha_5$ & $\delta \alpha_4$ & $\delta \alpha_0$ & $\delta \zeta_1$
\\[13mm]
\hlinewd{1.1pt}
\end{tabular}
\end{center}
\vspace{-5mm}
\end{normalsize}
\caption{Total set of independent counterterms contributing to $\mathcal{M}^{\mathrm{CT}}_{j}$ for the combinations $C_i$. For each process $j$, the counterterms are separated in two groups: those that are common to all the four combinations (second column), and those that are specific to each combination (last four columns).}
\label{Chap-Reno:tab:CTs-per-process}
\end{table}
\normalsize
It is clear that, in all three processes, the complete set of independent counterterms always depends on the combination $C_i$ (note that the last four columns of the table reflect eq. \ref{Chap-Reno:eq:CT-params-H}).
Moreover, processes $h_1 \to ZZ$ and $h_2 \to h_1 Z$ should have no scale dependence in $C_1$, $C_2$ and $C_3$, since they do not involve any counterterm fixed in $\overline{\text{MS}}$ in those combinations. By contrast, the combination $C_4$ in all processes is expected to lead to scale dependent results, as it includes the $\overline{\text{MS}}$-fixed $\delta \zeta_1$.
Finally, process $h_2 \to h_1 h_1$ depends on $\delta \mu^2$, so that we expect scale dependence in all combinations of that process.

\subsection{Decay widths}
\label{Chap-Reno:sec:decays}

Using \FMT, the expressions for the renormalized NLO decay width $\Gamma^{\mathrm{NLO}}_j$ for the process $j$ can be easily calculated from the renormalized NLO amplitude $\hat{\mathcal{M}}_j$. We assume that all external particles are OS. The final expressions are considerably clear, as long as they are written in terms of form factors.%
\fn{In what follows, we follow the conventions of \FMT; we omit the Feynman diagrams, as well as the expressions for the form factors.}

We start with the decay $h_2 \to ZZ$. Defining the momenta and Lorentz indices such that $h_2(p_1) \to Z^{\nu}(q_1) Z^{\sigma}(q_2)$, we have:
\bs
\label{eq:ff-h2ZZ}
\bea
\hs{-10mm}
&&
\mathcal{M}_{h_2 \to ZZ}^{\mathrm{tree}}
= \varepsilon^*_{\nu}(q_1) \, \varepsilon^*_{\sigma}(q_2) \, \,  f_3^{\mathrm{tree}} \, g^{\nu\sigma}, \\[2mm]
\hs{-10mm}
&&
\hat{\mathcal{M}}_{h_2 \to ZZ}^{\mathrm{loop}}
= \varepsilon^*_{\nu}(q_1) \, \varepsilon^*_{\sigma}(q_2)
\bigg(
f_3 \, g^{\nu\sigma} +
f_6 \, p_1^{\nu} \, p_1^{\sigma} +
f_9 \, p_1^{\nu} \, q_1^{\sigma} +
f_{24} \, p_1^{\sigma} \, q_2^{\nu} +
f_{27} \, q_1^{\sigma} \, q_2^{\nu} \nonumber\\[-3mm]
&&
\hs{40mm}
+ f_{33} \, p_1^{\omega} \, q_1^{\upsilon} \, \epsilon^{\nu\sigma\omega\upsilon} +
f_{15} \, p_1^{\sigma} \, q_1^{\nu} +
f_{18} \, q_1^{\nu} \, q_1^{\sigma} +
f_{21} \, q_1^{\nu} \, q_2^{\sigma}
\bigg), \hs{8mm}
\eea
\es
where the different $f_k$
correspond to form factors.%
\fn{We identify tree-level form factors with the superscript `tree'.
The form factor $f_{33}$ corresponds to the CP-violating component predicted in ref. \cite{Huang:2020zde}. It turns out that it does not contribute to $\Gamma^{\mathrm{NLO}}_{h_2 \to ZZ}$, as can be seen in eq. \ref{Chap-Reno:eq:GNLOh2ZZ}.}
These in general contain contributions from the counterterms, so that they will in general have different values in the different combinations.
The form factors $f_{15}$, $f_{18}$ and $f_{21}$ do not contribute, since the fact that the $Z$ bosons are OS implies $\varepsilon^*_{\nu}(q_1) \, q_1^{\nu} = \varepsilon^*_{\sigma}(q_2) \, q_2^{\sigma} = 0$.
The decay width is:
\bea
\label{Chap-Reno:eq:GNLOh2ZZ}
\Gamma^{\mathrm{NLO}}_{h_2 \to ZZ}
&=&
\dfrac{\sqrt{m_2^4 - 4 \, m_2^2 \, m_{\mathrm{Z}}^2}}{128\, \pi \, m_2^3 \, m_{\mathrm{Z}}^4} f_3^{\mathrm{tree}}
\bigg\{ \left( m_2^4 - 4 \, m_2^2 \, m_{\mathrm{Z}}^2 + 12 \, m_{\mathrm{Z}}^4 \right) \left( f_3^{\mathrm{tree}}  + 2 \, \mathrm{Re} [f_3] \right) \nonumber\\
&& \hs{17mm}
+ m_2^2 \,  \left( m_2^4 - 6 \, m_2^2 \, m_{\mathrm{Z}}^2 + 8 \, m_{\mathrm{Z}}^4 \right)  \, \operatorname{Re}[f_6 + f_9 + f_{24} + f_{27}] \bigg\}.
\eea
As for the decay $h_2 \to h_1 Z$, we define the momenta and Lorentz indices such that $h_2(p_1) \to h_1(q_1) Z^{\sigma}(q_2)$, so that
\be
\mathcal{M}_{h_2 \to h_1 Z}^{\mathrm{tree}}
=
\varepsilon^*_{\sigma}(q_2)
\Big(
f_{10}^{\mathrm{tree}} \, q_1^{\sigma} + f_{12}^{\mathrm{tree}} \, p_1^{\sigma}
\Big), \qquad
\hat{\mathcal{M}}_{h_2 \to h_1 Z}^{\mathrm{loop}}
= \varepsilon^*_{\sigma}(q_2)
\Big(
f_{10} \, q_1^{\sigma} +
f_{11} \, q_2^{\sigma} +
f_{12} \, p_1^{\sigma}
\Big),
\ee
but, as the $Z$ boson is OS, $f_{11}$ does not contribute. The decay width is:
\bea
\Gamma^{\mathrm{NLO}}_{h_2 \to h_1 Z}
&=&
\dfrac{1}{64 \, \pi \, m_2^3 \, m_{\textrm{Z}}^2}
\Big(
m_1^4 + m_2^4 + m_{\textrm{Z}}^4 - 2 \, m_1^2 \, m_2^2 - 2 \, m_1^2 \, m_{\textrm{Z}}^2 - 2 \, m_2^2 \, m_{\textrm{Z}}^2
\Big)^{3/2} \nonumber\\[-2mm]
&& \hs{20mm} \times 
\bigg\{ 
\left|f_{10}^{\mathrm{tree}} + f_{12}^{\mathrm{tree}}\right|^2
+
2 \, \mathrm{Re}
\Big[
\left(f_{10}^{\mathrm{tree}} + f_{12}^{\mathrm{tree}}\right)
\left(f_{10}^* + f_{12}^*\right)
\Big]
\bigg\}.
\eea
Finally, the decay $h_2 \to h_1 h_1$ is such that the renormalized NLO amplitude is simply given by:
\be
\hat{\mathcal{M}}_{h_2 \to h_1 h_1}
=
\mathcal{M}_{h_2 \to h_1 h_1}^{\mathrm{tree}}
+ 
\hat{\mathcal{M}}_{h_2 \to h_1 h_1}^{\mathrm{loop}}
=
f_1^{\mathrm{tree}} + f_1,
\ee
and the decay width is:
\be
\Gamma^{\mathrm{NLO}}_{h_2 \to h_1 h_1}
=
\dfrac{\sqrt{m_2^4 -4 \, m_1^2 \, m_2^2}}{32 \, \pi \, m_2^3} \, f_1^{\mathrm{tree}} (f_1^{\mathrm{tree}} + 2 \, \mathrm{Re}[f_1]).
\ee

\subsection{Numerical results}
\label{Chap-Reno:sec:num-res}

At last, we investigate the behaviour of the different combinations on the decay widths using numerical results.
First and foremost, we checked that all combinations in all processes lead to gauge independent decay widths.
This is in agreement with the discussion presented in section \ref{Chap-Reno:sec:gauge-dep}, and validates the simple prescription for the renormalization of mixing parameters proposed in section \ref{Chap-Reno:sec:RenoMixingParameters}.

We compare the different combinations by ascertaining how their results affect the numerical stability of the perturbative expansion. If perturbation theory is to be trusted, higher orders should contribute with smaller corrections. In particular, the NLO results should generate small corrections when compared to the LO ones. This leads us to define the quantity \cite{Krause:2016xku,Krause:2017mal}:
\be
\Delta \Gamma_j \equiv \dfrac{\Gamma^{\mathrm{NLO}}_j - \Gamma^{\mathrm{LO}}_j}{\Gamma^{\mathrm{LO}}_j},
\label{Chap-Reno:eq:def-Gamma}
\ee
where $\Gamma^{\mathrm{LO}}_j$ represents the decay width at LO for the process $j$.
\begin{figure}[h!]
\centering
\includegraphics[width=0.55\textwidth]{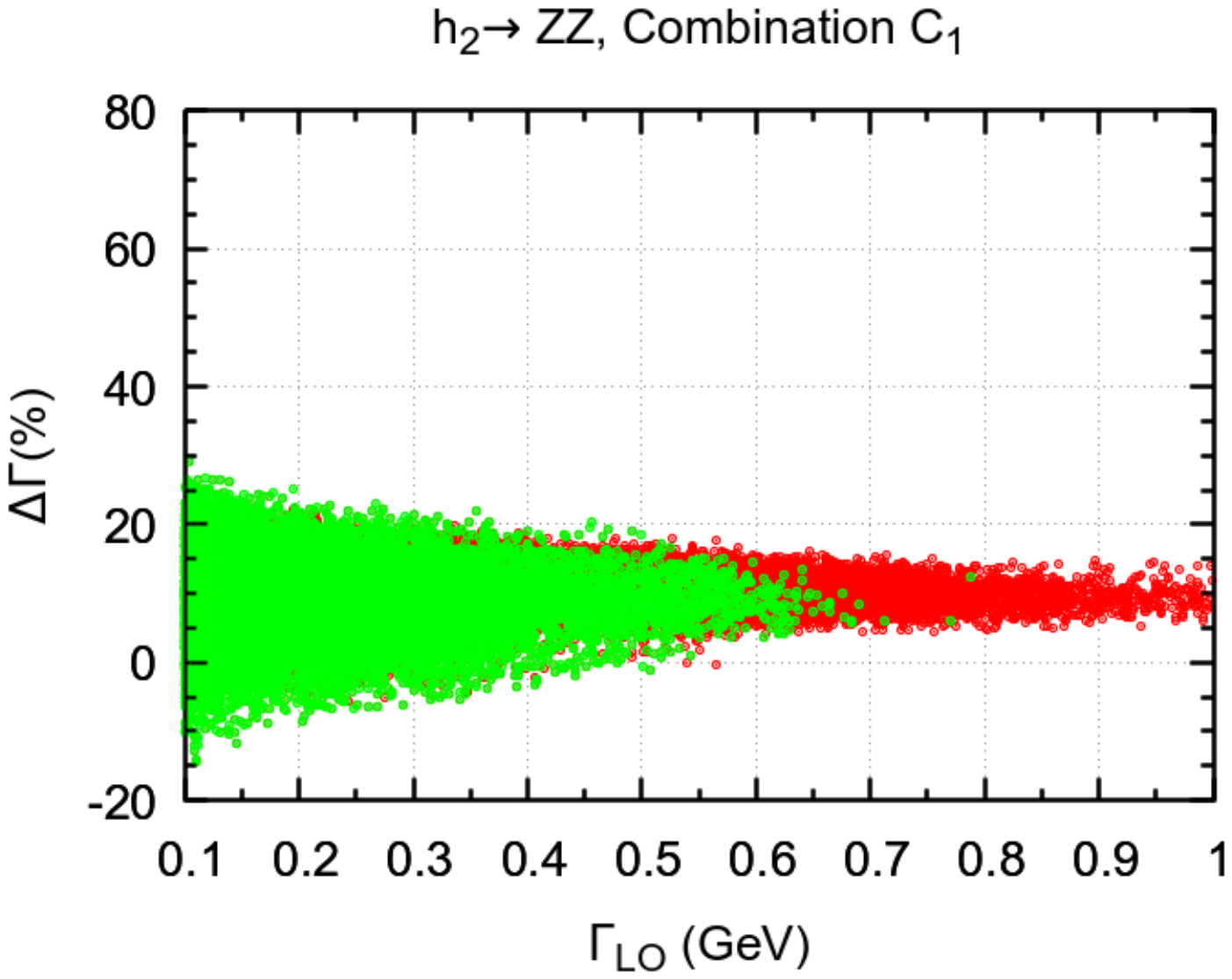}
\caption{
$\Delta \Gamma_{h_2 \to ZZ}$ in percentage as a function of $\Gamma^{\mathrm{LO}}_{h_2 \to ZZ}$, for the combination $C_1$.
Only the interval $ 0.1 \, \, \textrm{GeV} < \Gamma^{\mathrm{LO}} < 1 \, \, \textrm{GeV}$ is shown. 
In red, points passing all constraints except \textsc{HiggsBounds5}; in green, points passing all constraints.}
\label{Chap-Reno:fig:h2ZZ-C1}
\end{figure}	

In fig. \ref{Chap-Reno:fig:h2ZZ-C1}, we show $\Delta \Gamma_{h_2 \to ZZ}$ in percentage as a function of $\Gamma^{\mathrm{LO}}_{h_2 \to ZZ}$ for the combination $C_1$. Several aspects should be stressed here.
First if all, we checked numerically that the results are invariant under the renormalization scale, as expected from the discussion of section \ref{Chap-Reno:sec:influ}.%
\fn{Both scale invariance and finiteness were checked with the subroutines \t{setmudim} and \t{setdelta} from \ts{LoopTools}.}
Besides, the plot shows that \textsc{HiggsBounds5} hinders points with large values of $\Gamma^{\mathrm{LO}}$;
this is consistent with the fact that, had $h_2 \to ZZ$ a large value of $\Gamma^{\mathrm{LO}}$, $h_2$ would probably already have been discovered by now.
The majority of points passing all constraints thus obey $\Gamma^{\mathrm{LO}} < 0.7 \, \, \textrm{GeV}$; points with $\Gamma^{\mathrm{LO}}$ almost up to $0.8 \, \, \textrm{GeV}$ are still allowed, albeit very scarcely (a stringent fine tuning is required).
All in all, the points describe a smooth and well-behaved pattern; in particular, there is no region of parameter space generating singularities in $\Delta \Gamma$.
Finally, although there are points with small $\Delta \Gamma$ for all allowed values of $\Gamma^{\mathrm{LO}}$---indicating that, for those points, perturbation theory is valid---, larger values of $\Delta \Gamma$ are also possible for smaller values of  $\Gamma^{\mathrm{LO}}$.%
\fn{From eq. \ref{Chap-Reno:eq:def-Gamma}, larger values of $\Delta \Gamma$ are expected for smaller values of $\Gamma^{\mathrm{LO}}$.}
For those points, one would in principle need to calculate the following order in perturbation theory to check whether or not a perturbative description is possible.

\begin{figure}[htb]
\centering
\includegraphics[width=0.48\textwidth]{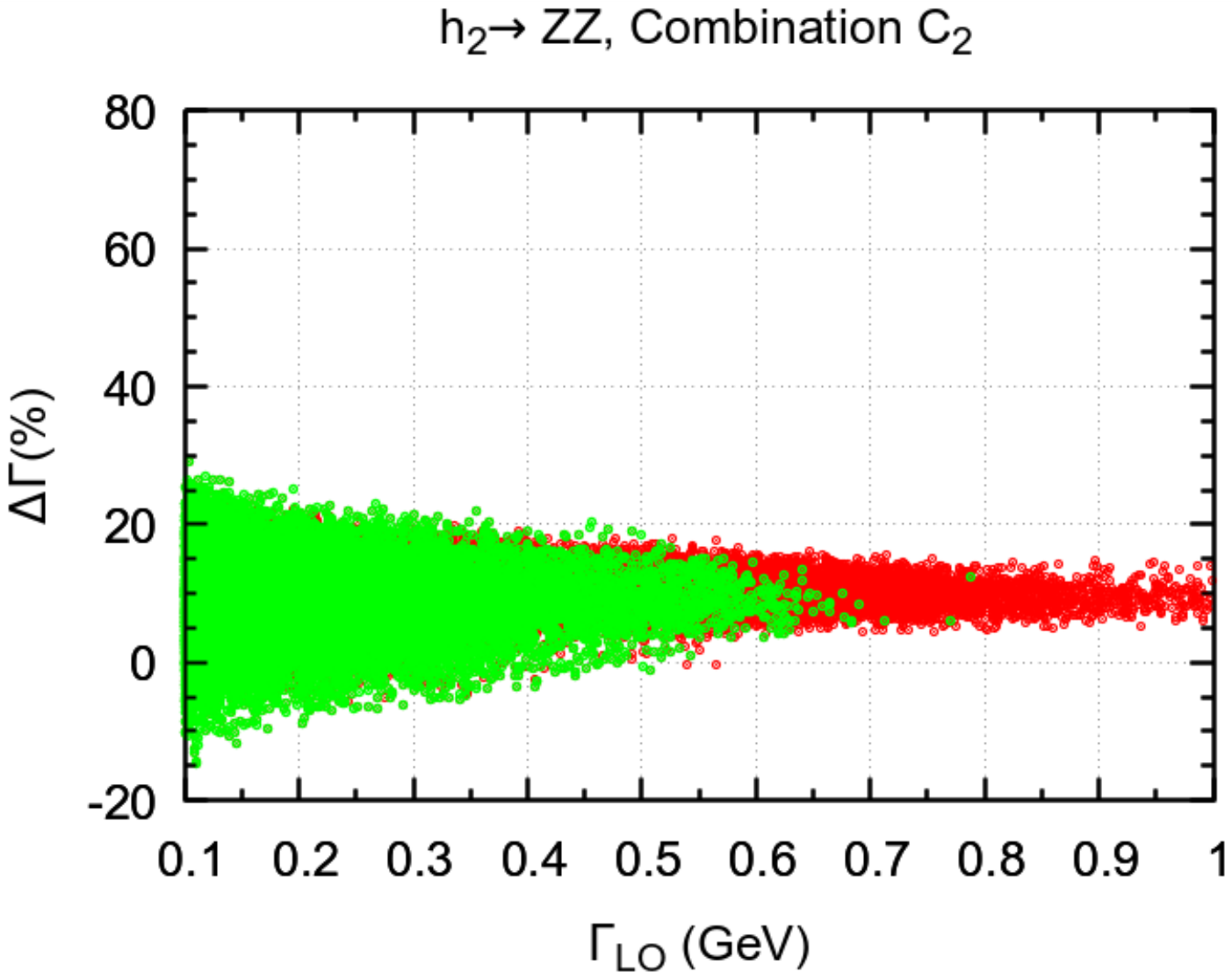}
\hfill
\includegraphics[width=0.48\textwidth]{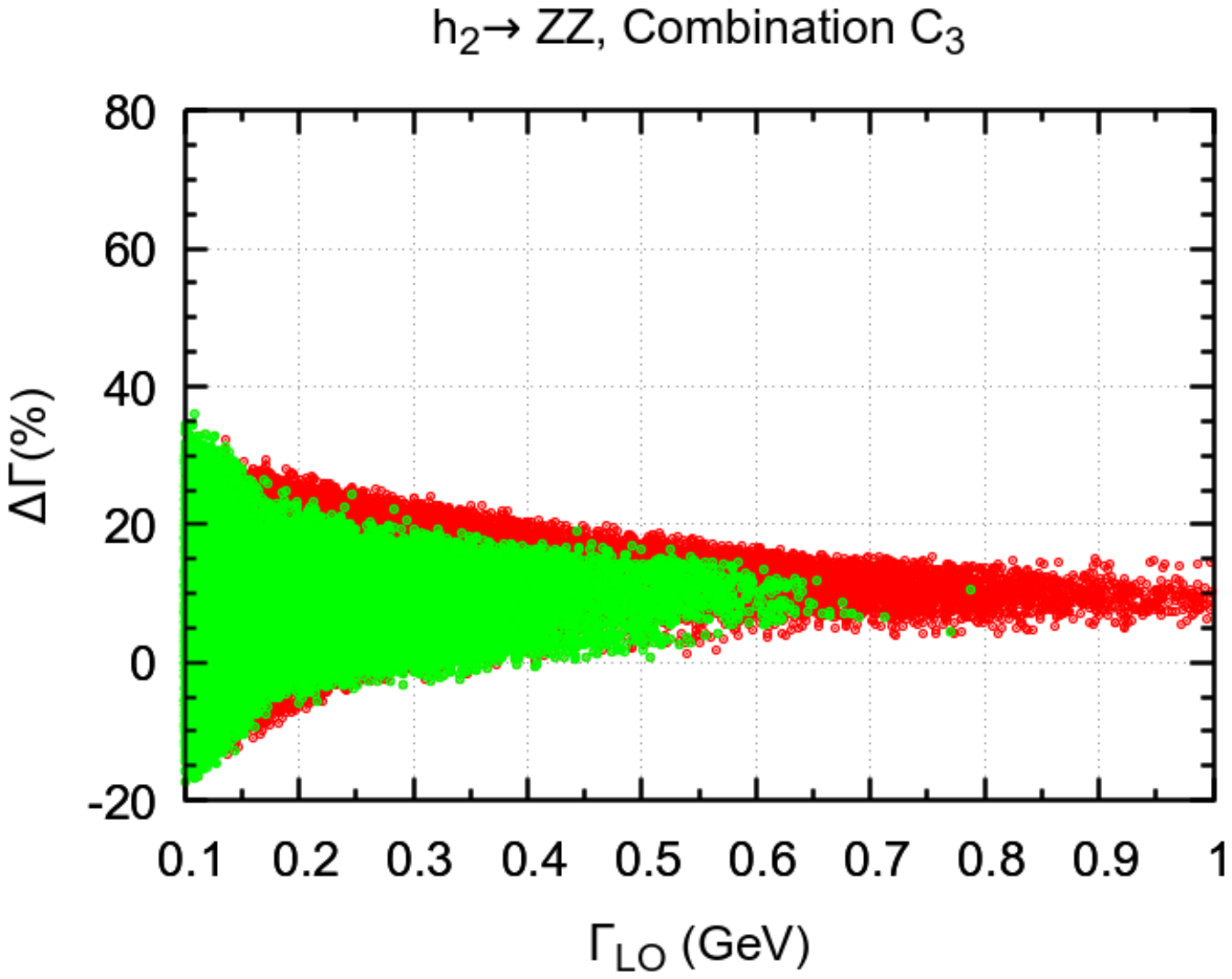}
\caption{
$\Delta \Gamma_{h_2 \to ZZ}$ in percentage as a function of $\Gamma^{\mathrm{LO}}_{h_2 \to ZZ}$, for the combinations $C_2$ (left) and $C_3$ (right).
Only the interval $ 0.1 \, \, \textrm{GeV} < \Gamma^{\mathrm{LO}} < 1 \, \, \textrm{GeV}$ is shown. 
The color conventions are those of fig. \ref{Chap-Reno:fig:h2ZZ-C1}.}
\label{Chap-Reno:fig:h2ZZ-C2-C3}
\end{figure}

Similar to fig. \ref{Chap-Reno:fig:h2ZZ-C1} are the two panels of fig. \ref{Chap-Reno:fig:h2ZZ-C2-C3}, where we show equivalent results for the combinations $C_2$ and $C_3$.
We checked that, as before, there is no scale dependence.
The values of $\Delta \Gamma_{h_2 \to ZZ}$ are essentially the same as in fig. \ref{Chap-Reno:fig:h2ZZ-C1}; moreover, we find the same smooth pattern in the points. This allows us to attest the quality of our prescription for the determination of the counterterms for the mixing parameters, proposed in section \ref{Chap-Reno:sec:RenoMixingParameters}.%
\fn{Except for $\delta \zeta_a$, which we do not investigate in this paper.}
Indeed, from table \ref{Chap-Reno:tab:CTs-per-process}, the combinations $C_1$, $C_2$ and $C_3$ for the process $h_2 \to ZZ$, besides depending all on $\delta \beta$, $\delta \alpha_1$, $\delta \alpha_2$ and $\delta \alpha_3$, depend individually on $\delta \alpha_5$, $\delta \alpha_4$ and $\delta \alpha_0$, respectively. Then, by considering fig.s \ref{Chap-Reno:fig:h2ZZ-C1} and \ref{Chap-Reno:fig:h2ZZ-C2-C3}, we conclude that all these counterterms lead to well-behaved results. Actually, not only are the results all well-behaved, but they are also very similar. 
This is consistent with the circumstance that $\delta \alpha_5$, $\delta \alpha_4$ and $\delta \alpha_0$ were all determined via symmetry relations, so that one would expect similar behaviours from the different lines of eqs. \ref{Chap-Reno:eq:indep-dalfas}.
%
All in all, then,
the combinations $C_1$, $C_2$ and $C_3$ are essentially equivalent for the process $h_2 \to ZZ$.

\begin{figure}[htb]
\centering
\includegraphics[width=0.55\textwidth]{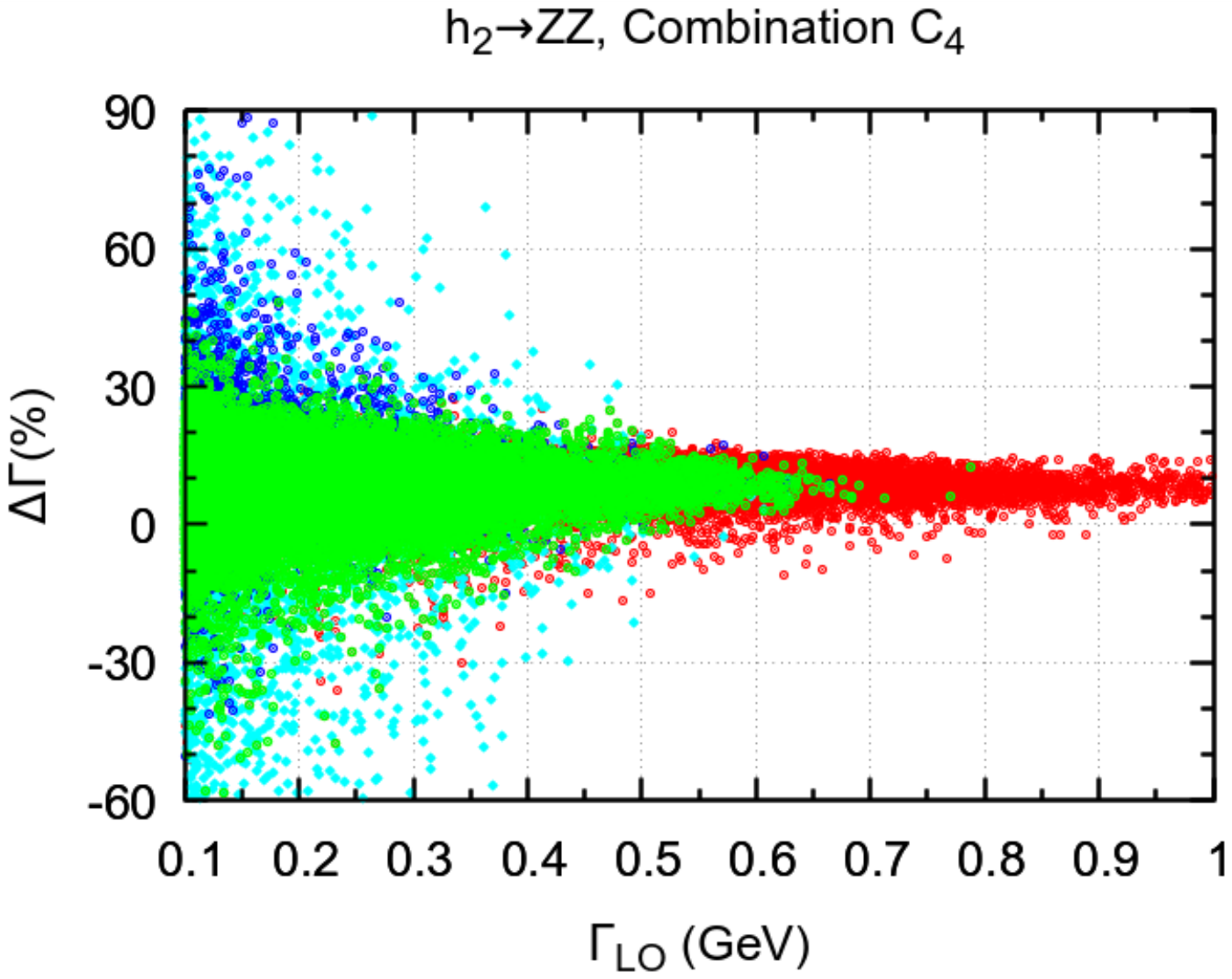}
\caption{
$\Delta \Gamma_{h_2 \to ZZ}$ in percentage as a function of $\Gamma^{\mathrm{LO}}_{h_2 \to ZZ}$, for the combination $C_4$.
Only the interval $ 0.1 \, \, \textrm{GeV} < \Gamma^{\mathrm{LO}} < 1 \, \, \textrm{GeV}$ is shown. 
In red, points with $\mu_{\mathrm{R}}= 350$ GeV passing all constraints except \textsc{HiggsBounds5}. The remaining colors represent different values of the renormalization scale for points passing all the constraints: in cyan, $\mu_{\mathrm{R}} = 125$ GeV; in green, $\mu_{\mathrm{R}}= 350$ GeV; in blue, $\mu_{\mathrm{R}}= 700$ GeV.}
\label{Chap-Reno:fig:h2ZZ-C4}
\vs{-3mm}
\end{figure}

The results for $C_4$ are shown in fig. \ref{Chap-Reno:fig:h2ZZ-C4}.
Here, as expected from the discussion of section \ref{Chap-Reno:sec:influ}, the results depend on the renormalization scale $\mu_{\mathrm{R}}$. We plot three different scales, for points passing all constraints. It is clear that the lighter scale leads to a big dispersion in the values of $\Delta \Gamma$, and that the intermediate scale leads to the most stable results. 
For $\Gamma^{\mathrm{LO}}_{h_2 \to ZZ} \gtrsim 0.4$ GeV, these turn out to be quite similar to those of the points passing all constraints in the remaining combinations: the ranges of values of $\Delta \Gamma$ are essentially the same, and the same well-behaved pattern is observed. This could hardly be expected from the expressions for the different counterterms; not only is the specific counterterm in $C_4$ ($\delta \zeta_1$) fixed through a method completely different from the one used to fix the specific counterterms of the combinations $C_1$, $C_2$ and $C_3$ ($\delta \alpha_5$, $\delta \alpha_4$ and $\delta \alpha_0$, respectively), but the role of counterterms is also different: $\delta \zeta_1$ is a counterterm for a phase of the potential, whereas $\delta \alpha_5$, $\delta \alpha_4$ and $\delta \alpha_0$ are counterterms for mixing parameters. For smaller values of $\Gamma^{\mathrm{LO}}_{h_2 \to ZZ}$, however, and whichever the scale chosen, $C_4$ leads to more unstable results than the remaining combinations.

\begin{figure}[h!]
\centering
\subfloat{\includegraphics[width=0.46\linewidth]{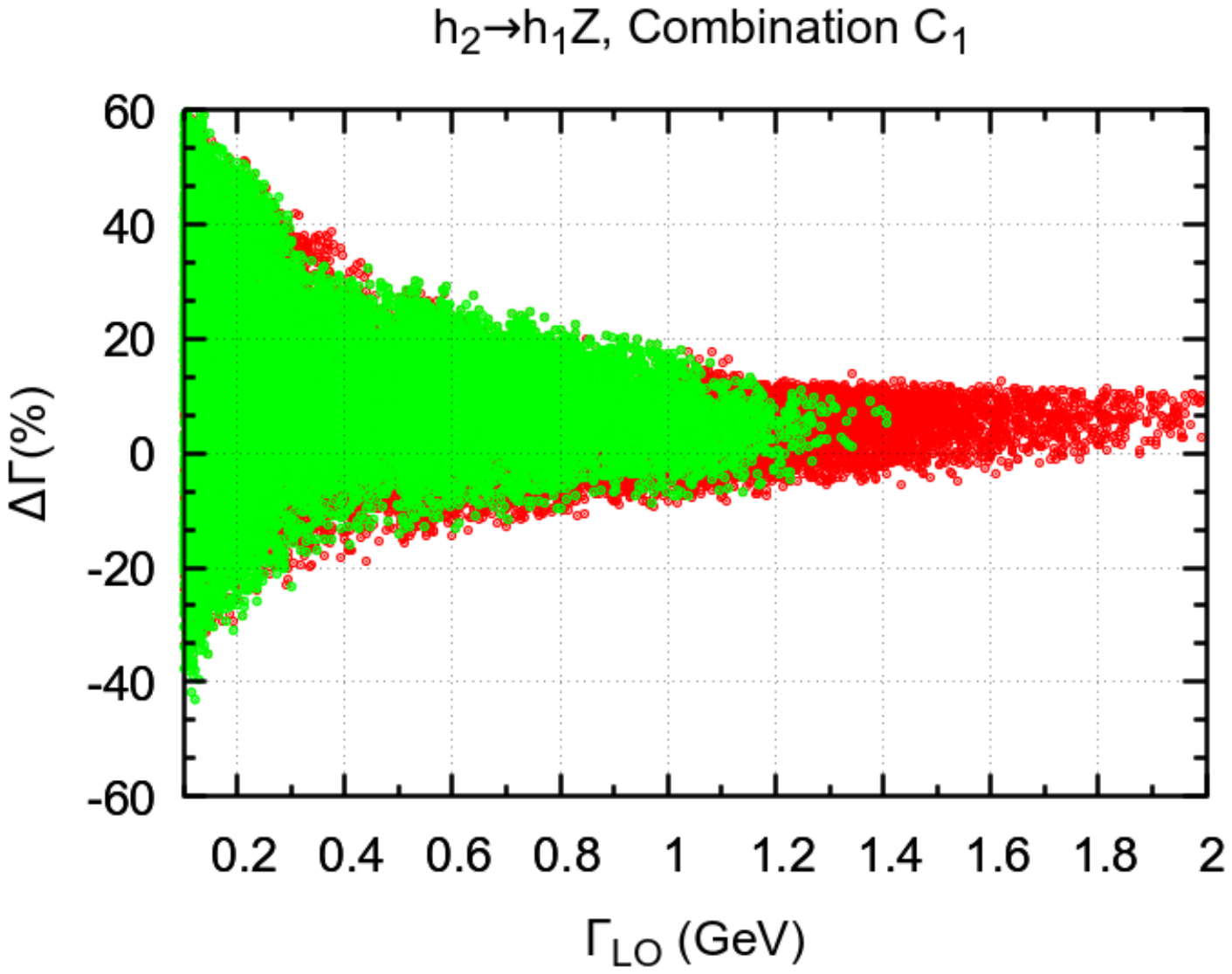}}\qquad
\subfloat{\includegraphics[width=0.46\linewidth]{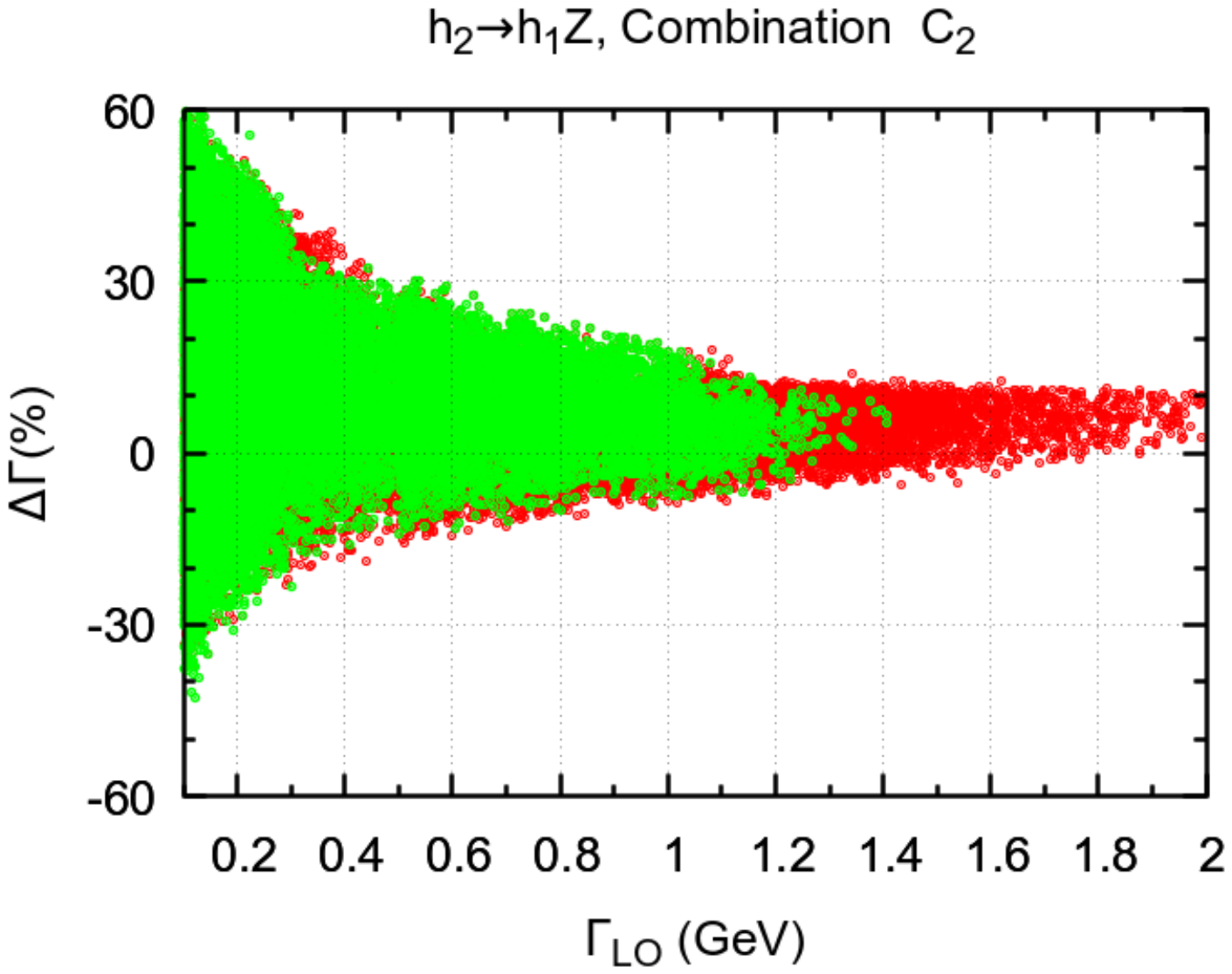}}\\
\subfloat{\includegraphics[width=0.46\textwidth]{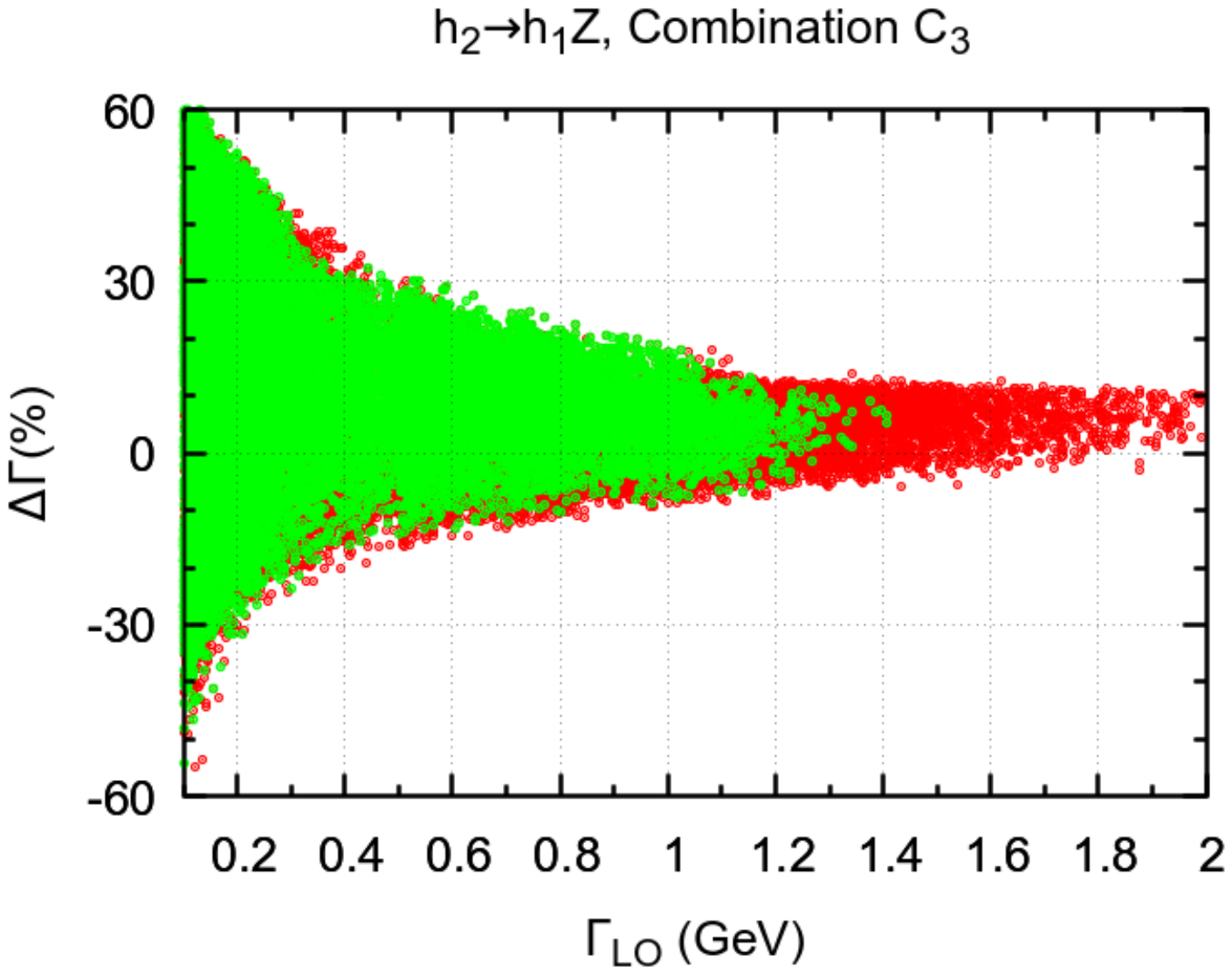}}\qquad%
\subfloat{\includegraphics[width=0.46\textwidth]{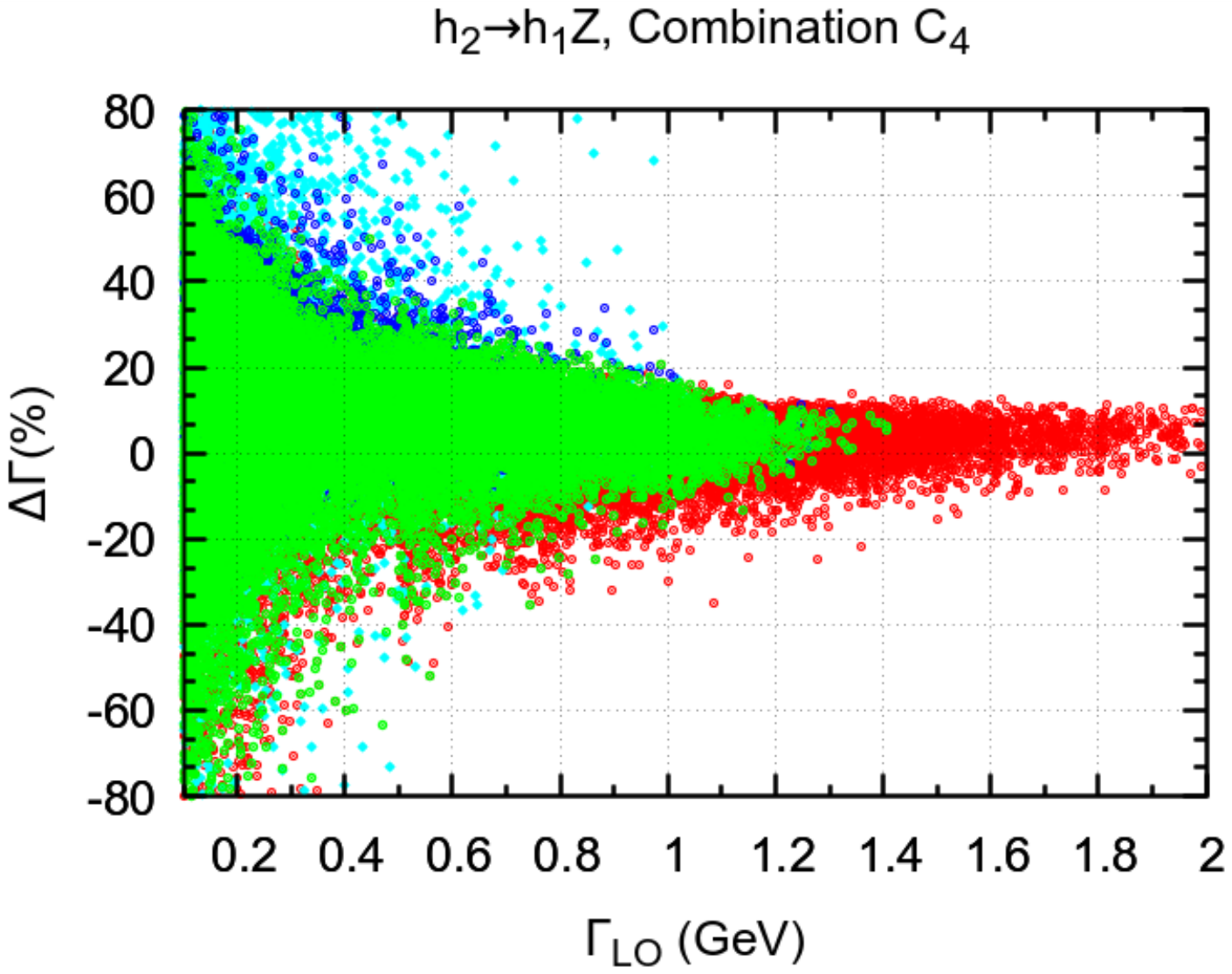}}%
\caption{
$\Delta \Gamma_{h_2 \to h_1Z}$ in percentage as a function of $\Gamma^{\mathrm{LO}}_{h_2 \to h_1Z}$, for the combinations $C_1$ (top left), $C_2$ (top right), $C_3$ (down left) and $C_4$ (down right).
Only the interval $ 0.1 \, \, \textrm{GeV} < \Gamma^{\mathrm{LO}} < 2 \, \, \textrm{GeV}$ is shown.
The color conventions for $C_1$, $C_2$ and $C_3$ are those of fig. \ref{Chap-Reno:fig:h2ZZ-C1}, whereas for $C_4$ are those of fig. \ref{Chap-Reno:fig:h2ZZ-C4}.
}
\label{Chap-Reno:fig:h2h1Z}
\end{figure}

In fig. \ref{Chap-Reno:fig:h2h1Z}, we show the results for the decay $h_2 \to h_1Z$, for the four combinations.
The conclusions are similar to the ones for $h_2 \to ZZ$, which we just discussed.
In particular, the combinations $C_1$, $C_2$ and $C_3$ are scale independent, whereas $C_4$ is scale dependent (in which case the intermediate scale is again the most stable one); \textsc{HiggsBounds5} hampers points with large values of $\Gamma^{\mathrm{LO}}$, and all points passing all constraints in the four combinations show a similar and well-behaved pattern. Broadly speaking, these points are similar to the equivalent ones in $h_2 \to ZZ$ (except that, for small values of $\Gamma^{\mathrm{LO}}_{h_2 \to h_1Z}$, there are valid points with larger values---in modulus---of $\Delta \Gamma$, especially for $C_4$).
This similarity is expected from table \ref{Chap-Reno:tab:CTs-per-process}, which shows that the only difference in the set of independent counterterms contributing to the two decays concerns field counterterms; but since all field counterterms are fixed through the same method (namely, OSS), it is not surprising that the two processes turn out to lead to similar NLO corrections.

\begin{figure}[h!]
\centering
\subfloat{\includegraphics[width=0.46\linewidth]{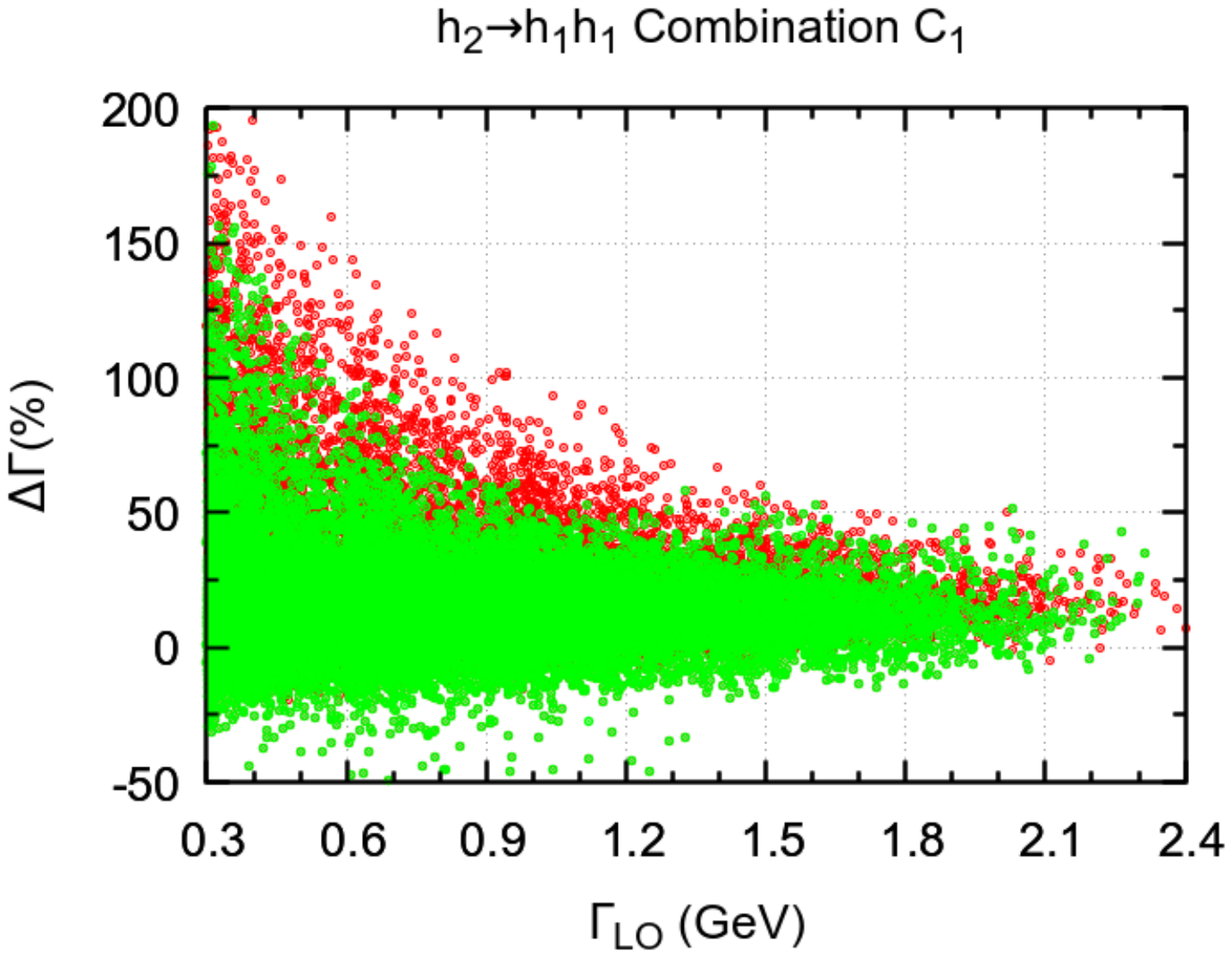}}\qquad
\subfloat{\includegraphics[width=0.46\linewidth]{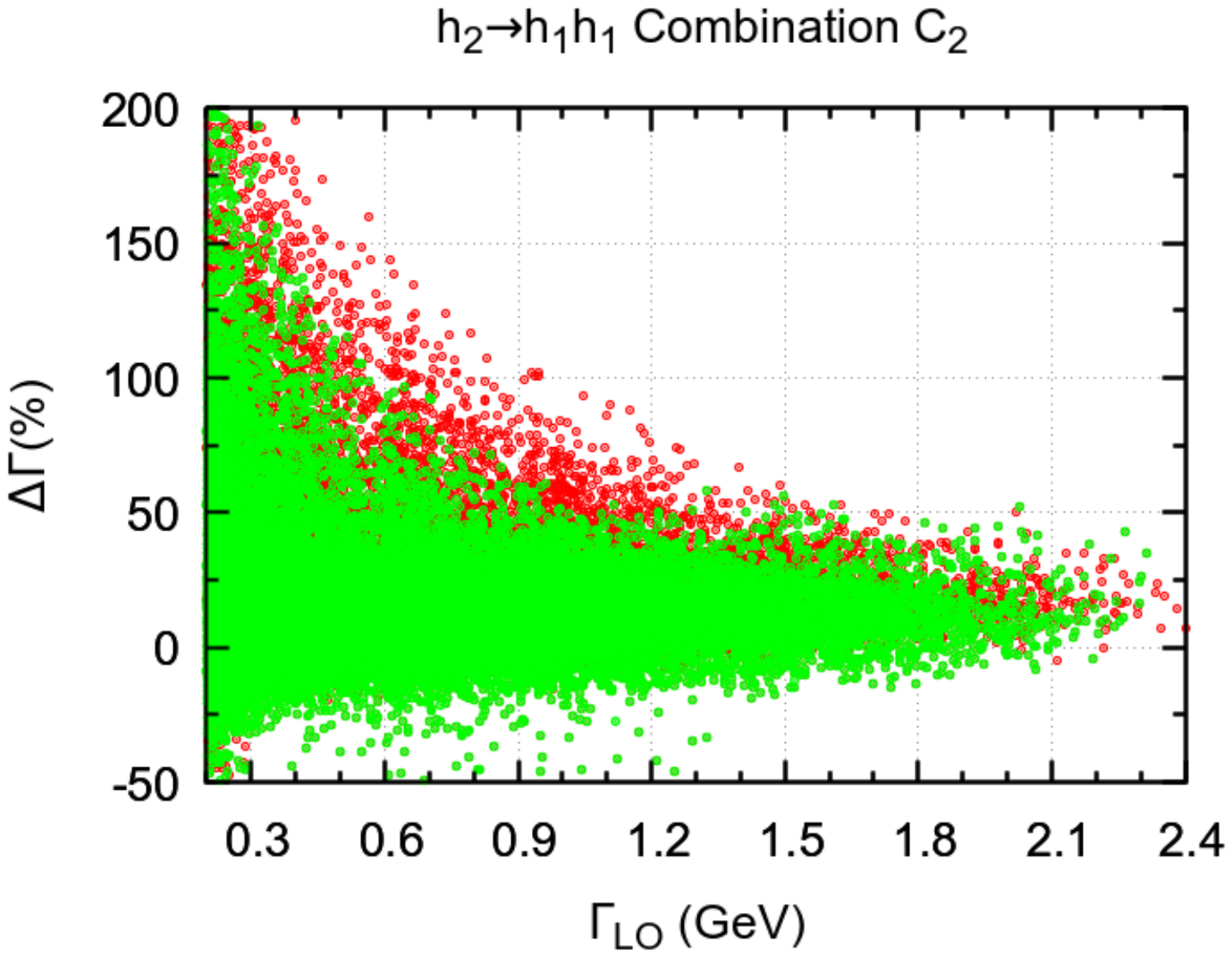}}\\
\subfloat{\includegraphics[width=0.46\textwidth]{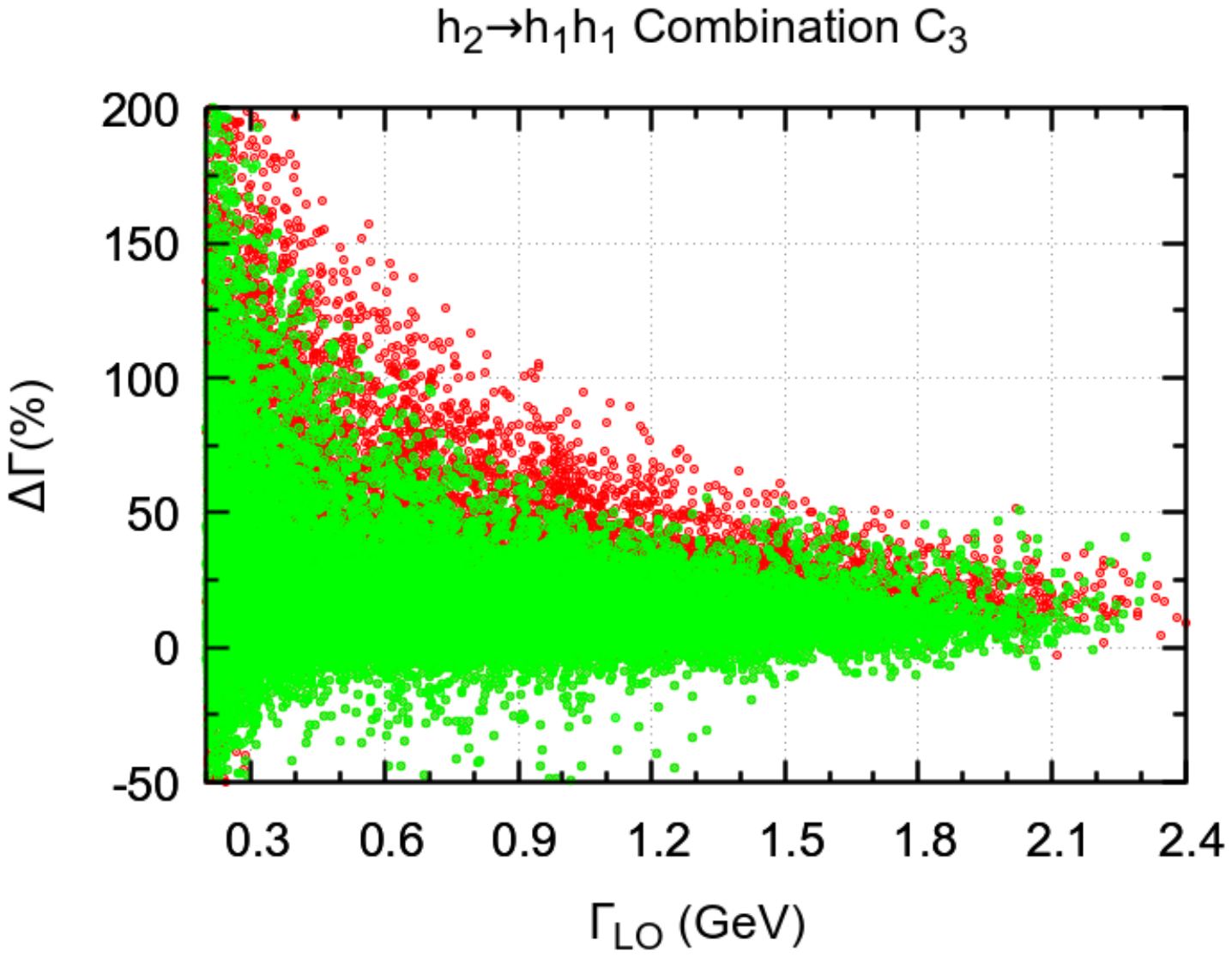}}\qquad%
\subfloat{\includegraphics[width=0.46\textwidth]{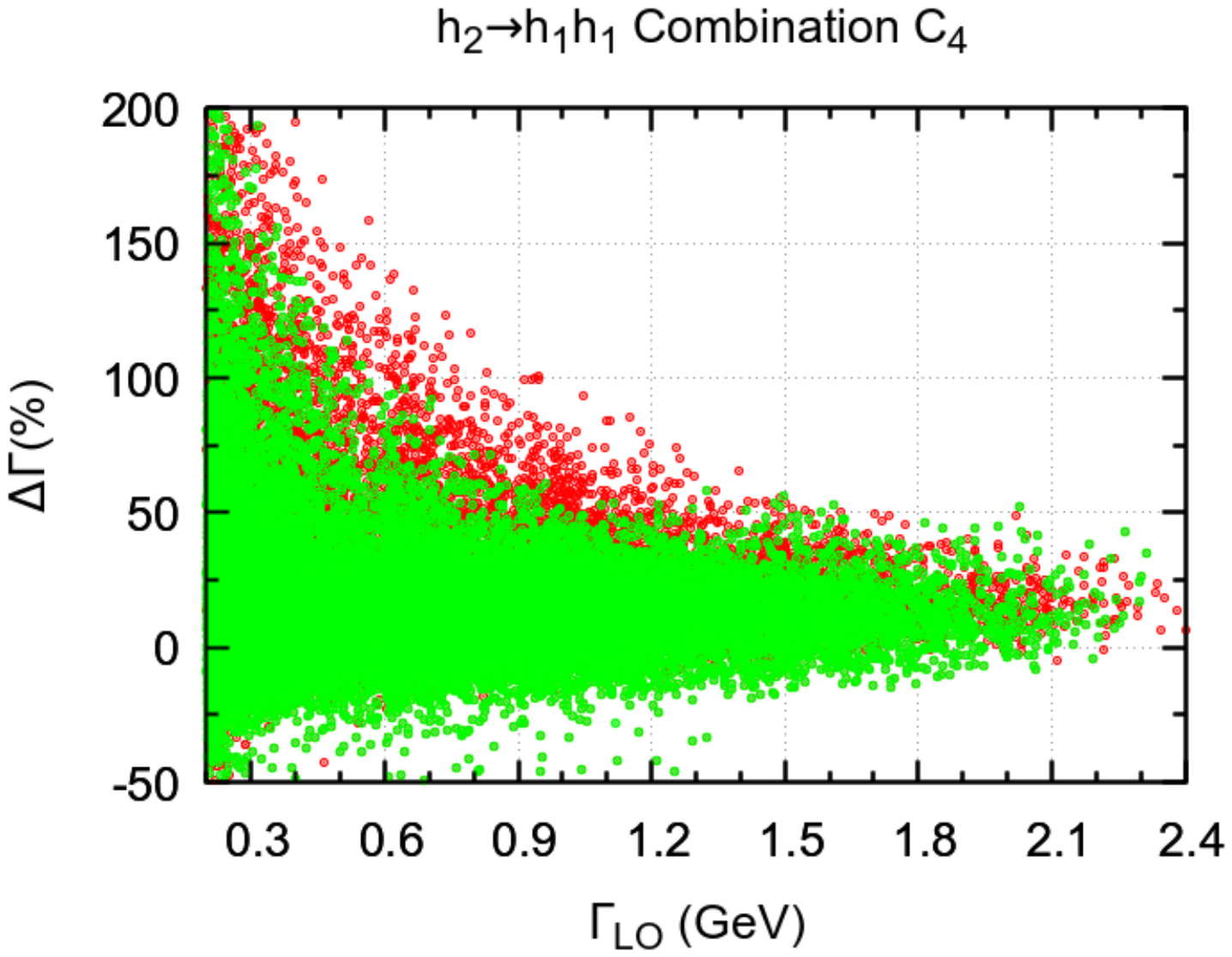}}%
\caption{
$\Delta \Gamma_{h_2 \to h_1h_1}$ in percentage as a function of $\Gamma^{\mathrm{LO}}_{h_2 \to h_1h_1}$, for the combinations $C_1$ (top left), $C_2$ (top right), $C_3$ (down left) and $C_4$ (down right).
Only the interval $ 0.1 \, \, \textrm{GeV} < \Gamma^{\mathrm{LO}} < 2.4 \, \, \textrm{GeV}$ is shown.
All points correspond to $\mu_{\mathrm{R}}= 350$ GeV; the color conventions are those of fig. \ref{Chap-Reno:fig:h2ZZ-C1}.
}
\label{Chap-Reno:fig:h2h1h1}
\end{figure}

\begin{figure}[htb]
\centering
\includegraphics[width=0.55\textwidth]{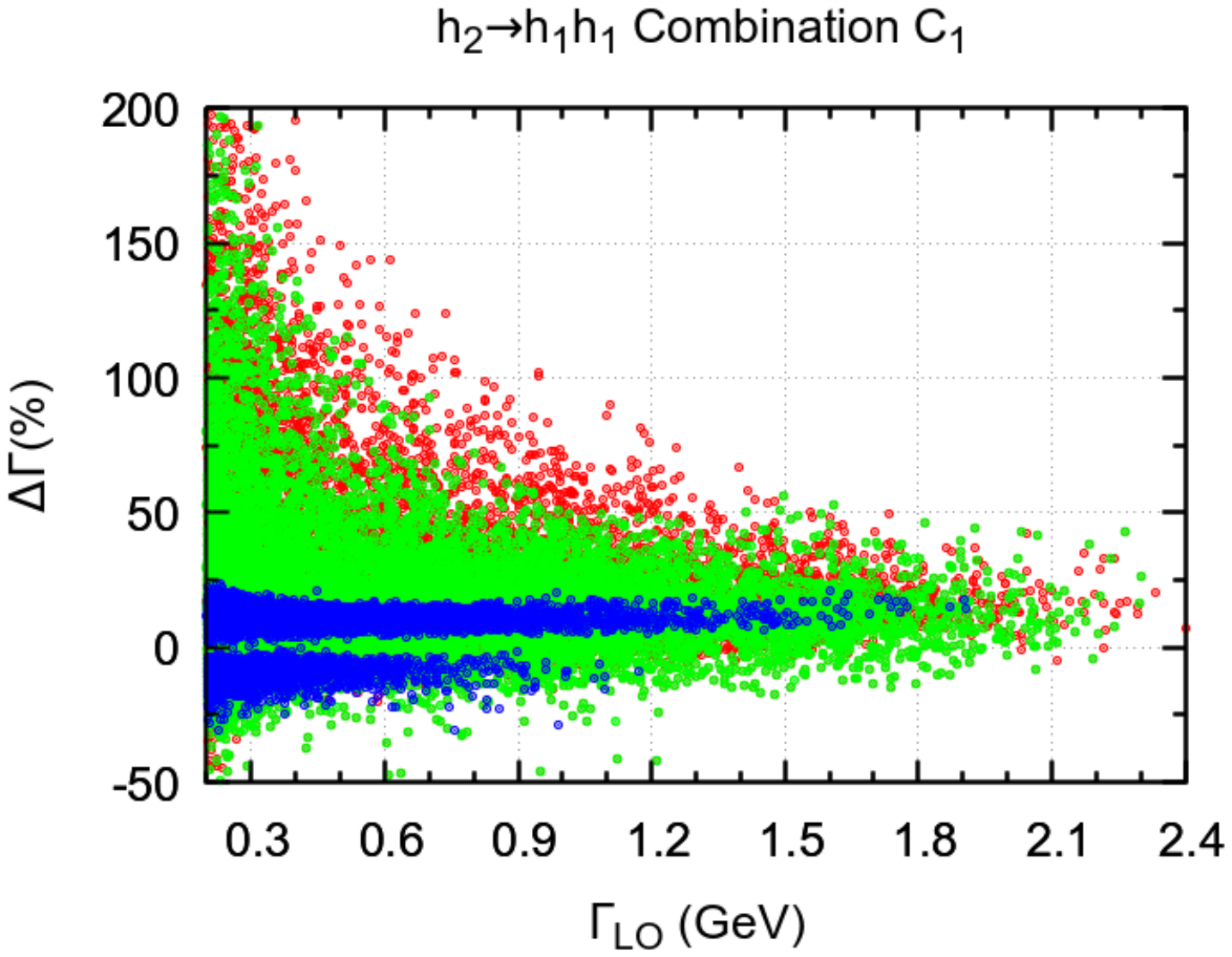}
\caption{
Identical to fig. \ref{Chap-Reno:fig:h2h1h1} top left, except that blue/black points are included, satisfying all constraints and, additionally, $|\lambda_i|<2$ (see text for details).}
\label{Chap-Reno:fig:h2h1h1-ll}
\end{figure}

In fig. \ref{Chap-Reno:fig:h2h1h1}, we show the results for $h_2 \to h_1 h_1$, for the four combinations.
Here, as predicted in section \ref{Chap-Reno:sec:influ} above, all combinations are scale dependent, due to the contribution of the counterterm $\delta \mu^2$; we plot the intermediate scale only, which we checked is again the most stable one.
The four combinations lead to similar results; a well-behaved pattern is once again to be observed in all of them. In contrast to what happened previously, \textsc{HiggsBounds5} does not preclude points with large values of $\Gamma^{\mathrm{LO}}_{h_2 \to h_1h_1}$; we checked that these points lead to small values of $\Gamma^{\mathrm{LO}}_{h_2 \to ZZ}$ and  $\Gamma^{\mathrm{LO}}_{h_2 \to h_1Z}$, and that only large values of these would be excluded by the experimental results.
Moreover, values of $\Delta \Gamma$ as high as $40\%$ are allowed for all the region of allowed points in all combinations. We found that more stable results can be obtained by requiring smaller values for the different $\lambda_i$ (parameters of the potential), as can be seen in fig. \ref{Chap-Reno:fig:h2h1h1-ll} for the combination $C_1$, where we constrained $|\lambda_i| < 2$ for all $\lambda_i$.
This can be explained by the circumstance that smaller values for the $\lambda_i$ correspond to a more stringent constraint regarding the tree-level perturbative unitarity. Hence, it is not surprising that such values lead to a more stable perturbative expansion \cite{Krause:2016xku,Denner:2016etu}.

\section{Summary}
\label{Chap-Reno:sec:conclu}

We presented the renormalization of the C2HDM, which is one of the simplest models containing an extended scalar sector with CP violation. We showed in detail how the presence of CP violation in the scalar sector leads to a rather peculiar program of renormalization, due to two main reasons.
First, several non-physical parameters must be introduced in order to ensure the generation of the necessary counterterms. This happens in such a way that, once the counterterms are obtained, the corresponding renormalized parameters are not needed and can be rephased away.
%
Second, for the same set of independent physical parameters, multiple combinations $C_i$ of independent counterterms can be chosen. It thus becomes relevant to study the impact of different combinations on observables.

We considered four combinations and applied them to three specific one-loop decays of $h_2$; the first three combinations take as independent a counterterm for a mixing parameter, and they fix it through symmetry relations, whereas $C_4$ takes a parameter of the potential as independent, and fixes it through $\overline{\text{MS}}$---thus yielding a dependence on the renormalization scale $\mu_{\mathrm{R}}$.
We checked that all combinations in all processes lead to gauge-independent and well-behaved results, which validates the prescriptions used to fix the counterterms.
For all processes, we found that the combination $C_4$ leads (at least for some values of $\mu_{\mathrm{R}}$ and of the LO decay width) to results very similar to those of the remaining combinations.
Among the points passing all constraints in all processes, one finds regions in the four combinations that lead to a numerical stability of the perturbative expansion; points leading to large NLO corrections are also allowed, in which case the results of higher orders should be investigated.
 
We stressed several aspects related to an adequate treatment of the C2HDM when considered up to one-loop level. 
We used the Fleischner-Jegerlehner tadpole scheme to ensure the selection of the true vev; this was the first time that this scheme was applied to a model with CP violation in scalar sector (and thus to a model with generally complex vevs); we confirmed that this scheme is equivalent to an approach where the tree-level vevs are used and where the (connected) GFs include one-loop tadpole insertions. 
We discussed the renormalization of a field whose mass is a dependent parameter.
We showed that the OSS conditions in the fermion sector leave some freedom, which can be used to define complex counterterms for the fermion masses.
We proposed a simple and easily generalizable prescription to calculate counterterms for mixing parameters, which ensures that observables are gauge independent.
%

Finally, \FMTS proved to be an ideal tool for renormalizing models such as the C2HDM and study them at NLO, since it calculates a multiplicity of crucial elements---Feynman rules, counterterms, one-loop amplitudes, decay widths, etc.---in a simultaneously automatic and flexible way. The \FMS model file for C2HDM can be found in the webpage of the program, and a detailed description of the advantages of using \FMTS to perform the renormalization of the model can be found in appendix \ref{App-FM-C2HDM}.

%% file: Chapters/Chapter_Lavoura.tex

\chapter{NLO corrections to $Z \to b\bar{b}$ in MHM}
\label{Chap-Lavou}

\vs{-5mm}


Multi-Higgs Models (MHM) can be investigated in a generic fashion; this is what we propose to do in this chapter, by considering a general model with an arbitrary number of scalar particles.
While light scalar particles may be detected directly through their production, heavy ones may be detected indirectly through their impact on the radiative corrections. Here, we investigate the second approach, focusing on one-loop corrections to $Z \to b\bar{b}$.
Such corrections can be connected to experiment through e.g. the forward--backward asymmetry measured in the process $e^- e^+ \to b \overline{b}$. The present values for these quantities are within 1$\sigma$ of the SM predictions~\cite{Tanabashi:2018oca}; therefore,
studying the one-loop corrections to  $Z \to b\bar{b}$ can be used to constrain New Physics.

Haber and Logan \cite{Haber:1999zh} have already addressed this case in models with extra scalars in any representation of the gauge group $\mathrm{SU(2)_L}$.
Their work has been used to constrain a multiplicity of models, e.g., 
2HDMs \cite{Dorsch:2013wja, Basler:2016obg,
Krause:2016xku, Fontes:2015gxa, Mader:2012pm, Belusca-Maito:2016dqe},
the Georgi--Machacek model\cite{Campbell:2016zbp, Hartling:2014aga,
Hartling:2014zca, Chiang:2014bia, Degrande:2015xnm},
scotogenic models \cite{Tang:2017rhv} and
models with $\mathrm{SU(2)_L}$ singlet scalars \cite{vonBuddenbrock:2018xar, Han:2017etg}; 
it was also used in fitting programs\cite{Flacher:2008zq, Haller:2018nnx}.
In this chapter, we improve their analysis
in several aspects: first, we consider CP-violating scalar sectors; second, we provide explicit expressions that a model should verify in order for the divergences of the one-loop $Z \to b\bar{b}$ to cancel; third, we introduce a practical formalism to write the results in models with singlets and doublets, which is possible due to a convenient parameterization that was introduced in refs.~\cite{Grimus:1989pu,Grimus:2002ux, Grimus:2007if, Grimus:2008nb}; finally, we apply the results to a particular CP-violating model (the C2HDM), where we ascertain the validity of common approximations---in particular, neglecting both $m_{\mathrm{Z}}$ and the contributions of the neutral scalars.


\section{\label{Chap-Lavou:sec:calc}The one-loop calculation}


We start by considering a generic MHM with $n$ complex charged scalars $H_a^\pm$ and $m$ real neutral scalars $S_l^0$, being in everything else equal to the SM.
In this notation, $H_1^{\pm}$ and $S_1^0$ are the would-be Goldstone bosons $G^{\pm}$ and $G^0$, respectively.
We approximate the CKM matrix such that $V_{tb} = 1$,
which leads us to consider only the quarks bottom with mass $m_b$ and
top with mass $m_t$. Besides, we neglect $m_b$ in the propagators and loop
functions, but not in the couplings (which will be justified below).
As usual in this thesis, we use the conventions of ref.~\cite{Romao:2012pq},
with all $\eta$'s positive.

\subsection{\label{Chap-Lavou:subsec:coup}Couplings}

The relevant interactions between gauge bosons and quarks can be parameterized as:
\bea
\mathcal{L}_{Zbb} &=& -
\frac{e}{s_{\mathrm{w}} c_{\mathrm{w}}}\, Z_\mu\, \bar b\, \gamma^\mu
\left( g_{\mathrm{L}b} \gamma_{\mathrm{L}} + g_{\mathrm{R}b} \gamma_{\mathrm{R}} \right) b,
\label{Chap-Lavou:Zbb}
\\[1mm]
\mathcal{L}_{Ztt} &=& -
\frac{e}{s_{\mathrm{w}} c_{\mathrm{w}}}\, Z_\mu \, \bar t\, \gamma^\mu
\left( g_{\mathrm{L}t} \gamma_{\mathrm{L}} + g_{\mathrm{R}t} \gamma_{\mathrm{R}} \right) t,
\label{Chap-Lavou:Ztt}
\\[1mm]
\mathcal{L}_{Wtb} &=&  -
\frac{e}{\sqrt{2} s_{\mathrm{w}}}\, 
\left( \bar{t} \gamma^\mu \gamma_{\mathrm{L}} b\, W_\mu^+
+ \bar{b} \gamma^\mu \gamma_{\mathrm{L}} t\, W_\mu^-
\right),
\label{Chap-Lavou:Wtb}
\eea
in such a way that the couplings $g_{\mathrm{L}b}$, $g_{\mathrm{R}b}$, $g_{\mathrm{L}t}$ and $g_{\mathrm{R}t}$ read:
\be
g_{\mathrm{L}b} = \frac{s_{\mathrm{w}}^2}{3} - \frac{1}{2},
\quad \quad
g_{\mathrm{R}b} = \frac{s_{\mathrm{w}}^2}{3},
\quad \quad
g_{\mathrm{L}t} = \frac{1}{2} - \frac{2 s_{\mathrm{w}}^2}{3},
\quad \quad
g_{\mathrm{R}t} = - \frac{2 s_{\mathrm{w}}^2}{3}.
\label{Chap-Lavou:difference}
\ee
To parameterize the interaction between scalars and quarks, we introduce the $n$-dimensional complex vector $c_a$, the $n$-dimensional complex vector $d_a$ and the $m$-dimensional complex vector $r_l$, such that:
\bea
\mathcal{L}_{Htb} &=&
\sum_{a=1}^n \left[
H_a^+\, \bar t \left( c_a^\ast \gamma_{\mathrm{L}} - d_a \gamma_{\mathrm{R}} \right) b + H_a^-\, \bar b \left( c_a \gamma_{\mathrm{R}} - d_a^\ast \gamma_{\mathrm{L}} \right) t
\right],
\label{Chap-Lavou:Htb}
\\ 
\mathcal{L}_{Sbb} &=& \sum_{l=1}^m S_l^0\,
\bar b \left( r_l \gamma_{\mathrm{R}} + r_l^\ast \gamma_{\mathrm{L}} \right) b.
\label{Chap-Lavou:Sbb}
\eea
Finally, we parameterize the interaction between scalars and the $Z$ gauge boson by introducing the $n \times n$ hermitian matrix $X$, the $m \times m$ real and antisymmetric matrix $Y$ and the $m$-dimensional real vector $y$, such that:
\bea
\mathcal{L}_{ZHH} &=&  -
\frac{e}{s_{\mathrm{w}} c_{\mathrm{w}}}\, Z_\mu
\sum_{a, a^\prime = 1}^n X_{a a^\prime}
\left( H_a^+\, i \partial^\mu H_{a^\prime}^-
- H_{a^\prime}^-\, i \partial^\mu H_a^+ \right),
\label{Chap-Lavou:ZHH}
\\ 
\mathcal{L}_{ZSS} &=&
\frac{ i e}{s_{\mathrm{w}} c_{\mathrm{w}}}\, Z_\mu
\sum_{l, l^\prime = 1}^m Y_{l l^\prime}
\left( S_l^0\, i \partial^\mu S_{l^\prime}^0
- S_{l^\prime}^0\, i \partial^\mu S_l^0 \right),
\label{Chap-Lavou:ZSS}
\\
\mathcal{L}_{ZZS} &=&
\frac{e \, m_{\mathrm{Z}}}{2 s_{\mathrm{w}} c_{\mathrm{w}}}\, Z_\mu Z^\mu \sum_{l= 1}^m y_l S_l^0.
\label{Chap-Lavou:ZZS}
\eea

\subsection{\label{Chap-Lavou:subsec:1loop}One-loop contribution: overview}

\begin{figure}[h!]
\centering
\begin{tabular}{cc}
\includegraphics[scale=0.8]{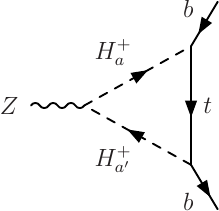}&\hskip 5mm
\includegraphics[scale=0.8]{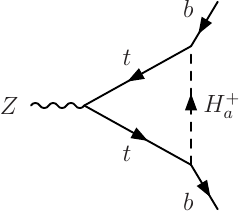}\\
a) & \hskip 5mm b)   
\end{tabular}
\caption{Two diagrams with charged scalars contributing to
$Z \to b\bar{b}$.}
\label{Chap-Lavou:fig:Generic-Charged}
\end{figure}

\begin{figure}[h!]
\centering
\begin{tabular}{cc}
\includegraphics[scale=0.8]{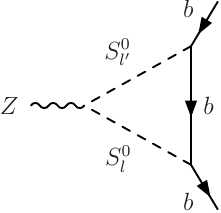}&\hskip 5mm
\includegraphics[scale=0.8]{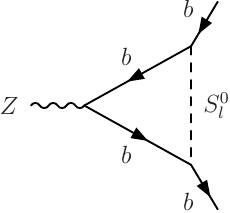}\\
a) & \hskip 5mm b)
\end{tabular}
\caption{Two diagrams with neutral scalars contributing to
$Z \to b\bar{b}$.}
\label{Chap-Lavou:fig:Generic-Neutral}
\end{figure}

\begin{figure}[h!]
\centering
\begin{tabular}{cccc}
\includegraphics[scale=0.8]{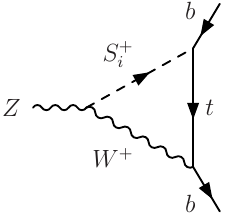}&\hskip 5mm
\includegraphics[scale=0.8]{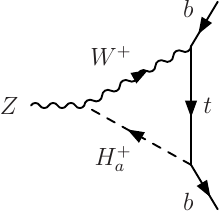}&\hskip 5mm
\includegraphics[scale=0.8]{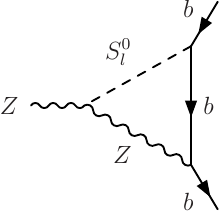}&\hskip 5mm
\includegraphics[scale=0.8]{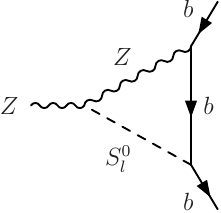} \\
i) & \hskip 5mm ii) & \hskip 5mm iii) & \hskip 5mm iv)
\end{tabular}
\caption[Diagrams referred as to ``type d)''.]{Diagrams referred as to ``type d)'' in  ref.~\cite{Haber:1999zh}.} 
\label{Chap-Lavou:fig:type_d)}
\end{figure}

We are looking for corrections---arising from scalar particles---that modify the tree-level couplings $g_{\mathrm{L}b}$ and $g_{\mathrm{R}b}$. Hence, we define the quantities $\Delta g_{\mathrm{L}b}$ and $\Delta g_{\mathrm{R}b}$ as the renormalized one-loop corrections that involve scalar particles, including the would-be Goldstone bosons and the already-observed neutral scalar with mass 125\,GeV. We thus write the renormalized effective couplings as:
\be
g_{\mathrm{L}b}^{\mathrm{eff}} = g_{\mathrm{L}b} + \Delta g_{\mathrm{L}b},
\qquad
g_{\mathrm{R}b}^{\mathrm{eff}} = g_{\mathrm{R}b} + \Delta g_{\mathrm{R}b}.
\label{Chap-Lavoura:eq:rutol}
\ee
In Figs. \ref{Chap-Lavou:fig:Generic-Charged}, \ref{Chap-Lavou:fig:Generic-Neutral} and \ref{Chap-Lavou:fig:type_d)}, we represent all the (non-renormalized) one-loop diagrams involving scalar particles.
We follow the terminology of Haber and Logan \cite{Haber:1999zh} closely. In particular, we identify the first diagram of figures \ref{Chap-Lavou:fig:Generic-Charged} and \ref{Chap-Lavou:fig:Generic-Neutral} as type a) and the second one as  type b). Haber and Logan also classify the diagrams of fig. \ref{Chap-Lavou:fig:type_d)} as type d), but neglected them.
Here, we shall ignore the diagrams \ref{Chap-Lavou:fig:type_d)} i) and ii), for two reasons: first, they are convergent; second, in models with only scalar singlets and doublets (which we are primarly interested in), they do not give new contributions beyond the SM, as there are no $Z W^\pm H_a^\mp$ couplings in those models (other than the $Z W^\pm G^\mp$ already present in the SM).
As for the diagrams \ref{Chap-Lavou:fig:type_d)} iii) and iv), they are proportional to $m_b$;%
\fn{This can be explained as follows. The coupling of the $Z$ to the bottom quarks conserves chirality (i.e. the $\bar{b}$ and the $b$ in eq. \ref{Chap-Lavou:Zbb} necessarily have the same chirality, due to the matrix $\gamma^{\mu}$); the same happens in the one-loop corrections to $Z \to b\bar{b}$, so that the incoming and the outgoing $b$ quarks have the same chirality. Now, in the diagrams in fig. \ref{Chap-Lavou:fig:type_d)} iii) and iv), besides the chirality-conserving coupling $Z \bar{b} b$, there is the coupling $S \bar{b} b$, which violates chirality (as there is no $\gamma^{\mu}$ in eq. \ref{Chap-Lavou:Sbb}). So, there must be a mass insertion in the internal $b$ propagator in order to change the chirality again.}
hence, and since they are also convergent, one may neglect them by taking $m_b=0$ (which is what was done in ref.~\cite{Haber:1999zh}).
Nevertheless, since $m_b$ could appear multiplied by a large coefficient,%
\fn{Such as $\tan{\beta}$ in the $\mathbbm{Z}_2$-symmetric 2HDM,
see for instance table 2 in ref.~\cite{Branco:2011iw}.}
we will also present their calculation in order to
check the validity of this approximation.

\begin{figure}[h!]
\centering
\begin{tabular}{ccc} 
\includegraphics[width=0.2\textwidth]{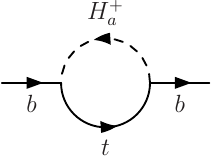}&\hskip 35mm&
\includegraphics[width=0.2\textwidth]{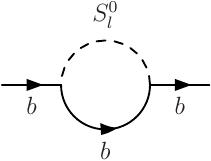}\\[+1mm]
i) && ii)
\end{tabular}
\caption{Contribution of the charged and neutral scalars
to the self-energy of the bottom quark,
leading to ``type c)'' contributions to the vertex.}
\label{Chap-Lavou:fig:Self-Energies}
\end{figure}

For the renormalization, we follow the usual procedure, which involves identifying the original quantities of the theory (parameters and fields) as bare quantities, and separating them into renormalized quantities and counterterms.%
\fn{For details, see chapters \ref{Chap-Selec} and \ref{Chap-Reno}.}
%
%
In the end, several counterterms contribute to the process $Z \to b\bar{b}$. However, in order to ensure the finiteness of the diagrams that interest us here, it is enough to consider the fermionic field counterterms; more specifically, it is enough to consider the contributions of the scalar particles to those counterterms.
%
%
We follow ref.~\cite{Haber:1999zh} in neglecting finite corrections coming from other counterterms.
We calculate the fermionic field counterterms using OSS, which fixes them in terms of one-loop fermionic 2-point functions.%
\fn{For details, cf. section \ref{Chap-Reno:sec:fermionsCT}.}
As we just mentioned, we consider only the contributions of the scalar particles (both charged and neutral), shown in fig. \ref{Chap-Lavou:fig:Self-Energies}.
We follow once again the classification of Haber and Logan, and identify the contribution of these diagrams to the effective vertex (via fermionic fields counterterms) as type c).
Using the conventions of section \ref{Chap-Reno:sec:fermionsCT}, we find:%
\fn{The second term on the r.h.s. of eqs. \ref{Chap-Reno:eq:reno-ferm-diag-L} and \ref{Chap-Reno:eq:reno-ferm-diag-R} is finite; therefore, and since it is proportional to $m_b$, we ignore it. Note, moreover, that we do not need to consider the operator $\widetilde{\operatorname{Re}}$ in this case, as there are no absorptive parts in the diagrams of fig. \ref{Chap-Lavou:fig:Self-Energies}.
Finally, although the quantities $\Delta g_{\aleph b}$ (with $\aleph = \{\mathrm{L},\mathrm{R}\}$) are finite (as they are the renormalized corrections), their different components ($\Delta g_{\aleph b}(i)$, for the diagrams of the generic type $i$) are in general divergent.}
\be
\Delta g_{\mathrm{L}b} \left( c \right)
=
- g_{\mathrm{L}b}\, \Sigma^{f,\mathrm{L}}_{bb} \left(m_b^2 \right),
\qquad
\Delta g_{\mathrm{R}b} \left( c \right)
=
- g_{\mathrm{R}b}\, \Sigma^{f,\mathrm{R}}_{bb} \left(m_b^2 \right).
\ee
Our calculations have been performed with \FM, by considering a model with a sufficiently large number of scalar particles, and by deriving general expressions from it. In this way, \FMS is a very useful tool also for investigating generic models.

\subsection{\label{Chap-Lavou:subsec:calc-Hpm}Calculation of the diagrams
involving charged scalars}

The diagrams in fig. \ref{Chap-Lavou:fig:Generic-Charged}a) lead to:%
\fn{We have set $m_b=0$ inside all the Passarino--Veltman functions;
when evaluating them numerically, however, it is convenient to keep $m_b \neq 0$ to avoid numerical instabilities.
Note, moreover, that the sums in eqs. \ref{Chap-Lavou:jdoipsa}
start at $a=1$ (i.e. they include the charged would-be Goldstone bosons $G^\pm$). However, one may show that $X_{1a}=X_{a1}=0$ for $a \neq 1$, so that the sum in eq. (\ref{Chap-Lavou:eq:15333}) may start at $a, a^\prime = 2$, while the term with $a = a^\prime = 1$ is separately included in the SM contribution.
Finally, the one-loop results for $\Delta g_{\mathrm{L}b}$ and $\Delta g_{\mathrm{R}b}$ have absorptive (i.e. imaginary) parts. If there are no scalars with mass below $m_{\mathrm{Z}}/2$, the imaginary parts only appear through cuts of the internal bottom-quark lines of fig. \ref{Chap-Lavou:fig:Generic-Neutral}b), thus affecting only the
contributions with neutral scalars. Yet, although those imaginary parts
may be of the same order of magnitude as the real parts, they
can be ignored here; indeed, the observables will depend on, for example,
\be
\left| g^{\mathrm{eff}}_{\mathrm{L}b} \right|^2
=
\left| g_{\mathrm{L}b} + \Delta g_{\mathrm{L}b} \right|^2
=
\left| g_{\mathrm{L}b} \right|^2
+ 2\, \textrm{Re} \left( g_{\mathrm{L}b}\, \Delta g_{\mathrm{L}b}^\ast \right)
+ \mathcal{O} \left( \Delta g_{\mathrm{L}b}^2 \right)
=
\left| g_{\mathrm{L}b} \right|^2
+ 2\, g_{\mathrm{L}b}\, \textrm{Re}\left( \Delta g_{\mathrm{L}b} \right)
+ \mathcal{O} \left( \Delta g_{\mathrm{L}b}^2 \right),
\ee
where the last equality follows from the fact that $g_{\mathrm{L}b}$ is real.  As a result, the impact of an imaginary $\Delta g_{\mathrm{L}b}$ on the observables effectively appears only at higher order.
}
\bs
\label{Chap-Lavou:jdoipsa}
\bea
\Delta g_{\mathrm{L}b} \left( a \right) &=&
\frac{1}{8\pi^2}
\sum_{a,a'=1}^{n} c_a X_{aa'} c_{a'}^\ast\,
C_{00} \left( m_{\mathrm{Z}}^2, 0, 0, m_{a'}^2, m_a^2, m_t^2 \right),
\label{Chap-Lavou:eq:15333}
\\
\Delta g_{\mathrm{R}b} \left( a \right) &=&
\Delta g_{\mathrm{L}b} \left( a \right) \left( c_a \rightarrow d_a^\ast \right),
\label{Chap-Lavou:eq:154}
\eea
\es
where $m_a$ denotes the mass of $H_a^\pm$.
The diagrams in fig. \ref{Chap-Lavou:fig:Generic-Charged}b) lead to:%
\fn{We took into account the fact that $\left( d - 2 \right) C_{00} \left( \dots \right) = 2\, C_{00} \left( \dots \right) - 1/2$, with
$d$ being the dimension of space-time.}
\bs
\bea
\Delta g_{\mathrm{L}b} \left( b \right) &=&
\frac{1}{16 \pi^2} \sum_{a=1}^{n} \left| c_a \right|^2
\bigg\{
{\vbox to 14 pt{}}
- m_t^2\, g_{\mathrm{L}t}\, C_0 \left( 0, m_{\mathrm{Z}}^2, 0, m_a^2, m_t^2, m_t^2 \right)
\nonumber \\ & &
+ g_{\mathrm{R}t} \bigg[
{\vbox to 12 pt{}}
2\, C_{00} \left( 0, m_{\mathrm{Z}}^2, 0, m_a^2, m_t^2, m_t^2 \right)
- \frac{1}{2} - m_{\mathrm{Z}}^2\, C_{12} \left( 0, m_{\mathrm{Z}}^2, 0, m_a^2, m_t^2, m_t^2 \right)
{\vbox to 12 pt{}}
\bigg]
{\vbox to 14 pt{}}
\bigg\} ,
\label{Chap-Lavou:eq:164a}
\\*[1mm] 
\Delta g_{\mathrm{R}b} \left( b \right) &=& \Delta g_{\mathrm{L}b} \left( b \right)
\left( c_a \rightarrow d_a,\ g_{\mathrm{L}t} \leftrightarrow g_{\mathrm{R}t} \right).
\label{Chap-Lavou:eq:164b}
\eea
\es
%

As for the type c) contributions,
arising from fig. \ref{Chap-Lavou:fig:Self-Energies} i), we find:
\bs
\label{Chap-Lavou:eq:158}
\bea
\Delta g_{\mathrm{L}b} \left( c \right) &=&
\frac{g_{\mathrm{L}b}}{16 \pi ^2}
\sum_{a=1}^{n} \left| c_a \right|^2 
B_1 \left( 0, m_t^2, m_a^2 \right),
\label{Chap-Lavou:eq:157}
\\
\Delta g_{\mathrm{R}b} \left( c \right) &=&
\Delta g_{\mathrm{L}b} \left( c \right) \left( c_a \rightarrow d_a,\
g_{\mathrm{L}b} \to g_{\mathrm{R}b} \right).
\label{Chap-Lavou:eq:15777}
\eea
\es
As mentioned before, we do not consider the contributions from the diagrams \ref{Chap-Lavou:fig:type_d)} i) and ii).
In the CP-conserving limit,
eqs. \ref{Chap-Lavou:jdoipsa}--\ref{Chap-Lavou:eq:158} agree with eqs.~(4.1)
of ref.~\cite{Haber:1999zh}.

The functions $B_1$ and $C_{00}$ are divergent;
all the other Passarino--Veltman functions appearing in this section are finite. Defining $\Delta_{\varepsilon}$ as in eq. \ref{Chap-Reno:Delta-eps}, one has \cite{Denner:2005nn}:
\be
B_1 \left( ... \right) = - \frac{\Delta_{\varepsilon}}{2} +
\mathrm{finite\ terms},
\qquad
C_{00} \left( ... \right) = \frac{\Delta_{\varepsilon}}{4} + \mathrm{finite\ terms}.
\label{Chap-Lavou:eq:4}
\ee
We can thus conclude that the terms in eqs. \ref{Chap-Lavou:jdoipsa}--\ref{Chap-Lavou:eq:158} proportional to $\Delta_{\varepsilon}$ are:
\bs
\label{Chap-Lavou:eq:166}
\begin{gather}
\Big\{
\Delta g_{\mathrm{L}b} \left( a \right) + \Delta g_{\mathrm{L}b} \left( b \right) + \Delta g_{\mathrm{L}b} \left( c \right)
\Big\}\Big|_{\Delta_{\varepsilon}}
=
\frac{\Delta_{\varepsilon}}{32 \pi^2}
\left[
\sum_{a,a'=1}^{n} c_a X_{aa'} c_{a'}^\ast
+ \left( g_{\mathrm{R}t} - g_{\mathrm{L}b} \right) \sum_{a=1}^{n} \left| c_a \right|^2
\right], \\
\Big\{
\Delta g_{\mathrm{R}b} \left( a \right) + \Delta g_{\mathrm{R}b} \left( b \right) + \Delta g_{\mathrm{R}b} \left( c \right)
\Big\}\Big|_{\Delta_{\varepsilon}}
=
\frac{\Delta_{\varepsilon}}{32 \pi^2} \left[
\sum_{a,a'=1}^{n} d_a^\ast X_{aa'} d_{a'}
+ \left( g_{\mathrm{L}t} - g_{\mathrm{R}b} \right) \sum_{a=1}^{n} \left| d_a \right|^2
\right].
\end{gather}
\es
Therefore, and using eq. \ref{Chap-Lavou:difference}, the divergences cancel if and only if:
\bs
\label{Chap-Lavou:eq:165}
\bea
\sum_{a,a'} c_a X_{aa'} c_{a'}^\ast
&=&
\frac{s_{\mathrm{w}}^2 - c_{\mathrm{w}}^2}{2} \sum_{a} \left| c_a \right|^2,
\label{Chap-Lavou:eq:165a}
\\
\sum_{a,a'} d_a^\ast X_{aa'} d_{a'}
&=&
\frac{s_{\mathrm{w}}^2 - c_{\mathrm{w}}^2}{2} \sum_{a} \left| d_a \right|^2.
\label{Chap-Lavou:eq:165b}
\eea
\es

\subsection{\label{Chap-Lavou:subsec:calc-S0}Calculation the diagrams
involving neutral scalars}

The diagrams in fig. \ref{Chap-Lavou:fig:Generic-Neutral}a) lead to:
\bs
\label{Chap-Lavou:eq:153}
\bea 
\Delta g_{\mathrm{L}b} \left( a \right) &=& 
\frac{i}{4\pi^2}
\sum_{l,l'=1}^{m} r_l Y_{ll'} r_{l'}^\ast\,
C_{00} \left( 0, m_{\mathrm{Z}}^2, 0, 0, m_{l'}^2, m_l^2 \right),
\label{Chap-Lavou:eq:153N}
\\
\Delta g_{\mathrm{R}b} \left( a \right) &=&
\Delta g_{\mathrm{L}b} \left( a \right) \left( r_l \rightarrow r_l^\ast \right),
\label{Chap-Lavou:eq:154N}
\eea
\es
where $m_l$ denotes the mass of $S^0_l$.
The diagrams in fig. \ref{Chap-Lavou:fig:Generic-Neutral}b) lead to:
\bs
\begin{align}
\Delta g_{\mathrm{L}b} \left( b \right) =&\, \frac{g_{\mathrm{R}b}}{16 \pi^2}
\sum_{l=1}^{m} \left| r_l \right|^2 
\bigg[ 2\, C_{00} \left( 0, m_{\mathrm{Z}}^2, 0, m_l^2, 0, 0 \right)  - \frac{1}{2} - m_{\mathrm{Z}}^2\, C_{12} \left( 0, m_{\mathrm{Z}}^2, 0, m_l^2, 0, 0 \right)
\bigg],
\label{Chap-Lavou:eq:155b}
\\
\Delta g_{\mathrm{R}b} \left( b \right) =&\,
\Delta g_{\mathrm{L}b} \left( b \right) \left( g_{\mathrm{R}b} \to g_{\mathrm{L}b} \right).
\label{Chap-Lavou:eq:156b}
\end{align}
\es
As for the type c) contributions, arising from fig. \ref{Chap-Lavou:fig:Self-Energies} ii), we find:
\bs
\label{Chap-Lavou:eq:158N}
\bea
\Delta g_{\mathrm{L}b} \left( c \right) &=&
\frac{g_{\mathrm{L}b}}{16 \pi^2}
\sum_{l=1}^{m} \left| r_l \right|^2  B_1 \left( 0, 0, m_l^2 \right),
\label{Chap-Lavou:eq:157N}
\\
\Delta g_{\mathrm{R}b} \left( c \right) &=&
\Delta g_{\mathrm{L}b} \left( c \right) \left( g_{\mathrm{L}b} \rightarrow g_{\mathrm{R}b} \right).
\label{Chap-Lavou:eq:159N}
\eea
\es
In the CP-conserving limit,
eqs. \ref{Chap-Lavou:eq:153}--\ref{Chap-Lavou:eq:158N} agree with eqs. (5.1)
of ref.~\cite{Haber:1999zh}.

Collecting all the divergent terms in eqs. \ref{Chap-Lavou:eq:153N},
\ref{Chap-Lavou:eq:155b} and \ref{Chap-Lavou:eq:157N}, we find:
\bea
& &
\Big\{
\Delta g_{\mathrm{L}b} \left( a \right)
+ \Delta g_{\mathrm{L}b} \left( b \right)
+ \Delta g_{\mathrm{L}b} \left( c \right)
\Big\}\Big|_{\Delta_{\varepsilon}}
=
\frac{\Delta_{\varepsilon}}{32 \pi^2}
\left[ 2 i \sum_{l,l'=1}^{m} r_l Y_{ll'} r_{l'}^\ast 
+ \left( g_{\mathrm{R}b} - g_{\mathrm{L}b} \right) \sum_{l=1}^{m} \left| r_l \right|^2
\right].
\label{Chap-Lavou:eq:166N}
\eea
Hence, given eq. \ref{Chap-Lavou:difference}, we conclude that a consistent theory requires:%
\fn{The same condition can also be obtained by collecting all the
divergent terms in eqs. \ref{Chap-Lavou:eq:154N}, \ref{Chap-Lavou:eq:156b},
and \ref{Chap-Lavou:eq:159N}.}
\be
\sum_{l,l'=1}^{m} r_l Y_{ll'} r_{l'}^*
=
\frac{i}{4}
\sum_{l} |r_l|^2.
\label{Chap-Lavou:eq:consistent_N}
\ee

Finally, we calculate the diagrams in fig. \ref{Chap-Lavou:fig:type_d)} iii) and iv). As mentioned above, although they are suppressed by $m_b$, they might be enhanced when the coupling of neutral scalars to the bottom quark gets enhanced. They lead to:
\bs
\bea
\Delta g_{\mathrm{L}b} \left( d \right) &=&
\frac{e \, m_b \, m_{\mathrm{Z}}}{8 \pi^2 s_{\mathrm{w}} c_{\mathrm{w}}}
\sum_{l=1}^{m} y_l\, \mathrm{Re} [r_l]
\bigg\{
{\vbox to 14 pt{}}
g_{\mathrm{L}b} \Big[ C_0 \left( m_{\mathrm{Z}}^2, 0, 0, m_{\mathrm{Z}}^2, m_l^2, 0 \right)
- C_1 \left( m_{\mathrm{Z}}^2, 0, 0, m_{\mathrm{Z}}^2, m_l^2, 0 \right) \Big] \nonumber \\
&& \hskip 36mm + g_{\mathrm{R}b}\, C_1 \left( m_{\mathrm{Z}}^2, 0, 0, m_l^2, m_{\mathrm{Z}}^2, 0 \right)
{\vbox to 14 pt{}} \bigg\}, \hspace*{5mm}
\label{Chap-Lavou:eq:16cc}
\\
\Delta g_{\mathrm{R}b} \left( d \right) &=&
\Delta g_{\mathrm{L}b} \left( d \right) \left( g_{\mathrm{L}b} \leftrightarrow
g_{\mathrm{R}b} \right).
\label{Chap-Lavou:eq:17cc}
\eea
\es
%

\section{\label{Chap-Lavou:sec:doubsing}Models with doublet and singlet scalars}

We now slightly restrict the spectrum of the previous section, and consider the more specific case of extensions of the SM with $n_d$ scalar doublets, $n_c$ singly-charged scalar $\mathrm{SU(2)_L}$ singlets and $n_n$ real scalar
gauge-invariant fields.%
\fn{The particle content is then $n \equiv n_d + n_c$ (complex) charged scalars $H_a^\pm$ and $m \equiv 2 n_d + n_n$ (real) neutral scalars $S_l^0$. As before, this counting includes the would-be Goldstone bosons $H_1^\pm = G^\pm$ and $S_1^0 = G^0$.  Without loss of generality, one may assume that the scalar with mass 125\,GeV found at the LHC is $S^0_2$; the generality is lost if one makes the further assumption that the masses are ordered, since there might be massive scalars below 125\,GeV.}

\subsection{Formalism}
\label{Chap-Lavoura:sec:form}

We use the formalism of refs. \cite{Grimus:2002ux, Grimus:2007if, Grimus:2008nb, Grimus:1989pu} (cf. also ref. \cite{Fontes:2019fbz}).
We thus parameterize the $n_d$ doublets as:
\be
\Phi_k = \left( \begin{array}{c} \varphi_k^+ \\*[1mm] \varphi_k^0 \end{array}
\right),
\quad \quad
\tilde{\Phi}_k \equiv i \sigma_2 \Phi_k^\ast
= \left( \begin{array}{c} {\varphi_k^0}^\ast \\*[1mm] - \varphi_k^- \end{array}
\right),
\ee
and we assume that the fields $\varphi_k^0$ have vevs $v_k \left/ \sqrt{2} \right.$, where the $v_k$ may be complex.
In order to write $\varphi_k^+$ and $\varphi_k^0$ as superpositions of the physical (i.e. eigenstates of mass) fields $H^+_a$ and $S_l^0$, we introduce the $n_d \times n$ matrix $\mathcal{U}$ and the $n_d \times m$ matrix $\mathcal{V}$, such that:%
\fn{Since $H_1^\pm$ and $S_1^0$ are would-be Goldstone bosons,
the first columns of $\mathcal{U}$ and $\mathcal{V}$ are fixed and given by
$\mathcal{U}_{k1} = v_k/v$ and $\mathcal{V}_{k1} = i v_k/v$, respectively, with $v^2 \equiv \sum_{k=1}^{n_d} |v_k|^2$
($v$ is real and positive by definition).}
\be
\varphi_k^+ = \sum_{a=1}^n \mathcal{U}_{ka} H_a^+,
\qquad
\varphi_k^0 = \frac{1}{\sqrt{2}}
\left( v_k + \sum_{l=1}^m \mathcal{V}_{kl} S_l^0 \right).
\label{Chap-Lavou:def:U&V}
\ee
Besides, we define an $n \times n$ unitary matrix $\tilde{\mathcal{U}}$,
\be
\tilde{\mathcal{U}} =
\left( \begin{array}{c} \mathcal{U} \\ \mathcal{T} \end{array} \right),
\label{Chap-Lavou:tildeU}
\ee
where the matrix $\mathcal{T}$ only exists when $n_c \neq 0$.
We also define an $m \times m$ real and orthogonal matrix $\tilde{\mathcal{V}}$,
\be
\tilde{\mathcal{V}} =
\left( \begin{array}{c} \mathrm{Re}\, \mathcal{V} \\
\mathrm{Im}\, \mathcal{V} \\
\mathcal{R} \end{array} \right),
\label{Chap-Lavou:jdisos}
\ee
where the matrix $\mathcal{R}$ only exists when $n_n \neq 0$.
Finally, to parameterize the Yukawa Lagrangian, we introduce the $n_d$-dimensional complex vectors $e$ and $f$, such that:%
\fn{It goes without saying that charged and neutral scalar singlets have no Yukawa couplings.}
\be
\begin{split}
\mathcal{L}_\textrm{Yukawa} = -
\left( \begin{array}{cc} \bar{t}_{\mathrm{L}} & \bar{b}_{\mathrm{L}} \end{array} \right)
\sum_{k=1}^{n_d} \Bigg[
& f_k
\left( \begin{array}{c} \varphi_k^+ \\*[2mm]
\varphi_k^0 \end{array} \right) b_{\mathrm{R}}
+ e_k
\left( \begin{array}{c} {\varphi_k^0}^\ast \\*[1mm]
- \varphi_k^-
\end{array} \right) t_{\mathrm{R}} \Bigg] + \textrm{h.c.}.
\label{Chap-Lavou:jcksps}
\end{split}
\ee

\subsection{Connection with the general description}
\label{Chap-Lavoura:sec:connec-gen}

We can compare the new formalism to the one introduced in section \ref{Chap-Lavou:subsec:coup} above.
For the Yukawa couplings, we can use eqs.~\ref{Chap-Lavou:Htb},
\ref{Chap-Lavou:Sbb}, and \ref{Chap-Lavou:def:U&V}--\ref{Chap-Lavou:jcksps} to conclude that:
\be
c_a = \sum_{k=1}^{n_d} \mathcal{U}_{ka}^\ast e_k,
\qquad
d_a = \sum_{k=1}^{n_d} \mathcal{U}_{ka} f_k,
\qquad
r_l = - \frac{1}{\sqrt{2}} \sum_{k=1}^{n_d} \mathcal{V}_{kl} f_k.
\ee
As for the couplings between $Z$ and scalars, one can show that, in this class of models \cite{Grimus:2007if},%
\fn{The last relation leads to $y_{l=1} = 0$, because $\mathcal{V}^\dagger \mathcal{V}$ is hermitian, which implies $\textrm{Im} \left( \mathcal{V}^\dagger \mathcal{V} \right)_{11} = 0$. Therefore, the sum in eq. \ref{Chap-Lavou:ZZS} really starts at $l=2$, as there is no vertex $Z Z G^0$.}
\bs
\bea
X_{a a^\prime} &=&  s_{\mathrm{w}}^2 \delta_{a a^\prime} -
\frac{\left( \mathcal{U}^{\mathrm{T}} \mathcal{U}^\ast \right)_{a a^\prime}}{2}
=
\frac{s_{\mathrm{w}}^2 - c_{\mathrm{w}}^2}{2}\, \delta_{a a^\prime}
+ \frac{\left( \mathcal{T}^{\mathrm{T}} \mathcal{T}^\ast \right)_{a a^\prime}}{2},
\label{Chap-Lavou:Xaa2} \\
Y_{l l^\prime} &=&  -
\frac{1}{4}\,
\textrm{Im} \left( \mathcal{V}^\dagger \mathcal{V} \right)_{l l^\prime},\\
y_l &=& - \textrm{Im} \left( \mathcal{V}^\dagger \mathcal{V} \right)_{1l}.
\label{Chap-Lavou:XY}
\eea
\es
%

\section{\label{Chap-Lavou:sec:exp}Connection with experiment}

Before we apply the new formalism to a particular case, we discuss how to relate the theoretical calculations with experimental observations (for details, cf. ref.~\cite{Tanabashi:2018oca} and references therein).
The effective couplings $g_{\mathrm{L}b}^{\mathrm{eff}}$ and $g_{\mathrm{R}b}^{\mathrm{eff}}$ in eq. \ref{Chap-Lavoura:eq:rutol}
may be determined experimentally from:
\begin{enumerate}
\item The rate $R_b$, given by:
\be
R_b =
\frac{\Gamma \left( Z \rightarrow b \bar{b} \right)}{\Gamma
\left( Z \rightarrow \textrm{hadrons} \right)}.
\label{Chap-Lavou:R_b}
\ee
\item Several asymmetries, including:
\begin{enumerate}
%
\item the $Z$-pole forward--backward asymmetry measured at LEP1,
\be
A_{\mathrm{FB}}^{(0,f)}
=
\frac{\sigma \left( e^- \rightarrow b_F \right)
- \sigma \left( e^- \rightarrow b_B \right)}{\sigma \left( e^-
\rightarrow b_F \right) + \sigma \left( e^- \rightarrow b_B \right)}
= \frac{3}{4} A_e  A_b,
\label{Chap-Lavou:A_FB}
\ee
where $b_F$ ($b_B$) stands for final-state bottom quarks
moving in the forward (backward) direction
with respect to the direction of the initial-state electron;
\item the left--right forward--backward asymmetry
measured by the SLD Collaboration,
\be
A_{\mathrm{LR}}^{\mathrm{FB}} \left( b \right) =  \frac{\sigma \! \left(e_L^- \! \rightarrow \! b_F \right)
\! - \sigma \left(e_L^- \! \rightarrow \! b_B \right)
\! - \sigma \left(e_R^- \! \rightarrow \! b_F \right)
\! + \sigma \left(e_R^- \! \rightarrow \! b_B \right)}{
\sigma \left(e_L^- \! \rightarrow \! b_F \right)
\! + \sigma \left(e_L^- \! \rightarrow \! b_B \right)
\! + \sigma \left(e_R^- \! \rightarrow \! b_F \right)
\! + \sigma \left(e_R^- \! \rightarrow \! b_B \right)}
= \frac{3}{4} A_b,
\label{Chap-Lavou:A_LRFB}
\ee
where $e_L^-$ ($e_R^-$) are initial-state left-handed (right-handed) electrons.
\end{enumerate}
\end{enumerate}
%



The recent fit of $R_b$ and $A_b$ to the electroweak data
by Erler and Freitas in ref.~\cite{Tanabashi:2018oca} finds:
\be
R_b^\textrm{fit} = 0.21629 \pm 0.00066,
\qquad
A_b^\textrm{fit} = 0.923 \pm 0.020,
\label{Chap-Lavou:uvido}
\ee
to be compared with the SM values:
\be
R_b^\textrm{SM} = 0.21582 \pm 0.00002,
\qquad
A_b^\textrm{SM} = 0.9347. 
\ee
Thus, the experimental $R_b$
is about $0.7\sigma$ above the SM value, while $A_b$ is about
$0.6\sigma$ below the SM value.  However, this good agreement only
applies to the overall fit of many observables producing
eqs. \ref{Chap-Lavou:uvido}; the measured values of the bottom-quark
asymmetries by themselves alone reveal a discrepancy close to $3\sigma$.%
\fn{As pointed out in ref.~\cite{Tanabashi:2018oca}, extracting $A_b$ from
$A_{\mathrm{FB}}^{(0,b)}$ and using $A_e = 0.1501 \pm 0.0016$ leads to a result which is $3.1 \sigma$ below the SM (the precise value of $A_b$ depends on which observables $A_e$ is extracted from), while combining
$A_{\mathrm{FB}}^{(0,b)}$ with $A_{\mathrm{LR}}^{\mathrm{FB}}$ leads to $A_b = 0.899 \pm 0.013$, which deviates from the SM value by $2.8 \sigma$.}
This suggests two possible approaches.
The first one consists in accepting the values in eq. \ref{Chap-Lavou:uvido} and using them as constraints on New Physics (NP).
The second approach, by contrast, seeks NP that might explain an $R_b$ just slightly above the SM, together with an $A_b$ that undershoots the SM by around $3\sigma$.

For what follows, it is convenient to switch from the parameterization defined in eq. \ref{Chap-Lavoura:eq:rutol} (which separates the tree-level contribution from the one-loop one) to a parameterization which separates the SM contribution from the BSM (or NP) one:
\be
g_{\mathrm{L} b}^{\mathrm{eff}} = g_{\mathrm{L} b}^\textrm{SM} + \delta g_{\mathrm{L} b},
\qquad
g_{\mathrm{R} b}^{\mathrm{eff}} = g_{\mathrm{R} b}^\textrm{SM} + \delta g_{\mathrm{R} b}.
\label{Chap-Lavou:delta_gLR}
\ee
The quantities in $\delta$ thus represent the NP contributions. This notation can also be applied to $R_b$ and $A_b$, thus yielding $\delta R_b$ and $\delta A_b$, which turn out to be related with $\delta g_{\mathrm{L}b}$ and $\delta g_{\mathrm{R}b}$ according to \cite{Haber:1999zh}:
\be
\delta R_b = - 0.7785\, \delta g_{\mathrm{L} b} + 0.1409\, \delta g_{\mathrm{R} b},
\qquad
\delta A_b = - 0.2984\, \delta g_{\mathrm{L} b} - 1.6234\, \delta g_{\mathrm{R} b}.
\label{Chap-Lavou:eq:truz}
\ee
%
%
Hence, if one wishes to follow the second approach mentioned above---using NP to keep $R_b$ close to its SM value and to reduce $A_b$ significantly---, one needs a small $\delta g_{\mathrm{L} b}$, together with a large (and positive) $\delta g_{\mathrm{R} b}$.


\section{A simple particular case: the C2HDM}
\label{Chap-Lavou:sec:particular}

We now investigate the application of the calculation described in section \ref{Chap-Lavou:sec:calc} to a particular case: the C2HDM, studied in detail in the previous chapters.%
\fn{Since the renormalization described in section \ref{Chap-Lavou:subsec:1loop} is simple, there is no need to resort to the unphysical parameters introduced in chapter \ref{Chap-Reno}.
}
In order to apply the one-loop corrections of sections \ref{Chap-Lavou:subsec:calc-Hpm} and \ref{Chap-Lavou:subsec:calc-S0}, we must determine the values that the quantities introduced in section \ref{Chap-Lavou:subsec:coup} take in the C2HDM. Such task can be simplified by determining first the values of the quantities of section \ref{Chap-Lavoura:sec:form} (which is the reason why the formalism of that section is so useful). This we do in section \ref{Chap-Lavoura:sec-exp} below; then, in section \ref{Chap-Lavoura:sec-results}, we present our numerical results.

\subsection{Expressions}
\label{Chap-Lavoura:sec-exp}

The first thing to notice is that, since there are no scalar singlets in the C2HDM, the description of section \ref{Chap-Lavoura:sec:form} becomes even more simplified, as the matrices $\mathcal{T}$ and $\mathcal{R}$ do not exist.
The matrix $\mathcal{U}$ can be easily obtained by comparing eqs. \ref{Chap-Real:eq:HiggsBasis}, \ref{Chap-Maggie:eq:2hdmdoubletexpansion}, \ref{Chap-Maggie:eq:Doublets} and \ref{Chap-Lavou:def:U&V}. To obtain $\mathcal{V}$, eq. \ref{Chap-Maggie:eq:c2hdmrot} also needs to be taken into account. The results are:
\bs
\label{Chap-Lavou:eq:700}
\bea
\mathcal{U} &=& \left( \begin{array}{cc}
c_{\beta} & - s_{\beta} \\ s_{\beta} & c_{\beta}
\end{array} \right),
\\
\mathcal{V} &=& \left( \begin{array}{cccc}
i c_\beta & R_{11} - i s_\beta R_{13} & R_{21} - i s_\beta R_{23} &
R_{31} - i s_\beta R_{33}
\\
i s_\beta & R_{12} + i c_\beta R_{13} & R_{22} + i c_\beta R_{23} &
R_{32} + i c_\beta R_{33}
\end{array} \right).
\eea
\es
To determine the Yukawa couplings $e$ and $f$, we assume the Type-II C2HDM.
Therefore, comparing eqs. \ref{Chap-Reno:eq:LYukawa-final} and \ref{Chap-Lavou:jcksps}, we directly find:
\be
e_1 = 0,
\qquad
f_2 = 0,
\qquad
e_2 =  \frac{\sqrt{2} m_t}{v_2},
\qquad
f_1 =  \frac{\sqrt{2} m_b}{v_1}.
\ee
Now that we have all the parameters of section \ref{Chap-Lavoura:sec:form}, we can trivially determine the general parameters of section \ref{Chap-Lavou:subsec:coup} using the relations of section \ref{Chap-Lavoura:sec:connec-gen}, and thus calculate the one-loop corrections $\Delta g_{\mathrm{L}b}$ and $\Delta g_{\mathrm{R}b}$ in the C2HDM. 
Particularly relevant for what follows are the values $c_2$ and $d_2$, given by:
\be
\label{Chap-Lavou:eq:207}
c_2 =  \frac{\sqrt{2} m_t}{v} \cot{\beta}, \quad \quad
d_2 = - \frac{\sqrt{2} m_b}{v} \tan{\beta}.
\ee
It is thus clear that $\left| c_2 \right|$ and $\left| d_2 \right|$ may be of vastly different orders of magnitude---in particular,
$\left| d_2 \right| \ll \left| c_2 \right|$ for $\tan{\beta} \sim 1$.
However, when $\tan{\beta} \gtrsim \sqrt{m_t / m_b} \approx 6$,
$\left| d_2 \right|$ becomes larger than $\left| c_2 \right|$,
which is the regime that we will be mostly interested in.

\subsection{Results}
\label{Chap-Lavoura:sec-results}

From the results for $\Delta g_{\mathrm{L}b}$ and $\Delta g_{\mathrm{R}b}$, we can easily obtain the NP contributions $\delta g_{\mathrm{L}b}$ and $\delta g_{\mathrm{R}b}$ introduced in eq. \ref{Chap-Lavou:delta_gLR}; as already seen, these allow a simple comparison with experiment, and are thus preferred in the results that are presented below.
We use superscripts $c$ and $n$ to denote the NP contributions to $\delta g_{\mathrm{L}b}$ and $\delta g_{\mathrm{R}b}$ coming from the charged scalar (in the C2HDM, just $H^{\pm}$) and neutral scalars, respectively. 
The calculations were performed with \FMS and converted to \textsc{LoopTools}~\cite{Hahn:1998yk} for numerical evaluation.
The numerical simulation follows the strategy described in section \ref{Chap-Maggie:sec:rest}---which selects, in particular, the range $0.8 \le \tan \beta  \le 35$.
We combine the results obtained from that run with those from a new run, characterized by the range $0 \le \tan \beta \le 100$. Such extreme (very low and very high) values of $\tan{\beta}$ may be in contradiction with certain Flavour Physics observables, notably $Z \rightarrow b \bar{b}$.%
\footnote{Cf. fig. \ref{Chap-Lavou:fig:RbAb-low_tb-2} below. Those extreme values of $\tan{\beta}$ (both high and low) also violate perturbative unitarity.}
Nevertheless, we will consider them in order to stress that the details of such a bound may require both the charged-scalar and the neutral-scalar contributions.

\begin{figure}[h!]
\centering
\includegraphics[width=0.47\textwidth]{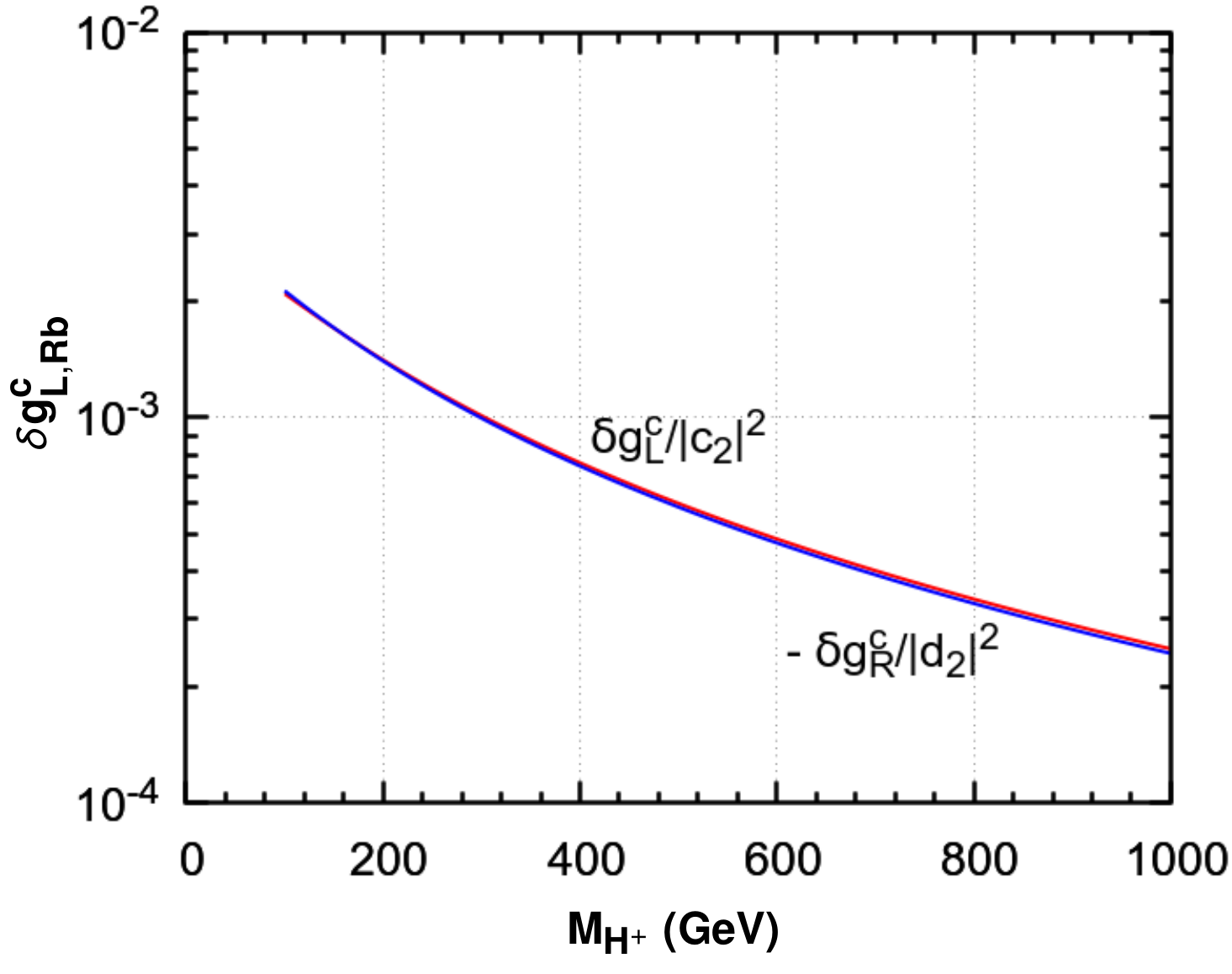}
\hs{2mm}
\includegraphics[width=0.47\textwidth]{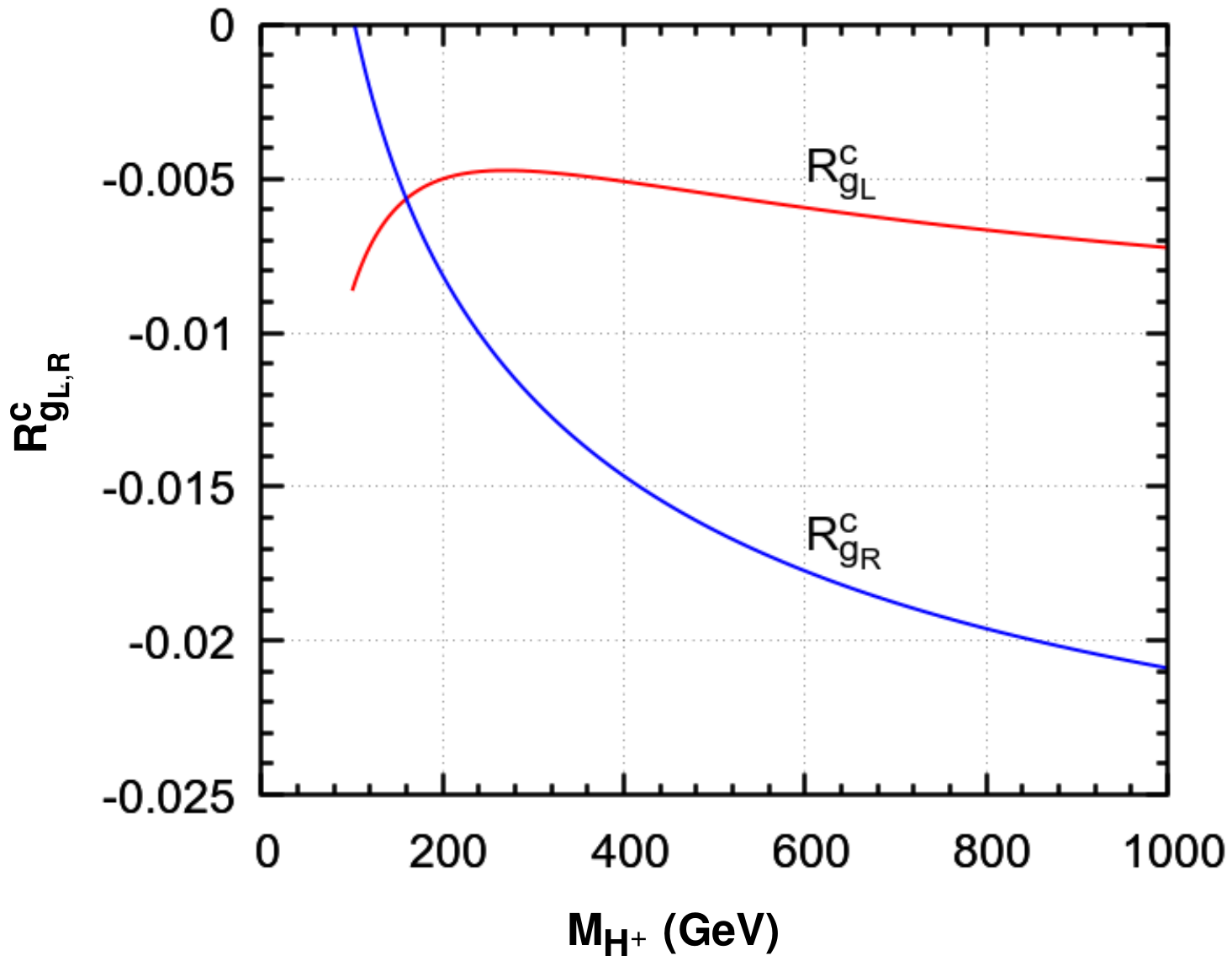}
\caption{Left panel: contribution of $H^{\pm}$
to $\delta g_{\mathrm{L}b}$ (red curve) and to $- \delta g_{\mathrm{R}b}$ (blue curve). Right panel: the assymetries $R_{g_{\mathrm{L}}}^c$ (red curve) and $R_{g_{\mathrm{R}}}^c$ (blue curve), defined in eq. \ref{Chap-Lavou:eq:1}.}
\label{Chap-Lavou:fig:2hdm-charged}
\end{figure}

We start by exploring $\delta g_{\mathrm{L}b}^c$ and $\delta g_{\mathrm{R}b}^c$, i.e. the contributions from the charged scalar.
More precisely, we want to ascertain the impact of the approximation $m_{\mathrm{Z}}=0$ on those contributions, since this approximation is usually employed in the literature.%
\fn{
The function $m_t^2\, C_0 \left( 0, 0, 0, m_a^2, m_t^2, m_t^2 \right)
= \frac{x}{1 - x} \left( 1 + \frac{\ln{x}}{1 - x} \right)$, with $x = \frac{m_t^2}{m_a^2}$, which shows up in the limit $m_{\mathrm{Z}}=0$,
has been given in eq.~(4.5) of ref. \cite{Haber:1999zh} and used by many authors (cf. e.g. ref. \cite{Ferreira:2019aps} and references therein).}
It is easy to prove that, with such approximation, we have:
\be
\dfrac{\delta g_{\mathrm{L}b}^c}{|c_2|^2}
=
-\dfrac{\delta g_{\mathrm{R}b}^c}{|d_2|^2}.
\label{Chap-Lavoura:eq:myapprox}
\ee
On the left panel of fig. \ref{Chap-Lavou:fig:2hdm-charged}, we show the exact contributions (i.e. without $m_{\mathrm{Z}}=0$) from the charged scalar to $\delta g_{\mathrm{L}b}$ and $\delta g_{\mathrm{R}b}$. We see that eq. \ref{Chap-Lavoura:eq:myapprox} holds very well.
This result can also be ascertained by the asymmetries $R_{g_{\mathrm{L,R}}}$, defined in terms of the $\delta$ quantities evaluated with $m_{\mathrm{Z}} \neq 0$ and with $m_{\mathrm{Z}} = 0$:
\be
R_{g_{\mathrm{L}}}^c = \frac{\delta g_{\mathrm{L}b}^c \left( m_{\mathrm{Z}} \right) -
\delta g_{\mathrm{L}b}^c \left( 0 \right)}{\delta g_{\mathrm{L}b}^c \left( m_{\mathrm{Z}} \right) + \delta g_{\mathrm{L}b}^c \left( 0 \right)},
\qquad
R_{g_{\mathrm{R}}}^c = \frac{\delta g_{\mathrm{R}b}^c \left( m_{\mathrm{Z}} \right) -
\delta g_{\mathrm{R}b}^c \left( 0 \right)}{\delta g_{\mathrm{R}b}^c \left( m_{\mathrm{Z}} \right) + \delta g_{\mathrm{R}b}^c \left( 0 \right)}.
\label{Chap-Lavou:eq:1}
\ee
We observe on the right panel of fig. \ref{Chap-Lavou:fig:2hdm-charged} that both asymmetries are very small, which thus confirms the validity of the approximation that neglects the $Z$ boson mass in the charged scalar contributions to NP.

\begin{figure*}[h!]
\centering
\includegraphics[width=0.47\textwidth]{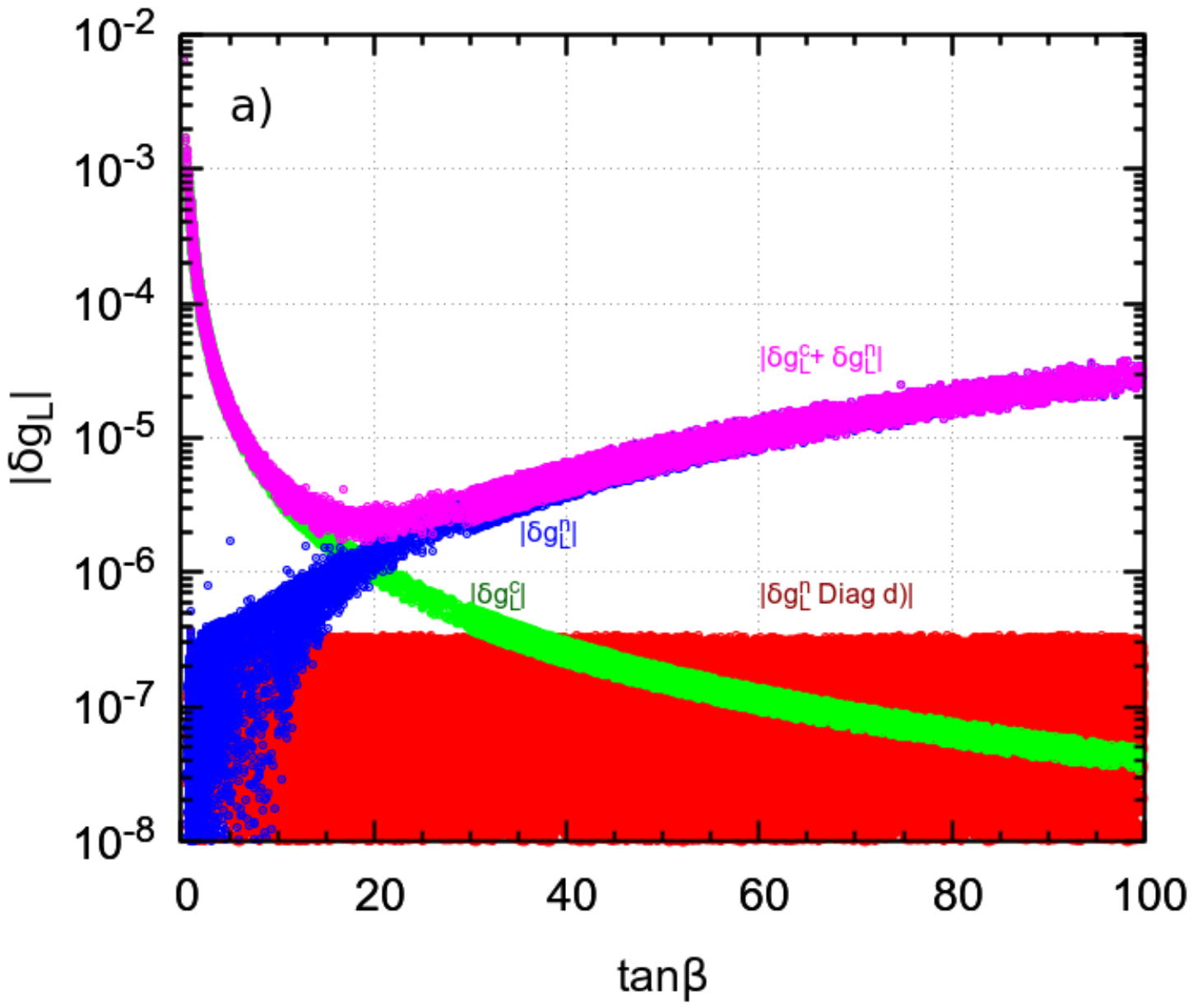}
\hs{2mm}
\includegraphics[width=0.47\textwidth]{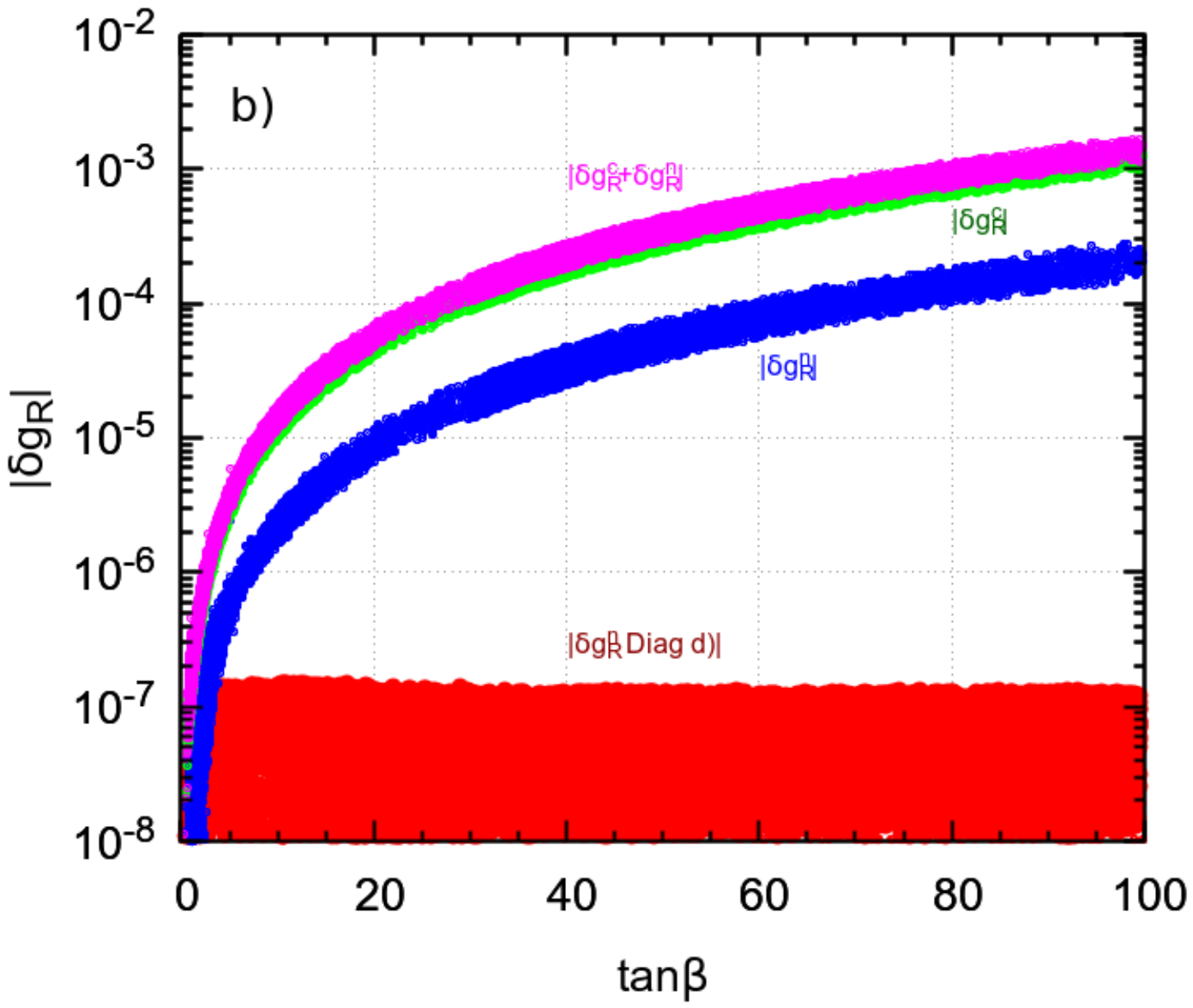}
\caption{Left panel: comparison of $|\delta g_{\mathrm{L}b}^n|$ (in blue)
and $|\delta g_{\mathrm{L}b}^c|$ (in green) with $|\delta g_{\mathrm{L}b}^n + \delta g_{\mathrm{L}b}^c|$ (in pink).
Right panel: comparison of $|\delta g_{\mathrm{R}b}^n|$ (in blue) and $|\delta g_{\mathrm{R}b}^c|$ (in green) with $|\delta g_{\mathrm{R}b}^n + \delta g_{\mathrm{R}b}^c|$ (in pink).
Also displayed (in red) are the contributions of the neutral type d) diagrams---Fig. \ref{Chap-Lavou:fig:type_d)} iii) and iv)---to both $\delta g_{\mathrm{L}b}^n$ (on the left panel) and $\delta g_{\mathrm{R}b}^n$ (on the right panel).}
\label{Chap-Lavou:fig:delg-compare-high_tb}
\end{figure*}

Note that $\delta g_{\mathrm{L}b}^c$ is positive, while $\delta g_{\mathrm{R}b}^c$ is negative. From eq. \ref{Chap-Lavou:eq:truz}, we see that a positive $\delta g_{\mathrm{L}b}$ tends to make $R_b$ smaller---whereas we expected it to become larger, recall eq. \ref{Chap-Lavou:uvido}.
From eq. \ref{Chap-Lavou:eq:truz}, we also conclude that the correction $\delta g_{\mathrm{R}b}^c$ is too small to have an impact on $R_b$.%
\fn{In principle, it could have a substantial impact on $A_b$
going in the wrong direction when compared with the
experimental measurements (cf. eq. \ref{Chap-Lavou:uvido}).
However, we will see below (on the right panel of fig. \ref{Chap-Lavou:fig:delg-compare-high_tb}) that this only happens for large values of $\tan\beta$, not allowed by perturbativity.}

\begin{figure*}[h!]
\centering
\includegraphics[width=0.47\textwidth]{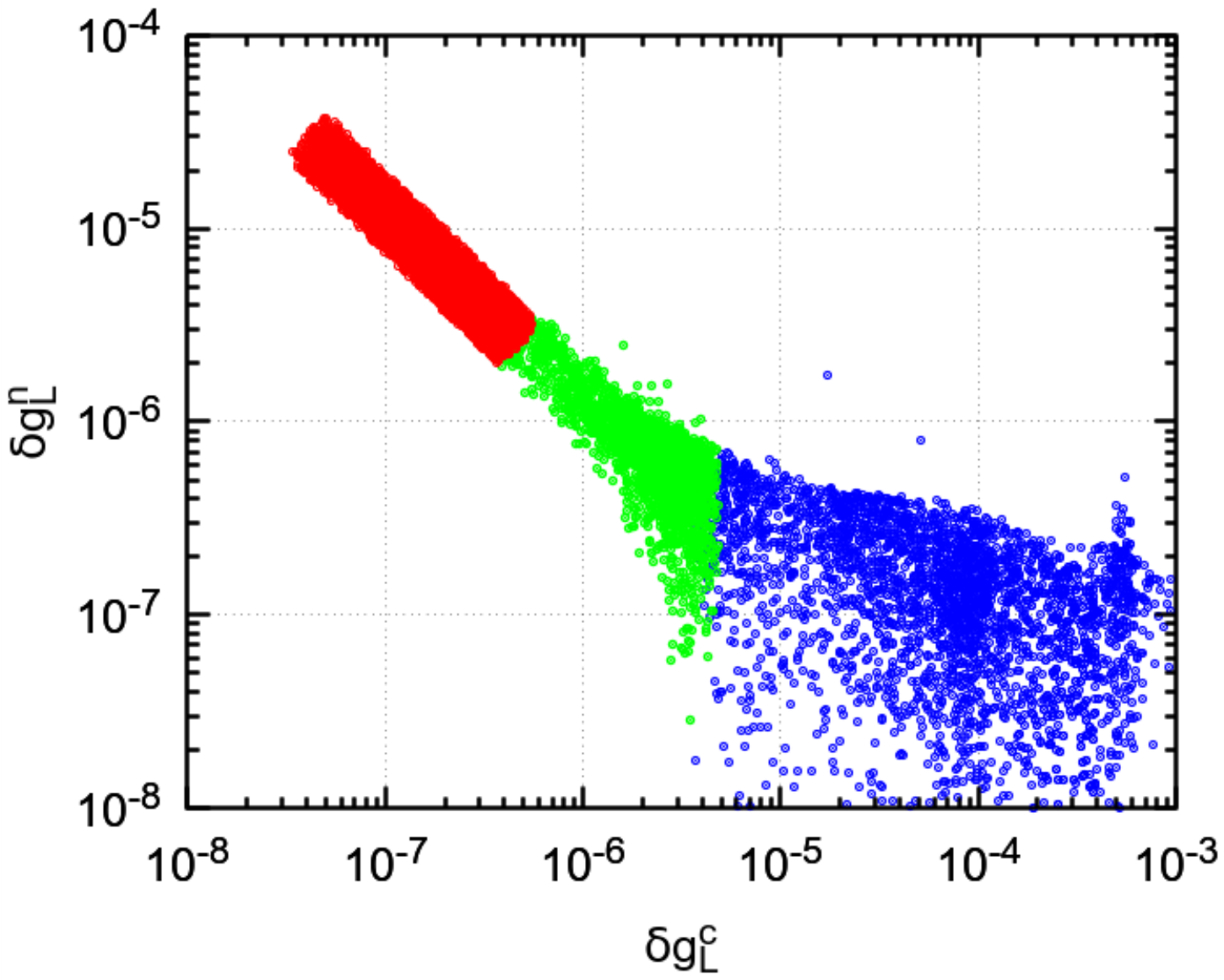}
\hs{2mm}
\includegraphics[width=0.47\textwidth]{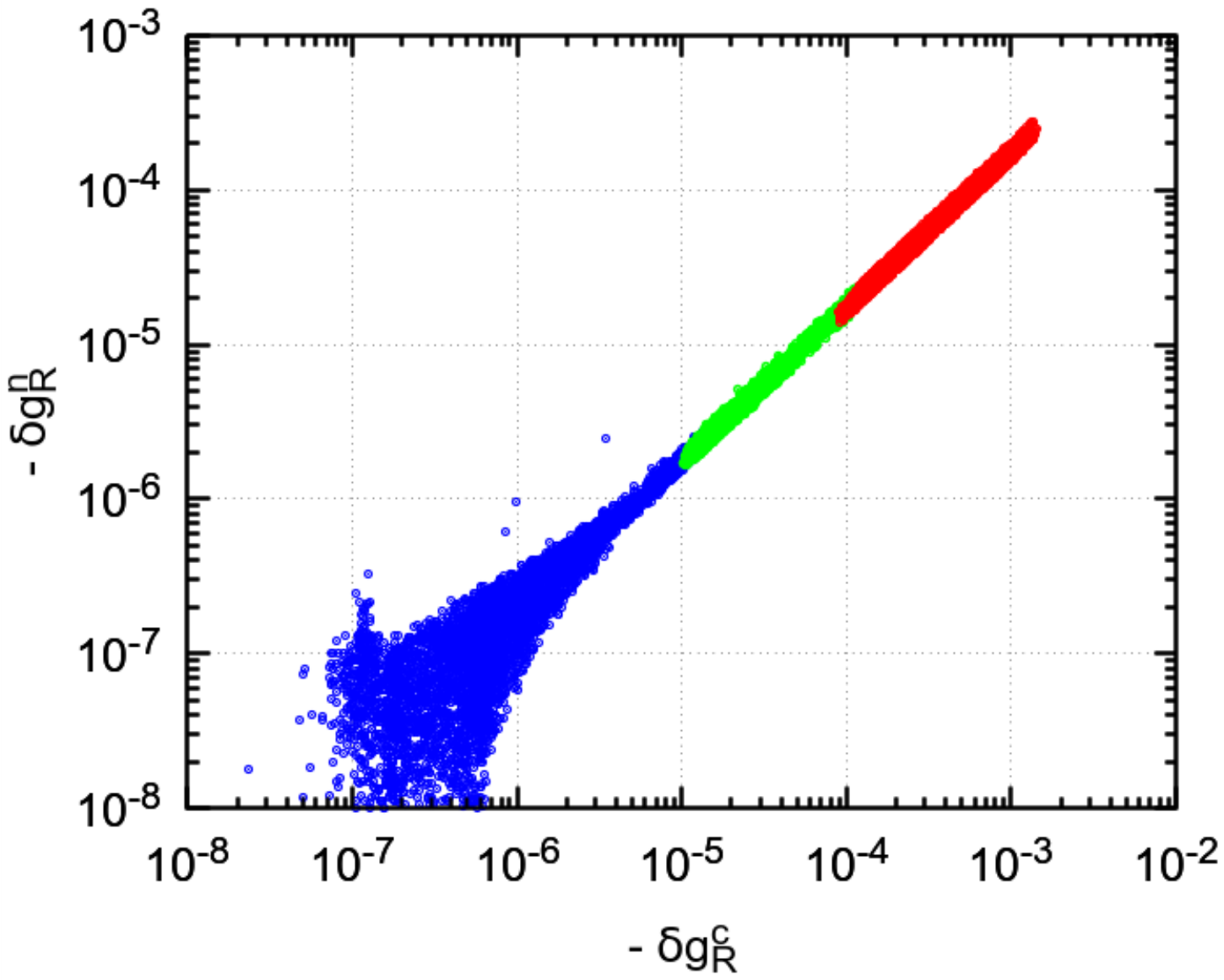}
\caption{Comparison of $\delta g_{\mathrm{L}b}^n$ with $\delta g_{\mathrm{L}b}^c$ (left) and of $\delta g_{\mathrm{R}b}^n$ with $\delta g_{\mathrm{R}b}^c$ (right). 
Blue points correspond to $0 \le \tan\beta \le 10$,
green points to $10 \le \tan\beta \le 30$ and 
red ones to $30 \le \tan\beta \le 100$.}
\label{Chap-Lavou:fig:delg-compare-low_tb}
\end{figure*}
We now compare the contributions from the charged scalar with those from the neutral scalars. The latter are frequently disconsidered in the literature (cf. e.g. refs. \cite{Deschamps:2009rh,Ferreira:2019aps}), which motivates us to investigate under which circumstances they can be large.
On the left panel of fig. \ref{Chap-Lavou:fig:delg-compare-low_tb},
we display $\delta g_{\mathrm{L}b}^n$ versus $\delta g_{\mathrm{L}b}^c$;
on the right panel, $-\delta g_{\mathrm{R}b}^n$ is displayed against $-\delta g_{\mathrm{R}b}^c$ (both $\delta g_{\mathrm{R}b}^n$ and $\delta g_{\mathrm{R}b}^c$ are negative).
We consider separately three different ranges of $\tan \beta$ and we represent each one by a different color.
We see in the right panel that $\left| \delta g_{\mathrm{R}b}^n \right|$
generally is of order $\left. \left| \delta g_{\mathrm{R}b}^c \right| \right/ 10$, but they may be comparable in the low-$\tan{\beta}$ regime.
The left panel shows a very different behaviour, namely: $\delta g_{\mathrm{L}b}^n \ll \delta g_{\mathrm{L}b}^c$ for low $\tan{\beta}$, $\delta g_{\mathrm{L}b}^n \sim \delta g_{\mathrm{L}b}^c$ for $\tan{\beta} \sim 30$ and $\delta g_{\mathrm{L}b}^n \gg \delta g_{\mathrm{L}b}^c$ for high $\tan{\beta}$. So,
one really cannot neglect the contributions from neutral scalars when $\tan{\beta} \gtrsim 10$. For low $\tb \sim 1$, $\delta g_{\mathrm{L}b}^c$ is much larger than $\delta g_{\mathrm{L}b}^n$, but $\delta g_{\mathrm{R}b}^n$ may not be much smaller than $\delta g_{\mathrm{R}b}^c$.

In fig. \ref{Chap-Lavou:fig:delg-compare-high_tb}, we provide a more explicit comparison between the contributions to NP arising from the charged and from the neutral scalars.
On the left panel, we see that the contribution from neutral scalars to $\delta g_{\mathrm{L}b}$ becomes larger than the contribution from the charged scalar---eventually by many orders of magnitude---as soon as $\tan{\beta} > 30$. We thus reinforce the same conclusion derived above, namely: one cannot neglect the contribution of the neutral scalars to $\delta g_{\mathrm{L}b}$.
%
In the same figure, we ascertain (in red) the importance of contributions from the diagrams \ref{Chap-Lavou:fig:type_d)} iii) and iv), calculated in section \ref{Chap-Lavou:subsec:calc-S0}.
We see that, when $\tb$ is low, they may constitute a substantial part of the NP contribution; yet, that is precisely the range in which $\delta g_{\mathrm{L} b}^n$ and $\delta g_{\mathrm{R} b}^n$ are too small to be of practical relevance (as can be seen by the blue points).
We conclude that, at least in the particular case of the C2HDM, it is correct to neglect the diagrams in fig. \ref{Chap-Lavou:fig:type_d)} iii) and iv), as was done in ref.~\cite{Haber:1999zh}.

The pink points in fig. \ref{Chap-Lavou:fig:delg-compare-high_tb} allow us to infer that a significant impact on $A_b$ and $R_b$
can only occur for either very low or very high values of $\tan\beta$;
specifically, for $\tb \lesssim 1$, $\delta g_{\mathrm{L}b}^c + \delta g_{\mathrm{L}b}^n \sim 10^{-3}$ and for $\tb \gtrsim 50$, $- \delta g_{\mathrm{R}b}^c - \delta g_{\mathrm{R}b}^n \gtrsim 10^{-3}$.
The impact on $A_b$ and $R_b$ is explored in more detail in fig. \ref{Chap-Lavou:fig:RbAb-low_tb-2}.
\begin{figure}[!htb]
\centering
\includegraphics[width=0.5\textwidth]{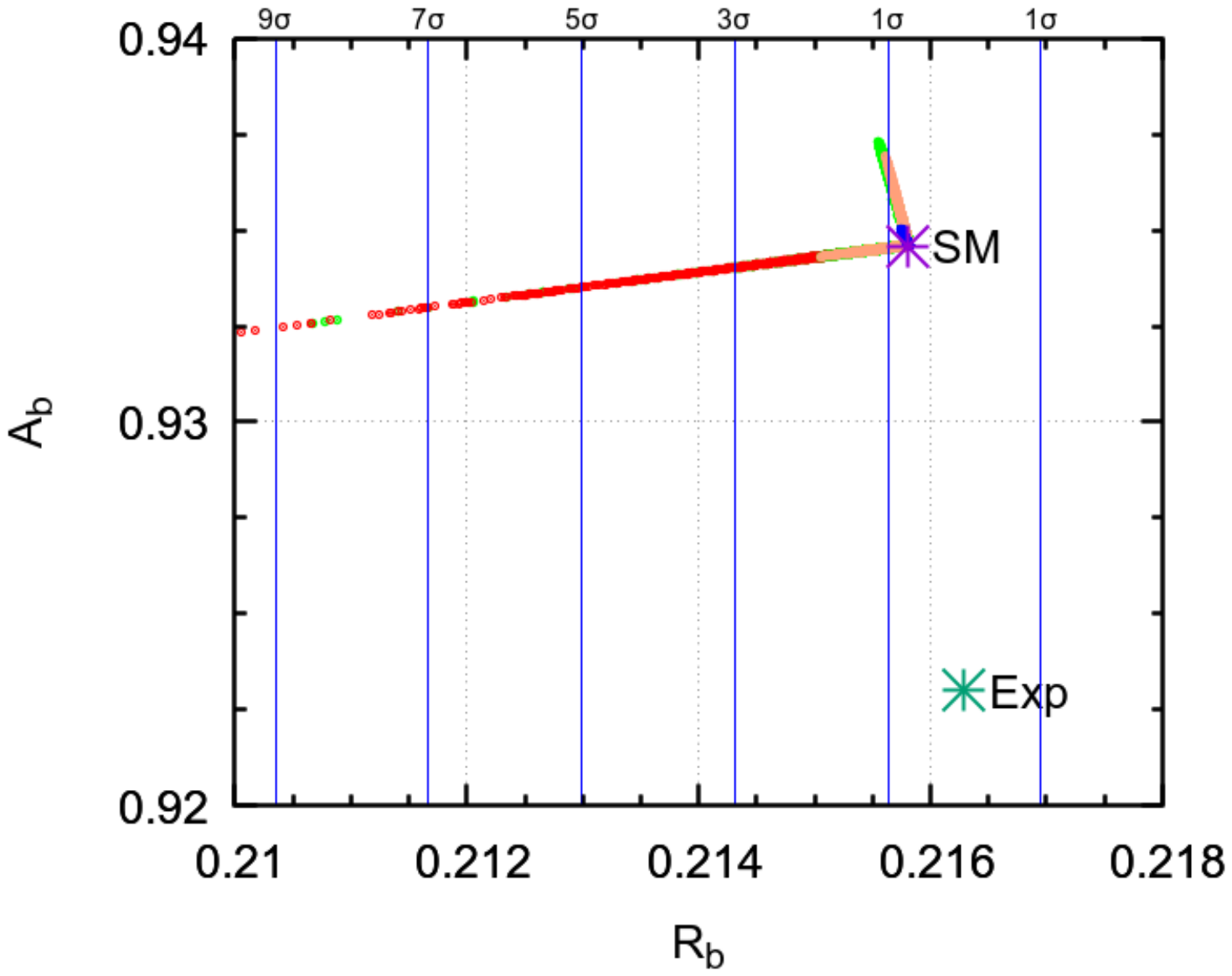}
\caption{$A_b$ versus $R_b$ in the C2HDM for all values of
$0<\tan\beta <100$.
The charged-scalar contribution is shown in
red for low $\tb$ and in orange for large $\tb$.
The contribution of the neutral scalars is shown in blue and lies very
close to the SM point.
In green (in background) the sum of the contributions.}
\label{Chap-Lavou:fig:RbAb-low_tb-2}
\end{figure}
%
%
%
There, we see that points deviate significantly from the experimental value of $R_b$ in the low $\tan\beta$ regime; we checked that points only lie within the 2$\sigma$ interval for $\tb > 0.8$, which is why this lower bound was imposed in chapter \ref{Chap-Maggie}. The contribution of the neutral scalars (in blue) is always very small; we verified that, for the neutral scalars to have meaningful impact in the C2HDM, one would have to consider values of $\tan\beta>250$, which would violate perturbativity. This is because the neutral contributions in the C2HDM are related to a ratio of vevs ($v_2/v_1 = \tan{\beta}$), which is limited by perturbative unitarity. However, such relation is not necessary in more general models, in which case the contribution from the neutral scalars may very well be important.%
\fn{It is a matter of course that, when studying any model, all the theoretical and experimental constraints need to be included, which may curtail a large part of the phase space for such extreme couplings. This will have to be evaluated in a case-by-case basis.}

\section{\label{Chap-Lavou:sec:conclusions}Summary}

We studied the one-loop contributions to $Z \rightarrow b \bar b$
in generic models with extra scalars. We started by performing the one-loop calculations for general couplings and we derived the conditions that must hold for the divergences to cancel.
We then concentrated on models with a generic number of scalar doublets and singlets, using the formalism of refs.~\cite{Grimus:2002ux, Grimus:2007if, Grimus:2008nb, Grimus:1989pu}, which is very useful to connect the general one-loop results to particular cases. We verified this explicitly in the case of the C2HDM; here, we also highlighted the possible importance of contributions arising from neutral scalars, which may in some cases be much larger than the contributions of the charged scalars. This imbalance must be considered when evaluating the limits on $A_b$ and $R_b$.

%% file: Chapters/Chapter_Valle.tex
\chapter{Simplest linear seesaw model}
\label{Chap-Valle}

\vs{-5mm}

\def\lfv{lepton flavour violation }
\def\lnv{lepton number violating }
\def\SM{$\mathrm{SU(3)_c \otimes SU(2)_L \otimes U(1)_Y}$ }
\def\su2{$\mathrm{SU(2)_L}$}
\newcommand{\sm}{{Standard Model }}
\def\vev#1{\left\langle #1\right\rangle}


In this final chapter, we investigate the phenomenology of a MHM motivated by non-zero neutrino masses. As is well known, neutrino masses constitute one of the most robust evidences for new physics. Ever since the discovery \cite{Kajita:2016cak,McDonald:2016ixn} and confirmation \cite{Eguchi:2002dm,Ahn:2002up} of neutrino oscillations, the efforts to understand the mechanism of neutrino mass have been intense.
A popular approach is the type-I seesaw mechanism, in which neutrinos get mass due to the exchange of heavy singlet mediators.

In its standard high-scale realization, the seesaw mechanism hardly leads to any phenomenological implication besides those associated to the neutrino masses themselves. However, the seesaw can arise from low-scale physics~\cite{Joshipura:1992hp,Boucenna:2014zba}, e.g. through the linear seesaw mechanism \cite{Akhmedov:1995ip,Akhmedov:1995vm,Malinsky:2005bi}. 
%
Here, we examine the scalar sector of the simplest variant of such mechanism, where the seesaw mechanism is realized using just the SM gauge structure associated to the \SM symmetry \cite{Schechter:1980gr,Chikashige:1980ui,Schechter:1981cv}.
%
In addition to the SM Higgs doublet, the model contains a second Higgs doublet, as well as a complex singlet scalar, both carrying lepton number charges. Spontaneous breaking of the lepton number (global) symmetry hence implies the existence of a Nambu-Golstone boson, a variant of the so-called majoron~\cite{Chikashige:1980ui,Schechter:1981cv}.
%
%

We start by presenting the scalar sector of the model in section 
\ref{Chap-Valle:sec:higgs-potential}, after which we provide our results.
The latter illustrate the phenomenology of the model,
giving special emphasis to the invisible Higgs decay channel with majoron emission \cite{Joshipura:1992hp}. This has implications for collider experiments \cite{Romao:1992zx,Eboli:1994bm,DeCampos:1994fi,Romao:1992dc,deCampos:1995ten,deCampos:1996bg,Diaz:1998zg,Hirsch:2004rw,Hirsch:2005wd,Bonilla:2015uwa,Bonilla:2015jdf}, and has indeed been searched by LEP \cite{LEPHiggsWorkingforHiggsbosonsearches:2001ypa} and LHC collaborations~\cite{Sirunyan:2018owy,Aaboud:2019rtt}.
The new signals can be studied in proton-proton collisions such as at the High-Luminosity LHC setup, as well as in the next generation of lepton collider experiments such as CEPC, FCC-ee, ILC and CLIC~\cite{CEPCStudyGroup:2018ghi, Abada:2019zxq,Bambade:2019fyw,deBlas:2018mhx}.

\section{The scalar sector}
\label{Chap-Valle:sec:higgs-potential}

As we just mentioned, besides the SM doublet $\Phi$, we consider an additional doublet $\chi_L$ and a (complex) singlet $\sigma$. We parameterize all these multiplets according to:
\begin{equation}
\label{Chap-Valle:eq:1}
\Phi=
\left(
\begin{array}{c}
\phi^+\\[+2mm]
\frac{1}{\sqrt{2}} \left( v_{\phi} + R_1 + i\, I_1\right)
\end{array}
\right),
~~~\chi_L=
\left(
\begin{array}{c}
\chi_L^+\\[+2mm]
\frac{1}{\sqrt{2}} \left( v_L + R_2 + i\, I_2\right)
\end{array}
\right),
~~~\sigma =     \frac{1}{\sqrt{2}} \left( v_\sigma + R_3 + i\, I_3\right),
\end{equation}
with $v_{\phi}$, $v_L$ and $v_\sigma$ real parameters, $\phi^+$ and $\chi_L^+$ complex fields, and $R_i$ and $I_i$ real fields.
We choose the following lepton number assignments,
\begin{equation}
\label{Chap-Valle:eq:2}
L[\Phi] =0,\quad L[\chi_L] =-2,\quad L[\sigma]=1.
\end{equation}
Therefore, the most general potential is:
\begin{align}
\label{Chap-Valle:eq:7}
V = &-\mu^2\, \Phi^\dagger \Phi + \lambda\, \left(\Phi^\dagger
\Phi\right)^2 -\mu_L^2\, \chi_L^\dagger \chi_L + \lambda_L\,
\left(\chi_L^\dagger   \chi_L\right)^2
- \mu_\sigma^2 \sigma^\dagger \sigma
+ \lambda_\sigma  \left(\sigma^\dagger \sigma\right)^2\nonumber\\[+2mm]
&
+ \beta_1\, \Phi^\dagger \Phi\, \chi_L^\dagger \chi_L
+ \beta_2\, \Phi^\dagger \Phi\,  \sigma^\dagger \sigma
+ \beta_3\,  \chi_L^\dagger \chi_L\,  \sigma^\dagger \sigma
+  \beta_5\,  \Phi^\dagger \chi_L\,  \chi_L^\dagger \Phi
- \left(\beta_4 \Phi \chi_L \sigma^2 + \text{h.c.}\right).
\end{align}
Although $\beta_4$ is in general complex, we will consider the region of parameter space where it takes real values only.%
\fn{\label{Chap-Valle:note_key}$\beta_4$ cannot be rendered real by a basis choice; more precisely, there is no basis where $v_{\phi}$, $v_L$, $v_\sigma$ and $\beta_4$ are all real. In principle, these quantities could be set real by imposing the symmetry CP on the theory; however, this is not a sound option, because this symmetry is violated by quark interactions. In other words, if $v_{\phi}$, $v_L$, $v_\sigma$ and $\beta_4$ are all taken to be real parameters, the model will be as theoretically unsound as the real 2HDM (recall the discussion in chapter \ref{Chap-Real}). Strictly speaking, then, and if we choose a basis where $v_{\phi}$, $v_L$, $v_\sigma$ are real, $\beta_4$ should be allowed to be complex, which would imply CP violation in the scalar sector, and the diagonalization of the neutral scalar states should be done accordingly. However, if we focus on the particular region of the parameter space of the (CP-violating) model where $\beta_4$ takes real values, then, in that region, CP can be well defined in the potential, so that CP-even states and CP-odd states can be separately diagonalized. This is what we do in the following.}
After writing the minimization equations, 
we can derive the following charged scalar mass matrix in the basis $(\phi^+,\chi^+_L)$:
\begin{equation}
\label{Chap-Valle:eq:9}
M^2_{\rm ch}=
\begin{pmatrix}
\dfrac{\beta_4 v_L v_{\sigma}^2-\beta_5 v_L^2 v_\phi}{2 v}
& \dfrac{\beta_5 v_L v_\phi - \beta_4 v_{\sigma}^2}{2} \\[+2mm]
\dfrac{\beta_5 v_L v_\phi -\beta_4 v_{\sigma}^2}{2}
& \dfrac{\beta_4 v_{\phi} v_{\sigma}^2-\beta_5 v_L v_\phi^2}{2 v_L} 
\end{pmatrix}.
\end{equation}
Diagonalizing this matrix, we find two mass states: the charged (massless) would-be Goldstone boson, $G^+$, as well as a massive state, $H^+$. The squared mass of the latter is:
\begin{equation}
\label{Chap-Valle:eq:11}
m_{\mathrm{H}^+}^2
= \frac{(\beta_4 v_{\sigma}^2 -\beta_5 v_L v_\phi) \left(v_{\phi}^2+v_L^2\right)}{2 v_{\phi} v_L} =
\frac{\beta_4 v_{\sigma}^2}{\sin 2\beta} - \frac{1}{2} \beta_5 v^2,
\end{equation}
where we defined:
\begin{equation}
\label{Chap-Valle:eq:16}
v=\sqrt{v_{\phi}^2+v_L^2},
\qquad
\tan\beta= \frac{v_L}{v_{\phi}}.
\end{equation}

The neutral scalar (CP-even) mass matrix in the basis $(R_1, R_2, R_3)$ is given by:
\begin{equation}
\label{Chap-Valle:eq:12}
M^2_{\rm ns} =
\begin{pmatrix}
2 \lambda v_{\phi}^2+\dfrac{\beta_4 v_L v_{\sigma}^2}{2 v_{\phi}} & \beta_1 v_{\phi}
v_L-\dfrac{\beta_4 v_{\sigma}^2}{2} + \beta_5 v_L v_\phi & (\beta_2 v_{\phi}-\beta_4
v_L) v_{\sigma} \\[+2mm] 
\beta_1 v_{\phi} v_L-\dfrac{\beta_4 v_{\sigma}^2}{2}+ \beta_5 v_L v_\phi & 2 \lambda_L
v_L^2+\dfrac{\beta_4 v_{\phi} v_{\sigma}^2}{2 v_L} & \beta_3 v_L
v_{\sigma}-\beta_4 v_{\phi} v_{\sigma} \\[+2mm]
(\beta_2 v-\beta_4 v_L) v_{\sigma} & \beta_3 v_L v_{\sigma}-\beta_4 v_{\phi} v_{\sigma} & 2
\lambda_\sigma v_{\sigma}^2 
\end{pmatrix}.
\end{equation}
We can thus define the scalar mass eigenstates $h_1$,
$h_2$ and $h_3$, as well as the orthogonal matrix $\mathcal{O}^{R}$, such that: 
\begin{equation}
\label{Chap-Valle:eq:20}
\begin{pmatrix}
h_1\\
h_2\\
h_3
\end{pmatrix}
= \mathcal{O}^{R}
\begin{pmatrix}
R_1\\
R_2\\
R_3
\end{pmatrix},
\qquad
\mathrm{with}
\qquad
\mathcal{O}^{R}  \cdot M_{\rm ns}^2\cdot \mathcal{O}^{R}{}^{\mathrm{T}} =
\text{diag}(m_{1}^2,m_{2}^2,m_{3}^2),
\end{equation}
where $m_1^2$, $m_2^2$ and $m_3^2$ are the squared masses of $h_1$,
$h_2$ and $h_3$, respectively, with $m_1 < m_2 < m_3$.
The matrix $\mathcal{O}^{R}$ can
be parameterized precisely as the matrix $R$ of the C2HDM (eq. \ref{Chap-Maggie:matrixR}):
\begin{equation}
\label{Chap-Valle:eq:21}
\mathcal{O}^{R}
=
\begin{pmatrix}
c_1c_2 & s_1 c_2 & s_2\\
-c_1s_2s_3-s_1c_3&c_1c_3-s_1s_2s_3&c_2s_3\\
-c_1s_2c_3+s_1s_3&-c_1s_3-s_1s_2c_3&c_2c_3
\end{pmatrix},
\end{equation}
where $c_i=\cos\alpha_i,s_i=\sin\alpha_i$.

The neutral pseudoscalar (CP-odd) mass matrix in the basis $(I_1, I_2, I_3)$ is given by:
\begin{equation}
\label{Chap-Valle:eq:13}
M^2_{\rm nps} =
\begin{pmatrix}
\dfrac{\beta_4 v_L v_{\sigma}^2}{2 v_{\phi}} & -\dfrac{\beta_4 v_{\sigma}^2}{2} & -\beta_4
v_L v_{\sigma} \\[+2mm]
-\dfrac{\beta_4 v_{\sigma}^2}{2} & \dfrac{\beta_4 v_{\phi} v_{\sigma}^2}{2 v_L} & \beta_4 v_{\phi} v_{\sigma}
\\[+2mm]
-\beta_4 v_L v_{\sigma} & \beta_4 v_{\phi} v_{\sigma} & 2 \beta_4 v_{\phi} v_L 
\end{pmatrix}.
\end{equation}
Similarly to what we did for the scalar states, we can define the pseudoscalar mass eigenstates $G_0$, $J$ and $A$, as well as the orthogonal matrix $\mathcal{O}^{I}$, such that:
\begin{equation}
\label{Chap-Valle:eq:24}
\begin{pmatrix}
G_0\\
J\\
A
\end{pmatrix}
=\mathcal{O}^{I}
\begin{pmatrix}
I_1\\
I_2\\
I_3
\end{pmatrix},
\qquad
\mathrm{with}
\qquad
\text{diag}(0,0,m_{A}^2) =
\mathcal{O}^{I} \cdot M_{\rm ns}^2\cdot\mathcal{O}^{I}{}^{\mathrm{T}}.
\end{equation}
$G_0$ is the neutral (massless) would-be Goldstone boson and $J$ is the majoron (which is also massless). The squared mass of $A$ is given by:
\begin{equation}
\label{Chap-Valle:eq:14}
m_A^2= \frac{\beta_4 \left(4 v_{\phi}^2 v_L^2+v_{\phi}^2 v_{\sigma}^2+v_L^2
v_{\sigma}^2\right)}{2 v_{\phi} v_L}= \beta_4 \left[ v^2
\sin 2\beta  + \frac{v_{\sigma}^2}{\sin2\beta} \right].
\end{equation}

Finally, we can use the relations of the model to rewrite the quartic parameters of the potential in terms of other parameters. Particularly relevant for what follows is the expression for $\lambda_L$, given by:
\begin{align}
\lambda_L=&-\frac{1}{4 v_{L}^3}\left[\vb{14}
\beta_{4} c_{1}^4 v_{\phi} v_{\sigma}^2+2
c_{1}^2 \left(\beta_{4} c_{3}^4 
s_{1}^2 v_{\phi} v_{\sigma}^2+c_{3}^2 \left(2 \beta_{4}
s_{1}^2 s_{3}^2 
v_{\phi} v_{\sigma}^2-m_{2}^2 v_{L}\right)+\beta_{4} s_{1}^2
s_{3}^4 
v_{\phi} v_{\sigma}^2
\right.\right.
\nonumber\\
&\hs{-5mm}\left.\left.
-m_{3}^2 s_{3}^2
v_{L}\right)+s_{1}^2 \left(\beta_{4} 
c_{2}^4 s_{1}^2 v_{\phi} v_{\sigma}^2-2 c_{2}^2 \left(m_{1}^2
v_{L}-\beta_{4} s_{1}^2 s_{2}^2 v_{\phi}
v_{\sigma}^2\right)+s_{2}^2 
\left(\beta_{4} c_{3}^4 s_{1}^2 s_{2}^2 v_{\phi}
v_{\sigma}^2
\right.\right.\right.
\nonumber\\
&\hs{-5mm}\left.\left.\left.
+c_{3}^2 \left(2 
\beta_{4} s_{1}^2 s_{2}^2 s_{3}^2 v_{\phi} v_{\sigma}^2-2 m_{3}^2
v_{L}\right)+\beta_{4} s_{1}^2 s_{2}^2 s_{3}^4
v_{\phi} v_{\sigma}^2-2 
m_{2}^2 s_{3}^2 v_{L}\right)\right)
+4 c_{1}
c_{3} s_{1} 
s_{2} s_{3} v_{L} (m_{2}^2-m_{3}^2) \vb{14}\right].
\label{Chap-Valle:eq:23b}
\end{align}
\section{Results}
\label{Chap-Valle:sec:profile-higgs-bosons}

\subsection{Simulation procedure and constraints}
\label{Chap-Valle:sec:theor-constr}

The generation of Feynman rules and the calculation of leading order (LO) decay widths was performed with \FM.
For the scatter plots below, we generated points in the parameter space of the model with $m_1$ fixed to 125 GeV, and we varied the other parameters in the following way:%
\begin{gather}
\label{Chap-Valle:eq:59}
m_{2}, m_{3},m_{A},m_{\mathrm{H}^+} \in [125, 800] \, \text{GeV},
\qquad
\alpha_1,\alpha_2,\alpha_3 \in [-\frac{\pi}{2},\frac{\pi}{2}],
\qquad
v_L \in [10^{-6},10^2] \, \text{GeV},
\nonumber\\
v_{\sigma} \in [10^3,1.2 \times 10^4] \, \text{GeV}.
\end{gather}
We imposed both theoretical and experimental constraints on the generated points; details can be found in ref. \cite{Fontes:2019uld}.
On the theoretical side, we considered BFB \cite{Klimenko:1984qx}, unitarity \cite{Bento:2017eti} and the oblique parameters S, T and U \cite{Grimus:2007if,Branco:2011iw}.
%
On the experimental side, we took into account both astrophysical constraints and contraints from the LHC. The first
affect the majoron $J$, which would be copiously produced in stars, leading to new mechanisms of stellar cooling.
In our model, that leads to the constraint \cite{Choi:1989hi,Valle:1990pka}:
\begin{equation}
  \label{Chap-Valle:eq:60}
  |\vev{ J|\Phi}|=
\frac{2 v_{\phi} v_L^2}{\sqrt{(v_{\phi}^2+v_L^2) \big(v_{\phi}^2 (4 v_L^2 +
  v_{\sigma}^2) + v_L^2 v_{\sigma}^2\big)}}
  \lesssim10^{-7},
\end{equation}
with $\vev{ J|\Phi}$ denoting the majoron projection into the SM doublet.
As for the LHC constraints, these concern both the experiments related to the 125 GeV scalar Higgs boson, $h_{1}$, as well as those related to additional scalar bosons. For the former, we used combined ATLAS and CMS analyses for 8 TeV \cite{TheATLASandCMSCollaborations:2015bln}, and ATLAS results for 13 TeV \cite{ATLAS:2018doi}; by default, we imposed these constraints at the $3 \sigma$ level. The information relative to other scalars was included using the \textsc{HiggsBounds4} package \cite{Bechtle:2013wla}.

\subsection{The impact of the astrophysical constraints on $v_L$}

\begin{figure}[h!]
\centering
\includegraphics[width=0.5\textwidth]{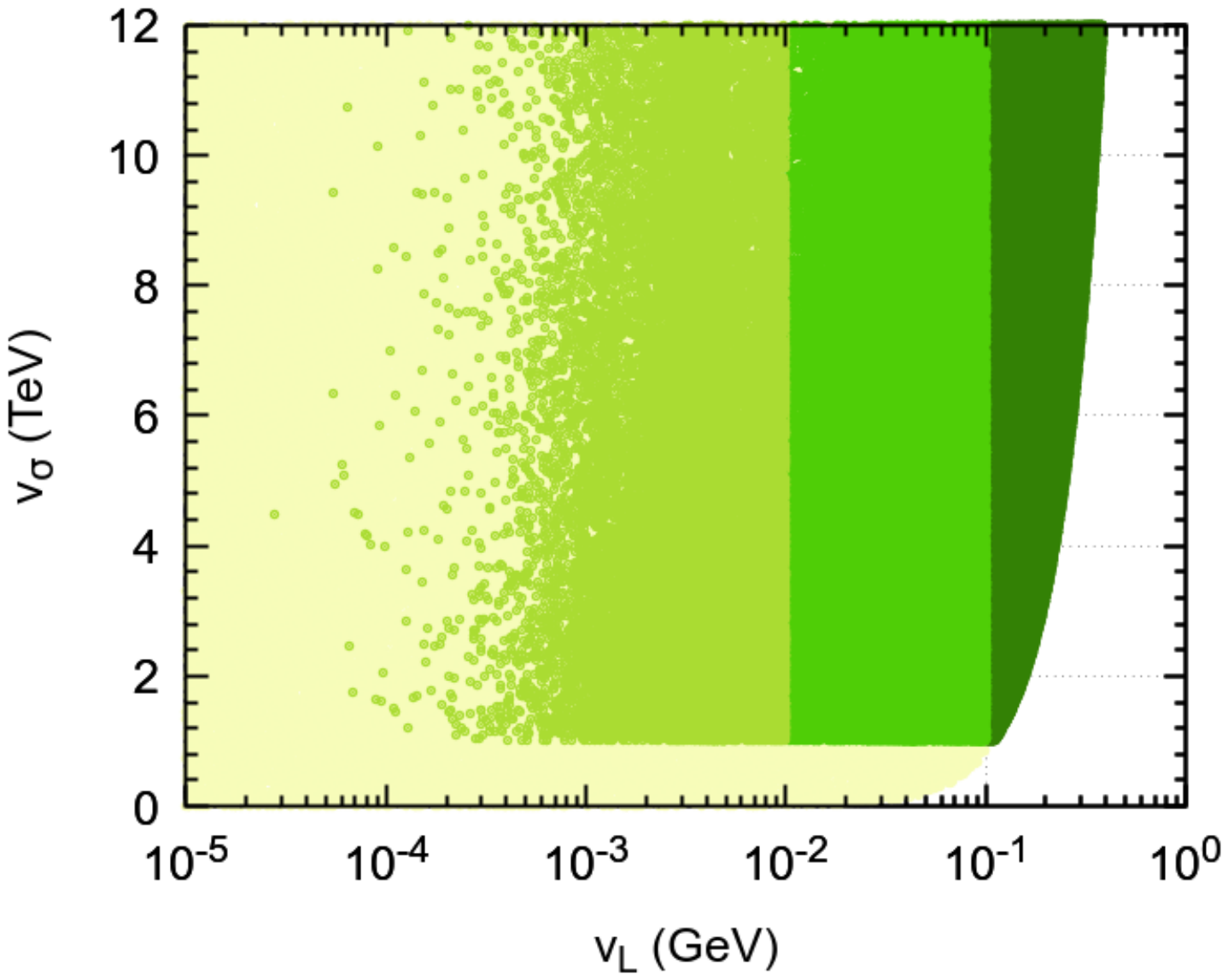}
\caption{$v_\sigma$ versus $v_L$. The yellow points satisfy eq.~(\ref{Chap-Valle:eq:60}) without any other constraint. The other points satisfy all the other theoretical and experimental constraints: in dark green, $v_L> 0.1$ GeV; in green, $v_L\in[0.01,01]$GeV; in yellow green, $v_L < 0.01$ GeV. Besides, a cut $v_\sigma > 1$ TeV was imposed.
}
\label{Chap-Valle:fig:stellar}
\end{figure}

We start by ascertaining the impact of the astrophysical constraint, eq.~(\ref{Chap-Valle:eq:60}).
In order to more easily comply with this constraint, we have sampled
$v_\sigma$ values above 1~TeV, aware that lower values could be possible. 
The allowed region in the ($v_\sigma,v_L$) plane is shown in fig.~\ref{Chap-Valle:fig:stellar}.
The yellow region is the set of values of
$(v_\sigma,v_L)$ that satisfy eq.~(\ref{Chap-Valle:eq:60}), ignoring all the remaining contraints. The green regions, in contrast, satisfy all constraints. We explicitly associate different shades of green to different ranges of $v_L$: points with $v_L > 0.1 \, \text{GeV}$ are shown in dark green, points with $v_L \in [0.01,0.1] \, \text{GeV}$ in green and points with $v_L < 0.01 \, \text{GeV}$ in yellow green. The figure shows that all points (in particular, the yellow ones) obey $v_L < 0.5$ GeV, which represents a much smaller range than that which we started with.
In other words, the astrophysical contraints by themselves force $v_L$ to be quite constrained.
%
%
Note that, besides eq.~(\ref{Chap-Valle:eq:60}), there is another expression relating the vevs, $m_{\mathrm{W}}=\dfrac{1}{2}\, \dfrac{e}{s_{\mathrm{w}}} \, \sqrt{v_L^2+v_\phi^2}$, which explains the boundary in the plane ($v_\sigma,v_L$).
%

\subsection{Mixing angles}

\begin{figure}[h!]
\centering
\includegraphics[width=0.45\textwidth]{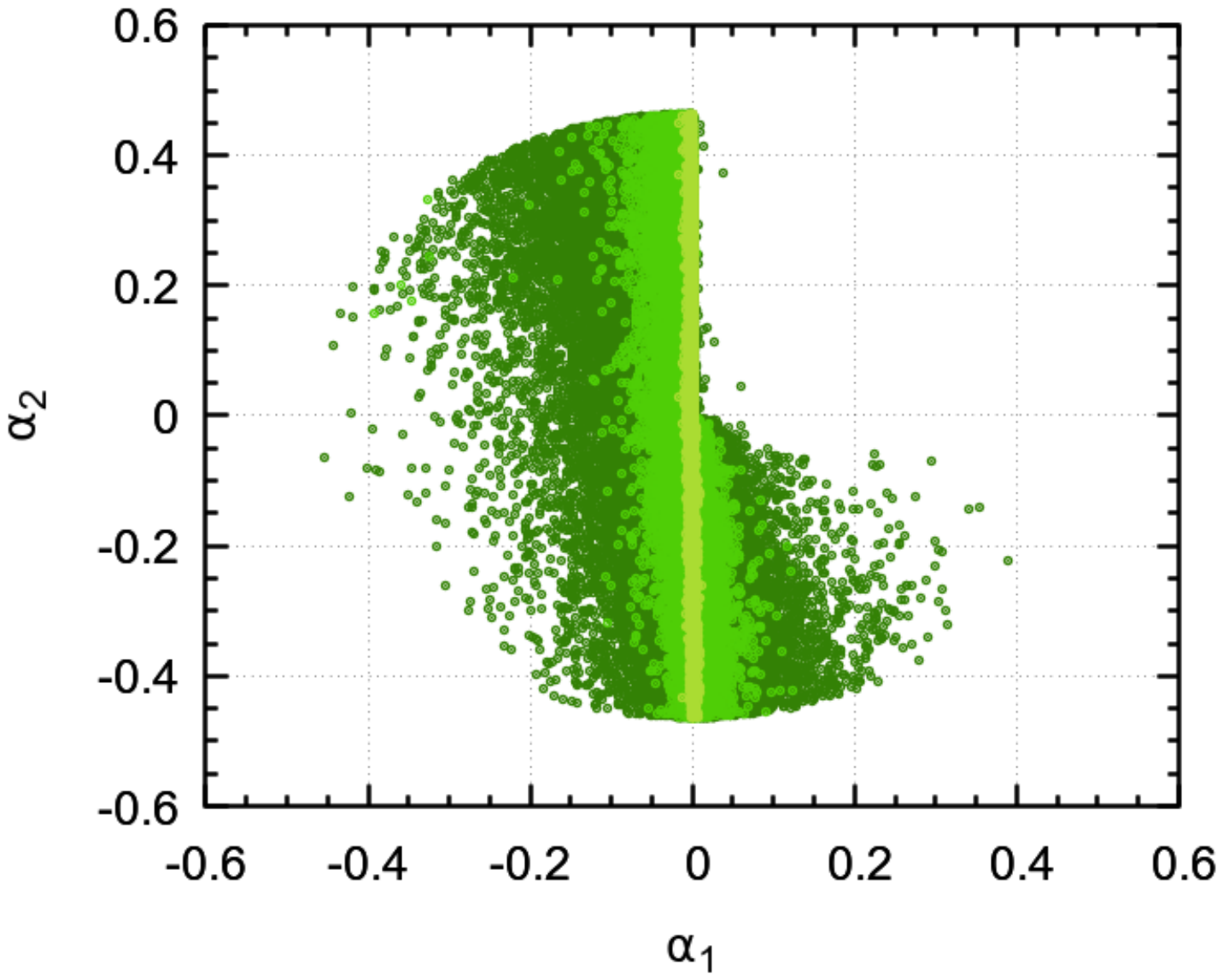}
\includegraphics[width=0.45\textwidth]{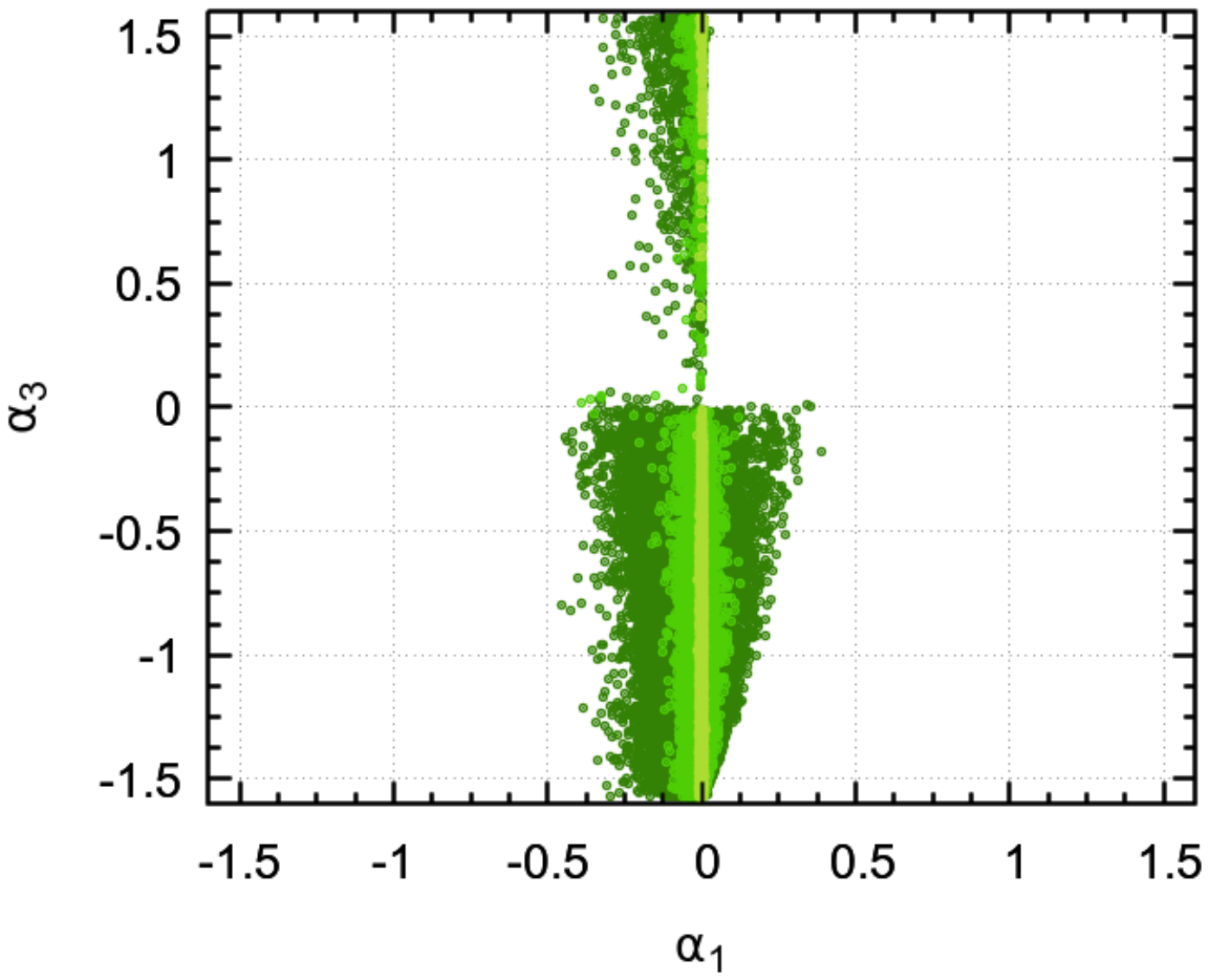}
\caption{$\alpha_2$ (left) and $\alpha_3$ (right) versus $\alpha_1$. Points obey to all theoretical and experimental constraints: in dark green, $v_L> 0.1$ GeV; in green, $v_L\in[0.01,01]$GeV; in yellow green, $v_L < 0.01$ GeV.}
\label{Chap-Valle:fig:alfas}
\end{figure}

\begin{figure}[h!]
\centering
\includegraphics[width=0.5\textwidth]{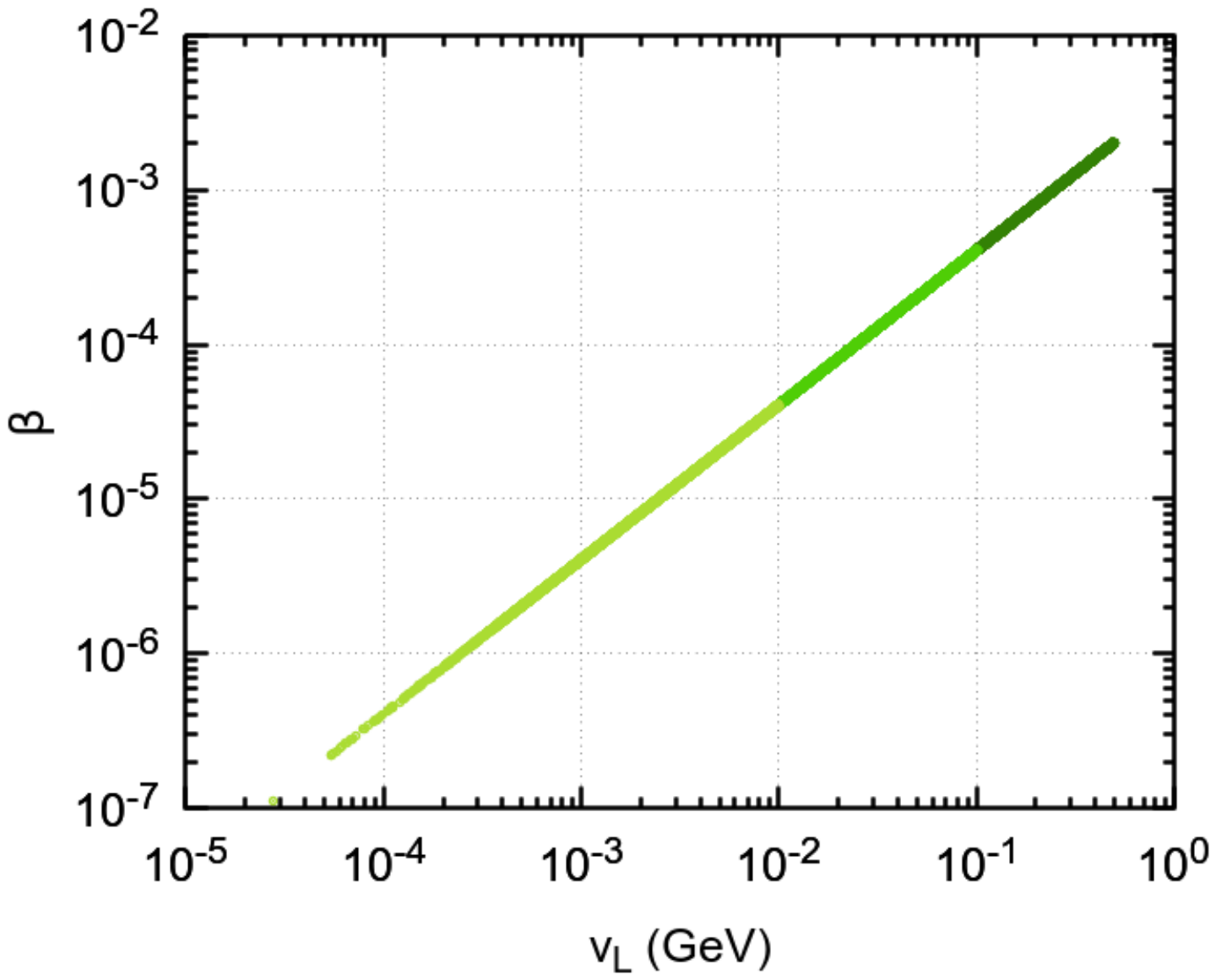}
\caption{Correlation between $\beta$ and $v_L$. The color
code is that of fig.~\ref{Chap-Valle:fig:alfas}.} 
\label{Chap-Valle:fig:betaalpha}
\end{figure}

In fig.~\ref{Chap-Valle:fig:alfas}, the mixing angles in the neutral CP-even rotation matrix are shown.
%
%
%
We see that $\alpha_1$ is close to zero for very small values of $v_L$.
The reason can be traced back to the perturbative unitarity constraints on the quartic coulings, in particular to those on $\lambda_L$, eq.~(\ref{Chap-Valle:eq:23b}). In fact, as $\lambda_L$ is inversely proportional to the third power of $v_L$, small values of $v_L$ demand a very small numerator, in order to prevent $\lambda_L$ violating the perturbative unitarity constraints. 
One can verify that the smallness of the numerator occurs for $\alpha_1$  close to zero, which thus explains the observed behaviour.
It turns out that it is not only $\alpha_1$ that almost vanishes for small values of $v_L$; the same should happen with $\beta$, according to eq. \ref{Chap-Valle:eq:16}. This is what is found in fig.~\ref{Chap-Valle:fig:betaalpha}.
%
%
%
%
%
%
%

\subsection{Higgs masses}

\begin{figure}[h!]
\centering
\includegraphics[width=0.31\textwidth]{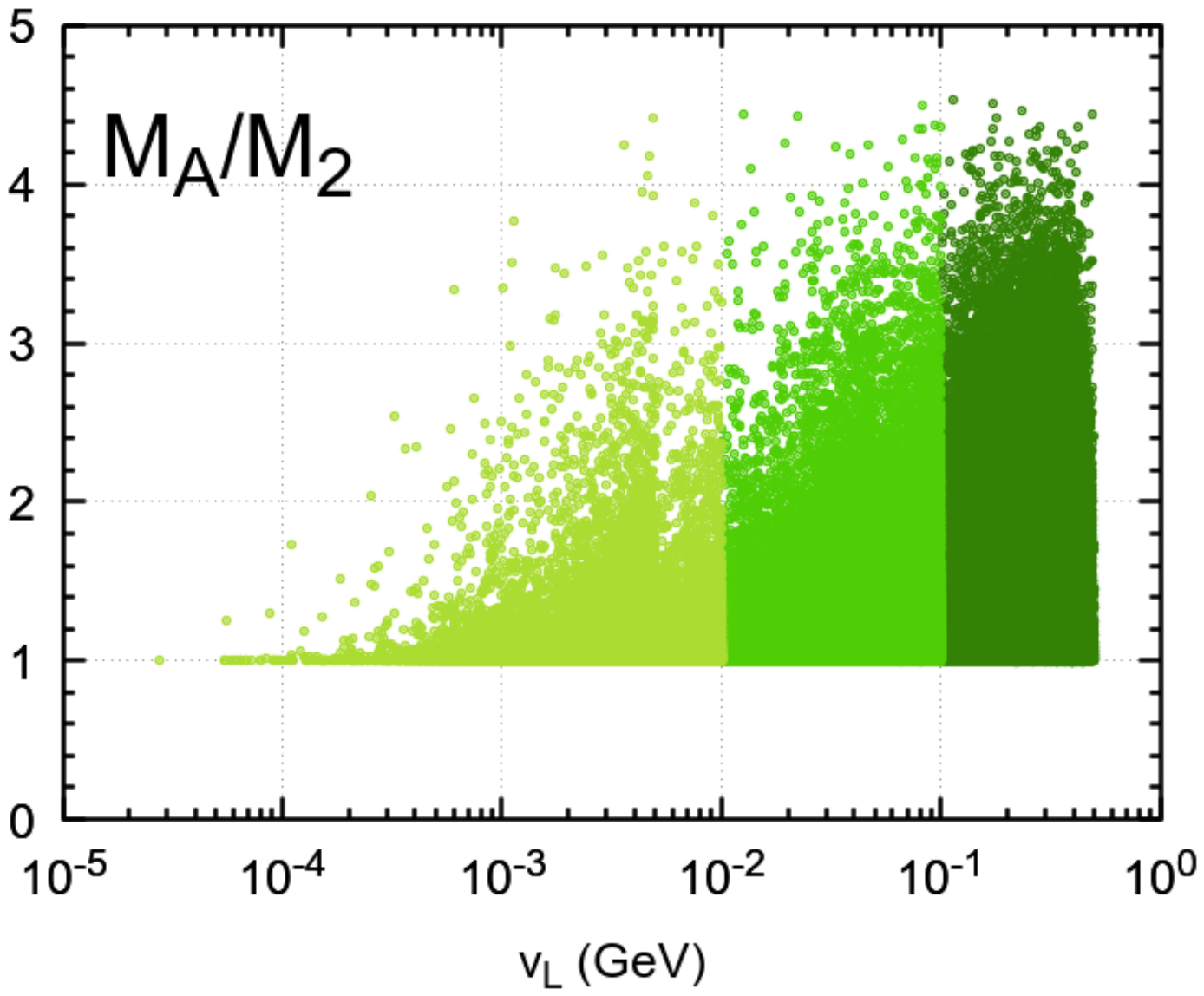}
\includegraphics[width=0.31\textwidth]{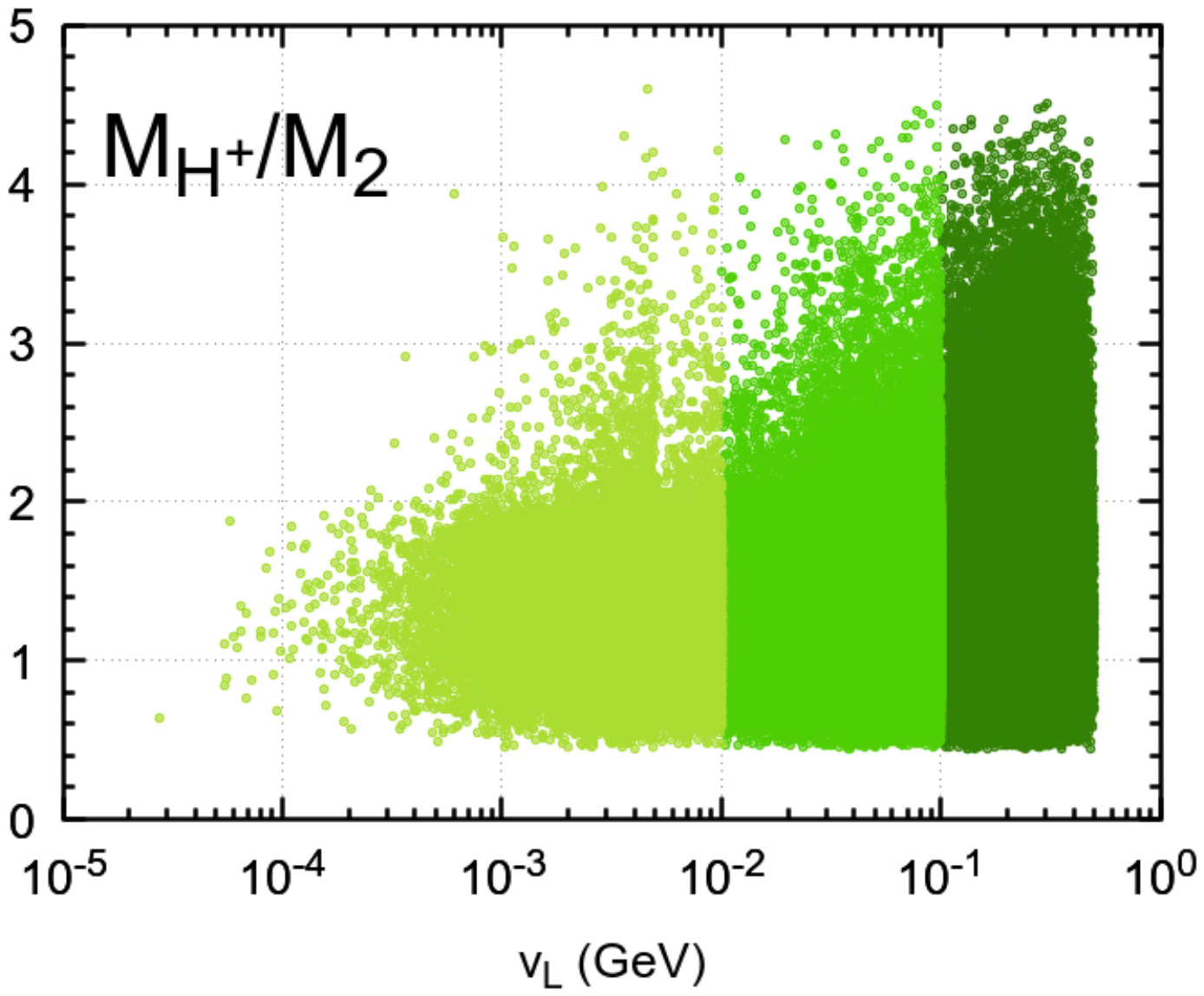} 
\includegraphics[width=0.31\textwidth]{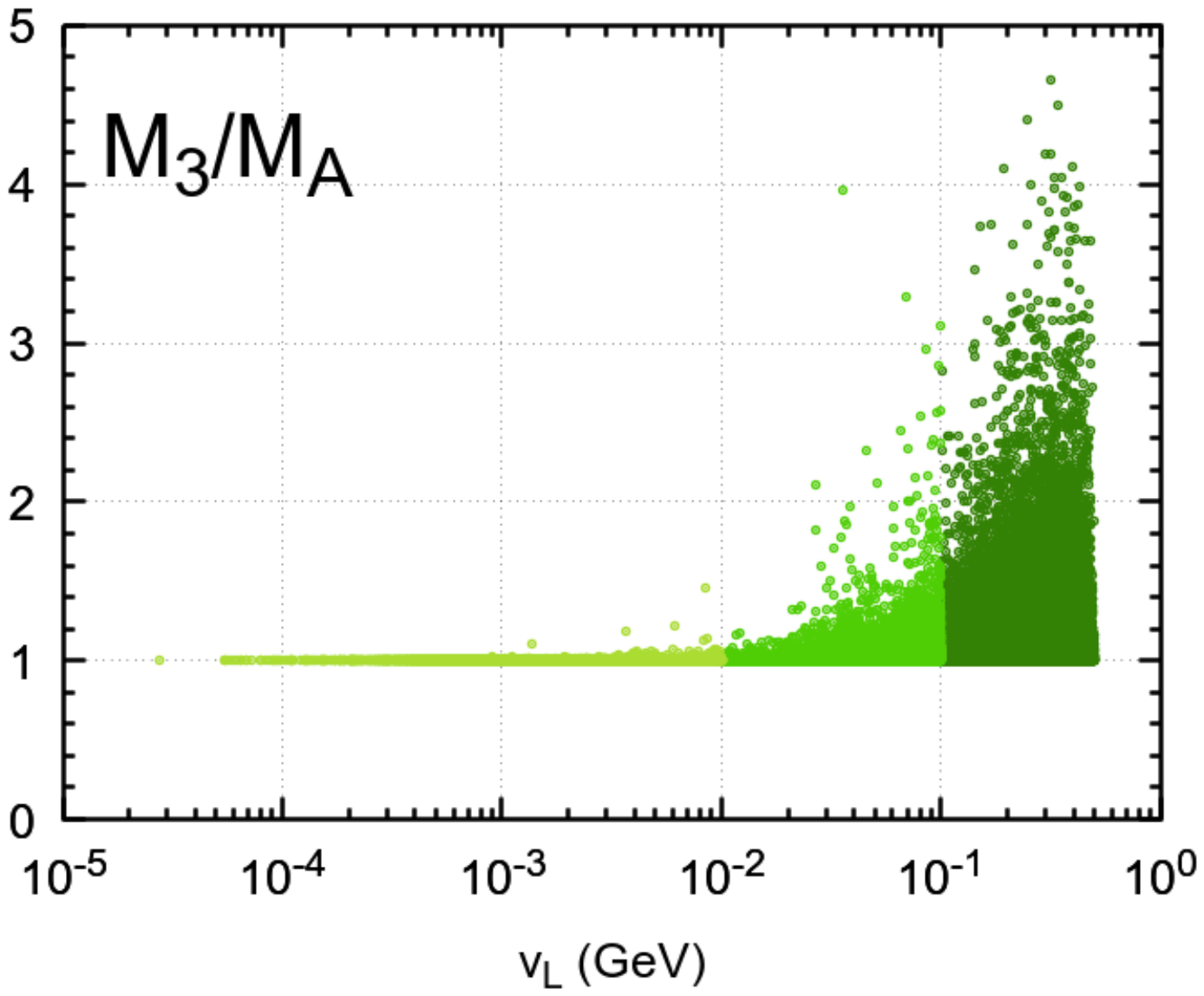}
\caption{Correlation between $m_{A}/m_{2}$ (left), $m_{\mathrm{H}^+}/m_{2}$ (middle), $m_{3}/m_{A}$ (right) and $v_L$. The color code is that of fig.~\ref{Chap-Valle:fig:alfas}.}
\label{Chap-Valle:fig:Ratios1}
\end{figure}

The argument of perturbative unitarity which we have just used to justify the smallness of $\alpha_1$ also leads to a compressed spectrum of the Higgs masses. Indeed, small values of $v_L$ require almost degenerate scalar masses in order for the unitarity constraints to be verified. This behaviour is illustrated in the plots in fig.~\ref{Chap-Valle:fig:Ratios1}. 
%
%
We should stress that this compression is much stronger than that usually imposed by the oblique parameters. On the other hand, the figures also show that, for larger values of $v_L$, the points satisfying all the constraints (including those from the oblique parameters) can have a sizeable splitting.

\subsection{Invisible Higgs decays}
\label{Chap-Valle:sec:invis-higgs-decays}

We now turn to the invisible decays of the CP-even neutral scalars to the majoron, $h_{i}\to JJ$ and $h_{i}\to 2h_{j} \to 4J$.%
\fn{For the latter, we must have $m_{i}> 2 m_{j}$.}
We investigate the branching ratios (BRs) corresponding to the total invisible decay of the scalar $h_i$ into $J$, $\textrm{BR}_{\textrm{Inv}}(h_{i})$.
We ascertain the impact of the LHC constraints on the 125 GeV Higgs boson by considering separately points complying with $3 \sigma$ (in green) and with $2 \sigma$ (in yellow green).

\begin{figure}[h!]
\centering
\includegraphics[width=0.45\textwidth]{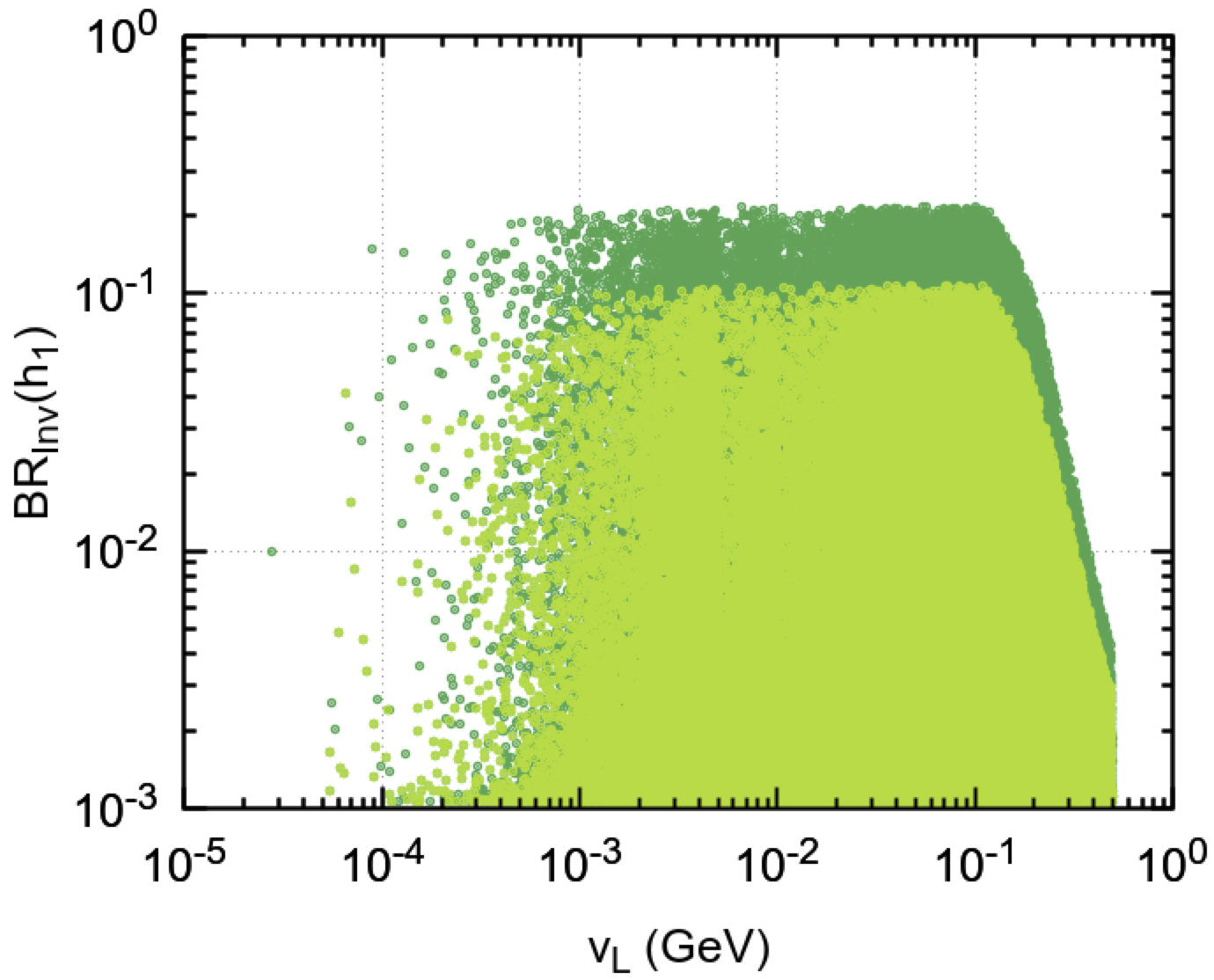}
\includegraphics[width=0.45\textwidth]{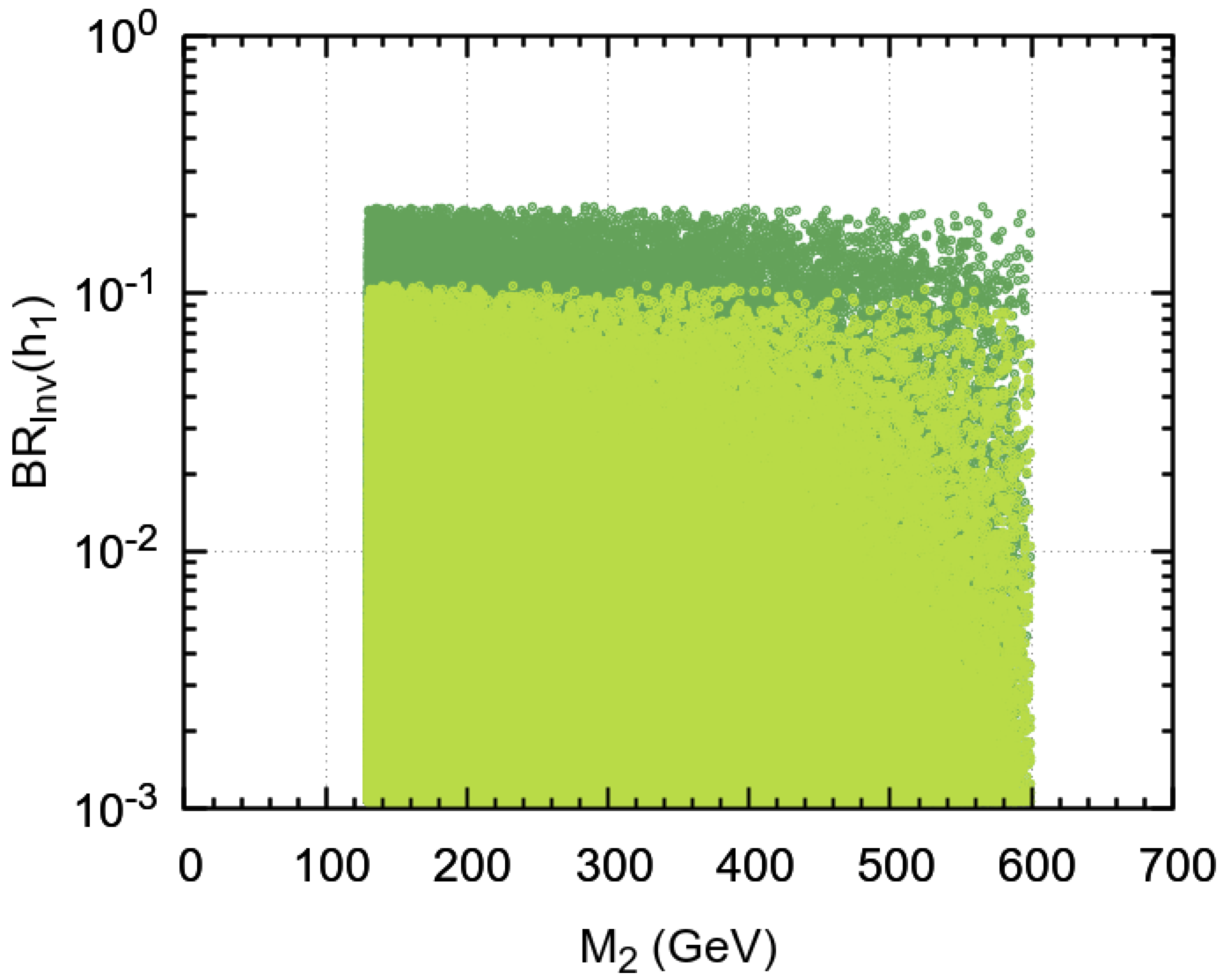}
\caption{Left panel: $\textrm{BR}_{\textrm{Inv}}(h_{1})$ versus $v_L$. Right panel: $\textrm{BR}_{\textrm{Inv}}(h_{1})$ versus $m_2$. The LHC constraints on the 125 GeV Higgs boson are imposed at 3$\sigma$ in green and at 2$\sigma$ in yellow green.}
\label{Chap-Valle:fig:BRinv}
\end{figure}

\begin{figure}[h!]
\centering
\includegraphics[width=0.50\textwidth]{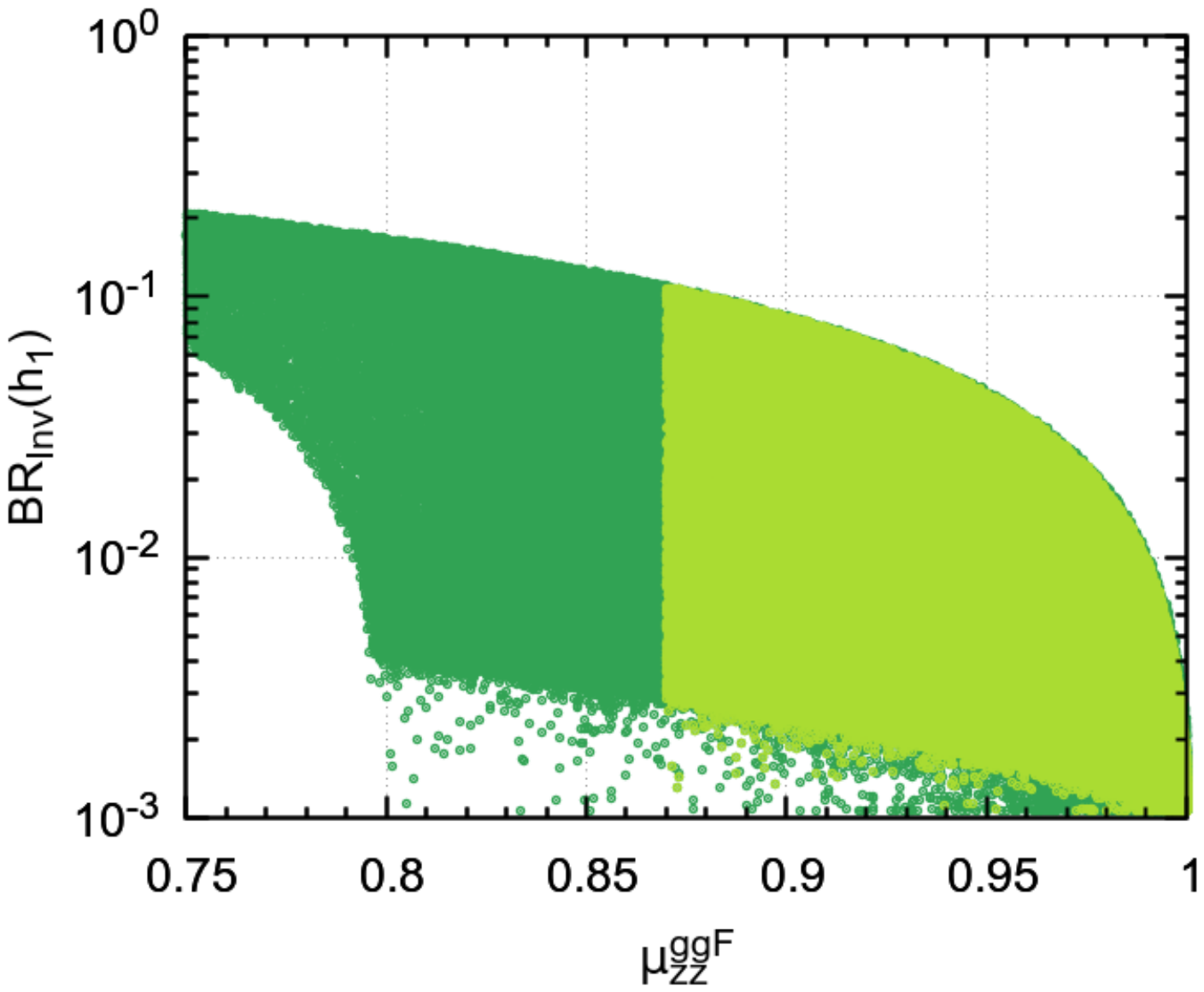}
\caption{$\textrm{BR}_{\textrm{Inv}}(h_{1})$  versus $\mu_{zz}^{ggF}$. The color code is that of fig. \ref{Chap-Valle:fig:BRinv}.}
\label{Chap-Valle:fig:BRinvb}
\end{figure}

\begin{figure}[h!]
\centering
\includegraphics[width=0.45\textwidth]{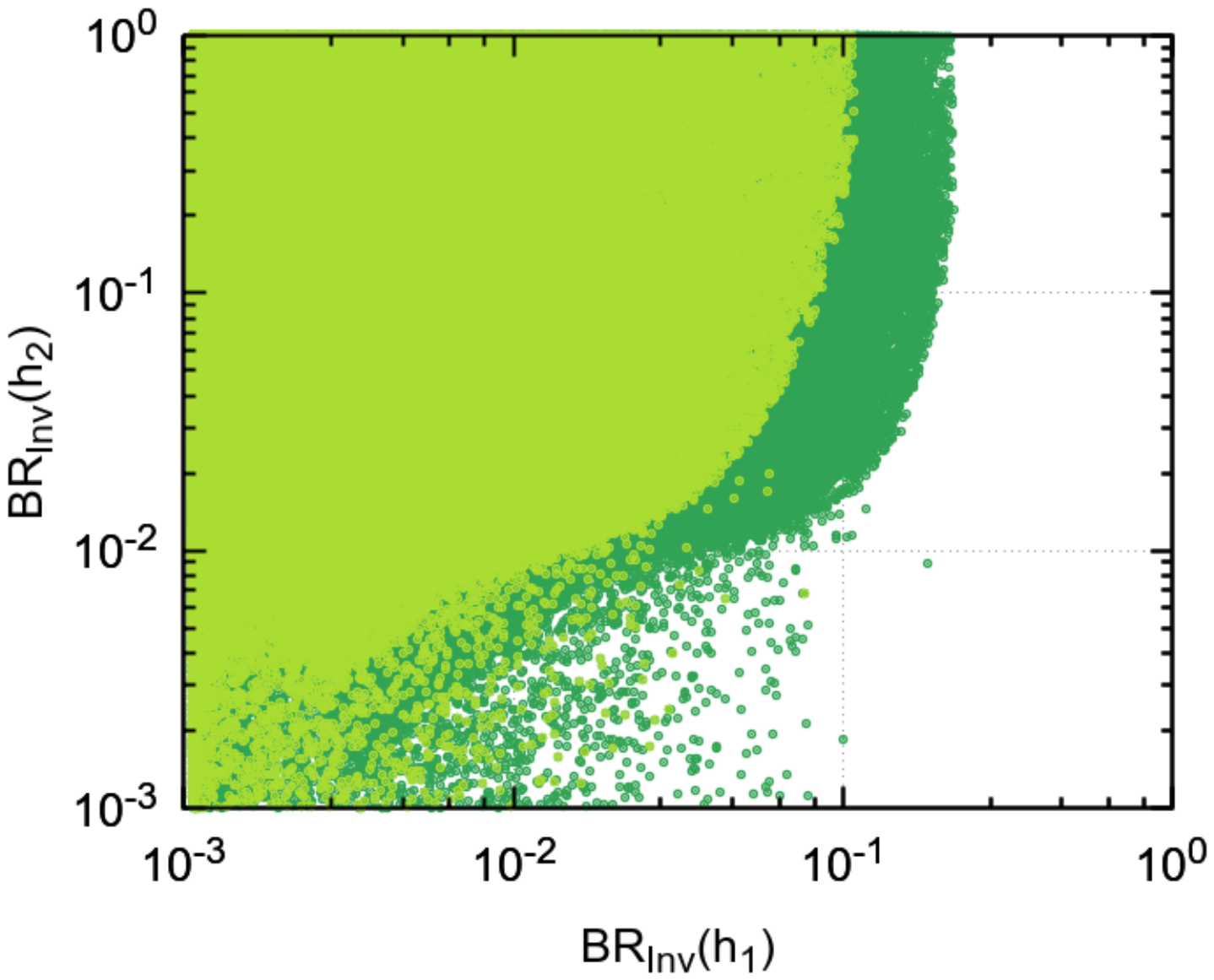}
\includegraphics[width=0.45\textwidth]{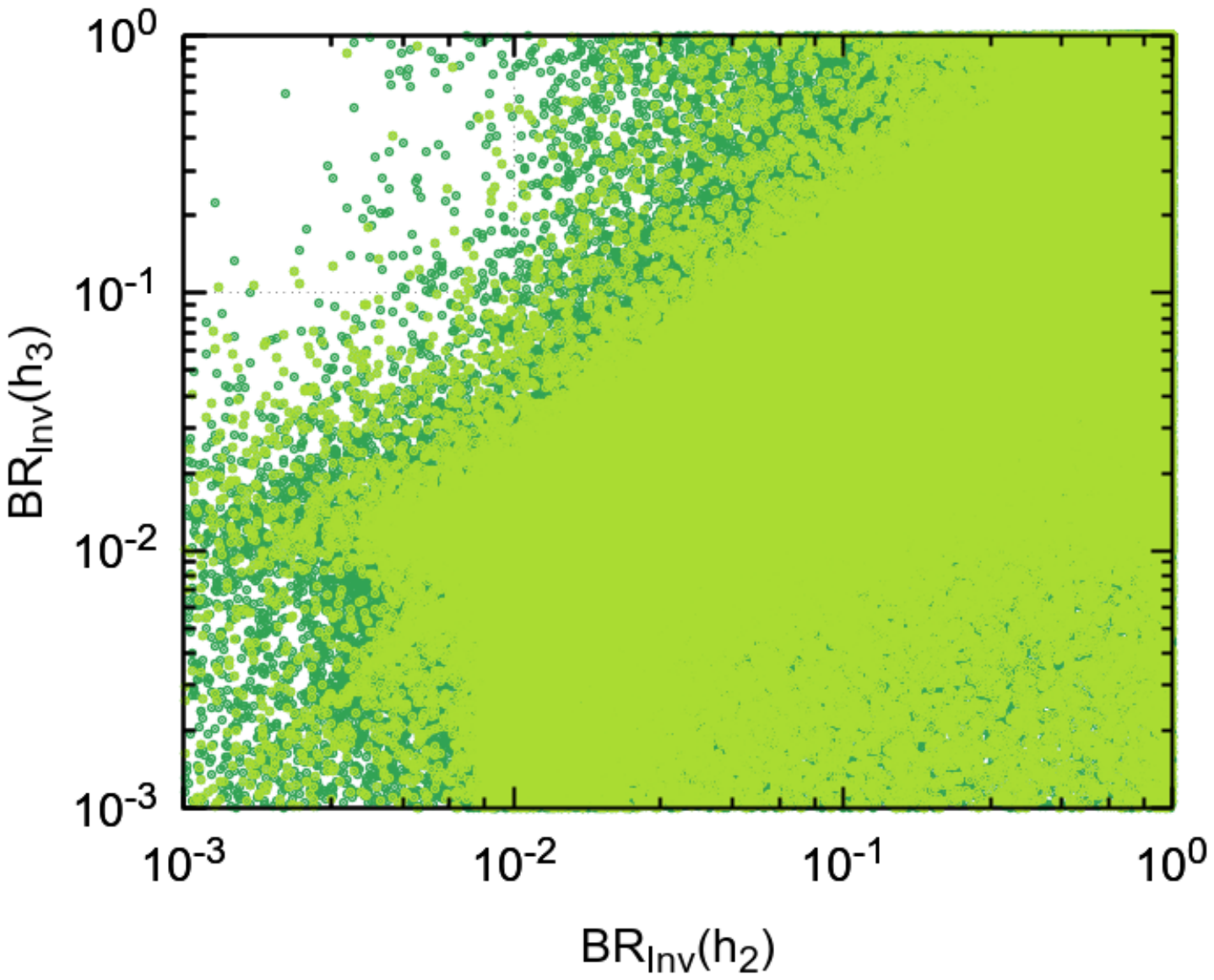}
\caption{Left panel: $\textrm{BR}_{\textrm{Inv}}(h_{2})$ versus $\textrm{BR}_{\textrm{Inv}}(h_{1})$. Right panel: $\textrm{BR}_{\textrm{Inv}}(h_{3})$ versus $\textrm{BR}_{\textrm{Inv}}(h_{2})$. The color code is that of fig. \ref{Chap-Valle:fig:BRinv}.} 
\label{Chap-Valle:fig:BRinvb2}
\end{figure}

In fig.~\ref{Chap-Valle:fig:BRinv}, we plot $\textrm{BR}_{\textrm{Inv}}(h_{1})$ as a function of $v_L$ on the left panel, and as a function of $m_2$ on the right panel.
We see that an invisible branching ratio around 20\%---close to the present upper bound \cite{Sirunyan:2018owy,Aaboud:2019rtt}---is possible for 3$\sigma$ constraints.
If 2$\sigma$ limits are applied, however, the ratio reduces to a maximum of 10\%. This is also illustrated in fig.~\ref{Chap-Valle:fig:BRinvb}, where we plot $\textrm{BR}_{\textrm{Inv}}(h_{1})$ versus the signal strength of the Higgs boson produced via gluon fusion and decaying in the $ZZ$ final state, $\mu_{zz}^{ggF}$.
We see that, if this signal strength becomes closer to one, $\textrm{BR}_{\textrm{Inv}}(h_{1})$ will be smaller, which can be important for LHC searches.
Finally, we show in fig. \ref{Chap-Valle:fig:BRinvb2} the correlation between the invisible branching ratios of different scalars. 
In the right panel, which shows $\textrm{BR}_{\textrm{Inv}}(h_{3})$ versus $\textrm{BR}_{\textrm{Inv}}(h_{2})$, it is clear that both BRs can have a wide range of values, from very small (hence visible) to close to 100\%.
\section{Summary}
\label{Chap-Valle:sec:Conclusions}

We investigated the phenomenology of the scalar sector of the simplest realization of the linear seesaw mechanism within the context of the SM gauge symmetry. In addition to the SM scalar doublet, it features two scalar multiplets charged under lepton-number: an extra doublet and a complex singlet. The physical content of the model includes a majoron $J$, which leads to severe constraints from astrophysical experiments.
This is reflected in the parameter space of the model, which nonetheless still allows a rich phenomenology. 
In particular, values of an invisible branching ratio up to $20\%$ are allowed for the current experimental precision. However, future
lepton colliders may play a decisive role here; in fact, $\textrm{BR}_{\textrm{Inv}}(h_{i})$ is
expected to be measured with precision better than 1\% level~\cite{Abada:2019zxq}, which will impose severe constraints on these decay modes.


%% file: Chapters/Conclusion.tex
\chapter*{Conclusion}
\addcontentsline{toc}{chapter}{Conclusion}

\vs{-5mm}

In this thesis, we discussed several aspects of Multi-Higgs models (MHM).
As a toolkit to more practically address the different topics, we presented \FM, a multi-tasking software for particle physics studies. By making use of already existing programs (\FR, \QG, \FC), \FMS automatically generates Feynman rules, generates and draws Feynman diagrams, generates amplitudes, performs both loop and algebraic calculations, and fully renormalizes models. In parallel with this automatic character, \FMS allows the user to manipulate the generated results in \t{\ts{Mathematica}} notebooks in a flexible and consistent way.

Throughout the thesis, the versatility of the software became manifest. 
It proved suitable for building and testing simple models, as was shown with the toy model discussed in chapter \ref{Chap-Real}; there, \FMS was used to generate Feynman rules and one-loop amplitudes, and to quickly check that the counterterms were not able to absorb all divergences.
The program also revealed quite useful to investigate the tree-level phenomenology of non-trivial models such as the one presented in chapter \ref{Chap-Valle}, by generating the Feynman rules and by calculating amplitudes and decay widths in a user-friendly way.
It also proved very convenient to ascertain more complex problems, such as the verification of both Ward identities in non-abelian theories (appendix \ref{App-WI}) and particular cases of the theorem proposed in appendix \ref{App-Theorem}, or the cancellations of UV divergences in generic models (chapter \ref{Chap-Lavou}).
It is fair to say that it even went beyond our initial expectations, by allowing to calculate the three-loop amplitudes involved in chapter \ref{Chap-Real}, which corresponds to a highly complex calculation with hundreds of Feynman diagrams.
But the most successful application of \FMS was perhaps illustrated in chapter \ref{Chap-Reno}, where the software was used to fully renormalize a non-trivial model such as the C2HDM and study it at NLO; this involved several NLO decay widths, dozens of counterterms, hundreds of Feynman diagrams and a myriad of calculations and manipulations that could hardly have been performed with the same flexibility and consistency with any other software.

As a prelude to the inclusion of higher orders in MHM, we devoted special attention to the topic which we dubbed \textit{selection of the true vev}.
We argued that this procedure is distinct from and prior to that of renormalization, despite the fact that the two are usually mixed together in the literature. We noted that such mixture can be misleading, precisely because it blurs the distinction between the two procedures.
We discussed two different methods to perform the selection of the true vev of the up-to-one-loop theory, PRTS and FJTS, and we highlighted the advantages of the latter over the former. We also showed that one does not need to select the true vev, even if that leads to divergent one-point functions. As a matter of fact, in what concerns the connected GFs with at least two external legs, the FJTS turns out to be equivalent to an approach where the bare vev is selected. 
In the end, then, as long as one includes one-loop tadpole insertions in the GFs (which can be trivially done with modern tools such as \FM), one can avoid the discussion that aims at the removal of 1-point functions and simply stick to the bare vev.

In our exploration of MHM, we put forward the daring claim according to which the real 2HDM---a variant of 2HDM which was studied in hundreds of papers---is not a consistent model. This inconsistency, however, turns out to be terribly difficult to prove. We started by resorting to a simple toy model, suffering from a similar pathology, and we showed that the problem there is undeniable. This ought to convince the community that, even if a thorough demonstration of the same problem in the real 2HDM is not yet available, this model will very likely suffer from the same unsoundness.
We approched the problem in the real 2HDM by calculating the most divergent part of a three-loop amplitude. Our undertaking is not yet complete, for a three-loop renormalization of the process at stake is still unavailable. Nonetheless, we hope that this venture spurs further interest in the problem, and that a full calculation will become possible in the future.

In face of the argument just described of inconsistency of the real 2HDM, we turned to the most simple model that simultaneously a) corrects such inconsistency and b) accounts for CP violation in Nature. This model is the C2HDM, which has played a major role in the present thesis.
We started by investigating its phenomenology at LO. We realized that several exciting scenarios---such as having large CP-odd Yukawa couplings, or having scalar bosons coupling in a fully CP-odd way to a fermion and in a fully CP-even way to another---are allowed by this model.

We then discussed the renormalization of the model. We showed that, despite the fact that such undertaking is inserted in a sequence of recent studies on renormalization, it nonetheless configures a very unique procedure. Indeed, the fact that CP violation shows up in several GFs at one-loop level forces the tree-level description to be more general than it otherwise would be; notably, several parameters that can usually be rephased away must be considered in the new tree-level description, so as to account for all the UV divergences. In the end, one ends up having more independent counterterms than independent renormalized parameters.
Consequently, the paradigm of renormalization is enlarged: instead of discussing merely which subtraction scheme should be employed to fix this and that particular counterterm, one is now faced with the possibility of choosing different combinations of independent parameters. Whereas in the usual paradigm it is just the choice of one or the other subtraction that has influence on the results, in the new paradigm the choice of one combination or the other will also contribute to the final predictions. Given its pioneering character, the renormalization of the C2HDM work opens the door to an exploration of precise predictions in this model, as well as to the renormalization of models with CP violation in the scalar sector.

There are good reasons to believe that such undertakings
will prove valuable.
The LHC Run-3 is anxiously expected for the period between 2022 and 2024, for an integrated luminosity of about $150 \, \mathrm{fb}^{-1}$. There, the collision statistics recorded until now will be doubled \cite{Bass:2021acr}. In addition, a factor of 20 in the increase of total statistics is expected for the high luminosity upgrade of the LHC, which shall begin before the end of this decade \cite{Bruning:2019aaa}. With such apparatus, several Higgs boson’s couplings are expected to be measured within a precision of a few $\%$, and the first direct indications of the Higgs self-interaction are hoped be detected \cite{Bass:2021acr}.
These projections are certainly stimulating for the community devoted to precise predictions.

But this is not all. In fact, plans for future colliders clearly aim at a better understanding of the scalar sector \cite{deBlas:2019rxi,DiMicco:2019ngk}.
Precision studies relative to the Higgs boson are selected by the recent European Particle Physics Strategy update as the main priority for the next high-energy collider \cite{Gianotti:2020aaa,European:2020aaa}.
After the LHC Run-3, several options are still open.
In context of the future of CERN, both a Future Circular Collider \cite{Benedikt:2019okv,Benedikt:2020ejr} (FCC-ee) and CLIC linear $e^+e^-$  collider \cite{Stapnes:2019aaa,Sicking:2020gjp} are being considered.
Other options, such as $e^+e^-$ collider options like the International Linear Collider (ILC) in Japan \cite{Michizono:2019aaa} and a circular collider CEPC in China \cite{Lou:2019aaa}, are equally under discussion.
The projects for the circular collider also consider other options, such as a proton-proton (FCC-hh) and proton-lepton (FCC-eh) colliders \cite{Bass:2021acr} 
(for a detailed discussion of the precision that is expected to be reached on measurements relative to properties of the Higgs boson with these different options, cf. ref. \cite{deBlas:2019rxi}).
All in all, the future of high-energy physics looks very promising.

%% file: Appendices/FM-Manual.tex
\chapter{FeynMaster 2 manual}
\label{App-FM-Manual}

\vs{-5mm}

\section{Introduction}
\label{Chap-FM:sec:Intro}

\n \textsc{FeynMaster} \cite{Fontes:2019wqh} was recently introduced as a multi-tasking software for particle physics studies. Combining \ts{FeynRules}~\cite{Christensen:2008py,Alloul:2013bka}, \QG~\cite{Nogueira:1991ex} and \ts{FeynCalc}~\cite{Mertig:1990an,Shtabovenko:2016sxi,Shtabovenko:2020gxv}, \FMS is able to perform the totality of the following list of tasks:
\vs{1.0mm}
\begin{center}
\quad a) generation and drawing of Feynman rules; \qquad b) generation and drawing of Feynman diagrams;\\
c) generation of amplitudes; \hs{3mm} d) loop calculations; \hs{3mm} e) algebraic calculations; \hs{3mm} f) renormalization.
\end{center}
\vs{1.0mm}
It was presented not as competing with other existing software that perform some of the elements of the above list (e.g., refs. \cite{Belanger:2003sd,Cullen:2011ac,Cullen:2014yla,Lorca:2004fg,Degrande:2014vpa,Shtabovenko:2016sxi,Mertig:1990an,Alloul:2013bka,Shtabovenko:2020gxv,Christensen:2008py,Hahn:1998yk,Hahn:2000kx,Kublbeck:1990xc,Pukhov:1999gg,Nogueira:1991ex,Semenov:1996es,Semenov:1998eb,Tentyukov:1999is,Wang:2004du,Alwall:2014hca}), but rather as an alternative---with four main advantages.
First, \ts{FeynMaster} has a hybrid character concerning automatization: not only does it automatically generate the results, but it also allows the user to act upon them through \t{\ts{Mathematica}} notebooks. Second, the complete set of analytical expressions for the counterterms in the modified minimal subtraction ($\overline{\text{MS}}$) scheme can be automatically calculated. Third, \ts{FeynMaster} includes a thorough interaction with numerical calculations through \t{\ts{Fortran}}, using \ts{LoopTools}~\cite{Hahn:1998yk} in one-loop calculations. Finally, all the printable outputs of \ts{FeynMaster}---the complete set of Feynman rules (tree-level and counterterms), the Feynman diagrams, as well as a list containing both the expressions and computed counterterms---are automatically written in \LaTeX \, files.

\n \FMS 2 has significant improvements over the previous version; we highlight three of them.
First of all, \FMS 2 is much more practical, since:
\begin{addmargin}[8mm]{0mm}
a) a model is now specified by a single model file,\\
b) the file controlling the \FMS run now contains just the data relevant for the run,\\
c) the presence of \QGS has virtually vanished.
\end{addmargin}
The last point means that there is no longer such thing as a \QGS style or convention, which in turn implies that there is a single set of conventions (the \FRS ones) used throughout the whole program.
A second major improvement in \FMS 2 is that the calculated results can now be automatically stored. That is, a certain process can be calculated once and for all, which considerably simplifies the manipulation of results and the use of \FMS as a whole. 
In the third place, the renormalization is significantly faster; for example, the generation and impression of the total set of Feynman rules (tree-level and counterterms) for the Standard Model is now completed in just 5 minutes in a regular laptop.
In addition, several functions were corrected, improved or simply created.

\n The general usage of \ts{FeynMaster} can be summarized in a few lines: after the user has defined the relevant directories and the model, the \ts{FeynMaster} run is controlled from a single file (\t{Control.m}). Here, the sequence of processes to study can be chosen, as well as many different options. \ts{FeynMaster} is then ready to run. The run automatically generates and opens several PDF files---according to the options chosen in \t{Control.m}---and creates the above-mentioned \t{\ts{Mathematica}} notebooks.

\n The present appendix is a complete and auto-sufficient manual of \ts{FeynMaster} 2, and is organized as follows. In section \ref{Chap-FM:sec:Instal}, we explain how to download and install \ts{FeynMaster}. Section \ref{Chap-FM:sec:Create} is devoted to the creation of models, and section \ref{Chap-FM:sec:Usage} to the detailed usage of \ts{FeynMaster}. Then, in section \ref{Chap-FM:sec:Examples}, we give some examples. Finally, in section \ref{Chap-FM:sec:Summary} we present a quick first usage of \ts{FeynMaster}.

\section{Installation}
\label{Chap-FM:sec:Instal}

\n \ts{FeynMaster} can be downloaded at:
\begin{center}
\url{https://porthos.tecnico.ulisboa.pt/FeynMaster/}.
\end{center}
\ts{FeynRules}, \QGS and \ts{FeynCalc}, essential to run \ts{FeynMaster}, can be downloaded in \url{https://feynrules.irmp.ucl.ac.be/}, \url{http://cfif.ist.utl.pt/~paulo/qgraf.html} and \url{https://feyncalc.github.io/}, respectively.%
\fn{The user of \ts{FeynMaster} is supposed to be familiar with both \ts{FeynRules} (in order to define new models) and \ts{FeynCalc} (in order to manipulate the final results).
There is no need to know \QG, since the non-trivial part of this program---the definition of the model---is automatically carried through by \ts{FeynMaster}.
We verified that \ts{FeynMaster} runs properly with the latest public versions of \ts{FeynRules}, \QGS and \ts{FeynCalc}---namely, versions 2.3.36, 3.4.2 and 9.3.0, respectively.
More instructions on how to download and install \QGS can be found in \url{https://porthos.tecnico.ulisboa.pt/CTQFT/node9.html} and in \url{https://porthos.tecnico.ulisboa.pt/CTQFT/node33.html}. Finally, when using Linux or Mac, the executable \QGS file should be named {\t{qgraf}}.

\n To run \ts{FeynMaster}, it is also necessary to have \t{\ts{Python}}, \t{\ts{Mathematica}} and \LaTeX \, installed; links to download are \url{https://www.python.org/downloads/} \url{http://www.wolfram.com/mathematica/} and \url{https://www.latex-project.org/get/}, respectively.
We tested \ts{FeynMaster} using version 3.6 of \t{\ts{Python}} and version {12.0} of \t{\ts{Mathematica}}. %
(Note that \ts{FeynMaster} will \textit{not} run if a \t{\ts{Python}} version prior to 3 is used, and it is only guaranteed to properly work if a version of \t{\ts{Mathematica}} not older than 10.3 is used.)
As for \LaTeX \,, the user is required to update the package database; note also that, in the first run of \ts{FeynMaster}, some packages (like \t{feynmp-auto} and \t{breqn}) may require authorization to be installed.
Finally, for the numerical calculations, the needed files are written in the \t{\ts{Fortran}} 77 format. A \t{\ts{Fortran}} compiler is needed, as well as an installation of \ts{LoopTools}. The programs were shown to work both with the gfortran and ifort compilers and with LoopTools-2.14 and above.}
After downloading \ts{FeynMaster}, the downloaded file should be extracted and the resulting folder (named \t{FeynMaster}) can be placed in any directory of the user's choice.
Then, the directories corresponding to \ts{FeynRules} and \QG, as well as the one corresponding to the \ts{FeynMaster} output (i.e., the directory where the outputs of \FMS shall be saved), should be defined. This must be done by writing the appropriate paths after \t{dirFR}, \t{dirQ} and \t{dirFMout}, respectively, in the beginning of the \t{RUN-FeynMaster} batch file lying inside the \t{FeynMaster} folder.%
\fn{\label{Chap-FM:note:instal}%
These lines are commented in the batch file, and must continue to be so; besides, the beginning of these lines (with \t{dirFR}, \t{dirFR} and \t{dirFMout}) should \textit{not} be erased.
By default, the folder with the models for \ts{FeynMaster} is inside the \t{FeynMaster} folder and is named \t{Models}; this can be changed by defining a variable \t{dirFRmod} in the beginning of \t{RUN-FeynMaster} and setting a path for it. In a similar way, the folder with the models for \QGS (that is, the folder that shall contain the \QGS models, which will be automatically generated by \ts{FeynMaster}) is automatically generated and set inside the \QGS folder with the default name \t{Models}; as before, this can be changed by defining a variable \t{dirQmod} in the beginning of \t{RUN-FeynMaster} and setting a path for it.}

\n To test \ts{FeynMaster}, jump to section \ref{Chap-FM:sec:Summary}, where instructions for a quick first usage are given.

\section{Creating a new model}
\label{Chap-FM:sec:Create}

\n \ts{FeynMaster} is a model dependent program: it cannot run without the specification of a model.
As already mentioned, one of the major simplifications of \ts{FeynMaster} 2 lies in model building: while in the previous version the specification of each model required two files, in \ts{FeynMaster} 2 only one file is required to fully define a model. 
This one file---which we now identify as \textit{the} \ts{FeynMaster} model file---is essentially a regular model file for \ts{FeynRules};
as such, it has the termination \t{.fr}.%
\fn{It is assumed here that the user is familiar with \ts{FeynRules} and knows how to create a \ts{FeynRules} model. If this is not the case, we refer to the \ts{FeynRules} website \url{https://feynrules.irmp.ucl.ac.be/}. Finally, it should be clear that the \ts{FeynMaster} model file can always be used as a \ts{FeynRules} model file.}
Its name is identified in the following as the \textit{internal name} of the model. This model file must be inside a folder with the same name---that is, with the internal name of the model---, which in turn must be inside the folder with the models for \ts{FeynMaster}.

\n \ts{FeynMaster} already comes with three models : QED, Scalar QED (SQED) and the Standard Model (SM).\fn{The SM model file is written in an arbitrary $R_{\xi}$ gauge and with the $\eta$ parameters of ref.~\cite{Romao:2012pq}, and it closely follows ref.~\cite{Denner:1991kt} for renormalization.}
These serve as prototypes, and we highly recommend them as guiding tools in the creation of a new model.
Since there are already models available, this section can be skipped in a first utilization of \ts{FeynMaster}.

\n We now describe in detail the creation of a \ts{FeynMaster} model file. As we just mentioned, this file is essentially a model for \ts{FeynRules}. Yet, it is characterized by special features: while some of the definitions and attributes are common to \FR, some were specifically designed for \FM, as we now show.

\vs{3mm}
\n \underline{\textbf{Lagrangian}}

\n The \ts{FeynMaster} model file must include the Lagrangian of the theory. The Lagrangian must be separated in different parts---each of them corresponding to a different type of interaction---and each of those parts should have a specific name: see table \ref{Chap-FM:tab:FRnames}.\fn{It is not mandatory to define all the 6 Lagrangian parts present in table \ref{Chap-FM:tab:FRnames} (for example, if the model does not have a ghost sector, there is no need to define LGhost); only, no other name besides the 6 ones specified in the right column of table \ref{Chap-FM:tab:FRnames} will be recognized by \ts{FeynMaster}.}
\begin{table}[!h]%
\begin{normalsize}
\normalsize
\begin{center}
\begin{tabular}
{@{\hspace{3mm}}>{\raggedright\arraybackslash}p{5cm}>{\raggedright\arraybackslash}p{2.9cm}@{\hspace{3mm}}}
\hlinewd{1.1pt}
Type of interactions & \ts{FeynRules} name \\
\hline
pure gauge & \t{LGauge} \\
fermion-gauge & \t{LFermions}\\
Yukawa & \t{LYukawa} \\
scalar-scalar and gauge-scalar & \t{LHiggs} \\
ghosts & \t{LGhost} \\
gauge fixing & \t{LGF} \\
\hlinewd{1.1pt}
\end{tabular}
\end{center}
\vspace{-5mm}
\end{normalsize}
\caption{Name of the different Lagrangian parts according to the type of interaction.}
\label{Chap-FM:tab:FRnames}
\end{table}
\normalsize
The Lagrangian parts involving fermions (\t{LFermions} and \t{LYukawa}) can be written either in terms of Dirac fermions or Weyl fermions; in both cases, the Feynman rules will be presented for Dirac fermions.

\vs{3mm}
\n \underline{\textbf{Parameters}}

\n Parameters are defined via the usual \FRS variable \t{M\$Parameters}.
Different attributes can be associated to a certain parameter; their order is irrelevant.
As in \FR, the parameters are by default real; otherwise, one must include the attribute \t{ComplexParameter} and set it to \t{True} (i.e., \t{ComplexParameter -> True}).
The \LaTeX \, name of a parameter should be set with the attribute \t{TeXName}\fn{Which does not exist in \FRS and should not be confused with the \FRS attribute \t{TeX}.} and written between double commas in \LaTeX \, style (ex: \verb|TeXName -> "\\lambda_2"|).\fn{The backslash should always be doubled in \t{TeXName}, as well as in \t{TeXAntiName} (see below).}
Parameters can have indices, which are defined as in \FRS (ex: \t{Indices -> \{Index[Gen], Index[Gen]\}}).%
\fn{In this example, the range of \t{Gen} must be defined in the model; for example,
\t{IndexRange[Index[Gen]]} \t{= Range[3]}.}
A numerical value can be given to a parameter through the attribute \t{NumValue} (ex: \verb|NumValue -> 0.0047|); in the case of parameters with indices, the set of values should be written as a \t{\ts{Mathematica}} list (ex: for a parameter with dimensions $2\times2$, one can set \t{NumValue -> \{0.5, 1, 1.5, 2\}}, for the matrix entries 11, 12, 21, 22, respectively).%
\fn{The attribution of numerical values is useful only in case the user wants to exploit the numerical interface of \ts{FeynMaster}---to be explained in section \ref{Chap-FM:sec:FC}. The SM model file includes numerical values according to \cite{Tanabashi:2018oca}.}
The renormalization rule for a parameter can be defined through the attribute \t{Renormalization}, in such a way that the rule must be written inside braces (ex: \verb|Renormalization -> {mf -> mf + dmf}|).\fn{It should be clear that it is not mandatory to define renormalization rules. That is, \FMS can be used even if no renormalization is performed.
Finally, note that, if a parameter is complex, the renormalization of its complex conjugate is automatically considered.}
Finally, parameters corresponding to counterterms should be flagged by setting the attribute \t{Counterterm} to \t{True}.\fn{There can be counterterms for both parameters and fields; since counterterms themselves are parameters, they are defined in the parameters sections \t{M\$Parameters}.}

\vs{3mm}
\n \underline{\textbf{Particle classes}}

\n Particles are defined via the usual \FRS variable \t{M\$ClassesDescription};
again, the order of attributes associated to a certain parameter is irrelevant.
The \t{ClassName} cannot correspond to \FMS internal indices, such as \t{J1}, \t{J2}, \t{J3}, \t{J4}, \t{p1}, \t{p2}, \t{q1}, \t{q2}.
The attribute \t{SelfConjugate} should always be defined, either to \t{True} or to \t{False}.
Unphysical particles (except Weyl fermions) should be flagged with the attribute \t{Unphysical} set to \t{True}.
Every propagating particle must have a mass assignment, either to a variable or to 0 (ex: \t{Mass -> MH} or \t{Mass -> 0}).
Propagating particles should also have a \LaTeX \, name (ex: \verb|TeXName -> "c_{W^-}"|); when a particle is not its own complex conjugate, a \LaTeX \, name for the antiparticle should also be specified, through the attribute \t{TeXAntiName} (ex: \verb|TeXAntiName -> "\\bar{c_{W^-}}"|).
The decay width can be defined through \t{DecayWidth} (ex: \verb|DecayWidth -> DWZ|).%
\fn{The decay width thus defined will show up in the denominator of the propagator of the particle at stake whenever such propagator mediates the $s$-channel of a 2 $\to$ 2 scattering process. This will also automatically happen for the corresponding would-be Goldstone boson of the particle at stake (in the case it corresponds to a gauge boson) if, in the definition of the particle corresponding to the would-be Goldstone, the attribute \t{Goldstone} is associated to the gauge boson at stake (ex: \t{Goldstone -> Z}).}
Weyl components must be specified for Dirac fermions whenever the latter are composed of Weyl particles defined in the model (ex: \verb|WeylComponents -> {uqL,uqR}|).
Neutrinos should be flagged with the attribute \t{Neutrino} as \t{True}.%
\fn{This is relevant only for the definition of number of polarizations in decay or scattering processes. In particular, the functions \t{DecayWidth} and \t{DiffXS} (cf. table \ref{Chap-FM:tab:FCfunc} below) use this information.}
Finally, renormalization is defined in a way similar to that of the parameters (ex: \t{Renormalization -> \{H -> H + 1/2 dZH H\}}).%
\fn{In the case of gauge bosons, generic Lorentz indices should be included. Renormalization of antiparticles is automatically included and needs not be specified by the user.}

\vs{3mm}
\n \underline{\textbf{Extra (optional) information}}

\n Besides \t{M\$Parameters} and \t{M\$ClassesDescription}, \FMS allows extra optional quantities to be defined.
These are presented in table \ref{Chap-FM:tab:FMmodel} and for each one we present a brief description, the default value and an example. A few of comments are in order.
%
%
\begin{table}[!h]%
\begin{center}
\begin{tabular}
{@{\hspace{3mm}}
>{\raggedright\arraybackslash}p{2.7cm}
>{}p{5.4cm}
>{}p{2.5cm}
>{}p{3.2cm}
@{\hspace{3mm}}}
\hlinewd{1.1pt}
Quantity & Description & Default value & Example\\
\hline\\[-2mm]
\t{M\$ModelExtName} &  model name to be printed in the \ts{FeynMaster} final documents & \textit{(internal name)} & \t{"Standard Model"}\\[7mm]
\t{M\$FCeqs} & identities for \ts{FeynCalc} & \textit{(empty)} & \t{\{xiA->1\}}\\[7mm]
\t{M\$FCsimp} & simplifications for \ts{FeynCalc} & \textit{(empty)} & \verb|{cw^2+sw^2->1}|\\[7mm]
\t{M\$PrMassFL} & when true, masses of propagators are extracted from the Lagrangian & \t{True} & \t{False}\\[7mm]
\t{M\$GFreno} & when true, renormalization rules are applied to the Gauge Fixing Lagrangian & \t{False} & \t{True}\\[10mm]
\t{M\$RestFile}  & restrictions file for the \ts{FeynRules} model & \textit{(empty)} & \t{"MyRest.rst"}\\[10mm]
\t{RenoPreRep} & function to rewrite the Lagrangian immediatly before renormalization & \textit{(empty)} & \textit{(cf. SM model file)}\\
\hlinewd{1.1pt}
\end{tabular}
\end{center}
\vspace{-5mm}
\caption{Optional variables for the \FMS model file. See text for details.}
\label{Chap-FM:tab:FMmodel}
\end{table}
\normalsize

\n First, \t{M\$FCeqs} should be distinguished \t{M\$FCsimp}: while both are defined in the \FMS model file as lists of replacement rules, the former ends up being converted into a list of equalities for the \FC, whereas the latter is only used as a set of replacement rules. So, considering the examples presented in table \ref{Chap-FM:tab:FMmodel}, the \FCS notebook will have \t{xiA} defined as 1 (i.e., \t{xiA:=1}), while it will use the rule \verb|cw^2+sw^2->1| to simplify the calculations (more details on section \ref{Chap-FM:sec:FC}).

\n Second, if \t{PrMassFL} is set to \t{True}, the poles of the propagators are extracted from the Lagrangian---i.e., they are defined as the bilinear terms of the field at stake in the Lagrangian---and the propagator is written in the most general form. If \t{PrMassFL} is set to \t{False}, the poles will match the variable corresponding to the mass of the propagator and the propagator will be written in the Feynman gauge. While setting \t{PrMassFL} to \t{True} is certainly the most faithful way to describe the propagator, this option may bring certain difficulties: on the one hand, it requires the definition of a Gauge Fixing term for gauge boson propagators; on the other hand, the bilinear terms can be very complicated expressions.

\n Finally, \t{RenoPreRep}, if used, must be defined as a \t{\ts{Mathematica}} function with a single argument. It is applied to the Lagrangian immediatly before the renormalization process, thus allowing to rewrite it (i.e., the Lagrangian) in a more convenient way. This may be useful to obtain simpler expressions for the Feynman rules of the counterterms.

\vs{-1mm}
\section{Usage}
\label{Chap-FM:sec:Usage}

\n Once the initial specifications are concluded (i.e., once the directories and the model are defined), \ts{FeynMaster} is ready to be used. In this section, we explain in detail how to use it. We start by showing how to edit the file that controls the \ts{FeynMaster} run. Then, after describing how to run \ts{FeynMaster}, we comment its outputs and explain how to use the notebooks we alluded to in the Introduction.

\subsection{\t{Control.m}}
\label{Chap-FM:sec:Control}

\n As mentioned before, the \ts{FeynMaster} run is uniquely controlled from the \t{Control.m} file (which lies inside the \t{FeynMaster} folder). In this section, we explain the different components of this file.

\vs{3mm}
\n \underline{\textbf{Model selection}}

\n First, the model must be chosen; this is done by editing the name after the variable \t{model} in the beginning of \t{Control.m}. This name must correspond to the internal name of the model (the same name given to the \FMS model file).

\vs{3mm}
\n \underline{\textbf{Process specification}}

\n Then, the user should start by specifying the desired process (or processes).%
\fn{
It is possible to define a sequence of processes---that is, a series of processes to be run in a single \ts{FeynMaster} run.
To define a sequence of processes, the user must copy the lines of the first process, paste them after it, and edit them to define the second process, and repeat the same procedure for more processes (an example of a sequence of processes will be given in section \ref{Chap-FM:sec:FullReno}).
}
One process is specified through the definition of the set of variables shown in table \ref{Chap-FM:tab:process}.%
\fn{As in table \ref{Chap-FM:tab:FMmodel}, all variables in table \ref{Chap-FM:tab:process} can be completely omitted from the \t{Control.m} file, in which case the default values are applied.}
\begin{table}[!h]%
\begin{center}
\begin{tabular}
{@{\hspace{3mm}}>{\raggedright\arraybackslash}p{2.5cm}>{}p{3.9cm}>{}p{2.3cm}>{}p{4.6cm}@{\hspace{3mm}}}
\hlinewd{1.1pt}
Variable & Description & Default value & Example\\
\hline\\[-2mm]
\t{inparticles} & incoming particles & \textit{(empty)} & \t{e,ebar} \\[2.5mm]
\t{outparticles} & outgoing particles & \textit{(empty)} & \t{H,Z} \\[2.5mm]
\t{loops} & number of loops & \t{0} & \t{1}\\[2.5mm]
\t{parsel} & intermediate particles & \textit{(empty)} &
\verb|{avoid,Z,1,3},{keep,e,1,1}|
\\[2.5mm]
\t{factor} & quantity to factor out & 1 & \verb|16 Pi^2/I| \\[2.5mm] 
\t{options} & options & \textit{(empty)} & \t{onepi}\\
\hlinewd{1.1pt}
\end{tabular}
\end{center}
\vspace{-5mm}
\caption{Variables that specify one process. See text for details.}
\label{Chap-FM:tab:process}
\end{table}
\normalsize
As before, for each variable, a brief description, the default value and an example are presented. We now describe them in more detail.

\n \t{inparticles} and \t{outparticles}, corresponding to the incoming and outgoing particles of the process, should contain only particles defined in the \ts{FeynMaster} model. Antiparticles are defined according to the \FRS convention (with the suffix \t{bar}).
Different particles should be separated by commas. Whenever both a particle and an antiparticle are considered, the particle should always be written first, and the antiparticle after it. Tadpoles are obtained by selecting a single incoming particle and no outgoing particles.

\n Concerning the \t{loops} variable, it should be clear that, whatever the number of loops, \ts{FeynMaster} will always correctly generate the amplitudes for every diagram involved, although it is only prepared to properly draw and compute diagrams with number of loops inferior to 2. The one-loop calculations are performed with the \FMS function \t{OneLoopTID} (cf. table \ref{Chap-FM:tab:FCfunc} below).

\n \t{parsel} allows the specification of intermediate particles contributing to the process.\fn{It is similar to (and actually based on) the \t{iprop} option in \QG.}
It applies not only to particles in loops, but to all intermediate particles. The selections must be written between braces and different selections should be separated by commas (see example in table \ref{Chap-FM:tab:process}, where two specific selections are given).
Each specific selection contains four arguments: the first should either be \t{avoid} or \t{keep}, the second should correspond to a particle of the model,\fn{Care should be taken not to select antiparticles, but only particles. This is because the propagator in \ts{FeynMaster} is defined through the particle, and not the antiparticle.} and the last two should be non-negative integer numbers such that the second is not smaller than the first.
We illustrate how it works by considering the example in table \ref{Chap-FM:tab:process}: \t{\{avoid,Z,1,3\}} discards all the diagrams with number of \t{Z} propagators between 1 and 3, while \t{\{keep,e,1,1\}} keeps only diagrams with number of \t{e} propagators between 1 and 1 (i.e., exactly equal to 1).

\n \t{factor} is a number, written in \t{\ts{Mathematica}} style that is to be factored out in the final calculations (more details on section \ref{Chap-FM:sec:FC}).

\n \t{options} allows all the \QGS options, as well as a couple of specific option for \FM---\t{ugauge}, which performs the calculation in the unitary gauge, and \t{noPVauto}, which employs the option \t{PaVeAutoReduce -> False} in \t{TID}.%
\fn{\label{Chap-FM:note:TID}
As shall be seen, loop integrals in \FMS are performed with the function \t{OneLoopTID}, which uses the \FCS function \t{TID}. By default, \t{OneLoopTID} uses \t{TID} with the option \t{PaVeAutoReduce -> True}. This simplifies some special cases of Passarino--Veltman functions. In some cases, however, it may be relevant to perform loop integrals in a more expanded way, in which case the option \t{PaVeAutoReduce -> False} must be used by selecting the option \t{noPVauto} in \t{Control.m}.}
. In table \ref{Chap-FM:tab:Qoptions}, besides these, we present some of the \QGS options; for each one, a brief description and the converse option are presented.

\begin{table}[!h]%
\begin{normalsize}
\normalsize
\begin{center}
\begin{threeparttable}
\begin{tabular}
{@{\hspace{3mm}}
>{\raggedright\arraybackslash}p{2.5cm}
>{\raggedright\arraybackslash}p{7.2cm}
>{\raggedright\arraybackslash}p{2.8cm}
@{\hspace{3mm}}}
\hlinewd{1.1pt}
Option & Description & Converse option \\
\hline\\[-1.5mm]
\t{ugauge} & unitary gauge & \textit{(none)}\\[3mm]
\t{noPVauto} & employs \t{PaVeAutoReduce -> False} in \t{TID} & \textit{(none)}\\[3mm]
\t{onepi} & 1-particle irreducible diagrams only\tnote{$\star$} & \t{onepr}\\[3mm]
\t{onshell} & no self-energy insertions on the external lines\tnote{$\star$} & \t{offshell}\\[3mm]
\t{nosigma} & no self-energy insertions (nowhere)\tnote{$\star$} & \t{sigma}\\[3mm]
\t{nosnail} & no snails (i.e., tadpoles or a collapsed tadpoles)\tnote{$\star$} & \t{snail}\\[3mm]
\t{notadpole} & no tadpole insertions, i.e. no 1-point insertions\tnote{$\star$} & \t{tadpole}\\[3mm]
\t{simple} & at most one propagator connecting any two different vertices, and no propagator starting and ending at the same vertex\tnote{$\star$} & \t{notsimple}\\
\hlinewd{1.1pt}
\end{tabular}
\begin{tablenotes}
{\small
\item[$\star$] Taken from the \QGS manual; please consult it for more information.
}
\end{tablenotes}
\end{threeparttable}
\end{center}
\vspace{-5mm}
\end{normalsize}
\vs{-1mm}
\caption{Options for \t{Control.m}.}
\label{Chap-FM:tab:Qoptions}
\end{table}

\vs{3mm}
\n \underline{\textbf{\FMS logical variables}}

\n Once the model and the process (or sequence of processes) are specified, the user must select the logical value of the 9 variables present in the end of \t{Control.m}. As logical variables, they admit only the values \t{True} (or, alternatively, \t{T} or \t{t}) and \t{False} (or, alternatively, \t{F} or \t{f}).
We describe them in table \ref{Chap-FM:tab:sel}. 
\begin{table}[!h]%
\begin{normalsize}
\normalsize
\begin{center}
\begin{tabular}
{@{\hspace{3mm}}>{\raggedright\arraybackslash}p{2.8cm}>{\raggedright\arraybackslash}p{7.2cm}@{\hspace{3mm}}}
\hlinewd{1.1pt}
Variable & Effect (when chosen as \t{True}) \\
\hline\\[-1.5mm]
\t{FRinterLogic} & establish an interface with \ts{FeynRules} \\[2.5mm] 
\t{RenoLogic} & perform renormalization \\[2.5mm]
\t{Draw} & draw and print the Feynman diagrams \\[2.5mm]
\t{Comp} & compute the final expressions \\[2.5mm]
\ \ \t{FinLogic} & print the final result of each diagram \\[2.5mm]
\ \ \t{DivLogic} & print the UV divergent part of each diagram \\[2.5mm]
\t{SumLogic} & compute the sum of the expressions\\[2.5mm]
\t{MoCoLogic} & apply momentum conservation \\[2.5mm]
\t{LoSpinors} & include spinors \\[2.5mm]
\hlinewd{1.1pt}
\end{tabular}
\end{center}
\vspace{-5mm}
\end{normalsize}
\caption{Logical variables of \t{Control.m}. See text for details.}
\label{Chap-FM:tab:sel}
\end{table}
\normalsize
Some remarks are in order, concerning the effect of these variables when set to \t{True}.

\n \t{FRinterLogic}, by establishing an interface with \ts{FeynRules}, performs several tasks. First, it runs \ts{FeynRules} (for the model selected in the initial variable \t{model}), prints the complete tree-level Feynman rules of the model in a PDF file and opens this file.\fn{\label{Chap-FM:nt:1st}The results are automatically written; this is especially challenging when it comes to (automatically) breaking the lines in a long equation. This challenge is in general surpassed with the \LaTeX\, \t{breqn} package, which \ts{FeynMaster} employs. However, \t{breqn} is not able to break a line whenever the point where the line is to be broken is surrounded by three or more parentheses; in those cases, unfortunately, the lines in the \ts{FeynMaster} PDF outputs simply go out of the screen. For documentation on the \t{breqn} package, cf.\url{https://www.ctan.org/pkg/breqn}.}
Second, it generates the \QGS model file, a crucial element in the generation of Feynman diagrams. Third, it generates the complete tree-level Feynman rules in \ts{FeynCalc} style, which will play a decisive role in all the calculations. Finally, it generates a \t{\ts{Mathematica}} notebook specifically designed to run \ts{FeynRules}---hereafter dubbed the \ts{FeynRules} notebook. This notebook is useful in case the user wants to have control over the generation of Feynman rules, and is the subject of section \ref{Chap-FM:sec:FR}.

\n Note that, even if all logical variables are set to \t{False}, \ts{FeynMaster} always performs some actions. First, it
generates a \t{\ts{Mathematica}} notebook specifically designed to run \ts{FeynCalc}---hereafter dubbed the \ts{FeynCalc} notebook. This notebook is very useful should the user want to have control over calculations, and is the subject of section \ref{Chap-FM:sec:FC}. Second, once there is a \QGS model, \ts{FeynMaster} always runs \QG, which writes in a symbolic form the total diagrams that contribute to the process at stake---the same process which was specified through the variables in table \ref{Chap-FM:tab:process}. Finally, \ts{FeynMaster} takes the \QGS output and writes the amplitude for each diagram in a file that the \ts{FeynCalc} notebook shall have access to.

\n \t{RenoLogic} concerns the renormalization of the model. If \t{FRinterLogic} is set to \t{True}, \t{RenoLogic} prints the complete set of Feynman rules for the counterterms interactions in a PDF file and opens this file; moreover, it stores those interactions in a file which the \FCS notebook shall have access to. A second important feature of \t{RenoLogic} is described below, in the context of the \t{Comp} variable.

\n \t{Draw} takes the \QGS output, draws the Feynman diagrams in a \LaTeX \,file, prints them in a PDF file and opens this file. This operation is achieved with the help of \t{feynmf} \cite{Ohl:1995kr}, a \LaTeX \,package to draw Feynman diagrams. Since the diagrams are written in a \LaTeX \,file, they can not only be edited by the user, but also directly copied to the \LaTeX \,file of the user's paper.\fn{As already suggested, \t{Draw} is at present only guaranteed to properly draw the diagrams up to one-loop. Moreover, diagrams with more than two particles in the initial or final states, as well as some reducible diagrams, are also not warranted.}

\n \t{Comp} computes the final expressions using \ts{FeynCalc} and stores them in a file.%
\fn{By `final expressions' we mean the analytical expressions for the diagrams; these are written in terms of Passarino--Veltman integrals in case the number of loops equals 1; more details on section \ref{Chap-FM:sec:FC}. It is normal that the warning `\textit{front end is not available}' shows up when \t{Comp} is set to \t{True}.}

\n \t{FinLogic} and \t{DivLogic} are nothing but options for \t{Comp}, so that their value is only relevant if \t{Comp} is set to \t{True}.
If at least one of them is set to \t{True}, a PDF file is generated and opened: when \t{FinLogic} is selected, the file includes the (total) final analytical expression for each diagram; when \t{DivLogic} is selected, it includes the analytical expression for the UV divergent part of each diagram.\fn{The limitation we alluded to in note \ref{Chap-FM:nt:1st} applies here too.}

\n At this point, we should clarify the difference between UV divergences and infrared (IR) divergences. It is well known that, while the former are in general present in loop integrals, the latter can only show up when there is a massless particle running inside the loop (in which case the IR divergence comes from the integration region near $k^2=0$, with $k$ the loop momentum). In the present version of \ts{FeynMaster}, we restrict the treatment of divergences to the UV ones. Indeed, we assume that the IR divergences can be regulated by giving the massless particle a fake mass---which one shall eventually be able to set to zero in physical processes, after considering real emission graphs. With that assumption, IR divergences will never show up explicitly (only implicitly through the fake mass). In the following, unless in potentially dubious statements, we will stop writing UV explicitly: it is assumed that, whenever we mention divergences, we shall be referring to UV divergences.

\n \t{SumLogic} is relevant both when \t{Comp} is \t{True} and when it is \t{False}. In the first case, it calculates the sum of the analytical expressions (when the PDF file with the analytical expressions is generated, \t{SumLogic} prints their sum).%
\fn{More specifically: if \t{FinLogic} is \t{True}, \t{SumLogic} includes in the PDF file the sum of the total final expressions; if \t{DivLogic} is \t{True}, it includes the sum of the expressions for the divergent parts; if both are \t{True}, it includes both the sum of the total expressions and the sum of the expressions for the divergent parts.}
The situation where \t{SumLogic} is \t{True} and \t{Comp} is \t{False} shall be described in section \ref{Chap-FM:sec:FC}.

\n We now explain the effect of \t{RenoLogic} when \t{Comp} is set to \t{True}.
In case the user defined a single process in \t{Control.m}, \t{RenoLogic} causes \ts{FeynMaster} to look for counterterms that might absorb the divergences of the process at stake, and to calculate those counterterms in $\overline{\text{MS}}$.%
\fn{That is, calculates them in such a way that the counterterms are precisely equal to the divergent part they absorb (except for the $\ln(4\pi)$ and the Euler-Mascheroni  constant $\gamma$, which are also absorbed in the $\overline{\text{MS}}$ scheme). By `calculating' we mean here writing the analytical expression.}
Such counterterms are then stored in a file (\t{CTfin.m}, described below) and,
should the PDF file with the analytical expressions be generated, printed in such file.
In case the user defined a sequence of processes, the subsequently computed counterterms are added to \t{CTfin.m}; however, what is particularly special about the sequence is that, for a certain process of the sequence, \ts{FeynMaster} will compute the counterterms by making use of the counterterms already computed in the previous processes.%
\fn{This is, in fact, the major advantage of writing a series of processes in a single \ts{FeynMaster} run (as opposed to one process per run); indeed, this feature is not possible with one process per run, since \FMS rewrites \t{CTfin.m} for the model at stake everytime \t{Comp} and \t{RenoLogic} are both set to \t{True}.}
In the end of the run, \t{CTfin.m} contains all the counterterms that were computed (in $\overline{\text{MS}}$) to absorb the divergences of the processes of the sequence. In this way, and by choosing an appropriate sequence of processes, it is possible to automatically renormalize the whole model in $\overline{\text{MS}}$ with a single \ts{FeynMaster} run.

\n As \t{SumLogic} described above, the last two logical variables of table \ref{Chap-FM:tab:sel}---\t{MoCoLogic} and \t{LoSpinors}---are relevant both when \t{Comp} is \t{True} and when it is \t{False}. \t{LoSpinors}---only relevant with external fermions---is actually prior to any calculation, since it modifies the amplitudes themselves, including spinors in them. When \t{Comp} is \t{True}, \t{MoCoLogic} is such that the calculations use momentum conservation. The situation where \t{MoCoLogic} is \t{True} and \t{Comp} is \t{False} is also postponed to section \ref{Chap-FM:sec:FC}.

\subsection{Run}
\label{Chap-FM:sec:Run}

\n After editing the \t{Control.m} file, everything is set. To run \ts{FeynMaster}, just run the \t{RUN-\ts{FeynMaster}} batch file inside the \t{FeynMaster} folder.\fn{Care should be taken not to run \ts{FeynMaster} when the relevant notebooks are open. More specifically, if \t{FRinterLogic} is set to \t{True}, and if the \ts{FeynRules} notebook created for the process at stake already exists, this notebook cannot be open during the run; in the same way, if \t{Comp} is set to \t{True}, and if the \ts{FeynCalc} notebook designed for the process at stake already exists, such notebook cannot be open during the run.}

\subsection{Outputs}
\label{Chap-FM:sec:Output}

\n Depending on the logical value of the variables of table \ref{Chap-FM:tab:sel}, \ts{FeynMaster} can have different outputs. We now list the total set of outputs, assuming that all those variables are set to \t{True}.%
\fn{Actually, when \t{Comp} and \t{RenoLogic} are both \t{True} and there are external fermions, \t{LoSpinors} should be \t{False}. This is irrelevant for what follows, since \t{LoSpinors} has no influence on the outputs as a whole.}
First, in the directory where the \FMS model is, two files are generated: the \ts{FeynRules} notebook (\t{Notebook.nb}) and an auxiliary file for it (\t{PreControl.m}). Second, the \QGS model with the internal name of the model is created, and placed inside the folder with the models for \QGS (cf. note \ref{Chap-FM:note:instal}); besides, a file named \t{last-output} (with the last output from \QG) is created inside the directory corresponding to \QG. Then, if it does not exist yet, a directory with the internal name of the model is created inside the directory for the \FMS output. Inside it, and if they do not exist yet, three directories are created, \t{Counterterms}, \t{FeynmanRules} and \t{Processes}, which we now describe.

\n \t{Counterterms} contains one folder, \t{TeXs-drawing}, and two files, \t{CTini.m} and \t{CTfin.m}. \t{TeXs-drawing} is where the PDF file with the complete set of Feynman rules for the counterterms interactions is stored, as well as the \LaTeX \, file that creates it. \t{CTini.m} is the file which the \ts{FeynCalc} notebook has access to and where the Feynman rules for the counterterms interactions are stored. \t{CTfin.m}, in turn, is the aforementioned file containing the counterterms that were computed (in $\overline{\text{MS}}$) to absorb the divergences of the processes of the sequence at stake.

\n \t{FeynmanRules}, besides several auxiliary files to be used in the \ts{FeynCalc} notebook, contains yet another \t{TeXs-drawing} folder, where the PDF file with the complete  set of Feynman rules for the
tree-level interactions is stored, as well as the \LaTeX \, file that creates it.

\n \t{Processes} contains a folder for each of the different processes studied. These folders are named with the index (in the sequence of processes) corresponding to the process at stake, as well as with a string containing the names of the incoming and the outgoing particles joined together.
Inside each folder, there are three other folders, \t{Lists}, \t{TeXs-drawing} and \t{TeXs-expressions}, as well as three files, \t{Amplitudes.m}, \t{Helper.m} and the \ts{FeynCalc} notebook, \t{Notebook.nb}.
In order:
\t{Lists} contains files where the analytical expressions for the process at staked are stored (more details below);
\t{TeXs-drawing} contains the PDF file with the printed Feynman diagrams, as well as the \LaTeX \,file that creates it;
\t{TeXs-expressions} contains the PDF file with the printed expressions, as well as the \LaTeX \,file that creates it;
\t{Amplitudes.m} contains the amplitudes for the diagrams (written in \ts{FeynCalc} style);
\t{Helper.m} is an auxiliary file for the \ts{FeynCalc} notebook.

\n Finally, recall that, in case there is already a \QGS model, \ts{FeynMaster} will run even if all variables of table \ref{Chap-FM:tab:sel} are set to \t{False}. This is relevant since it generates not only the \QGS output (\t{last-output}), but also the folder (or folders) for the specific process (or processes) selected, containing the files described above.\fn{While the \QGS output is not overwritten when \QGS is run on its own, it is overwritten when \QGS is run inside \ts{FeynMaster}.}

\subsection{The notebooks}
\label{Chap-FM:sec:NB}

\n As previously mentioned, a major advantage of \ts{FeynMaster} is its hybrid character concerning automatization. Indeed, not only does it automatically generate the results, but it also allows the user to handle them. This is realized due to the automatic creation of the \ts{FeynRules} notebook and the \ts{FeynCalc} notebook. We now describe them in detail.

\subsubsection{The \ts{FeynRules} notebook}
\label{Chap-FM:sec:FR}

\n We mentioned in section \ref{Chap-FM:sec:Control} that, when \ts{FeynMaster} is run with the logical variable \t{FRinterLogic} set to \t{True}, the \ts{FeynRules} notebook \t{Notebook.nb} is automatically created in the directory where the \FMS model is. By running it, the user can access the vertices for the different Lagrangian parts, according to table \ref{Chap-FM:tab:FRverts}.\fn{The run will generate several \ts{FeynMaster} internal files, among which is \t{built-model}, the \QGS model file.}
\begin{table}[!h]%
\begin{normalsize}
\normalsize
\begin{center}
\begin{tabular}
{@{\hspace{3mm}}>{\raggedright\arraybackslash}p{3.2cm}>{\raggedright\arraybackslash}p{3.8cm}@{\hspace{3mm}}}
\hlinewd{1.1pt}
Lagrangian part & vertices \\
\hline
\t{LGauge} & \t{vertsGauge} \\
\t{LFermions} & \t{vertsFermionsFlavor} \\
\t{LYukawa} & \t{vertsYukawa} \\
\t{LHiggs} & \t{vertsHiggs} \\
\t{LGhost} & \t{vertsGhosts} \\
\hlinewd{1.1pt}
\end{tabular}
\end{center}
\vspace{-5mm}
\end{normalsize}
\caption{Names of the different vertices according to the Lagrangian part (compare with table \ref{Chap-FM:tab:FRnames}).}
\label{Chap-FM:tab:FRverts}
\end{table}
\normalsize
Besides the usual \ts{FeynRules} instructions, two useful functions---\t{GetCT} and \t{MyTeXForm}---are available.
\t{GetCT} is a function that, for a certain Lagrangian piece given as argument, yields the Feynman rules for the respective counterterms.%
\fn{In order to clarify the particles involved in the Feynman rule at stake, each term includes a factor $\eta_i$ for each particle $i$.}
\t{MyTeXForm} is \ts{FeynMaster}'s version of \t{\ts{Mathematica}}'s \t{TeXForm}; it is a function that uses \t{\ts{Python}} (as well as inner \ts{FeynMaster} information concerning the \LaTeX \,form of the parameters of the model) to write expressions in a proper \LaTeX \,form.%
\fn{\t{MyTeXForm} prints the \LaTeX \,form of the expression at stake not only on the screen, but also in an external file named \t{MyTeXForm-last-output.tex} in the directory where the notebook lies.}

\subsubsection{The \ts{FeynCalc} notebook}
\label{Chap-FM:sec:FC}

\n Whenever \ts{FeynMaster} is run, and independently of the logical values of the variables of table \ref{Chap-FM:tab:sel}, the \ts{FeynCalc} notebook is automatically created. This notebook, as already the \ts{FeynRules} one just described, is totally ready-to-use: the user does not have to define directories, nor import files, nor change conventions.
Just by running the notebook, there is immediate access to a whole set of results: not only to some basic elements---such as the total Feynman rules for the model and amplitudes for the diagrams---, but also to 
the totality of the results obtained should the \t{Comp} logical variable had been turned on. This last feature is made possible in \FMS 2 due to the creation of lists with analytical expressions. Indeed, when \t{Comp} is set to true, \FMS now stores the analytical expressions inside the aforementioned folder \t{Lists}, in such a way that, after the run, the \ts{FeynCalc} notebook has immediate access to those expressions (more details below). 

\n But this is just part of the flexibility involved in the \ts{FeynCalc} notebook. In fact, since all the referred results are written in a \t{\ts{Mathematica}} notebook, the user has great control over them, as he or she can operate algebraically on them, or select part of them, or print them into files, etc.
Moreover, since the \ts{FeynCalc} package is loaded, and since all the results are written in a \ts{FeynCalc}-readable style, the control at stake is even greater, for the user can apply all the useful tools of that package: operate on the Dirac algebra, perform contractions, solve loop integrals, etc.\fn{In the following, we assume the user to be familiar with \ts{FeynCalc}. For more informations, consult the \ts{FeynCalc} website: \url{https://feyncalc.github.io/}.} 

\n We now present some useful features introduced by \ts{FeynMaster} in the \ts{FeynCalc} notebook. We start with the variables related to the analytical expressions for the Feynman diagrams: see table \ref{Chap-FM:tab:FCvars}.
\begin{table}[!h]%
\begin{normalsize}
\normalsize
\begin{center}
\begin{tabular}
{@{\hspace{3mm}}>{\raggedright\arraybackslash}p{1.8cm}>{\raggedright\arraybackslash}p{8.0cm}@{\hspace{3mm}}}
\hlinewd{1.1pt}
Variable & Meaning \\
\hline\\[-1.5mm]
\t{amp} & list with all the amplitudes \\[2.5mm]
\t{amp}\textit{i} & amplitude for diagram \textit{i} \\[2.5mm]
\t{res} & list with all the final expressions \\[2.5mm]
\t{res}\textit{i} & final expression for diagram \textit{i} \\[2.5mm]
\t{resD} & list with all the expressions for the divergent parts \\[2.5mm]
\t{resD}\textit{i} & expression for the divergent part of diagram \textit{i} \\[2.5mm]
\t{restot} & sum of all the final expressions \\[2.5mm]
\t{resDtot} & sum of all the expressions for the divergent parts \\[2.5mm]
\hlinewd{1.1pt}
\end{tabular}
\end{center}
\vspace{-5mm}
\end{normalsize}
\caption{Useful variables concerning expressions for the diagrams. See text for details.}
\label{Chap-FM:tab:FCvars}
\end{table}
\normalsize
We should clarify the meaning of final expression, corresponding to the \t{res} list: for a certain index $j$, \t{res}\textit{j} (or \t{res[[j]]}) takes the amplitude \t{amp}\textit{j}, divides it by $i$, rewrites the loop integral in terms of Passarino--Veltman integrals (in case it is a one-loop process), writes it in 4 dimensions---including possible finite parts coming from this conversion%
\fn{As is well known, in the {dimensional regularization scheme}, the infinities are tamed by changing the dimensions of the integrals from 4 to $d$, in such a way that the divergences are regulated by the parameter $\varepsilon = 4 - d$. When solving the integrals in terms of Passarino--Veltman integrals, the result will in general depend explicitly on the dimension $d$, as well as on the Passarino--Veltman integrals themselves---which usually diverge, with divergence proportional to $1/\varepsilon$. But since $d = 4-\varepsilon$, there will in general be finite terms (order $\varepsilon^0$) coming from the product between $d$ and the divergent parts in the Passarino--Veltman integrals. Hence, when converting the result back to 4 dimensions (since the final result is written in 4 dimensions), one cannot forget to include such terms. {Finally, recall that IR divergences will never show up explicitly if the potentially IR divergent integrals are tamed by giving the massless particle a fake mass.}\label{Chap-FM:note:DimReg}}%
---and factorizes the previously selected \t{factor}.
Note that, due to the division by $i$, the final expressions---that is, \t{res}---correspond to $\mathcal{M}$, and not to $i \mathcal{M}$.
The divergent parts are written in terms of the variable \t{div}, defined as:%
\fn{In the PDF file with the printed expressions, we change the name \t{div} to $\omega_{\varepsilon}$.}
\begin{equation*}
\t{div} = \dfrac{1}{2} \left( \dfrac{2}{\varepsilon} - \gamma + \ln 4\pi \right).
\end{equation*}

\n Let us now consider the details related to the \t{Lists} folder. When \t{Comp} is set to \t{True}, the lists \t{res} and \t{resD} are calculated and stored in \t{Lists} in the files \t{res.in} and \t{resD.in}, respectively. If, in addition, \t{SumLogic} is set to \t{True}, \t{restot} and \t{resDtot} are also calculated and stored in \t{Lists} in the files \t{restot.in} and \t{resDtot.in}, respectively.
The access of the \ts{FeynCalc} notebook to these files depends on the variable \t{compNwrite}, which is visibly defined in the middle of the notebook: when it is \t{False}, the notebook will load the files stored in \t{Lists} (should they exist); when it is \t{True}, it will calculate them anew.
By default, \t{compNwrite} is \t{False}, which means that, after having been calculated with \t{Comp}, the analytical expressions are by default immediatly accessible to the \ts{FeynCalc} notebook.

\n Next, we consider useful functions to manipulate the results: see table \ref{Chap-FM:tab:FCfunc}.
\begin{table}[!h]%
\begin{normalsize}
\normalsize
\begin{center}
\begin{threeparttable}
\begin{tabular}
{@{\hspace{3mm}}
>{\raggedright\arraybackslash}p{2.3cm}
>{\raggedright\arraybackslash}p{11.0cm}@{\hspace{3mm}}}
\hlinewd{1.1pt}
Function & Action \\
\hline\\[-2.5mm]
\t{ChangeTo4} & change to 4 dimensions\\[2.5mm]
\t{DecayWidth} & calculate the decay width\tnote{$\star$}\\[2.5mm]
\t{DiffXS} & calculate the differential cross section\tnote{$\dagger$}\\[2.5mm]
\t{GetDirac} & yield the Dirac structures as a list\\[2.5mm]
\t{GetDiv} & get the divergent part of an expression \\[2.5mm]
\t{GetFinite} & get the finite part of an expression in dimensional regularization\tnote{$\ddagger$}\\[2.5mm]
\t{FacToDecay} & rewrite expressions (with form factors) for decays\tnote{$\star$} \tnote{$\diamond$}\\[2.5mm]
\t{FCtoFT} & convert expressions to \t{\ts{Fortran}} \\[2.5mm]
\t{MyTeXForm} & write expressions in a proper \LaTeX \,form  \\[2.5mm]
\t{MyPaVeReduce} & apply \ts{FeynCalc}'s \t{PaVeReduce} and convert to 4 dimensions\\[2.5mm]
\t{OneLoopTID} & solve one-loop integrals with \FCS TID decomposition method\tnote{$\#$}\\[2.5mm]
\t{RepDirac} & replace the Dirac structures with elements \t{ME[j]}\\[2.5mm]
\t{TakeReal} & applies the operator Re to the Passarino--Veltman functions\\[2.5mm]
\t{TrG5} & calculate the trace for expressions with $\gamma_5$\\[0.5mm]
\hlinewd{1.1pt}
\end{tabular}
\begin{tablenotes}
{\small
\item[$\star$] Only applicable to processes with 1 incoming and 2 outgoing particles.
\item[$\dagger$] Only applicable to processes with 2 incoming and 2 outgoing particles.
\item[$\ddagger$] Cf. note \ref{Chap-FM:note:DimReg}.
\item[$\diamond$] Not yet available for three external gauge bosons.
\item[$\#$] Note the arguments: \t{OneLoopTID(k,amp)}, with \t{k} the loop momentum and \t{amp} the amplitude. Cf. also note \ref{Chap-FM:note:TID}.
}
\end{tablenotes}
\end{threeparttable}
\end{center}
\vspace{-5mm}
\end{normalsize}
\vs{-1mm}
\caption{Useful functions. See text for details.}
\label{Chap-FM:tab:FCfunc}
\end{table}
\normalsize
We now explain some of them in more detail (note that a description of a certain function \t{Func} shows up in the notebook by writing \t{?Func}).

\n Concerning \t{DecayWidth} and \t{DiffXS}, while the former is written solely in terms of masses, the latter is written also in terms of the center of momentum energy \t{S} as well of the scattering angle \t{Theta}.

\n \t{GetDirac} and \t{RepDirac} are useful functions to handle expressions with external fermions. The former yields a list containing the different Dirac structures of the expression given as argument; the latter replaces the Dirac structures of the expression given as argument with non-matricial elements \t{ME[j]}---that is, the first Dirac structure is replaced by \t{ME[1]}, the second by \t{ME[2]}, and so on---, thus simplying the manipulation of the expression. \t{RepDirac} admits a second (optional) argument, consisting of a list of Dirac structures (ex: \t{\{GA[p1],GS[q1]\})}; this can be relevant to establish a correspondance between \t{ME[j]} elements and Dirac structures. Indeed, such correspondance can obey one of two rules:
either only one argument is given to \t{RepDirac}, in which case the \t{ME[j]} elements follow the order of the Dirac structures yielded by \t{GetDirac}, or a list of Dirac structures is given as a second argument of \t{RepDirac}, in which case the \t{ME[j]} elements follow the order of that list.

\n \t{GetDiv} yields the UV divergent part of an expression; care should be taken if IR divergences are not regulated via a fake mass, for in that case, although they can show up as poles of the Passarino--Veltman functions, they will not be detected by \t{GetDiv}. Moreover, \t{GetDiv} only yields the UV divergent part of the Passarino--Veltman functions \cite{Denner:2005nn} that \ts{FeynCalc} and \ts{FeynMaster} can handle---namely, integrals whose power of the loop momenta in the numerator is at most 3 (except for the $D$ functions, where we extended \ts{FeynCalc} up to the fourth power of the loop momenta).

\n \t{FacToDecay} simplifies expressions to render the calculation of the decay width easier. It yields a two-element list: the first element is the expression given as argument, but rewritten in terms of form factors; the second is a list of replacements, associating to each form factor the corresponding analytical expression (written using the kinematics of the process).%
\fn{The form factors are defined as complex parameters in the options of the \ts{FeynCalc} function \t{ComplexConjugate}. This will be relevant to calculate the DecayWidth.}

\n \t{FCtoFT} is the function that allows the numerical interface of \ts{FeynMaster};  when applied to an expression, it generates at least five files: \t{MainFT.F}, \t{MyFunctionFT.F}, \t{MyVariables.h}, \t{MyParameters.h} and \t{Makefile-template}. The first one, \t{MainFT.F}, is the beginning of a main \t{\ts{Fortran}} program, which must be completed according to the user's will.
\t{MainFT.F} calls the function \t{MyFunction}, which is the \t{\ts{Fortran}} version of the expression \t{FCtoFT} was applied to, and which is written in the file \t{MyFunctionFT.F}; in turn, \t{MyVariables.h} and \t{MyParameters.h} respectively contain the variables and the numerical values associated to the different parameters in the \ts{FeynMaster} model file. Finally, \t{Makefile-template} consists of a skeleton of a makefile. In the case of cross sections, the integration routine \t{IntGauss.f} is also generated and called in \t{MainFT.F}, which in this case is completed with a concrete example.%
\fn{\label{Chap-FM:note:FCtoFT}
By default, the Passarino--Veltman functions will be converted as complex; this can be modified by defining a new variable in the \ts{FeynCalc} notebook \t{retil}, and define it as \t{True} (i.e. \t{retil=True}).
\t{FCtoFT} admits a second (optional) argument, as we now explain. Suppose the expression to be converted to \t{\ts{Fortran}} (let us call it \t{expr}) is a complicated expression, but such that it can be written in a simple way using three form factors, \t{F1}, \t{F2} and \t{F3}; that is, \t{expr = f(F1,F2,F3)}, where \t{f} is a simple function, while \t{F1}, \t{F2} and \t{F3}
correspond to complicated expressions. In cases like this, it is convenient to write \t{FCtoFT} with two arguments: the first one is the expression \t{expr}, but written in terms of auxiliary variables \t{F1aux}, \t{F2aux} and \t{F3aux} instead of the complicated expressions \t{F1}, \t{F2} and \t{F3} (that is,
\t{expr = f(F1aux,F2aux,F3aux)}); the second argument is a list of the replacements between the auxiliary variables and their corresponding form
factor (in this example, \t{\{F1aux -> F1, F2aux -> F2, F3aux -> F3\}}).}

\n \t{MyTeXForm} is the same function as the one described in section \ref{Chap-FM:sec:FR}. 

\n \t{MyPaVeReduce} is \ts{FeynMaster}'s version of \ts{FeynCalc}'s \t{PaVeReduce}; it applies \t{PaVeReduce} and writes the result in 4 dimensions---again, not without including possible finite parts coming from this conversion.

\n Finally, we consider \t{TrG5}. Since \ts{FeynMaster} is prepared to compute divergent integrals---and, more specifically, to compute them via dimensional regularization---, it defines Dirac and Lorentz structures (like $g^{\mu\nu}$ or $\gamma^{\mu}$) in dimension $d$, not in dimension 4. However, the definition of $\gamma_5$ in dimension $d$ is not trivial, as chiral fermions are a property of four dimensions. In fact, the treatment of $\gamma_5$ in dimensional regularization is still an open problem (see e.g. refs. \cite{Akyeampong:1973xi,Chanowitz:1979zu,Barroso:1990ti,Kreimer:1993bh,Jegerlehner:2000dz,Greiner:2002ui,Zerf:2019ynn}). By default, \ts{FeynMaster} assumes the so-called naive dimensional regularization scheme~\cite{Jegerlehner:2000dz}, which takes the relation $\{\gamma_5,\gamma^{\mu}\}=0$ to be valid in dimension $d$. This naive approach is applied both when the loop diagram has external fermions, and when it has only inner fermions (forming a closed loop). In the second case, in order to calculate the corresponding trace, \ts{FeynMaster} uses \t{TrG5}. This function starts by separating the expression it applies to in two terms: one with $\gamma_5$, another without $\gamma_5$. It then computes the trace of the former in $4$ dimensions, while keeping the dimension of the latter in its default value $d$.\fn{It is a matter of course that the calculation of the term with $\gamma_5$ in dimension 4 can only be an issue when the integral multiplying it is divergent. This is simply because one does not need to regularize an integral that is not divergent. In particular, there is no need to use dimensional regularization for a finite integral, so that all the calculations can be made in dimension 4. Note also that \ts{FeynCalc} already includes different schemes to handle $\gamma_5$, and is expected to improve the treatment of $\gamma_5$ in dimensional regularization in future versions.}

\n Having clarified the functions in table \ref{Chap-FM:tab:FCfunc}, we must consider a new table, table \ref{Chap-FM:tab:FCreno}, which contains some useful variables concerning renormalization. 
\begin{table}[!h]%
\begin{normalsize}
\normalsize
\begin{center}
\begin{tabular}
{@{\hspace{3mm}}>{\raggedright\arraybackslash}p{2.5cm}>{\raggedright\arraybackslash}p{12.0cm}@{\hspace{3mm}}}
\hlinewd{1.1pt}
Function & Meaning \\
\hline\\[-1.5mm]
\t{CT}\textit{process} & expression containing the total counterterm for the \textit{process} at stake \\[2.5mm]
\t{PreResReno} & sum of the total divergent part and \t{CT}\textit{process} \\[2.5mm]
\t{CTfinlist} & list with all the counterterms computed so far in $\overline{\text{MS}}$\\[2.5mm]
\t{ResReno} & the same as \t{PreResReno}, but using counterterms previously stored in CTfinlist\\[6.5mm]
\t{PosResReno} & the same as \t{ResReno}, but using also the counterterms calculated for the process at stake\\[2.5mm]
\hlinewd{1.1pt}
\end{tabular}
\end{center}
\vspace{-5mm}
\end{normalsize}
\caption{Useful variables concerning renormalization. See text for details.}
\label{Chap-FM:tab:FCreno}
\end{table}
\normalsize
Two notes should be added. First, in the \t{CT}\textit{process} variable, \textit{process} corresponds to the names of the incoming and the outgoing particles joined together (for example, in the SM, for the process $h \to Z\gamma$, \t{CT}\textit{process} is \t{CTHZA}). Second, \t{PosResReno} should always be zero, since in the $\overline{\text{MS}}$ scheme the divergents parts are exactly absorbed by the counterterms.

\n Some final comments on the \ts{FeynCalc} notebook.
First, the indices of the particles are described in table \ref{Chap-FM:tab:FCconv},
\begin{table}[!h]%
\begin{normalsize}
\normalsize
\begin{center}
\begin{tabular}
{@{\hspace{3mm}}
>{}m{1.7cm}
>{\raggedright\arraybackslash}m{1.5cm}
>{\raggedright\arraybackslash}m{1.3cm}
>{\raggedright\arraybackslash}m{1.7cm}
>{\raggedright\arraybackslash}m{1.3cm}
>{\raggedright\arraybackslash}m{1.3cm}
>{\raggedright\arraybackslash}m{1.7cm}
>{\raggedright\arraybackslash}m{1.6cm}
@{\hspace{3mm}}}
\hlinewd{1.1pt}
& \small order in \t{Control.m} & \small \QGS index & \small Lorentz index in \ts{FeynCalc} & \small Lorentz index in \LaTeX \,& \small Color index in \LaTeX \,& \small Momentum index in \ts{FeynCalc} & \small Momentum index in \LaTeX \,\\
\hline\\[-2.5mm]
\multirow{2}{2.5cm}{\\[-3.5mm]Incoming particles} & 
\multirow{2}{2.5cm}{1\\[1.5mm]2} &
\multirow{2}{2.5cm}{-1\\[1.5mm]-3} & \multirow{2}{2.5cm}{\t{-J1}\\[1.5mm]\t{-J3}} & \multirow{2}{2.5cm}{$\mu$\\[1.5mm]$\rho$} & 
\multirow{2}{2.5cm}{$a$\\[1.5mm]$c$} & 
\multirow{2}{2.5cm}{\t{p1}\\[1.5mm]\t{p2}} & \multirow{2}{2.5cm}{$p_1$\\[1.5mm]$p_2$}\\[8mm]
\multirow{2}{2.5cm}{\\[-3.5mm]Outgoing particles}  &
\multirow{2}{2.5cm}{1\\[1.5mm]2} &
\multirow{2}{2.5cm}{-2\\[1.5mm]-4} & \multirow{2}{2.5cm}{\t{-J2}\\[1.5mm]\t{-J4}} & \multirow{2}{2.5cm}{$\nu$\\[1.5mm]$\sigma$} &
\multirow{2}{2.5cm}{$b$\\[1.5mm]$d$} &
\multirow{2}{2.5cm}{\t{q1}\\[1.5mm]\t{q2}} & \multirow{2}{2.5cm}{$q_1$\\[1.5mm]$q_2$}\\[8mm]
\hlinewd{1.1pt}
\end{tabular}
\end{center}
\vspace{-5mm}
\end{normalsize}
\caption{Particle indices for the \ts{FeynCalc} notebook. }
\label{Chap-FM:tab:FCconv}
\end{table}
\normalsize
and momentum conservation can be applied through the replacement rule \t{MomCons}.\fn{Whenever there is one incoming particle and two outgoing particles, \t{MomCons} replaces \t{p1} by the remaining momenta; in all the other cases, MomCons replaces \t{q1} by the remaining momenta.}
Second, the \t{Helper.m} file contains, among other definitions, both the \t{factor} (in case it was defined in \t{Control.m}) as well as the \ts{FeynCalc} identities (in case they were defined as \t{M\$FCeqs} in the \ts{FeynMaster} model).
Third, even if \t{Comp} is set to \t{False} in \t{Control.m}, setting \t{SumLogic} and \t{MoCoLogic} to \t{True} has consequences for the \ts{FeynCalc} notebook:
when the notebook is run with \t{compNwrite} set to \t{True}, the total expressions will be calculated and momentum conservation will be applied, respectively.
Finally, the replacement rule \t{FCsimp} contains the simplifications for \ts{FeynCalc} (in case they were defined as \t{M\$FCsimp} in the \ts{FeynMaster} model) and is applied in the calculation of \t{res}, \t{resD}, \t{restot} and \t{resDtot}.

\section{Examples}
\label{Chap-FM:sec:Examples}

\subsection{Creation and complete automatic renormalization of a toy model}
\label{Chap-FM:sec:FullReno}

\n Here we exemplify how to create a model, on the one hand, and how to completely renormalize it, on the other. The model will be very simple: QED with an extra fermion. We first show how to create such a toy model.

\n Probably the simplest way to create any model whatsoever is to copy and modify an already existing model. Given the similarity between our toy model and QED, we go to the directory with models for \FMS and
duplicate the directory \t{QED}, after which we name the duplicate \t{QED2}. We get inside \t{QED2} and change the name of the model file from \t{QED.fr} to \t{QED2.fr}. We then open \t{QED2.fr} and edit the model in three steps:
first, we modify the parameter list to:
\vs{2mm}
\begin{small}
\begin{addmargin}[8mm]{0mm}
\begin{verbatim}
(***** Parameter list *****)
M$Parameters = {
  m1 == {TeXName -> "m_1", Renormalization -> {m1 -> m1 + dm1}},
  m2 == {TeXName -> "m_2", Renormalization -> {m2 -> m2 + dm2}},
  ee == {TeXName -> "e", Renormalization -> {ee -> ee + de ee}},
  xiA == {TeXName -> "\\xi_A"},
  de == {Counterterm -> True, TeXName -> "\\delta e"},
  dZ3 == {Counterterm -> True, TeXName -> "\\delta Z_3"},
  dZ1L == {Counterterm -> True, TeXName -> "\\delta Z_1^L"},
  dZ1R == {Counterterm -> True, TeXName -> "\\delta Z_1^R"},
  dZ2L == {Counterterm -> True, TeXName -> "\\delta Z_2^L"},
  dZ2R == {Counterterm -> True, TeXName -> "\\delta Z_2^R"},  
  dm1 == {Counterterm -> True, TeXName -> "\\delta m_1"},
  dm2 == {Counterterm -> True, TeXName -> "\\delta m_2"}};
\end{verbatim}
\end{addmargin}
\end{small}
\vs{3mm}
Then, in the particle classes list, we slightly modify what we had, and we add a second fermion:\fn{The fermions are defined both in terms of Weyl spinors (the \t{W} variables) and Dirac spinors (the \t{F} variables). It is certainly true that, in models with no parity violation (like the present one), there is no need to define the fermions in terms of Weyl spinors. Nevertheless, we consider them for illustrative purposes.}
\vs{5mm}
\begin{small}
\begin{addmargin}[8mm]{0mm}
\begin{verbatim}
(***** Particle classes list *****)
M$ClassesDescription = {
  W[1] == {
		ClassName -> psi1L,
		SelfConjugate -> False,
		QuantumNumbers -> {Q-> Q},
		Renormalization -> {psi1L -> psi1L + 1/2 dZ1L psi1L},
		Chirality -> Left},
  W[2] == {
		ClassName -> chi1R,
		SelfConjugate -> False,
		QuantumNumbers -> {Q-> Q},
		Renormalization -> {chi1R -> chi1R + 1/2 dZ1R chi1R},
		Chirality -> Right},
  W[3] == {
		ClassName -> psi2L,
		SelfConjugate -> False,
		QuantumNumbers -> {Q-> Q},
		Renormalization -> {psi2L -> psi2L + 1/2 dZ2L psi2L},
		Chirality -> Left},
  W[4] == {
		ClassName -> chi2R,
		SelfConjugate -> False,
		QuantumNumbers -> {Q-> Q},
		Renormalization -> {chi2R -> chi2R + 1/2 dZ2R chi2R},
		Chirality -> Right},		
  F[1] == {
		ClassName -> f1,
		TeXName -> "f_1",
		TeXAntiName -> "\\bar{f_1}",
		SelfConjugate -> False,
		QuantumNumbers -> {Q-> Q},
		Mass -> m1,
		WeylComponents -> {psi1L, chi1R}},
  F[2] == {
		ClassName -> f2,
		TeXName -> "f_2",
		TeXAntiName -> "\\bar{f_2}",
		SelfConjugate -> False,
		QuantumNumbers -> {Q-> Q},
		Mass -> m2,
		WeylComponents -> {psi2L, chi2R}},			
  V[1] == {
		ClassName -> A,
		TeXName -> "\\gamma",
		Renormalization -> {A[mu_] -> A[mu] + 1/2 dZ3 A[mu]},
		Mass -> 0,
		SelfConjugate -> True}};
\end{verbatim}
\end{addmargin}
\end{small}
\vs{3mm}
We modify the Lagrangian to include a second fermion:
\vs{3mm}
\begin{small}
\begin{addmargin}[8mm]{0mm}
\begin{verbatim}
LGauge := -1/4 FS[A, \[Mu], \[Nu]] FS[A, \[Mu], \[Nu]]
LFermions := I psi1Lbar.sibar[mu].del[psi1L, mu] + I chi1Rbar.si[mu].del[chi1R, mu] \
		   - m1 (psi1Lbar.chi1R + chi1Rbar.psi1L) \
		   + ee psi1Lbar.sibar[mu].psi1L A[mu] + ee chi1Rbar.si[mu].chi1R A[mu] \
		   + I psi2Lbar.sibar[mu].del[psi2L, mu] + I chi2Rbar.si[mu].del[chi2R, mu] \
		   - m2 (psi2Lbar.chi2R + chi2Rbar.psi2L) \
		   + ee psi2Lbar.sibar[mu].psi2L A[mu] + ee chi2Rbar.si[mu].chi2R A[mu]
LGF := -1/2/xiA del[A[mu], mu] del[A[nu], nu]			 
\end{verbatim}
\end{addmargin}
\end{small}
\vs{3mm}
This completes the model. Now, we want to proceed to its complete automatic renormalization---that is, to the determination of the analytical expressions for the complete set of counterterms (in the $\overline{\text{MS}}$ scheme).
To do so, we open \t{Control.m}; we start by setting \t{model:\,\,QED2}. Then, we must choose a sequence of processes such that all the counterterms are computed. To do so, note that the total set of counterterms is:
\be
\delta Z_3, \quad \delta Z^{L}_1, \, \delta Z^{R}_1, \, \delta m_1, \quad \delta Z^{L}_2, \, \delta Z^{R}_2, \, \delta m_2, \quad \delta e.
\ee
However, from the renormalization of QED, we know that the first one, $\delta Z_3$, can be determined by the vacuum polarization of the photon; the following three, $\delta Z^{L}_1, \, \delta Z^{R}_1, \, \delta m_1$, can be determined by the self-energy of $f_1$; by the same token, $\delta Z^{L}_2, \, \delta Z^{R}_2, \, \delta m_2$ can be determined by the self-energy of $f_2$; finally, $\delta e$ can be determined by one of the vertices (either $f_1 \bar{f}_1 \gamma$ or $f_2 \bar{f}_2 \gamma$) at 1 loop. Therefore, we write:
\vs{2mm}
\begin{small}
\begin{addmargin}[10mm]{0mm}
\begin{verbatim}
inparticles: A
outparticles: A
loops: 1

inparticles: f1
outparticles: f1
loops: 1
options: onepi

inparticles: f2
outparticles: f2
loops: 1
options: onepi

inparticles: A
outparticles: f1,f1bar
loops: 1
options: onepi
\end{verbatim}
\end{addmargin}
\end{small}
\vs{1mm}
Finally, concerning the logical variables of \t{Control.m}, we set them all to \t{True}, except \t{LoSpinors}, which we set to \t{False}. This being done, everything is set. We then go to the \t{FeynMaster} folder and run batch the file \t{RUN-\ts{FeynMaster}}. In total, 10 PDF files are automatically and subsequentially generated and opened: one for the tree-level Feynman rules, another one for the counterterms Feynman rules, and two files per process---one with the Feynman diagrams, another with the respective expressions. In the last file for the expressions, we read ``\textit{This completes the renormalization of the model}'', and the list of the full set of counterterms is presented.

\subsection{$h \to \gamma\gamma$ in the Standard Model}
\label{Chap-FM:sec:HAA}

\n In this example, we use the $h \to \gamma\gamma$ in the SM as an illustration of several features of \ts{FeynMaster}.
We use the SM model file that comes with \FM. As for \t{Control.m}, we set it as:%
\fn{\label{Chap-FM:note:TrueFalse} We are setting \t{FRinterLogic} to \t{True}, which only needs to be done in case it was not yet done before. Actually, generating all Feynman rules for both the tree-level interactions and the counterterms in the SM may take around 5 minutes. Therefore, if we have already performed that operation, we can save time by setting \t{FRinterLogic} and \t{RenoLogic} to \t{False}.}
\vs{2mm}
\begin{small}
\begin{addmargin}[10mm]{0mm}
\begin{verbatim}
model: SM

inparticles: H
outparticles: A,A
loops: 1

FRinterLogic: T
RenoLogic: T
Draw: T
Comp: F
FinLogic: F
DivLogic: F
SumLogic: T
MoCoLogic: F
LoSpinors: F
\end{verbatim}
\end{addmargin}
\end{small}
%
We then run the batch file \t{RUN-FeynMaster}. In total, 3 PDF files will automatically be generated and opened: one for the tree-level Feynman rules, another one for the counterterms Feynman rules and a third one for the Feynman diagrams. We go to the directory \t{SM/Processes/1-HAA} (meanwhile generated inside the directory for the \FMS output) and open the \ts{FeynCalc} notebook \t{Notebook.nb}. We then run the \t{Notebook.nb}, after which we are ready to test some relevant features.

\subsubsection{Notebook access to Feynman rules}

\n First, we want to gain some intuition on how the notebook has access to the SM Feynman rules and to the amplitudes of the $h \to \gamma\gamma$ decay. We write
\vs{1mm}
\begin{small}
\begin{addmargin}[8mm]{0mm}
\begin{Verbatim}[commandchars=\\\{\}]
\textcolor{mygray}{In[14]:=} amp1
\end{Verbatim}
\end{addmargin}
\end{small}
\vs{-2mm}
which should yield the expression:
\be
\dfrac{2 \, e^3 \, {m_W} \, g^{-J2-J4}}{s_w \left(k_1^2 - {{m_W}}^2\right) \left((p_1-k_1)^2 - {{m_W}}^2\right)} \,  \,  - \dfrac{2 \, D \, e^3 \, {m_W} \, g^{-J2-J4}}{s_w \left(k_1^2 - {{m_W}}^2\right) \left((p_1-k_1)^2 - {{m_W}}^2\right)}.
\ee
This is the amplitude for the first Feynman diagram, where $D$ represents the dimension. Now, where does the notebook get this information from? To answer the question, we open \t{Amplitudes.m} inside \t{1-HAA}. If we check the first line, we realize that \t{amp1} is essentially a product of Feynman rules such as \t{propWP[...]} and \t{vrtxAAWPWPbar[...]}.\fn{Amplitudes 13 to 22 have an extra factor 3, which corresponds to the color number. This factor is automatically added by \FMS for diagrams with closed loops of quarks.} These rules are defined in the auxiliary files inside the directory \t{SM/FeynmanRules} (inside the directory for the \FMS output). Although they have been automatically generated, they can always be edited for particular purposes.

\subsubsection{Finiteness and gauge invariance}
\label{Chap-FM:sec:FT}

\n Next, we use some of the features described in section \ref{Chap-FM:sec:FC} to test two important properties of $h \to \gamma\gamma$: finiteness and gauge invariance. We start with the former; by writing
\vs{1mm}
\begin{small}
\begin{addmargin}[8mm]{0mm}
\begin{Verbatim}[commandchars=\\\{\}]
\textcolor{mygray}{In[15]:=} resD
\end{Verbatim}
\end{addmargin}
\end{small}
\vs{-1mm}
we obtain the list with all the expressions for the divergents parts. It is a non-trivial list: although some of its elements are zero, most of them are not. However, when we sum the whole list, we find:
\vs{1mm}
\begin{small}
\begin{addmargin}[8mm]{0mm}
\begin{Verbatim}[commandchars=\\\{\}]
\textcolor{mygray}{In[16]:=} resDtot
\end{Verbatim}
\vs{-3mm}
\begin{Verbatim}[commandchars=\\\{\}]
\textcolor{mygray}{Out[16]=} 0
\end{Verbatim}
\end{addmargin}
\end{small}
\vs{-2mm}
so that the process as a whole is finite, as expected for this decay mode.

\n Let us now check gauge invariance. First of all, note that the total amplitude $M$ for $h \to \gamma\gamma$ can be written as
\be
M = \epsilon_1^{\nu} \epsilon_2^{\sigma} \, M_{\nu\sigma},
\ee
where we are just factoring out the polarization vectors $\epsilon_1^{\nu}$ and $\epsilon_2^{\sigma}$ of the two photons. Then, it is easy to show that gauge invariance forces $M^{\nu\sigma}$ to have the form
\be
M^{\nu\sigma} = ( g^{\nu\sigma} q_1 . q_2 - q_1^{\sigma}  q_2^{\nu}) F,
\label{Chap-FM:eq:invgaugefinal}
\ee
where $q_1$ and $q_2$ are the 4-momenta of the two photons, and $F$ is a scalar function of the momenta and the masses. In other words, it is a consequence of gauge invariance that, in the total process, the coefficient of $g^{\nu\sigma} q_1 . q_2$ must be exactly opposite to that of $q_1^{\sigma}  q_2^{\nu}$. To test this, we define some replacement rules:
\begin{small}
\vs{1mm}
\begin{addmargin}[8mm]{0mm}
\begin{Verbatim}[commandchars=\\\{\}]
\hs{12mm} \textcolor{uglyblue}{(* momentum conservation in scalar products and four-vectors *)}
\end{Verbatim}
\vs{-3mm}
\begin{Verbatim}[commandchars=\\\{\}]
\textcolor{mygray}{In[17]:=} dist = \{SP[p1, x_] -> SP[q1, x] + SP[q2, x], FV[p1, x_] -> FV[q1, x] + FV[q2, x]\};
\end{Verbatim}
\vs{-3mm}
\begin{Verbatim}[commandchars=\\\{\}]
\hs{12mm} \textcolor{uglyblue}{(* external particles on-shell *)}
\end{Verbatim}
\vs{-3mm}
\begin{Verbatim}[commandchars=\\\{\}]
\textcolor{mygray}{In[18]:=} onshell = \{SP[q1, q1] -> 0, SP[q2, q2] -> 0, SP[p1, p1] -> mH^2\};
\end{Verbatim}
\vs{-3mm}
\begin{Verbatim}[commandchars=\\\{\}]
\hs{12mm} \textcolor{uglyblue}{(* kinematics *)}
\end{Verbatim}
\vs{-3mm}
\begin{Verbatim}[commandchars=\\\{\}]
\textcolor{mygray}{In[19:=} kin = \{SP[q1, q2] -> MH^2/2, SP[p1, q1] -> MH^2/2, SP[p1, q2] -> MH^2/2\};
\end{Verbatim}
\vs{-3mm}
\begin{Verbatim}[commandchars=\\\{\}]
\hs{12mm} \textcolor{uglyblue}{(* transversality of the external photons polarizations *)}
\end{Verbatim}
\vs{-3mm}
\begin{Verbatim}[commandchars=\\\{\}]
\textcolor{mygray}{In[20]:=} transv = \{FV[q1, -J2] -> 0, FV[q2, -J4] -> 0\};
\end{Verbatim}
\end{addmargin}
\vs{1mm}
\end{small}
which we use to define a new \t{res} list:
\vs{3mm}
\begin{small}
\begin{addmargin}[8mm]{0mm}
\begin{Verbatim}[commandchars=\\\{\}]
\textcolor{mygray}{In[21]:=} resnew = (res /. dist /. onshell /. kin /. transv) // Simplify;
\end{Verbatim}
\end{addmargin}
\end{small}
\vs{1mm}
Finally, we write the coefficients of $g^{\nu\sigma} q_1 . q_2$ and $q_1^{\sigma}  q_2^{\nu}$ as
\vs{3mm}
\begin{small}
\begin{addmargin}[8mm]{0mm}
\begin{Verbatim}[commandchars=\\\{\}]
\textcolor{mygray}{In[22]:=} resnewT = (Coefficient[resnew, MT[-J2, -J4]] // MyPaVeReduce) 
\hs{12mm}  /(MH^2/2) // Simplify // FCE;
\end{Verbatim}
\vs{-3mm}
\begin{Verbatim}[commandchars=\\\{\}]
\textcolor{mygray}{In[23]:=} resnewL = (Coefficient[resnew, FV[q1, -J4]*FV[q2,-J2]] // MyPaVeReduce)
\hs{12mm}  // Simplify // FCE;
\end{Verbatim}
\end{addmargin}
\end{small}
\vs{1mm}
respectively, to conclude that
\vs{3mm}
\begin{small}
\begin{addmargin}[8mm]{0mm}
\begin{Verbatim}[commandchars=\\\{\}]
\textcolor{mygray}{In[24]:=} Total[resnewT] + Total[resnewL] // Simplify
\end{Verbatim}
\vs{-3mm}
\begin{Verbatim}[commandchars=\\\{\}]
\textcolor{mygray}{Out[24]:=} 0
\end{Verbatim}
\end{addmargin}
\end{small}
in accordance with gauge invariance. For what follows, it is convenient to save the expressions for the total transverse and longitudinal part. We write
\vs{3mm}
\begin{small}
\begin{addmargin}[8mm]{0mm}
\begin{Verbatim}[commandchars=\\\{\}]
\textcolor{mygray}{In[25]:=} FT = Total[resnewT] // Simplify;
\end{Verbatim}
\vs{-3mm}
\begin{Verbatim}[commandchars=\\\{\}]
\textcolor{mygray}{In[26]:=} FL = Total[resnewL] // Simplify;
\end{Verbatim}
\end{addmargin}
\end{small}

\subsubsection{\t{MyTeXForm}}

\n We now illustrate how to use \t{MyTeXForm} inside the \ts{FeynCalc} notebook. Suppose we want to write the sum of final results for the diagrams with quartic vertices (diagrams 1 to 6)  in a \LaTeX \, document. We define the variable \t{toprint1} as
\vs{3mm}
\begin{small}
\begin{addmargin}[8mm]{0mm}
\begin{Verbatim}[commandchars=\\\{\}]
\textcolor{mygray}{In[27]:=} toprint1 = Sum[res[[i]], {\{i, 1, 6\}}] // Simplify
\end{Verbatim}
\end{addmargin}
\end{small}
after which we write
\vs{3mm}
\begin{small}
\begin{addmargin}[8mm]{0mm}
\begin{Verbatim}[commandchars=\\\{\}]
\textcolor{mygray}{In[28]:=} toprint1 // MyTeXForm
\end{Verbatim}
\end{addmargin}
\end{small}
If we now copy the outcome as plain text and paste it in a \LaTeX \, document like the present one, we get:
\be
\begin{split}
& - \Big(( e^3 \,  \big(  \left( m_h^2 + 6 \, m_W^2 \right)  \, B_0\left(p_1^2, m_W^2, m_W^2\right) \\
& + m_W^2 \,  \left( -4 + B_0\left(q_1^2, m_W^2, m_W^2\right) + B_0\left(q_2^2, m_W^2, m_W^2\right) \right)  \big)  \, g^{\nu \sigma} \Big) / \left( 16 \, {m_W} \, \pi^2 \, {s_w} \right)
\end{split}
\ee
Note that we did not need to break the line manually in the \LaTeX \, equation. This is because we are using the \t{breqn} package, which automatically breaks lines in equations.\fn{For documentation, cf. \url{https://www.ctan.org/pkg/breqn} . Recall that the line breaking does not work when the point where the line is to be broken is involved in three or more parentheses.}

\subsubsection{\t{\ts{Fortran}} interface}

\n We mentioned in the Introduction that \FMS includes a numerical interface with \t{\ts{Fortran}}.
We now show how it works in the context of $h \to \gamma\gamma$.
Suppose we want to plot the decay width as a function of the Higgs mass; we could start by computing the total $h \to \gamma\gamma$ decay width:
\vs{3mm}
\begin{small}
\begin{addmargin}[8mm]{0mm}
\begin{Verbatim}[commandchars=\\\{\}]
\textcolor{mygray}{In[27]:=} X0 = restot // DecayWidth
\end{Verbatim}
\end{addmargin}
\end{small}
\vs{-0mm}
However, although this works, it takes a long time and produces large expressions. It is simpler to exploit the generic form of eq.~\ref{Chap-FM:eq:invgaugefinal} and use a form factor; that is,
\vs{3mm}
\begin{small}
\begin{addmargin}[8mm]{0mm}
\begin{Verbatim}[commandchars=\\\{\}]
\textcolor{mygray}{In[27]:=} X0 = FTaux (MT[-J2, -J4] MH^2/2 -  FV[q1, -J4] FV[q2, -J2]) // DecayWidth 
\end{Verbatim}
\end{addmargin}
\end{small}
where \t{FTaux} is an auxiliary variable that, in the end, must be replaced by the absolute value of the total transverse part \t{FT}, defined above. Then, we write
\vs{3mm}
\begin{small}
\begin{addmargin}[8mm]{0mm}
\begin{Verbatim}[commandchars=\\\{\}]
\textcolor{mygray}{In[28]:=} (X0 /. FTaux -> Abs[FT] // Simplify) // FCtoFT
\end{Verbatim}
\end{addmargin}
\end{small}
where we performed the referred replacement.%
\fn{Alternatively, following the instructions in note \ref{Chap-FM:note:FCtoFT}, we could also have written \t{FCtoFT[X0, \{FTaux -> Abs[FT]\}]}.}
As explained in section \ref{Chap-FM:sec:FC}, the command \t{FCtoFT} generates five files: \t{MainFT.F}, \t{MyFunctionFT.F}, \t{MyVariables.h}, \t{MyParameters.h} and \t{Makefile-template}. We open \t{MainFT.F} and, immediately after the comments \textit{Write now the rest of the program},
we write%
\footnote{The parameters loaded from the file \t{MyParameter.h} cannot be changed inside the \ts{Fortran} program (\t{MainFT.F}). Hence, since we define the parameter \t{MH} as the Higgs boson mass, we name \t{xMH} the variable we are using to make the plot; in doing so, we must be careful to replace \t{MH} for \t{xMH} in the arguments of \t{MyFunction} inside the loop.}
\vs{1mm}
\begin{small}
\begin{addmargin}[8mm]{0mm}
\begin{Verbatim}[commandchars=\\\{\}]
xMH=38d0
do i=1,162
xMH=xMH+1d0
write(50,98)xMH,MyFunction(..., xMH, ...)
enddo
\end{Verbatim}
\end{addmargin}
\end{small}
%
where \t{50} and \t{98} represent the output file and the impression format, respectively. We are varying the Higgs mass from 38 GeV to 200 GeV in steps of 1 GeV. The result is presented in fig.~{\ref{Chap-FM:fig:HAA}.
\begin{figure}[htb]
\centering
\includegraphics[width=0.5\textwidth]{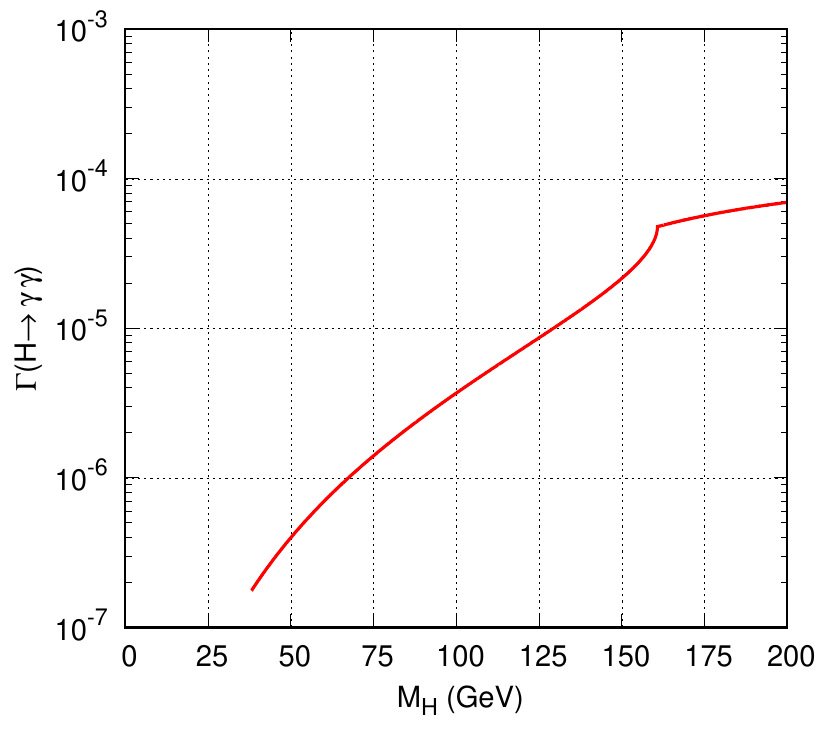}
\vs{-5mm}
\caption{Width of the process $h \to \gamma \gamma$ as a function of the Higgs boson mass.}
\label{Chap-FM:fig:HAA}
\end{figure}	

\vs{-3mm}
\subsubsection{Edition of Feynman diagrams}

\n Finally, we briefly explain how to edit the Feynman diagrams. Recall that they were written in a \LaTeX \, file inside \t{SM/1-HAA/TeXs-drawing} folder. We open the file \t{diagrams.tex} and consider the first diagram; the original code produces the original diagram:

\vs{3mm}
\hspace{7mm} 
\begin{minipage}[h]{.80\textwidth}
\begin{small}
\begin{verbatim}
(...)
\fmflabel{$\gamma$}{...} 
\fmflabel{$\gamma$}{...} 
\fmf{dashes,tension=3}{...} 
\fmf{photon,tension=3}{...} 
\fmf{photon,tension=3}{...} 
\fmf{photon,label=$W^{+}$,right=1}{...}
\fmf{phantom_arrow,tension=0,right=1}{...}
\fmf{photon,label=$W^{+}$,right=1}{...}
\fmf{phantom_arrow,tension=0,right=1}{...}
(...)
\end{verbatim}
\end{small}
\end{minipage}
\hspace{-45mm} 
\begin{minipage}[h]{.40\textwidth}
\begin{picture}(0,80)
\begin{fmffile}{1} 
\begin{fmfgraph*}(100,56) 
\fmfset{arrow_len}{3mm} 
\fmfset{arrow_ang}{20} 
\fmfleft{nJ1} 
\fmflabel{$h$}{nJ1} 
\fmfright{nJ2,nJ4} 
\fmflabel{$\gamma$}{nJ2} 
\fmflabel{$\gamma$}{nJ4} 
\fmf{dashes,tension=3}{nJ1,nJ1J3J2} 
\fmf{photon,tension=3}{nJ2,nJ2nJ4J1J4} 
\fmf{photon,tension=3}{nJ4,nJ2nJ4J1J4} 
\fmf{photon,label=$W^+$,right=1}{nJ1J3J2,nJ2nJ4J1J4} 
\fmf{phantom_arrow,tension=0,right=1}{nJ1J3J2,nJ2nJ4J1J4} 
\fmf{photon,label=$W^+$,right=1}{nJ2nJ4J1J4,nJ1J3J2} 
\fmf{phantom_arrow,tension=0,right=1}{nJ2nJ4J1J4,nJ1J3J2} 
\end{fmfgraph*} 
\end{fmffile}
\end{picture}
\end{minipage}
\vs{1mm}

\n However, we can modify the code in order to change the aspect of the diagram. In particular, we can change the labels, the tensions and the curvatures.\fn{The tensions represent the strength of the lines: the larger the tension, the tighter the line will be. The default tension is 1. The curvature is represented by the variable \t{right}. Note that tensions, labels and curvatures are just a few examples of variables that can be changed to generate a different diagram. For more informations, please consult the \t{feynmf} manual.} For example:

\vs{3mm}
\hspace{7mm} 
\begin{minipage}[h]{.80\textwidth}
\begin{small}
\begin{verbatim}
(...)
\fmflabel{$\gamma_1$}{...} 
\fmflabel{$\gamma_2$}{...} 
\fmf{dashes,tension=1}{...} 
\fmf{photon,tension=3}{...} 
\fmf{photon,tension=3}{...} 
\fmf{photon,label=$W^{-}$,right=1}{...} 
\fmf{phantom_arrow,tension=0,right=1}{...} 
\fmf{photon,label=$W^{-}$,right=1}{...}
\fmf{phantom_arrow,tension=0,right=1}{...} 
(...)
\end{verbatim}
\end{small}
\end{minipage}
\hspace{-45mm} 
\begin{minipage}[h]{.40\textwidth}
\begin{picture}(0,80)
\begin{fmffile}{2} 
\begin{fmfgraph*}(100,56) 
\fmfset{arrow_len}{3mm} 
\fmfset{arrow_ang}{20} 
\fmfleft{nJ1} 
\fmflabel{$h$}{nJ1} 
\fmfright{nJ2,nJ4} 
\fmflabel{$\gamma_1$}{nJ2} 
\fmflabel{$\gamma_2$}{nJ4} 
\fmf{dashes,tension=1}{nJ1,nJ1J3J2} 
\fmf{photon,tension=3}{nJ2,nJ2nJ4J1J4} 
\fmf{photon,tension=3}{nJ4,nJ2nJ4J1J4} 
\fmf{photon,label=$W^{+}$,right=1}{nJ1J3J2,nJ2nJ4J1J4} 
\fmf{phantom_arrow,tension=0,right=1}{nJ1J3J2,nJ2nJ4J1J4} 
\fmf{photon,label=$W^+$,right=1}{nJ2nJ4J1J4,nJ1J3J2} 
\fmf{phantom_arrow,tension=0,right=1}{nJ2nJ4J1J4,nJ1J3J2} 
\end{fmfgraph*} 
\end{fmffile}
\end{picture}
\end{minipage}
\vs{8mm}

\vs{-5mm}
\subsection{QED Ward identity}
\label{Chap-FM:sec:WI}

\n In the previous example, we showed how to use \ts{FeynMaster} to manipulate the results of a single process. Here, we illustrate how it can also be used to combine information of several processes. For that purpose, we consider a simple task: prove the QED Ward identity.

\n It is easy to show that the Ward identity at one-loop order in QED can be written as:
\be
p_1^{\nu} \, \Gamma_{\nu}(p_1,p_2,p_3) = e \left( \vphantom{\dfrac{A^B}{A^B}} \Sigma(p_2) - \Sigma(p_3) \right),
\label{Chap-FM:eq:provanda}
\ee
where
\be
\Gamma_{\nu}(p_1,p_2,p_3) =
\hs{8mm}
\begin{minipage}{0.35\textwidth}
\begin{fmffile}{QEDB11}
\begin{fmfgraph*}(100,100) 
\fmfset{arrow_len}{3mm} 
\fmfset{arrow_ang}{20} 
\fmfleft{nJ1}  
\fmfright{nJ2,nJ4} 
\fmf{photon,label=$p_1$,tension=4}{nJ1,J2J3nJ1} 
\fmf{fermion,label=$p_2$,label.side=left,tension=4}{nJ2J1J5,nJ2} 
\fmf{fermion,label=$p_3$,label.side=left,tension=4}{nJ4,J4nJ4J6} 
\fmf{fermion,tension=1,label.dist=3thick}{J2J3nJ1,nJ2J1J5} 
\fmf{fermion,tension=1,label.dist=3thick}{J4nJ4J6,J2J3nJ1} 
\fmf{photon,label.side=left,tension=1,label.dist=3thick}{J4nJ4J6,nJ2J1J5} 
\end{fmfgraph*}
\end{fmffile}
\end{minipage}
\hs{-17mm},
\hs{15mm}
\Sigma(p_i) = 
\begin{minipage}{0.35\textwidth}
\begin{fmffile}{QEDB12} 
\begin{fmfgraph*}(100,57) 
\fmfset{arrow_len}{3mm} 
\fmfset{arrow_ang}{20} 
\fmfleft{nJ1} 
\fmfright{nJ2} 
\fmf{fermion,label=$p_i$,tension=3}{nJ1,J2nJ1J3} 
\fmf{fermion,label=$p_i$,label.side=right,tension=3}{nJ2J1J4,nJ2} 
\fmf{fermion,right=1}{J2nJ1J3,nJ2J1J4} 
\fmf{photon,right=1}{nJ2J1J4,J2nJ1J3} 
\end{fmfgraph*} 
\end{fmffile} 
\end{minipage}
\hs{-15mm},
\label{Chap-FM:eq:pic}
\ee
and where the momenta $p_1$ and $p_3$ are incoming, while $p_2$ is outgoing.
In order to prove eq. \ref{Chap-FM:eq:provanda} with \ts{FeynMaster}, we need to consider the two processes depicted in eq. \ref{Chap-FM:eq:pic}: the QED vertex and the fermion self-energy. Hence, we open and edit \t{Control.m} according to:
\vs{2mm}
\begin{small}
\begin{addmargin}[10mm]{0mm}
\begin{Verbatim}[commandchars=\\\{\}]
model: QED

inparticles: A
outparticles: f,fbar
loops: 1
options: onepi

inparticles: f
outparticles: f
loops: 1
options: onepi

FRinterLogic: T
RenoLogic: F
Draw: F
Comp: T
FinLogic: F
DivLogic: F
SumLogic: T
MoCoLogic: F
LoSpinors: F
\end{Verbatim}
\end{addmargin}
\end{small}
%
We then run \ts{FeynMaster}.
After this, we go to the directory \t{QED/Processes/1-Affbar} (meanwhile generated inside the directory for the \FMS output), we copy the notebook lying there to a different directory and we rename it \t{Notebook-Global.nb}. This is going to be the notebook where we shall combine the information of both processes. We open it, and delete most of the lines there: in a first phase, we only want to load the general files. So it must look like this:
\vs{1mm}
\begin{small}
\begin{addmargin}[8mm]{0mm}
\begin{Verbatim}[commandchars=\\\{\}]
\textcolor{mygray}{In[1]:=} << FeynCalc`
\textcolor{mygray}{In[2]:=} dirNuc = "(...)"; 
\textcolor{mygray}{In[3]:=} dirFey = "(...)"; 
\textcolor{mygray}{In[4]:=} dirCT = "(...)"; 
\textcolor{mygray}{In[5]:=} Get["Feynman-Rules-Main.m", Path -> \{dirFey\}] 
\textcolor{mygray}{In[6]:=} Get["FunctionsFM.m", Path -> \{dirNuc\}]
\textcolor{mygray}{In[7]:=} compNwrite = False; 
\end{Verbatim}
\end{addmargin}
\end{small}
\vs{-1mm}
Now, we want to load the first process. To do so, we write:
\vs{1mm}
\begin{small}
\begin{addmargin}[8mm]{0mm}
\begin{Verbatim}[commandchars=\\\{\}]
\textcolor{mygray}{In[8]:=} dirHome = "(...)"; 
\textcolor{mygray}{In[9]:=} SetDirectory[dirHome];
\textcolor{mygray}{In[10]:=} << Helper.m;
\textcolor{mygray}{In[11]:=} Get["Definitions.m", Path -> \{dirNuc\}];
\textcolor{mygray}{In[12]:=} Get["Finals.m", Path -> \{dirNuc\}]; 
\end{Verbatim}
\end{addmargin}
\end{small}
\vs{-1mm}
where \t{dirHome} should be set to the directory corresponding to \t{1-Affbar}. Next, we define new variables: \t{X0} as $\Gamma$ of eq. \ref{Chap-FM:eq:provanda}, with the above-mentioned momentum definitions, and \t{X1} as the whole l.h.s. of eq. \ref{Chap-FM:eq:provanda}.
\vs{1mm}
\begin{small}
\begin{addmargin}[8mm]{0mm}
\begin{Verbatim}[commandchars=\\\{\}]
\textcolor{mygray}{In[13]:=} X0 = res[[1]] /. \{p1 -> p2 - p3, q1 -> p2, q2 -> -p3\};
\textcolor{mygray}{In[14]:=} X1 = Contract[X0 FV[p2 - p3, -J1]];
\end{Verbatim}
\end{addmargin}
\end{small}
\vs{-1mm}
We now load the second process:
\vs{1mm}
\begin{small}
\begin{addmargin}[8mm]{0mm}
\begin{Verbatim}[commandchars=\\\{\}]
\textcolor{mygray}{In[15]:=} dirHome = "(...)"; 
\textcolor{mygray}{In[16]:=} SetDirectory[dirHome];
\textcolor{mygray}{In[17]:=} << Helper.m ;
\textcolor{mygray}{In[18]:=} Get["Definitions.m", Path -> \{dirNuc\}];
\textcolor{mygray}{In[19]:=} Get["Finals.m", Path -> \{dirNuc\}]; 
\end{Verbatim}
\end{addmargin}
\end{small}
\vs{-1mm}
where \t{dirHome} should now be set to the directory corresponding to \t{2-ff}. From this, and recalling that the default momentum of a self-energy is \t{p1} (cf. table \ref{Chap-FM:tab:FCconv}), we can obtain the r.h.s. of eq. \ref{Chap-FM:eq:provanda} by writing:
\vs{1mm}
\begin{small}
\begin{addmargin}[8mm]{0mm}
\begin{Verbatim}[commandchars=\\\{\}]
\textcolor{mygray}{In[20]:=} Y0a = res[[1]] /. {p1 -> p2}; 
\textcolor{mygray}{In[21]:=} Y0b = res[[1]] /. {p1 -> p3};
\textcolor{mygray}{In[22]:=} Y1 = (e (Y0a - Y0b) // DiracSimplify) // Simplify;
\end{Verbatim}
\end{addmargin}
\end{small}
\vs{-1mm}
Finally, we prove the Ward identity by showing that both sides of eq. \ref{Chap-FM:eq:provanda} are equal:
\vs{1mm}
\begin{small}
\begin{addmargin}[8mm]{0mm}
\begin{Verbatim}[commandchars=\\\{\}]
\textcolor{mygray}{In[23]:=} WI = (Y1 - X1) // Simplify;
\textcolor{mygray}{In[24]:=} CheckWI = MyPaVeReduce[WI] // Calc
\end{Verbatim}
\end{addmargin}
\end{small}
\vs{-1mm}
which yields 0, thus completing the proof.

\section{Quick first usage}
\label{Chap-FM:sec:Summary}

\n For a quick first usage of \ts{FeynMaster}, the user should follow this sequence of steps:
\vs{1.0mm}
\begin{center}
\begin{addmargin}[8mm]{0mm}
1) Make sure you have installed \t{\ts{Python}}, \t{\ts{Mathematica}} and \LaTeX \,, on the one hand, and \ts{FeynRules}, \QGS and \ts{FeynCalc}, on the other;\\[2mm]
2) Download \ts{FeynMaster} in \url{https://porthos.tecnico.ulisboa.pt/FeynMaster/};\\[2mm]
3) Extract the downloaded file and place the resulting folder in a directory at will;\\[2mm]
4) Edit the file \t{RUN-FeynMaster} as explained in section \ref{Chap-FM:sec:Instal};\\[2mm]
5) Run \t{RUN-FeynMaster}.
\end{addmargin}
\end{center}
This should generate and open 4 PDF files relative to QED: the Feynman rules for the tree-level interactions, the Feynman rules for the counterterms
interactions, the Feynman diagram for the one-loop vacuum polarization, and a document containing not only the expressions for the vacuum polarization, but also the expression for the associated counterterm in $\overline{\text{MS}}$.

%% file: Appendices/Pilaftsis.tex
\chapter{A note on a model by Pilaftsis}
\label{App-Pilaftsis}

In this appendix, we discuss a model by Pilaftsis \cite{Pilaftsis:1998pe}, and show that it is a theoretically sound model.
As shall be seen, CP is not a symmetry of the model. However, at tree-level, CP can be well-defined in a certain sector of the model, since there is a specific basis where all the phases of that sector vanish. In particular, the imaginary part of a certain complex parameter vanishes (in that specific basis).
For renormalization purposes, however, and since the parameter at stake is in general complex, its imaginary part is renormalized. We show that the counterterm of that imaginary part is the key to understand Pilaftsis's claims about the renormalization of the mixing between CP-even and CP-odd states.

\section{Tree-level}
\label{App-Pilaf:sec:tree}

We start by presenting Pilaftsis's model at tree-level.
Suppose two abelian symmetries, $\mathrm{U(1)_Q}$ and $\mathrm{U(1)_Y}$, the former being global, the latter gauge.
%
%
Suppose also four complex (singlet) scalars, $\Phi_1$, $\Phi_2$, $\chi_L$ and $\chi_{R}$, with charges \cite{Pilaftsis:1998pe}:%
\fn{Pilaftsis considers $\chi_L^{\pm}$ and $\chi_{R}^{\pm}$ instead of $\chi_L$ and $\chi_{R}$, respectively, and claims that they are charged fields \cite{Pilaftsis:1998pe}. However, it is not clear what he means by ``charged'' in this context, as there not seems to be a symmetry associated with electric charge. In any event, this is not relevant for the argument here.}
\bs
\bea
&&
Q(\Phi_1) = 2,\quad
Q(\Phi_2) = 1,\quad
Q(\chi_L) = -2,\quad
Q(\chi_{\mathrm{R}}) = 0, \\
&&
Y(\Phi_1) = 1,\quad
Y(\Phi_2) = 1,\quad
Y(\chi_L) = -\dfrac{1}{2},\quad
Y(\chi_{\mathrm{R}}) = \dfrac{1}{2},
\eea
\es
where $Q$ and $Y$ represent the operators of $\mathrm{U(1)_Q}$ and $\mathrm{U(1)_Y}$, respectively.
The complete Lagrangian includes the ``Higgs potential'' $\mathcal{L}_V^{\mu}$ and the sector $\mathcal{L}_Y^{\chi}$, such that \cite{Pilaftsis:1998pe}:%
\fn{We use $g$ instead of $h$ to describe the parameter associated with the interaction $\Phi_2 \chi_L \chi_{\mathrm{R}}^*$, to avoid any confusion with the $h$ field defined below. The ellipses in eq. \ref{App-Pilaftsis:eq:Ppot} represent (according to Pilaftsis) new interactions which are allowed by $\mathrm{U(1)_Q}$ and $\mathrm{U(1)_Y}$ \cite{Pilaftsis:1998pe}.}
\bs
\begin{flalign}
\label{App-Pilaftsis:eq:Ppot}
\mathcal{L}_V^{\mu} &= \mu_1^2 \Phi_1^* \Phi_1 + \mu_2^2 \Phi_2^* \Phi_2 + \left(\mu^2 \Phi_1^* \Phi_2 + \mathrm{h.c.} \right)
+ \lambda_1 {\left ( \Phi_1^{*} \Phi_1 \right )}^2 + \lambda_2 {\left ( \Phi_2^{*} \Phi_2 \right )}^2 + \lambda_{34} \, \Phi_1^{*} \Phi_1 \Phi_2^{*} \Phi_2 + \, ... \, ,\\[2mm]
- \mathcal{L}_Y^{\chi}
&=
m_L^2 \, \chi_L \chi_L^* +
m_{\mathrm{R}}^2 \, \chi_{\mathrm{R}} \chi_{\mathrm{R}}^* +
\left(
f \, \Phi_1 \chi_L \chi_{\mathrm{R}}^* +
g \, \Phi_2 \chi_L \chi_{\mathrm{R}}^* +
\mathrm{h.c.} \right).
\label{App-Pilaftsis:eq:Yuk}
\end{flalign}
\es
The parameters $\mu^2$, $f$ and $g$ are in general complex, while the remaining ones are real by construction. The terms involving $\mu^2$ and $g$ break the symmetry $\mathrm{U(1)_Q}$ softly.
Moreover, the model violates CP, as the condition that would ensure CP conservation, $\operatorname{Im} \left[ \mu^2 f g^* \right] = 0$, in general does not hold  \cite{Pilaftsis:1998pe}.
The fields $\chi_L$ and $\chi_R$ do not develop vevs, whereas $\Phi_1$ and $\Phi_2$ do \cite{Pilaftsis:1998pe}.
These can be parameterized in a general basis as:
\be
\Phi_1 =
\dfrac{1}{\sqrt{2}}
\left(v_1 \, e^{i \zeta_1}  + H_1^{\prime} + i A_1^{\prime} \right),
\qquad
\Phi_2 =
\dfrac{1}{\sqrt{2}}
\left( v_2 \, e^{i \zeta_2} + H_2^{\prime} + i A_2^{\prime}\right),
\label{App-Pilaftsis:eq:66}
\ee
with $v_1$ and $v_2$ are the (real) vevs, $\zeta_1$ and $\zeta_2$ (real) phases, and $H_1^{\prime}$, $H_2^{\prime}$, $A_1^{\prime}$ and $A_2^{\prime}$ real fields.
The minimization (or tadpole) equations read:
\bs
\label{App-Pilaftsis:eq:min-eqs-prime}
\begin{gather}
0= - t_{H_{1}^{\prime}} \left.\equiv \dfrac{\partial \mathcal{L}_{V}^{\mu}}{\partial H_{1}^{\prime}}\right|_{\left\langle H_{i}^{\prime}\right\rangle=\left\langle A_{i}^{\prime}\right\rangle=0}
=
c_{\zeta_1} v_1 \left(\mu_1^2 + \lambda_1 \, v_1^2 + \dfrac{1}{2} \lambda_{34} \, v_2^2 \right) +
c_{\zeta_2} v_2 \, \mu_{\mathrm{R}}^{2} -
s_{\zeta_2} v_2 \, \mu_{\mathrm{I}}^{2}, \\
0= - t_{H_{2}^{\prime}} \left.\equiv \dfrac{\partial \mathcal{L}_{V}^{\mu}}{\partial H_{2}^{\prime}}\right|_{\left\langle H_{i}^{\prime}\right\rangle=\left\langle A_{i}^{\prime}\right\rangle=0}
=
c_{\zeta_2} v_2 \left(\mu_2^2 + \lambda_2 \, v_2^2 + \dfrac{1}{2} \lambda_{34} \, v_1^2\right) +
c_{\zeta_1} v_1 \, \mu_{\mathrm{R}}^{2} +
s_{\zeta_1} v_1 \, \mu_{\mathrm{I}}^{2}, \\
0= - t_{A_{1}^{\prime}} \left.\equiv \dfrac{\partial \mathcal{L}_{V}^{\mu}}{\partial A_{1}^{\prime}}\right|_{\left\langle H_{i}^{\prime}\right\rangle=\left\langle A_{i}^{\prime}\right\rangle=0}
=
s_{\zeta_1} v_1 \left(\mu_1^2 + \lambda_1 \, v_1^2 + \dfrac{1}{2} \lambda_{34} \, v_2^2 \right) +
s_{\zeta_2} v_2 \, \mu_{\mathrm{R}}^{2} +
c_{\zeta_2} v_2 \, \mu_{\mathrm{I}}^{2},
\\
0= - t_{A_{2}^{\prime}} \left.\equiv \dfrac{\partial \mathcal{L}_{V}^{\mu}}{\partial A_{2}^{\prime}}\right|_{\left\langle H_{i}^{\prime}\right\rangle=\left\langle A_{i}^{\prime}\right\rangle=0}
=
s_{\zeta_2} v_2 \left(\mu_2^2 + \lambda_2 \, v_2^2 + \dfrac{1}{2} \lambda_{34} \, v_1^2\right) +
s_{\zeta_1} v_1 \, \mu_{\mathrm{R}}^{2} -
c_{\zeta_1} v_1 \, \mu_{\mathrm{I}}^{2},
\end{gather}
\es
where we are using the short notation $s_x = \sin x, c_x = \cos x$,
and we are defining:
\be
\mu_{\mathrm{R}}^2 \equiv \operatorname{Re} \mu^2,
\qquad
\mu_{\mathrm{I}}^2 \equiv \operatorname{Im} \mu^2.
\ee
Equations \ref{App-Pilaftsis:eq:min-eqs-prime} define the true vevs $v_1$ and $v_2$ and the true phases $\zeta_1$ and $\zeta_2$ at tree-level (i.e. the true vevs and the true phases are those for which eqs. \ref{App-Pilaftsis:eq:min-eqs-prime} hold).%
\fn{\label{App-Pilaftsis:note:the-note}Actually, eqs. \ref{App-Pilaftsis:eq:min-eqs-prime} are just a necessary condition to ensure that $v_1$, $v_2$, $\zeta_1$ and $\zeta_2$ are the true vevs and phases; they are not a sufficient one. In fact, they only guarantee that a local extreme of the potential is selected: they do not ensure that such extreme a global one, not even that it is a minimum. In what follows, we assume that the true vevs and phases are all non-zero.}
Now, one is not required to work in this general basis: one has the freedom to choose the particular basis for $\Phi_1$ and $\Phi_2$ where $\zeta_1$ and $\zeta_2$ are rephased away. This is the basis chosen by Pilaftsis; consequently, he writes \cite{Pilaftsis:1998pe}:
\be
\Phi_1 =
\dfrac{1}{\sqrt{2}}
\left(v_1 + H_1 + i A_1 \right),
\qquad
\Phi_2 =
\dfrac{1}{\sqrt{2}}
\left( v_2 + H_2 + i A_2\right),
\label{App-Pilaftsis:eq:6}
\ee
where $H_1$, $H_2$, $A_1$ and $A_2$ are real fields.
%
%
%
%
%
%
%
%
%
In this particular basis, the tadpole equations become\cite{Pilaftsis:1998pe}:%
\fn{We use the notation $t_{\phi}$ (instead of $T_{\phi}$) to represent the tree-level tadpole of the generic field $\phi$; moreover, since $\mathcal{L} \ni - \mathcal{L}_{V}^{\mu}$, we use a minus sign before $t_{\phi}$.}
\bs
\label{App-Pilaftsis:eq:min-eqs}
\bea
&& 0 = - t_{H_{1}} \left.\equiv \dfrac{\partial \mathcal{L}_{V}^{\mu}}{\partial H_{1}}\right|_{\left\langle H_{i}\right\rangle=\left\langle A_{i}\right\rangle=0}=v_{1}\left(\mu_{1}^{2}+ \mu_{\mathrm{R}}^{2} \dfrac{v_{2}}{v_{1}}+\lambda_{1} v_{1}^{2}+\dfrac{1}{2} \lambda_{34} v_{2}^{2}\right), \\
&& 0= - t_{H_{2}} \left.\equiv \dfrac{\partial \mathcal{L}_{V}^{\mu}}{\partial H_{2}}\right|_{\left\langle H_{i}\right\rangle=\left\langle A_{i}\right\rangle=0}=v_{2}\left(\mu_{2}^{2}+ \mu_{\mathrm{R}}^{2} \dfrac{v_{1}}{v_{2}}+\lambda_{2} v_{2}^{2}+\dfrac{1}{2} \lambda_{34} v_{1}^{2}\right), \\
&& 0= - t_{A_{1}} \left.\equiv \dfrac{\partial \mathcal{L}_{V}^{\mu}}{\partial A_{1}}\right|_{\left\langle H_{i}\right\rangle=\left\langle A_{i}\right\rangle=0}=v_{2} \mu_{\mathrm{I}}^{2},
\label{App-Pilaftsis:eq:3}
\\
&& 0= - t_{A_{2}} \left.\equiv \dfrac{\partial \mathcal{L}_{V}^{\mu}}{\partial A_{2}}\right|_{\left\langle H_{i}\right\rangle=\left\langle A_{i}\right\rangle=0}=-v_{1} \mu_{\mathrm{I}}^{2},
\label{App-Pilaftsis:eq:4}
\eea
\es
in such a way that the last two equations (which hold in the basis \ref{App-Pilaftsis:eq:6}) imply:
\be
\mu_{\mathrm{I}}^{2} = 0.
\label{App-Pilaftsis:eq:joker}
\ee
It is crucial to realize that this equation holds in the \textit{particular} basis choice implicit in eqs. \ref{App-Pilaftsis:eq:min-eqs} (that is, in the particular basis defined in eq. \ref{App-Pilaftsis:eq:6}), but does \textit{not} hold in general. That is to say, in the general basis \ref{App-Pilaftsis:eq:66}, where the minimization equations are eqs. \ref{App-Pilaftsis:eq:min-eqs-prime}, $\mu_{\mathrm{I}}^{2}$ is in general different from zero. Eq. \ref{App-Pilaftsis:eq:joker} is a consequence of a specific basis choice, and is only ensured to be true in that particular basis. Or, as we could also say it, $\mu^{2}$ is in general a complex parameter, which turns out to be real in a particular basis.

Now, that particular basis is very peculiar; the reason is that, in such basis, there are no phases in the potential ($\mathcal{L}_V^{\mu}$). Indeed, the phases $\zeta_1$ and $\zeta_2$ were rephased away (by the basis choice), and the only remaining phase left---the phase of $\mu^{2}$---is forced to be zero (by the minimization conditions of that basis).
This implies that ``CP is a good symmetry of the Higgs potential'' \cite{Pilaftsis:1998pe}. Indeed, despite the fact that CP is violated in the theory (in the sector $ \mathcal{L}_Y^{\chi}$, as $f$ and $g$ are in general complex), the circumstance that there is a basis where the sector $\mathcal{L}_V^{\mu}$ has no phases means that CP can be well-defined in that sector.
Therefore, one can attribute a CP value to the fields of such basis: $H_1$ and $H_2$ are CP-even fields, and $A_1$ and $A_2$ are CP-odd fields \cite{Pilaftsis:1998pe}.

It should be clear that this attribution of CP values to the fields only makes sense in the particular basis which we just described.
In a general basis, as we saw, there will be phases in $\mathcal{L}_V^{\mu}$; one of the consequences is that the bilinear interactions between the $H_i^{\prime}$ fields and the $A_i^{\prime}$ fields will not vanish; hence, the $H_i^{\prime}$ fields and the $A_i^{\prime}$ fields must be diagonalized together, in a $4 \times 4$ matrix.
By contrast, in the particular basis chosen by Pilaftsis, there are no phases in $\mathcal{L}_V^{\mu}$. In that basis, then, the fields $H_i$ and $A_i$ do not interact at the bilinear level; therefore, the former and the latter can be separately diagonalized. That is to say, instead of considering a $4 \times 4$ matrix of bilinear interactions, one can consider two $2 \times 2$ matrices: one for the $H_i$ fields ($H_1$ and $H_2$) and another one for the $A_i$ fields ($A_1$ and $A_2$).
This is what Pilaftsis does: he introduces 
the angles $\theta$ and $\beta$ such that \cite{Pilaftsis:1998pe}
\bs
\label{App-Pilaftsis:eq:12}
\bea
\left(\begin{array}{c}
H_{1} \\
H_{2}
\end{array}\right)
&=&
\left(\begin{array}{cc}
c_{\theta} & -s_{\theta} \\
s_{\theta} & c_{\theta}
\end{array}\right)\left(\begin{array}{l}
h \\
H
\end{array}\right),
\\
\left(\begin{array}{c}
A_{1} \\
A_{2}
\end{array}\right)
&=&
\left(\begin{array}{cc}
c_{\beta} & -s_{\beta} \\
s_{\beta} & c_{\beta}
\end{array}\right)\left(\begin{array}{c}
G_{0} \\
A
\end{array}\right),
\eea
\es
where $h$ and $H$ are the CP-even mass states, and $A$ and $G_0$ are the CP-odd mass states, with $G_0$ being the massless would-be Goldstone boson.
As before, the attribution of well-defined CP values to the mass states only makes sense in the particular basis used by Pilaftsis. 
In that basis, the tadpole equations could have been written
for the mass states as well (and not only for the symmetry states $H_1$, $H_2$, $A_1$ and $A_2$, as in eqs. \ref{App-Pilaftsis:eq:min-eqs}). In particular, we would find \cite{Pilaftsis:1998pe}:
\be
0 = -t_{A} \left.\equiv \dfrac{\partial \mathcal{L}_{V}^{\mu}}{\partial A}\right|_{\left\langle h\right\rangle=\left\langle H\right\rangle=\left\langle A\right\rangle=\left\langle G_0\right\rangle=0}= - v \, \mu_{\mathrm{I}}^2,
\label{App-Pilaftsis:eq:13}
\ee
with $v = \sqrt{v_1^2 + v_2^2}$.
%
%
%
%
%
As for the $\chi_L$ and $\chi_R$ fields, Pilaftsis diagonalizes them by introducing the angle $\phi$, such that \cite{Pilaftsis:1998pe}:%
\fn{Again, we omit the $+$ and $-$ superscripts on the fields.}
\be
\left(\begin{array}{c}
\chi_L \\
\chi_R
\end{array}\right)
=
\left(\begin{array}{cc}
c_{\phi} & s_{\phi} \\
-s_{\phi} & c_{\phi}
\end{array}\right)\left(\begin{array}{c}
\chi_1 \\
\chi_2
\end{array}\right),
\ee
where $\chi_1$ and $\chi_2$ are the (complex) diagonalized states, with (real) masses $M_1$ and $M_2$. 

\section{Up to one-loop level}

When considering the theory up to one-loop level, as we saw in detail in chapter \ref{Chap-Selec}, one should select the vev, and then renormalize the theory in order to obtain finite $S$-matrix elements.
In the usual procedure, the last point means taking an independent set of parameters, identifying them with bare quantities, identifying these in turn with the sum of a renormalized parameter and a counterterm, and finally fixing the counterterms. 
But this is not what Pilaftsis does---or, at least, this is not something he alludes to. He never mentions bare parameters, let alone separating them into renormalized parameters and counterterms.
In what follows, we employ the usual renormalization procedure to show that Pilaftsis' model is theoretically sound.

Concerning the selection of the vev, Pilaftsis requires that the quantum corrections do ``not shift the true ground state of the effective potential'' \cite{Pilaftsis:1998pe}. This means that he wants to select the true vev. To this end, we use the Parameter-Renormalized tadpole scheme (PRTS), discussed in detail in chapter \ref{Chap-Selec}, which is the tadpole scheme which Pilaftsis himself uses (although he does not use this terminology).

Concerning the renormalization, and as we just noted, the parameters of the model are identified as bare parameters
and they equated to the sum of the renormalized parameter
and the corresponding counterterm. In particular,
\be
\mu_{\mathrm{I}(0)}^2 = \mu_{\mathrm{I}}^2 + \delta \mu_{\mathrm{I}}^2.
\label{App-Pilaftsis:eq:mui-exp}
\ee
A clarification is in order here.
The circumstance that $\mu_{\mathrm{I}}^2$ is zero in a particular basis at tree-level (recall eq. \ref{App-Pilaftsis:eq:joker}) does not mean that it does not need to be renormalized.
Renormalization must always start from the most general basis at tree-level, since the inclusion of high-orders in general modifies the bases (as we explicitly saw in chapter \ref{Chap-Reno}).
Actually, if one does not start from the most general basis (i.e. if one does not consider the parameters in all their generality), one is not ensured to be able to absorb all the divergences that may show up at higher orders.
Consequently, since $\mu_{\mathrm{I}}^2$ is not zero in the most general basis, it must be renormalized; or, which is the same, since $\mu^2$ is in general complex, it must be renormalized as complex.

We start by showing that the one-loop mixing between CP-even and CP-odd states is finite; then, we discuss the results obtained.

\subsection{Renormalization of the mixing between CP-even and CP-odd states}

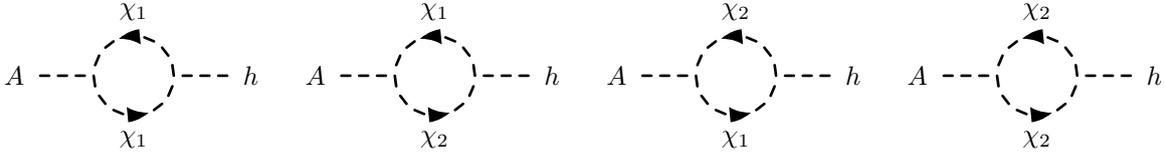
\begin{figure}[htp] 
\centering 
\subfloat{
\begin{fmffile}{App-Pilaftsis-3}
\begin{fmfgraph*}(70,40) 
\fmfset{arrow_len}{3mm} 
\fmfset{arrow_ang}{20} 
\fmfleft{nJ1} 
\fmflabel{$A$}{nJ1} 
\fmfright{nJ2} 
\fmflabel{$h$}{nJ2} 
\fmf{dashes,tension=3}{nJ1,nJ1J1J4} 
\fmf{dashes,tension=3}{nJ2,nJ2J3J2} 
\fmf{scalar,label=$\chi_1$,right=1}{nJ2J3J2,nJ1J1J4} 
\fmf{scalar,label=$\chi_1$,right=1}{nJ1J1J4,nJ2J3J2} 
\end{fmfgraph*} 
\end{fmffile} 
}
\hspace{0.8cm} 
\subfloat{ 
\begin{fmffile}{App-Pilaftsis-4}
\begin{fmfgraph*}(70,40) 
\fmfset{arrow_len}{3mm} 
\fmfset{arrow_ang}{20} 
\fmfleft{nJ1} 
\fmflabel{$A$}{nJ1} 
\fmfright{nJ2} 
\fmflabel{$h$}{nJ2} 
\fmf{dashes,tension=3}{nJ1,nJ1J1J4} 
\fmf{dashes,tension=3}{nJ2,nJ2J2J3} 
\fmf{scalar,label=$\chi_1$,right=1}{nJ2J2J3,nJ1J1J4} 
\fmf{scalar,label=$\chi_2$,right=1}{nJ1J1J4,nJ2J2J3} 
\end{fmfgraph*} 
\end{fmffile} 
} 
\hspace{0.8cm} 
\subfloat{ 
\begin{fmffile}{App-Pilaftsis-5}
\begin{fmfgraph*}(70,40)
\fmfset{arrow_len}{3mm} 
\fmfset{arrow_ang}{20} 
\fmfleft{nJ1} 
\fmflabel{$A$}{nJ1} 
\fmfright{nJ2} 
\fmflabel{$h$}{nJ2} 
\fmf{dashes,tension=3}{nJ1,nJ1J2J3} 
\fmf{dashes,tension=3}{nJ2,nJ2J1J4} 
\fmf{scalar,label=$\chi_1$,right=1}{nJ1J2J3,nJ2J1J4} 
\fmf{scalar,label=$\chi_2$,right=1}{nJ2J1J4,nJ1J2J3} 
\end{fmfgraph*} 
\end{fmffile} 
} 
\hspace{0.8cm} 
\subfloat{ 
\begin{fmffile}{App-Pilaftsis-6}
\begin{fmfgraph*}(70,40) 
\fmfset{arrow_len}{3mm} 
\fmfset{arrow_ang}{20} 
\fmfleft{nJ1} 
\fmflabel{$A$}{nJ1} 
\fmfright{nJ2} 
\fmflabel{$h$}{nJ2} 
\fmf{dashes,tension=3}{nJ1,nJ1J1J4} 
\fmf{dashes,tension=3}{nJ2,nJ2J3J2} 
\fmf{scalar,label=$\chi_2$,right=1}{nJ2J3J2,nJ1J1J4} 
\fmf{scalar,label=$\chi_2$,right=1}{nJ1J1J4,nJ2J3J2} 
\end{fmfgraph*} 
\end{fmffile} 
}
\caption{Feynman diagrams contributing to the one-loop mixing between $A$ and $h$ in the PRTS.}
\label{App-Pilaftsis:fig:myf}
\end{figure}

In the PRTS, to recap what we saw in chapter \ref{Chap-Selec}, the true vevs and the true phases up to one-loop are determined by a modified version of the tree-level tadpole equations;
such version identifies all parameters as bare parameters (except the vevs and the phases, which will be the true ones) and replaces the tree-level tadpoles $t_{\phi}$ by tadpole counterterms $\delta t_{\phi}$.
The latter are such that they cancel the corresponding one-loop tadpoles $T_{\phi}$:%
\fn{Pilaftsis explicitely writes this equation for the mass state $A$ in his eq. 4.8 (in a different notation) \cite{Pilaftsis:1998pe}.}
\be
\delta t_{\phi} = - T_{\phi},
\label{App-Pilaftsis:eq:mytad}
\ee
which are in general different from zero.
The true quantities up to one-loop level (the true vevs $v_1$ and $v_2$ and the true phases $\zeta_1$ and $\zeta_2$) are then those for which the up-to-one-loop tadpole equations,
\bs
\label{App-Pilaftsis:eq:tad-eqs-prime}
\begin{gather}
T_{H_{1}^{\prime}}
=
c_{\zeta_1} v_1 \left(\mu_{1(0)}^2 + \lambda_{1(0)} \, v_1^2 + \dfrac{1}{2} \lambda_{34(0)} \, v_2^2 \right) +
c_{\zeta_2} v_2 \, \mu_{\mathrm{R}(0)}^{2} -
s_{\zeta_2} v_2 \, \mu_{\mathrm{I}(0)}^{2}, \\
T_{H_{2}^{\prime}}
=
c_{\zeta_2} v_2 \left(\mu_{2(0)}^2 + \lambda_{2(0)} \, v_2^2 + \dfrac{1}{2} \lambda_{34(0)} \, v_1^2\right) +
c_{\zeta_1} v_1 \, \mu_{\mathrm{R}(0)}^{2} +
s_{\zeta_1} v_1 \, \mu_{\mathrm{I}(0)}^{2}, \\
T_{A_{1}^{\prime}}
=
s_{\zeta_1} v_1 \left(\mu_{1(0)}^2 + \lambda_{1(0)} \, v_1^2 + \dfrac{1}{2} \lambda_{34(0)} \, v_2^2 \right) +
s_{\zeta_2} v_2 \, \mu_{\mathrm{R}(0)}^{2} +
c_{\zeta_2} v_2 \, \mu_{\mathrm{I}(0)}^{2},
\\
T_{A_{2}^{\prime}} 
=
s_{\zeta_2} v_2 \left(\mu_{2(0)}^2 + \lambda_{2(0)} \, v_2^2 + \dfrac{1}{2} \lambda_{34(0)} \, v_1^2\right) +
s_{\zeta_1} v_1 \, \mu_{\mathrm{R}(0)}^{2} -
c_{\zeta_1} v_1 \, \mu_{\mathrm{I}(0)}^{2},
\end{gather}
\es
are verified. 
Now, the true vevs $v_1$ and $v_2$ and the true phases $\zeta_1$ and $\zeta_2$ do not need to be finite. Actually, they are in general divergent, and thus in general can be renormalized.%
\fn{Recall section \ref{Chap-Selec:sec:CPV}. They only need to be renormalized if they chosen as independent parameters; if they are chosen as dependent parameters, there are two alternatives: either they are replaced (before renormalization) by the parameters they depend on, and the latter are renormalized; or they are renormalized by convenience. See also section \ref{Chap-Selec:sec:conv}.}
It turns out that we will not need the counterterms for the phases; as a consequence, before we expand eqs. \ref{App-Pilaftsis:eq:tad-eqs-prime} in terms of renormalized parameters and counterterms, we can go to the basis where $\zeta_1$ and $\zeta_2$ were rephased away.%
\fn{We checked that the counterterms for the phases are not needed to absorb the divergences of the 2-point functions for the mixing between CP-even and CP-odd states. We did not check other cases; yet, the GFs for the mixing between CP-even and CP-odd states and their renormalizability are the focus of Pilaftsis's (and thus our) discussion; Pilaftsis does not even show the whole model (recall the ellipses in eq. \ref{App-Pilaftsis:eq:Ppot}). Hence, we are not interested in studying the full renormalization of the model, but only the renormalization of the referred mixing.}
In that particular basis, the tadpole equations for the symmetry states read:
\bs
\label{App-Pilaftsis:eq:tad-eqs-prime-particular}
\bea
&&
T_{H_{1}}
=
v_{1}\left(\mu_{1(0)}^{2}+ \mu_{\mathrm{R}(0)}^{2} \dfrac{v_{2}}{v_{1}}+\lambda_{1(0)} v_{1}^{2}+\dfrac{1}{2} \lambda_{34(0)} v_{2}^{2}\right), \\
&&
T_{H_{2}}
=
v_{2}\left(\mu_{2(0)}^{2}+ \mu_{\mathrm{R}(0)}^{2} \dfrac{v_{1}}{v_{2}}+\lambda_{2(0)} v_{2}^{2}+\dfrac{1}{2} \lambda_{34(0)} v_{1}^{2}\right), \\
&&
T_{A_{1}}
=
v_{2} \, \mu_{\mathrm{I}(0)}^{2},
\label{App-Pilaftsis:eq:lo}
\\
&&
T_{A_{2}}
=
-v_{1} \, \mu_{\mathrm{I}(0)}^{2}.
\label{App-Pilaftsis:eq:po}
\eea
\es
We can also derive the tadpole equations for the mass states; we find, in particular:
\be
T_A = - v \, \mu_{\mathrm{I}(0)}^2.
\label{App-Pilaftsis:eq:Tkey}
\ee
Let us consider the different components of this equation in detail.
Due to the CP-violating interactions coming from the $\mathcal{L}_Y^{\chi}$ sector, $T_A$ is such that:
\begin{equation}
i T_A
=
\begin{minipage}[h]{.40\textwidth}
\hspace{4mm}
\vspace{-2mm}
\begin{picture}(0,42)
\begin{fmffile}{App-Pilaftsis-1}
\begin{fmfgraph*}(80,48) 
\fmfset{arrow_len}{3mm} 
\fmfset{arrow_ang}{20} 
\fmfleft{nJ1} 
\fmflabel{$A$}{nJ1} 
\fmfright{nJ2} 
\fmf{dashes,tension=2}{nJ1,nJ1J1J4} 
\fmf{phantom,tension=3}{nJ2,nJ2J3J2} 
\fmf{scalar,label=$\chi_1$,right=1}{nJ2J3J2,nJ1J1J4} 
\fmf{scalar,label=$\chi_1$,right=1}{nJ1J1J4,nJ2J3J2} 
\end{fmfgraph*} 
\end{fmffile}
\end{picture}
\end{minipage}
\hspace{-32mm}
+
\hspace{4mm}
\begin{minipage}[h]{.40\textwidth}
\vspace{2mm}
\begin{picture}(0,42)
\begin{fmffile}{App-Pilaftsis-2}
\begin{fmfgraph*}(80,48) 
\fmfset{arrow_len}{3mm} 
\fmfset{arrow_ang}{20} 
\fmfleft{nJ1} 
\fmflabel{$A$}{nJ1} 
\fmfright{nJ2} 
\fmf{dashes,tension=2}{nJ1,nJ1J1J4} 
\fmf{phantom,tension=3}{nJ2,nJ2J2J3} 
\fmf{scalar,label=$\chi_2$,right=1}{nJ2J2J3,nJ1J1J4} 
\fmf{scalar,label=$\chi_2$,right=1}{nJ1J1J4,nJ2J2J3} 
\end{fmfgraph*} 
\end{fmffile} 
\end{picture}
\end{minipage}
\hspace{-35mm},
\label{App-Pilaftsis:eq:my-tads-drawn}
\end{equation}
which is not zero and divergent.
On the other hand, as we saw, the true vevs $v_1$ and $v_2$ are in general divergent, so that $v=\sqrt{v_1^2+v_2^2}$ is also in general divergent. Hence, similarly to eq. \ref{Chap-Selec:eq:vev-ren-PR}, we define:
\be
v = v_{\mathrm{R}} + v_{\mathrm{R}} \, \delta Z_v,
\label{App-Pilaftsis:eq:my-true-vev}
\ee
where $v_{\mathrm{R}}$ is the renormalized vev. Finally, we can use eq. \ref{App-Pilaftsis:eq:mui-exp} to expand eq. \ref{App-Pilaftsis:eq:Tkey} into:
\be
T_A
=
- \left( v_{\mathrm{R}} + v_{\mathrm{R}} \, \delta Z_v \right) \left( \mu_{\mathrm{I}}^2 + \delta \mu_{\mathrm{I}}^2 \right) 
=
- \left( v_{\mathrm{R}} \, \mu_{\mathrm{I}}^2 + v_{\mathrm{R}} \, \delta \mu_{\mathrm{I}}^2 + \mu_{\mathrm{I}}^2 \, v_{\mathrm{R}} \, \delta Z_v \right),
\label{App-Pilaftsis:eq:19}
\ee
where we neglected second order terms. In that case, we can always fix $\delta \mu_{\mathrm{I}}^2$ to be:
\be
\delta \mu_{\mathrm{I}}^2 = - \dfrac{T_A}{v_{\mathrm{R}}},
\label{App-Pilaftsis:eq:magis}
\ee
in which case eq. \ref{App-Pilaftsis:eq:19} implies:%
\fn{Just as at tree-level (cf. note \ref{App-Pilaftsis:note:the-note}), we assume that the true vev $v$ in eq. \ref{App-Pilaftsis:eq:my-true-vev} is non-zero.}
\be
\mu_{\mathrm{I}}^2 = 0.
\label{App-Pilaftsis:eq:magis-new}
\ee
We will discuss all these results at the end. For now, let us consider the one-loop mixing between $A$ and $h$. The renormalized GF is given by:
\be
i \hat{\Sigma}^{A h}(k^2)
=
i \Sigma^{A h}(k^2) + i \Sigma^{A h}_{\mathrm{CT}},
\label{App-Pilaftsis:eq:Reno}
\ee
where $i \Sigma^{A h}$ is the non-renormalized one-loop 2-point function, which in the PRTS is composed of the Feynman diagrams represented in figure \ref{App-Pilaftsis:fig:myf}.%
$i \Sigma^{A h}_{\mathrm{CT}}$ represents the totality of counterterms for that mixing. Taking the limit $\theta \to 0$ (as Pilaftsis does), we find:%
\fn{It is worth mentioning in passing that \textsc{FeynMaster 2} is the ideal tool to tackle these problems. It is very simple to create the \textsc{FeynMaster} model file of Pilaftsis's model; this automatically generates and draws the complete set of Feynman rules (for both the tree-level interactions and the counterterms). Then, one can in no time calculate one-loop processes (and, of course, obtain the corresponding Feynman diagrams), and ascertain whether the counterterms render them finite.}
\be
i \Sigma^{A h}_{\mathrm{CT}} = i \, c_{\beta} \, \delta \mu^2_{\mathrm{I}},
\label{App-Pilaftsis:eq:my-funny-CT}
\ee
so that, using eq. \ref{App-Pilaftsis:eq:magis}, eq. \ref{App-Pilaftsis:eq:Reno} becomes:
\be
i \hat{\Sigma}^{A h}(k^2) = i \Sigma^{A h} (k^2) - i \, c_{\beta} \dfrac{T_{A}}{v_{\mathrm{R}}}.
\label{App-Pilaftsis:eq:Reno2}
\ee
Finally, using again  \textsc{FeynMaster 2}, it is very easy to show that:%
\fn{We used the relations of the theory to replace the masses $M_1$ and $M_2$ by parameters of the potential and mixing angles.}
\be
i \Sigma^{A h} \big|_{\varepsilon}
=
-\dfrac{i}{8 \pi^2}  \dfrac{1}{\varepsilon} \operatorname{Im}(g) \Big[ c_{\beta} f +  s_{\beta} \, g \Big],
\qquad
T_A \big|_{\varepsilon}
=
-\dfrac{1}{8 \pi^2} \dfrac{1}{\varepsilon} \dfrac{v_{\mathrm{R}}}{c_{\beta}} \operatorname{Im}(g) \Big[ c_{\beta} f +  s_{\beta} \, g \Big],
\ee
where $|_{\varepsilon}$ represents as usual the terms proportional to $1/\varepsilon$, so that:
\be
\hat{\Sigma}^{A h} \big|_{\varepsilon} = 0.
\label{App-Pilaftsis:eq:end}
\ee
Therefore, the process is renormalized.
Following a similar reasoning, we would also conclude that the remaning 2-point functions for the mixing between CP-even and CP-odd states (namely, $\hat{\Sigma}^{A H}$, $\hat{\Sigma}^{G_0 h}$ and $\hat{\Sigma}^{G_0 H}$) are finite as well.

\subsection{Discussion}

There are several aspects that should be mentioned.
First and foremost, CP no longer is a symmetry of the potential.
Recall that, at tree-level, we were able to choose a basis where the $H_i$ fields did not mix with the $A_i$ fields. Hence, we were able to identify the former as CP-even and the latter as CP-odd, and we rotated the former to $h$ and $H$, and the latter to $A$ and $G_0$. So, all these states had well-defined CP value.
However, when we considered the theory up to one-loop level, we found out that such scenario no longer holds, because there is a mixing between the purported CP-odd state $A$ and the purported CP-even state $h$. In fact, although that mixing is finite (eq. \ref{App-Pilaftsis:eq:end}), it is in general non-zero.

At this point, one might be led to think that maybe we just picked the wrong basis; in other words, maybe there is another basis where there are no phases.
But this is not possible, for the simple reason that all the parameters in the potential are already real. Indeed, in eqs. \ref{App-Pilaftsis:eq:tad-eqs-prime-particular} we chose to rephase $\zeta_1$ and $\zeta_2$ away; then, in eq. \ref{App-Pilaftsis:eq:magis} we chose the renormalization condition for $\delta \mu_{\mathrm{I}}^2$ in such a way that the renormalized parameter $\mu_{\mathrm{I}}^2$ vanished, eq. \ref{App-Pilaftsis:eq:magis-new}. Hence, according to our choices for the renormalized up-to-one-loop theory, there are no phases in the potential.%
\fn{We are not referring to the effective potential (which includes loop effects), but the the potential before the inclusion of loop effects.}
Had we chosen another basis for the parameters, there would be phases.

With our choices, then, if we just take into account the tree-level interactions, there are no bilinear interactions between CP-odd and CP-states. However, and as we suggested in chapter \ref{Chap-Selec}, what defines the states is no longer just the tree-level terms, but the effective theory. This includes, besides the tree-level interactions, also the one-loop interactions. And what we find is that, when we include these elements, there is a mixing between CP-odd and CP-states. We insist that, had we chosen another basis for the parameters, this mixing would be verified already the tree-level.

The impossibility to define CP in the effective potential comes from the fact that, at one-loop level, the CP-violating interactions from the $\mathcal{L}_Y^{\chi}$ sector contaminate the interactions of the potential (as we saw in eq. \ref{App-Pilaftsis:eq:my-tads-drawn} and in figure \ref{App-Pilaftsis:fig:myf}). In the tree-level description of section \ref{App-Pilaf:sec:tree}, there was no contamination, so that we could choose a basis where CP was well-defined in the potential; in the up-to-one-loop theory, by contrast, that is no longer possible. This does not mean that the theory is unsound; it only means that the particular basis which was considered at tree-level and where CP could be well-defined no longer exists, so that the states no longer can be given a CP value.

This leads us to another aspect, related to the imaginary part of $\mu^2$. At tree-level, we found that, in the basis where the phases $\zeta_1$ and $\zeta_2$ were rephased away, $\mu_{\mathrm{I}}^{2}$ vanished (eq. \ref{App-Pilaftsis:eq:joker}).
When going to one-loop level, we noted that $\mu_{\mathrm{I}}^{2}$ needed in general to be renormalized, so that we identified it with a bare parameter, which was equated to a renormalized parameter and a counterterm (eq. \ref{App-Pilaftsis:eq:mui-exp}). Then, although the up-to-one-loop true phases $\zeta_1$ and $\zeta_2$ could in general be renormalized, we decided to rephase them away (so that no renormalized phases nor their counterterms show up). Even despite that decision, we found that $\mu_{\mathrm{I}(0)}^{2}$ was not zero, because the CP-violating interactions from the $\mathcal{L}_Y^{\chi}$ sector generate one-loop tadpoles in the CP-odd fields (eqs. \ref{App-Pilaftsis:eq:Tkey} and \ref{App-Pilaftsis:eq:my-tads-drawn}). This means, in particular, that $\mu_{\mathrm{I}(0)}^{2}$ is divergent; therefore, even if we choose the finite parts in such a way that the renormalized parameter $\mu_{\mathrm{I}}^{2}$ vanishes (as we did), the counterterm $\delta \mu_{\mathrm{I}}^2$ does not vanish.

The circumstance that it does not vanish is the key to understand the renormalizability of the one-loop mixing between CP-even and CP-odd states. Indeed, $\delta \mu_{\mathrm{I}}^2$ is the only counterterm that contributes to that mixing (eq. \ref{App-Pilaftsis:eq:my-funny-CT}). It turns out that, by virtue of the up-to-one-loop tadpole equations in the basis where $\zeta_1$ and $\zeta_2$ were rephased away, $\delta \mu_{\mathrm{I}}^2$ can be chosen as proportional to $T_A$ only (eq. \ref{App-Pilaftsis:eq:magis}). As a consequence, the counterterm for the one-loop mixing between $A$ and $h$ (the last term of eq. \ref{App-Pilaftsis:eq:Reno2}) is proportional to the one-loop tadpole of $A$. In the end, then, $T_A$ `saves the day', but only due to a set of three reasons:
\begin{center}
\begin{addmargin}[8mm]{0mm}
i) the parameter $\mu^2$ is in general complex, so that $\mu^2_{\mathrm{I}}$ is renormalized,\\
ii) its counterterm ($\delta \mu^2_{\mathrm{I}}$) is the only one contributing to $\Sigma^{A h}_{\mathrm{CT}}$, \\
iii) that counterterm is intimately related to $T_A$ due to the tadpole equations.
\end{addmargin}
\end{center}
In the context of the usual description of renormalization, this is the explanation for why the Lagrangian ``fortunately (...) provides the necessary renormalization constant'' \cite{Pilaftsis:1998pe}.
It also shows that the model is theoretically sound, in the sense that the divergences that show up at one-loop level can be removed by the counterterms. Finally, note that although in general there are counterterms for the true phases, as well as field counterterms for the mixing between CP-odd and CP-even states, we did not need to consider them, as $\delta \mu_{\mathrm{I}}^{2}$ ensures that those mixings are finite.

%% file: Appendices/OSS.tex
\chapter{OSS and dependent masses}
\label{App-OSS}

\vs{-5mm}

\n The on-shell subtraction (OSS) scheme, originally proposed in ref.~\cite{Ross:1973fp}, has become one of the most common subtraction schemes in the renormalization of gauge theories, having been recurrently described and used by many authors \cite{Aoki:1982ed,Hollik:1988ii,Bohm:1986rj,Denner:1991kt,Schwartz:2013pla,Grimus:2016hmw,Denner:2019vbn}.
In this appendix, we present a clear and simple description of the one-loop renormalization conditions that characterize the OSS scheme, explaining how they allow to derive the mass and field counterterms. We consider both the scenario with no mixing of fields, as well as that with mixing; we also study the scenarios where a dependent mass prevents the application of the OSS conditions, and we show how the counterterms are determined in those cases.
This appendix can thus be seen as an introduction to sections \ref{Chap-Reno:sec:CT-gauge} to \ref{Chap-Reno:sec:CT-scalar} above.%

\n We will assume scalars particles, for simplicity; in the case of fermions and gauge bosons, the 2-point functions must be projected onto (i.e. multiplied by) physical states: the spinor $u(p)$ for fermions and the polarization vector $\varepsilon_{\mu}(k)$ for gauge bosons.

\section{No mixing, independent mass}
\label{App-OSS:sec-first}

\n We start with the simplest case: a generic single scalar neutral field of a generic theory.
Suppose a real scalar field $\phi$ with mass $m$, such that:
\be
\mathcal{L} \ni - \dfrac{1}{2} \phi \Box \phi - \dfrac{1}{2} m^2 \phi^2.
\label{App-OSS:eq:LagToy}
\ee
We assume that $m$ is an independent parameter, i.e. that it is not written in terms of other parameters.
When considering the theory up to one-loop level, we follow the usual procedure, namely: we identify the tree-level quantities as bare ones, and split them into renormalized quantities and counterterms,
\be
m_{(0)}^2 = m_{\mathrm{R}}^2 + \delta m^2,
\quad
\phi_{(0)} = \phi + \dfrac{1}{2} \delta Z_{\phi} \phi.
\label{App-OSS:eq:ExpanToy}
\ee
Note that we explicitely represent the renormalized mass $m_{\mathrm{R}}$ with an index $\mathrm{R}$. Moreover, no mixed fields counterterms are introduced, since we are assuming that $\phi$ mixes with no other field at one-loop.
Identifying the up-to-one-loop 2-point function with $\Gamma$ and the one-loop 2-point function with $\Sigma$, we have:%
\fn{Since the 2-point GFs can only depend on the squared momentum $p^2$ (as we are considering scalar functions), we write it explicitly in their argument (instead of just $p$).}
\be
\hat{\Gamma}(p^2) = p^2 - m_{\mathrm{R}}^2 + \hat{\Sigma}(p^2),
\label{App-OSS:eq:OSaux1}
\ee
where the caret means `renormalized'. Then, expanding the bare version of eq. \ref{App-OSS:eq:LagToy} with eq. \ref{App-OSS:eq:ExpanToy}, we find:%
\fn{Recall that, as seen in eq. \ref{Chap-Reno:eq:FJTS-2p}, the non-renormalized one-loop 2-point function $\Sigma(p^2)$ in general includes diagrams with one-loop tadpoles if the FJTS is adopted.}
\be
\hat{\Sigma}(p^2) = \Sigma(p^2) + (p^2 - m_{\mathrm{R}}^2) \delta Z_{\phi} - \delta m^2.
\label{App-OSS:eq:OSaux2}
\ee
Now, the  pole mass $m_{\mathrm{P}}$ is defined as the value of the momentum for which $\hat{\Gamma}(p^2)$ is zero, that is:%
\fn{We are using the definition of \textit{real} pole mass (also known as Breit-Wigner mass), where $m_{\mathrm{P}}$ is real as a consequence of the fact that we are considering the real part of $\hat{\Sigma}(m_{\mathrm{P}}^2)$. This is what is done in the traditional OSS scheme \cite{Aoki:1982ed, Denner:2019vbn}.
We are using the operator $\widetilde{\operatorname{Re}}$ (and not $\mathrm{Re}$), which neglects the imaginary---also called absorptive---parts of loop integrals, while keeping the imaginary parts of complex parameters. The use of $\widetilde{\operatorname{Re}}$ in the renormalization conditions is also common practice in the traditional OSS scheme.
%
}
\be
\widetilde{\operatorname{Re}} \, \hat{\Gamma}(m_{\mathrm{P}}^2) = m_{\mathrm{P}}^2 - m_{\mathrm{R}}^2 + \widetilde{\operatorname{Re}} \,  \hat{\Sigma}(m_{\mathrm{P}}^2) = 0.
\label{App-OSS:eq:def:polemass}
\ee
This allows us to see how the OSS conditions fix the counterterms $\delta m^2$ and $\delta Z_{\phi}$. In the case we are considering (a single field), there are only two such conditions. The first one simply states that the renormalized mass is equal to the pole mass,
\be
m_{\mathrm{R}} \stackrel{\mathrm{OSS}}{=} m_{\mathrm{P}},
\label{App-OSS:eq:my-first-OSS}
\ee
thus giving a physical meaning to the renormalized mass.
In that case, eq. \ref{App-OSS:eq:def:polemass} implies:
\be
\widetilde{\operatorname{Re}} \, \hat{\Gamma}(m_{\mathrm{R}}^2) \stackrel{\mathrm{OSS}}{=} 0,
\label{App-OSS:eq:OSSfirst}
\ee
which, together with eqs. \ref{App-OSS:eq:OSaux1} and \ref{App-OSS:eq:OSaux2}, fixes $\delta m^2$ to:
\be
\delta m^2 \stackrel{\mathrm{OSS}}{=} \widetilde{\operatorname{Re}} \, \Sigma(m_{\mathrm{R}}^2).
\label{App-OSS:eq:dm-OSS}
\ee
One should not confuse eqs. \ref{App-OSS:eq:def:polemass} and \ref{App-OSS:eq:OSSfirst}: the former is the definition of the pole mass, valid in every subtraction scheme.
By contrast, eq. \ref{App-OSS:eq:OSSfirst} is specific of OSS, and results from the application of the OSS condition \ref{App-OSS:eq:my-first-OSS} to eq. \ref{App-OSS:eq:def:polemass}.%
%

\n The second OSS condition takes the (renormalized, real) residue of the renormalized propagator at the pole mass,
\be
\hat{R}
\equiv
\lim_{p^2 \to m_{\mathrm{P}}^2} (-i) (p^2 - m_{\mathrm{P}}^2) \, \widetilde{\operatorname{Re}} \, \hat{G}(p^2)
=
\lim_{p^2 \to m_{\mathrm{P}}^2} (-i) (p^2 - m_{\mathrm{P}}^2) \, \widetilde{\operatorname{Re}} \left[ i \hat{\Gamma}^{-1}(p^2) \right],
\label{App-OSS:eq:resi}
\ee
and sets it to one. Actually, one usually sets the inverse of the residue equal to one, which means:%
\fn{In this equation, $m_{\mathrm{P}}$ and $m_{\mathrm{R}}$ can be used indifferently, as they are chosen to be the same in OSS.}
\be
\lim_{p^{2} \to m_{\mathrm{R}}^2} \frac{1}{p^2 - m_{\mathrm{R}}^{2}} \widetilde{\operatorname{Re}} \, \hat{\Gamma}(p^2) \stackrel{\mathrm{OSS}}{=} 1.
\ee
Then, using eqs. \ref{App-OSS:eq:OSaux1} and \ref{App-OSS:eq:OSaux2}, as well as L'Hôpital's rule, the expression for $\delta Z_{\phi}$ follows:
\be
\delta Z_{\phi} \stackrel{\mathrm{OSS}}{=} -\left.\widetilde{\operatorname{Re}} \, \frac{\partial \Sigma(p^2)}{\partial p^{2}}\right|_{p^{2}=m_{\mathrm{R}}^{2}}.
\label{App-OSS:eq:dZ-OSS}
\ee

\section{No mixing, dependent mass}
\label{App-OSS:sec:dep-mass}

\n If we suppose instead that the bare mass $m_{(0)}$ is a dependent parameter, then both the renormalized mass $m_{\mathrm{R}}$ and the counterterm $\delta m^2$ are also dependent, i.e. fixed.%
\fn{The case of dependent masses is common in the minimal supersymmetric extension of the SM; see e.g. ref. \cite{Diaz:1991ki}.}
Now, since $m_{\mathrm{R}}$ is fixed, it cannot be set to be equal to the pole mass $m_{\mathrm{P}}$, so that the first OSS condition cannot be used. In that case, then, the two masses will in general be different, and eq. \ref{App-OSS:eq:def:polemass} (which is always valid) ensures that their difference is of one-loop order. But this means that, to first order in perturbation theory, $m_{\mathrm{P}}^2$ can be replaced by $m_{\mathrm{R}}^2$ when $m_{\mathrm{P}}^2$ is an argument of a one-loop function. In particular, the relation
\be
\widetilde{\operatorname{Re}} \, \hat{\Sigma}(m_{\mathrm{P}}^2) = \widetilde{\operatorname{Re}} \, \hat{\Sigma}(m_{\mathrm{R}}^2)
\ee
in valid to first order, so that eqs. \ref{App-OSS:eq:def:polemass} and \ref{App-OSS:eq:OSaux2} imply, to that order,
\be
m_{\mathrm{P}}^2 = m_{\mathrm{R}}^2 - \widetilde{\operatorname{Re}} \, \Sigma(m_{\mathrm{R}}^2) + \delta m^2.
\label{App-OSS:eq:def:polemass2}
\ee
Hence, the pole mass is completely determined.%
\fn{Note that $m_{\mathrm{P}}^2$ is finite, as it should be. 
A simple way to see this is to recall that the divergent parts of counterterms must be the same whichever the subtraction scheme chosen. Then, the divergent part of the l.h.s. of eq. \ref{App-OSS:eq:dm-OSS} is equal to that of the r.h.s., for all subtraction schemes. Therefore, the divergent parts of the r.h.s. of eq. \ref{App-OSS:eq:def:polemass2} cancel.}

\n Just as in the case of the independent mass, $\delta Z_{\phi}$ can be fixed by setting the residue equal to one. Only, one must be careful with the fact that $m_{\mathrm{P}}^2$ is different from $m_{\mathrm{R}}^2$. Inserting eqs. \ref{App-OSS:eq:OSaux1} and \ref{App-OSS:eq:OSaux2} inside eq. \ref{App-OSS:eq:resi}, equating the latter to one, using L'Hôpital's rule and the relation
\be
\left.\widetilde{\operatorname{Re}} \, \frac{\partial \hat{\Sigma}(p^2)}{\partial p^{2}}\right|_{p^{2}=m_{\mathrm{P}}^{2}}
=
\left.\widetilde{\operatorname{Re}} \, \frac{\partial \hat{\Sigma}(p^2)}{\partial p^{2}}\right|_{p^{2}=m_{\mathrm{R}}^{2}},
\ee
which holds to first order, we find:
\be
\delta Z_{\phi}
=
-\left.\widetilde{\operatorname{Re}} \, \frac{\partial \Sigma(p^2)}{\partial p^{2}}\right|_{p^{2}=m_{\mathrm{R}}^{2}},
\ee
which is precisely the same as the field counterterm fixed through OSS in the case of the independent mass, eq. \ref{App-OSS:eq:dZ-OSS}.

\section{Mixing, all masses independent}
\label{Chap-OSS:sec:mix-ind}

\n We now consider a generalization of eqs. \ref{App-OSS:eq:LagToy} to \ref{App-OSS:eq:OSaux2} for two fields, $\phi_1$ and $\phi_2$:
\begin{gather}
\mathcal{L} \ni
- \dfrac{1}{2} \phi_1 \Box \phi_1 - \dfrac{1}{2} m_1^2 \phi_1^2
- \dfrac{1}{2} \phi_2 \Box \phi_2 - \dfrac{1}{2} m_2^2 \phi_2^2,
\label{App-OSS:eq:LagToy2}
\\
m_{1(0)}^2 = m_{1\mathrm{R}}^2 + \delta m_1^2,
\quad
m_{2(0)}^2 = m_{2\mathrm{R}}^2 + \delta m_2^2, \\
\begin{pmatrix}
\phi_{1(0)} \\ \phi_{2(0)}
\end{pmatrix}
=
\begin{pmatrix}
1 + \dfrac{1}{2} \delta Z_{11} & \dfrac{1}{2} \delta Z_{12} \\
\dfrac{1}{2} \delta Z_{21} & 1 + \dfrac{1}{2} \delta Z_{22}
\end{pmatrix}
\begin{pmatrix}
\phi_1 \\ \phi_2
\end{pmatrix},
\label{App-OSS:eq:ExpanToy2}
\\
\hat{\Gamma}_{ij} (p^2) =
\begin{pmatrix}
p^2 - m_{1\mathrm{R}}^2 + \hat{\Sigma}_{11}(p^2) &
\hat{\Sigma}_{12}(p^2) \\
\hat{\Sigma}_{21}(p^2) & p^2 - m_{2\mathrm{R}}^2 + \hat{\Sigma}_{22}(p^2)
\end{pmatrix},
\label{App-OSS:eq:OSaux3}
\\
\hat{\Sigma}_{ij}(p^2) = \Sigma_{ij}(p^2) + \frac{1}{2}\left(p^{2}-m_{j\mathrm{R}}^{2}\right) \delta Z_{ji}
+ \frac{1}{2}\left(p^{2}-m_{i\mathrm{R}}^{2}\right) \delta Z_{ij}
-\delta_{ij} \, \delta m_{i}^{2},
\label{App-OSS:eq:OSaux4}
\end{gather}
where $i,j = \{1,2\}$, and where both $m_{1(0)}$ and $m_{2(0)}$ are assumed to be independent parameters.
In OSS, the counterterms are once again calculated by reference to the pole masses and the residues.

\n But, now that there is mixing, what is the definition of the pole masses and the residues?
Because $\hat{\Gamma}(p^2)$ is now a matrix (eq. \ref{App-OSS:eq:OSaux3}), the pole masses are determined as the values of the momentum for which its eigenvalues are zero.
And while that leads to complicated expressions, these can be simplified by fixing the off-diagonal field counterterms in an appropriate way. More specifically, if we choose those counterterms to be such that:
\be
\widetilde{\operatorname{Re}} \, \hat{\Gamma}_{ij} (m_{i\mathrm{P}}^2)
\stackrel{j \neq i}{=}
0,
\qquad
\widetilde{\operatorname{Re}} \, \hat{\Gamma}_{ji} (m_{i\mathrm{P}}^2)
\stackrel{j \neq i}{=}
0,
\label{App-OSS:eq:aux3OSS}
\ee
then the definition of the pole masses $m_{i\mathrm{P}}$ is a trivial generalization of eq. \ref{App-OSS:eq:def:polemass}, namely,
\be
\widetilde{\operatorname{Re}} \, \hat{\Gamma}_{ii}(m_{i\mathrm{P}}^2) = m_{i\mathrm{P}}^2 - m_{i\mathrm{R}}^2 + \widetilde{\operatorname{Re}} \,  \hat{\Sigma}_{ii}(m_{i\mathrm{P}}^2) = 0,
\label{App-OSS:eq:def:polemass-i}
\ee
in which case the residues become:
\be
\hat{R}_i
=
\lim_{p^2 \to m_{i\mathrm{P}}^2} (p^2 - m_{i\mathrm{P}}^2) \, \widetilde{\operatorname{Re}} \, \hat{\Gamma}_{ii}^{-1}(p^2).
\label{App-OSS:eq:def:residue-i}
\ee
Eq. \ref{App-OSS:eq:aux3OSS} ensures that OS particles do not mix with each other, and allows to calculate the off-diagonal field counterterms ($\delta Z_{12}$ and $\delta Z_{21}$) through eqs. \ref{App-OSS:eq:OSaux3} and \ref{App-OSS:eq:OSaux4}. 
Now, OSS assumes not only eq. \ref{App-OSS:eq:aux3OSS}, but also that the renormalized masses are equal to the pole masses, $m_{i\mathrm{R}} \stackrel{\mathrm{OSS}}{=} m_{i\mathrm{P}}$ (which we can always do, since the renormalized masses $m_{i\mathrm{R}}$ are assumed to be free parameters). In that case, eq. \ref{App-OSS:eq:aux3OSS} becomes:
\be
\widetilde{\operatorname{Re}} \, \hat{\Gamma}_{ij} (m_{i\mathrm{R}}^2)
\stackrel{j \neq i, \, \mathrm{OSS}}{=}
0,
\qquad
\widetilde{\operatorname{Re}} \, \hat{\Gamma}_{ji} (m_{i\mathrm{R}}^2)
\stackrel{j \neq i, \, \mathrm{OSS}}{=}
0,
\label{App-OSS:eq:aux3OSS-OSS}
\ee
while eq. \ref{App-OSS:eq:def:polemass-i} becomes:
\be
\widetilde{\operatorname{Re}} \, \hat{\Gamma}_{ii}(m_{i\mathrm{R}}^2) \stackrel{\mathrm{OSS}}{=} 0,
\label{App-OSS:eq:OS-mixing-2}
\ee
which allows to determine the mass counterterms. Besides, and also in a trivial generalization of the case with no mixing, OSS fixes the diagonal field counterterms by requiring the different inverse residues to be equal to one:
\be
\lim_{p^{2} \to m_{i\mathrm{R}}^2} \frac{1}{p^2 - m_{i\mathrm{R}}^{2}} \widetilde{\operatorname{Re}} \, \hat{\Gamma}_{ii}(p^2) \stackrel{\mathrm{OSS}}{=} 1.
\label{App-OSS:eq:OS-mixing-3}
\ee
The expressions for the complete set of mass and field counterterms are, thus,
\be
\delta m_i^2 \stackrel{\mathrm{OSS}}{=} \widetilde{\operatorname{Re}} \, \Sigma_{ii}(m_{i\mathrm{R}}^2),
\quad
\delta Z_{ij} \stackrel{j \neq i, \, \mathrm{OSS}}{=} 2 \, \frac{\widetilde{\operatorname{Re}}  \, \Sigma_{ij}(m_{j\mathrm{R}}^{2})}{m_{i\mathrm{R}}^2 - m_{j\mathrm{R}}^2},
\quad
\delta Z_{ii} \stackrel{\mathrm{OSS}}{=} -\left.\widetilde{\operatorname{Re}} \, \frac{\partial \Sigma_{ii}(p^2)}{\partial p^{2}}\right|_{p^{2}=m_{i\mathrm{R}}^{2}}.
\label{App-OSS:eq:total-set-OSS}
\ee

\section{Mixing, one dependent mass}

\n Finally, we consider the situation where $m_{2(0)}$ is a dependent parameter (while $m_{1(0)}$ is still independent). It follows that $m_{2\mathrm{R}}^2$ and $\delta m_2^2$ are dependent. Nonetheless, we can always impose eq. \ref{App-OSS:eq:aux3OSS}, which not only fixes the counterterms $\delta Z_{12}$ and $\delta Z_{21}$, but also implies that eqs. \ref{App-OSS:eq:def:polemass-i} and \ref{App-OSS:eq:def:residue-i} are valid. Now, while we have the freedom to choose $m_{1\mathrm{R}} \stackrel{\mathrm{OSS}}{=} m_{1\mathrm{P}}$ (so that eqs. \ref{App-OSS:eq:OS-mixing-2} and \ref{App-OSS:eq:OS-mixing-3} hold for $i=1$, and these respectively allow to calculate $\delta m_1^2$ and $\delta Z_{11}$), we do not have the freedom to fix $m_{2\mathrm{R}}^2$.
In this case, though, we have a trivial generalization of section \ref{App-OSS:sec:dep-mass}: the expression for $m_{2\mathrm{P}}^2$ follows from eq. \ref{App-OSS:eq:def:polemass-i}, 
while $\delta Z_{22}$ can be determined by fixing the residue of the OS propagator of $\phi_2$ to one.%
\fn{As a consequence, the LSZ factors become trivial; cf. appendix \ref{App-LSZ}.}
The expressions then become:
\begin{gather}
m_{2\mathrm{P}}^2 = m_{\mathrm{2R}}^2 - \widetilde{\operatorname{Re}} \, \Sigma_{22}(m_{2\mathrm{R}}^2) + \delta m_2^2, \\
\delta m_1^2 = \widetilde{\operatorname{Re}} \, \Sigma_{11}(m_{1\mathrm{R}}^2),
\quad
\delta Z_{ij} \stackrel{j \neq i}{=} 2 \, \frac{\widetilde{\operatorname{Re}}  \, \Sigma_{ij}(m_{j\mathrm{R}}^{2})}{m_{i\mathrm{R}}^2 - m_{j\mathrm{R}}^2},
\quad
\delta Z_{ii} = -\left.\widetilde{\operatorname{Re}} \, \frac{\partial \Sigma_{ii}(p^2)}{\partial p^{2}}\right|_{p^{2}=m_{i\mathrm{R}}^{2}},
\label{App-OSS:eq:total-set}
\end{gather}
where we neglected higher order terms.
%
%
Hence, 
eq. \ref{App-OSS:eq:total-set} only differs from eq. \ref{App-OSS:eq:total-set-OSS} in the fact that $\delta m_2^2$ is \textit{a priori} fixed.

\n The generalization to the case with four fields with one dependent mass (as in section \ref{Chap-Reno:sec:CT-scalar} above) is straightforward.

%% file: Appendices/LSZ.tex
\chapter{The LSZ reduction formula revisited}
\label{App-LSZ}

\vs{-5mm}

\n The Lehmann–Symanzik–Zimmermann (LSZ) \cite{Lehmann:1954rq} reduction formula is one of the most fundamental expressions in quantum field theory, establishing a relation between $S$-matrix elements and GFs.
So central is the role it plays that it is presented in virtually every textbook on quantum field theory. However, such presentation is usually done at the tree-level only, and scarcely applied to the renormalized theory as well. In particular, and to the best of our knowledge, no textbook really discusses the specific case (inside the renormalized theory) of field mixing.%
\fn{This does not mean that there is no literature discussing non-trivial consequences of the LSZ reduction formula (in particular, non-trivial renormalized LSZ factors); such discussion is common e.g. in the cMSSM \cite{Frank:2006yh}.}

\n In this appendix, we present a very brief description of this case, considered only up to one-loop level. We show that, although field mixing in general leads to a complicated expression for the LSZ formula, a simple renormalization condition applied to the off-diagonal elements of one-loop bilinear interactions implies a simple expression. As we will show, such renormalization condition can be applied even to fields whose renormalized masses are dependent.

\section{Tree-level}

\n We start by discussing the formula applied to the tree-level case. We consider the case for external scalar fields, following ref. \cite{Bohm:2001yx} closely.

\n If we define the GF in coordinate space as in eq. \ref{Chap-Selec:eq:GF-vev}, 
%
%
%
then the LSZ formula states that the $S$-matrix element for $s$ incoming particles (with momenta $p_1, ..., p_s$) and $n-s$ outgoing particles (with momenta $-p_{s+1}, ..., -p_n$) is given by:
\be
\left\langle -p_{s+1}, ... , -p_{n}  | S | p_{1}, ... ,  p_{s}\right\rangle
=
R_{1}^{-1/2} ... \, \, R_{n}^{-1/2} \, \, (-i)^n \, \left(p_1^2 - m_{1}^2\right) ... \left( p_n^2 - m_{n}^2\right) \, \mathcal{G} (p_1, ... , p_n) \Big|_{p_i^2 = m_{i}^2},
\label{App-LSZ:eq:LSZ-original}
\ee
where $\mathcal{G} (p_1, ... , p_n)$ is defined as the Fourier transform of $G^{\phi_1...\phi_n}(x_1, ... , x_n)$, that is,
\bea
\mathcal{G} (p_1, ... , p_n) 
&\equiv&
\int d^4x_1 ... \, d^4x_n \, e^{-i (p_1 x_1 + ... + p_n x_n)} G^{\phi_1...\phi_n}(x_1, ... , x_n) \no
&=&
(2 \pi)^4 \delta^{(4)} (p_1 + ... + p_n) \, G(p_1, ... , p_n),
\eea
and where the different $R_j$ ($j=1, ..., n$) are the wave-function renormalization constants, also known as \textit{LSZ factors} (or Z-factors).
These are given by the residues of the propagators at the corresponding mass,
\be
R_j
=
i^{-1} \left(p_j^2 - m_j^2\right)
G(p_j, -p_j)
\big|_{p_j^2 = m_j^2}.
\ee

\n Moreover, if we define the \textit{amputated} GF as:%
\fn{Also known as \textit{truncated }GF \cite{Bohm:2001yx,Denner:2019vbn}.}
\be
\mathcal{G}_{\mathrm{ampu.}} (p_1, ... , p_n)
=
G^{-1}(p_1, -p_1) ... \,  G^{-1}(p_n, -p_n) \, \, \mathcal{G}(p_1, ... , p_n),
\label{App-LSZ:eq:ampu-GF}
\ee
we can re-write the LSZ reduction formula as:
\be
\left\langle -p_{s+1}, ... , -p_{n}  | S | p_{1}, ... ,  p_{s}\right\rangle
=
R_{1}^{1/2} ... \, \, R_{n}^{1/2} \, \, \mathcal{G}_{\mathrm{ampu.}} (p_1, ... , p_n) \Big|_{p_i^2 = m_{i}^2}.
\label{App-LSZ:eq:LSZ-simple}
\ee
The amputated GF is just the normal GF, but without the external propagators---which are removed (or amputated) by the factors $G^{-1}(p_1, -p_1) ... \,  G^{-1}(p_n, -p_n)$. Then, the LSZ reduction formula of eq. \ref{App-LSZ:eq:LSZ-simple} says that the $S$-matrix elements is obtained by considering the amputated GF with OS external fields and by multiplying it by the square root of product of the LSZ factors.

\section{Renormalized theory}

\n We now consider the case of a renormalized theory. As mentioned above, we restrict ouselves to the theory up to one-loop level. We start with the simple scenario where fields do not mix at loop level, and consider afterwards the one where they are allowed to mix.

\subsection{No mixing}

\n The LSZ reduction formula for the renormalized theory is a trivial generalization of eq. \ref{App-LSZ:eq:LSZ-simple}, namely, 
\be
\left\langle -p_{s+1}, ... , -p_{n}  | S | p_{1}, ... ,  p_{s}\right\rangle
=
\hat{R}_{1}^{1/2} ... \, \, \hat{R}_{n}^{1/2} \, \, \hat{\mathcal{G}}_{\mathrm{ampu.}} (p_1, ... , p_n) \Big|_{p_i^2 = m_{i\mathrm{P}}^2},
\label{App-LSZ:eq:Sm-reno-no-mix}
\ee
where $\hat{R}_j$ are the renormalized LSZ factors, which read
\be
\hat{R}_j
=
i^{-1} \left(p_j^2 - m_{j\mathrm{P}}^2\right)
\hat{G}(p_j, -p_j)
\Big|_{p_j^2 = m_{j\mathrm{P}}^2},
\label{App-LSZ:eq:reno-resi}
\ee
and $\hat{\mathcal{G}}_{\mathrm{ampu.}}$ represents the renormalized amputated GF, which is such that:
\be
\hat{\mathcal{G}}_{\mathrm{ampu.}} (p_1, ... , p_n) \Big|_{p_i^2 = m_{i\mathrm{P}}^2}
=
\hat{R}_{1}^{-1} ... \, \, \hat{R}_{n}^{-1} \, \, (-i)^n \, \left(p_1^2 - m_{1\mathrm{P}}^2\right) ... \left( p_n^2 - m_{n\mathrm{P}}^2\right) \, \hat{\mathcal{G}} (p_1, ... , p_n) \Big|_{p_i^2 = m_{i\mathrm{P}}^2}.
\label{App-LSZ:eq:G-reno-ampu}
\ee
Note that the masses that appear in these equations are the physical masses, $m_{i\textrm{P}}$; these correspond to the pole masses (i.e. the values of the momentum for which the complete propagator at stake has a pole) and are in general different from the renormalized masses $m_{i\textrm{R}}$ (recall section \ref{App-OSS:sec-first}).
Moreover, $\hat{\mathcal{G}}$ in the r.h.s. of eq. \ref{App-LSZ:eq:G-reno-ampu} reads:
%
%
\be
\hat{\mathcal{G}} (p_1, ... , p_n) 
=
\int d^4x_1 ... \, d^4x_n \, e^{-i (p_1 x_1 + ... + p_n x_n)}
\langle \Omega | T \{ \phi_{1}(x_1) ... \, \phi_{n}(x_n) \} \Omega \rangle,
\label{App-LSZ:eq:G-reno-physical}
\ee
%
%
%
%
where $\phi_{1}(x_1), ..., \, \phi_{n}(x_n)$ correspond to the renormalized fields, characterized by the renormalized masses $m_{1\mathrm{R}}, ..., m_{n\mathrm{R}}$. 
It is worth stressing that, just as at tree-level the external propagators of $\mathcal{G}$ in eq. \ref{App-LSZ:eq:LSZ-original} were removed by the remaining factors of the r.h.s of that equation, so the external propagators of $\hat{\mathcal{G}}$ are removed by the remaining factors of the r.h.s. of eq. \ref{App-LSZ:eq:G-reno-ampu}.%
\fn{While in the first case (eq. \ref{App-LSZ:eq:LSZ-original}) there were loops nowhere, in the second case (eq. \ref{App-LSZ:eq:G-reno-ampu}) there are loops everywhere: in the residues $\hat{R}$, in the pole masses $m_{\mathrm{P}}$ and in the GF $\hat{\mathcal{G}}$.
Note that Schwartz \cite{Schwartz:2013pla} makes an incorrect description of this topic in section 18.3.2 of his book. Indeed, his eq. 18.53 is correct (and not just almost correct, as he claims), whereas eq. 18.54 is incorrect: the structure $(\slashed{p_f} - m_{\mathrm{P}}) . . . (\slashed{p_i} - m_{\mathrm{P}})$ should not be there. Cf. also ref. \cite{Bohm:2001yx}.}
This is the reason why only the amputated diagrams contribute to the $S$-matrix element (eq. \ref{App-LSZ:eq:LSZ-simple} at tree-level and eq. \ref{App-LSZ:eq:Sm-reno-no-mix} at loop level).%
\fn{It goes without saying that the amputation (which removes corrections to external propagators) does not remove one-loop tadpole insertions.}

Before considering the case with mixing, we should also emphasize that nowhere did we need to specify a subtraction scheme. Eqs. \ref{App-LSZ:eq:Sm-reno-no-mix} to \ref{App-LSZ:eq:G-reno-physical} are general definitions, valid for every subtraction scheme. In particular, the fact that only amputated diagrams contribute to $S$-matrix elements does not depend on the renormalization conditions imposed on the fields.
Hence, and contrary to what is often claimed in the literature, the amplitudes---which correspond to the non-trivial part of the $S$-matrix elements---never receive contributions from corrections to external propagators.%
\fn{What this means is that the expression for the $S$-matrix elements is always given by eq. \ref{App-LSZ:eq:Sm-reno-no-mix}, where $\hat{\mathcal{G}}_{\mathrm{ampu.}}$ is the (renormalized) \textit{amputated} GF.
Now, as shall be seen in appendix \ref{App-Theorem},
there are a) particular cases where $\hat{\mathcal{G}}_{\mathrm{ampu.}}$ can be calculated by considering reducible diagrams with non-renormalized corrections to external legs instead of counterterms, and b) particular cases where the renormalized residues $\hat{R}_i$ (in eq. \ref{App-LSZ:eq:Sm-reno-no-mix}) can be calculated by considering renormalized corrections to external propagators.
In any event, this does not change the fact that the expression for the $S$-matrix elements includes the amputated GF only.}

As suggested in appendix \ref{App-OSS}, the OSS scheme will further simplify the results, by setting the renormalized residues equal to one (so that they can be ignored), and by setting the renormalized masses $m_{i\mathrm{R}}$ equal to the pole masses $m_{i\mathrm{P}}$. In that particular case, then, eq. \ref{App-LSZ:eq:Sm-reno-no-mix} becomes:
\be
\left\langle -p_{s+1}, ... , -p_{n}  | S | p_{1}, ... ,  p_{s}\right\rangle
\OSSeq
\hat{\mathcal{G}}_{\mathrm{ampu.}} (p_1, ... , p_n) \Big|_{p_i^2 = m_{i\mathrm{R}}^2}.
\label{App-LSZ:eq:Sm-reno-no-mix-OSS}
\ee

\subsection{Mixing}

\n When fields mix at one-loop level, eqs. \ref{App-LSZ:eq:Sm-reno-no-mix} to \ref{App-LSZ:eq:G-reno-physical} are still in principle valid; only, the mixing between the renormalized fields $\phi_{1}(x_1), ...,  \, \phi_{n}(x_n)$ complicates the quantities therein involved. For example, as we saw in section \ref{Chap-OSS:sec:mix-ind}, the pole masses $m_{i\textrm{P}}$ (which are still equivalent to the poles of the complete propagators) are now given by more complicated expressions. Indeed, they correspond to the eigenvalues of the mass matrix of the effective (i.e. the renormalized up-to-one-loop level) propagator; but since fields mix, such matrix will have off-diagonal entries, so that 
the determination of the eigenvalues requires a diagonalization of the matrix, which ends up leading to complicated expressions.

\n Fortunately, the strategy followed in section \ref{Chap-OSS:sec:mix-ind} allows a simple way around. Recall that, since there is field mixing, there are in general off-diagonal field counterterms. Then, we can always choose those counterterms to be such that eq. \ref{App-OSS:eq:aux3OSS} holds, that is,
\be
\widetilde{\operatorname{Re}} \, \hat{\Gamma}_{ij} (m_{i\mathrm{P}}^2)
\stackrel{j \neq i}{=}
0,
\qquad
\widetilde{\operatorname{Re}} \, \hat{\Gamma}_{ji} (m_{i\mathrm{P}}^2)
\stackrel{j \neq i}{=}
0.
\label{App-LSZ:eq:aux3OSS}
\ee
These relations ensure that, in the limit $p^2=m_{i\mathrm{P}}^2$, the field $\phi_i$ does not mix with other fields.
%
%
%
Thus, although in general (i.e. for a general momentum) the field $\phi_i$ mixes with other fields at one-loop level, it mixes with no other field when the momentum is such that $p^2 = m_{i\mathrm{P}}^2$.

\n The renormalization condition \ref{App-LSZ:eq:aux3OSS} simplifies to a great extent the LSZ formula for the case with mixing. For a thing, 
and as noted in appendix \ref{App-OSS}, it allows a simple definition of both the pole mass $m_{i\mathrm{P}}$, which becomes
\be
\widetilde{\operatorname{Re}} \, \hat{\Gamma}_{ii}(m_{i\mathrm{P}}^2) = m_{i\mathrm{P}}^2 - m_{i\mathrm{R}}^2 + \widetilde{\operatorname{Re}} \,  \hat{\Sigma}_{ii}(m_{i\mathrm{P}}^2) = 0,
\label{App-LSZ:eq:def:mP-i}
\ee
as well as of the renormalized residue $\hat{R}_i$, given by
\be
\hat{R}_i
=
\lim_{p^2 \to m_{i\mathrm{P}}^2} (p^2 - m_{i\mathrm{P}}^2) \, \widetilde{\operatorname{Re}} \left[\hat{\Gamma}_{ii}^{-1}(p^2) \right].
\label{App-LSZ:eq:def:res-i}
\ee
Moreover, eq. \ref{App-LSZ:eq:aux3OSS} also ensures that one does not have to consider GFs with mixing in the calculation of $S$-matrix elements. The reason is that $S$-matrix elements are calculated for OS fields (i.e. for squared momenta equal to the squared pole masses);
for example, if the field $\phi_i$ with momentum $p_i$ shows up in the GF of eq. \ref{App-LSZ:eq:G-reno-physical},
the LSZ formula \ref{App-LSZ:eq:Sm-reno-no-mix} will evaluate the expression for $p_i^2 = m_{i\mathrm{P}}^2$. 
But this is precisely the limit in which eq. \ref{App-LSZ:eq:aux3OSS} holds; as a consequence, the field $\phi_i$ does not mix with other fields in the LSZ formula.
Hence, if one uses one the renormalization condition \ref{App-LSZ:eq:aux3OSS} to fix the off-diagonal counterterms, one can ignore one-loop mixing in the calculation of $S$-matrix elements.

Finally, recall that, as we saw in section \ref{Chap-OSS:sec:mix-ind}, although the renormalization condition \ref{App-LSZ:eq:aux3OSS} is necessary for OSS, it is not sufficient.
In other words, eq. \ref{App-LSZ:eq:aux3OSS} can be used even if not the OSS conditions cannot be used as a whole. This aspect is particularly relevant for the case of a dependent mass---like $m_3$ in the C2HDM. Indeed, even if the renormalized mass $m_{3\mathrm{R}}$ cannot be set equal to the pole mass $m_{3\mathrm{P}}$, 
the off-diagonal field counterterms can still be fixed so that 
eq. \ref{App-LSZ:eq:aux3OSS} holds. In that case, the mixing of $h_3$ with other fields can be ignored in the LSZ formula. Moreover, if the diagonal field counterterm $\delta Z_{33}$ is used to set the residue of the OS propagator of $h_3$ to one, then LSZ becomes further simplified, as that residue can be ignored.

%% file: Appendices/Fermions.tex
\chapter{CP violation in fermionic 2-point functions}
\label{App-Fermions}

\vs{-5mm}

\n In this appendix, we study CP violation in fermionic 2-point functions. As we will show, this is particularly relevant for renormalization purposes---and hence for the renormalization of the C2HDM. However, we do not restrict ourselves to a particular theory; in fact, the appendix can be applied to every model with Dirac fermions.
We organize it as follows:
in a first part, we show how a complex mass shows up in the Dirac representation of the Lorentz group;
then, we consider CP violation in fermionic 2-point functions and prove that a complex mass is intrinsically related with it;
after that, we investigate the consequences of a CP-violating one-loop 2-point function on counterterms;
finally, we ascertain how to handle CP violation using the OSS scheme.

\section{Complex mass}
\label{App-Fermions:sec:D1}

\n Let us consider two fields: a left-handed Weyl spinor $f^{\text{w}}_{\mathrm{L}}$---transforming in the $\left(\frac{1}{2}, 0\right)$ representation of the Lorentz group---and a right-handed Weyl spinor $f^{\text{w}}_{\mathrm{R}}$---transforming in the $\left(0, \frac{1}{2}\right)$ representation of the Lorentz group.%
\fn{We consider a single flavour for simplicity.}
The Lagrangian for the mass term is such that:
\be
-\mathcal{L}_{\text{mass},f} = m_f \, f_{\mathrm{L}}^{\text{w}\dagger} f_{\mathrm{R}}^{\text{w}} + \text{h.c.}
=
m_f \, f_{\mathrm{L}}^{\text{w}\dagger} f_{\mathrm{R}}^{\text{w}} + m_f^* \, f_{\mathrm{R}}^{\text{w}\dagger} f_{\mathrm{L}}^{\text{w}}.
\label{App-Fermions:eq:apD:mass-base}
\ee
Note that the parameter $m_f$ is in general complex.
Moreover, $f_{\mathrm{L}}^{\text{w}}$ and $f_{\mathrm{R}}^{\text{w}}$ are two separate and unrelated fields, which implies in particular that they can be rotated in independent ways. Later on, they (or some rotated versions of them) will be embedded in a Dirac spinor as its left-handed and right-handed components. But until then, $f_{\mathrm{L}}^{\text{w}}$ and $f_{\mathrm{R}}^{\text{w}}$ are not connected with each other in any way. Now, the fact that we can redefine them differently means that the phase of $m_f$ can be absorbed in the fields. Indeed, defining $m_f = |m_f| e^{i \theta_m}$, we can for example rephase $f_{\mathrm{R}}^{\text{w}}$ as
\be
f_{\mathrm{R}}^{\text{w}} \rightarrow e^{-i \theta_m} f_{\mathrm{R}}^{\text{w}}
\label{App-Fermions:eq:apD:7}
\ee
to render $m_f$ real.
Such rephasing (or another one equivalent to it) is usually performed, since $m_f$ is usually identified with the (real) pole mass.
Yet, it is illustrative to keep $m_f$ complex. In particular, we want to ascertain how a complex mass shows up in the Dirac representation $\left(\frac{1}{2}, 0\right) \oplus \left(0, \frac{1}{2}\right)$ of the Lorentz group. Thus, we now embed the Weyl spinors in a Dirac spinor $f$, such that:
\be
f = \begin{pmatrix}
f_{\mathrm{L}}^{\text{w}} \\
f_{\mathrm{R}}^{\text{w}}
\end{pmatrix} = f_{\mathrm{L}} + f_{\mathrm{R}},
\qquad
\text{with}
\quad
f_{\mathrm{L}} = \gamma_{\mathrm{L}} \, f
\equiv \dfrac{1-\gamma_5}{2} f,
\quad 
f_{\mathrm{R}} = \gamma_{\mathrm{R}} \, f
\equiv \dfrac{1+\gamma_5}{2} f.
\ee
%
In the Weyl representation, 
\be
f_L
=
\begin{pmatrix}
f_{\mathrm{L}}^{\text{w}} \\
0
\end{pmatrix},
\quad
f_R =
\begin{pmatrix}
0 \\
f_{\mathrm{R}}^{\text{w}}
\end{pmatrix},
\ee
which means that eq. \ref{App-Fermions:eq:apD:mass-base} can be rewritten as:
\be
-\mathcal{L}_{\text{mass},f} = m_f \, \bar{f}_{\mathrm{L}} f_{\mathrm{R}} + m_f^* \, \bar{f}_{\mathrm{R}} f_{\mathrm{L}}.
\label{App-Fermions:eq:apD:my11}
\ee
Separating $m_f$ into its real and imaginary parts, we find:
\be
\begin{split}
-\mathcal{L}_{\text{mass},f} 
&=  \Big(\mathrm{Re} [m_f] + i \,  \mathrm{Im}[m_f]\Big) \bar{f}_{\mathrm{L}} f_{\mathrm{R}} + \Big(\mathrm{Re} [m_f] - i \,  \mathrm{Im}[m_f]\Big) \bar{f}_{\mathrm{R}} f_{\mathrm{L}} \\
&= \mathrm{Re} [m_f] \Big(\bar{f}_{\mathrm{L}} f_{\mathrm{R}} + \bar{f}_{\mathrm{R}} f_{\mathrm{L}}\Big) + i \, \mathrm{Im} [m_f] \Big(\bar{f}_{\mathrm{L}} f_{\mathrm{R}} -  \bar{f}_{\mathrm{R}} f_{\mathrm{L}}\Big).
\label{App-Fermions:eq:apD:mass-2}
\end{split}
\ee
Using the properties $\gamma_5^{\dagger}=\gamma_5$ and $\{\gamma_5, \gamma_{\mu}\}=0$, it is straightforward to conclude that
\be
\bar{f}_{\mathrm{L}} f_{\mathrm{R}} + \bar{f}_{\mathrm{R}} f_{\mathrm{L}} = \bar{f} f,
\qquad
\bar{f}_{\mathrm{L}} f_{\mathrm{R}} - \bar{f}_{\mathrm{R}} f_{\mathrm{L}} = \bar{f} \gamma_5 f,
\ee
so that eq. \ref{App-Fermions:eq:apD:mass-2} becomes:
\be
-\mathcal{L}_{\text{mass},f} = \mathrm{Re} [m_f] \, \bar{f} f + i \, \mathrm{Im} [m_f] \, \bar{f} \gamma_5 f.
\label{App-Fermions:eq:apD:13}
\ee
In the Dirac representation, then, a mass $m_f$ with non-null imaginary part is equivalent to the presence of a term of the form $\bar{f} \gamma_5 f$.%
\fn{This term has no physical meaning whatsoever, as it depends on the basis chosen: given eq. \ref{App-Fermions:eq:apD:7}, that term will not show up if, before we embed the Weyl spinors $f_{\mathrm{L}}^{\text{w}}$ and $f_{\mathrm{R}}^{\text{w}}$ in the Dirac spinor $f$, we rotate them to render $m_f$ real.}
We now show that such term violates CP.

\section{CP violation in 2-point functions}

\n We start by considering the 2-point function for $f$. Following eq. \ref{Chap-Reno:eq:GammaRenoFermion}, we can parameterize it as:
\be
\Gamma^{\bar{f} f}(p) = \slashed{p} \frac{1-\gamma_{5}}{2} \Gamma^{f,\mathrm{L}}(p^2)+ \slashed{p} \frac{1+\gamma_{5}}{2} \Gamma^{f,\mathrm{R}}(p^2)+\frac{1-\gamma_{5}}{2} \Gamma^{f,\mathrm{l}}(p^2)+\frac{1+\gamma_{5}}{2} \Gamma^{f,\mathrm{r}}(p^2),
\label{App-Fermions:eq:apD:17}
\ee
or equivalently as
\be
\Gamma^{\bar{f} f} (p)
=
\slashed{p} \, \Gamma^{f,{\mathrm{V}}}(p^2)
+ \slashed{p} \gamma_5 \, \Gamma^{f,{\mathrm{A}}}(p^2)
+ \Gamma^{f,{\mathrm{S}}}(p^2)
+ \gamma_5 \, \Gamma^{f,{\mathrm{P}}}(p^2),
\label{App-Fermions:eq:apD:1}
\ee
with the correspondence
\be
\Gamma^{f,{\mathrm{V}}} = \dfrac{\Gamma^{f,\mathrm{L}}+\Gamma^{f,\mathrm{R}}}{2},
\quad 
\Gamma^{f,{\mathrm{A}}} = \dfrac{\Gamma^{f,\mathrm{R}}-\Gamma^{f,\mathrm{L}}}{2},
\quad 
\Gamma^{f,{\mathrm{S}}} = \dfrac{\Gamma^{f,\mathrm{l}}+\Gamma^{f,\mathrm{r}}}{2},
\quad 
\Gamma^{f,{\mathrm{P}}} = \dfrac{\Gamma^{f,\mathrm{r}}-\Gamma^{f,\mathrm{l}}}{2}.
\label{App-Fermions:eq:apD:conv}
\ee
Now, eq. \ref{App-Fermions:eq:apD:1} corresponds to the effective Lagrangian
\be
\mathcal{L}_{\text{eff}}^{\bar{f} f} =  i \, \bar{f} \, \slashed{\partial} f \, \Gamma^{f,{\mathrm{V}}}(\partial^2)  + i \, \bar{f} \, \slashed{\partial} \gamma_5 f \, \Gamma^{f,{\mathrm{A}}}(\partial^2) + \bar{f} f \, \Gamma^{f,{\mathrm{S}}} (\partial^2)+ \bar{f} \gamma_5 f \, \Gamma^{f,{\mathrm{P}}}(\partial^2).
\label{App-Fermions:eq:apD:2}
\ee
In table \ref{App-Fermions:tab:apD:CPtable},
\newcolumntype{?}{!{\vrule width 0.005mm}}
\begin{table}[!h]%
\begin{normalsize}
\normalsize
\begin{center}
\begin{tabular}
{@{\hspace{3mm}}
>{\centering\arraybackslash}p{1.5cm}
?
>{\centering\arraybackslash}p{2.0cm}
>{\centering\arraybackslash}p{2.0cm}
>{\centering\arraybackslash}p{2.0cm}
@{\hspace{3mm}}}
\hlinewd{1.1pt}
& C & P & CP\\
\hline\\[-3mm]
$\bar{\psi}_i \, \psi_j$ & $\bar{\psi}_j \, \psi_i$ & $\bar{\psi}_i \, \psi_j$ & $\bar{\psi}_j \, \psi_i$\\[1mm]
$\bar{\psi}_i \, \gamma_5 \, \psi_j$ & $\bar{\psi}_j \, \gamma_5 \, \psi_i$ & $- \bar{\psi}_i \, \gamma_5 \, \psi_j$ & $- \bar{\psi}_j \, \gamma_5 \, \psi_i$ \\[1mm]
$\bar{\psi}_i \, \slashed{\partial} \, \psi_j$ & $\bar{\psi}_j \, \slashed{\partial} \, \psi_i$ & $\bar{\psi}_i \, \slashed{\partial} \, \psi_j$ & $\bar{\psi}_j \, \slashed{\partial} \, \psi_i$ \\[1mm]
$\bar{\psi}_i \, \slashed{\partial} \, \gamma_5 \psi_j$ & $- \bar{\psi}_j \, \slashed{\partial} \, \gamma_5 \, \psi_i$ & $- \bar{\psi}_i \, \slashed{\partial} \, \gamma_5 \, \psi_j$ & $\bar{\psi}_j \, \slashed{\partial} \, \gamma_5 \, \psi_i$ \\[1mm]
\hlinewd{1.1pt}
\end{tabular}
\end{center}
\vspace{-5mm}
\end{normalsize}
\caption{Results of the application of the symmetries C, P and CP to the terms in the left column.}
\label{App-Fermions:tab:apD:CPtable}
\end{table}
\normalsize
we show the action of the operators C, P and CP on the four bilinear structures of eq. \ref{App-Fermions:eq:apD:2}, but for general fields $\psi_i$ and $\psi_j$.%
\fn{We have omitted possible free phases present in each discrete symmetry (see page 35 of ref. \cite{Branco:1999fs}).
Note also that, in the last two lines of the table, the Dirac fields are exchanged with one another under the symmetry C; as a consequence, the derivative (which was acting on the field $\psi_j$) will now act on $\psi_i$. Now, assuming that the total derivative of the terms in the left collumn are zero---i.e. $\partial_{\mu} (\bar\psi_i \gamma^{\mu} \psi_j) = 0$
and
$\partial_{\mu} (\bar{\psi}_i \gamma^{\mu} \gamma_5 \psi_j) = 0$---, then the term with the derivative acting on one of the fields has the opposite sign relative to the term with the derivative acting on the other field---i.e.
$\slashed{\partial} \, \bar\psi_i \psi_j = - \bar\psi_i \slashed{\partial} \, \psi_j$
and
$\slashed{\partial} \, \bar\psi_i \gamma_5 \psi_j = - \bar\psi_i \slashed{\partial} \, \gamma_5 \psi_j$.
Therefore, after applying the symmetry C, if we want to leave the derivative between the two fields (so that it still acts on the field that currently stands to its right), we obtain a minus sign. In this way, when the derivative is ensandwiched between Dirac fields, it is as if acquires a minus sign under C.}
In the particular case $\psi_i=\psi_j=f$, there is CP violation in eq. \ref{App-Fermions:eq:apD:2} if and only if $\Gamma^{f,{\mathrm{P}}} \neq 0$. Using eq. \ref{App-Fermions:eq:apD:conv}, we thus conclude:
\be 
\text{CP violation in} \,  \mathcal{L}_{\text{eff}}^{\bar{f} f}
\quad \Leftrightarrow
\quad 
\Gamma^{f,{\mathrm{P}}} \neq 0
\quad \Leftrightarrow
\quad 
\Gamma^{f,\mathrm{l}} \neq \Gamma^{f,\mathrm{r}}.
\label{App-Fermions:eq:apD:box}
\ee
Comparing eqs.  \ref{App-Fermions:eq:apD:13} and \ref{App-Fermions:eq:apD:2}, it is clear that the term in eq.  \ref{App-Fermions:eq:apD:13} proportional to $\mathrm{Im} [m_f]$ violates CP.
One might wonder why this conclusion is relevant; after all, and as we showed, one can always avoid such term by a rephasing of the Weyl spinors. But in that case, what happens at loop level? In particular, if we perform such rephasing and if CP violation contributes to fermionic 2-point functions at one-loop, how are we ensured that there will be a counterterm for such contribution?
It is to these questions that we now turn.

\section{Fermionic counterterms and CP violation}

\n When renormalizing the theory, the tree-level parameters are identified as usual with bare parameters. Since the tree-level parameter $m_f$ was in general complex,
the bare parameter $m_{f(0)}$ is also in general complex. Writing eq. \ref{App-Fermions:eq:apD:my11} in terms of bare quantities, we have:
\be
-\mathcal{L}_{\text{mass},f}
= m_{f(0)} \, \bar{f}_{\mathrm{L}(0)} \, f_{\mathrm{R}(0)} + m_{f(0)}^* \, \bar{f}_{\mathrm{R}(0)} \, f_{\mathrm{L}(0)},
\label{App-Fermions:eq:bare-mass-Lag}
\ee
such that
\be
\label{App-Fermions:eq:m-expa}
m_{f(0)} = m_f + \delta m_f,
\qquad
f_{\mathrm{L}(0)} = f_{\mathrm{L}} + \frac{1}{2} \delta Z^{f,\mathrm{L}} f_{\mathrm{L}},
\qquad
f_{\mathrm{R}(0)} = f_{\mathrm{R}} + \frac{1}{2} \delta Z^{f,\mathrm{R}} f_{\mathrm{R}}.
\ee
Here, $m_f$ and $\delta m_f$ are in general complex.
However, based on what we showed one could do at tree-level, we can take the freedom in the basis choice of the (renormalized) chiral spinors $f_{\mathrm{L}}$ and $f_{\mathrm{R}}$ to render $m_f$ real. After that, $\delta m_f$ is still in general complex.%
\fn{Cf. section \ref{Chap-Reno:sec:simple} for an explicit example of a case where a counterterm remains complex even after the corresponding renormalized parameter is rendered real.}
%
%
%
Separating the renormalized terms and the counterterms, eq. \ref{App-Fermions:eq:bare-mass-Lag} then becomes:
\be
-\mathcal{L}_{\text{mass},f}
=  
\mathcal{L}^{\text{reno.}}_{\text{mass},f}
+ \mathcal{L}^{\text{CT}}_{\text{mass},f},
\ee
where
\bs
\label{App-Fermions:eq:reno-and-CT}
\begin{flalign}
\label{App-Fermions:eq:f-reno}
&\mathcal{L}^{\text{reno.}}_{\text{mass},f}
=
m_f \left( \bar{f}_{\mathrm{L}} f_{\mathrm{R}} + \bar{f}_{\mathrm{R}} f_{\mathrm{L}}\right) 
= m_f \bar{f} f,
\\
\label{App-Fermions:eq:f-CT}
&\mathcal{L}^{\text{CT}}_{\text{mass},f}
=
\delta M_f \, \bar{f}_{\mathrm{L}} f_{\mathrm{R}}
+ \delta M_f^* \, \bar{f}_{\mathrm{R}} f_{\mathrm{L}}
=
\mathrm{Re} [\delta M_f] \, \bar{f} f + i \, \mathrm{Im} [\delta M_f] \, \bar{f} \gamma_5 f,
\end{flalign}
\es
with
\be
\delta M_f \equiv \left(\delta m_f + \frac{1}{2} m_f \, \delta Z^{f,\mathrm{L} *} + \frac{1}{2} m_f \, \delta Z^{f,\mathrm{R}}\right).
\label{App-Fermions:eq:delta-Mf}
\ee
Eqs. \ref{App-Fermions:eq:reno-and-CT} clarify the importance of the discussion of the previous section: we see that, due to our basis choice for the chiral spinors, there is no CP violation in the renormalized terms, eq. \ref{App-Fermions:eq:f-reno}. But it is also clear that, even after that basis choice, there \textit{is} CP violation in the counterterms, eq. \ref{App-Fermions:eq:f-CT}, as long as $\mathrm{Im} [\delta M_f] \neq 0$ (recall eqs. \ref{App-Fermions:eq:apD:2} and \ref{App-Fermions:eq:apD:box}). Therefore, if CP-violating effects show up at one-loop level, there will in general be counterterms to absorb their divergences.

\section{CP violation in the OSS scheme}

\n We now focus on the consequences of the OSS renormalization conditions on the different terms of $\delta M_f$.
We start by noting that, motivated by the fact that
\be
\delta m_f \, \bar{f}_{\mathrm{L}} f_{\mathrm{R}}
+\delta m_f^*  \, \bar{f}_{\mathrm{R}} f_{\mathrm{L}} =
\delta m_f \, \bar{f} \gamma_{\mathrm{R}} f
+\delta m_f^* \, \bar{f} \gamma_{\mathrm{L}} f,
\ee
we make the identification:%
\fn{Left-handed and right-handed components of fermion mass counterterms have already been considered by Kniehl and Sirlin in the case of fermion mixing \cite{Kniehl:2006bs,Kniehl:2006rc}.}
\be
\delta m_f^{\mathrm{R}} \equiv \delta m_f,
\qquad
\delta m_f^{\mathrm{L}} \equiv \delta m_f^*,
\ee
which implies, in particular,
\be
\delta m_f^{\mathrm{R}} = \delta m_f^{\mathrm{L} *},
\qquad
\mathrm{Im} \left[\delta m_f\right] = \dfrac{\delta m_f^{\mathrm{R}} - \delta m_f^{\mathrm{L}}}{2 i}.
\label{App-Fermions:eq:apD:reldeltams}
\ee
Then, inserting eqs. \ref{Chap-Reno:eq:GammaRenoDecompFermion} in the two eqs. \ref{Chap-Reno:eq:AuxFermionsA} for $i=j$, we obtain respectively:%
\fn{We omit the indices $i$. Note that the case $i \neq j$ is irrelevant for this discussion, as the expressions for $\delta Z_{ij}^{f,\mathrm{L}}$ and $\delta Z_{ij}^{f,\mathrm{R}}$ (with $i \neq j$, eqs. \ref{Chap-Reno:eq:reno-fermB-pre} and \ref{Chap-Reno:eq:reno-fermB}) directly follow from the insertion of eqs. \ref{Chap-Reno:eq:GammaRenoDecompFermion} in the two equations \ref{Chap-Reno:eq:AuxFermionsA} with $i \neq j$. In that case, then, the OSS conditions leave no freedom.}
\bs
\label{App-Fermions:eq:apD:setMasses}
\bea
\delta m_{f}^{\mathrm{R}}
&=&
\dfrac{1}{2} \widetilde{\operatorname{Re}}
\bigg[
2 \, m_{f} \Sigma^{f,\mathrm{L}}(m_{f}^2) +  2 \, \Sigma^{f,\mathrm{r}}(m_{f}^2) + m_{f} \left(\delta Z^{f, \mathrm{L}} - \delta Z^{f, \mathrm{R}} \right)
\bigg], \label{App-Fermions:eq:apD:my4} \\
\delta m_{f}^{\mathrm{L}}
&=&
\dfrac{1}{2} \widetilde{\operatorname{Re}}
\bigg[
2 \, m_{f} \Sigma^{f,\mathrm{R}}(m_{f}^2) +  2 \, \Sigma^{f,\mathrm{l}}(m_{f}^2) + m_{f} \left(\delta Z^{f, \mathrm{R}} - \delta Z^{f, \mathrm{L}} \right)
\bigg].
\label{App-Fermions:eq:apD:my3}
\eea
\es
Inserting eqs. \ref{Chap-Reno:eq:GammaRenoDecompFermion} in eq. \ref{Chap-Reno:eq:AuxFermionsB} (again for $i=j$), and subtracting the complex conjugate of eq. \ref{App-Fermions:eq:apD:my4} to eq. \ref{App-Fermions:eq:apD:my3} while using eqs. \ref{App-Fermions:eq:apD:reldeltams} and \ref{Chap-Reno:eq:apD:30c} (which assumes hermiticity of the up-to-one-loop action, recall section \ref{Chap-Reno:sec:fermionsCT}), we get:
\bs
\label{App-Fermions:eq:apD:setFields}
\begin{flalign}
\dfrac{\delta Z^{f, \mathrm{L}} + \delta Z^{f, \mathrm{L} *}}{2} &=  -\widetilde{\operatorname{Re}} \, \Sigma^{f, \mathrm{L}}\left(m_{f}^{2}\right)-  m_{f} \frac{\partial}{\partial p^{2}} \widetilde{\operatorname{Re}}\bigg[m_{f}\left(\Sigma^{f, \mathrm{L}}(p^2)+\Sigma^{f, \mathrm{R}}(p^2)\right) \nonumber
\\[-4mm]
& \hs{65mm} + \Sigma^{f, \mathrm{l}}(p^2)+\Sigma^{f, \mathrm{r}}(p^2)\bigg]\bigg|_{p^{2}=m_{f}^{2}} , \\[-2mm]
\dfrac{\delta Z^{f, \mathrm{R}} + \delta Z^{f, \mathrm{R}  *}}{2} &=  -\widetilde{\operatorname{Re}} \, \Sigma^{f, \mathrm{R}}\left(m_{f}^{2}\right)- m_{f} \frac{\partial}{\partial p^{2}} \widetilde{\operatorname{Re}}\bigg[m_{f}\left(\Sigma^{f, \mathrm{L}}(p^2)+\Sigma^{f, \mathrm{R}}(p^2)\right)
\nonumber \\[-4mm]
&  \hs{65mm} +\Sigma^{f, \mathrm{l}}(p^2)+\Sigma^{f, \mathrm{r}}(p^2)\bigg]\bigg|_{p^{2}=m_{f}^{2}} .
\end{flalign}
\es
There are thus 6 real degrees of freedom (dof)---two for each of the three complex counterterms $\delta m_{f}, \delta Z^{f, \mathrm{L}}$ and $\delta Z^{f, \mathrm{R}}$---, but the diagonal OSS conditions only imply the 4 relations of eqs. \ref{App-Fermions:eq:apD:setMasses} and \ref{App-Fermions:eq:apD:setFields}. Hence, there are two free dof, so that many solutions satisfy those conditions. Before considering them, note that, for the specific case of the non-renormalized diagonal one-loop 2-point function,
eq. \ref{App-Fermions:eq:apD:box} implies:
\be
\text{CP violation in} \, \widetilde{\operatorname{Re}} \,  \Sigma^{\bar{f} f}
\quad
\Leftrightarrow
\quad
\widetilde{\operatorname{Re}} \, \Sigma^{f, \mathrm{l}}
\neq
\widetilde{\operatorname{Re}} \, \Sigma^{f, \mathrm{r}}.
\label{App-Fermions:eq:my-CP-one-loop}
\ee

\n Now, among the many possible solutions, we restrict ourselves to two.
The first chooses $\mathrm{Im} \left[\delta Z^{f, \mathrm{L}}\right] = \mathrm{Im} \left[\delta Z^{f, \mathrm{R}}\right] = 0$.
In this case, eq. \ref{App-Fermions:eq:my-CP-one-loop} together with eqs. \ref{App-Fermions:eq:apD:setMasses} and \ref{App-Fermions:eq:apD:setFields} leads to:
\be
\text{CP violation in} \, \widetilde{\operatorname{Re}} \,  \Sigma^{\bar{f} f}
\quad
\stackrel{\scriptsize\mathrm{Im} \big[\delta Z^{f, {\mathrm{L}/\mathrm{R}}}\big] = 0}{\Leftrightarrow}
\quad 
\delta m_f^{\mathrm{R}} \neq \delta m_f^{\mathrm{L}} .
\label{App-Fermions:eq:apD:box2}
\ee
This relation shows that, whenever there is no CP violation in  one-loop fermionic diagonal 2-point functions, there is no need to introduce $\delta m_{f}^{\mathrm{L}}$ nor $\delta m_{f}^{\mathrm{R}}$: they would be the same, and both equal to $\delta m_{f}$, which is thus preferred. Conversely, if there is CP violation in $\widetilde{\operatorname{Re}} \, \Sigma^{\bar{f} f}$, the choice $\mathrm{Im} \left[\delta Z^{f, \mathrm{L}}\right] = \mathrm{Im} \left[\delta Z^{f, \mathrm{R}}\right] = 0$ forces   $\delta m_{f}^{\mathrm{L}}$  and  $\delta m_{f}^{\mathrm{R}}$ to be different (or, what is equivalent, forces $\delta m_{f}$ to be complex). This is the solution we adopted in eqs. \ref{Chap-Reno:eq:reno-ferm} for the case $i=j$.\fn{These follow from eqs. \ref{App-Fermions:eq:apD:setMasses} and \ref{App-Fermions:eq:apD:setFields} when imposing $\mathrm{Im} \left[\delta Z^{f, \mathrm{L}}\right] = \mathrm{Im} \left[\delta Z^{f, \mathrm{R}}\right] = 0$.}

\n Another simple solution uses one of the two free dofs to impose $\delta m_{f}^{\mathrm{L}} = \delta m_{f}^{\mathrm{R}}$. In this case, by subtracting eq. \ref{App-Fermions:eq:apD:my4} to eq. \ref{App-Fermions:eq:apD:my3}, taking the imaginary part of the result and using eqs. \ref{Chap-Reno:eq:apD:30c} and \ref{App-Fermions:eq:my-CP-one-loop}, we get:
\be
\text{CP violation in} \, \widetilde{\operatorname{Re}} \,  \Sigma^{\bar{f} f}
\quad
\stackrel{\delta m_{f}^{\mathrm{L}} = \delta m_{f}^{\mathrm{R}}}{\Leftrightarrow}
\quad 
\mathrm{Im} \left[\delta Z^{f, \mathrm{L}} \right] \neq \mathrm{Im} \left[\delta Z^{f, \mathrm{R}}\right].
\label{App-Fermions:eq:apD:casedmLdmR}
\ee
It is then clear that, in the context of OSS, CP violation in fermionic diagonal 2-point functions does not necessarily force the relation $\delta m_{f}^{\mathrm{L}} \neq \delta m_{f}^{\mathrm{R}}$. However, by choosing $\delta m_{f}^{\mathrm{L}} = \delta m_{f}^{\mathrm{R}}$, it follows that $\delta Z^{f, \mathrm{L}}$ and $\delta Z^{f, \mathrm{R}}$ cannot be both real. This is the solution usually adopted in the cMSSM (cf. e.g. refs.~\cite{Heinemeyer:2010mm,Fritzsche:2011nr}).

%% file: Appendices/WI.tex
\chapter{Ward identity for $A \bar{f} f$ and the $\delta Z_e$ counterterm}
\label{App-WI}

\vs{-5mm}

In this appendix, we derive the Ward identity (WI) for the one-loop photon-fermion-fermion vertex function, for a general $R_{\xi}$ gauge and for on-shell (OS) fermions and vanishing photon momentum.
More exactly, we consider the general case $A \bar{f}_r f_s$, where the flavor $r$ is in general different from $s$.
The particular case of flavor conservation ($r=s$) is especially useful to determine the counterterm $\delta Z_e$ for the electric charge in the on-shell subtraction (OSS) scheme, as shall be discussed at the end of the appendix.
For that reason, the WI in the flavor-conserving case plays an important role in the renormalization of the SM---and in all its extensions that keep its gauge sector, as will be seen---, wherefore it is often cited in the literature (e.g. \cite{Aoki:1982ed, Denner:1991kt, Bohm:2001yx, Denner:2016etu}).
In spite of this, it was derived only recently, in an appendix of a report by Denner and Dittmaier \cite{Denner:2019vbn}.%
\fn{Even more recently, Dittmaier showed how to determine $\delta Z_e$ to all orders in perturbation theory based on the concept of charge universality \cite{Dittmaier:2021loa}.}

In their paper, Denner and Dittmaier derived the WI using the Lee identities (cf. ref. \cite{Denner:2019vbn} and references therein). Here, we follow an alternative method (also suggested in ref. \cite{Denner:2019vbn}), which starts from the BRST \cite{Becchi:1975nq,Tyutin:1975qk} invariance of complete Green's functions (GFs). 
Despite being longer than (and posterior to) the derivation of ref. \cite{Denner:2019vbn}, the derivation that we present here is not deprived of relevance, for several reasons. First, it is an alternative derivation to that of ref. \cite{Denner:2019vbn}---which, given the complexity of both derivations, is not trivial. Second, by not containing external sources, the derivation presented here contains less unphysical quantities than that of ref. \cite{Denner:2019vbn}. Moreover, as mentioned above, we consider the case with general flavors, which has interest beyond the determination of $\delta Z_e$. Finally, we perform the derivation with the general $\eta$'s of ref. \cite{Romao:2012pq}, the main advantage of which is that the final WI can be easily adapted to different sign conventions. 

In this appendix, we start by introducing our conventions for the GFs, as well as the techniques used in the derivation. Then, we derive the WI. Finally, we use the WI in the flavor-conserving case to derive $\delta Z_e$ in OSS.%
\fn{I am indebted to Dittmaier for useful suggestions on the derivation of the WI.}

\section{Introduction of techniques and conventions}

In what follows, we use lowercase roman letters for flavour indices, greek letters for Lorentz indices, natural numbers for specific GFs, and we omit the Dirac indices.
Repeated flavour indices are only to be summed whenever an explicit sum symbol is shown.
The WI derived below corresponds to a relation between vertex (or one-particle irreducible, 1PI) GFs in momentum space. To obtain them, we start with complete GFs in position space. For example, in the photon-fermion-fermion case, the complete GF is:%
\be
G^{Af\bar{f}}_{\mu ji}(x,y,z) 
=
\ppo \Omega | T \{A_{\mu}(x) \, f_j(y) \, \bar{f}_i(z) \} | \Omega \ppc.
\label{App-WI:eq:WIbase}
\ee
In what follows, we abbreviate $\ppo \Omega | T \{ ... \} | \Omega \ppc$ with $\ppo ... \ppc$ for simplicity, as usual.
One obtains the equivalent to eq. \ref{App-WI:eq:WIbase} in momentum space by applying the Fourier Transform (FT).
We define the complete GF in momentum space with momenta $k$ and $p$ incoming and $p^{\prime}$ outgoing, $G^{Af\bar{f}}_{\mu ji}(k,-p^{\prime},p)$, such that:
\be
\begin{split}
(2 \pi)^4 \delta^4(k-p^{\prime}+p)
\, G^{Af\bar{f}}_{\mu ji}(k,-p^{\prime},p)
& =
\mathrm{FT}\Big[\ppo A_{\mu} (x) \, f_j(y) \, \bar{f}_i (z) \ppc \Big] \\
& = 
\int d^4x \,  d^4y \, d^4z \, e^{-i \big( k x - p' y + p z \big)} \ppo A_{\mu} (x) \, f_j(y) \, \bar{f}_i (z) \ppc, 
\end{split}
\ee
and we represent it with a dark circle:%
\fn{We suppose an outgoing anti-fermion $\bar{f}_i$, which is equivalent to an incoming fermion ${f}_i$. As a consequence, when it is set OS, we shall use a spinor $v$ instead of a spinor $u$.}
\be
G^{Af\bar{f}}_{\mu ji}(k,-p^{\prime},p)
=
\hs{3mm}
\begin{minipage}{0.35\textwidth}
\begin{fmffile}{conv1} 
\begin{fmfgraph*}(70,70) 
\fmfset{arrow_len}{3mm} 
\fmfset{arrow_ang}{20} 
\fmfleft{nJ1} 
\fmfright{nJ2,nJ4}
\fmf{phantom,label=$A_{\mu}$,label.side=left,tension=0}{nJ1,nJ1nJ2nJ4}
\fmf{photon,label=$\Large \xxrightarrow[k]{}$,label.dist=3,tension=3}{nJ1,nJ1nJ2nJ4}
\fmf{fermion,label=\rotatebox{57}{$\Large\xxleftarrow[p]{}$},label.side=left,label.dist=-5,label.angle=30,tension=3}{nJ4,nJ1nJ2nJ4}
\fmf{phantom,label=$\bar{f}_i$,label.side=right,tension=0}{nJ4,nJ1nJ2nJ4}
\fmf{fermion,label=\rotatebox{-57}{$\Large\xxrightarrow[p']{}$},label.side=right,label.dist=-5,label.angle=30,tension=3}{nJ1nJ2nJ4,nJ2}
\fmf{phantom,label=$f_j$,label.side=left,tension=0}{nJ1nJ2nJ4,nJ2}
\fmfv{decor.shape=circle,decor.filled=70,decor.size=11thick}{nJ1nJ2nJ4}
\end{fmfgraph*} 
\end{fmffile}
\end{minipage}
\hs{-28mm}
.
\ee
We are following the conventions of ref. \cite{Denner:2019vbn} for the arguments and the indices of GFs, namely: the $n^{\mathrm{th}}$ argument denotes the incoming momentum of the field corresponding to the $n^{\mathrm{th}}$ superscript (in the case of 2-point functions, however, we exploit momentum conservation to write only one argument, always positive);
as for the superscripts, they depend on the type of GFs: in complete and connected GFs, they correspond to outgoing fields, whereas in 1PI GFs they correspond to the incoming fields.
%
%

%

Now, the complete GF is the sum of connected and disconnected ones. With 3 external particles, though, the only way to form a disconnected GF would be through a tadpole of one of them. As such scenario is excluded for the particles at stake, the complete GF is simply equal to the connected one, which we represent with a gray circle:
\vs{-4mm}
\be
\hs{10mm}
G^{Af\bar{f}}_{\mu ji}(k,-p^{\prime},p)_{\text{con.}} =  
\hs{3mm}
\begin{minipage}{0.35\textwidth}
\begin{fmffile}{conv2} 
\begin{fmfgraph*}(70,70) 
\fmfset{arrow_len}{3mm} 
\fmfset{arrow_ang}{20} 
\fmfleft{nJ1} 
\fmfright{nJ2,nJ4}
\fmf{phantom,label=$A_{\mu}$,label.side=left,tension=0}{nJ1,nJ1nJ2nJ4}
\fmf{photon,label=$\Large \xxrightarrow[k]{}$,label.dist=3,tension=3}{nJ1,nJ1nJ2nJ4}
\fmf{fermion,label=\rotatebox{57}{$\Large\xxleftarrow[p]{}$},label.side=left,label.dist=-5,label.angle=30,tension=3}{nJ4,nJ1nJ2nJ4}
\fmf{phantom,label=$\bar{f}_i$,label.side=right,tension=0}{nJ4,nJ1nJ2nJ4}
\fmf{fermion,label=\rotatebox{-57}{$\Large\xxrightarrow[p']{}$},label.side=right,label.dist=-5,label.angle=30,tension=3}{nJ1nJ2nJ4,nJ2}
\fmf{phantom,label=$f_j$,label.side=left,tension=0}{nJ1nJ2nJ4,nJ2}
\fmfv{decor.shape=circle,decor.filled=30,decor.size=11thick}{nJ1nJ2nJ4}
\end{fmfgraph*} 
\end{fmffile}
\end{minipage}
\hs{-28mm}.
\ee
We shall identify the cases of the connected GF at tree-level (tr) and one-loop (1L) with the label \textbf{tr} or \textbf{1L} inside the gray circle.
One can obtain 1PI GFs from connected ones by factorizing the external propagators:
\be
G^{Af\bar{f}}_{\mu ji}(k,-p^{\prime},p)_{\text{con.}} =  
\hs{3mm}
\begin{minipage}{0.35\textwidth}
\begin{fmffile}{SMAff2} 
\begin{fmfgraph*}(80,80) 
\fmfset{arrow_len}{3mm} 
\fmfset{arrow_ang}{20} 
\fmfleft{nJ1} 
\fmfright{nJ2,nJ4} 
\fmf{photon,label=$\Large \xxrightarrow[k]{}$,label.side=right,label.dist=3,tension=1.8}{nJ1,x1}
\fmf{phantom,label=$A_{\mu}$,label.side=left,tension=0}{nJ1,x1}
\fmfv{decor.shape=circle,decor.filled=30,decor.size=6thick}{x1}
\fmf{photon,tension=1,label=$A_{\nu}$,label.side=left}{x1,X}
\fmf{fermion,tension=0.9}{x4,X}
\fmf{phantom,label=$\bar{f}_l$,label.side=right,label.dist=4,tension=0}{x4,X}
\fmfv{decor.shape=circle,decor.filled=30,decor.size=6thick}{x4}
\fmf{fermion,label=\rotatebox{57}{$\Large\xxleftarrow[p]{}$},label.side=left,label.dist=-3,label.angle=30,tension=1.5}{nJ4,x4}
\fmf{phantom,label=$\bar{f}_i$,label.side=right,label.dist=4,tension=0}{nJ4,x4}
\fmf{fermion,label=\rotatebox{-57}{$\Large\xxrightarrow[p']{}$},label.side=right,label.dist=-5,label.angle=30,tension=1.5}{x2,nJ2}
\fmf{phantom,label=$f_j$,label.side=left,label.dist=4,tension=0}{x2,nJ2}
\fmfv{decor.shape=circle,decor.filled=30,decor.size=6thick}{x2}
\fmf{fermion,tension=0.9}{X,x2} 
\fmf{phantom,label=$f_m$,label.side=right,label.dist=4,tension=0}{x2,X}
\fmfv{decor.shape=circle,decor.filled=shaded,decor.size=11thick}{X}
\end{fmfgraph*} 
\end{fmffile}
\end{minipage}
\hs{-25mm}
+
\hs{6mm}
\begin{minipage}{0.35\textwidth}
\begin{fmffile}{SMAff3} 
\begin{fmfgraph*}(80,80) 
\fmfset{arrow_len}{3mm} 
\fmfset{arrow_ang}{20} 
\fmfleft{nJ1} 
\fmfright{nJ2,nJ4} 
\fmf{photon,label=$\Large \xxrightarrow[k]{}$,label.dist=3,tension=1.8}{nJ1,x1}
\fmf{phantom,label=$A_{\mu}$,label.side=left,tension=0}{nJ1,x1}
\fmfv{decor.shape=circle,decor.filled=30,decor.size=6thick}{x1}
\fmf{photon,tension=0.8,label=$Z_{\nu}$,label.side=left}{x1,X}
\fmf{fermion,tension=0.9}{x4,X}
\fmf{phantom,label=$\bar{f}_l$,label.side=right,label.dist=4,tension=0}{x4,X}
\fmfv{decor.shape=circle,decor.filled=30,decor.size=6thick}{x4}
\fmf{fermion,label=\rotatebox{57}{$\Large\xxleftarrow[p]{}$},label.side=left,label.dist=-3,label.angle=30,tension=1.5}{nJ4,x4}
\fmf{phantom,label=$\bar{f}_i$,label.side=right,label.dist=4,tension=0}{nJ4,x4}
\fmf{fermion,label=\rotatebox{-57}{$\Large\xxrightarrow[p']{}$},label.side=right,label.dist=-5,label.angle=30,tension=1.5}{x2,nJ2}
\fmf{phantom,label=$f_j$,label.side=left,label.dist=4,tension=0}{x2,nJ2}
\fmfv{decor.shape=circle,decor.filled=30,decor.size=6thick}{x2}
\fmf{fermion,tension=0.9}{X,x2} 
\fmf{phantom,label=$f_m$,label.side=right,label.dist=4,tension=0}{x2,X}
\fmfv{decor.shape=circle,decor.filled=shaded,decor.size=11thick}{X}
\end{fmfgraph*} 
\end{fmffile}
\end{minipage}
\hs{-25mm}
+
\hs{6mm}
\begin{minipage}{0.35\textwidth}
\begin{fmffile}{SMAff3b} 
\begin{fmfgraph*}(80,80) 
\fmfset{arrow_len}{3mm} 
\fmfset{arrow_ang}{20} 
\fmfleft{nJ1} 
\fmfright{nJ2,nJ4} 
\fmf{photon,label=$\Large \xxrightarrow[k]{}$,label.dist=3,tension=1.8}{nJ1,x1}
\fmf{phantom,label=$A_{\mu}$,label.side=left,tension=0}{nJ1,x1}
\fmfv{decor.shape=circle,decor.filled=30,decor.size=6thick}{x1}
\fmf{dashes,tension=0.8,label=$S$,label.side=left}{x1,X}
\fmf{fermion,tension=0.9}{x4,X}
\fmf{phantom,label=$\bar{f}_l$,label.side=right,label.dist=4,tension=0}{x4,X}
\fmfv{decor.shape=circle,decor.filled=30,decor.size=6thick}{x4}
\fmf{fermion,label=\rotatebox{57}{$\Large\xxleftarrow[p]{}$},label.side=left,label.dist=-3,label.angle=30,tension=1.5}{nJ4,x4}
\fmf{phantom,label=$\bar{f}_i$,label.side=right,label.dist=4,tension=0}{nJ4,x4}
\fmf{fermion,label=\rotatebox{-57}{$\Large\xxrightarrow[p']{}$},label.side=right,label.dist=-5,label.angle=30,tension=1.5}{x2,nJ2}
\fmf{phantom,label=$f_j$,label.side=left,label.dist=4,tension=0}{x2,nJ2}
\fmfv{decor.shape=circle,decor.filled=30,decor.size=6thick}{x2}
\fmf{fermion,tension=0.9}{X,x2} 
\fmf{phantom,label=$f_m$,label.side=right,label.dist=4,tension=0}{x2,X}
\fmfv{decor.shape=circle,decor.filled=shaded,decor.size=11thick}{X}
\end{fmfgraph*} 
\end{fmffile}
\end{minipage}
\hs{-22mm},
\label{App-WI:eq:my109}
\ee
where the shaded circle represents the 1PI GFs 
(in the particular case of one-loop 1PI GFs, we use a hatched circle, instead of a shaded one),
and where $S$ represents a generic neutral scalar.%
\fn{Using a generic field is not only more economical, but also keeps the description general. In the SM, there are only two neutral scalars ($G_0$ and $H$, with masses $\xi_Z m_{\mathrm{Z}}$ and $m_H$, respectively), but in BSM models there are in general more. As we will see, these terms will end up not contributing to the WI.}
If we consider the first term of the r.h.s., for example, the 1PI GF is:
\be
i \Gamma^{A\bar{f}f}_{\nu ml}(k,-p^{\prime},p) \, = 
\hspace{3mm}
\begin{minipage}{0.35\textwidth}
\begin{fmffile}{conv5} 
\begin{fmfgraph*}(60,60) 
\fmfset{arrow_len}{3mm} 
\fmfset{arrow_ang}{20} 
\fmfleft{nJ1} 
\fmfright{nJ2,nJ4} 
\fmf{photon,label=$\Large \xxrightarrow[k]{}$,label.dist=3,tension=3}{nJ1,X}
\fmf{phantom,label=$A_{\nu}$,label.side=left,tension=0}{nJ1,X}
\fmf{fermion,label=\rotatebox{57}{$\Large\xxleftarrow[p]{}$},label.side=left,label.dist=-5,label.angle=30,tension=3}{nJ4,X}
\fmf{phantom,label=$f_l$,label.side=right,tension=0}{nJ4,X}
\fmf{fermion,label=\rotatebox{-57}{$\Large\xxrightarrow[p']{}$},label.side=right,label.dist=-5,label.angle=30,tension=3}{X,nJ2}
\fmf{phantom,label=$\bar{f}_m$,label.side=left,tension=0}{X,nJ2}
\fmfv{decor.shape=circle,decor.filled=shaded,decor.size=11thick}{X}
\fmfv{decor.shape=square,decor.size=3thick}{nJ1}
\fmfv{decor.shape=square,decor.size=3thick}{nJ2}
\fmfv{decor.shape=square,decor.size=3thick}{nJ4}
\end{fmfgraph*} 
\end{fmffile}
\end{minipage}
\hs{-30mm},
\vs{3mm}
\ee
%
%
where the black squares indicate amputated (or truncated) propagators. This allows us to write:
\be
\begin{minipage}{0.35\textwidth}
\begin{fmffile}{SMAff2b} 
\begin{fmfgraph*}(80,80) 
\fmfset{arrow_len}{3mm} 
\fmfset{arrow_ang}{20} 
\fmfleft{nJ1} 
\fmfright{nJ2,nJ4} 
\fmf{photon,label=$\Large \xxrightarrow[k]{}$,label.dist=3,tension=1.8}{nJ1,x1}
\fmf{phantom,label=$A_{\mu}$,label.side=left,tension=0}{nJ1,x1}
\fmfv{decor.shape=circle,decor.filled=30,decor.size=6thick}{x1}
\fmf{photon,tension=0.8,label=$A_{\nu}$,label.side=left}{x1,X}
\fmf{fermion,tension=0.9}{x4,X}
\fmf{phantom,label=$\bar{f}_l$,label.side=right,label.dist=4,tension=0}{x4,X}
\fmfv{decor.shape=circle,decor.filled=30,decor.size=6thick}{x4}
\fmf{fermion,label=\rotatebox{57}{$\Large\xxleftarrow[p]{}$},label.side=left,label.dist=-3,label.angle=30,tension=1.5}{nJ4,x4}
\fmf{phantom,label=$\bar{f}_i$,label.side=right,label.dist=4,tension=0}{nJ4,x4}
\fmf{fermion,label=\rotatebox{-57}{$\Large\xxrightarrow[p']{}$},label.side=right,label.dist=-5,label.angle=30,tension=1.5}{x2,nJ2}
\fmf{phantom,label=$f_j$,label.side=left,label.dist=4,tension=0}{x2,nJ2}
\fmfv{decor.shape=circle,decor.filled=30,decor.size=6thick}{x2}
\fmf{fermion,tension=0.9}{X,x2} 
\fmf{phantom,label=$f_m$,label.side=right,label.dist=4,tension=0}{x2,X}
\fmfv{decor.shape=circle,decor.filled=shaded,decor.size=11thick}{X}
\end{fmfgraph*} 
\end{fmffile}
\end{minipage}
\hs{-25mm}
= \sum\limits_{m,l} G^{AA}_{\mu\nu}(k) \, G^{f\bar{f}}_{jm}(p^{\prime}) \, i \Gamma^{A\bar{f}f}_{\nu ml}(k,-p^{\prime},p) \, G^{f\bar{f}}_{li}(p).
\vs{2mm}
\ee
Finally, the Slavnov operator sometimes generates two fields in the same point in position space (cf. eq. \ref{App-WI:eq:SMAff,slav} below). We represent the composite operator in momentum space with a white star; for example,%
\fn{In this way, the white star includes possible factors, sums and matrices (both in the flavour space and in the Dirac space) included in the composite operator.}
\vs{-2mm}
\be
\hs{15mm}
\sum\limits_{l}
\dfrac{i g}{\sqrt{2}}  V_{il}^{\dagger}  \, \gamma_{\mathrm{R}} \int d^4z \, e^{-i p z} \ppo \bar{d}_l (z) \, c^{-}(z) \ppc \, \, =
\hs{12mm}
\begin{minipage}{0.35\textwidth}
\begin{fmffile}{comb1} 
\begin{fmfgraph*}(80,80) 
\fmfset{arrow_len}{3mm} 
\fmfset{arrow_ang}{20} 
\fmfleft{nJ1}
\fmfright{nJ2}
\fmflabel{$\vs{-1mm}{\Large \xxrightarrow[p]{}}$ {\footnotesize$il$} \hs{-2mm}}{nJ1}
\fmf{phantom,tension=20}{nJ1,x1} 
\fmf{phantom,tension=3}{x2,nJ2} 
\fmf{fermion,right=0.8,label=$d_l$}{x1,x2} 
\fmf{ghost,label=$c^-$,right=0.8}{x2,x1}
\fmfv{decor.shape=pentagram,decor.filled=empty,decor.size=6thick}{x1}
\fmfv{decor.shape=circle,decor.filled=70,decor.size=12thick}{x2}
\end{fmfgraph*} 
\end{fmffile}
\end{minipage}
\hs{-33mm}
\times 
(2 \pi)^4 \delta^4(p).
\ee

\section{Derivation of the Ward identity}

We start by considering BRST transformation ($\delta_{\mathrm{BRST}}$) and the Slavnov operator ($s$), related through
$ \delta_{\mathrm{BRST}} (\phi) = \theta \, s(\phi)$,
for any field $\phi$ and any Grassman number $\theta$.
As suggested above, complete GFs are invariant under a BRST transformation;
we want to choose the GF that, when subject to a BRST transformation, ends up yielding the 1PI GF for $A \bar{f}_i f_j$. Now, we know that \cite{Romao:2012pq,Denner:2019vbn}%
\fn{These relations hold in every theory with the same gauge sector as that of the SM.
Ref. \cite{Romao:2012pq} omitted CKM elements, but we must consider them here explicitly.}
%
\bs
\label{App-WI:eq:SMAff,slav}
\bea
\label{App-WI:eq:8a}
&& s \, \bar{c}_A(x) = \dfrac{\eta_G}{\xi_A}  \partial_{\mu}^x A_{\mu}(x) , \\
&& s \, f_j(y) = \sum\limits_m \Bigg(i \eta \dfrac{e}{\sqrt{2} s_{\text{w}}} F_{jm} c_{\pm}(y) \gamma_{\mathrm{L}} \psi_m(y) + i \eta_e e Q_m \delta_{jm} c_A(y) f_m(y) \nonumber \\[-5mm]
&& \hspace{50mm} - i \eta \eta_Z \dfrac{e}{s_{\text{w}} c_{\text{w}}} \delta_{jm} c_Z(y) \Big[ Q_m s_{\text{w}}^2 - T_3^m \gamma_{\mathrm{L}} \Big] f_m(y) \Bigg),
\label{App-WI:eq:comp1}
\\
&& s \, \bar{f}_i(z) = \sum\limits_l \Bigg(i \eta  \dfrac{e}{\sqrt{2} s_{\text{w}}} F^{\dagger}_{il} \bar{\psi}_l (z) \gamma_{\mathrm{R}}  c_{\mp}(z) + i \eta_e e Q_l \delta_{il} \bar{f}_l(z) c_A(z) \nonumber \\[-5mm]
&& \hspace{50mm}  -  i \eta \eta_Z \dfrac{e}{s_{\text{w}} c_{\text{w}}} \delta_{il} \bar{f}_l(z) \Big[ Q_l s_{\text{w}}^2 - T_3^l \gamma_{\mathrm{R}} \Big] c_Z(z) \Bigg).
\label{App-WI:eq:comp2}
\eea
\es
%
Here, the fields $c_a$ and $\bar{c}_a$ (with $a = A, Z, \pm$) represent ghosts and anti-ghosts,
respectively;
\begin{table}[!h]
\begin{normalsize}
\normalsize
\begin{center}
\begin{tabular}
{@{\hspace{3mm}}
>{\centering\arraybackslash}p{2.7cm}
>{\centering\arraybackslash}p{2.7cm}
>{\centering\arraybackslash}p{1cm}
>{\centering\arraybackslash}p{0.8cm}
>{\centering\arraybackslash}p{0.8cm}@{\hspace{3mm}}}
\hlinewd{1.1pt}
$f$ & $\psi$ & $F_{ij}$ & $c_{\pm}$ & $c_{\mp}$ \\
\hline\\[-3.7mm]
up-type quark & down-type quark & $V_{ij}$ & $c_+$ & $c_-$ \\[1.4mm]
down-type quark & up-type quark & $V_{ij}^{\dagger}$ & $c_-$ & $c_+$ \\[1.4mm]
charged lepton & neutrino & $\delta_{ij}$ & $c_-$ & $c_+$ \\[1.4mm]
neutrino & charged lepton & $\delta_{ij}$ & $c_+$ & $c_-$ \\
\hlinewd{1.1pt}
\end{tabular}
\end{center}
\vspace{-5mm}
\end{normalsize}
\caption{Possible meanings of the general quantities $f$, $\psi$, $F_{ij}$, $c_{\pm}$ and $c_{\mp}$ introduced in eqs. \ref{App-WI:eq:SMAff,slav}.}
\label{App-WI:table:code}
\end{table}
\normalsize
we are using a general notation to account for all types of fermions, as is explained in table \ref{App-WI:table:code} (where $V_{ij}$ represents the CKM matrix).
Then, we choose:
\vs{-3mm}
\bs
\bea
\hs{-3mm} && \hs{-3mm} \delta_{\mathrm{BRST}} \,\ppo \bar{c}_A(x) \, f_j(y) \, \bar{f_i}(z) \ppc = 0 \\[2mm]
\hs{-3mm} \Leftrightarrow 
&& \hs{-3mm} \Big \langle \theta \Big[ s \, \bar{c}_A(x) \Big] \, f_j(y) \, \bar{f_i}(z) \Big \rangle
+
\Big \langle \bar{c}_A(x) \, \theta \Big[ s \, f_j(y) \Big] \, \bar{f_i}(z) \Big \rangle
+ 
\Big \langle \bar{c}_A(x) \, f_j(y) \, \theta \Big[ s \, \bar{f_i}(z)\Big]  \Big \rangle = 0 \\[2mm]
\hs{-3mm} \Leftrightarrow &&
\hs{-3mm} \Big \langle \Big[ s \, \bar{c}_A(x) \Big] f_j(y) \, \bar{f_i}(z) \Big \rangle
=
\Big \langle \bar{c}_A(x) \Big[ s \, f_j(y) \Big] \bar{f_i}(z) \Big \rangle
- \, \Big \langle \bar{c}_A(x) \, f_j(y) \Big[ s \, \bar{f_i}(z)\Big] \Big \rangle.
\label{App-WI:eq:SMAffmaster}
\eea
\es

Replacing eqs. \ref{App-WI:eq:SMAff,slav} in eq. \ref{App-WI:eq:SMAffmaster} and going to momentum space, we get:
\be
\dfrac{\eta_G}{\xi_A} i k_{\mu} G^{Af\bar{f}}_{\mu ji} = G^{\bar{c}_Asf\,\bar{f}}_{1ji} + G^{\bar{c}_Asf\,\bar{f}}_{2ji} + G^{\bar{c}_Asf\,\bar{f}}_{3ji}
- \Big(G_{1ji}^{\bar{c}_Af\,s\bar{f}} +G_{2ji}^{\bar{c}_Af\,s\bar{f}} + G_{3ji}^{\bar{c}_Af\,s\bar{f}} \Big), 
\label{App-WI:eq:70}
\ee
where we omitted the momentum arguments. As explained before, $G^{Af\bar{f}}_{\mu ji}$ is given by eq. \ref{App-WI:eq:my109} and\fn{Note the appearance of a minus sign, due to an anticommutation of ghost fields, which is required in order to have a consistent definition of the ghost propagator as $(2 \pi)^4 \delta^4(p) G^{c_a\bar{c}_{a^{\prime}}}(p) = \mathrm{FT} \big[\ppo c_a(x) \bar{c}_{a^{\prime}}(y)\ppc\big] = - \mathrm{FT} \big[\ppo \bar{c}_{a^{\prime}}(y) c_a(x) \ppc\big]$.
}
%
%
%
\begingroup\makeatletter\def\f@size{9.5}\check@mathfonts
\def\maketag@@@#1{\hbox{\m@th\normalsize\normalfont#1}}
\makeatother
\bs
\label{App-WI:eq:my11}
\begin{flalign}
\label{App-WI:eq:C11a}
&(2 \pi)^4 \delta^4(k-p^{\prime}+p) G^{\bar{c}_Asf\,\bar{f}}_{1ji}(k,-p^{\prime},p) = \sum\limits_m (- i) \eta \dfrac{e}{\sqrt{2} s_{\text{w}}}  V_{jm} \, \gamma_{\mathrm{L}} \, \mathrm{FT}\Big[\ppo c^{+}(y) \,  \bar{c}_A(x) \, \psi_m(y) \, \bar{f}_i (z) \ppc \Big],& \\
&(2 \pi)^4 \delta^4(k-p^{\prime}+p) G^{\bar{c}_Asf\,\bar{f}}_{2ji}(k,-p^{\prime},p) = \sum\limits_m (- i) \eta_e e Q_m \, \delta_{jm}  \, \mathrm{FT}\Big[\ppo c_A(y) \, \bar{c}_A(x) \,  f_m(y) \, \bar{f}_i (z) \ppc \Big],&
\label{App-WI:eq:my116b} \\
&(2 \pi)^4 \delta^4(k-p^{\prime}+p) G^{\bar{c}_Asf\,\bar{f}}_{3ji}(k,-p^{\prime},p) = \sum\limits_m i \eta \eta_Z \dfrac{e}{s_{\text{w}} c_{\text{w}}} \, \delta_{jm} \big(Q_m s_{\text{w}}^2 - T_3^m \gamma_{\mathrm{L}}\big) \, \mathrm{FT}\bigg[\Big \langle c_Z(y) \, \bar{c}_A(x) \, f_m(y) \, \bar{f}_i (z) \Big \rangle  \bigg],& \\
\label{App-WI:eq:C11d}
&(2 \pi)^4 \delta^4(k-p^{\prime}+p) G_{1ji}^{\bar{c}_Af\,s\bar{f}}(k,-p^{\prime},p) = \sum\limits_{l} (-i) \eta \dfrac{e}{\sqrt{2} s_{\text{w}}}  V_{il}^{\dagger} \, \mathrm{FT} \Big[\ppo c^{-}(z) \, \bar{c}_A(x) \, f_j(y) \, \bar{\psi}_l (z) \ppc \Big] \gamma_{\mathrm{R}},& \\
&(2 \pi)^4 \delta^4(k-p^{\prime}+p) G_{2ji}^{\bar{c}_Af\,s\bar{f}}(k,-p^{\prime},p) = \sum\limits_{l} (-i) \eta_e e Q_l \, \delta_{il} \, \mathrm{FT}\Big[\ppo   c_A(z) \, \bar{c}_A(x) \, f_j(y) \, \bar{f}_l (z)\ppc \Big],&
\label{App-WI:eq:my116e} \\
&(2 \pi)^4 \delta^4(k-p^{\prime}+p) G_{3ji}^{\bar{c}_Af\,s\bar{f}}(k,-p^{\prime},p) = \sum\limits_{l} i \eta \eta_Z \dfrac{e}{s_{\text{w}} c_{\text{w}}}  \, \delta_{il} \,  \mathrm{FT}\bigg[\Big \langle c_Z(z) \, \bar{c}_A(x) \, f_j(y) \, \bar{f}_l (z) \Big \rangle \bigg] \big(Q_l s_{\text{w}}^2 - T_3^l \gamma_{\mathrm{R}}\big).&
\end{flalign}
\es
\endgroup
Just as in eq. \ref{App-WI:eq:my109}, we can factorize the external propagators. For example,
%
\bs
\bea
&&
\hs{-18mm}
G^{\bar{c}_Asf\,\bar{f}}_{2ji}(k,-p^{\prime},p)
=
\hs{3mm}
\begin{minipage}{0.35\textwidth}
\begin{fmffile}{SMAff13} 
\begin{fmfgraph*}(70,70) 
\fmfset{arrow_len}{3mm} 
\fmfset{arrow_ang}{20} 
\fmfleft{nJ3} 
\fmfright{nJ2,nJ4}
\fmflabel{{\small$jm$}\hspace{-2mm}\rotatebox{-52}{$\Large\xxrightarrow[p']{}$}\hspace{-1mm}}{nJ2}
\fmf{ghost,label=$\Large\xxrightarrow[k]{}$,label.side=right,tension=3}{nJ3,nJ1nJ2nJ4}
\fmf{phantom,label=$c_A$,label.side=left,label.dist=3}{nJ3,nJ1nJ2nJ4}
\fmf{fermion,label=\rotatebox{52}{$\Large\xxleftarrow[p]{}$},label.dist=-5,label.angle=30,label.side=left,tension=3}{nJ4,nJ1nJ2nJ4}
\fmf{phantom,label=$\bar{f}_i$,label.side=right,tension=0}{nJ4,nJ1nJ2nJ4}
\fmf{ghost,label=$c_A$,label.side=left,label.dist=3,left=0.8,tension=0}{nJ1nJ2nJ4,nJ2}
\fmf{fermion,label=$f_m$,label.side=right,label.dist=3,tension=3}{nJ1nJ2nJ4,nJ2}
\fmfv{decor.shape=circle,decor.filled=70,decor.size=11thick}{nJ1nJ2nJ4}
\fmfv{decor.shape=pentagram,decor.filled=empty,decor.size=6thick}{nJ2}
\end{fmfgraph*} 
\end{fmffile}
\end{minipage}
\hs{-26mm}
=
\hs{2mm}
\begin{minipage}{0.35\textwidth}
\begin{fmffile}{SMAff14} 
\begin{fmfgraph*}(80,68) 
\fmfset{arrow_len}{3mm} 
\fmfset{arrow_ang}{20} 
\fmfleft{nJ3} 
\fmfright{nJ2,nJ4}
\fmflabel{{\small$jm$}\hspace{-2mm}\rotatebox{-52}{$\Large\xxrightarrow[p']{}$}\hspace{-1mm}}{nJ2}
\fmf{ghost,label=$\Large\xxrightarrow[k]{}$,label.side=right,tension=3}{nJ3,x1}
\fmf{phantom,label=$c_A$,label.side=left,label.dist=3}{nJ3,x1}
\fmfv{decor.shape=circle,decor.filled=30,decor.size=6thick}{x1}
\fmf{ghost,label=$c_A$,label.side=left,label.dist=3,tension=2}{x1,nJ1nJ2nJ4}
\fmf{fermion,label=\rotatebox{52}{$\Large\xxleftarrow[p]{}$},label.dist=-5,label.angle=30,label.side=left,tension=3}{nJ4,x2}
\fmf{phantom,label=$\bar{f}_i$,label.side=right,tension=0}{nJ4,x2}
\fmfv{decor.shape=circle,decor.filled=30,decor.size=6thick}{x2}
\fmf{fermion,tension=2,label=$\bar{f}_l$}{x2,nJ1nJ2nJ4}
\fmf{ghost,label=$c_A$,label.side=left,label.dist=3,left=0.8,tension=1.2}{nJ1nJ2nJ4,nJ2}
\fmf{fermion,label=$f_m$,label.side=right,label.dist=3,tension=0.1}{nJ1nJ2nJ4,nJ2}
\fmfv{decor.shape=circle,decor.filled=shaded,decor.size=11thick}{nJ1nJ2nJ4}
\fmfv{decor.shape=pentagram,decor.filled=empty,decor.size=6thick}{nJ2}
\end{fmfgraph*} 
\end{fmffile}
\end{minipage}
\hs{-24mm}
+
\hs{3mm}
\begin{minipage}{0.35\textwidth}
\begin{fmffile}{SMAff15} 
\begin{fmfgraph*}(80,68) 
\fmfset{arrow_len}{3mm} 
\fmfset{arrow_ang}{20} 
\fmfleft{nJ3} 
\fmfright{nJ2,nJ4}
\fmflabel{{\small$jm$}\hspace{-2mm}\rotatebox{-52}{$\Large\xxrightarrow[p']{}$}\hspace{-1mm}}{nJ2}
\fmf{ghost,label=$\Large\xxrightarrow[k]{}$,label.side=right,tension=3}{nJ3,x1}
\fmf{phantom,label=$c_A$,label.side=left,label.dist=3}{nJ3,x1}
\fmfv{decor.shape=circle,decor.filled=30,decor.size=6thick}{x1}
\fmf{ghost,label=$c_Z$,label.side=left,label.dist=3,tension=2}{x1,nJ1nJ2nJ4}
\fmf{fermion,label=\rotatebox{52}{$\Large\xxleftarrow[p]{}$},label.dist=-5,label.angle=30,label.side=left,tension=3}{nJ4,x2}
\fmf{phantom,label=$\bar{f}_i$,label.side=right,tension=0}{nJ4,x2}
\fmfv{decor.shape=circle,decor.filled=30,decor.size=6thick}{x2}
\fmf{fermion,tension=2,label=$\bar{f}_l$}{x2,nJ1nJ2nJ4}
\fmf{ghost,label=$c_A$,label.side=left,label.dist=3,left=0.8,tension=1.2}{nJ1nJ2nJ4,nJ2}
\fmf{fermion,label=$f_m$,label.side=right,label.dist=3,tension=0.1}{nJ1nJ2nJ4,nJ2}
\fmfv{decor.shape=circle,decor.filled=shaded,decor.size=11thick}{nJ1nJ2nJ4}
\fmfv{decor.shape=pentagram,decor.filled=empty,decor.size=6thick}{nJ2}
\end{fmfgraph*} 
\end{fmffile}
\end{minipage}
\hs{-25mm},
\\[16mm]
&&
\hs{-18mm}
G_{2ji}^{\bar{c}_Af\,s\bar{f}}(k,-p^{\prime},p) =
\hs{3mm}
\begin{minipage}{0.35\textwidth}
\begin{fmffile}{SMAff4} 
\begin{fmfgraph*}(70,70) 
\fmfset{arrow_len}{3mm} 
\fmfset{arrow_ang}{20} 
\fmfleft{nJ3} 
\fmfright{nJ2,nJ4}
\fmflabel{{\small$il$} \vspace{-1mm}\hspace{-2mm}\rotatebox{52}{$\Large\xxleftarrow[p]{}$}\hspace{-2mm}}{nJ4}
\fmf{ghost,label=$\Large\xxrightarrow[k]{}$,label.side=right,tension=3}{nJ3,nJ1nJ2nJ4}
\fmf{phantom,label=$c_A$,label.side=left,label.dist=3}{nJ3,nJ1nJ2nJ4}
\fmf{fermion,label=$\bar{f}_l$,label.side=left,label.dist=3,tension=3}{nJ4,nJ1nJ2nJ4}
\fmf{ghost,label=$c_A$,label.side=left,label.dist=4,left=0.8,tension=0}{nJ1nJ2nJ4,nJ4}
\fmf{fermion,label=\rotatebox{-52}{$\Large\xxrightarrow[p']{}$},label.side=right,label.dist=-5,label.angle=30,tension=3}{nJ1nJ2nJ4,nJ2}
\fmf{phantom,label=$f_j$,label.side=left,tension=0}{nJ1nJ2nJ4,nJ2}
\fmfv{decor.shape=circle,decor.filled=70,decor.size=11thick}{nJ1nJ2nJ4}
\fmfv{decor.shape=pentagram,decor.filled=empty,decor.size=6thick}{nJ4}
\end{fmfgraph*} 
\end{fmffile}
\end{minipage}
\hs{-26mm}
=
\hs{2mm}
\begin{minipage}{0.35\textwidth}
\begin{fmffile}{SMAff5} 
\begin{fmfgraph*}(80,68) 
\fmfset{arrow_len}{3mm} 
\fmfset{arrow_ang}{20} 
\fmfleft{nJ3} 
\fmfright{nJ2,nJ4}
\fmflabel{{\small$il$} \vspace{-1mm}\hspace{-2mm}\rotatebox{52}{$\Large\xxleftarrow[p]{}$}\hspace{-2mm}}{nJ4}
\fmf{ghost,label=$\Large\xxrightarrow[k]{}$,label.side=right,tension=3}{nJ3,x1}
\fmf{phantom,label=$c_A$,label.side=left,label.dist=3}{nJ3,x1}
\fmfv{decor.shape=circle,decor.filled=30,decor.size=6thick}{x1}
\fmf{ghost,label=$c_A$,label.side=left,label.dist=3,tension=1.9}{x1,nJ1nJ2nJ4}
\fmf{fermion,label=$\bar{f}_l$,label.side=left,label.dist=3,tension=1.2}{nJ4,nJ1nJ2nJ4}
\fmf{ghost,label=$c_A$,label.side=left,label.dist=4,left=0.8,tension=0}{nJ1nJ2nJ4,nJ4}
\fmf{fermion,tension=2,label=$f_m$}{nJ1nJ2nJ4,x2}
\fmfv{decor.shape=circle,decor.filled=30,decor.size=6thick}{x2}
\fmf{fermion,label=\rotatebox{-52}{$\Large\xxrightarrow[p']{}$},label.side=right,label.dist=-5,label.angle=30,tension=3}{x2,nJ2}
\fmf{phantom,label=$f_j$,label.side=left,tension=0}{x2,nJ2}
\fmfv{decor.shape=circle,decor.filled=shaded,decor.size=11thick}{nJ1nJ2nJ4}
\fmfv{decor.shape=pentagram,decor.filled=empty,decor.size=6thick}{nJ4}
\end{fmfgraph*} 
\end{fmffile}
\end{minipage}
\hs{-24mm}
+
\hs{3mm}
\begin{minipage}{0.35\textwidth}
\begin{fmffile}{SMAff6} 
\begin{fmfgraph*}(80,68) 
\fmfset{arrow_len}{3mm} 
\fmfset{arrow_ang}{20} 
\fmfleft{nJ3} 
\fmfright{nJ2,nJ4}
\fmflabel{{\small$il$} \vspace{-1mm}\hspace{-2mm}\rotatebox{52}{$\Large\xxleftarrow[p]{}$}\hspace{-2mm}}{nJ4}
\fmf{ghost,label=$\Large\xxrightarrow[k]{}$,label.side=right,tension=3}{nJ3,x1}
\fmf{phantom,label=$c_A$,label.side=left,label.dist=3}{nJ3,x1}
\fmfv{decor.shape=circle,decor.filled=30,decor.size=6thick}{x1}
\fmf{ghost,label=$c_Z$,label.side=left,label.dist=3,tension=1.9}{x1,nJ1nJ2nJ4}
\fmf{fermion,label=$\bar{f}_l$,label.side=left,label.dist=3,tension=1.2}{nJ4,nJ1nJ2nJ4}
\fmf{ghost,label=$c_A$,label.side=left,label.dist=4,left=0.8,tension=0}{nJ1nJ2nJ4,nJ4}
\fmf{fermion,tension=2,label=$f_m$}{nJ1nJ2nJ4,x2}
\fmfv{decor.shape=circle,decor.filled=30,decor.size=6thick}{x2}
\fmf{fermion,label=\rotatebox{-52}{$\Large\xxrightarrow[p']{}$},label.side=right,label.dist=-5,label.angle=30,tension=3}{x2,nJ2}
\fmf{phantom,label=$f_j$,label.side=left,tension=0}{x2,nJ2}
\fmfv{decor.shape=circle,decor.filled=shaded,decor.size=11thick}{nJ1nJ2nJ4}
\fmfv{decor.shape=pentagram,decor.filled=empty,decor.size=6thick}{nJ4}
\end{fmfgraph*} 
\end{fmffile}
\end{minipage}
\hs{-25mm},
\\[-1mm] \nonumber
\eea
\es
and in an equivalent way for the remaining GFs. Finally, defining:
\vs{2mm}
\bea
i \Gamma_{2jl}^{c_As\bar{f}\,f}(k,-p',p) \, \equiv
\quad 
\begin{minipage}{0.35\textwidth}
\begin{fmffile}{SMAff22}
\begin{fmfgraph*}(60,60) 
\fmfset{arrow_len}{3mm} 
\fmfset{arrow_ang}{20} 
\fmfleft{nJ3} 
\fmfright{nJ2,nJ4}
\fmflabel{{\small$jm$}\hspace{-2mm}\rotatebox{-52}{$\Large\xxrightarrow[p']{}$}\hspace{-1mm}}{nJ2}
\fmf{ghost,label=$\Large\xxrightarrow[k]{}$,label.side=right,tension=3}{nJ3,nJ1nJ2nJ4}
\fmf{phantom,label=$c_A$,label.side=left,label.dist=3}{nJ3,nJ1nJ2nJ4}
\fmf{fermion,label=\rotatebox{52}{$\Large\xxleftarrow[p]{}$},label.dist=-5,label.angle=30,label.side=left,tension=3}{nJ4,nJ1nJ2nJ4}
\fmf{phantom,label=$f_l$,label.side=right,tension=0}{nJ4,nJ1nJ2nJ4}
\fmf{ghost,label=$\hs{1mm}c_A\hs{1mm}$,label.side=left,label.dist=1,left=0.8,tension=0}{nJ1nJ2nJ4,nJ2}
\fmf{fermion,label=$\bar{f}_m$,label.side=right,label.dist=3,tension=3}{nJ1nJ2nJ4,nJ2}
\fmfv{decor.shape=circle,decor.filled=shaded,decor.size=11thick}{nJ1nJ2nJ4}
\fmfv{decor.shape=pentagram,decor.filled=empty,decor.size=6thick}{nJ2}
\fmfv{decor.shape=square,decor.size=3thick}{nJ3}
\fmfv{decor.shape=square,decor.size=3thick}{nJ4}
\end{fmfgraph*} 
\end{fmffile}
\end{minipage}
\hs{-25mm}
,
\\[1mm] \nonumber
\eea
and in a similar way for the other GFs, we can rewrite eq. \ref{App-WI:eq:70} as:
%
\bea
& \hs{-35mm} \dfrac{\eta_G}{\xi_A} i k_{\mu} \sum\limits_{m,l} \bigg[ G^{AA}_{\mu\nu}(k) \, G^{f\bar{f}}_{jm}(p^{\prime})\, i \Gamma^{A\bar{f}f}_{\nu ml} \, G^{f\bar{f}}_{li}(p)
+ G^{AZ}_{\mu\nu}(k) \, G^{f\bar{f}}_{jm}(p^{\prime}) \, i \Gamma^{Z\bar{f}f}_{\nu ml} \, G^{f\bar{f}}_{li}(p) \nonumber \\[-1mm]
& \hs{50mm} + G^{AS}_{\mu}(k) \, G^{f\bar{f}}_{jm}(p^{\prime}) \, i \Gamma^{S\bar{f}f}_{ml}\, G^{f\bar{f}}_{li}(p) \bigg] \nonumber \\[-1mm]
& \hs{-20mm} = \sum\limits_{V}
G^{c_V\bar{c}_A}(k) \bigg[
\sum\limits_{l} 
\Big( i \Gamma_{1jl}^{c_Vs\bar{f}\,f} + i \Gamma_{2jl}^{c_Vs\bar{f}\,f} + i \Gamma_{3jl}^{c_Vs\bar{f}\,f} \Big) G^{f\bar{f}}_{li}(p) \nonumber \\[-1mm]
& \hs{40mm} - \sum\limits_{m} G^{f\bar{f}}_{jm}(p^{\prime}) \Big( i \Gamma_{1mi}^{c_V\bar{f}\,sf} + i \Gamma_{2mi}^{c_V\bar{f}\,sf} + i \Gamma_{3mi}^{c_V\bar{f}\,sf} \Big) \bigg],
\eea
%
where $V=\{A,Z\}$ and we ommited the arguments of the 3-point 1PI GFs. Multiplying from the left by ${(G^{f\bar{f}}_{rj})}^{-1}(p^{\prime})$ and from the right by ${(G^{f\bar{f}}_{is})}^{-1}(p)$, and using $ G^{f\bar{f}}_{ij}(p) \, {(G^{f\bar{f}}_{jm})}^{-1}(p) = \delta_{im}$, we get:
\bea
&& \hs{-18mm} \dfrac{\eta_G}{\xi_A} i k_{\mu} \bigg[
G^{AA}_{\mu\nu}(k) \, i \Gamma^{A\bar{f}f}_{\nu rs} +
G^{AZ}_{\mu\nu}(k) \, i \Gamma^{Z\bar{f}f}_{\nu rs} +
G^{AS}_{\mu}(k) \, i \Gamma^{S\bar{f}f}_{rs} \bigg]
\nonumber
\\[-1mm]
&& \hs{-12mm} = \sum\limits_{V}  G^{c_V\bar{c}_A}(k) \bigg[ \sum\limits_j
{(G^{f\bar{f}}_{rj})}^{-1}(p^{\prime}) \Big( i \Gamma_{1js}^{c_Vs\bar{f}\,f} + i \Gamma_{2js}^{c_Vs\bar{f}\,f}  + i \Gamma_{3js}^{c_Vs\bar{f}\,f} \Big)
\nonumber\\[-3mm]
&& \hs{40mm}
- \sum\limits_i \Big( i \Gamma_{1ri}^{c_V\bar{f}\,sf} + i \Gamma_{2ri}^{c_V\bar{f}\,sf} + i \Gamma_{3ri}^{c_V\bar{f}\,sf}\Big) {(G^{f\bar{f}}_{is})}^{-1}(p)
\bigg].
\label{App-WI:eq:basis}
\eea
Now, we restrict this expression by imposing several conditions. In the first place, we consider it at one-loop order only; this leads us to expand the different terms in eq. \ref{App-WI:eq:basis}. For example,
%
%
\be
\bigg(G^{AA}_{\mu\nu}(k) \, i \Gamma^{A\bar{f}f}_{\nu rs} \bigg)\Big|_{\text{1L}} = 
{G^{AA}_{\mu\nu}(k)}{\Big|_{\text{tr}}} \, i \Gamma^{A\bar{f}f}_{\nu rs} \Big|_{\text{1L}} +
{G^{AA}_{\mu\nu}(k)}{\Big|_{\text{\scriptsize 1L}}} \, i \Gamma^{A\bar{f}f}_{\nu rs}\Big|_{\text{\scriptsize tr}},
\ee
where `1L' and `tr'  represent one-loop level and tree-level, respectively. Eq. \ref{App-WI:eq:basis} is thus unfolded in a multiplicity of terms. Yet, some of them vanish; for example, there are no mixed propagators at tree-level, so that ${G^{AZ}_{\mu\nu}(k)}{\Big|_{\text{tr}}} = {G^{c_Z\bar{c}_A}(k)}{\Big|_{\text{tr}}}  =
{G^{AS}_{\mu}(k)}{\Big|_{\text{tr}}} =
0$; what is more, some of the 1PI GFs vanish, as can be seen in table \ref{App-WI:table:diags},
\begin{table}[!h]
\begin{normalsize}
\normalsize
\begin{center}
\begin{tabular}
{@{\hspace{3mm}}
>{\raggedright\arraybackslash}p{2.5cm}
>{\centering\arraybackslash}p{1.5cm}
>{\centering\arraybackslash}p{1.3cm}@{\hspace{3mm}}}
\hlinewd{1.1pt}
 & tree-level & one-loop\\
\hline\\[-3.7mm]
$\Gamma_{1js}^{c_As\bar{f}\,f}, \, \Gamma_{1ri}^{c_A\bar{f}\,sf}$ & 0 & 6\\[1.4mm]
$\Gamma_{2js}^{c_As\bar{f}\,f}, \, \Gamma_{2ri}^{c_A\bar{f}\,sf}$ & 1 & 0\\[1.4mm]
$\Gamma_{3js}^{c_As\bar{f}\,f}, \, \Gamma_{3ri}^{c_A\bar{f}\,sf}$ & 0 & 0\\[1.4mm]
$\Gamma_{1js}^{c_Zs\bar{f}\,f}, \, \Gamma_{1ri}^{c_Z\bar{f}\,sf}$ & 0 & 6\\[1.4mm]
$\Gamma_{2js}^{c_Zs\bar{f}\,f}, \, \Gamma_{2ri}^{c_Z\bar{f}\,sf}$ & 0 & 0\\[1.4mm]
$\Gamma_{3js}^{c_Zs\bar{f}\,f}, \, \Gamma_{3ri}^{c_Z\bar{f}\,sf}$ & 1 & 0\\
\hlinewd{1.1pt}
\end{tabular}
\end{center}
\vspace{-5mm}
\end{normalsize}
\caption{Number of Feynman diagrams corresponding to the different 1PI GFs from the r.h.s. of eq. \ref{App-WI:eq:basis} at tree-level and one-loop (the one-loop diagrams of the first row are depicted in eq. \ref{App-WI:eq:my-diagrams}).}
\label{App-WI:table:diags}
\end{table}
\normalsize
where we show the number of diagrams corresponding to the different 1PI GFs from the r.h.s. of eq. \ref{App-WI:eq:basis} in both orders. Then, the first condition transforms eq. \ref{App-WI:eq:basis} into:
%
\begin{equation}
\begin{split}
\dfrac{\eta_G}{\xi_A} i k_{\mu} 
& 
\Bigg[ {G^{AA}_{\mu\nu}(k)}{\Big|_{\text{tr}}} \, i \Gamma^{A\bar{f}f}_{\nu rs} \Big|_{\text{1L}} +
{G^{AA}_{\mu\nu}(k)}{\Big|_{\text{1L}}} \, i \Gamma^{A\bar{f}f}_{\nu rs} \Big|_{\text{tr}} +
{G^{AZ}_{\mu\nu}(k)}{\Big|_{\text{1L}}} \,  i \Gamma^{Z\bar{f}f}_{\nu rs} \Big|_{\text{tr}} + {G^{AS}_{\mu}(k)}{\Big|_{\text{1L}}} \,  {\Gamma^{S\bar{f}f}_{rs}}{\Big|_{\text{tr}}} \Bigg] \\[-1mm]
\hs{-5mm} = & \hs{3mm} \sum\limits_{j} \Bigg[
{G^{c_A\bar{c}_A} (k)}{\Big|_{\text{tr}}} \, {{(G^{f\bar{f}}_{rj})}^{-1}(p^{\prime})}{\Big|_{\text{tr}}} \, {i \Gamma_{1js}^{c_As\bar{f}\,f}}{\Big|_{\text{1L}}}
+ {G^{c_A\bar{c}_A} (k)}{\Big|_{\text{tr}}} \, {{(G^{f\bar{f}}_{rj})}^{-1}(p^{\prime})}{\Big|_{\text{1L}}} \, {i \Gamma_{2js}^{c_As\bar{f}\,f}}{\Big|_{\text{tr}}} \\[-2mm]
& 
\hs{6.5mm} 
+ {G^{c_A\bar{c}_A} (k)}{\Big|_{\text{1L}}} \, {{(G^{f\bar{f}}_{rj})}^{-1}(p^{\prime})}{\Big|_{\text{tr}}} \, {i \Gamma_{2js}^{c_As\bar{f}\,f}}{\Big|_{\text{tr}}}
+ {G^{c_Z\bar{c}_A} (k)}{\Big|_{\text{1L}}} \, {{(G^{f\bar{f}}_{rj})}^{-1}(p^{\prime})}{\Big|_{\text{tr}}} \, {i \Gamma_{3js}^{c_Zs\bar{f}\,f}}{\Big|_{\text{tr}}} \Bigg] \\[-2mm]
& 
- \sum\limits_{i} \Bigg[
{G^{c_A\bar{c}_A} (k)}{\Big|_{\text{tr}}} \, {i \Gamma_{1ri}^{c_A\bar{f}\,sf}}{\Big|_{\text{1L}}} \, {{(G^{f\bar{f}}_{is})}^{-1}}(p){\Big|_{\text{tr}}}
+ {G^{c_A\bar{c}_A} (k)}{\Big|_{\text{tr}}} \, {i \Gamma_{2ri}^{c_A\bar{f}\,sf}}{\Big|_{\text{tr}}} \, {{(G^{f\bar{f}}_{is})}^{-1}}(p){\Big|_{\text{1L}}} \\[-2mm]
& \hs{6.5mm}
+ {G^{c_A\bar{c}_A} (k)}{\Big|_{\text{1L}}} \, {i \Gamma_{2ri}^{c_A\bar{f}\,sf}}{\Big|_{\text{tr}}} \, {{(G^{f\bar{f}}_{is})}^{-1}}(p){\Big|_{\text{tr}}}
+ {G^{c_Z\bar{c}_A} (k)}{\Big|_{\text{1L}}} \, {i \Gamma_{3ri}^{c_Z\bar{f}\,sf}}{\Big|_{\text{tr}}} \, {{(G^{f\bar{f}}_{is})}^{-1}(p)}{\Big|_{\text{tr}}} \Bigg] .
\end{split}
\label{App-WI:eq:torestrict}
\end{equation}
%
The second condition consists in setting the fermion with momentum $p$ OS, which involves multiplying eq. \ref{App-WI:eq:torestrict} by $v_s(-p)$ from the right and using:%
\fn{The momentum in the argument of the spinor $v$ corresponds to the momentum of the (outgoing) physical particle.}
\be
{(G^{f\bar{f}}_{is})}^{-1}(p) \Big|_{\text{tr}} v_s(-p) = (-i) (\slashed{p}-m_{f,s}) \, \delta_{is} \, v_s(-p) = 0.
\label{App-WI:eq:OS-spinors}
\ee
Therefore, eq. \ref{App-WI:eq:torestrict} becomes:
%
%
%
%
\begin{flalign}
&\dfrac{\eta_G}{\xi_A} i k_{\mu}
\Bigg[ {G^{AA}_{\mu\nu}(k)}{\Big|_{\text{tr}}} \, i \Gamma^{A\bar{f}f}_{\nu rs} \Big|_{\text{1L}} +
{G^{AA}_{\mu\nu}(k)}{\Big|_{\text{1L}}} \, i \Gamma^{A\bar{f}f}_{\nu rs} \Big|_{\text{tr}} +
{G^{AZ}_{\mu\nu}(k)}{\Big|_{\text{1L}}} \,  i \Gamma^{Z\bar{f}f}_{\nu rs} \Big|_{\text{tr}} + {G^{AS}_{\mu}(k)}{\Big|_{\text{1L}}} \,  {\Gamma^{S\bar{f}f}_{rs}}{\Big|_{\text{tr}}} \Bigg] v_s(-p) & \nonumber \\[-1mm]
= & \sum\limits_{j} \Bigg[
{G^{c_A\bar{c}_A} (k)}{\Big|_{\text{tr}}} \, {{(G^{f\bar{f}}_{rj})}^{-1}(p^{\prime})}{\Big|_{\text{tr}}} \, {i \Gamma_{1js}^{c_As\bar{f}\,f}}{\Big|_{\text{1L}}}
+ {G^{c_A\bar{c}_A} (k)}{\Big|_{\text{tr}}} \, {{(G^{f\bar{f}}_{rj})}^{-1}(p^{\prime})}{\Big|_{\text{1L}}} \, {i \Gamma_{2js}^{c_As\bar{f}\,f}}{\Big|_{\text{tr}}} & \nonumber \\[-1mm]
& 
\hs{4.5mm} 
+ {G^{c_A\bar{c}_A} (k)}{\Big|_{\text{1L}}} \, {{(G^{f\bar{f}}_{rj})}^{-1}(p^{\prime})}{\Big|_{\text{tr}}} \, {i \Gamma_{2js}^{c_As\bar{f}\,f}}{\Big|_{\text{tr}}}
+ {G^{c_Z\bar{c}_A} (k)}{\Big|_{\text{1L}}} \, {{(G^{f\bar{f}}_{rj})}^{-1}(p^{\prime})}{\Big|_{\text{tr}}} \, {i \Gamma_{3js}^{c_Zs\bar{f}\,f}}{\Big|_{\text{tr}}} \Bigg] \, v_s(-p) & \nonumber \\[-1mm]
& 
\hs{30mm} 
- \sum\limits_{i} {G^{c_A\bar{c}_A} (k)}{\Big|_{\text{tr}}} \, {i \Gamma_{2ri}^{c_A\bar{f}\,sf}}{\Big|_{\text{tr}}} \, {{(G^{f\bar{f}}_{is})}^{-1}}(p){\Big|_{\text{1L}}} \, v_s(-p). & 
\label{App-WI:eq:myC19}
\end{flalign}
%
%
%
Next, we perform the following sequence of tasks:
%
\begin{center}
\begin{addmargin}[8mm]{0mm}
a) take the derivative of eq. \ref{App-WI:eq:myC19} with respect to $k^{\mu}$ (for fixed $p^{\mu}$ and $p^{\prime \mu} = k^{\mu} + p^{\mu}$);\\
b) multiply by $k^2$;\\
c) set $k=0$;\\
d) multiply by $\bar{u}_r(p)$ from the left.\\
\end{addmargin}
\end{center}
Note that, with the step c), $p^{\prime}$ automatically becomes equal to $p$, so that it also becomes OS, which allows us to perform the last step.
Before implementing this sequence of four tasks, we must consider some identities that shall prove useful later.

\subsubsection*{Auxiliary identities}

First of all, we have the following tree-level definitions \cite{Romao:2012pq}:%
\fn{We omit the $i \varepsilon $ in the denominators of propagators. Once again, these relations are valid in all extensions of the SM that do not modify its gauge structure.}
\bs
\label{App-WI:eq:rels}
\begin{gather} 
{G^{c_A\bar{c}_A} (k)}{\Big|_{\text{tr}}} = \dfrac{\eta_G i}{k^2},
\qquad
{G^{c_Z\bar{c}_Z} (k)}{\Big|_{\text{tr}}} = \dfrac{\eta_G i}{k^2 - \xi_Z m_{\textrm{Z}}^2},
\\[2mm]
G^{SS}(k){\Big|_{\text{tr}}} = \dfrac{i}{k^2 - m_{\mathrm{S}}^2},
\qquad
{{(G^{f\bar{f}}_{rj})}^{-1}(p^{\prime})}{\Big|_{\text{tr}}} = -i (\slashed{k} + \slashed{p} - m_{f,r}) \, \delta_{rj},
\\[2mm]
{G^{AA}_{\mu\nu}(k)}{\Big|_{\text{tr}}} = -i \bigg[ \dfrac{g_{\mu\nu}}{k^2} - \left(1- \xi_A \right) \dfrac{k_{\mu} k_{\nu}}{(k^2)^2}\bigg],
\qquad
{G^{ZZ}_{\mu\nu}(k)}{\Big|_{\text{tr}}} = \dfrac{-i}{k^2 -  m_{\mathrm{Z}}^2} \bigg[ g_{\mu\nu}  - \dfrac{1 - \xi_Z}{k^2 - {\xi_Z}  m_{\mathrm{Z}}^2} \, k_{\mu} k_{\nu} \bigg],
\label{App-WI:eq:phot-tree-prop}
\\[2mm]
i \Gamma^{A\bar{f}f}_{\nu rs} \Big|_{\text{tr}} = i \eta_e e Q_r \, \gamma_{\nu} \,  \delta_{rs}, 
\qquad
i \Gamma_{2js}^{c_As\bar{f}\,f}\Big|_{\text{tr}} = - i \eta_e e Q_s \delta_{js},
\qquad
i \Gamma_{2ri}^{c_A\bar{f}\,sf} \Big|_{\text{tr}} = - i \eta_e e Q_r \delta_{ri},
\label{App-WI:eq:aux-A}
\\[2mm]
i \Gamma^{Z\bar{f}f}_{\nu rs} \Big|_{\text{tr}} = -i \eta \eta_Z \dfrac{e}{s_{\text{w}} c_{\text{w}}} \gamma_{\nu} \, \delta_{rs} \, \left(g_V^r - g_A^r \gamma_5 \right)
=
i \eta \eta_Z \, \delta_{rs} \left( \dfrac{e \, s_{\text{w}}}{c_{\text{w}}} Q_r \, \gamma_{\nu} - \dfrac{e}{s_{\text{w}} c_{\text{w}}} T_3^r \, \gamma_{\nu} \gamma_{\mathrm{L}}\right),
\label{App-WI:eq:Zrs}
\\[2mm]
i \Gamma_{3js}^{c_Zs\bar{f}\,f} \Big|_{\text{tr}}
=
i \eta \eta_Z \, \delta_{js} \left( \dfrac{e \, s_{\text{w}}}{c_{\text{w}}} Q_j - \dfrac{e}{s_{\text{w}} c_{\text{w}}} T_3^j \, \gamma_{\mathrm{L}}\right),
\label{App-WI:eq:G3}
\end{gather}
\es
where we used:
\be
g_V^r = \dfrac{1}{2} T_3^r - s_{\text{w}}^2 Q_r,
\qquad
g_A^r = \dfrac{1}{2} T_3^r.
\label{App-WI:eq:gVf-gAf}
\ee

Besides, we define two new $\eta$ variables, $\eta_f$ and $\eta_g$ (not included in ref. \cite{Romao:2012pq}), which are either $1$ or $-1$ according to different conventions. They are such that:%
\fn{As shall be clear below, it is irrelevant for the present  derivation whether reducible diagrams with one-loop tadpoles are considered in $i \hat{\Sigma}^{\bar{f}f}_{ij}$ and $i \hat{\Sigma}_{\mu \nu}^{V V^{\prime}}$ or not.}
\bs
\bea
\label{App-WI:eq:my-ferm-trick}
-{(\hat{G}^{f\bar{f}}_{ij})}^{-1}(p){\Big|_{\text{1L}}} = i \hat{\Gamma}^{\bar{f}f}_{ij}(p){\Big|_{\text{1L}}} &=& 
\eta_f \, i \hat{\Sigma}^{\bar{f}f}_{ij}(p), \\[2mm]
i \hat{\Gamma}^{V V^{\prime}}_{\mu \nu} (k) {\Big|_{\text{1L}}} &=& \eta_{g} \, i \hat{\Sigma}_{\mu \nu}^{V V^{\prime}} (k),
\label{App-WI:eq:etag}
\eea
\es
in such a way that:
\bs
\bea
i \hat{\Sigma}^{\bar{f}f}_{ij}(p) \hs{-2mm} &=& \hs{-2mm} i \Sigma^{\bar{f}f}_{ij}(p) + 
i \, \eta_f 
\Big(
F^{\mathrm{V}}_{\mathrm{L}ij} \, \, \slashed{p} \gamma_{\mathrm{L}}
+ F^{\mathrm{V}}_{\mathrm{R}ij} \, \, \slashed{p} \gamma_{\mathrm{R}}
-  F^{\mathrm{S}}_{\mathrm{L}ij} \, \gamma_{\mathrm{L}}
-  F^{\mathrm{S}}_{\mathrm{R}ij} \, \gamma_{\mathrm{R}}
\Big),
\label{App-WI:eq:ffCT}
\\
i \hat{\Sigma}_{\mathrm{T}}^{V V^{\prime}}(k^2) \hs{-2mm} &=& \hs{-2mm} i \Sigma_{\mathrm{T}}^{V V^{\prime}}(k^2) + i \, \eta_{g} \left[ - \frac{1}{2}\left(k^{2}- m_{\mathrm{V}}^{2}\right) \delta Z_{V V^{\prime}}   - \frac{1}{2}\left(k^{2}-m_{V^{\prime}}^{2}\right) \delta Z_{V^{\prime} V} + \delta_{V V^{\prime}} \delta  m_{\mathrm{V}}^{2} \right], \hspace{10mm}
\label{App-WI:eq:ZACT}
\eea
\es
with%
\fn{On the definition of $\delta m_{f, i}^{\mathrm{L}}$ and $\delta m_{f, i}^{\mathrm{R}}$, cf. appendix \ref{App-Fermions}.}
\be
\begin{aligned}
&& F_{\mathrm{L}ij}^{\mathrm{V}}=\frac{1}{2}\left(\delta Z_{i j}^{f, L}+\delta Z_{i j}^{f, L \dagger}\right), \\
&& F_{\mathrm{R}ij}^{\mathrm{V}}=\frac{1}{2}\left(\delta Z_{i j}^{f, R}+\delta Z_{i j}^{f, R \dagger}\right),
\end{aligned}
\qquad
\begin{aligned}
&& F_{\mathrm{L}ij}^{\mathrm{S}}=m_{f, i} \frac{1}{2} \delta Z_{i j}^{f, L}+m_{f, j} \frac{1}{2} \delta Z_{i j}^{f, R \dagger}+\delta_{i j} \, \delta m_{f, i}^{\mathrm{L}}, \\
&& F_{\mathrm{R}ij}^{\mathrm{S}}=m_{f, i} \dfrac{1}{2} \delta Z_{i j}^{f, R}+m_{f, j} \frac{1}{2} \delta Z_{i j}^{f, L \dagger}+\delta_{i j} \, \delta m_{f, i}^{\mathrm{R}}.
\end{aligned}
\label{App-WI:eq:ferm-aux}
\ee

Finally, we consider an auxiliary Ward identity. We start by noting that:
\be
\delta_{\mathrm{BRST}} \,\ppo A_{\mu}(x) \, \bar{c}_A(y) \ppc = 0,
\qquad
\delta_{\mathrm{BRST}} \,\ppo Z_{\mu}(x) \, \bar{c}_A(y) \ppc = 0,
\ee
with \cite{Romao:2012pq}
\bea
s A_{\mu}(x) &=& -\partial_{\mu} c_{A}(x) -i \eta_{e} e \Big[W_{\mu}^{+}(x) c^{-}(x) - W_{\mu}^{-}(x) c^{+}(x)\Big], \\
s Z_{\mu}(x) &=& -\partial_{\mu} c_{Z}(x) -i \eta \eta_{Z} \, e \dfrac{c_{\textrm{w}}}{s_{\textrm{w}}} \Big[W_{\mu}^{+}(x) c^{-}(x) - W_{\mu}^{-}(x) c^{+}(x)\Big].
\eea
So, using eq. \ref{App-WI:eq:8a}, going to momentum space, combining the two relations, considering the expressions at one-loop level and using the well-known fact that the photon propagator does not get longitudinal corrections at higher orders \cite{Bohm:2001yx}, we find the following WI:
\be
k_{\mu} \, G^{c_A\bar{c}_A}(k){\Big|_{\text{1L}}} = 
\eta \eta_Z \eta_e \dfrac{s_{\textrm{w}}}{c_{\textrm{w}}}  \left( k_{\mu} G^{c_Z\bar{c}_A}(k){\Big|_{\text{1L}}} - \eta_G \dfrac{1}{\xi_A} \, k_{\nu} \, G_{\mu\nu}^{AZ}(k){\Big|_{\text{1L}}} \right).
\label{App-WI:eq:mae}
\ee
Now, we have:
\bea
k_{\mu} \, G^{c_Z\bar{c}_A}(k) \Big|_{\text{1L}}
\!\!\!\!&=&\!\!\!\!
k_{\mu} \hs{1mm}
\begin{minipage}{0.35\textwidth}
\begin{fmffile}{SMdecom3} 
\begin{fmfgraph*}(70,50) 
\fmfset{arrow_len}{3mm} 
\fmfset{arrow_ang}{20}
\fmfleft{nJ1} 
\fmfright{nJ2} 
\fmf{ghost,label=$\Large \xxrightarrow[k]{}$,label.side=right,label.dist=2,tension=2}{nJ1,x}
\fmf{phantom,label=$c_A$,label.side=left,tension=0}{nJ1,x}
\fmf{ghost,label=$c_Z$,label.side=left,tension=2}{x,nJ2} 
\fmfv{decor.shape=circle,decor.filled=30,decor.size=11thick,label=$\hs{-4mm}\text{1L}$}{x}
\end{fmfgraph*} 
\end{fmffile}
\end{minipage}
\hs{-30mm}
=
k_{\mu} \hs{1mm}
\begin{minipage}{0.35\textwidth}
\begin{fmffile}{SMdecom4} 
\begin{fmfgraph*}(130,50) 
\fmfset{arrow_len}{3mm} 
\fmfset{arrow_ang}{20}
\fmfleft{nJ1} 
\fmfright{nJ2} 
\fmf{ghost,label=$\Large \xxrightarrow[k]{}$,label.side=right,label.dist=2,tension=3}{nJ1,x1}
\fmf{phantom,label=$c_A$,label.side=left,tension=0}{nJ1,x1}
\fmf{ghost,label=$c_A$,label.side=left,tension=2}{x1,x}
\fmf{phantom,label=$\hs{-12mm}\text{\small tr}$,label.side=left,label.dist=0,tension=0}{x1,x}
\fmf{ghost,label=$c_Z$,label.side=left,tension=2}{x,x2} 
\fmf{ghost,label=$c_Z$,label.side=left,tension=3}{x2,nJ2}
\fmfv{decor.shape=circle,decor.filled=hatched,decor.size=11thick}{x}
\fmfv{decor.shape=circle,decor.filled=30,decor.size=6thick}{x1}
\fmfv{decor.shape=circle,decor.filled=30,decor.size=6thick,label=$\hs{-3.5mm}\text{\small tr}$}{x2}
\end{fmfgraph*} 
\end{fmffile}
\end{minipage}
\nonumber
\\
\!\!\!\!&=&\!\!\!\!
k_{\mu} \, {G^{c_Z\bar{c}_Z} (k)}{\Big|_{\text{tr}}}
\, i \Sigma^{\bar{c}_Zc_A}(k) \, {G^{c_A\bar{c}_A} (k)}{\Big|_{\text{tr}}} = -k_{\mu}  \dfrac{i \Sigma^{\bar{c}_Zc_A}(k)}{k^2 (k^2 - \xi_Z m_{\mathrm{Z}}^2)},
\label{App-WI:eq:my-AG0-conex}
\\
k_{\nu} \, G^{AZ}_{\mu\nu}(k) \Big|_{\text{1L}}
\!\!\!\! &=& \!\!\!\!
k_{\nu} \, \,
\begin{minipage}{0.35\textwidth}
\begin{fmffile}{SMdecom1} 
\begin{fmfgraph*}(70,50) 
\fmfset{arrow_len}{3mm} 
\fmfset{arrow_ang}{20}
\fmfleft{nJ1} 
\fmfright{nJ2} 
\fmf{photon,label=$\Large \xxrightarrow[k]{}$,label.side=right,label.dist=2,tension=2}{nJ1,x}
\fmf{phantom,label=$A_{\mu}$,label.side=left,tension=0}{nJ1,x}
\fmf{photon,label=$Z_{\nu}$,label.side=left,tension=2}{x,nJ2} 
\fmfv{decor.shape=circle,decor.filled=30,decor.size=11thick,label=$\hs{-4mm}\text{1L}$}{x}
\end{fmfgraph*} 
\end{fmffile}
\end{minipage}
\hs{-30mm}
=
k_{\nu} 
\hs{1mm}
\begin{minipage}{0.35\textwidth}
\begin{fmffile}{SMdecom2} 
\begin{fmfgraph*}(100,50) 
\fmfset{arrow_len}{3mm} 
\fmfset{arrow_ang}{20}
\fmfleft{nJ1} 
\fmfright{nJ2} 
\fmf{photon,label=$\Large \xxrightarrow[k]{}$,label.side=right,label.dist=2,tension=3}{nJ1,x1}
\fmf{phantom,label=$A_{\mu}$,label.side=left,tension=0}{nJ1,x1}
\fmf{photon,label=$A_{\alpha}$,label.side=left,tension=2}{x1,x}
\fmf{phantom,label=$\hs{-9mm}\text{\small tr}$,label.side=left,label.dist=0,tension=0}{x1,x}
\fmf{photon,label=\hs{1mm}$Z_{\beta}$,label.side=left,tension=2}{x,x2} 
\fmf{photon,label=$Z_{\nu}$,label.side=left,tension=3}{x2,nJ2}
\fmfv{decor.shape=circle,decor.filled=hatched,decor.size=11thick}{x}
\fmfv{decor.shape=circle,decor.filled=30,decor.size=6thick}{x1}
\fmfv{decor.shape=circle,decor.filled=30,decor.size=6thick,label=$\hs{-3.5mm}\text{\small tr}$}{x2}
\end{fmfgraph*} 
\end{fmffile}
\end{minipage}
\hs{-20mm}
\nonumber
\\
\!\!\!\!&=&\!\!\!\!
k_{\nu} {G^{AA}_{\mu\alpha}(k)}{\Big|_{\text{tr}}} \eta_{g} i \Sigma^{AZ}_{\alpha \beta}(k) {G^{ZZ}_{\beta\nu}(k)}{\Big|_{\text{tr}}}
=
- \eta_{g} \, \xi_A \xi_Z \, k_{\mu} \dfrac{i\Sigma^{AZ}_{\mathrm{L}}(k)}{k^2(k^2 - \xi_Z m_{\mathrm{Z}}^2)},
\label{App-WI:eq:my-AZ-conex}
\eea
where we used the decomposition of eq. \ref{Chap-Reno:eq:SigmaRenoGaugeNeutral} for non-renormalized functions. Hence, it follows from eq. \ref{App-WI:eq:mae} the following particular case:
\be
\lim_{k \to 0} \, k^2 \, G^{c_A\bar{c}_A}(k){\Big|_{\text{1L}}} = 
\eta \eta_Z \eta_e \dfrac{s_{\textrm{w}}}{c_{\textrm{w}}} \dfrac{1}{m_{\mathrm{Z}}^2} \left[\dfrac{1}{\xi_Z} i \Sigma^{\bar{c}_Zc_A}(0) - \eta_G \eta_g i\Sigma^{AZ}_{\mathrm{L}}(0)\right],
\label{App-WI:eq:use-later}
\ee
which shall be useful later.

Having seen this, we apply to the different sides of eq. \ref{App-WI:eq:myC19} the sequence of four tasks introduced immediatly after it. In what follows, we write explicitely all momentum dependences.

\subsubsection*{Left-hand side of eq. \ref{App-WI:eq:myC19}}

Here, we have four terms. We consider each one separately:

\begin{itemize}
\item The first one can be simplified if we note that, due to eq. \ref{App-WI:eq:phot-tree-prop}, we have:
\be
i k_{\mu} {G^{AA}_{\mu\nu}(k)}{\Big|_{\text{tr}}}
= \xi_A \dfrac{k_{\nu}}{k^2}.
\ee
As a consequence, when the derivative in order to $k^{\mu}$  applies to $i \Gamma^{A\bar{f}f}_{\nu rs} (k,-p',p) \Big|_{\text{1L}}$, this contribution ends up vanishing when we multiply by $k^2$ and set $k=0$. In the end, then, the total contribution of the first term is:
\be
- \bar{u}_r(p) \, \eta_G \, i \Gamma^{A\bar{f}f}_{\mu rs} (0,-p,p) \Big|_{\text{1L}} v_s(-p).
\label{App-WI:eq:lhs-first}
\ee

\item The second term vanishes, because, as already mentioned, the photon propagator does not get longitudinal corrections at higher orders.

\item As for the third term, eq. \ref{App-WI:eq:my-AZ-conex} implies that, when the derivative in order to $k^{\mu}$  applies to $i \Sigma^{AZ}_{\mathrm{L}}(k)$, this contribution ends up vanishing when we multiply by $k^2$ and set $k=0$. Using eq. \ref{App-WI:eq:Zrs}, we find that the final contribution of the third term is:
\be
\bar{u}_r(p) \, \eta \eta_G \eta_Z \eta_{g} \dfrac{1}{m_{\textrm{Z}}^2} \delta_{rs} i \Sigma^{AZ}_{\mathrm{L}}(0)
\Big[ \dfrac{e \, s_{\text{w}}}{c_{\text{w}}} Q_r  \gamma_{\mu} - \dfrac{e}{s_{\text{w}} c_{\text{w}}} T_3^r \gamma_{\mu} \gamma_{\textrm{L}} \Big]
v_s(-p).
\label{App-WI:eq:lhs-third}
\ee

\item Finally, for the fourth term,
we can expand $k_{\mu} G^{A{S}}_{\mu}(k)$ as in eq. \ref{App-WI:eq:my-AZ-conex} and use the fact that Lorentz invariance implies
$
i \Sigma^{AS}_{\mu}(k) = k_{\mu} \, i \Sigma_{AS}(k).
$
In that case, the contribution of the fourth term vanishes after the sequence of four tasks.
\end{itemize}

Therefore, the total contribution of the l.h.s of eq. \ref{App-WI:eq:myC19} is the sum of eqs. \ref{App-WI:eq:lhs-first} and \ref{App-WI:eq:lhs-third}:
\be
\bar{u}_r(p)
\bigg\{
- \eta_G \, i \Gamma^{A\bar{f}f}_{\mu rs} (0,-p,p) \Big|_{\text{1L}}
+
\eta \eta_G \eta_Z \eta_{g} \dfrac{1}{m_{\textrm{Z}}^2} \, \delta_{rs} \, i \Sigma^{AZ}_{\mathrm{L}}(0)
\Big[ \dfrac{e \, s_{\text{w}}}{c_{\text{w}}} Q_r  \gamma_{\mu} - \dfrac{e}{s_{\text{w}} c_{\text{w}}} T_3^r \gamma_{\mu} \gamma_{\textrm{L}} \Big]
\bigg\}
v_s(-p).
\label{App-WI:eq:total-lhs}
\ee

\subsubsection*{Right-hand side of eq. \ref{App-WI:eq:myC19}}

In this case, there are five terms:

\begin{itemize}
\item The first one contains ${{(G^{f\bar{f}}_{rj})}^{-1}(p^{\prime})}{\Big|_{\text{tr}}}$, with $p^{\prime} = p+k$. As we saw, when we set $k=0$, $p^{\prime}$ will become OS. This means that, after we perform the four tasks, all the terms containing ${{(G^{f\bar{f}}_{rj})}^{-1}(p^{\prime})}{\Big|_{\text{tr}}}$ will vanish. In that case, we can write:
\be
\begin{split}
&\sum\limits_j 
\dfrac{\partial}{\partial k^{\mu}}
\left[
{G^{c_A\bar{c}_A} (k)}{\Big|_{\text{tr}}} \, {{(G^{f\bar{f}}_{rj})}^{-1}(p+k)}{\Big|_{\text{tr}}} \, {i \Gamma_{1js}^{c_As\bar{f}\,f}}(k,-p^{\prime},p)  {\Big|_{\text{1L}}}
\right] \\
& \hs{50mm}
=
{G^{c_A\bar{c}_A} (k)}{\Big|_{\text{tr}}}
\left(-i \, \gamma_{\mu} \right)
\, {i \Gamma_{1rs}^{c_As\bar{f}\,f}}(k,-p^{\prime},p) {\Big|_{\text{1L}}}
+ (...),
\end{split}
\ee
where the ellipses represent terms that will not contribute. Now, as shown in table \ref{App-WI:table:diags}, ${i \Gamma_{1rs}^{c_As\bar{f}\,f}}{\Big|_{\text{1L}}}$ corresponds to 6 diagrams; more specifically, and using the conventions of  table \ref{App-WI:table:code},
\be
\hs{-15mm}
i \Gamma_{1rs}^{c_As\bar{f}\,f}{\Big|_{\text{1L}}} 
=
\hs{2mm}
\begin{minipage}{0.35\textwidth}
\begin{fmffile}{SMAff334} 
\begin{fmfgraph*}(80,68) 
\fmfset{arrow_len}{3mm} 
\fmfset{arrow_ang}{20} 
\fmfleft{nJ3} 
\fmfright{nJ2,nJ4}
\fmflabel{{\small$rk$}}{nJ2}
\fmf{ghost,label=$c_A$,label.side=right,tension=7}{nJ3,y0}
\fmf{ghost,label=$c_{\pm}$,label.side=right,tension=5}{y0,nJ2}
\fmf{fermion,label=$f_s$,label.side=left,tension=3}{nJ4,z0}
\fmf{fermion,label=$\psi_m$,label.side=left,tension=2}{z0,nJ2}
\fmf{photon,label=$W^{\pm}$,label.dist=3,label.side=left,tension=2}{y0,z0}
\fmf{phantom_arrow,tension=1}{z0,y0}
\fmfv{decor.shape=pentagram,decor.filled=empty,decor.size=6thick}{nJ2}
\end{fmfgraph*} 
\end{fmffile}
\end{minipage}
\hs{-21mm}
+
\hs{5mm}
\begin{minipage}{0.35\textwidth}
\begin{fmffile}{SMAff335} 
\begin{fmfgraph*}(80,68) 
\fmfset{arrow_len}{3mm} 
\fmfset{arrow_ang}{20} 
\fmfleft{nJ3} 
\fmfright{nJ2,nJ4}
\fmflabel{{\small$rk$}}{nJ2}
\fmf{ghost,label=$c_A$,label.side=right,tension=7}{nJ3,y0}
\fmf{ghost,label=$c_{\pm}$,label.side=right,tension=5}{y0,nJ2}
\fmf{fermion,label=$f_s$,label.side=left,tension=3}{nJ4,z0}
\fmf{fermion,label=$\psi_m$,label.side=left,tension=2}{z0,nJ2}
\fmf{scalar,label=$G^{\pm}$,label.dist=3,label.side=right,tension=2}{z0,y0}
\fmfv{decor.shape=pentagram,decor.filled=empty,decor.size=6thick}{nJ2}
\end{fmfgraph*} 
\end{fmffile}
\end{minipage}
\hs{-20mm}.
\vspace{5mm}
\label{App-WI:eq:my-diagrams}
\ee
However, instead of calculating these diagrams explicitly, we go back to eq. \ref{App-WI:eq:myC19}, multiply by $k^2$, set $p$ OS and set $k=0$. The l.h.s. vanishes, so that, recalling eqs. \ref{App-WI:eq:rels}, \ref{App-WI:eq:my-AG0-conex} and \ref{App-WI:eq:use-later}, we get:
\begin{gather}
i \, \eta_G \, (-i) (\slashed{p}-m_{f,r}) \, i \Gamma_{1rs}^{c_As\bar{f}\,f}(0,-p,p){\Big|_{\text{1L}}} \, v_s(-p)
\ \ + \ \ 
i \eta_G {(G^{f\bar{f}}_{rs})}^{-1}(p)\Big|_{\text{1L}} \, (-i) \eta_e e Q_s \, v_s(-p) \no
+ \, \, \eta \eta_Z \eta_e \dfrac{s_{\textrm{w}}}{c_{\textrm{w}}} \dfrac{1}{m_{\mathrm{Z}}^2} \left[\dfrac{1}{\xi_Z} i \Sigma^{\bar{c}_Zc_A}(0) - \eta_G \eta_g i\Sigma^{AZ}_{\mathrm{L}}(0)\right]  \, (-i) (\slashed{p}-m_{f,s}) \, (-i) \, \eta_e \, e \, Q_s \, \delta_{rs} \, v_s(-p)\no
+ \dfrac{i \Sigma^{\bar{c}_Zc_A}(0)}{\xi_Z m_{\mathrm{Z}}^2} \, (-i) (\slashed{p}-m_{f,r}) \, i \eta \eta_Z \, \delta_{rs} \left( \dfrac{e \, s_{\text{w}}}{c_{\text{w}}} Q_s - \dfrac{e}{s_{\text{w}} c_{\text{w}}} T_3^s \, \gamma_{\mathrm{L}}\right) \, v_s(-p)\no
- \, \, i \eta_G \, (-i) \eta_e e Q_r \, {(G^{f\bar{f}}_{rs})}^{-1}(p)\Big|_{\text{1L}} \, v_s(-p)  = 0
.
\hs{5mm}
\label{App-WI:eq:muitos}
\end{gather}
The second and the fifth terms cancel exactly; the third term vanishes, due to eq. \ref{App-WI:eq:OS-spinors}. As for the first term, we can use eq. \ref{App-WI:eq:comp1} to write:
\be
i \Gamma_{1rs}^{c_As\bar{f}\,f}(0,-p,p){\Big|_{\text{1L}}} \, v_s(-p)
=
\gamma_{\mathrm{L}} \, i \Lambda_1^{c_As\bar{f}\,f} \, \delta_{rs} \, v_s(-p),
\ee
where $i \Lambda_{1}^{c_As\bar{f}\,f}$ is a scalar constant, since $i \Gamma_{1rs}^{c_As\bar{f}\,f}(0,-p,p){\Big|_{\text{1L}}}$ can only depend on $\slashed{p}$ and is multiplied by $v_s(-p)$.%
\fn{Also, whatever the particular choice for $F_{ij}$ (cf. table \ref{App-WI:table:code}), that choice always appears twice in each of the diagrams of eq. \ref{App-WI:eq:my-diagrams}, in the combination $\sum\limits_m F_{sk}^* F_{rk}$, which is equal to $\delta_{rs}$ whichever the choice taken.}
Hence, using eq. \ref{App-WI:eq:OS-spinors}, eq. \ref{App-WI:eq:muitos} leads to:
\be
i \Lambda_{1}^{c_As\bar{f}\,f} = \eta \eta_Z \eta_G \, \dfrac{i \Sigma^{\bar{c}_Zc_A}(0)}{\xi_Z m_{\textrm{Z}}^2} \dfrac{e}{s_{\textrm{w}} c_{\textrm{w}}} T_3^r .
\ee
Therefore, the final contribution of the first term of the r.h.s. of eq. \ref{App-WI:eq:myC19} is:
\be
\bar{u}_r(p) \, \eta \eta_Z 
\,
\dfrac{i \Sigma^{\bar{c}_Zc_A}(0)}{\xi_Z m_{\textrm{Z}}^2}
\dfrac{e}{s_{\textrm{w}} c_{\textrm{w}}} T_3^r \, \gamma_{\mu} \gamma_{\mathrm{L}} \, \delta_{rs} \, v_s(-p).
\ee

\item In the second term, the derivative in order to $k^{\mu}$ generates two non-trivial terms:
\be
\begin{split}
\sum\limits_j
\bigg\{
\dfrac{\partial}{\partial k^{\mu}}
\left[
{G^{c_A\bar{c}_A} (k)}{\Big|_{\text{tr}}}\right] \,
&
{{(G^{f\bar{f}}_{rj})}^{-1}(p+k)}{\Big|_{\text{1L}}} \, {i \Gamma_{2js}^{c_As\bar{f}\,f}} (k,-p^{\prime},p)  {\Big|_{\text{tr}}}\\
&+
{G^{c_A\bar{c}_A} (k)}{\Big|_{\text{tr}}} \,
\dfrac{\partial}{\partial k^{\mu}}
\left[
{{(G^{f\bar{f}}_{rj})}^{-1}(p+k)}{\Big|_{\text{1L}}} \,
\right]
{i \Gamma_{2js}^{c_As\bar{f}\,f}} (k,-p^{\prime},p) {\Big|_{\text{tr}}}
\bigg\}.
\label{App-WI:eq:cancelanda}
\end{split}
\ee
The first term will cancel with another term, to be introduced below. So, we use eqs. \ref{App-WI:eq:aux-A} and \ref{App-WI:eq:my-ferm-trick}, as well as the chain rule, to obtain the final contribution:
\be
\bar{u}_r(p) \, \eta_e \eta_G \eta_f \, e Q_s \dfrac{\partial \, i \Sigma^{\bar{f}f}_{rs}(p)}{\partial p^{\mu}} \, v_s(p).
\ee
\item In the third term, similarly to what happened in the first, the only term contributing after we take the derivative in order to $k^{\mu}$ is:
\be
{G^{c_A\bar{c}_A} (k)}{\Big|_{\text{1L}}}
\, \left(-i \, \gamma_{\mu} \right)
\, {i \Gamma_{2rs}^{c_As\bar{f}\,f}}(k,-p^{\prime},p){\Big|_{\text{tr}}}.
\ee
Consequently, after the remaining two tasks, we can use eqs. \ref{App-WI:eq:aux-A} and \ref{App-WI:eq:use-later} to conclude that the final contribution of the third term is:
\be
- \bar{u}_r(p) \, \eta \eta_Z \, e Q_s \, \dfrac{s_{\textrm{w}}}{c_{\textrm{w}}} \dfrac{1}{m_{\mathrm{Z}}^2} \left[\dfrac{1}{\xi_Z} i \Sigma^{\bar{c}_Zc_A}(0) - \eta_G \eta_g i\Sigma^{AZ}_{\mathrm{L}}(0)\right] \delta_{rs}
\, \gamma_{\mu} \, v_s(-p).
\ee

\item The fourth term is similar to the first and the third, in the sense that only one term ends up contributing. 
Using eqs. \ref{App-WI:eq:G3} and \ref{App-WI:eq:my-AG0-conex}, we conclude that its final contribution is:
\be
\bar{u}_r(p) \, \eta \eta_Z \, \dfrac{i \Sigma^{\bar{c}_Zc_A}(0)}{\xi_Z m_{\mathrm{Z}}^2} 
\left[
\dfrac{e \, s_{\textrm{w}}}{c_{\textrm{w}}} Q_r \, \gamma_{\mu}
-
\dfrac{e}{s_{\textrm{w}} c_{\textrm{w}}} T_3^r \, \gamma_{\mu} \gamma_{\mathrm{L}} \right] \delta_{rs} \, v_s(-p).
\ee

\item Finally, in the fifth term, the derivative in order to $k^{\mu}$ generates only one non-trivial term, which precisely cancels the first term of eq. \ref{App-WI:eq:cancelanda}.

\end{itemize}

Therefore, the total contribution of the r.h.s. of eq. \ref{App-WI:eq:myC19} is:
\be
\bar{u}_r(p)
\left\{
\eta_e \eta_G \eta_f \, e Q_s \dfrac{\partial \, i \Sigma^{\bar{f}f}_{rs}(p)}{\partial p^{\mu}} 
+ \eta \eta_G \eta_Z \eta_g \, \dfrac{e \, s_{\textrm{w}}}{c_{\textrm{w}} m_{\mathrm{Z}}^2} \, i\Sigma^{AZ}_{\mathrm{L}}(0) \,
Q_s \, \delta_{rs}
\, \gamma_{\mu} 
\right\} v_s(-p).
\label{App-WI:eq:total-rhs}
\ee

\subsubsection*{Joining sides and discussion}

If we join eqs. \ref{App-WI:eq:total-lhs} and \ref{App-WI:eq:total-rhs},
and if we use the fact that $\Sigma^{AZ}_{\mathrm{L}}(0) = \Sigma^{AZ}_{\mathrm{T}}(0)$, which results from the analiticity of $\Sigma_{\mu\nu}^{AZ}(k)$ for $k^2=0$ (from the non-renormalized version of eq. \ref{Chap-Reno:eq:SigmaRenoGaugeNeutral}), we find:
\begin{equation}
\bar{u}_r(p) \left\{ i \Gamma^{A\bar{f}f}_{\mu rs}(0,-p,p) \Big|_{\text{1L}}
+ \eta \eta_Z \eta_{g}
\dfrac{e \, T_3^r}{m_{\mathrm{Z}}^2 s_{\text{w}} c_{\text{w}}}  \, \gamma_{\mu}  \gamma_{\mathrm{L}}  \, \delta_{rs} \, i \Sigma^{AZ}_{\mathrm{T}}(0)
+
\eta_e \eta_f \, e Q_s \dfrac{\partial \, i \Sigma^{\bar{f}f}_{rs}(p)}{\partial p^{\mu}}
\right\} v_s(-p) = 0.
\label{App-WI:eq:SMAtheWI-pre}
\end{equation}
This is the WI we wanted to derive: a relation for the one-loop vertex function $i \Gamma^{A\bar{f}f}_{\mu rs} \Big|_{\text{1L}}$ in terms of other one-loop vertex functions, in the case of OS fermions and vanishing photon momentum.
In the particular case of flavor conservation ($s=r$), the identity reads:
\begin{equation}
\bar{u}_r(p) \left\{ i \Gamma^{A\bar{f}f}_{\mu rr}(0,-p,p) \Big|_{\text{1L}}
+ \eta \eta_Z \eta_{g}
\dfrac{e \, T_3^r}{m_{\mathrm{Z}}^2 s_{\text{w}} c_{\text{w}}}  \, \gamma_{\mu}  \gamma_{\mathrm{L}} \, i \Sigma^{AZ}_{\mathrm{T}}(0)
+
\eta_e \eta_f \, e Q_r \dfrac{\partial \, i \Sigma^{\bar{f}f}_{rr}(p)}{\partial p^{\mu}}
\right\} v_r(-p) = 0.
\label{App-WI:eq:SMAtheWI}
\end{equation}
Some comments are in order.

First, all we assumed in the derivation was invariance of complete GFs under BRST transformations in the context of a general $R_{\xi}$ gauge and Lorentz invariance. The number of fermions or scalars in the theory was not relevant for the derivation of the identities \ref{App-WI:eq:SMAtheWI-pre} and \ref{App-WI:eq:SMAtheWI}, which thus also hold for models with more fermions and scalars than the SM. In fact, as long as the gauge sector of the SM is not modified, the identities are valid.

Second, these identities are manifestly gauge independent, since the dependences on gauge parameters of intermediate results cancelled one another.
However, they do depend on the conventions for signs, as the presence of the $\eta$'s in the final results testifies. And while this has no physical meaning whatsoever, it may be useful when one wants to use the WI in a different convention.
As mentioned above, besides the $\eta$'s of ref. \cite{Romao:2012pq}, we introduced $\eta_f$ and $\eta_g$, which concern the definition of the one-loop contribution to the 2-point vertex functions of fermions and gauge bosons, respectively. This introduction is relevant, since both $\eta_f$ and $\eta_g$ end up showing up in the final form of the WIs, on the one hand, and since different works in the literature use different values for these variables, on the other.
Therefore, the consideration of the identities with general $\eta$'s allows a simple application to different conventions; for example, in the case of flavor conservation, eq. (C.30) of ref. \cite{Denner:2019vbn} is obtained from eq. \ref{App-WI:eq:SMAtheWI} above by setting $\eta_e = \eta_Z = \eta_f = -\eta = -\eta_{g}=1$; or, in the case of flavor violation ($r \neq s$), eq. (33) of ref. \cite{Soares:1988fj} is obtained from eq. \ref{App-WI:eq:SMAtheWI-pre} above by setting $\eta_e = - \eta_f = 1$ in the appropriate limit.

Third, eqs. \ref{App-WI:eq:SMAtheWI-pre} and \ref{App-WI:eq:SMAtheWI} do not depend on the treatment of tadpoles. Indeed, nowhere in the derivation did we need to specify which tadpole scheme or which vev we were considering. Among the differents GFs used, the only one where one-loop tadpoles could contribute is the 2-point function of fermions; yet, the derivative in order to $p^{\mu}$ ensures that, even if they are included, their contribution vanishes. 

Finally, note the similarity with QED, where the WI is the same as that in eq. \ref{App-WI:eq:SMAtheWI} above, except that it does not include the second term on the l.h.s. \cite{Bohm:2001yx}. This extra term is a consequence of the fact the SM has a more complicated gauge symmetry than QED, which leads in particular to a mixing between the $Z$ boson and the photon.

\section{The counterterm $\delta Z_e$}

The WI \ref{App-WI:eq:SMAtheWI} is very useful in OSS. In fact, as suggested in section \ref{Chap-Reno:sec:charge}, the original expression for $\delta Z_e$ in OSS is quite complicated, but it can be simplified thanks to eq. \ref{App-WI:eq:SMAtheWI}.
To see how this is done, we start by rewriting eq. \ref{Chap-Reno:eq:OSS-vert-cond} for a spinor $v(p)$ and with diagonal flavour indices $r$:%
\fn{Due to charge universality, the OSS condition for the electric charge can be written for any charged particle \cite{Denner:2019vbn}. Note also that, in this equation as well as in the following ones, we do not need to consider the operator $\widetilde{\operatorname{Re}}$ (usually employed in the OSS scheme), as it is irrelevant for these particular cases. More specifically, the absorptive parts of $i \hat{\Gamma}^{A\bar{f}f}_{\mu rr}(0,-p,p) \Big|_{\text{1L}}$, $i\hat{\Sigma}_{\mathrm{T}}^{AZ}(0)$ and $\frac{\partial \, i \hat{\Sigma}^{\bar{f}f}_{rr}(p)}{\partial p^{\mu}}$ all vanish (this also explains the absence of that operator in eq. \ref{Chap-Reno:eq:OSS-vert-cond}).}
\be
\bar{u}_r(p) \, \,
i \hat{\Gamma}^{A\bar{f}f}_{\mu rr}(0,-p,p) \Big|_{\text{1L}}
\, v_r(-p) \OSSeq 0.
\label{App-WI:eq:OSS-vert-cond}
\ee
Next, we recall two other OSS renormalization conditions. The first one concerns the mixing between the photon and the $Z$ boson, and states that such mixing vanishes for an OS photon (eq. \ref{Chap-Reno:eq:OSS-gn1}); then, using eqs. \ref{Chap-Reno:eq:GammaRenGaugeNeutral}
and \ref{Chap-Reno:eq:SigmaRenoGaugeNeutral}, we obtain:%
\be
i\hat{\Sigma}_{\mathrm{T}}^{AZ}(0)  \OSSeq 0.
\label{App-WI:eq:my-gauge-mix-cond}
\ee
The second one refers to the residue of the OS fermion propagator, which is fixed to one (eq. \ref{Chap-Reno:eq:fermion-res-cond}); then, using eq. \ref{Chap-Reno:eq:GammaRenFermion}, we obtain:
\be
\dfrac{\partial \, i \hat{\Sigma}^{\bar{f}f}_{rr}(p)}{\partial p^{\mu}} v_{r}(-p) \OSSeq 0.
\label{App-WI:eq:fermion-deriv}
\ee
Now, eq. \ref{App-WI:eq:SMAtheWI} (which involves non-renormalized GFs) considers the limit of OS fermions and vanishing photon momentum. But, and as we just saw, eqs. \ref{App-WI:eq:OSS-vert-cond}, \ref{App-WI:eq:my-gauge-mix-cond} and \ref{App-WI:eq:fermion-deriv} (which involve renormalized GFs) all hold in the OSS scheme in that limit. As all of them are equal to zero, they can subtracted (with appropriate factors) to eq. \ref{App-WI:eq:SMAtheWI} without modifying it. Therefore, and since each of them has a non-renormalized equivalent in eq. \ref{App-WI:eq:SMAtheWI}, it follows that the WI also holds for the counterterms in OSS; that is:
\begin{equation}
\bar{u}_r(p) \left\{ i \Gamma^{A\bar{f}f}_{\mu rr}(0,-p,p) \Big|_{\text{CT}}
+ \eta \eta_Z \eta_{g}
\dfrac{e \, T_3^r}{m_{\mathrm{Z}}^2 s_{\text{w}} c_{\text{w}}}  \, \gamma_{\mu}  \gamma_{\mathrm{L}} \, i \Sigma^{AZ}_{\mathrm{T}}(0)\Big|_{\text{CT}}
+
\eta_e \eta_f \, e Q_r \dfrac{\partial \, i \Sigma^{\bar{f}f}_{rr}(p)}{\partial p^{\mu}}\Big|_{\text{CT}}
\right\} v_r(-p) \OSSeq 0.
\label{App-WI:eq:SMAtheWICTs}
\end{equation}
This relation allows us to obtain a simple expression for $\delta Z_e$ in terms of other counterterms; all we need is the expressions for the total counterterms involved in eq. \ref{App-WI:eq:SMAtheWICTs}. Using \textsc{FeynMaster}, we find:
\be
i \Gamma^{A\bar{f}f}_{\mu ij}(k,-p^{\prime},p) \Big|_{\text{CT}}
=
\hs{5mm}
\begin{minipage}[h]{.40\textwidth}
\begin{fmffile}{Juca}
\begin{fmfgraph*}(50,60)
\fmfset{arrow_len}{3mm}
\fmfset{arrow_ang}{20}
\fmfleft{nJ1}
\fmfright{nJ2,nJ4}
\fmf{photon,label=$\gamma_{\mu} \hs{5mm}$,label.side=left,tension=3}{nJ1,nJ2nJ4nJ1}
\fmf{fermion,tension=3,label=$\bar{f}_i$,label.side=right}{nJ2nJ4nJ1,nJ2}
\fmf{fermion,tension=3,label=$f_j$}{nJ4,nJ2nJ4nJ1}
\fmfv{decor.shape=pentagram,decor.filled=full,decor.size=6thick}{nJ2nJ4nJ1}
\end{fmfgraph*}
\end{fmffile}
\end{minipage}
\hs{-43mm}
=
\hs{3mm}
i e \gamma_{\mu} \left(F_{\mathrm{L}ij} \, \gamma_{\mathrm{L}} + F_{\mathrm{R}ij} \, \gamma_{\mathrm{R}}\right),
\quad 
\vs{2mm}
\label{App-WI:eq:vert:boneco}
\ee
where
%
\bs
\begin{gather}
F_{\mathrm{L}ij}=-\eta_e \, Q_i \left[\delta_{i j}\left(\delta Z_{e}+\frac{1}{2} \delta Z_{A A}\right)+\frac{1}{2}\left(\delta Z_{i j}^{f, L}+\delta Z_{i j}^{f, L \dagger}\right)\right] - \eta \, \eta_Z  \, \delta_{i j} \, g^{j}_{\mathrm{L}} \frac{1}{2} \delta Z_{Z A}, \\
F_{\mathrm{R}ij}=-\eta_e \, Q_i \left[\delta_{i j}\left(\delta Z_{e}+\frac{1}{2} \delta Z_{A A}\right)+\frac{1}{2}\left(\delta Z_{i j}^{f, R}+\delta Z_{i j}^{f, R \dagger}\right)\right] - \eta \, \eta_Z \, \delta_{i j} \, g^{j}_{\mathrm{R}} \frac{1}{2} \delta Z_{Z A},
\end{gather}
\label{App-WI:eq:vert:conta}
\es
with
\be
g^r_{\mathrm{L}} = \dfrac{g_V^r+g_A^r}{s_{\text{w}} c_{\text{w}}},
\hs{10mm}
g^r_{\mathrm{R}} = \dfrac{g_V^r-g_A^r}{s_{\text{w}} c_{\text{w}}},
\ee
and $g_V^r$ and $g_A^r$ are given in eq. \ref{App-WI:eq:gVf-gAf}.
As for the mixing between the photon and the $Z$, we find from eq. \ref{App-WI:eq:ZACT} that:
%
%
%
\be
i \Sigma^{AZ}_{\mathrm{T}}(0) \Big|_{\text{CT}} = \dfrac{i}{2} \eta_{g} m_{\mathrm{Z}}^2 \delta Z_{ZA}.
\ee

Finally, from eq. \ref{App-WI:eq:ffCT}, we find:
\be
\dfrac{\partial \, i \Sigma^{\bar{f}f}_{rr}(p)}{\partial p^{\mu}}\Big|_{\text{CT}} = 
 i \eta_f 
\Big(
F^{\mathrm{V}}_{\mathrm{L}rr} \, \gamma_{\mu} \, \gamma_{\mathrm{L}}
+ F^{\mathrm{V}}_{\mathrm{R}rr} \, \gamma_{\mu} \, \gamma_{\mathrm{R}}
\Big),
\ee
%


with $F^{\mathrm{V}}_{\mathrm{L}rr}$ and $F^{\mathrm{V}}_{\mathrm{R}rr}$ given in eq. \ref{App-WI:eq:ferm-aux}.
Inserting these equations in eq. \ref{App-WI:eq:SMAtheWICTs}, we obtain at last:
\begin{equation}
\delta Z_e \OSSeq -\frac{1}{2} \delta Z_{A A} + \eta \eta_Z \eta_e \dfrac{s_{\text{w}}}{c_{\text{w}}} \dfrac{1}{2} \delta Z_{Z A}.
\end{equation}
Due to the WI, then, $\delta Z_e$ in OSS is related through a very simple expression with other counterterms. The fact that such expression does not depend on the fermion species reflects the universality of the electric charge \cite{Dittmaier:2021loa}. As before, this expression is written for general $\eta$'s, so that it can be easily applied to different sign conventions. Setting all $\eta$'s positive, we obtain the result presented in eq. \ref{Chap-Reno:eq:deltaZe}.

%% file: Appendices/Sym.tex
\chapter{Symmetry relations}
\label{App-Sym}

\vs{-7mm}

\n To our knowledge, the procedure of fixing counterterms through symmetry relations has been originally proposed in ref.~\cite{Kanemura:2004mg}, having been later worked out and used by several authors \cite{Krause:2016oke,Krause:2017mal,Altenkamp:2017ldc,Denner:2018opp}.
As mentioned in section \ref{Chap-Reno:sec:UTOL}, the theory at stake can be renormalized in its symmetric form (before SSB); in that case, it is enough to consider one field counterterm for each multiplet in order to cancel UV divergences \cite{Denner:2019vbn}.
On the other hand, the theory can also be renormalized in the broken phase, where counterterms for mixing parameters and mixed fields counterterms in general appear.
The two methods are not independent whenever fields in the symmetry basis are related with fields in the physical basis through mixing parameters.

\n In this appendix, we start by showing that one can exploit such relation to write the divergent parts of the counterterms for mixing parameters in terms of the divergent parts of field counterterms; this we do in section \ref{App-Sym:sec:proc-sym}, considering a simple example. Then, in section \ref{App-Sym:sec:C2-case}, we apply the same method to the specific case of the mixing parameters of the scalar sector of the C2HDM, and discuss the limit of CP conservation.

\n The symmetry relations as such only have to be true for the divergent parts, since the common ground between the renormalization in the two forms (symmetric and broken) was the elimination of UV divergences.
Yet, as we will see in section \ref{App-Sym:sec:dep-sec}, in the case of \textit{independent} counterterms for mixing parameters (which need to be fixed), one can decide to fix them by requiring that the symmetry relations also hold for the finite parts. This is the reason why (independent) counterterms for mixing parameters can be fixed through symmetry relations. In that final section, we discuss the generality of the method, and consider the consequences of symmetry relations on dependent counterterms.

\vs{-3mm}

\section{A simple example}
\label{App-Sym:sec:proc-sym}

\n Let us consider an example involving SM particles only. From the bare version of eq. \ref{Chap-Reno:eq:gauge-rot-tree}, we have:
\be
\begin{pmatrix}
B_{\mu(0)} \\
W^3_{\mu(0)}
\end{pmatrix}
=
\begin{pmatrix}
c_{{\text{w}}(0)} & -s_{{\text{w}}(0)} \\
s_{{\text{w}}(0)} & c_{{\text{w}}(0)} 
\end{pmatrix}
\begin{pmatrix}
A_{\mu(0)} \\
Z_{\mu(0)}
\end{pmatrix}.
\label{App-Sym:eq:sym-bare}
\ee
According to the previous paragraphs, from the point of view of the cancellation of divergences, $W_{\mu}^a$ and $B_{\mu}$ can be renormalized with a single counterterm each. Hence,
\be
W^a_{\mu(0)}
\big|_{\varepsilon}
=
\left(1 + \dfrac{1}{2} \delta Z_{W}^{\prime}\right)\Big|_{\varepsilon} W^a_{\mu},
\qquad
B_{\mu(0)}\big|_{\varepsilon}
=
\left(1 + \dfrac{1}{2} \delta Z_{B}\right)\Big|_{\varepsilon} B_{\mu},
\ee
where $|_{\varepsilon}$ represents the terms proportional to $1/\varepsilon$, as usual.%
\fn{We define the counterterm $\delta Z_{W}^{\prime}$ with a prime to distinguish it from the counterterm for the physical $W^{+}$ field.}
It follows, in particular,
\be
\left.
\begin{pmatrix}
B_{\mu(0)} \\
W^3_{\mu(0)}
\end{pmatrix}
\right|_{\varepsilon}
=
\left.
\begin{pmatrix}
1 + \dfrac{1}{2} \delta Z_{B} & 0 \\
0 & 1 + \dfrac{1}{2} Z_{W}^{\prime}
\end{pmatrix}
\right|_{\varepsilon}
\begin{pmatrix}
B_{\mu} \\
W^3_{\mu}
\end{pmatrix}.
\label{App-Sym:eq:sym-expa-left}
\ee
On the other hand, from eqs. \ref{Chap-Reno:eq:field-gauge-expa} and \ref{Chap-Reno:eq:unnec},
\be
\begin{pmatrix}
A_{\mu(0)} \\ Z_{\mu(0)}
\end{pmatrix}
=
\begin{pmatrix}
1 + \dfrac{1}{2} \delta Z_{AA} & \dfrac{1}{2} \delta Z_{AZ} \\
\dfrac{1}{2} \delta Z_{ZA} & 1 + \dfrac{1}{2} \delta Z_{ZZ}
\end{pmatrix}
\begin{pmatrix}
A_{\mu}
\\
Z_{\mu}
\end{pmatrix},
\quad 
s_{{\text{w}}(0)} = s_{\text{w}} + \delta s_{\text{w}}, 
\quad
c_{{\text{w}}(0)} = c_{\text{w}} + \delta c_{\text{w}},
\label{App-Sym:eq:sym-expa-right}
\ee
and the renormalized quantities are defined such that:
\be
\begin{pmatrix}
B_{\mu} \\
W^3_{\mu}
\end{pmatrix}
=
\begin{pmatrix}
c_{\text{w}} & -s_{\text{w}} \\
s_{\text{w}} & c_{\text{w}} 
\end{pmatrix}
\begin{pmatrix}
A_{\mu} \\
Z_{\mu}
\end{pmatrix}.
\label{App-Sym:eq:sym-reno}
\ee
As a consequence, eq. \ref{App-Sym:eq:sym-expa-left} becomes:
\be
\left.
\begin{pmatrix}
B_{\mu(0)} \\
W^3_{\mu(0)}
\end{pmatrix}
\right|_{\varepsilon}
=
\left.
\begin{pmatrix}
1 + \dfrac{1}{2} \delta Z_{B} & 0 \\
0 & 1 + \dfrac{1}{2} Z_{W}^{\prime}
\end{pmatrix}
\right|_{\varepsilon}
\begin{pmatrix}
c_{\text{w}} & -s_{\text{w}} \\
s_{\text{w}} & c_{\text{w}} 
\end{pmatrix}
\begin{pmatrix}
A_{\mu} \\
Z_{\mu}
\end{pmatrix}.
\label{App-Sym:eq:sym-left}
\ee
On the other hand, we can apply eq. \ref{App-Sym:eq:sym-expa-right} to the r.h.s. of eq. \ref{App-Sym:eq:sym-bare} and consider the divergent parts only to get:
\be
\left.
\begin{pmatrix}
B_{\mu(0)} \\
W^3_{\mu(0)}
\end{pmatrix}
\right|_{\varepsilon}
=
\left.
\left[
\begin{pmatrix}
c_{\text{w}} + \delta c_{\text{w}} & -s_{\text{w}}  - \delta s_{\text{w}}\\
s_{\text{w}} + \delta s_{\text{w}} & c_{\text{w}} + \delta c_{\text{w}} 
\end{pmatrix}
\begin{pmatrix}
1 + \dfrac{1}{2} \delta Z_{AA} & \dfrac{1}{2} \delta Z_{AZ} \\
\dfrac{1}{2} \delta Z_{ZA} & 1 + \dfrac{1}{2} \delta Z_{ZZ}
\end{pmatrix}
\right]
\right|_{\varepsilon}
\begin{pmatrix}
A_{\mu}
\\
Z_{\mu}
\end{pmatrix}.
\label{App-Sym:eq:sym-right}
\ee
Finally, equating eqs. \ref{App-Sym:eq:sym-left} and \ref{App-Sym:eq:sym-right} to first order and noting that $\delta \theta_{\text{w}} =  - \delta c_{\text{w}}/s_{\text{w}}$, one concludes that:
\be
\delta \theta_{\text{w}}
\big|_{\varepsilon}
=
\dfrac{1}{4} \left(\delta Z_{AZ} - \delta Z_{Z A}\right)\big|_{\varepsilon}.
\label{App-Sym:eq:relWein}
\ee
This is the symmetry relation; it fixes the divergent part of the counterterm for the mixing parameter $\theta_{\text{w}}$ in terms of the divergent parts of the counterterms for the physical fields associated with $\theta_{\text{w}}$.
This relation is also valid in the SM, as well as in all its extensions that keep its gauge structure.%
\fn{The same is true for another relation that also follows from equating eqs. \ref{App-Sym:eq:sym-left} and \ref{App-Sym:eq:sym-right} to first order, namely,
\be
\delta Z_{ZZ}
\big\rvert_{\varepsilon}
=
\left.
\bigg[
\delta Z_{AA}
+
\cot\left(2 \, \theta_{\text{w}}\right)
\Big(
\delta Z_{AZ} + \delta Z_{ZA}
\Big)
\bigg]
\right\rvert_{\varepsilon}.
\ee
}

\n The procedure just used is generalizable to other mixing parameters (or sets of mixing parameters) of different models and can be summarized in four steps.
First, one considers the relation involving the mixing parameters at stake (i.e. transforming the fields in the symmetry basis into the mass basis)
in its bare form; in the above example, this is simply eq. \ref{App-Sym:eq:sym-bare}. 
Second, the bare fields in the symmetry basis are expanded---aiming at the elimination of divergences, so that only one field counterterm per multiplet is taken and only the divergent parts are considered---and, after that, the renormalized relation between the two bases is used; in the above example, the result of this step is eq. \ref{App-Sym:eq:sym-left}. 
In the third step, one expands instead the bare mixing parameters and bare physical fields and takes the divergent parts; in the above example, the result is eq. \ref{App-Sym:eq:sym-right}.
Finally, one equates the two previous steps to first order, which yields the divergent parts of the counterterms for the mixing parameters in terms of the divergent parts of field counterterms.

\section{The case of the C2HDM}
\label{App-Sym:sec:C2-case}

\n Applying this procedure to the scalar sector of the C2HDM, we get, for the counterterms for mixing parameters of the charged fields,%
\fn{
Besides, we also find the following relations for the field counterterms:
\bs
\begin{gather}
\mathrm{Im} \, \delta Z_{G^+H^+}
\big\rvert_{\varepsilon}
=
\mathrm{Im} \, \delta Z_{H^+G^+}
\big\rvert_{\varepsilon},
\\
\delta Z_{H^+H^+}\big\rvert_{\varepsilon}
=
\left.
\bigg[
\delta Z_{G^+G^+}
+
\cot\left(2 \beta\right)
\mathrm{Re}
\Big(
\delta Z_{G^+H^+} + \delta Z_{H^+G^+}
\Big)
\bigg]
\right\rvert_{\varepsilon}.
\label{App-Sym:eq:dZHPHP}
\end{gather}
\es
Had we started with a parameterization of $X$ that included an overall phase, the symmetry relations would have forced the divergent part of its counterterm to be zero. This means that such phase is really not needed, which explains why we ignored it in the first place (cf. eq. \ref{Chap-Reno:eq:charged-param-original} and note \ref{Chap-Reno:note:overall}).
}
\bs
\label{App-Sym:eq:charged-div-rel}
\bea
\label{App-Sym:eq:dbeta-div-rel}
{\delta \beta}
\big\rvert_{\varepsilon}
&=& \dfrac{1}{4}
\operatorname{Re}
\left.\Big[\delta Z_{G^+H^+} - \delta Z_{H^+G^+}\Big]
\right|_{\varepsilon},
\\
\label{App-Sym:eq:dzetaa-div-rel}
\delta \zeta_a
\big|_{\varepsilon}
&=&
-\dfrac{1}{2} \cot(2 \beta) \operatorname{Im} \left[\delta Z_{G^+H^+} \right]\big|_{\varepsilon},
\eea
\es
and, for those of the neutral fields,
%
%
\bs
\label{App-Sym:eq:dalphas}
\begin{flalign}
&
\delta \alpha_0
\big|_{\varepsilon}
=
\dfrac{1}{4} \sec(\alpha_2) \sec(\alpha_3)  \big( \delta Z_{G_0h_3} - \delta Z_{h_3G_0} \big)
\Big|_{\varepsilon},
\\[3mm]
&
\delta \alpha_1
\big|_{\varepsilon}
=
\dfrac{1}{4} \sec(\alpha_2)  \Big[  \cos(\alpha_3) \left( \delta Z_{h_1h_2} - \delta Z_{h_2h_1} \right)  +  \sin(\alpha_3) \left( \delta Z_{h_3h_1} -\delta Z_{h_1h_3} \right)  \Big]
\Big|_{\varepsilon},
\\[3mm]
&
\delta \alpha_2
\big|_{\varepsilon}
=
\dfrac{1}{4}  \sin(\alpha_3) \Big[ \delta Z_{h_1h_2} - \delta Z_{h_2h_1} + \cot(\alpha_3) \left( \delta Z_{h_1h_3} - \delta Z_{h_3h_1} \right)  \Big]
\Big|_{\varepsilon},
\\[3mm]
&
\delta \alpha_3
\big|_{\varepsilon}
=
\dfrac{1}{4} \bigg[\delta Z_{h_2h_3} - \delta Z_{h_3h_2} -   \cos(\alpha_3) \tan(\alpha_2)  \left( \delta Z_{h_1h_2} - \delta Z_{h_2h_1} \right) \nonumber\\[-3mm]
& \hs{55mm} + \sin(\alpha_3) \tan(\alpha_2) \left( \delta Z_{h_1h_3} - \delta Z_{h_3h_1} \right)  \bigg]
\bigg|_{\varepsilon},
\\[3mm]
&
\delta \alpha_4
\big|_{\varepsilon}
=
\dfrac{1}{4} \bigg[\delta Z_{h_1G_0} -\delta Z_{G_0h_1} + \sec(\alpha_3) \tan(\alpha_2) \left( \delta Z_{G_0h_3} - \delta Z_{h_3G_0} \right)  \bigg]
\bigg|_{\varepsilon},
\\[3mm]
&
\delta \alpha_5
\big|_{\varepsilon}
=
\dfrac{1}{4} \bigg[\delta Z_{h_2G_0} -\delta Z_{G_0h_2} + \tan(\alpha_3) \left( \delta Z_{G_0h_3} - \delta Z_{h_3G_0} \right)
\bigg]
\bigg|_{\varepsilon}.
\end{flalign}
\es
%
Note that the expressions for the divergent parts of $\delta \alpha_0$, $\delta \alpha_4$ and $\delta \alpha_5$ depend only on the divergent parts of fields counterterms for the mixing between $G_0$ and other neutral scalar fields. This means that, should there be no such mixing at one-loop, the divergent parts of the corresponding mixed fields counterterms would vanish---which would then imply the vanishing of the divergent parts of $\delta \alpha_0$, $\delta \alpha_4$ and $\delta \alpha_5$. In other words, if $G_0$ did not mix with the fields $h_1$, $h_2$ and $h_3$ at one-loop, the counterterms $\delta \alpha_0$, $\delta \alpha_4$ and $\delta \alpha_5$ would not be needed. This is consistent with description of the theory considered solely at tree-level, where the angles $\alpha_0$, $\alpha_4$ and $\alpha_5$ are not introduced. However, since $G_0$ does mix with the remaining neutral scalar fields at one-loop, the divergent parts of $\delta \alpha_0$, $\delta \alpha_4$ and $\delta \alpha_5$ are in general not zero, so that these counterterms must be considered---which in turn means that $\alpha_{0}$, $\alpha_{4}$ and $\alpha_{5}$ have to be included at tree-level when aiming at the one-loop renormalization of the theory.

\n It is also interesting to consider the limit of CP conservation; one possible limit is \cite{ElKaffas:2006gdt}:
%
%
\be
\alpha_1 \stackrel{\mathrm{CP}}{=} \alpha + \dfrac{\pi}{2},
\qquad
\alpha_2 \stackrel{\mathrm{CP}}{=} 0,
\qquad
\alpha_3 \stackrel{\mathrm{CP}}{=} 0
\qquad
(\text{with}
\ \
\alpha_0 = \beta,
\quad
\alpha_4 = 0,
\quad
\alpha_5 = 0),
\ee
in which case \cite{ElKaffas:2006gdt}:
%
\be
h_1 \stackrel{\mathrm{CP}}{=} h, 
\qquad
h_2 \stackrel{\mathrm{CP}}{=} -H,
\qquad
h_3 \stackrel{\mathrm{CP}}{=} A_0,
\ee
with $h$ and $H$ being CP-even and $A_0$ CP-odd. Then, given that there would be no one-loop mixing between fields with different CP values, eqs. \ref{App-Sym:eq:dalphas} would become:
\begin{flalign}
&
\delta \alpha_0
\big|_{\varepsilon}
\stackrel{\mathrm{CP}}{=}
\dfrac{1}{4} \big( \delta Z_{G_0A_0} - \delta Z_{A_0G_0} \big)
\big|_{\varepsilon},
\\[3mm]
&
\delta \alpha_1
\big|_{\varepsilon}
\stackrel{\mathrm{CP}}{=}
\dfrac{1}{4} \left( \delta Z_{Hh} - \delta Z_{hH} \right)
\big|_{\varepsilon},
\\[3mm]
&
\delta \alpha_2\big|_{\varepsilon}
\stackrel{\mathrm{CP}}{=}
\delta \alpha_3\big|_{\varepsilon}
\stackrel{\mathrm{CP}}{=}
\delta \alpha_4\big|_{\varepsilon}
\stackrel{\mathrm{CP}}{=}
\delta \alpha_5\big|_{\varepsilon}
\stackrel{\mathrm{CP}}{=}
0.
\end{flalign}
This is consistent with the symmetry relations in a 2HDM with CP conservation (cf. e.g. ref.~\cite{Krause:2016oke}).
Moreover, still in the context of CP conservation, $Z_{G^+H^+}$ would be real, which would imply that
$
\delta \zeta_a
\big|_{\varepsilon}
$
would vanish; consequently, if CP were a symmetry of the theory, there would also be no need to consider $\delta \zeta_a$.

\section{Dependent and independent mixing parameters}
\label{App-Sym:sec:dep-sec}

\n The symmetry relations have interesting consequences both when the counterterm for the mixing parameter at stake is dependent and when it is independent.
In the first case, one can use the symmetry relations to establish relations between the divergent parts of different independent counterterms. For example, given eq. \ref{Chap-Selec:eq:cwCT},
one can use eq. \ref{App-Sym:eq:relWein} to conclude that:
\be
\dfrac{1}{4} \left(\delta Z_{A Z} - \delta Z_{Z A}\right)\Big|_{\varepsilon}
=
\left.
\left(\dfrac{m_{\mathrm{W}} \, \delta m_{\mathrm{Z}}^2}{2 s_{\text{w}} m_{\mathrm{Z}}^3} -  \dfrac{\delta m_{\mathrm{W}}^2}{2 s_{\text{w}} m_{\mathrm{W}} m_{\mathrm{Z}}} \right)
\right|_{\varepsilon}.
\label{App-Sym:eq:eq:my-rel-thetaw}
\ee
Note that the same relation is in general not valid for the finite parts.%
\fn{This is the reason why the counterterms involved are still independent (even if their divergent parts are related).
Again, eq. \ref{App-Sym:eq:eq:my-rel-thetaw} is also valid in the SM, as well as in all extensions of this model that do not modify its gauge structure.}

\n In the case of counterterms for independent mixing parameters, the finite parts are \textit{a priori} not fixed. In fact, even if we are able to determine their divergent parts, the finite parts are free, and must be fixed by renormalization conditions.
But here is the trick: we can choose as renormalization conditions the symmetry relations \textit{including finite parts}.
Said otherwise, we can decide that the way through which we fix the independent counterterms is requiring that the symmetry relations hold not only for the divergent parts, but also for the finite parts \cite{Denner:2018opp}.
In particular, we can decide that the relations involved in eqs. \ref{App-Sym:eq:charged-div-rel} and \ref{App-Sym:eq:dalphas} also hold for the finite parts---whenever the counterterm for mixing parameters at stake is independent (which depends on the combination $C_i$).
This fixes the counterterms involved, thus justifying the expressions presented in section \ref{Chap-Reno:sec:RenoMixingParameters}.%
\fn{As discussed in section \ref{Chap-Reno:sec:gauge-dep}, the independent counterterms for the mixing parameters are calculated in the Feynman gauge.}

\n The prescription we just described---of promoting symmetry relations for independent mixing parameter counterterms to renormalization conditions---is completely general, not being restricted to the C2HDM. In fact, just as the procedure presented in section \ref{App-Sym:sec:proc-sym} is easily generalizable to other models, so is this prescription to fix independent counterterms for mixing parameters. Actually, not only it is easily generalizable, as it is very simple, leading to gauge independent observables (as discussed in section \ref{Chap-Reno:sec:gauge-dep}). 

%% file: Appendices/FM-C2HDM.tex
\chapter{FeynMaster 2: an ideal tool for renormalizing the C2HDM}
\label{App-FM-C2HDM}

\vs{-5mm}

In this appendix, we describe the application of \FMTS to the renormalization of the C2HDM. We start by summarizing its usefulness in the renormalization of models. Then, we explain how it can be exploited to renormalize a model such as the C2HDM, providing illustrations of the different steps.

%

\section{An ideal tool for renormalizing models}

Given the list of tasks that it is able to perform (cf. chapter \ref{Chap-FM} and appendix \ref{App-FM-Manual}), it is easy to conclude that \FMTS is an extremely useful tool in the renormalization of a model. Such usefulness is manifest, first and foremost, in the fact that the different steps required for the renormalization of a model---definition of the model, generation of Feynman rules, generation of counterterms, one-loop calculations and renormalization conditions---can all be performed with \FMT.
One starts by writing the model, with the desired conventions. This allows to automatically generate the Feynman rules not only for the tree-level interactions, but also for the counterterms. With the former, one-loop processes can be automatically calculated and stored, which allows to fix all the counterterms in a desired subtraction scheme, thus completing the renormalization of the model.

But this is not all. One can then study different one-loop processes (meanwhile renormalized), disposing not only of all the tools of \FC, but also of some specific functions of \FMT. These can also be combined with the numerical interface with \t{\ts{Fortran}}, thus allowing a more complete analysis. In the end of the day, renormalization and the study of NLO processes has become quite simple.

A constant element in the chain of processes just described should be stressed: \textit{flexibility}.
This property begins in the moment when, in the writing of the model, the user defines the conventions. Not only he or she has the freedom to define those conventions at will, but---and what is perhaps more relevant---they are kept throughout the whole process: in Feynman rules, one-loop calculations, counterterms, etc. Indeed, by combining all the required steps in a single software, \FMTS ensures that the conventions do not change when changing from one step to the other.%
\fn{This tends to be a problem when one needs to combine different software (e.g., one software for the generation of Feynman rules, and another one for the one-loop calculations), since different software usually define different quantities and variables in different ways.}

Flexibility is present in many other ways. Intermediate and final results can be manipulated: not only due to the intuitive structure of \FMT, but also to the user-friendly character of \FC \, and \t{\ts{Mathematica}}. Notebooks for processes are automatically written, in which all of the results already calculated (Feynman rules, counterterms, one-loop amplitudes, etc.) are immediately available. The writing of Feynman diagrams and analytical expressions in \LaTeX \hs{0.1mm} is immediate. The treatment of Dirac structures and the calculation of (one-loop) decay widths and cross sections is remarkably simple given the functions available in \FMT. The numerical interface includes the beginning of a \t{\ts{Fortran}} main file (adapted to the process at stake), as well as a template of a makefile.

These examples suffice to make our point: \FMTS is very convenient for renormalizing a model and to study it at NLO. We now illustrate these features in the particular case of the C2HDM.

\section{The C2HDM}

We organize this section in three parts. In a first one, we briefly describe some particularities concerning the implementation of the C2HDM as a model for \FMT. Then, we explain how to calculate the counterterms. In the end, we show how to use \FMTS to study processes at NLO.

\subsection{The model}

The C2HDM as a model for \FMTS is available online in the webpage of \FMTS (in the Type II version, cf. section \ref{Chap-Reno:section:Yukawa}). 
Here, we do not describe in detail all the different components of the model (the manual---in appendix \ref{App-FM-Manual}---can be checked for details). Rather, we stress some aspects which may be less obvious.

First, the model for the C2HDM is an extension of the model for the SM, in the sense that it was built upon the model file for the SM. This implies, in particular, that it also contains the $\eta$'s from ref.~\cite{Romao:2012pq}, which are very useful to compare different conventions.
Now, the theory (the Lagrangian, with its particles and parameters) is originally written as a bare theory. And since we intend to renormalize it, we need to write it in a general basis. The bare quantities are identified with the sum of a renormalized quantity and its corresponding counterterm, as usual. The renormalized quantities obey eqs. \ref{Chap-Reno:eq:key} to \ref{Chap-Reno:eq:X-reno}, which are enforced through the restrictions file.
%
%
%
%
%
Counterterms such as $\delta \lambda_1$ will show up because the model is written in such a way that most of the dependent parameters are renormalized.%
\fn{Recall note \ref{Chap-Reno:eq:note-dep-CT} and eq. \ref{Chap-Reno:eq:unnec}.
}
Although we did not need to renormalize them (they are dependent, so that one could rewrite them in terms of independent parameters, and renormalize the latter), we decided to renormalize them for convenience. This is indeed very convenient: although it introduces several unnecessary counterterms, it renders the expressions considerably more compact.%
\fn{
Actually, if no unnecessary counterterms were used, the renormalization would probably be unfeasible: as it is, it takes about 50 minutes in our laptops to generate the total set of Feynman rules for the tree-level as well as for the counterterm interactions. It would take much longer if unnecessary counterterms were avoided.} 

\begin{figure}[htb]
\centering
\includegraphics[width=0.85\textwidth]{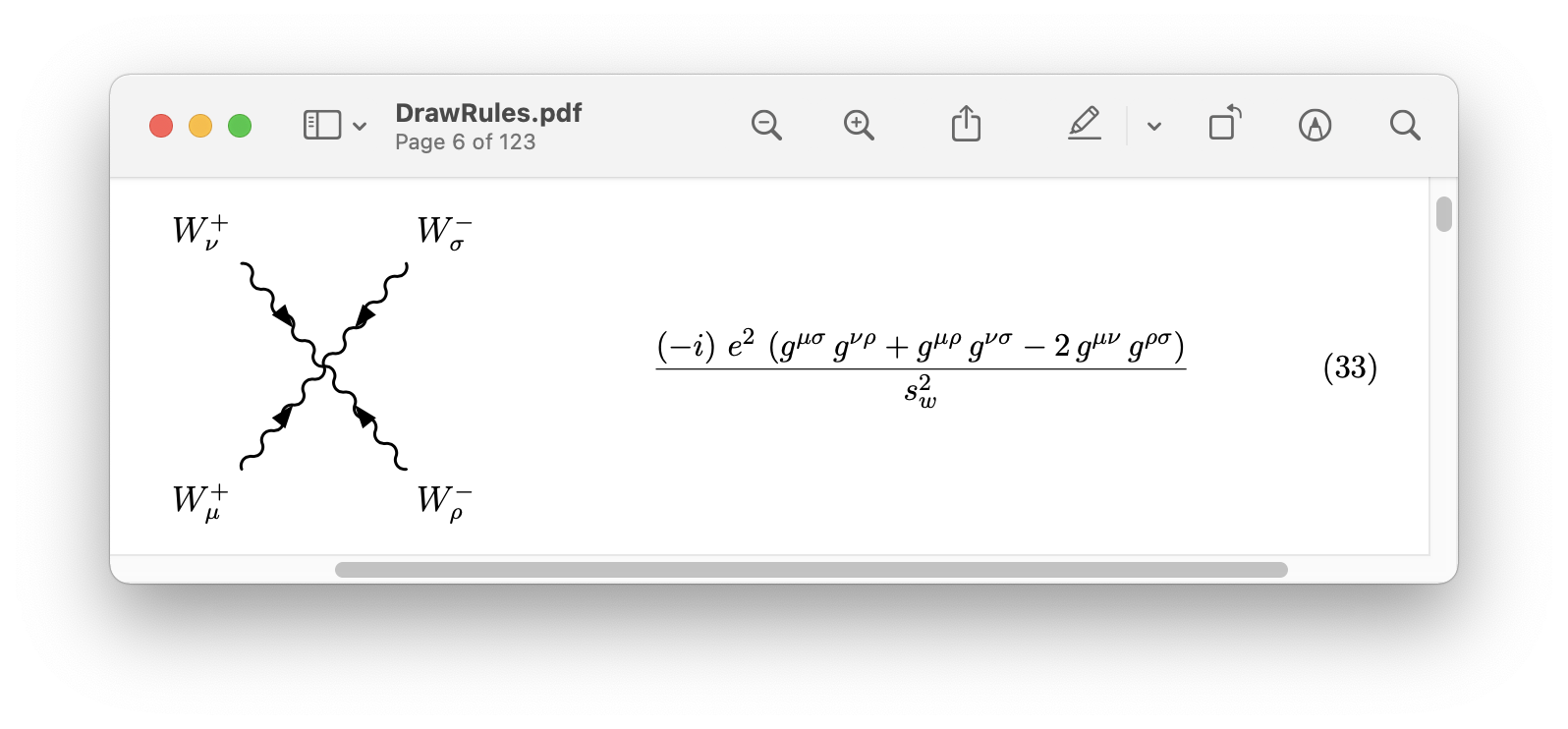}
\vs{-8mm}
\caption{Excerpt from the \t{DrawRules.pdf} file, containing the Feynman diagrams and rules for the tree-level interactions.}
\label{App-FM-C2HDM:fig:FM:rule-ext}
\vs{-8mm}
\end{figure}

\begin{figure}[htb]
\centering
\includegraphics[width=0.85\textwidth]{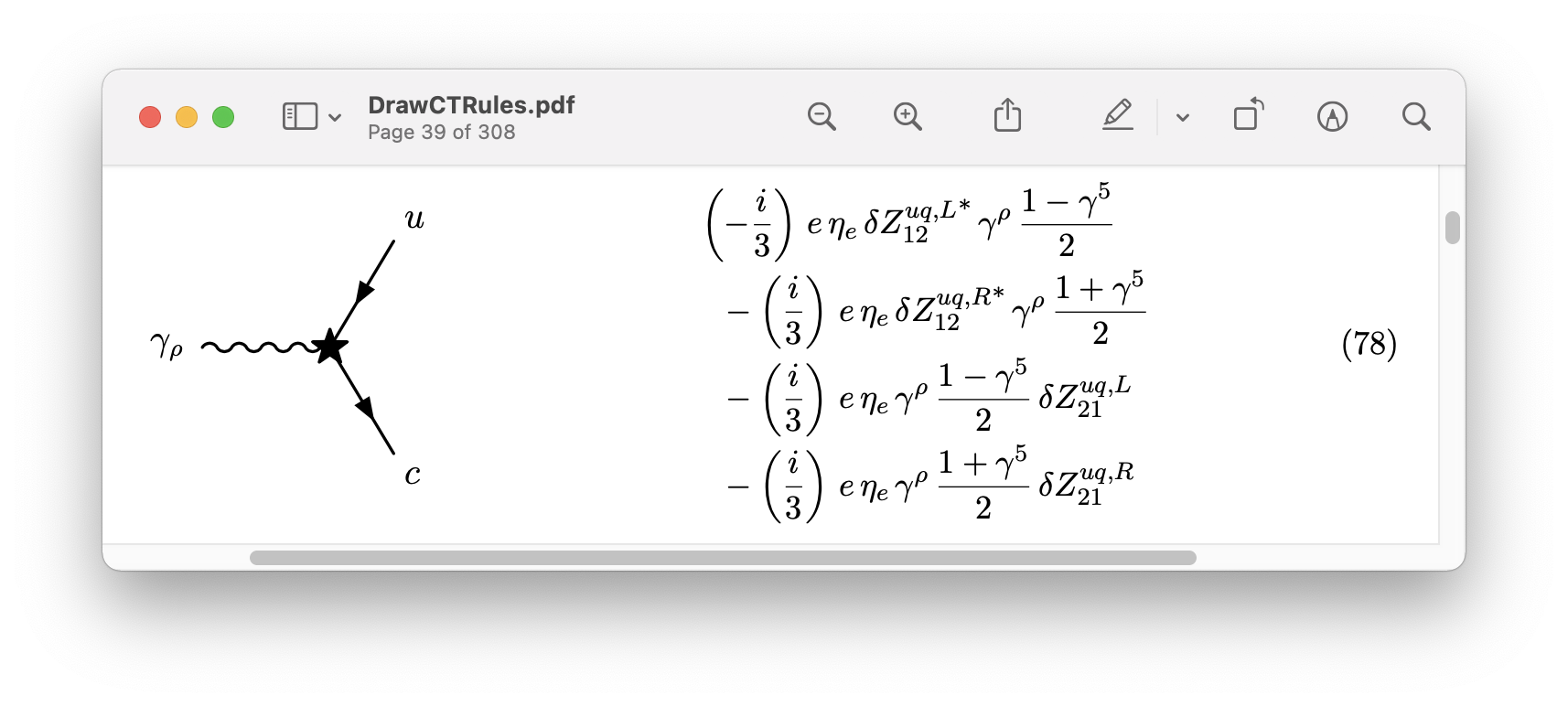}
\vs{-8mm}
\caption[Excerpt from the \t{DrawCTRules.pdf} file, containing the Feynman diagrams and rules for the counterterm interactions]{Excerpt from the \t{DrawCTRules.pdf} file, containing the Feynman diagrams and rules for the counterterm interactions ($\eta_e$ is one of the $\eta$'s from ref.~\cite{Romao:2012pq}).}
\label{App-FM-C2HDM:fig:FM:rule-CT-ext}
\vs{-2mm}
\end{figure}

\begin{figure}[htb]
\centering
\includegraphics[width=0.95\textwidth]{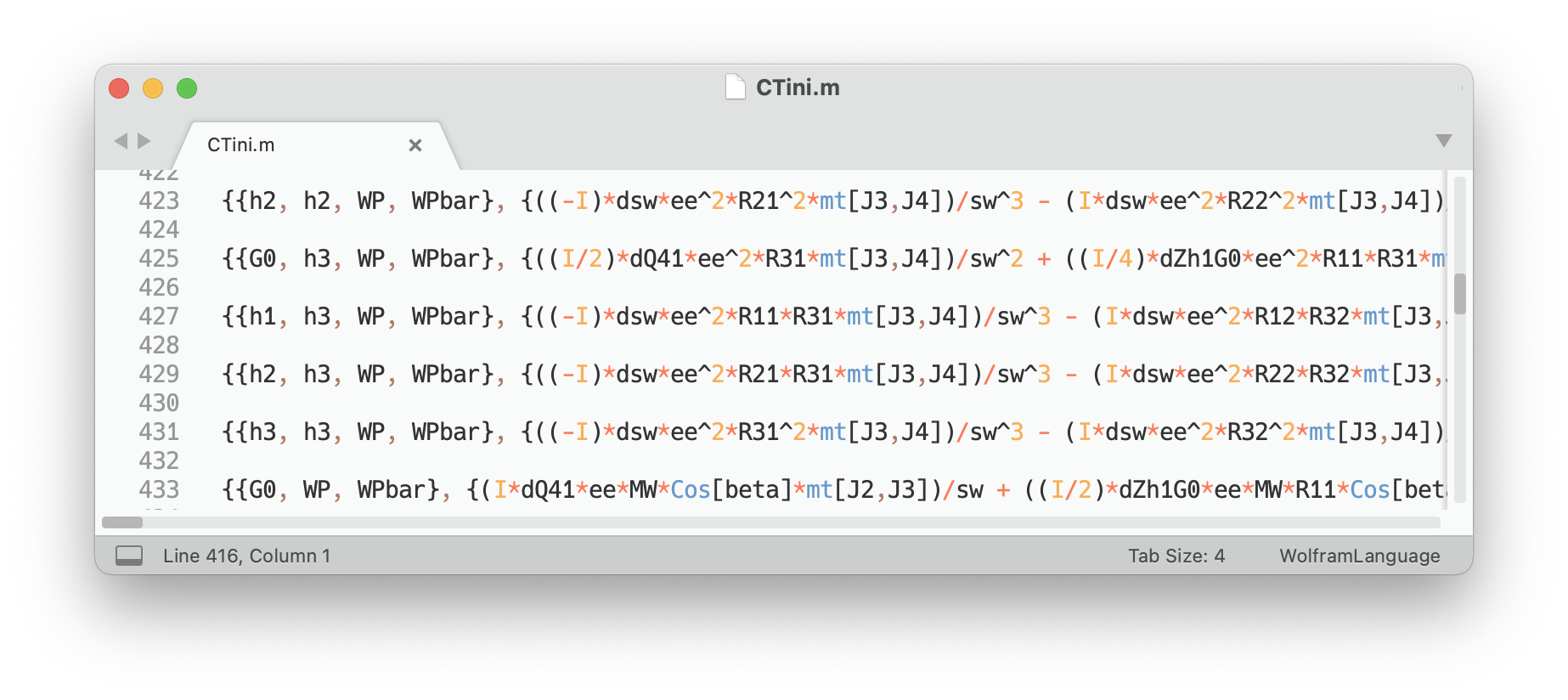}
\vs{-8mm}
\caption{Excerpt from the \t{CTini.m} file, containing the Feynman rules for the counterterms, written in a \FC-readable format.}
\label{App-FM-C2HDM:fig:FM:rule-CT-int}
\vs{-2mm}
\end{figure}

By running \FMTS with the variables \t{FRinterLogic} and \t{RenoLogic} (of the \t{Control.m} file) set to \t{True}, the Feynman diagrams and rules for the tree-level and counterterm interactions are automatically generated. In figs. \ref{App-FM-C2HDM:fig:FM:rule-ext} and \ref{App-FM-C2HDM:fig:FM:rule-CT-ext},
we present excerpts of the files containing the diagrams and rules written in \LaTeX. The corresponding files with the internal rules (i.e. the rules written in a \FC-readable format, to be used in the calculations) are also automatically generated. We show in fig. \ref{App-FM-C2HDM:fig:FM:rule-CT-int} 
an excerpt of one of them, \t{CTini.m}, which is the one concerned with the counterterms.
There, each non-empty line begins with the interaction at stake, containing after it the total set of counterterms contributing to it. Note the presence of dependent counterterms, such as $\delta Q_{41}$ and $\delta s_{\mathrm{w}}$ (represented by \t{dQ41} and \t{dsw}, respectively). Note
also the presence of renormalized parameters such as \t{R21}, corresponding to the entry $21$ of the matrix $R$.%
\fn{We used eqs. \ref{Chap-Reno:eq:Qsimple1} and \ref{Chap-Reno:eq:Qsimple2} to replace the elements of matrix $Q$ by those of matrix $R$. We could also have used eq. \ref{Chap-Maggie:matrixR} to rewrite the entries of the matrix $R$, but the Feynman rules would then become longer.}
%

\begin{figure}[htb]
\centering
\includegraphics[width=0.95\textwidth]{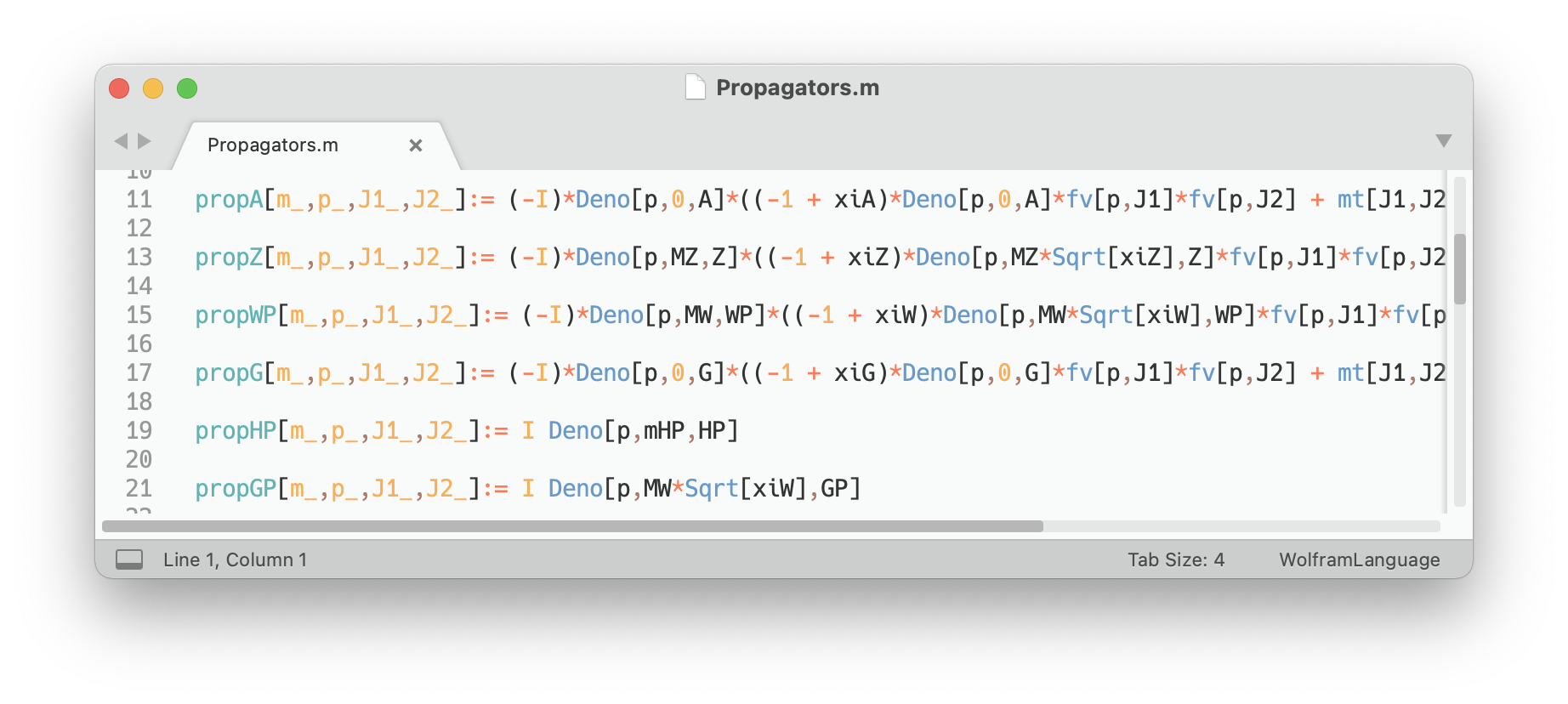}
\vs{-8mm}
\caption{Excerpt from the modified \t{Propagators.m} file of the C2HDM, containing the Feynman rules for the propagators written in a \FC-readable format.}
\label{App-FM-C2HDM:fig:FM:rule-int}
\end{figure}

Finally, we have set the option \t{M\$PrMassFL} of the \FMS model file (defined in the \t{Extra.fr} file) to \t{False}. This means that \FMS does not extract the mass of a certain propagator from the term in the Lagrangian which is bilinear in the corresponding field, but rather writes a simplified version of the propagator.%
\fn{Otherwise, the masses in the propagators of the scalar particles would be written in terms of parameters in the potential. For more details, cf. the manual.}
But it also means that the gauge boson propagators are written in the Feynman gauge. In order to study gauge-dependences, however, they must be written in an arbitrary $R_{\xi}$ gauge. As the totality of propagators that depend on the gauge exist also in the SM, we can copy them---assuming the user has already generated the Feynman rules for SM---from the file \t{Propagators.m} of the SM (which lies inside the folder \t{FeynmanRules}, contained in the \FMS output of the SM), and paste it in the equivalent file for the C2HDM. The final version should look like fig. \ref{App-FM-C2HDM:fig:FM:rule-int}

\subsection{Calculation of counterterms}

\begin{figure}[htb]
\centering
\includegraphics[width=0.5\textwidth]{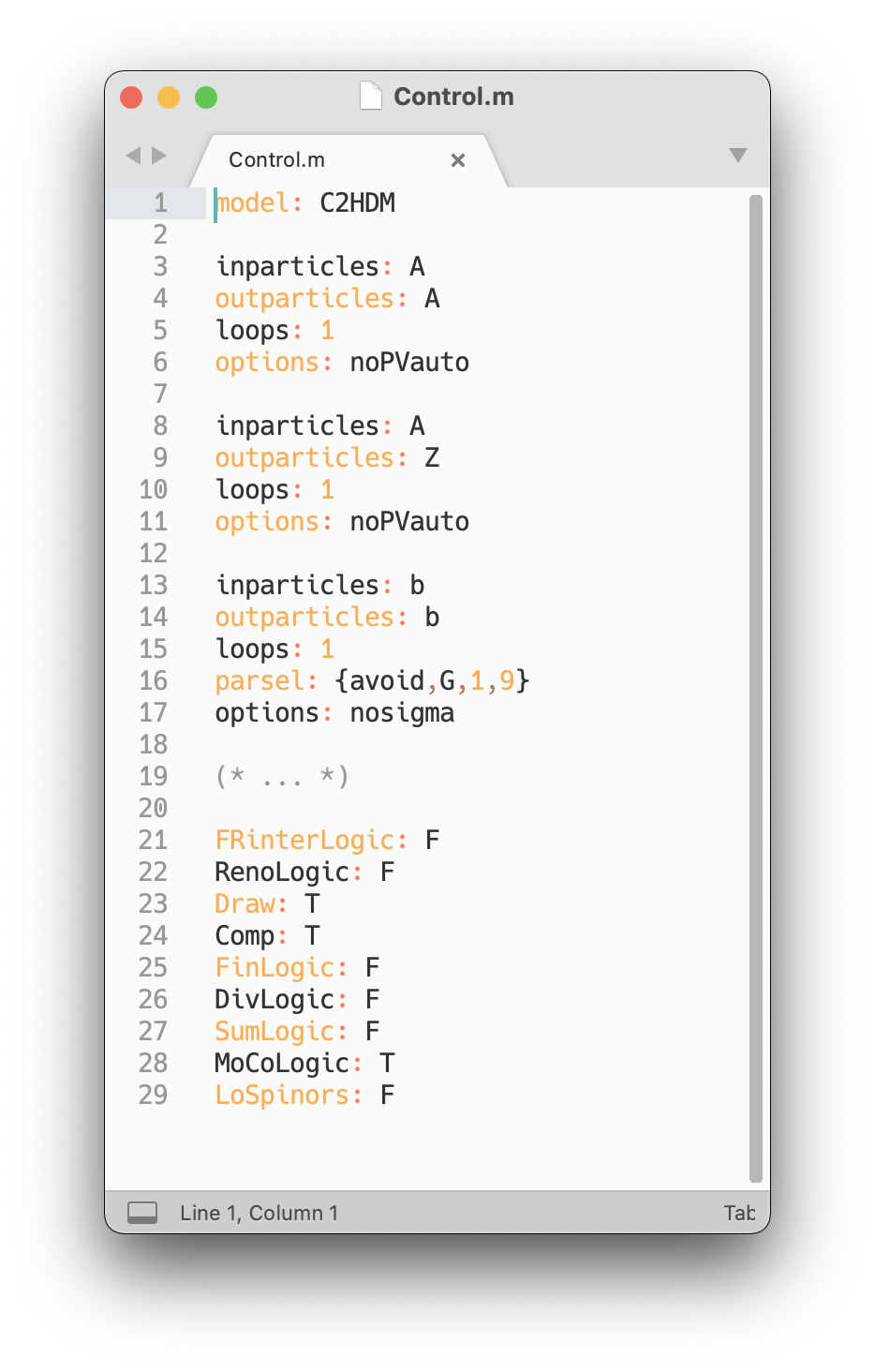}
\vs{-8mm}
\caption{Sketch of the \t{Control.m} file used to generate all the one-loop GFs required to renormalize the model. See text for details.}
\label{App-FM-C2HDM:fig:FM:Control}
\vs{-1mm}
\end{figure}

As we just saw, the running of \FMS already generated the totality of Feynman rules for the counterterm interactions. That is, for each process, the total set of counterterms contributing to it is already known. Now, those counterterms need to be calculated, which can be done in different ways, i.e. through different subtraction schemes. Here, we adopt the schemes defined in section \ref{Chap-Reno:sec:calculation-CTs}. We thus need to calculate several GFs: mostly 2-point functions, but also some 3-point functions (due to $\overline{\text{MS}}$ renormalization). This can be done in a single \FMS run, as illustrated in fig. \ref{App-FM-C2HDM:fig:FM:Control}.
Some clarifications are in order, concerning this figure.

First, we only show some of the GFs involved in section \ref{Chap-Reno:sec:calculation-CTs}; the remaining ones should be added in the place of the ellipses ``\t{(* ...\,\,\,*)}''. 
Second, in the calculation of $\Sigma^{\bar{b} b}$, we exclude diagrams with gluons via \t{\{avoid,G,1,9\}}; the same applies for all GFs with external fermions.
Third, note the options \t{noPVauto} and \t{nosigma}.
The former must be employed in some GFs to prevent meaningless divergences.%
\fn{The conditions of section \ref{Chap-Reno:sec:calculation-CTs} predict that some counterterms are calculated taking the limit of zero external momentum in some 2-point functions. Now, as explained in the manual of \FMT, loop integrals are performed with the function \t{OneLoopTID}, which employs the \ts{FeynCalc} function \t{TID}. By default, \t{OneLoopTID} uses \t{TID} with the option \t{PaVeAutoReduce -> True}, which simplifies some special cases of Passarino--Veltman functions. But it turns out that, in a general $R_{\xi}$ gauge, that option causes spurious divergences to appear in some amplitudes when the limit of zero external momentum is taken. We verified this phenomenon for the cases $\Sigma^{AA}$, $\Sigma^{AZ}$, $\Sigma^{G_0G_0}$ and $\Sigma^{G^{+}G^{+}}$. This can be solved by making \t{PaVeAutoReduce -> False}, which in turn can be obtained by selecting the option \t{noPVauto} in the \t{Control.m} file. Finally, note that nothing of this happens in the Feynman gauge: all the counterterms are well behaved in the limit of zero external momentum in that case.} 
The latter ensures two things: on the one hand, that no one-loop corrections to the external legs are included (such corrections are not needed for renormalizing purposes);\fn{Actually, this aspect is only relevant in the case of GFs with more than two external particles.} on the other hand, that reducible diagrams with broad tadpoles are included. We thus prove what we claimed in sections \ref{Chap-Selec:sec:FJTS} and  \ref{Chap-Reno:sec:FJTS-C2HDM}, namely, that the inclusion of broad tadpoles in \FMS is trivial.

Once all the GFs required in section \ref{Chap-Reno:sec:calculation-CTs} are calculated, the total set of counterterms can be directly determined in an analytical form through the results derived in that section.

\vs{-1mm}

\subsection{NLO processes}

As suggested above, the usefulness of \FMS does not end when the counterterms are determined. After that, multiple tasks are enabled by the tools provided in \FM. We now briefly illustrate three of them.

A first one consists in verifying the finiteness of a certain one-loop process. Once in the possession of the expressions for the counterterms, this task is almost immediate. One starts by selecting the process at stake in \t{Control.m} with the options \t{Comp} and \t{SumLogic} set to True. By running \FM, the total divergent part of the process is calculated, and a notebook for the process is automatically generated. One then opens the latter, having thus immediate access not only to the total divergent part (through the variable \t{resDtot}), but also to the total counterterms contributing to the process at stake (through the variable \t{CT}\textit{process}, where \textit{process} corresponds to the names of the incoming and the outgoing particles joined together). The expressions for the counterterms can be imported and replaced inside \t{CT}\textit{process}. The divergent part of the resulting expression can be obtained with the function \t{GetDiv} and compared with \t{resDtot}.%
\fn{As a matter of fact, the counterterm should be compared with $i$ times \t{resDtot}, since the amplitude calculated corresponds to $\mathcal{M}$, and not to $i \mathcal{M}$.}
If the process is finite, the difference must be zero.%
\fn{The comparison can be made analytically or numerically. While the former is more robust, it can take some time. The numerical method, on the other hand, requires a software to generate points in the parameter space of the model.}

\begin{figure}[htb]
\centering
\includegraphics[width=0.9\textwidth]{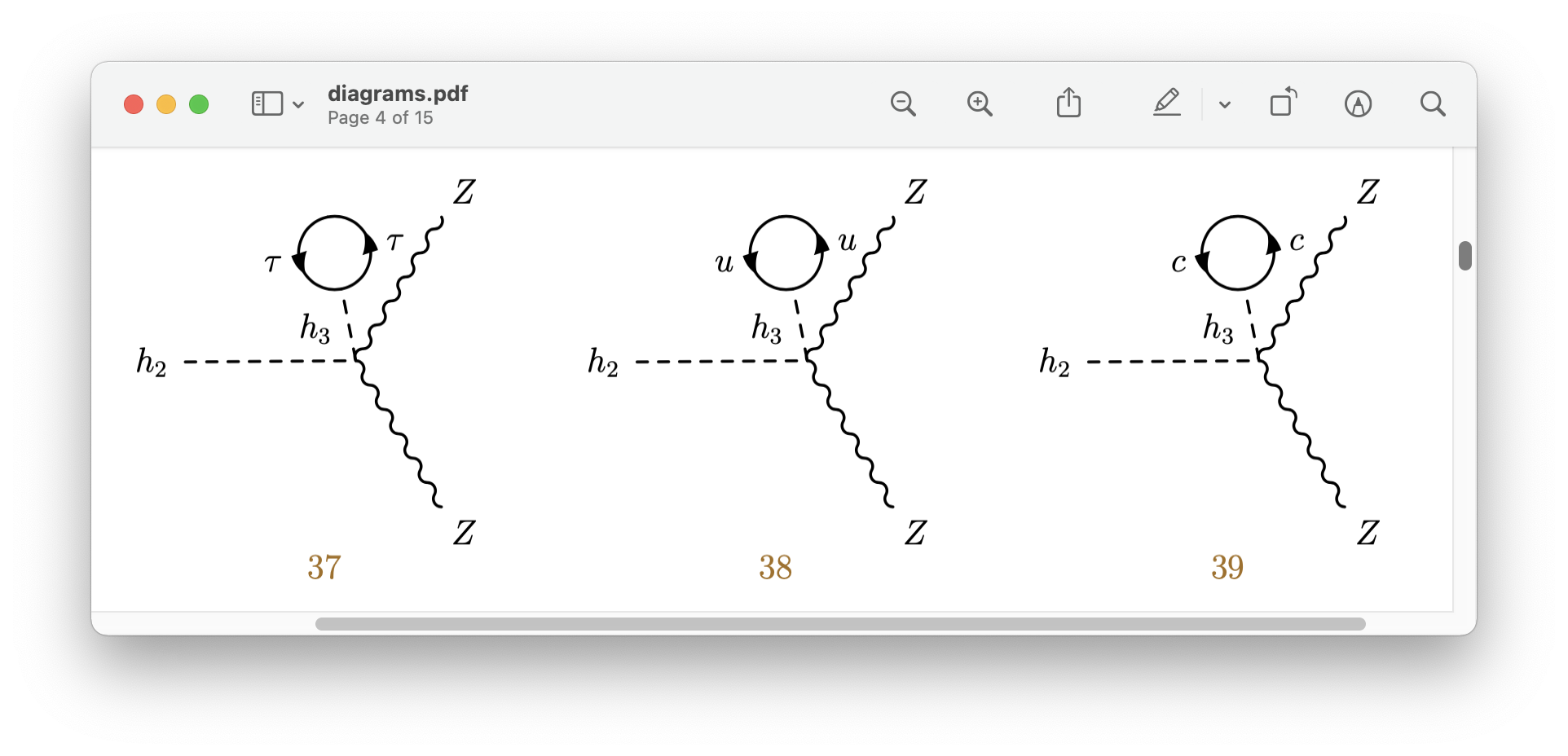}
\vs{-6mm}
\caption{Excerpt from \t{diagrams.pdf} for $h_2 \to ZZ$ at one-loop (in the FJTS).}
\label{App-FM-C2HDM:fig:Feynman-diagram}
\end{figure}

A second task that becomes quite simple when using \FMS is the calculation of a decay width at up-to-one-loop level. This requires the calculation of the tree-level amplitude, as well as of the renormalized one-loop one. The user can start with the former (choosing the process at stake at tree-level in \t{Control.m} with \t{Comp} set to True). Then, in a similar way, he or she calculates the non-renormalized one-loop amplitude (if the variable \t{Draw} in \t{Control.m} is set to True, \FMS automatically draws the Feynman diagrams, as in fig. \ref{App-FM-C2HDM:fig:Feynman-diagram}).
Here, as in the previous paragraph, the total counterterm should be added to obtain the (finite) renormalized one-loop amplitude. 
After summing up the two amplitudes (tree-level and renormalized one-loop), one will in general have a complicated expression. In this case, although the \FMS  function \t{DecayWidth} can be directly applied to it, it will probably take quite some time; it is thus convenient to rewrite the expression in terms of form factors. This can be done automatically using another \FMS function, \t{FacToDecay}, which yields a two-element list: the first one is precisely the expression rewritten in terms of form factors; the second one is a list of replacements, associating to each form factor the corresponding analytical expression (written using the kinematics of the process). The function \t{DecayWidth} can now be easily applied to the first element of the output of \t{FacToDecay}.%
\fn{The user must in general define which parameters are complex.
Note, however, that the form factors used in \t{FacToDecay} are by default set as complex.}
In this way, one obtains in a couple of steps a simple formula for the NLO decay width, as well as a list with the different form factors contained in that formula. 

A final example concerns the numerical interface of \FMT, enabled by yet another \FMS function, \t{FCtoFT}. This function automatically converts the expression it was applied to into \t{\ts{Fortran}} format and generates several files to run a \t{\ts{Fortran}} program (see the manual for details). \t{FCtoFT} is especially convenient in cases where the user has a simple expression written in terms of form factors and a list of replacement rules for each of them---precisely the case described in the previous paragraph. The reason is that \t{FCtoFT} can be used with two arguments, namely: a simple expression in terms of form factors, and the list of replacement rules for them. Finally, note that \t{FCtoFT} turns out to be virtually indispensable: the expressions which this function is applied to can involve millions of lines (see table \ref{App-FM-C2HDM:tab:lines}), which means that a correct conversion is far from being a trivial task.%
\fn{According to the process, the compilation takes between one and two hours in a computer with CPU i7.}

\begin{table}[h!]
\centering
\begin{tabular}
{
@{\hspace{-0.8mm}}
>{\centering}p{2.4cm}
>{\centering\arraybackslash}p{2cm}@{\hspace{3mm}}
}
\hlinewd{1.1pt}
Decay & Lines of code\\
\hline\\[-4mm]
$h_2 \to ZZ$ & 2074699 \\[1mm]
$h_2 \to h_1 Z$ & 4462302 \\[1mm]
$h_2 \to h_1 h_1$ & 6809331 \\[0.2mm]
\hlinewd{1.1pt}
\end{tabular}
\vspace{-1mm}
\caption{Number of lines in the  \t{\ts{Fortran}} codes associated to the decays studied in chapter \ref{Chap-Reno}.}
\label{App-FM-C2HDM:tab:lines}
\end{table}
\normalsize
%
%
%



%% file: Appendices/Theorem.tex
\chapter{A theorem for renormalization}
\label{App-Theorem}

\vs{-5mm}

It is often mentioned that some processes can be renormalized (i.e. rendered finite) simply by considering all the diagrams (1PI and reducible) that contribute to it, and by taking the on-shell (OS) limit. The $h \to Z \gamma$ one-loop decay in the SM is one of those cases; indeed,
\vspace{-3mm}
\begin{equation}
\begin{minipage}[h]{.25\textwidth}
\vspace{6mm}
\begin{fmffile}{App-Theorem-apphZA-1}
\begin{fmfgraph*}(66,66)
\fmfset{arrow_len}{3mm}
\fmfset{arrow_ang}{20}
\fmfleft{nJ1} 
\fmfright{nJ2,nJ4}
\fmflabel{$h$}{nJ1}
\fmflabel{$\gamma$}{nJ2}
\fmflabel{$Z$}{nJ4}
\fmf{dashes,tension=3}{nJ1,X}
\fmf{photon,tension=3}{nJ2,X}
\fmf{photon,tension=3}{nJ4,X}
\fmfv{decor.shape=circle,decor.filled=hatched,decor.size=9thick}{X}
\end{fmfgraph*}
\end{fmffile}
\vspace{6mm}
\end{minipage}
\hs{-12mm}
+
\hs{9mm}
\begin{minipage}[h]{.25\textwidth}
\vspace{6mm}
\begin{fmffile}{App-Theorem-apphZA-2}
\begin{fmfgraph*}(66,66)
\fmfset{arrow_len}{3mm}
\fmfset{arrow_ang}{20}
\fmfleft{nJ1} 
\fmfright{nJ2,nJ4}
\fmflabel{$h$}{nJ1}
\fmflabel{$\gamma$}{nJ2}
\fmflabel{$Z$}{nJ4}
\fmf{dashes,tension=5}{nJ1,X}
\fmf{photon,tension=2.6}{nJ4,X}
\fmf{photon,label=$Z$,label.dist=3,tension=5}{X,y}
\fmf{photon,tension=5}{y,nJ2}
\fmfv{decor.shape=circle,decor.filled=hatched,decor.size=9thick}{y}
\end{fmfgraph*}
\end{fmffile}
\vspace{6mm}
\end{minipage}
\hs{-12mm}
\OSeq
\hs{3mm}
\mathrm{finite},
\label{App-Theorem:eq:intro-hZA}
\vspace{-3mm}
\end{equation}
where the hatched circles represent one-loop non-renormalized GFs and $\OSeq$ means that the external particles are OS. As we now show, this is just a particular case of a general theorem.

\section{The theorem}

In any renormalizable theory:
\be
\mywbox{
\hs{-3mm}
\parbox{3.2cm}{\centering Process $P$ does not exist at order $n$ of perturbation theory}
\hs{-3mm}
}
\,
\Leftrightarrow
\,
\mywbox{
\hs{-1mm}
\parbox{10.2cm}{\centering Let $X$ and $Y$ be two sets of diagrams that contribute to $P$ at order $n+1$  of perturbation theory: $X$ the totality of irreducible diagrams, and $Y$ all the possible reducible diagrams that are obtained exclusively by inserting one-loop 2-point functions into external legs of irreducible diagrams that exist at lower orders; then the sum of $X$ and $Y$ must be finite if and only if the external particles in the one-loop 2-point functions in $Y$ are OS}
\hs{-1mm}
}
\label{App-Theorem:eq:apT:theorem}
\ee

\section{Proof}

\n
We start by showing that the l.h.s. of eq. \ref{App-Theorem:eq:apT:theorem} is a sufficient condition for the r.h.s. Then, we prove that it is also a necessary condition.

\subsection{Sufficiency}
\label{App-Theorem:sec:suf}

We start with $n=0$.
Suppose $\phi_i, \phi_j, \phi_k, \phi_l, \phi_m$ generic scalar fields with well-defined mass, such that $\phi_l$ and $\phi_m$ share the same quantum numbers.\fn{Scalar fields are considered for convenience. The generalization to other particles is trivial.}
Suppose also that the Lagrangian does not include a term for $\phi_i \phi_j \phi_k \phi_m$,
\be
\left\langle\dfrac{\partial^4 \mathcal{L}}{\partial \phi_i \partial \phi_j \partial \phi_k \partial \phi_m}\right\rangle = 0,
\ee
but does include a term for $\phi_i \phi_j \phi_k \phi_l$,
\be
\mathcal{L} \ni \lambda \, \phi_i \phi_j \phi_k \phi_l,
\label{App-Theorem:eq:apT:myL}
\ee
with $\lambda$ a generic parameter.%
\fn{As an example, one can consider the model described in appendix \ref{App-Pilaftsis}, where there is no coupling $A \, h \, h \, h $, although there is the coupling $h \, h \, h \, h$.}
The Feynman rule corresponding to eq. \ref{App-Theorem:eq:apT:myL} is:
\be
\begin{minipage}{0.35\textwidth}
\vs{5mm}
\begin{fmffile}{App-Theorem-appT1}
\begin{fmfgraph*}(42,42) 
\fmfset{arrow_len}{3mm} 
\fmfset{arrow_ang}{20} 
\fmfleft{nJ1,nJ3} 
\fmfright{nJ2,nJ4}
\fmflabel{$\phi_i$}{nJ1}
\fmflabel{$\phi_j$}{nJ3}
\fmflabel{$\phi_k$}{nJ2}
\fmflabel{$\phi_l$}{nJ4}
\fmf{dashes,tension=3}{nJ1,X}
\fmf{dashes,tension=3}{nJ3,X}
\fmf{dashes,tension=3}{nJ2,X}
\fmf{dashes,tension=3}{nJ4,X}
\end{fmfgraph*} 
\end{fmffile}
\vs{5mm}
\end{minipage}
\hspace{-35mm}
=
i \, n_{\text{o}} \, \lambda,
\ee
where $n_{\text{o}}$ is the symmetry factor of the vertex $\phi_i \phi_j \phi_k \phi_l$ (as there may be equal fields). Although there is no term in $\mathcal{L}$ for $\phi_i \phi_j \phi_k \phi_m$---so that there is no such interaction at tree-level---, the theory may be such that the interaction is non-null at one-loop. In that case, and since the theory is renormalizable, there will be a counterterm for it, which by definition obeys:
\begin{equation}
\left.
\begin{bmatrix}
\hs{7mm}
\begin{minipage}[h]{.25\textwidth}
\vspace{6mm}
\begin{fmffile}{App-Theorem-appT2}
\begin{fmfgraph*}(42,42)
\fmfset{arrow_len}{3mm}
\fmfset{arrow_ang}{20}
\fmfleft{nJ1,nJ3} 
\fmfright{nJ2,nJ4}
\fmflabel{$\phi_i$}{nJ1}
\fmflabel{$\phi_j$}{nJ3}
\fmflabel{$\phi_k$}{nJ2}
\fmflabel{$\phi_m$}{nJ4}
\fmf{dashes,tension=3}{nJ1,X}
\fmf{dashes,tension=3}{nJ3,X}
\fmf{dashes,tension=3}{nJ2,X}
\fmf{dashes,tension=3}{nJ4,X}
\fmfv{decor.shape=circle,decor.filled=hatched,decor.size=9thick}{X}
\end{fmfgraph*}
\end{fmffile}
\vspace{6mm}
\end{minipage}
\hs{-15mm}
+
\hs{9mm}
\begin{minipage}[h]{.25\textwidth}
\vspace{6mm}
\begin{fmffile}{App-Theorem-appT3}
\begin{fmfgraph*}(42,42)
\fmfset{arrow_len}{3mm}
\fmfset{arrow_ang}{20}
\fmfleft{nJ1,nJ3} 
\fmfright{nJ2,nJ4}
\fmflabel{$\phi_i$}{nJ1}
\fmflabel{$\phi_j$}{nJ3}
\fmflabel{$\phi_k$}{nJ2}
\fmflabel{$\phi_m$}{nJ4}
\fmf{dashes,tension=3}{nJ1,X}
\fmf{dashes,tension=3}{nJ3,X}
\fmf{dashes,tension=3}{nJ2,X}
\fmf{dashes,tension=3}{nJ4,X}
\fmfv{decor.shape=pentagram,decor.filled=full,decor.size=6thick}{X}
\end{fmfgraph*}
\end{fmffile}
\vspace{6mm}
\end{minipage}
\hs{-18mm}
\hphantom{.}
\end{bmatrix}
\right|_{\varepsilon}
=0,
\label{App-Theorem:eq:apT:base}
\end{equation}
%
where $|_{\varepsilon}$ represents the terms proportional to $1/\varepsilon$, as usual.
The reason why this counterterm in general exists is that, since $\phi_l$ and $\phi_m$ have the same quantum numbers, there is in general a counterterm for their mixing:%
\fn{This holds even if there is no bilinear term that mixes $\phi_l$ and $\phi_m$ at tree-level. Note also that there is no summation on repeated indices.}
\be
\phi_{l(0)} = ... + \dfrac{1}{2} \delta Z_{lm} \, \phi_m + ... \, , 
\hs{10mm}
\phi_{m(0)} = ... + \dfrac{1}{2} \delta Z_{ml} \, \phi_l + ... \, , 
\label{App-Theorem:eq:apT:Reno}
\ee
where the ellipses represent other possible terms. This means that the term in eq. \ref{App-Theorem:eq:apT:myL} generates the following term in the Lagrangian of the counterterms, $\mathcal{L}_{\mathrm{CT}}$:
\be
\mathcal{L}_{\mathrm{CT}} \ni n_{\text{s}} \, \lambda \, \dfrac{1}{2} \delta Z_{lm} \, \phi_i \phi_j \phi_k \phi_m,
\label{App-Theorem:eq:apT:LCT}
\ee
where $n_{\text{s}}$ is the total number of particles in eq. \ref{App-Theorem:eq:apT:myL} equal to $\phi_l$. The Feynman rule for eq. \ref{App-Theorem:eq:apT:LCT} is thus:
\be
\begin{minipage}{0.35\textwidth}
\vs{5mm}
\begin{fmffile}{App-Theorem-appT4}
\begin{fmfgraph*}(42,42) 
\fmfset{arrow_len}{3mm} 
\fmfset{arrow_ang}{20} 
\fmfleft{nJ1,nJ3} 
\fmfright{nJ2,nJ4}
\fmflabel{$\phi_i$}{nJ1}
\fmflabel{$\phi_j$}{nJ3}
\fmflabel{$\phi_k$}{nJ2}
\fmflabel{$\phi_m$}{nJ4}
\fmf{dashes,tension=3}{nJ1,X}
\fmf{dashes,tension=3}{nJ3,X}
\fmf{dashes,tension=3}{nJ2,X}
\fmf{dashes,tension=3}{nJ4,X}
\fmfv{decor.shape=pentagram,decor.filled=full,decor.size=6thick}{X}
\end{fmfgraph*} 
\end{fmffile}
\vs{5mm}
\end{minipage}
\hspace{-35mm}
\ni
\, \, 
i \, n_{\text{s}} \, n_{\text{c}} \, \lambda  \, \dfrac{1}{2} \delta Z_{lm},
\label{App-Theorem:eq:apT:myCT}
\ee
where $n_{\text{c}}$ is the symmetry factor of the vertex $\phi_i \phi_j \phi_k \phi_m$.

On the other hand, we can consider all the reducible diagrams that are obtained by inserting a one-loop mixing 2-point function $i\Sigma_{lm}(p)$ in the vertex $\phi_i \phi_j \phi_k \phi_l$, thus contributing to $\phi_i \phi_j \phi_k \phi_m$. We dub this set of diagrams $Y_1$.
As $\phi_i, \phi_j, \phi_k$ may be equal to $\phi_m$, but in such a way that each one has its own momentum $p_a$, there can be several configurations in  $Y_1$.%
\fn{The different configurations all have the 2-point function $i\Sigma_{lm}$, which means that, in all of them, the corrected leg has an inner $\phi_l$ propagator and an external $\phi_m$ field. What distinguishes the different configurations is the momentum flowing through the corrected leg.}
The sum of all possible configurations is:
\be
\begin{minipage}{0.35\textwidth}
\vs{5mm}
\begin{fmffile}{App-Theorem-appT6}
\begin{fmfgraph*}(80,80) 
\fmfset{arrow_len}{3mm} 
\fmfset{arrow_ang}{20} 
\fmfleft{nJ1,nJ3} 
\fmfright{nJ2,nJ4}
\fmflabel{$\phi_i$}{x1}
\fmflabel{$\phi_j$}{x3}
\fmflabel{$\phi_k$}{x2}
\fmflabel{$\phi_m$}{nJ4}
\fmf{phantom,tension=4}{nJ1,x1}
\fmf{dashes,tension=2}{x1,X}
\fmf{phantom,tension=4}{nJ3,x3}
\fmf{dashes,tension=2}{x3,X}
\fmf{phantom,tension=4}{nJ2,x2}
\fmf{dashes,tension=2}{x2,X}
\fmf{dashes,tension=3,label=\rotatebox{50}{$\Large\xxrightarrow[p_a]{}$},label.side=left,label.dist=-6,label.angle=30}{nJ4,x4}
\fmf{dashes,tension=2.5,label=$\phi_l$,label.side=left,label.dist=0.5}{x4,X}
\fmfv{decor.shape=circle,decor.filled=hatched,decor.size=9thick}{x4}
\end{fmfgraph*} 
\end{fmffile}
\vs{5mm}
\end{minipage}
\hspace{-25mm}
+
\hs{3mm}
. . .
\hs{3mm}
=
\,
\sum_{a=1}^{n_{Y_1}} i \, n_{\text{o}} \, \lambda \, \dfrac{i}{p_a^2 - m_l^2} \, i \Sigma_{lm}(p_a^2),
\label{App-Theorem:eq:apT:red}
\ee
where the ellipses represents all the other (besides the one depicted) possible configurations in $Y_1$, and $n_{Y_1}$ is the total number of configurations.
Using combinatorial arguments, one concludes that:
\be
n_{Y_1} = \dfrac{n_{\text{c}} \, n_{\text{s}}}{{n_{\text{o}}}}.
\label{App-Theorem:eq:apT:frac}
\ee
Now, the renormalized one-loop mixing 2-point function between $\phi_l$ and $\phi_m$, $i\hat{\Sigma}_{lm}(p_a^2)$, is by definition finite:
\be
i\hat{\Sigma}_{lm}(p_a^2) \Big|_{\varepsilon} = 0
\quad
\Leftrightarrow
\quad
\left. \Big[i \Sigma_{lm}(p_a^2) + i \Sigma^{\text{CT}}_{lm}(p_a^2) \Big] \right|_{\varepsilon} = 0,
\label{App-Theorem:eq:apT:sum-pre}
\ee
where $i\Sigma^{\text{CT}}_{lm}(p_a^2)$ represents the counterterm, which is composed exclusively of counterterms for the mixing of fields.%
\fn{This is because there is no bilinear term at tree-level mixing $\phi_l$ and $\phi_m$, as we assumed that both fields are physical fields, in the sense that they have well defined mass.}
More specifically,%
\fn{
Note the appearence of the structure $p^2-m^2$, which is a consequence of the structure of the kinetic terms.}
\be
i\Sigma^{\text{CT}}_{lm}(p_a^2) = 
\hs{8mm}
\begin{minipage}{0.35\textwidth}
\begin{fmffile}{App-Theorem-appT7} 
\begin{fmfgraph*}(60,50) 
\fmfset{arrow_len}{3mm} 
\fmfset{arrow_ang}{20}
\fmfleft{nJ1}
\fmflabel{$\phi_l$}{nJ1} 
\fmfright{nJ2} 
\fmflabel{$\phi_m$}{nJ2} 
\fmf{dashes,tension=2}{nJ1,x}
\fmf{phantom,label=$\overrightarrow{p_a}$,label.side=right,tension=0}{nJ1,x}
\fmf{dashes,tension=2}{x,nJ2} 
\fmfv{decor.shape=pentagram,decor.filled=full,decor.size=6thick}{x}
\end{fmfgraph*} 
\end{fmffile}
\end{minipage}
\hs{-25mm}
=
\, \, 
i \Bigg[\dfrac{1}{2} \delta Z_{lm} \Big(p_a^2-m_l^2\Big) + \dfrac{1}{2} \delta Z_{ml} \Big( p_a^2-m_m^2\Big) \Bigg].
\label{App-Theorem:eq:mixCTpre}
\ee
Inserting this expression inside eq. \ref{App-Theorem:eq:apT:sum-pre} and taking $\phi_m$ OS ($p_a^2 = m_m^2$), we find:
\be
i \Sigma_{lm}(m_m^2) \Big|_{\varepsilon}
=
-  i \Big(m_m^2-m_l^2\Big) \dfrac{1}{2} \delta Z_{lm} \Big|_{\varepsilon},
\label{App-Theorem:eq:apT:sum}
\ee
so that, going back to eq. \ref{App-Theorem:eq:apT:red}, taking only the divergent parts and setting the external particle of the corrected leg (in each configuration) OS, we get:
\be
\sum_{a=1}^{n_{Y_1}} i \, n_{\text{o}} \, \lambda \, \dfrac{i}{m_m^2 - m_l^2} \, i \Sigma_{lm}(m_m) \Big|_{\varepsilon}
=
i \, n_{Y_1} \, n_{\text{o}} \, \lambda \,
\dfrac{1}{2} \delta Z_{lm} \Big|_{\varepsilon}
=
i \, n_{\text{s}} \, n_{\text{c}} \, \lambda  \,
\dfrac{1}{2} \delta Z_{lm} \Big|_{\varepsilon}
,
\ee
where in the last equality we used eq. \ref{App-Theorem:eq:apT:frac}. This result equals precisely the divergent part of the r.h.s. of eq. \ref{App-Theorem:eq:apT:myCT}. Such equality is true if and only if the external particle of the corrected leg in each configuration is OS, for otherwise eq. \ref{App-Theorem:eq:apT:sum} will not have the structure required to cancel the inner propagators in eq. \ref{App-Theorem:eq:apT:red}.
Now, on the one hand, since the only way for the theory to generate a counterterm for a process that does not exist at tree-level is through the counterterm for the mixing of fields (as in eq. \ref{App-Theorem:eq:apT:Reno}), then the total counterterm for $\phi_i \phi_j \phi_k \phi_m$ at one-loop equals the sum of all the terms in the Lagrangian of counterterms obtained through mixing field renormalization.
On the other hand, although the counterterm in eq. \ref{App-Theorem:eq:apT:myCT} and the diagrams in eq. \ref{App-Theorem:eq:apT:red} were obtained from the term in the Lagrangian in eq. \ref{App-Theorem:eq:apT:myL}, the fact that this term is completely general implies that the reasoning followed applies to any other term.
%
%
In other words, for every other term $t$ in the Lagrangian that also generates a counterterm for $\phi_i \phi_j \phi_k \phi_m$, the set of diagrams $Y_t$---defined as the set of diagrams that contribute to $\phi_i \phi_j \phi_k \phi_m$ and were obtained by inserting a one-loop mixing 2-point function in the vertex involved in $t$---will be equivalent to such counterterm if and only if the external particle of the corrected leg in each configuration is OS.
It then follows that the total counterterm for $\phi_i \phi_j \phi_k \phi_m$ at one-loop equals the sum of all $Y_t$ if and only if the external particle of the corrected leg in each configuration is OS. Finally, given eq. \ref{App-Theorem:eq:apT:base} and the fact that the sum of all $Y_t$ corresponds to the set of diagrams identified with $Y$ in the statement of the theorem, the sufficiency condition is proven for $n=0$.

Let us now consider $n=1$. Suppose the process $P$ does not exist at one-loop---which in turn necessarily means that it does not exist at tree-level either. 
The first point to stress is that all the processes $P'$ that may generate a counterterm for $P$ at two-loop (i.e. a counterterm to absorb the divergences of the two-loop irreducible diagrams of $P$) do not exist \textit{themselves} at tree-level. 
Indeed, suppose that a process $P'$ generates a counterterm for $P$ at two-loop; as $P$ exists neither at tree-level nor at one-loop, such generation is exclusively due mixing field renormalization; but if $P'$ existed at tree-level, it would generate a counterterm for $P$ at one-loop precisely through mixing field renormalization, which contradicts the assumption according to which $P$ does not exist at one-loop. We thus conclude that $P'$ does not exist at tree-level, and this conclusion holds for any process $P'$ generating a counterterm for $P$ at two-loop.

Let us now consider one of those processes, which we identify as $P_1'$. If $P_1'$ does not exist at tree-level, we can apply the theorem for $n=0$ to conclude that the sum of one-loop irreducible diagrams and one-loop $Y$ diagrams with OS external particles in the 2-point functions is finite. We dub this sum $Z_1'$. Now, $Z_1'$ generates counterterms for $P$ at two-loop exclusively through mixing field renormalization (again, as in eq. \ref{App-Theorem:eq:apT:Reno}). So, for each term included in $Z_1'$, the reasoning followed for $n=0$ applies. That is to say, the divergent parts of the counterterm (for $P$) generated by each term of $Z_1'$ are precisely equal to the divergent parts of the sum of reducible diagrams obtained by insertion of one-loop 2-point functions in the term of $Z_1'$ at stake if and only if the external fields in the 2-point functions are OS.%
\fn{The demonstration for this claim is identical to the one followed for $n=0$.}
Since the same is valid for all the other terms $Z_t'$ associated to different processes $P'$, the sufficiency condition is proven for $n=1$.

Having seen this, one can generalize the same reasoning to $n>1$.

\vs{3mm}

\subsection{Necessity}

It is simple to prove that the l.h.s. of eq. \ref{App-Theorem:eq:apT:theorem} is a necessary condition to the r.h.s.

We start with $n=0$.
Suppose once again that the Lagrangian includes a term for $\phi_i \phi_j \phi_k \phi_l$, as described in eq. \ref{App-Theorem:eq:apT:myL}.
The renormalization of the theory requires the renormalization of the parameters; in particular, the bare parameter $\lambda_{(0)}$ is identified with:
\be
\lambda_{(0)} = \lambda + \delta \lambda,
\ee
where $\lambda$ is the renormalized parameter and $\delta \lambda$ the counterterm, as usual. This means that, while the total counterterm for $\phi_i \phi_j \phi_k \phi_m$ in section \ref{App-Theorem:sec:suf} above was composed exclusively of (mixing) field counterterms, the total counterterm for $\phi_i \phi_j \phi_k \phi_l$ receives a contribution from:
\be
\mathcal{L}_{\mathrm{CT}} \ni \delta \lambda \, \phi_i \phi_j \phi_k \phi_l.
\ee
That is, there is a parameter counterterm ($\delta \lambda$) contributing to the total counterterm for $\phi_i \phi_j \phi_k \phi_l$. As it is obvious, this means that such parameter counterterm is required to absorb the divergences of the irreducible diagrams of $\phi_i \phi_j \phi_k \phi_l$ at one-loop order. However, one cannot in general find reducible diagrams that mimic the action of $\delta \lambda$.
As a consequence, the set of diagrams  identified in the statement of the theorem as $Y$ will in general not cancel the divergences of the irreducible diagrams (independently of whether or not some of the states are OS), so that the necessity condition is proven for $n=0$.

The proof for $n>0$ follows a similar reasoning. Suppose $n=1$; if a process $P$ exists at one-loop, there will be parameter counterterms contributing to the total counterterm of $P$ at 2-loop, which once again cannot be mimicked by reducible diagrams. The same reasoning can be generalized to $n>1$.

\section{Discussion}

Several points concerning the theorem just presented are worth mentioning.
First of all, the generality of the theorem should be stressed. The theorem is not bound to a particular theory; rather, it is valid for \textit{any} renormalizable quantum field theory. Besides the requirement of renormalizability, indeed, the only other assumption made was that the fields involved in the process at stake had well-defined mass, in the sense of being eigenstates of mass (which is always true for particles in physical processes). At last, note that, although we derived the proof for four scalar fields, the theorem also holds for processes with three particles, for other types of particles, as well as for processes including different types of particles.

Second, the only reducible diagrams mentioned in the theorem are those involving 2-point functions in the external legs. In particular, reducible diagrams with tadpole insertions or with 2-point functions in internal legs do not belong to the set identified as $Y$.

This leads us to another relevant conclusion, concerning the special role of processes with 3 particles. In this case, the set of diagrams $Y$ constitute the totality of reducible diagrams contributing to the process.
Hence, a corollary of the original theorem is:
\be
\mywbox{
\hs{-1mm}
\parbox{3.2cm}{\centering Process $P$ with 3 particles does not exist at order $n$  of perturbation theory
\hs{-1mm}
}
}
\,
\Leftrightarrow
\,
\mywbox{
\hs{-2mm}
\parbox{9.4cm}{\centering At order $n+1$, the sum of the irreducible and reducible diagrams must be finite if and only if the external particles in the 2-point functions in the reducible diagrams are OS}
\hs{-2mm}
}
\label{App-Theorem:eq:apT:theorem3}
\ee

Yet another aspect that is worth stressing is that the only particles that need to be OS for the sum of $X$ and $Y$ diagrams to be finite are the external particles of the 2-point functions involved in the $Y$ diagrams. In particular, the remaining particles in the $Y$ diagrams, as well as all the particles in the irreducible diagrams, need not be OS. For example, the Higgs boson and the external $Z$ boson in eq. \ref{App-Theorem:eq:intro-hZA} do not need to be OS: in order for the sum of diagrams in that equation to be finite, it is enough that the photon is OS.

Related to this topic is one last important point. Despite appearances, eqs. \ref{App-Theorem:eq:apT:theorem} and \ref{App-Theorem:eq:apT:theorem3} are independent of the substraction scheme used: although they require the external particles in the 2-point functions in $Y$ to be OS (in order for the sum of $X$ and $Y$ to be finite), their validity is independent of whether one decides to renormalize fields using the on-shell subtraction (OSS) scheme, or any other subtraction scheme. 
What is particular about OSS, however, is that the renormalized mixing 2-point functions are set to zero when the external particles are OS (recall appendix \ref{App-OSS}). This means that, in OSS, eq. \ref{App-Theorem:eq:apT:sum} is valid not only for the divergent parts, but also for the finite parts. In that particular subtraction scheme, therefore, there is an equality not only between the divergent parts of the counterterm for $P$ (on the one hand) and those of the sum of diagrams of $Y$ with external particles with OS 2-point mixing functions (on the other), but also between their finite parts.
The consequence is thus clear: by summing the set of $X$ and $Y$ diagrams, and by setting the one-loop mixing 2-point functions in the external legs of $Y$ OS, one obtains not only a finite result, but precisely \textit{the same} result that one would obtain should one renormalize the theory in OSS.
For example, eq. \ref{App-Theorem:eq:intro-hZA} is not only finite, but actually equal to the result for $h \to Z \gamma$ that one would obtain if one would calculate the process using counterterms and renormalizing them through OSS.

\section{The theorem in the literature}

In the literature, one can find references that derive results corresponding to particular cases of the theorem \cite{Fontes:2014xva, Soares:1988fj, Grimus:2002ux}.%
%
\fn{Although ref. \cite{Soares:1988fj} states that ``no explanation can be provided within the context of OSS'' for a particular case of the corollary \ref{App-Theorem:eq:apT:theorem3}, we have just provided such an explanation.}
It is also very common to find references where the theorem is applied to processes that do not exist at tree-level, in such a way that, instead of using the counterterm procedure and fixing the counterterms through OSS, the authors calculate reducible diagrams with one-loop 2-point mixing functions; for example, this is the usual procedure in processes with flavour-changing neutral currents \cite{Eilam:1990zc,HoKim:1999bs,Bejar:2000ub,He:2009rz,Abbas:2015cua,Bardhan:2016txk}.

One can also find articles in the literature that, despite considering reducible diagrams with one-loop corrections to external legs, are actually performing something quite different from what is at stake in the theorem proposed above.
The procedure we are alluding to can be found in several papers by Hollik, among other authors \cite{Bohm:1986rj,Hollik:1988ii,Beenakker:1991ca,Grzadkowski:1991nj,Hollik:1993cg,Hollik:1993cf}.
As we now clarify, Hollik's approach is not directed to render processes finite by considering reducible diagrams instead of counterterms; rather, it aims at the calculation of (renormalized) residues of propagators. To achieve this goal, he considers renormalized corrections to external legs  (i.e. the non-renormalized corrections and the corresponding counterterms), whereas in our case we only considered
non-renormalized corrections.

To understand Hollik's procedure, we start by noting that he performs a minimalist renormalization, in the sense that the minimal number of field renormalization constants is introduced. In fact, Hollik renormalizes the fields in the symmetric form of the theory, so that he considers only one field counterterm for each multiplet of the original theory \cite{Hollik:1988ii}.
Now, one of the advantages of renormazing fields in the mass basis (as we did in chapter \ref{Chap-Reno}) is that one has enough field counterterms to fix all the (renormalized) residues of the propagators of the physical fields to one. Precisely because he has fewer field counterterms, Hollik cannot do this for all physical fields. In other words, he can only set the residue of some of the fields to unity. For example, he usually chooses field counterterms so that charged leptons have residue one; but this means that he cannot set the residue of neutrinos to one \cite{Hollik:1988ii}. 

When the renormalized residue of a field is different from one, the renormalized LSZ factors (or wave function renormalization factors) must be considered in the LSZ formula. Actually, as discussed in appendix \ref{App-LSZ}, those factors are in general present in the LSZ formula for the renormalized theory. Only, they are equal to the renormalized residues (eq. \ref{App-LSZ:eq:Sm-reno-no-mix}), which means that they can be ignored when the latter are equal to one. Whenever that is not the case, however, they must be calculated in order to obtain the correct $S$-matrix element.

To see how this is related to renormalized corrections to external legs,
suppose a process where the renormalized residue $\hat{R}$ of a certain field is not set to one (in the following, we work with $\hat{R}^{-1}$ instead of $\hat{R}$, for convenience).
Suppose also that the mass of that field is renormalized through OSS (implying that the pole mass $m_{\mathrm{P}}$ is equal to the renormalized mass $m_{\mathrm{R}}$).
We can split $\hat{R}^{-1}$ into tree-level and renormalized one-loop components as $\hat{R}^{-1} = \hat{R}^{-1}_{\mathrm{tree}} + \hat{R}^{-1}_{\mathrm{loop}}$, with $\hat{R}_{\mathrm{tree}}^{-1} = 1$ by construction.%
\fn{This follows from the definition of the renormalized residue in eq. \ref{App-LSZ:eq:reno-resi}, in the case where the mass is renormalized in OSS.
Concerning $\hat{R}^{-1}_{\mathrm{loop}}$, and as we suggested, Hollik must calculate the renormalized LSZ factor (i.e. the renormalized residue) of external neutrino lines in processes with external neutrinos. This he does explicitly in eq. 68 of ref. \cite{Hollik:1993cg}, where (with his conventions) he finds $\hat{R}_{\mathrm{loop}} = - \left[\Sigma_{\mathrm{L}}^{\nu}(0) + \delta Z_{\mathrm{L}}\right]$, where $\Sigma_{\mathrm{L}}^{\nu}(0)$ represents a part of a non-renormalized GF and $\delta Z_{\mathrm{L}}$ a counterterm.}
So, $\hat{R}^{-1} = 1 + \hat{R}^{-1}_{\mathrm{loop}}$, which means that $\hat{R}^{-1}_{\mathrm{loop}}$ represents the deviation of $\hat{R}^{-1}$ to one.
Since $\hat{R}^{-1}_{\mathrm{loop}}$ is assumed to be a small quantity (it is a higher order correction), one can expand $\hat{R}^{1/2}$ in eq. \ref{App-LSZ:eq:Sm-reno-no-mix} in series, thus obtaining $1 - \frac{1}{2} \hat{R}^{-1}_{\mathrm{loop}}$ to first order. Then, according to the expression for the $S$-matrix element in eq. \ref{App-LSZ:eq:Sm-reno-no-mix}, this expansion multiplies $\hat{\mathcal{G}}_{\mathrm{ampu.}}$. If we separate the latter into tree-level component ($\hat{\mathcal{G}}_{\mathrm{ampu.}}^{\mathrm{tree}}$) and renormalized one-loop component ($\hat{\mathcal{G}}_{\mathrm{ampu.}}^{\mathrm{loop}}$), we conclude that the $S$-matrix element is such that:
\be
\begin{split}
&\left\langle -p_{s+1}, ... , -p_{n}  | S | p_{1}, ... ,  p_{s}\right\rangle
\, \propto \, 
\Big(1 - \frac{1}{2} \hat{R}^{-1}_{\mathrm{loop}} \Big) \, \hat{\mathcal{G}}_{\mathrm{ampu.}}(p_1, ... , p_n)
\Big|_{p_i^2 = m_{i\mathrm{P}}^2}
\\
& \hs{10mm}
=
\bigg[
\hat{\mathcal{G}}_{\mathrm{ampu.}}^{\mathrm{tree}}(p_1, ... , p_n)
+
\hat{\mathcal{G}}_{\mathrm{ampu.}}^{\mathrm{loop}}(p_1, ... , p_n)
-
\frac{1}{2} \hat{R}^{-1}_{\mathrm{loop}} \,  \hat{\mathcal{G}}_{\mathrm{ampu.}}^{\mathrm{tree}}(p_1, ... , p_n)
+
(...)
\bigg]
\Big|_{p_i^2 = m_{i\mathrm{P}}^2}
,
\end{split}
\label{App-Theorem:eq:S-matrix}
\ee
where the last term represents higher order contributions.
Now, considering the definition of the renormalized residue (eq. \ref{App-LSZ:eq:reno-resi}), noting that $\hat{G}(p)^{-1} = -i \hat{\Gamma}(p)$ and expanding $\hat{\Gamma}(p)$ as in eqs. \ref{Chap-Reno:eq:GammaRenScalar}, one finds:%
\fn{Note that both the numerator and the denominator vanish in the limit at stake (the vanishing of the numerator is due to the fact that we decided to renormalize the mass in OSS, recall eqs. \ref{App-OSS:eq:OSaux1} and \ref{App-OSS:eq:OSSfirst}).}
\be
\hat{R}^{-1}_{\mathrm{loop}}
=
\dfrac{\hat{\Sigma}(p^2)}{p^2 - m_{\mathrm{R}}^2} \Big|_{p^2 = m_{\mathrm{R}}^2}.
\ee
In this case, the third term of the last line of eq. \ref{App-Theorem:eq:S-matrix} (the term which results from the circumstance that we did not set $\hat{R}=1$) can be written as:
\be
- \frac{1}{2} \hat{R}^{-1}_{\mathrm{loop}} \,  \hat{\mathcal{G}}_{\mathrm{ampu.}}^{\mathrm{tree}}(p_1, ... , p_n)
\Big|_{p_i^2 = m_{i\mathrm{P}}^2}
=
\dfrac{1}{2}
\bigg[
i \hat{\Sigma}(p^2)
\,
\dfrac{i}{p^2 - m_{\mathrm{R}}^2}
\,
\hat{\mathcal{G}}_{\mathrm{ampu.}}^{\mathrm{tree}}(p_1, ... , p_n)
\bigg]
\Big|_{p_i^2 = m_{i\mathrm{P}}^2}.
\ee
It turns out that the quantities inside the squared brackets can be obtained by a Feynman diagram. For example, in the case of a 3-particle process, 
\vs{-5mm}
\be
\bigg[
i \hat{\Sigma}(p^2)
\,
\dfrac{i}{p^2 - m_{\mathrm{R}}^2}
\,
\hat{\mathcal{G}}_{\mathrm{ampu.}}^{\mathrm{tree}}(p_1, ... , p_n)
\bigg]
\Big|_{p_i^2 = m_{i\mathrm{P}}^2}
=
\hs{3mm}
\begin{minipage}[h]{.25\textwidth}
\vspace{6mm}
\begin{fmffile}{App-Theorem-hocus-pocus}
\begin{fmfgraph*}(90,90)
\fmfset{arrow_len}{3mm}
\fmfset{arrow_ang}{20}
\fmfleft{nJ1} 
\fmfright{nJ4,nJ2}
\fmf{dashes,tension=5}{nJ1,X}
\fmf{dashes,tension=2.6}{nJ4,X}
\fmf{dashes,label=\rotatebox{45}{$\Large\xxrightarrow[p]{}$},label.side=left,label.dist=-4,label.angle=30,tension=5}{X,y}
\fmf{dashes,label=\rotatebox{45}{$\Large\xxrightarrow[p]{}$},label.side=left,label.dist=-4,label.angle=30,tension=5}{y,nJ2}
\fmfv{decor.shape=circle,decor.filled=30,decor.size=11thick,label=$\hs{-2.5mm}\vs{-2.7mm}\text{R}$}{y}
\end{fmfgraph*}
\end{fmffile}
\vspace{6mm}
\end{minipage},
\label{App-Theorem:eq:diag-resi}
\vs{-5mm}
\ee
where the gray circle with R represents a renormalized correction to the external leg, corresponding to $i \hat{\Sigma}(p^2)$.
We insist: it is the \textit{renormalized} one-loop 2-point function, and not the non-renormalized one (which we represent with a hatched circle, as in eq. \ref{App-Theorem:eq:intro-hZA}).

In summary, the fact that the renormalized residue $\hat{R}$ is different from one leads to a new contribution to the $S$-matrix element. Such contribution is given by the third term of the last line of eq. \ref{App-Theorem:eq:S-matrix}, which in turn is given by diagram of eq. \ref{App-Theorem:eq:diag-resi} multiplied by a factor of $1/2$.%
\fn{More precisely, the diagram is only that of eq. \ref{App-Theorem:eq:diag-resi} in the case of 3-particle processes involving scalar particles only; other cases are a trivial generalization of this case.
Besides, this whole argument---according to which the deviations of $\hat{R}$ from one can be given by a Feynman diagram---requires the mass to be renormalized in the OSS scheme, which is what is usually done (even when the field counterterms are not sufficient to set all residues to one), in particular by Hollik; for example, in the sequence of three Feynman diagrams immediatly after eq. 68 of ref. \cite{Hollik:1993cg} (where, in his notation, the shaded circles represent renormalized GFs), the last two diagrams are equivalent to eq. \ref{App-Theorem:eq:diag-resi}. Recall that, as we noted, one has to multiply such diagrams by a factor of $1/2$ in order to obtain the correct contribution to the $S$-matrix elements; Hollik mentions this explicitly e.g. in the caption of fig. 3 of ref. \cite{Hollik:1993cf}. Cf. also eq. 4.12 of ref. \cite{Hollik:1988ii}.}

We described this topic in detail to clarify that, although it also involves reducible diagrams with one-loop corrections to external legs, it configures a situation  very different from that of the theorem we introduced in this appendix. Indeed, the one-loop corrections to external legs implied in the theorem above are always non-renormalized 2-point functions, which involve field mixing, and are only considered in the particular case where the process at stake does not exist at lower orders.
By contrast, the one-loop corrections to external legs involved in the calculation of non-trivial contributions of $\hat{R}$ to the $S$-matrix elements are renormalized 2-point functions, which are only considered when the renormalized residue of the field at stake is different from one (and they are considered precisely as a way to calculate such residue).